%% file: thesis.tex
\documentclass[a4paper,	10pt,OPENANY,oneside]{book}
\usepackage{subcaption}

\usepackage[total={16cm,25.4cm},top=2.5	cm, left=3cm]{geometry} 
\usepackage{bookman}                                       
\usepackage{amsmath,amssymb,amsfonts,latexsym}             
\usepackage[pdftex]{graphicx}                              
\usepackage{graphicx}
\usepackage{anyfontsize}                                   
\usepackage{cancel}                                        
\usepackage{multicol}                                      
\usepackage{caption}                               	        
\usepackage{cases}                                                                                         
\usepackage{tabulary}                                      
\usepackage{times} 	                                        
\usepackage[utf8]{inputenc}                                
\usepackage[T1]{fontenc}
\usepackage{lmodern}
\usepackage[english]{babel}
\usepackage{acronym}
\usepackage{float}
\usepackage{color}                                        
\usepackage{mathtools}

\usepackage[pdfpagemode={UseOutlines},bookmarks=false,bookmarksopen=false,
   bookmarksopenlevel=0,bookmarksnumbered=false,hypertexnames=false,
   colorlinks,linkcolor={blue	},citecolor={green},urlcolor={red},
   pdfstartview={FitV},unicode,breaklinks=false,colorlinks=false]{hyperref}   
\usepackage{braket}
\usepackage{fancyhdr}                                     
\usepackage{textcomp}                                     
\usepackage{setspace}                                     
\usepackage{slashed}
\usepackage{cleveref}
\crefformat{section}{\S#2#1#3} 
\crefformat{subsection}{\S#2#1#3}
\crefformat{subsubsection}{\S#2#1#3}
\usepackage[force]{feynmp-auto}

\usepackage{ytableau}
\usepackage{cite}
\hypersetup{
    colorlinks=true, 
    pdfborder = {0 0 0.5 [3 3]},
    anchorcolor=black,
    citecolor=blue,
    linktoc=all,    
    linktocpage=true,
    linkcolor=blue,
    urlcolor=blue
}
\DeclareUnicodeCharacter{00A0}{~}

\def \12x12 {$\frac{1}{2}\otimes \frac{1}{2}$ }
\def \0x1 {$0\otimes1$ }

\newcommand{\bs}{\boldsymbol}

\newcommand{\mc}{\mathcal}

\newcommand{\fb}[1]{{\color{black} { #1}}}

\hypersetup{urlcolor=blue, colorlinks=true}               
\graphicspath{{Figures/}}                                 
\normalfont

\newcommand{\Rlm}[4]{{}_{#1}R_{#2 #3 #4}}
\newcommand{\Rin}[4]{{}_{#1}R^{\rm in}_{#2 #3 #4}}
\newcommand{\sSlm}[3]{{}_{#1} S_{#2 #3}}
\newcommand{\sYlm}[3]{{}_{#1} Y_{#2 #3}}

\newcommand{\rs}{r_*}
\newcommand{\Binc}[3]{B^{\rm inc}_{#1 #2 #3}}
\newcommand{\Bref}[3]{B^{\rm ref}_{#1 #2 #3}}
\newcommand{\Btrans}[3]{B^{\rm trans}_{#1 #2 #3}}

\newcommand{\diff}[2]  {\frac{d #1}{d #2}}

\newcommand{\pdiff}[2]  {\frac{\partial #1}{\partial #2}}
\newcommand{\spdiff}[2] {\frac{\partial^2 #1}{\partial #2^2}}

\newcommand\portada{
	\begin{titlepage}
		\begin{center}

            \vspace{4cm}
            {\Huge{} }
			{\Huge{Scattering Amplitude Techniques in Classical Gauge Theories and Gravity} } \\ 
			\vspace{2cm}
 			{\it\Large{by}\par}
 			\Large{Yilber Fabian Bautista Chivata}
		\vspace*{2.5cm}

        \Large
        A Dissertation Submitted to \\
        The  Faculty of Graduate Studies\\ 
        In Partial Fulfilment of the Requirements\\
       for the Degree of \\
        Doctor of Philosophy  

        \vspace*{2.5cm}
        Graduate Program in Physics
        \\
        York University\\
        Toronto, Ontario \\
        \vspace*{2.0cm}
        July 2022

        \vspace*{2.0cm}

        \copyright\ Yilber Fabian Bautista Chivata, 2022 \\

		\end{center}
\end{titlepage}
}

\begin{document}

\addtocontents{toc}{\protect\setstretch{0.6}}
\frontmatter    
\portada
\pagestyle{plain}
\setcounter{page}{2}

\include{examiners}
\include{declaration}
\include{statement_of_contributions}

\include{abstract}

\include{dedication}
\include{acknowledgments}


\tableofcontents
\listoftables
\listoffigures
\include{acro_list}
\include{introduction}

\mainmatter                                              
\pagestyle{fancy}
\fancyfoot[C]{}

\include{./Chapters/chapter_preliminaries}
\include{./Chapters/chapter_electromagnetic_radiation}
\include{Chapters/soft_constraints}

\include{Chapters/multiple_double_copy}

\include{Chapters/bounded}

\include{./Chapters/chapter_double_copy}	
\include{./Chapters/chapter_higher_spins}
\include{./Chapters/conclusions}
\appendix	
\include{./appendices/soft_constraints}
\include{./appendices/appendix_multipoles}
\include{./appendices/spin_2}
\include{./appendices/app_bounded}
\include{./appendices/app_dc}

\include{./appendices/teukolsky}
\backmatter  
\bibliographystyle{JHEP}
\bibliography{Bibliography}  
\end{document}

%% file: examiners.tex
\cleardoublepage 

\begin{center}\textbf{Examining Committee Membership}\end{center}
  \noindent
The following served on the Examining Committee for this thesis. The decision of the Examining Committee is by majority vote.
  \bigskip
  
  \noindent
\begin{tabbing}
Internal-External Member: \=  \kill 
External Examiner: \>  Donal O'Connell \\ \> Professor, School of Physics and Astronomy\\ \>University of Edinburgh \\ 
\end{tabbing} 
  \bigskip
  
  \noindent
\begin{tabbing}
Internal-External Member: \=  \kill 
Supervisor(s): \> Sean Tulin \\
\> Associate Professor, Department of Physics and Astronomy \\
\> York  University\\
\end{tabbing}
  \bigskip
  
  \noindent
  \begin{tabbing}
Internal-External Member: \=  \kill 
Committee Members: \> Matthew C Johnson \\
\>Associate Professor, Department of Physics and Astronomy \\
\> York  University\\
\>Associate Faculty, Perimeter Institute for Theoretical Physics
\\
\\
\>Tom Kirchner  \\
\>Professor of Physics, Department of Physics and Astronomy \\
\> York  University\\
\end{tabbing}
  \bigskip
  
  \noindent
\begin{tabbing}
Internal-External Member: \=  \kill 
Internal-External Member: \> EJ Janse van Rensburg\\
\> Professor of Mathematics \& Statistics, Department of Mathematics and Statistics \\
\> York University\\
\end{tabbing}

%% file: declaration.tex
\cleardoublepage

  \noindent
I hereby declare that I am the sole author of this thesis. This is a true copy of the thesis, including any required final revisions, as accepted by my examiners.

  \bigskip
  
  \noindent
I understand that my thesis may be made electronically available to the public.

%% file: statement_of_contributions.tex
\begin{center}\textbf{Statement of Contributions}\end{center}

Yilber Fabian Bautista was the sole author of  \cref{ch:electromagnetism}, \cref{conclusions} as well as the Preliminaries \cref{ch_preliminaries} and the general Introduction , which were not written for publication.

This thesis consists in part of five manuscripts written for publication, as well as  current work in progress by the author and his collaborators.

\begin{enumerate}
    \item \textit{Y.F. Bautista and A. Guevara: From Scattering Amplitudes to Classical Physics: Universality, Double Copy and Soft Theorems.} \href{https://arxiv.org/abs/1903.12419}{ [hep-th:1903.12419]}
    
    \item \textit{Y.F. Bautista and A. Guevara: On the Double Copy for Spinning Matter. JHEP 11 (2021) 184} \href{https://arxiv.org/abs/1908.11349}{ [hep-th:1908.11349]}

    \item \textit{Y.F. Bautista, A. Guevara, C. Kavanagh and J. Vines: From Scattering in Black Hole Backgrounds to Higher-Spin Amplitudes: Part I.} \href{https://arxiv.org/abs/2107.10179}{ [hep-th:2107.10179]}
    
    \item \textit{Y.F. Bautista and N. Siemonsen: Post-Newtonian Waveforms from Spinning Scattering Amplitudes. JHEP 01 (2022) 006} \href{https://arxiv.org/abs/2110.12537}{ [hep-th:2110.12537]}
    
    \item \textit{Y.F. Bautista and A. Laddha: Soft Constraints on KMOC Formalism.} \href{https://arxiv.org/abs/2111.11642}{ [hep-th:2111.11642]}
    
    \item \textit{Y.F. Bautista, A. Guevara, C. Kavanagh and J. Vines: From Scattering in Black Hole Backgrounds to Higher-Spin Amplitudes: Part II.}  In preparation. 
    
    \item \textit{Y.F. Bautista, A. Laddha and Y. Zhang: Subleading-Soft Constraints on KMOC Formalism.} In preparation. 
    
\end{enumerate}
Each of the chapters of this thesis takes elements of the different manuscripts as follows:

\textbf{Research presented in  \cref{ch:electromagnetism}} takes elements of works 1,3 and 5.

\textbf{Research presented in  \cref{ch:soft_constraints}} is based in  works  5 and 7.

\textbf{Research presented in  \cref{sec:spin_in_qed}} takes elements of works 1,3,4.

\textbf{Research presented in  \cref{ch:bounded}} is based in  work  4.

\textbf{Research presented in  \cref{ch:double_copy}} is based in  work  2.

\textbf{Research presented in  \cref{ch:GW_scattering}} takes elements of works 3 and 6.

%% file: abstract.tex
\chapter{\centering Abstract}

In this thesis we present a study of the computation of   classical observables in gauge theories and gravity directly from scattering amplitudes. In particular, we discuss the  direct application of modern amplitude techniques  in  the one, and two-body problems for both, scattering and bounded scenarios,  and in both, classical electrodynamics and gravity, with particular emphasis on spin effects in general, and in four spacetime dimensions. Among these observables we have the conservative linear impulse and  the radiated waveform in the two-body problem, and the differential cross section for the scattering of waves off classical spinning compact objects.
Implication of classical soft theorems in the computation of classical radiation is also discussed. Furthermore, formal aspects of the double copy for massive spinning matter, and its application in a classical two-body context are  considered. Finally, the relation between the minimal coupling gravitational Compton amplitude and the scattering of gravitational waves off the Kerr black hole is presented.

%% file: dedication.tex
\clearpage
\begin{center}
    \thispagestyle{empty}
    \vspace*{\fill}
  \textit{\hspace{1.6cm} To Cindy:\\ \hspace{6.5cm} For her unconditional support\\  \hspace{7.5cm}  through the years we  spent together. }
    \vspace*{\fill}
\end{center}
\clearpage

%% file: acknowledgments.tex
\chapter{\centering Acknowledgments}

I would like to express my sincere gratitude to my collaborator and friend, Alfredo Guevara for teaching me modern amplitudes methods, and for his guidance and advise during several stages of my Ph.D. program; without him, this work would have not been possible.  
In the same way, I would like to thank my advisor Sean Tulin for teaching me Dark Matter Physics and Computing Programming, but specially for trusting   me and giving me the opportunity to explore several research avenues on my own. I would further like to thank the members of the evaluation committee for valuable inputs on the thesis which helped to  improve the final version of the text. 

I would also like to thank my amplitudes/gravity collaborators, Chris Kavanagh, Alok Laddha, Nils Siemonsen, Justin Vines and Yong Zhang for enlightening discussions, they were very helpful  for my formation as a scientist. In the same way, I would like to express my appreciation to my Dark Physics collaborators,  Brian Colquhoun, Andrew Robertson, Laura Sagunski and  Adam Smith-Orlik; collaborating with theme has dotted me with interdisciplinary skills that will be of great value in my professional life.

In the same way, I want to thank the  unconditionally support from my family. My mother Ilda Chivata, my father Abelardo Bautista, my siblings Nilson, Mauricio, Camilo, Julian and Estefany.

To my Waterloo friends Shanming Ruan, Jinxiang Hu, Ramiro Cayuso, Sara Bogojević, Aiden Suter, Alexandre Homrich, Francisco Borges, Bruno Jimenez,  Benjamin Perdomo, Laura Blanco and Tato, I have nothing but deep appreciation  for all the shared moments; they have made my time in the city more enjoyable.
I would also like to thank  Debbie Guenther, and the staff from the Bistro Black Hole at the Perimeter Institute. They have made of PI the best place to  be a student. Likewise, I want to thank Cristalina Carmela Del Biondo for all of her help and document preparation during my Ph.D. program at York University.

I am also grateful to my friends back home, Camilo Gonzalez, Walter Fonseca, Fabio Peña, John  Mateus, Andrea Daza, Diego Hernandez, Andrea Campos, William Aldana, Juan Calos Martinez, Camilo Ochoa and Natalia Escobar, Lilith Escalante and Tatiana Salcedo, for always receiving me with warm open arms; despite the distance, they all have become a second family to me. 

I would  specially like to express my most sincere gratitude to Cindy Castiblanco, for  her unconditional support during the  last 8 years of my life, for all the adorable moments we shared.

%% file: acro_list.tex
\section*{List of Acronyms}
\begin{acronym}[MPC] 
\acro{BHPT}{Black Hole Perturbation Theory}
\acro{GWs}{Gravitational Waves}
\acro{GW}{Gravitational-Wave}
\acro{BBH}{Binary Black Hole}
\acro{BNS}{Binary Neutron Star}
\acro{GR}{General Relativity}
\acro{BH-NS}{Black Hole-Neutron Star}
\acro{NR}{Numerical Relativity}
\acro{GSF}{Gravitational Self-Force}
\acro{PN}{post-Newtonian}
\acro{EOB}{Effective One Body}
\acro{BH}{Black Hole}
\acro{PM}{Post-Minkowskian}
\acro{KMOC}{Kosower, Maybee and O'connell}
\acro{EFT}{Effective Field Theory}
\acro{EoM}{Equations of Motion}
\acro{PL}{post-Lorentzian}
\acro{YM}{Yang-Mills}
\acro{SYM}{Super Yang-Mills}
\acro{KLT}{Kawai-Lewellen-Tye}
\acro{BCJ}{Bern-Carrasco-Johansson}
\acro{QFT}{Quantum Field Theory}
\acro{SQED}{Scalar Quantum Electrodynamics}
\acro{QED}{Quantum Electrodynamics}
\acro{QCD}{Quantum Chromodynamics}
\acro{NLO}{Next to Leading Order}
\acro{LO}{Leading Order}
\acro{irreps.}{irreducible representations}
\acro{SSC}{Spin Supplementary Condition}
\acro{KK}{Kaluza-Klein}
\acro{DoF}{Degrees of Freedom}
\acro{CoM}{Center of Mass}
\acro{LHS}{Left Hand Side}
\acro{RHS}{Right Hand Side}
\acro{HCL}{Holomorphic Classical Limit}
\acro{GME}{Gravitational Memory Effect}
\acro{PW}{Plane Wave}
\end{acronym}

%% file: introduction.tex
\chapter{Introduction}\label{sec:introduction}

The more than 100 years old prediction made by Einstein for the existence of  \ac{GWs}  \cite{1918SPAW154E}\footnote{Although see \href{https://www.americanscientist.org/article/the-secret-history-of-gravitational-waves}{The Secret History of Gravitational Waves},  for an interesting  narrative on the development of the theory of Gravitational waves. }, and the recent direct confirmation by the LIGO and VIRGO collaborations \cite{LIGOScientific:2016aoc}, started the so called era  of gravitational wave astronomy. This  new window into the universe not only allows us to test \ac{GR} to an unprecedented degree of accuracy, but also permits  observational  investigation  of theories of modified gravity \cite{Mastrogiovanni_2020}, while adding important new  elements to the  multi-messenger astronomy club, the latter of which aims to look for the existence physics beyond the standard model  \cite{Chia:2020dye}. In  a nutshell, 
\ac{GWs} are  perturbations  of space and time that propagate  through the universe carrying energy, linear  and angular momentum which can be measured in terrestrial detectors.    Since the first event detected by the LIGO collaboration in the fall of $2015$, an order of  $100$ binary  events have been subsequently detected including events from \ac{BBH}  \cite{LIGOScientific:2016aoc}, \ac{BNS}\cite{Abbott_2017}, and the more exotic, \ac{BH-NS}  system \cite{Abbott_2021}. 

LIGO/VIRGO successful direct detection of \ac{GWs}  accounts for just the beginning of the gravitational wave era. Indeed,  it is of common knowledge an upgrade of the LIGO/VIRGO detectors will take place within the next decade; this will be  known as the  era of the advance LIGO and VIRGO detectors, A+/Virgo+, and as  a result, earth base gravitational wave instruments  expect to observe an order of $10$ binary events every two weeks \cite{Shoemaker:2019bqt}, increasing the statistical power in the measurement of classical gravitational observables in terrestrial  detectors.  Furthermore,  the near future space-based \href{https://www.lisamission.org/articles/lisa-mission/lisa-mission-gravitational-universe}{LISA} mission is expected to join the \ac{GW}  instruments club in the couple of  decades, bringing  into the table access to   binary merges of super massive  black holes happening at  large   red-shift  values  ($z\sim 7$)   \cite{Favata:2010zu}; such events will be further added to the  \ac{BBH}  gravitational wave catalog. Additional \ac{GW} observatories such as KAGRA \cite{KAGRA:2018plz}, LIGO-India \cite{Saleem:2021iwi}, the Einstein Telescope \cite{Maggiore:2019uih} and the Cosmic Explorer \cite{Reitze:2019iox}, will make of \ac{GW} astronomy a highly active area of research in the coming decades. These will be instruments aiming to prove larger portions of the \ac{GW}  spectrum, ranging from frequencies  of $10^3Hz$ (Sound frequencies) to $10^{-3}Hz$ (the m-sound )  \footnote{For a related discussion see  \href{https://www.youtube.com/watch?v=M_bbZttuJeE}{Salam Distinguished Lectures 2022: Lecture 1: "What Gravitational Waves tell us about the Universe"}, by  A. Buonanno.}.

In order to analyze data obtained from these different observatories, more refined theoretical predictions -- which are the basis of  \ac{GW}  templates production -- will be needed.   Traditionally, the production of  \ac{GW} template  has been  a collaborative effort that takes elements from \ac{NR} 
 \cite{Pretorius:2005gq}, \ac{BHPT} and \ac{GSF}  \cite{1973Teukolsky,Kokkotas:1999bd}, the \ac{PN} formalism  \cite{Blanchet:2013haa,Futamase:2007zz} and the  \ac{EOB} method  \cite{PhysRevD.59.084006,Buonanno:2000ef,Santamaria:2010yb,Damour2011}.
 More recently, however, efforts have been focused on the \ac{BBH} scattering problem, in order to connect classical computations performed in the context of the \ac{PM} theory \cite{Damour_2016,Porto:2016pyg,Goldberger:2017vcg,Goldberger:2017ogt,Vines:2018gqi,Damour:2019lcq,Damour:2020tta,Kalin:2019rwq,Kalin:2019inp,Kalin:2020mvi,Kalin:2020fhe,Goldberger:2020fot,Brandhuber:2021kpo,Brandhuber:2021eyq}, with those approaches based on the classical limit of \ac{QFT} scattering amplitudes \cite{Cheung:2020gyp,Cheung:2018wkq,Bern:2019nnu,Bern:2019crd,Bern:2021dqo,Bjerrum-Bohr:2018xdl,Cristofoli:2019neg,Bjerrum-Bohr:2019kec,Bjerrum-Bohr:2021vuf,DiVecchia:2020ymx,DiVecchia:2021ndb,Bern:2020buy,Chung:2019duq,Chung:2020rrz,Cachazo:2017jef,Guevara:2017csg,Guevara:2018wpp,Guevara:2019fsj,Aoude:2020onz,Bautista:2019evw, Blumlein:2020pyo, Blumlein:2020znm}. 
 
\begin{figure}[h!]
\begin {center}
\includegraphics[width=12truecm]{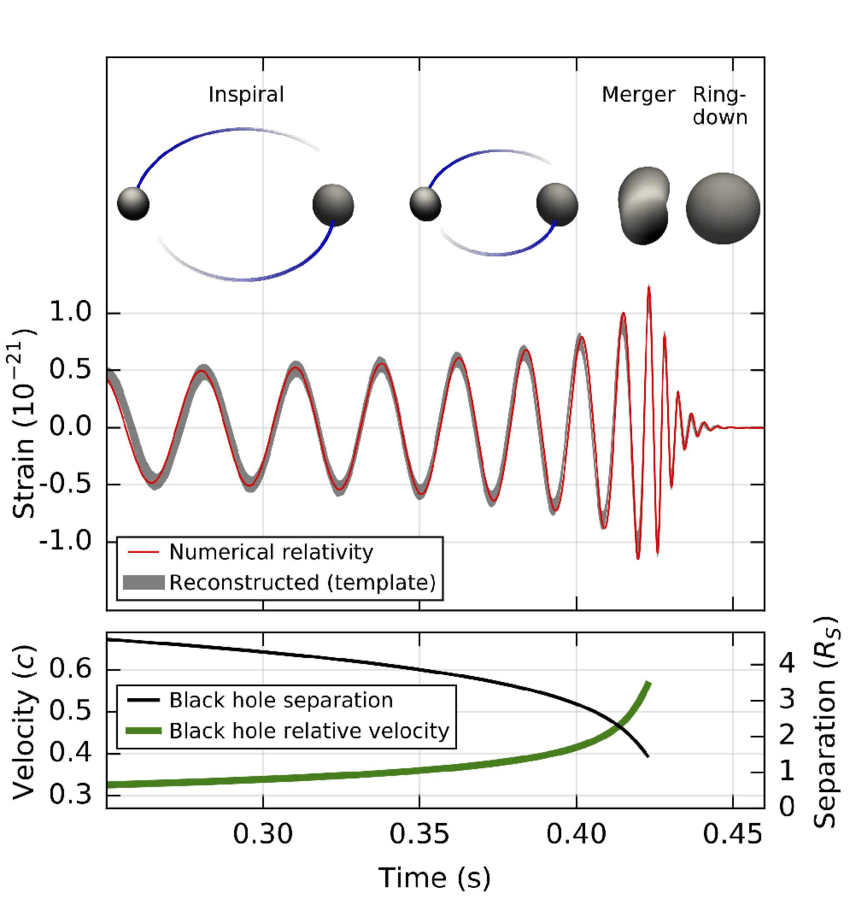} 
\end{center}
\caption{   GW150914 signal interpretation as  as seen at Hanford observatory. The three stages of the coalescing process are indicated. The lower plot shows the velocity of the components as function of their spatial separation. Figure reproduced from \cite{PhysRevLett.116.061102}.}
\label{fig:bbh}
\end{figure}

 The theoretical  predictions relevant  for \ac{GW} observatories fall into the category formed by the  so called two-body. 
 The two-body  problem for coalescing compact objects -- describing events such as  those observed in  LIGO/VIRGO detectors --  is customary divided into three stages. The earliest one is known as the inspiral stage. This phase comprehends the majority of the coalescing process and is characterized by the non-relativistic motion of the binary components; in addition, the  gravitational attraction between the two bodies falls into the  weak regime which makes of perturbation theory a suitable candidate to deal with  the problem.  
 The next stage is the merge. In this phase the two body collapse due to their 
 strong gravitational pull; the objects move with relativistic velocities and the problem becomes  non-perturbative, making of \ac{NR}  the, so far, suitable  tool to study the complex dynamics of the system. The final phase is known as the  ring down, where a Kerr \ac{BH} is formed from the combination of the two coalescing compact bodies. This \ac{BH}  radiates \ac{GWs} product of the excitation of its quasi-normal modes until a static configuration is reached. The main tool tho study this final stage is BHPT. See Figure \ref{fig:bbh} for a reconstruction example of  the coalescing process for the first \ac{GW} detection  GW150914.

\begin{figure}[h!]
\begin {center}
\includegraphics[width=12truecm]{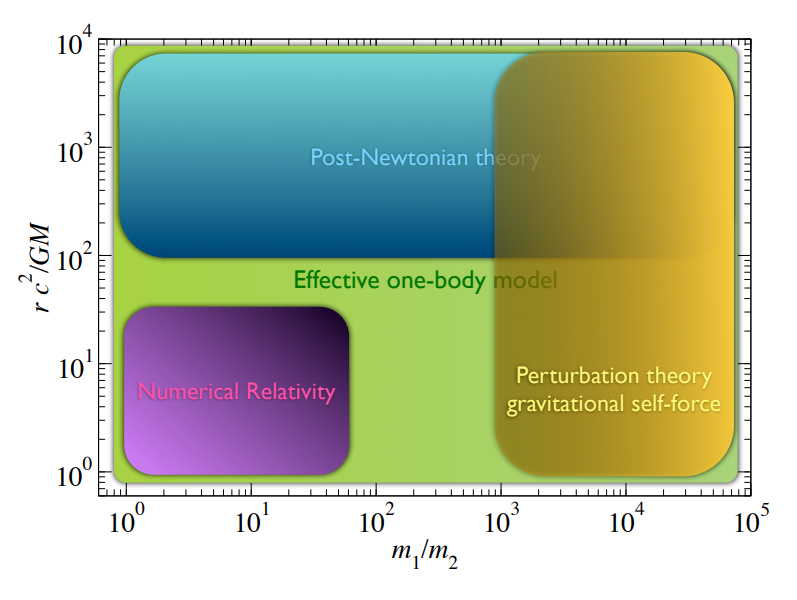} 
\end{center}
\caption{ Validity of the different methods producing gravitational wave templates in a typical \ac{BBH} system. Figure reproduced from     \cite{Buonanno:2014aza}.}
\label{fig:methods}
\end{figure}

\begin{figure}
\begin {center}
\includegraphics[width=12truecm]{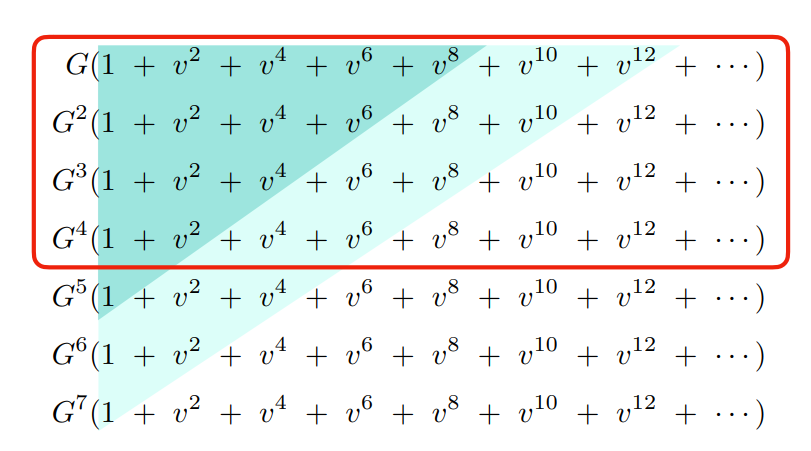} 
\end{center}
\caption{State of the art results for the \ac{GR} two-body conservative potential. The horizontal lines in the red box  indicate the  state of the art \ac{PM}   results, whereas the vertical lines correspond then to the \ac{PN}  information currently  available from \ac{PM}  methods. The dark blue triangle indicate the  state of the art   \ac{PM}-\ac{PN} overlap analytic information  available for gravitational wave template production, whereas the light triangle indicates the required orders in $v$ and $G$, needed  by future detectors.
Figure reproduced from     \cite{Buonanno:2022pgc}.}
\label{fig:PMpn}
\end{figure}

To be more  precise, the let us consider  in more detail several  of  the scales involved in the two-body problem. As already mentioned, a  significant part of the observed \ac{GW} signal  is encapsulated by the  inspiral phases where  there are  $\sim 10^5$ cycles for the two bodies going around each other. This makes the use of \ac{NR}  computationally expensive and therefore, analytic methods are  more suitable  to study such a phase. The traditional method to deal with such endeavor has been the \ac{PN} formalism. It  assumes both a weak field approximation (expanding  in powers of $G$, the Newton constant, or more precisely $GM/c^2b$, with $b$ the separation between the bodies), as well as a non-relativistic approximation (expansion in powers of $v^2/c^2$, with $v$ the typical velocities of the coalescing bodies). The current state of art results for  the two-body (conservative) dynamics is 5\ac{PN} order \cite{Bini:2019nra} (See also Figure \ref{fig:PMpn}). Let us now imagine  the  scenario  where one of the compact objects is much more massive than its companion, i.e. where the  mass ratio condition   $m_1/m_2\gg1$ is satisfied. Systems with such a property   are known as extreme mass ratio systems. A more suitable tool to study them  is  \ac{GSF},  where the problem is effectively reduced to solve for the geodesic motion of the small massive object in
the gravitational background field of its massive companion. One of the advantages of \ac{GSF} over the \ac{PN}  approximation is that  in principle \ac{GSF} is non-perturbative in the sense it only assumes an expansion in powers of $m_1/m_2$, but can keep all orders in $G$ if desired, therefore accounting for parts of higher \ac{PN} orders.  See Figure \ref{fig:methods} for typical systems where these methods are used. An alternative analytic approach to the two-body problem in the inspiral stage is provided by the \ac{PM}  approximation, which assumed   the problem can be treated using an expansion in powers of $GM/c^2b$ only. In reality, this approach is more suitable for the scattering problem as opposite to the bounded orbits scenario. Nevertheless, since the \ac{PM}  approximation  contains all powers in the velocity expansion, it naturally encapsulates \ac{PN} information for the bounded scenario by means of the virial theorem $GM/c^2b=v^2/c^2$ (See Figure \ref{fig:PMpn}). \ac{PM}  methods include worldline and classical methods, as well as the more recent \ac{QFT} approach as mentioned above. Let us finally mention the  \ac{EOB}  method (now days enlarged by the  Tutti-Frutti method \cite{Bini:2019nra}) is  the formalism that allows to  put together the information provided by the different  approximations to the problem.

Of special interest for this thesis is the scattering amplitudes treatment of the two-body problem. This approach has   recently gained  attention since it provides with a scalable way of dealing with  the two-body problem to very high orders in perturbation theory, while including spin \cite{Bern:2020buy,Guevara:2017csg,Chung:2019duq} and tidal  effects  \cite{Bern:2020uwk}. This is possible due to having  at  hand all of the 
 \ac{QFT} machinery developed for particle colliders such as double copy \cite{1986NuPhB.2691K,Bern:2008qj}, unitarity methods \cite{Bjerrum-Bohr:2021vuf}, leading singularity computations \cite{Cachazo:2017jef}, the spinor helicity formalism \cite{Arkani-Hamed:2017jhn}, integration by parts identities \cite{CHETYRKIN1981159,TKACHOV198165}  and differential equations \cite{KOTIKOV1991158,Bern:1992em,2000,Henn:2013pwa,Henn:2014qga} for loop integration, among some others. This arsenal of tools  makes of scattering amplitude methods a great candidate for hard core computations in gravity, and although these methods are  valid only in scattering scenarios, extrapolations to bounded scenarios are partially   understood  \cite{Kalin:2019rwq,Kalin:2019inp,Saketh:2021sri}; we will get back to this in a moment.
 
 At first it might seem  a bit odd to use the \ac{QFT} machinery to deal with problems which are of a purely classical nature.  Let us however remember  the correspondence principle  states  classical physics should emerge from quantum physics in the limit of large quantum numbers; that is, in the limit of macroscopic conserved charges such as  mass, electric charge, orbital angular momentum, spin  angular momentum, etc. In the context of the two-body problem, the transition from quantum to classical physics has been extensively studied \cite{Bern:2019crd,DiVecchia:2021bdo}, and with the introduction of the \ac{KMOC} formalism \cite{Kosower:2018adc}, a more precise map from the classical limit of scattering amplitude to classical observables in gauge theories an gravity has been established. Among the objectives of the amplitudes program in the two-body problem we have \cite{Buonanno:2022pgc}:
\begin{itemize}
    \item The production of state of the art predictions for the inspiral stage of the two-body problem in General Relativity and its possible modifications.
    \item Unraveling of  hidden theoretical structures in the gravity, while looking for a scalable framework for computing classical observables beyond the inspiral phase.
    \item The connection of  non-perturbative solutions in classical gravity, to perturbative scattering amplitudes realizations. 
\end{itemize}
 
From a \ac{QFT} setup, classical compact objects are understood as point particles dotted with a spin multipole structure. Additional finite size effects such as tidal deformability  can be taken into account by including higher dimensional (non-minimal coupling)  operators in the \ac{QFT} description \cite{Bern:2020uwk,Aoude:2020ygw}. Then, the amplitudes formulation of the two-body problem  relies mostly (but not only) on the computation of the $2\to2$ and $2\to3$ scattering amplitudes for spinning massive particles interchanging and radiating gravitons:
\begin{equation}\label{eq:4_and_5_point}
    \vcenter{\hbox{\includegraphics[width=80mm,height=23mm]{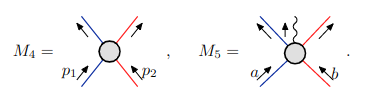}}} 
\end{equation}
The classical limit of these  amplitudes are  then  associated to conservative and radiative effects in the two-body problem, respectively \cite{Goldberger:2016iau,Kosower:2018adc}. These will be  some of the main objects of study for the present thesis. Here it is precise to mention, for \ac{BH}s, this \ac{EFT}  description can so far only  account for physics happening away from the \ac{BH}'s horizon. In fact, the amplitudes description of a Kerr \ac{BH} actually describes a naked singularity rather than an actual \ac{BH}, whose radius $a$, agrees  with the spin vector of the Kerr \ac{BH} \cite{Vines:2017hyw}. This means dissipation effects at the  horizon  are yet to be understood in a \ac{QFT} formulation of the problem, although  there are already some hints from the classical worldline approach \cite{Goldberger:2019sya,Goldberger:2020geb,Goldberger:2020fot}, as well as the amplitudes formulation of  the scattering of waves off the Kerr \ac{BH} \cite{Bautista:2021wfy,BGKV}.

In practice, the two-body dynamics   is studied in two separated sectors as given by the two amplitudes in \eqref{eq:4_and_5_point}: The first one corresponds to the  conservative sector where not radiative effects are accounted for (although radiation reaction effects are encapsulated in the conservative amplitude\footnote{ Radiation reaction reefers to  radiation  that is emitted by the binary system, but subsequently reabsorbed by the same system.}).  One of the greatest achievements from the amplitudes computation is the solution  for the conservative  dynamics at 4\ac{PM} order (3 loops), for binary systems composed of  scalar objects \cite{Bern:2021yeh,Dlapa:2021vgp}, and at 2\ac{PM} including spin effects  up to quartic order  \cite{Chen:2021qkk}, with recent new results at 3\ac{PM} at the spin quadrupole level \cite{Jakobsen:2022fcj}. Preliminary results at 2\ac{PM} but up to fifth \cite{Bern:2022kto} and  seventh  \cite{Aoude:2022trd} order in spin have recently appeared. 
The second sector  corresponds to the  radiative sector which has into account radiation effects encoded in   the   energy and angular momentum  emitted from the binary towards future null infinity in the form of  GWs. The current state of art from the amplitudes approach to the radiative dynamics is 3\ac{PM} order for scattering scenarios \cite{Herrmann:2021tct,Herrmann:2021lqe,Manohar:2022dea}.

In the conservative sector, the transition from scattering to bounded systems can be made in several ways. Without a particular hierarchy, the first path one could take  is  by computing the Hamiltonian of the two-body system. The   instantaneous potential for the gravitational interaction between the two bodies can be  calculated for instance via a  UV-EFT scattering amplitude matching procedure \cite{Cheung:2018wkq,Bern:2019crd}, or via  the scattering angle and the radial action \cite{Kalin:2019rwq,Kalin:2019inp},  or  by using a relativistic Lippmann-Schwinger equation \cite{Cristofoli:2019neg}, or the spinor helicity variables in conjunction with the \textit{holomorphic classical limit} \cite{Guevara:2017csg,Chung:2019duq}.
The second  way for transitioning from scattering to bounded orbit systems is  via a  direct  analytically continuation of  the scattering observables \cite{Kalin:2019rwq,Kalin:2019inp,Saketh:2021sri}. 

The radiative sector is a bit more complicated. The traditional way of including radiation effects for coalescing compact objects in a Hamiltonian is via the \ac{EOB} method. In this method, radiation reaction forces are included "by hand" in the particles \ac{EoM}  \cite{Khalil:2021txt}. This is an effect entering at the  2.5-\ac{PN} order at the level of the \ac{EoM}, and a 5-\ac{PN} effect at the level of the  radiated energy flux -- one of the important radiation observable --. These terms added "by hand" in the \ac{EoM} are dictated by  balance equation \cite{Bini:2012ji}, which have into account the lost of linear and angular momentum in the form of radiation emitted in the coalescing process. Once such terms are included in the \ac{EoM}, they can also be added to the two-body Hamiltonian, the latter of which  is the object used to compute all other  observables for a given -- bounded or unbounded -- system. From an amplitude perspective, for the radiative dynamics, analytic continuation methods applied to scattering observables  seem to still be valid when including radiation effects in the \ac{PM}  worldline \ac{EFT}  \cite{Cho:2021arx}. However, at 5\ac{PN} order,  back reaction makes non universal the unbounded and bounded problems, leading to non local in time terms; these are also known as  tail terms \cite{RevModPhys.52.299,Blanchet_1993,Khalil:2021fpm}, and it still  needs to be understood how to account for them  from an amplitudes approach. This then motivates to look for alternative continuation methods that can deal with the radiative bounded scenario directly from the scattering amplitudes. In this thesis we will take the first steps towards finding one of such methods, following the ideas of the authors in   \cite{Bautista:2021inx}; we will come back to this discussion below. 

\begin{table}
\begin{centering}
\begin{tabular}{|c|c|c|}
\cline{2-3} \cline{3-3} 
\multicolumn{1}{c|}{} & \ac{GR} & QED\tabularnewline
\hline 
quantum: & $\lambda_{c}\sim\frac{\hbar}{m}$ & $\lambda_{c}\sim\frac{\hbar}{m}$\tabularnewline
\hline 
classical particle size: & $r_{S}=2GM$ & $r_{Q}=\frac{e^{2}Q_{i}^{2}}{4\pi m}$\tabularnewline
\hline 
particle separation: & $b$ & $b$\tabularnewline
\hline 
\end{tabular}
\par\end{centering}
\caption{Parameter comparison for the two-body problem in \ac{GR} and electrodynamics. Here we have used units in which $c=1$. Table adapted from \cite{Bern:2021xze}.}
\label{tab:grvsqed}
\end{table}

From the discussion above it might seen  as if the amplitude methods are useful only for the two body problem in gravity.  Let us take the opportunity to stress however, amplitude computations indeed extended beyond the two-body problem. In  fact, with the introduction of the \ac{KMOC} formalism,  a variety of classical problems in  gauge theories and gravity can now be approached from a pure \ac{QFT} perspective. For instance, computation of related problems in classical electrodynamics as a toy model for gravity are now doable in a \ac{QFT} setup \cite{Kosower:2018adc,Bautista:2019tdr,Bern:2021xze,Herrmann:2021tct}. 
Perhaps  the closest scenario to the discussion above is the  relativistic two-body problem now in classical  Electrodynamics. That is, one can compute   classical observables for the relativistic scattering of two point-charges, in what is  been called the \ac{PL} expansion by the authors of ref. \cite{Bern:2021xze}. In Table \ref{tab:grvsqed} we have drawn a parallel of the relevant scales available for the two-body problem in \ac{GR} and Electrodynamics, when approached from a \ac{QFT} perspective.
The scale controlling the quantum effects is the Compton wavelength of the participating particles $\lambda_c$, which is related to the Plank constant and the particle's mass. The  typical classical particle size $r_S$ and $r_Q$ corresponding to the Schwarzschild radius and  charge radius, respectively. Finally, we have the particles  separation $b$. Extracting  classical information from a \ac{QFT} scattering process in the   \ac{PM}  approximation requires  $\lambda_C\ll r_S\ll b$ as mentioned above. The first inequality corresponds to take the point particle limit, whereas the second inequality is equivalent to a large angular momentum expansion, which effectively  permits to deal with  the problem in a perturbative fashion.
The electrodynamics analog to the \ac{PM}  expansion  is then the \ac{PL} expansion, which corresponds to the regime where $\lambda_c\ll r_Q\ll b$. Additional considerations have to be made when including radiation and spin effects. For the former, one requires the  wavelength of the emitted wave to be much bigger than the size of the system; this in turn allows to recover the source multipole expansion for the radiation field. For the latter, combinations of the \ac{BH} spin and the frequency of the emitted wave of the form $\omega a$ should remain finite. This then translates  to take the large spin limit, as required by the correspondence principle. 

Other problems in \ac{GR} with immediate analog in classical electrodynamics include Gravitational and electromagnetic Bremsstrahlung radiation in a $2\to3$ scattering process \cite{Kosower:2018adc,Bautista:2019tdr,Bern:2021xze,Herrmann:2021tct}. The map of the 3-particle amplitude to the linearized effective  Kerr metric  \cite{Chung:2018kqs,Guevara:2018wpp} and  the root Kerr charge configuration \cite{Arkani-Hamed:2019ymq}, the Thomson scattering \cite{Jackson:100964}, and the scattering of waves off the Schwarzschild/Kerr black hole \cite{Bautista:2021wfy,BGKV} in bounded scenarios,   the computation of   the  Maxwell dipole and  the Einstein quadrupole radiation formulas directly from scattering amplitudes \cite{Goldberger:2017vcg,Bautista:2021inx} , the memory effect in gravity and electrodynamics  \cite{Bautista:2019tdr,Bautista:2021llr,Manu:2020zxl},  among some others. In this thesis we will approach several of these problems, with particular interests in spin effects both, in electrodynamics and in classical gravity. 

Let us now take  the opportunity to summarize the content of this thesis, while highlighting  the contributions made by the author towards approaching some of the  aforementioned problems. We however stress that if it is true a vast majority of the content of this thesis will be aimed to provide results relevant to  the two-body problem, this thesis also aims to provide a more general understanding of the \ac{QFT} description of purely classical problems, but at the same time, to provide some formal derivations in pure \ac{QFT} scenarios, specially in the context of the double copy. 

This thesis is organized as follows: 
In   \cref{ch_preliminaries} we present a preliminary compilation of several modern amplitude methods that are relevant for understanding the  main body of this thesis. In particular, in \cref{sec:KMOC} we review the \ac{KMOC} formalism  in the context of the  two body problem. This will provide us with  a robust framework for computing  observables in (classical) gauge theories and gravity directly from the (classical limit of) \ac{QFT}  amplitudes. In this section we provide a detail discussion on how to take the classical limit of \ac{QFT} formulas in order to obtain the desired classical information. We focus on two main observables: The first one is the linear impulse acquired by a classical object in a $2\to2$ scattering process, at  generic order in perturbation theory. This observable is directly related to the scattering angle and therefore to the Hamiltonian of the system, as discussed above, so it is of main importance for the two-body problem. The second observable we discuss is the radiated classical electromagnetic/gravitational field in an inelastic $2\to3$ scattering process, similarly, to generic order in perturbation theory. This will give us directly the waveform emitted from the scatter objects towards future null infinity. This waveform  can be used to compute the (gravitational) wave energy flux, which is  one of the main observables measured in a (gravitational) wave observatory. We then move   to \cref{sec:double-copy-preliminaires} where we introduce some generalities of the Double copy for massless particles. In particular, we introduce the concept of  \ac{YM}   partial amplitudes  and discuss their double copy formulation in the \ac{KLT} form. We also discuss the color-kinematics duality and the  \ac{BCJ} formulation of the double copy of \ac{YM} amplitudes. We provide simple examples for the double copy of the 3 and 4-point amplitudes. 
Understanding of the double copy for massless particles will be of special use when formulating the double copy for massive particles with spin, specially in  \cref{sec:spin_in_qed},  \cref{ch:double_copy}, and \cref{ch:GW_scattering}. We move then to \cref{sec:spinor-helicity} where we review  the spinor helicity formalism  for both massless and massive particles in 4 dimensions. We discuss the spinor helicity representation of massless and massive momenta, as well as polarization vectors. In the massless case we discuss how little group arguments fix completely the all helicity 3-point amplitude. In the   massive case we discuss the exponential representation of the minimal coupling 3-point and the Compton amplitudes for spinning particles. Spinor-helicity variables will be of special use in \cref{ch:double_copy}, \cref{ch:GW_scattering} and \cref{app_B}. We conclude in \cref{sec:outlook_preliminaries} with a small outlook of the chapter. 

Having  acquired some preliminary knowledge of several of the modern amplitude techniques introduced in \cref{ch_preliminaries}, as a warm up in \cref{ch:electromagnetism} we begin the study of   classical observables in \ac{SQED}  from  an amplitudes setup. This will provide some flavour on the amplitude formalism when dealing with classical observables, while avoiding the complications introduced by spin or higher Lorentz index structures. We start in \cref{sec:scalarqedlag} by  deriving from the \ac{SQED} Lagrangian the three level  amplitudes for a  scalar matter line emitting one or two photons. We give  this amplitudes special names $A_n, \,\,n=3,4$, since they will be the building blocks for more complex amplitudes, as well as  the topic of extensive studies in the  proceeding  chapters. We discuss immediate application of these amplitudes in a classical context. In particular, for $A_3$ we discuss how despite this being an amplitude with photon emissions, it does not carries any  radiative content in Lorentz signature. For the case of   $A_4$, we connect its classical limit to  the Thomson scattering process in classical Electrodynamics (The analogous process in \ac{GR} will be studied in \cref{ch:GW_scattering}).  Additional properties  of these amplitudes such as soft exponentiation and the definition of orbit multipole moments are discussed. The latter correspond to  the amplitude analog of the multipolar expansion in classical electrodynamics. It is then argue that $A_4$ has indeed non trivial orbit multipoles as opposite to $A_3$, which makes  $A_4$ carry radiative degrees of freedom that $A_3$ does not possess. Soft exponentiation then allows us to argue these amplitudes can be derived directly from soft theorems and Lorentz invariance,  without the need of a Lagrangian formulation. We move then to \cref{sec:examplesa_tree_level} where a first application of the $A_n$ amplitudes in the two-body problem in \ac{SQED} is introduced. At leading order in perturbation theory, we show how the classical content of the conservative and radiative two-body amplitudes is controlled by $A_n$ from the factorization properties:
\begin{equation}\label{cuts m4 m5}
    \vcenter{\hbox{\includegraphics[width=96mm,height=23mm]{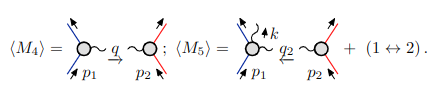}}} 
\end{equation}

These factorization properties are indeed more universal, and holds for spinning particles in both \ac{QED} and Gravity, which unify the computation of leading order radiation in classical electrodynamics and gravity  in the compact formula \eqref{eq:newM5clas}. One can  then  obtain many of the physical features of the two-body problem from understanding the universality (and double copy ) properties of $A_n$ amplitudes. For instance, we show how the soft exponentiation of $A_n$ induces an all orders soft exponentiation of the two-body radiative amplitude, whose leading soft piece reproduces the memory waveform in  \ac{SQED} This memory waveform is  universal (independent of the spin of the massive matter), for both \ac{QED} and GR, as it is dictated  only by the Weinberg soft theorem (In \cref{sec:spin_in_qed} we argue how this universality can be seen from the spin multipole expansion for both \ac{QED} and \ac{GR}).  We illustrate the computation of the leading order (\cref{sec:examplesa_tree_level}) and Next to leading order (\cref{sec:2PLimpulse}) classical impulse in a $2\to2$ scattering process, using the \ac{KMOC} formalism, and  introducing some integration techniques that will be used in \cref{ch:soft_constraints}. This allows us to  shows how one can  recover the classical result of Saketh et al \cite{Saketh:2021sri} for the 2\ac{PL} linear impulse, purely from  amplitudes arguments. 
We also show how 3\ac{PL} radiation results  reproduce known classical results for colorless radiation  computed from the worldline formalism by Goldberger and Ridgway \cite{Goldberger:2016iau}. Finally, we conclude in \cref{sec:outlook_sqed} with an outlook of the chapter.

Continuing with the \ac{SQED} theme, in \cref{ch:soft_constraints} we study low energy Bremsstrahlung  radiation for the scattering two-body problem, from an amplitudes perspective. It is well known classical soft theorems predict the form of the wave emitted in a $N$-particle scattering process in the limit in which the frequency of the emitted wave is much smaller than the momenta of the other objects involved. Classical soft theorems are non perturbative statements and to prove them from perturbative approaches becomes a highly  non-trivial task. However, they can also be  used to probe perturbative approaches to the computation of classical radiation, in particular, on the \ac{KMOC} formula for classical two-body  radiation as discussed in \cref{ch_preliminaries}  and \cref{ch:electromagnetism}. In this chapter we show that classical soft theorems impose an infinite series of constraints on \ac{KMOC} formula. These constraints relate the expectation value of certain monomials of exchange momenta, to the linear impulse classical objects acquire due to the exchange of photons/gravitons in the scattering process, at arbitrary order in perturbation theory. We start in   \cref{sec:introduction_soft_constraints} by  reviewing some facts from classical soft theorems, and summarizing the main results of the chapter. Next we move to  \cref{sec:classical_review} where we show explicit the prediction form  classical soft theorem  for the form of the radiated field in a $2\to 3$ scattering process to leading order in the soft expansion, and subleading order in perturbation theory.  In \cref{sec:soft_constraints_deriv} we provide a formal derivation of the  constraints imposed by the Weinberg soft theorem on the \ac{KMOC} formula for the radiated field, which we  then verify in  \cref{sec:lo-soft_costraitns} up to  \ac{NLO} in the perturbative expansion, matching the expected results introduced earlier in \cref{sec:classical_review}. In \cref{sec:discusion} we provide an outlook of the chapter. Here we argue that although  the soft constrains presented in this thesis were derived in the context of \ac{SQED} , and to leading order in perturbation theory, analogous constraints follow for the gravitational case \cite{Manu:2020zxl,Bautista:2019tdr}, both at leading and subleading orders in the soft expansion\footnote{See also \href{https://www.youtube.com/watch?v=b0dyWOgwWbc&t=1698s}{“Soft theorems and classical radiation”}, where the author argues non-linear Christodoulou memory effect \cite{PhysRevLett.67.1486},  can be obtained directly from the \ac{KMOC} formula in Gravity. Non linear memory originates from gravitational waves that are sourced by the previously emitted waves \cite{Favata:2010zu}. From the amplitudes approach, this is a  two-loop effect under current investigation by the author \cite{BLZ}.} \cite{BLZ}. In fact, in \cref{sec:spin_in_qed} we show how the leading soft constrains in the gravitational context at \ac{LO}  in perturbation theory recover the  burst memory waveform of Braginsky and Thorne \cite{Braginsky1987}. Soft theorem are non perturbative statements and in principle can inform about radiation to higher orders in perturbation theory, in fact, they are already used to compute radiation reaction effects in the high energy (eikonal) approximation of the two-body problem \cite{DiVecchia:2021ndb,DiVecchia:2021bdo,DiVecchia:2022owy,DiVecchia:2022nna}.
Finally in \cref{app:radiation_NLO} we provide some computational details on the verification of the soft constraints at \ac{NLO} in perturbation theory.

By then, the reader should had gained some familiarity with the amplitudes approach to obtaining  classical physics from  \ac{SQED}. The natural 
thing to do next is to use the amplitude machinery to  approach more complicated problems. There are several directions one could follow. For instance, one could  introduce spin effects from \ac{QED},  or study classical observables in gravitational  physics involving scalar and spinning\footnote{ Spin effects are important since they encode information regarding the formation mechanism of the binary system (see for instance \cite{Gerosa:2013laa,Vitale:2015tea, Biscoveanu:2021nvg}). For nearly extremal Kerr BHs, the individual spins of the binary's components   are expected to be measured with great precision by LISA  \cite{Burke:2020vvk}, and therefore, it is important to have perturbative results for both conservative and radiative dynamics to high powers in the spin expansion.} compact objects (minimally) coupled to gravity. These will be  in fact the topics of study of \cref{sec:spin_in_qed}. Continuing the study of our favorite $A_n$ amplitudes, in \cref{sec:spinin_matter} we show  when introducing spin effects, these amplitudes can be written in a spin multipole decomposition in generic spacetime dimensions.  We differentiate two types of spin multipole moments:  \textit{covariant}, and \textit{rotation} multipole. The former corresponds to  \ac{irreps.} of the Lorentz group in general dimension, SO$(D-1,1)$, whereas the latter are  \ac{irreps.} of the rotation subgroup SO$(D-1)$; these  are the ones describing actual classical rotating objects. We compute the multipole decomposition for amplitudes involving particles of spin $0,\frac{1}{2},\,\,\text{and}\,\,1$, which are computed from the \ac{SQED} , \ac{QED} and Maxwell-Proca Lagrangians respectively.  We show  $A_n$ amplitudes can be written in terms of the Lorentz generators $J_s$ in the spin $s$ representation, with the multipole coefficients being universal functions (independent of the spin of the scattered particles). For $A_3$, we show how for spin $1/2$, \ac{QED} predicts the electron ( tree level) gyromagnetic factor $g=2$, whereas for spin $1,$ the Proca Lagrangian predicts $g=1$ (We will revisit the $g$-factor in \cref{ch:double_copy}, from double copy arguments, and argue  $g=2$ for spin $1$ particles, where the massive vector particles are actually  W-bosons). From unitary arguments, we show $A_4$ can be constructed from $A_3$ in a spin multipole form, with an exponential structure analog to the soft exponentiation of the scalar amplitude. These amplitudes can then be used to compute two-body observables in electrodynamics from the factorization properties \eqref{cuts m4 m5}. One can easily recover known linear in spin results \cite{Li:2018qap}. In \cref{sec:spin_dc} we introduce a \textit{covariant} spin multipole double copy in generic space-time dimensions for the $A_n$ amplitudes. This double copy prescription has the property of preserving the spin multipole structure of the gravitational amplitudes, which can be used to compute two-body radiation from \eqref{cuts m4 m5}. This in turn implies leading order gravitational radiation can be computed from the double copy of photon radiation avoiding the complications form colour radiation.  Using double copy arguments and universality of the coupling of matter to gravity, we show the 3-point amplitude for a massive spinning particle of generic spin takes an exponential structure (this exponential will be matched to the linearized Kerr metric in \cref{ch:GW_scattering}). For the case of $A_4$ in gravity, we compute its covariant multipole decomposition up to quartic order in spin and show it agrees with the more lengthy Feynman diagrammatic computation from minimal coupling Lagrangians. We further decompose the  Compton amplitude in terms of \ac{irreps.} of SO$(D-1,1)$ by  introducing the Ricci decomposition method, which allows to decompose the products of two Lorentz generators into the correspondent \ac{irreps.} In order to make contact with actual classical rotating compact objects,  we write the amplitudes  in terms of the multipole moments for the rotation subgroup SO$(D-1)$. We show this is achieved by aligning spinning particles polarization tensors -- which have different little group transformation properties --  towards canonical polarizations with the same little group scaling for incoming and outgoing matter. This is done by fixing a condition on the spin tensor that goes by the name of the  \ac{SSC} \cite{Barker:1975ae,Barker1979}. This alignment also goes by the name of Hilbert space matching \cite{Chung:2020rrz}. In $D=4$, we obtain up to quadratic in spin, a vector representation of the classical gravitational Compton amplitude, which has the property of  factorizing into the product of the scalar amplitude and the spinning $s$-amplitude in \ac{QED},  as dictated by the equivalence principle (In \cref{ch:GW_scattering} we show it reproduces results for Gravitational wave scattering off Kerr \ac{BH} up to quadratic order in spin). Having understood the double copy properties of the $A_n$ amplitudes when including spin effects,  as well as how to take  their classical limit, in \cref{sec:M4M5_spin} we show double copy of $A_n$ amplitudes induce a classical double copy formula for the two-body amplitudes \eqref{cuts m4 m5}, including spin effects. We use this formulas to compute radiation for scalar, linear and quadratic in spin, recovering known result in the literature  \cite{Goldberger:2016iau,Luna:2017dtq,Bautista:2019evw,Li:2018qap,Jakobsen:2021lvp}. 
In addition, we show for scalar matter an exponential soft theorem for the two-body radiation amplitude in gravity can be obtained, analog to the electromagnetic case of \cref{ch:electromagnetism}, whose leading order allows to recover the  memory waveform derived by Braginsky and Thorne \cite{Braginsky1987}. We also show that spin effects in $M_5$ are subleading in the soft expansion, and therefore recovering the universality of the Weinberg soft theorem. We conclude in \cref{sec:outlook_spin} with an outlook of the chapter. In \cref{app_B} we provide some spinor helicity formulas to connect vectors results in this chapter to those given in spinor form in the literature.

In \cref{ch:bounded} we do a transition from scattering to bounded scenarios. In particular,  we show the inspiral waveform for two Kerr black holes orbiting in general (and quasi-circular orbits), whose spins are aligned with the direction of the system's angular moment, and to leading and subleading order in the velocity expansion, can be obtained directly from the spinning amplitudes derived in \cref{sec:spin_in_qed}. Using an empiric  formula for the waveform inspired by previous computations \cite{HariDass:1980tq,Goldberger:2017vcg}, we propose such formula could  modifies \ac{KMOC} formula for the radiated field \eqref{eq:rad_field_full_n},  to allow objects to move in generic trajectories. Particles \ac{EoM} at leading order in velocity, but to all orders in spin, are analogously obtained from the conservative 4-point  amplitude, via a modification of  the   \ac{KMOC} formula for the linear impulse \eqref{eq:impulse_final}. We start this chapter in \cref{sec:intro_bounded} with a small introduction and summary of our results. We  then move to \cref{sec:classical_derivation} where we provide a classical derivation of the gravitational waveform using the  multipolar \ac{PM} formalism  \cite{Thorne:1980ru, Blanchet:1985sp, Blanchet:1995fg, Blanchet:1998in, Blanchet:2013haa}. In this section, we review the classical Lagrangian description of a spinning \ac{BH} focusing on the conservative sector, where object's \ac{EoM} are derived, and provide explicit solutions to the \ac{EoM} for the quasi-circular orbits scenario. We then use them into the multipolar expansion of the radiated field, obtaining solutions for the leading and subleading in velocity contributions to the  waveform for binary systems in  both, general and quasi-circular orbits, to all orders in spin. We move then to the amplitudes formulation of the problem in \cref{sec:scatteringampl}. Introducing the general formalism, we write the formulas for the radiated field as well as particles equations of motion in terms of the non-relativistic limit of two-body scattering amplitudes. We review how to obtain the scalar waveform, reproducing the well known Einstein quadrupole radiation formula, and then provide spin corrections to it. For generic orbits, we show there is a one to one correspondence between the scalar amplitude and the source mass quadrupole moment, and in the same way, the linear in spin amplitude is in direct correspondence with the current quadrupole moment. We then show that at quadratic order in spin, and leading order in velocity, the radiated field acquires a vanishing  contribution from the spin quadrupole radiation amplitude. The all orders in spin waveform at leading order in velocity is then obtained from the solutions to the particles \ac{EoM}. We obtain the first subleading order in velocity correction to the quadrupole formula in the spinless limit, and show it agrees with the classical derivation obtained in \cref{sec:classical_derivation}. We argue that although in general waveforms derived from different methods can differed one from the other by a time independent constant, physical observables such as  the gravitational wave energy flux or the radiation scalar are insensitive to such a constant, as they can be computed from time derivatives of the waveform. This is a manifestation of a residual gauge freedom present in the waveform, which can be eliminated for physical observables. We conclude  in \cref{sec:discusion} with an outlook of the chapter. In \cref{sec:app_bounded}  we include some useful integrals and identities used for several computations in this chapter.

In \cref{ch:double_copy} we start a more formal study of the double copy for amplitudes involving massive  spinning matter in generic space-time dimensions. In this chapter we  aim to, on the one hand, provide a formal derivation of the  double copy prescriptions  introduced in \cref{sec:spin_in_qed}, and on the other hand, to derive the gravitational Lagrangians for the  theories obtained from such double copies. We star in \cref{sec:intro_dc} with a small introduction and a summary of the main results of the chapter. We argue double copy of spin $s$  with spin  $\tilde{s}$ matter, leads to universal coupling of the resulting $(s,\tilde{s})$ massive particles to the graviton -- as required by the equivalence principle -- but in general the coupling to the dilaton and the two-form  (axion) potential are not universal. For interactive spin 1 particles in gravity, this allows us to define two independent gravitational  theories which we name the $0\otimes1$ and the $\frac{1}{2}\otimes \frac{1}{2}$ theories. We provide general dimension tree-level Lagrangians in the Einstein frame for one and two spinning matter lines. Theory $\frac{1}{2}\otimes \frac{1}{2}$ is a simpler theory as compared to the  $0\otimes1$ counterpart, since on the one hand, it does not include quartic  terms in the two matter lines Lagrangian, and on the other hand, one can consistently truncate the double copy spectrum to remove the coupling of matter to the two-form potential.  On the other hand, in $D=4$ and in the massless limit,  the $0\otimes1$ theory reproduces the  bosonic interaction of $\mathcal{N}=4$ Supergravity, which arises from the double copy $\mathcal{N}=4$ Super Yang-Mills and \ac{YM} theories. In general dimensions,  this theory is the \ac{QFT} version of the worldline double copy model constructed by Goldberger and Ridgway in \cite{Goldberger:2016iau,Goldberger:2017ogt} and extended to include spin effects in \cite{Goldberger:2017vcg,Li:2018qap}. In \cref{sec:sec2} we derive the massive  double copy formulas, for the theories consider,  from dimensional reduction and compactification of the massless counterparts. We provide a variety of examples of amplitudes derived from such double copy formulas for one matter line emitting radiation. Furthermore, we show explicitly how to obtain the multipole double copy prescription introduced in \cref{sec:spin_in_qed} from these dimensionally reduced double copy formulas. In the same section we discuss how setting $g=2$ for the gyromagnetic factor removes the  divergences of the Compton amplitude in the massless limit. Such amplitude coincides with the minimal coupling\footnote{Following \cite{Arkani-Hamed:2017jhn}, minimal coupling amplitudes are   those which have a well defined high energy limit. This definition of minimal coupling differs from the  usual definition of minimal coupling of promoting partial derivatives to covariant derivatives. } Compton amplitude written in spinor helicity variables in \cref{sec:spinor-helicity}. We continue in \cref{constructing the lagrangians} where we construct massive Lagrangians both for \ac{QCD}  and the gravitational theories from  \ac{KK} reduction and  compactification. For spin 1 in \ac{QCD}, we introduce a  modification of the Proca Lagrangian to set $g=2$ which is characteristic of the W-boson. We then show that \ac{QCD}  amplitudes $A_n$ for generic $n$, entering in the double copy formulas derived in \cref{sec:sec2}, are obtained from the compactification of their massless counterpart. This is the reason these amplitudes possessa well defined high energy limit. In \cref{sec:grav_theories} we derive the Lagrangians for one matter line for  the $0\otimes1$ and $\frac{1}{2}\otimes\frac{1}{2}$ gravitational theories. In \cref{two matter lines} we study the massive double copy construction for spinning amplitudes including two matter lines. We use the massive version of the \ac{BCJ} prescription introduced in \cref{sec:double-copy-preliminaires}, providing the two-matter lines gravitational Lagrangians for the different double copy prescriptions. For inelastic scattering, we probe there is a Generalized Gauge Transformation that allow to recover the classical double copy formula for the radiation amplitude obtained from the factorization \eqref{cuts m4 m5}, this time directly from the quantum \ac{BCJ} double copy. We finalize in \cref{sec:discusion_dc} with an outlook of the chapter. In \cref{4 d double copy} we prove our general dimensional $\frac{1}{2}\otimes\frac{1}{2}$ gravitational Lagrangian agrees with the $D=4$ derivation obtained in \cite{Johansson:2019dnu}. In \cref{residues at 4 points} we study the unitarity properties of the  $\frac{1}{2}\otimes\frac{1}{2}$ amplitudes at four points. 

Up to this point, we would have claimed the classical limit of amplitudes for massive spinning  matter minimally couple to gravity actually describes the  Kerr \ac{BH}. Perhaps the strongest hint is given by the computation of the waveforms for bounded systems described in \cref{ch:bounded}. However, the  spin structure of the non-relativistic waveforms derived there  follows mostly from $A_3$ whose classical limit  now days is well known encode all the spin multipoles of  the linearized Kerr metric. The natural question to ask is whether $A_4$ has actually anything to do with Kerr. In \cref{ch:GW_scattering} we show that $A_4$ is indeed very related to Kerr as it  describes the low energy regime for the scattering of  gravitational waves off the Kerr BH. We start this chapter with a small introduction and a summary of the results in  \cref{sec:intro_gwscatt}. We stress finding the connection of $A_n$ amplitudes to Kerr is important since they are the building blocks for the two-body amplitudes. In particular, it is important to prove the  2\ac{PM} scattering angle for aligned spin computed in \cite{Guevara:2018wpp} actually describes the scattering of two Kerr BHs and not other classical compact objects. In \cref{sec:esponential34} we show how to take the  classical limit of $A_n$ amplitudes written in spinor helicity form. For $n=3$ we indeed recover the Linearized Kerr metric, whereas for $n=4$, up to quartic order in spin, the gravitational Compton amplitude can be written in an exponential form for both, same and opposite helicity configurations of the external graviton legs, in agreement with the classical heavy particle effective theory derivation of \cite{Aoude:2020onz}. In \cref{sec:gw_Scatt} we use $A_4$ amplitude to study the scattering of gravitational  waves off Kerr, obtaining the differential cross section for generic spin orientation of the \ac{BH},  recovering the linear in spin results of \cite{Dolan:2008kf} for polar scattering. Spin induced polarization of the waves is discussed in \cref{sec:polapp}, which to linear order in spin recovers the \ac{BHPT} results of  \cite{Dolan:2008kf} and therefore clarifying the mismatch from the Feynman diagrammatic computation of \cite{Guadagnini:2008ha,Barbieri:2005kp}. The solution of the discrepancy comes by including all the Feynman diagrams contributing to the Compton amplitude and not just the graviton exchange diagram, as done by the authors  in \cite{Guadagnini:2008ha,Barbieri:2005kp} \cite{Guadagnini:2008ha,Barbieri:2005kp}. The quartic in spin result for the differential cross section provides a highly non-trivial prediction, pushing the linear in spin state of the art result of   \cite{Dolan:2008kf} since 2008, while providing a way to resum the partial infinite sums appearing from \ac{BHPT}  for generic  orientation of the spin of the BH. In \cref{ch:teukolsky} we provide a detail derivation of the differential cross section up to quartic\footnote{Higher order in spin results required a more careful analysis but nevertheless will be shown in \cite{BGKV}. } order in spin from \ac{BHPT}, finding perfect agreement with the amplitudes computation, therefore showing the Compton amplitude indeed possessthe same spin multipole structure as that of  the Kerr \ac{BH} when perturbed by a gravitational wave.
In \cref{ap:hcl} we show the classical limit of the Compton amplitude derived in here indeed can be used to compute the 2\ac{PM} aligned spin scattering angle for the scattering of two Kerr BHs, therefore confirming the validity of the  predictions of \cite{Guevara:2018wpp}. We close with an outlook of this chapter in \cref{sec:outlook_gw}.

We finalize this thesis with at general discussion in \cref{conclusions}.


%% file: Chapters/chapter_preliminaries.tex
\chapter{Prelimimaries}\label{ch_preliminaries}
\section{Introduction}
In this chapter we will introduce  some aspect of scattering  amplitudes  that will be of great use for the present  thesis. 
We start  in  \cref{sec:KMOC}   reviewing some features of the \ac{KMOC}  formalism \cite{Kosower:2018adc}, which is a robust frame for the computation of (classical) observable directly from the (classical limit of the) scattering amplitudes.  In this thesis we will be interested in 2 \ac{KMOC}  observables: 1) The linear impulse in a $2\to2$ elastic scattering process, and 2) The radiated field at future null infinity from a $2\to3$ inelastic scattering process. This section will be of great use for most of the content of the present thesis, specially for \cref{ch:electromagnetism},\cref{ch:soft_constraints}, \cref{sec:spin_in_qed}.  In \cref{ch:bounded} we motivate a modification of \ac{KMOC}  formalism to study two-body systems for bounded orbits. Next, we move to  \cref{sec:double-copy-preliminaires} where we introduce some general aspects of the double copy \cite{1986NuPhB.2691K,Bern:2008qj}. In particular, we  focus on the massless double copy of Yang Mills amplitudes in both, the \ac{KLT} and the \ac{BCJ}  representations.
Intuition from the massless double copy will be of great use when formulating double copy  prescriptions for  massive particles with and without spin, presented in  \cref{sec:spin_in_qed} and \cref{ch:double_copy}. Finally, in \cref{sec:spinor-helicity} we introduce  the spinor-helicity formalism for massless and massive particles. In particular, we will review how helicity arguments fix the 3 and 4 point amplitudes for massive/massless spinning particles. Knowledge of  this formalism will be of great use through several chapters of this thesis, in particular when discussing higher spin amplitudes in \cref{ch:GW_scattering}.

\section{The Kosower, Maybee and O'Connell formalism (KMOC)}\label{sec:KMOC}

In this section we start by introducing  the \ac{KMOC} formalism. As already mentioned, the \ac{KMOC}  formalism  \cite{Kosower:2018adc,Maybee:2019jus,Guevara:2019fsj,Cristofoli:2021vyo,Aoude:2021oqj,Herrmann:2021lqe}, provides us with a robust framework for the computation of (classical) observables in gauge theories and gravity, directly from (the classical limit of )  \ac{QFT}  scattering amplitudes.  It has become one of the cornerstone in the amplitude program in classical physics, and is directly relevant to understand the content of this thesis.  In this formalism, classical compact objects are described in an effective way as point particles, whose finite size effects  can be mapped into intrinsic properties of the elementary particles used in the \ac{EFT} description. 
In what follows we review some of the most relevant aspects of this formalism. For a nice review, the reader is recommended to consult the original \ac{KMOC}  work, as well as the recent reference  \cite{Kosower:2022yvp}.

In the \ac{KMOC}  formalism,  the expectation value for the change of a   quantum mechanical observable, $\Delta\hat{\mathcal{O}}$, due to a scattering process is computed from the scattering matrix through the formula 
\begin{equation}\label{eq:kmoc_formula}
     \Delta \hat{\mathcal{O} } =\, _{\rm{in}}\langle\Psi|S^\dagger \hat{\mathcal{O}}S|\Psi\rangle_{\rm{in}} -\, _{\rm{in}}\langle\Psi| \hat{\mathcal{O}}|\Psi\rangle_{\rm{in}} \,,
\end{equation}
This corresponds to the  difference of the measurement of the given operator in the final and initial state, where we have relied on $S$ as a time evolution operator determining the form of the asymptotic  final state of the system $\ket{\Psi}_{\text{out}} = S\ket{\Psi}_{\text{in}}$. 
The connection of $ \Delta \hat{\mathcal{O}}$ to the a \ac{QFT}  scattering amplitude is done in two steps: First, we need to  split the $S$-operator in  the usual way, $S=1+iT$, after which,  exploding   the unitarity condition, $SS^\dagger=1$,  allows us to rewrite \eqref{eq:kmoc_formula} in the  form:
\begin{equation}\label{eq:Kmocds}
    \Delta \hat{\mathcal{O}} =  \,_{\rm{in}}\langle\psi|[T, i \hat{\mathcal{O}}]|\Psi\rangle_{\rm{in}} +\, _{\rm{in}}\langle\psi| T^\dagger[\hat{\mathcal{O}},T]|\Psi\rangle_{\rm{in}}\,.
\end{equation}
Second, we need to specify  the system's  initial state. For the moment let us assume it can be decompose into  multi-particle plane wave states, which in momentum space are proportional to  $\ket{p_1,\cdots,p_n}$. These states are the  tensor product of individual momentum eigenstates  $a_p^\dagger \ket{0}$, where $a_p^\dagger$ is the creation operator for state of momentum $p$. The conjugate states are labeled by  $ \bra{p_1',\cdots p_m'}$, and together with $T$, define the \ac{QFT}  scattering amplitude via 
\begin{equation}\label{eq:amp_def}
    \mathcal{A}(p_1,\cdots,p_n\to p_1',\cdots,p_m')\hat{\delta}^d(p_1+\cdots p_n-p_1'-\cdots p_m') = \bra{p_1',\cdots,p_m'}T\ket{p_1,\cdots,p_n}\,.
\end{equation}
where $\hat{\delta}^d(p_1+\cdots p_n-p_1'-\cdots p_m')$ is the momentum conserving delta function in general dimension (we will specialize to $d=4$ in several parts of this thesis below, for the moment let us keep the generic dimension approach). 

The extraction of classical information in this formalism has two main  ingredients to be taken in mind: 1) A parameter that controls the classical expansion, and 2) The  choice of  suitable wave functions describing the multi-particle  initial state   of the system. For the former, it is natural to use $\hbar$ as the parameter that controls the classical expansion. It appears in two main places in the computations: First, in the  coupling constants, which by reintroducing $\hbar\neq1$,  are to be re-scaled via $g\to g/\sqrt{\hbar}$, and second, the wave numbers associated to massless momenta for the  force carriers, which are introduced  as $q=\hbar \bar{q}$. We will discuss in detail below how to extract the classical limit of \eqref{eq:Kmocds}, as well as the choice of suitable on-shell initial state, for a given observable. For the moment, the classical classical piece 
 $\langle  \mathcal{O} \rangle $\footnote{We use $\langle...\rangle$ to imply that the classical limit for the given observable is taken. } of the observable, can be formally defined as 
\begin{equation}\label{eq:KMOCfin}
    \langle \Delta \hat{\mathcal{O}} \rangle =  \lim_{\hbar\to 0}\hbar^{\beta_\mathcal{O}} \left[\,_{\rm{in}}\langle\Psi|[T, i \hat{\mathcal{O}}]|\Psi\rangle_{\rm{in}} +\, _{\rm{in}}\langle\Psi| T^\dagger[\hat{\mathcal{O}},T]|\Psi\rangle_{\rm{in}}\,\right]\,,
\end{equation}
here  $-\beta_{\mathcal{O}}$ is the power of the LO-piece in the $\hbar$-expansion of the quantities inside the square brackets,  which depends on the specific observable, as well as on the theory considered. Then, the factor of $\hbar^{\beta_\mathcal{O}}$ in this formula then ensures  $ \langle  \mathcal{O} \rangle\sim \hbar^0$, i.e. classical scaling. For instance,   for the radiated photon field, we have $\beta_{\mathcal{O}}=\frac{3}{2}$, whereas for the linear impulse we use $\beta_{\mathcal{O}}=1$.

In this thesis we are interested in two observables: 1) The conservative linear impulse $\langle \Delta\hat{ p }\rangle$ (a global observable, i.e. independent of the particles positions), acquired by classical compact objects  in a $2\to2$ scattering process in  Electrodynamics/Gravity. 2) The classical radiated electromagnetic/gravitational field $\langle \hat{A}^\mu \rangle /\langle \hat{h}^{\mu\nu}\rangle$ (a local "observable", i.e. dependent on the particles positions ) in a non-conservative $2\to3$ scattering process.  

\subsection{Linear impulse in   $2\to2$ scattering}
At the classical level, the linear impulse dictates  the total change
in the momentum of one of the particles after the scattering process. At the quantum level, the impulse corresponds to the difference  between the expected outgoing and the incoming momenta of such particle, as given by the  \ac{KMOC}  formula \eqref{eq:KMOCfin}. For this observable it is convenient to choose the initial state of the system   $ |\Psi\rangle_{\text{in}}$,   as follows
\begin{equation}\label{eq:state}
    |\Psi\rangle_{\text{in}} =\int \prod_i \big[\hat{d}^dp_i
   \hat{\delta}^{(+)}(p_i^2-m_i^2)\phi_i(p_i)e^{ib_i{\cdot}p_i/\hbar}\big] |p_1p_2\rangle
\end{equation}
where we have employed the  notation of the original reference \cite{Kosower:2018adc}, however, unlike for the original work, and to be more general,  we have move to a frame where both particles are displaced by the positions $b_i$, with respect to such reference frame. Then, the  difference $b_2-b_1=b$, corresponds then to the impact parameters, which is the  distance of closest approach  between the  particles during the scattering process. Notice  $|\Psi\rangle_{\text{in}}$  is built from  on-shell states, of positive energy, as dictated by $\hat{\delta}^{(+)}(p_i^2-m_i^2) = (2\pi) \delta(p_i^2-m_i^2)\Theta(p^0)$ , there $\Theta(x)$ is the heaviside step function.   $\phi_i(p_i)$, corresponds to relativistic wave functions associated to the incoming massive particles, whose classical limit shall result into the point particle description of the compact objects. We will come on this below. 

The system's initial states is assumed to be normalized to the unit $\,_{\text{in}}\langle\psi|\psi\rangle_{\text{in}} = 1$. From this, it follows the normalization condition for the wave functions, 
\begin{equation}
    \int \hat{d}\Phi_i(p_i) |\phi_i(p_i)|^2 =1\,.
\end{equation}
Here we have written the on-shell phase-space measure as   $d\Phi(p_i) =\hat{d}^d p_i \hat{\delta}^{+}(p_i^2-m_i^2) $.

The next task is to relate the observable \eqref{eq:KMOCfin} to the scattering amplitude  using \eqref{eq:amp_def}, together with the initial two-particle state \eqref{eq:state}. Notice in general the computation of an  observable will have the contribution of two terms, one which is  linear in the amplitude, whereas the second one is quadratic. This is in general true to all orders in perturbation theory, except for the leading order, where the latter is subleading. Let us for the moment  focus on the contribution linear in the amplitude. At $(n)$-order in perturbation theory it reads explicitly 
\begin{equation}\label{eq:imp1}
    I_{(1)}^{(n)\,\mu}=\int d\Phi(p_{1},p_{2})d\Phi(p_{1}',p_{2}')\phi_{1}(p_{1})\phi_{2}(p_{2})\phi_{1}^{*}(p_{1}')\phi_{2}^{*}(p_{2}')i(p_{1}^{'\mu}-p_{1}^{\mu})e^{i(p_1{\cdot}b_1+p_2{\cdot}b_2)}\bra{p_{1}'p_{2}'}T\ket{p_{1}p_{2}}^{(n)}
\end{equation}
Here we have used $\hat{p}_i\ket{p_i,p_j} = p_i\ket{p_i,p_j}$, and 
$d\Phi(p_{i},p_{j})= d\Phi(p_{i})d\Phi(p_{j})$. 
In addition, we have  labeled the conjugate states with primed variables as mentioned above.
Next we can replace $\bra{p_{1}'p_{2}'}T\ket{p_{1}p_{2}}$ in terms of the  scattering amplitude as given by \eqref{eq:amp_def}, this will introduce a $d$-fold delta function that will allow us to perform $d$-integrals in the previous formula. 
 Introducing the  momentum miss-match $q_i=p_i'-p_i$, and changing the  integration variables from $p_i'\to q_i$, and  further using the momentum conserving delta function to do perform the integration in $q_2$, followed by the relabel $q_1\to q$, \eqref{eq:imp1} becomes
\begin{equation}\label{eq:i11}
\begin{split}
     I_{(1)}^{(n)\,\mu}&=\int d\Phi(p_{1},p_{2})\hat{d}^{d}q\hat{\delta}(-2p_{1}{\cdot}q+q^{2})\hat{\delta}(2p_{2}{\cdot}q+q^{2})\Theta(p_{1}^{0}+q^{0})\Theta(p_{2}^{0}-q^{0})\\
     &\qquad\qquad\qquad
     \phi_{1}(p_{1})\phi_{1}^{*}(p_{1}-q)\phi_{2}(p_{2})\phi_{2}^{*}(p_{2}+q)
     iq^{\mu}e^{-iq{\cdot}b}\mathcal{A}^{(n)}(p_{1},p_{2}\to p_{1}-q,p_{2}+q)
\end{split}
\end{equation}

Before discussing how to take the  classical limit of this expression, let us analyze the analogous expression for the term quadratic in the amplitude, entering in the \ac{KMOC}  formula \eqref{eq:KMOCfin}. Since there are two factors of $T$ in this term, we need to introduce a complete set momentum eigenstates between the two factors of $T$ in such way we can extract a momentum eigenvalue when the momentum operator hits the momentum eigenstates. At $(n)$ order in perturbation this can be done as follows 
\begin{equation}
\begin{split}
    I_{(2)}^{(n)\,\mu}&=\sum_{X=0}^{n-1}\int\prod_{m=0}^{X}d\Phi(r_{m})\prod_{i=1}^{2}d\Phi(R_{i})d\Phi(p_{i})d\Phi(p_{i}')\phi_{i}(p_{i})\phi_{i}^{*}(p'_{i})e^{ip_{i}{\cdot}b_{i}}(R_{1}^{\mu}-p_{1}^{\mu})\\
    &\times \sum_{a=0}^{n-1-X}\bra{p_{1}',p_{2}'}T\ket{R_{1},R_{2},r_{X}}^{(a)}\bra{R_{1},R_{2},r_{X}}T^{\dagger}\ket{p_{1}p_{2}}^{(n-a-X-1)}\,,
\end{split}
\end{equation}
Here we have used $\hat{p_1}\ket{R_1,R_2,r_X} = R_1\ket{R_1,R_2,r_X}$, where  $r_X$ represent additional massless states propagating through the cut. They only appear at sub-sub-leading (two loops) order in perturbation theory. We now proceed in an analogous way to the linear in amplitude computation, that is. we need to replace the dependence of  the scattering amplitude via \eqref{eq:amp_def}. Defining the momentum mismatch $q_i=p_i'-p_i$, as well as the momentum transfer $w_i=R_i-p_i$,  allows us to change the integration variables $p_i'\to q_i$ and $R_i\to w_i$. In addition, we can use the momentum conserving delta function for each amplitude to perform the integration in $q_2$ and $w_2$, which followed by the relabeling $w_1\to w$ and $q_1\to q$ results into
\begin{equation}
     \begin{split}I_{(2)}^{(n)\,\mu} & =\sum_{X=0}^{n-1}\int\hat{d}^{d}w\hat{d}^{d}q\prod_{m=0}^{X}d\Phi(r_{m})\prod_{i=1}^{2}d\Phi(p_{i})\hat{\delta}(-2p_{1}{\cdot}q+q^{2})\hat{\delta}(2p_{2}{\cdot}q+q^{2})e^{-iq{\cdot}b}w^{\mu}\\
 & \hat{\delta}(-2p_{1}{\cdot}w+w^{2})\hat{\delta}(2p_{2}{\cdot}w+w^{2})\phi_{1}(p_{1})\phi_{1}^{*}(p_{1}-q)\phi_{2}(p_{2})\phi_{2}^{*}(p_{2}+q)\\
 & \times\Theta(p_{1}+w)\Theta(p_{2}-w)\Theta(p_{1}+q)\Theta(p_{2}-q)\\
 & \times\sum_{a_{1}=0}^{n-1-X}\mathcal{A}^{(a_{1})\,\mu}(p_{1},p_{2}{\rightarrow}p_{1}{-}w,p_{2}{+}w,r_{X})\,,\\
 & \times\mathcal{A}^{(n-a_{1}-X-1)\star}(p_{1}{-}w,p_{2}{+}w,r_{X}{\rightarrow}p_{1}-q,p_{2}+q)
\end{split}
\end{equation}
This is the analog to \eqref{eq:i11}. 
With these two contributions at hand, the quantum mechanical impulse particle 1 acquires during the scattering process, at  $(n)$-order in perturbation theory is simply given by the sum 
\begin{equation}\label{eq:qmimpulse}
    \Delta p_1^\mu = I_{(1)}^{(n)\,\mu}+I_{(2)}^{(n)\,\mu}
\end{equation}

\subsubsection{Classical limit}
We now proceed  to extract the classical piece in the QM-impulse \eqref{eq:qmimpulse}. This is done through a series of steps: 1) The factors of $\hbar$ are restored in the formulas through the rescaling of the coupling constant $g\to g/\sqrt{\hbar}$ and the massless momenta $q\to \bar{q}\hbar$ and $w\to \bar{w}\hbar$. 2) There are 3 length scales to consider in the problem. The first one is defined by the size of massive particles, given by the   Compton wavelength $\lambda_c=\hbar/m$ (which in the classical context traduces to the radius of the classical charge/Black hole given by $r_Q=e^2 Q^2/(4\pi m)$ / or $r_S=2GM$). The second scale corresponds to the spread of the relativistic wave function $l_s$, and third, the separation between the particles $b$. In the classical limit, the following approximation should hold $\lambda_c \ll l_s\ll b $, which holds true if $b$ scales as $b\to b/\hbar$. The first part of the inequality simply imposes the effective point particle description of the classical objects, the second on the other hand ensures a non-overlapping of the particles' wave functions (typical of the long range scattering in classical physics), finally, the approximation $\lambda_c\ll b$  in the classical context becomes
$e^2Q^2/4(\pi m)\ll b$ the Post-Lorentzian  (PL) approximation, which allows us to compute observables order by order in perturbation theory (In the gravitational context this is $2GM\ll b $, which corresponds to the \ac{PM} approximation  ). 3) In the case in which there is the emission of external radiation (as will be the case for the waveform emission), the massless momenta of the photon/graviton need to also be re-scaled analogously as  $k\to \bar{k}\hbar$. This is equivalent to ask for long wavelength radiation (which in the bounded system scenario allows to recover the source multipole expansion). 

After this considerations, the previous discussion is equivalent to approximate the wave functions $\phi_i(p_i+\hbar \bar{q})\approx \phi_i(p_i)$, followed by a Laurent-expansion of  all of the components of the  integrands, in powers of $\hbar$. At this stage, the explicit dependence of the wavefunction can be integrated out, leaving us with the classical observable

\begin{equation}\label{eq:impulse_final}
\begin{split}
    \langle \Delta p_1^{(n)\,\mu} \rangle &= \lim_{\hbar\to 0}\Big[ \int \hat{d}^dq\hat{\delta}(-2p_1{\cdots}q+q^2) \hat{\delta}(2p_2{\cdots}q+q^2) i q^\mu e^{-i q{\cdot}b}\mathcal{A}(p_1,p_2\to p_1-q,p_2+q)\\
  &+  \sum_{X=0}^{n-1}\int\hat{d}^{d}w\hat{d}^{d}q\prod_{m=0}^{X}d\Phi(r_{m})\hat{\delta}(-2p_{1}{\cdot}q+q^{2})\hat{\delta}(2p_{2}{\cdot}q+q^{2})e^{-iq{\cdot}b}w^{\mu}
  \hat{\delta}(-2p_{1}{\cdot}w+w^{2})\hat{\delta}(2p_{2}{\cdot}w+w^{2})\\
 & \times\sum_{a_{1}=0}^{n-1-X}\mathcal{A}^{(a_{1})\,\mu}(p_{1},p_{2}{\rightarrow}p_{1}{-}w,p_{2}{+}w,r_{X})
  \times\mathcal{A}^{(n-a_{1}-X-1)\star}(p_{1}{-}w,p_{2}{+}w,r_{X}{\rightarrow}p_{1}-q,p_{2}+q)\Big]
 \end{split}
\end{equation}

As mentioned, in general we will have to Laurent-expand in $\hbar$ both, the on-shell delta functions, as well as the reduced amplitudes. At leading order $(n=0)$, only the term linear in the amplitude contributes to the impulse, in that case we can drop the $q^2$ factor inside the delta functions, since no singular terms in $\hbar$ appear in the amplitude. However, at higher orders in the perturbative expansion, possible singular terms arise in the amplitude. This singular terms are expected to be  cancelled  between the linear and quadratic in amplitudes terms. In \cref{ch:soft_constraints} we will see and explicit example of this cancellation. There is however no formal proof this cancellation happens to any order in perturbation theory.  

We conclude then that the classical  linear impulse is controlled basically by the (n)-Loop 4-point amplitude $\mathcal{A}(p_1,p_2\to p_1-q,p_2+q)$, as well as the $4+X$-cut amplitudes from the iterated piece.

\subsection{The radiated field in  $2\to 3 $ scattering}
We now move to the analysis of the computation of the radiated photon/graviton field in a $2\to3$ scattering process. This will be  analog to the previous example, whit a few interesting features arising from the non-conservative dynamics. Here we will be interested in computing the expectation value of the photon/graviton field operator $\hat{A}^\mu(x)/\hat{h}^{\mu\nu}(x)$. This unlike the case for the impulse is a local observable, depending on the position $x$ at which the field is measured. In particular, we are interested in the asymptotic for of the radiated field at future null infinity\footnote{The radiative field is an observable as it is defined at null infinity where (small) spatial gauge transformations vanish. There could still be some residual gauge due to time integration of the source \eqref{eq:rad_field_full_n}. That is, in general two waveforms $A_1^\mu(R,T_R,\hat{n})$ and $A_2^\mu(R,T_R,\hat{n})$ can differ by a time independent constant  $A_1^\mu(R,T_R,\hat{n})-A_2^\mu(R,T_R,\hat{n})=C^\mu(R,\hat{n})$. Observables such as the field strength tensor,  the Newman-Penrose scalar, or the wave energy flux can be computed from time derivatives of the waveform, therefore  insensitive to $C^\mu(R,\hat{n})$. We see this explicitly in \cref{ch:bounded}.}, which scales as $1/R$, with $R=|\vec{x}|$. This scaling follows naturally from the mode integration of the field operator. In what follows  we focus on the electromagnetic case, but the results can be easily generalized to the gravitational case.

We want to compute the expectation value of the field operator, whose mode expansion is 
\begin{equation}\label{eq:radi_f}
    \hat{A}^\mu(x)=\text{Re}\sum_{\eta =\pm 1}\int d\Phi(k) \epsilon_\eta^\mu e^{-ik{\cdot}x}a_\eta^\dagger(k)   
\end{equation}
here the sum over $\eta$ is a sum over the photon  polarization, and $a_\eta^\dagger(k)$ are creation operators for photons of momentum $k$ and helicity $\eta$. The next step is to put this operator inside our favorite \ac{KMOC}  formula \eqref{eq:KMOCfin}. Since this is a $2\to3$ scattering process, we can reuse  \eqref{eq:state} as our two-particle initial state. Although no initial radiation is present in the initial state, $a_\eta^\dagger(k)$  creates a particle of momentum $k$ and helicity $\eta$, when acting on the conjugate states $\bra{p_1,p_2',k_\eta }$. 

We then have as usual two contributions to the radiated field. The first one linear in the amplitude, however, since there is the creation of such massless momenta state, the controlling amplitude in this case at $(n)$-Loop order is  the 5-point amplitude $\mathcal{A}(p_1,p_2\to p_1-q_1,p_2-q_2,k_\eta)$. Analogously, the quadratic in amplitude part will be controlled by the $5+X$-amplitudes as shown below. At this stage, the classical limit outlined in previous section can be implemented straightforwards. This in turn integrates out the dependence on the particles wave functions, leaving us with an expression analog to \eqref{eq:impulse_final}, inside the radiated photon phase space. That is, writing the radiated field in \eqref{eq:radi_f} as an effective source integrated over the massless photon phase space, 
\begin{equation}\label{eq:rad_field_full}
   \langle  \hat{A}^\mu(x)\rangle=\text{Re}\int d\Phi(k) e^{-ik{\cdot}x} \langle J^\mu(k)\rangle 
\end{equation}
where the angular brackets indicate the classical limit has been taken. 
At $(n)$-order in perturbation theory, we naturally identify the source as given by the sum of two terms as follows 
 
\begin{equation}\label{eq:radi_field}
 \langle J^{(n)\,\mu} (k)\rangle = \mathcal{R}^{(n)\,\mu}(k)+\mathcal{C}^{(n)\,\mu}(k),
\end{equation}
which   have the explicit recursive form 
\begin{equation}\label{eq:r-kernel}
    \mathcal{R}^{(n)\,\mu}(k)=i\lim_{\hbar\to0}\hbar^{\frac{3}{2}}\int\prod_{i=1}^{2}\hat{d}^{4}q_{i}\hat{\delta}(2p_{i}{\cdot}q_{i}-q_{i}^{2})e^{ib_{i}{\cdot}q_{i}}\hat{\delta}^{4}(q_{1}{+}q_{2}-k)\mathcal{A}^{(n)\,\mu}\left(p_{1},p_{2}{\rightarrow}p_{1}{-}q_{1},p_{2}{-}q_{2},k\right)\,,
\end{equation}
and 
\textcolor{black}{
\begin{equation}\label{eq:c-kernel}
\begin{split}\mathcal{C}^{(n)\,\mu}(k) & ={\color{black}}\lim_{\hbar\to0}\hbar^{\frac{3}{2}}\sum_{X=0}^{n-1}\int\prod_{m=0}^{X}d\Phi(r_{m})\prod_{i=1}^{2}\hat{d}^{4}w_{i}\hat{d}^{4}q_{i}\hat{\delta}(2p_{i}{\cdot}q_{i}{-}q_{i}^{2})\hat{\delta}(2p_{i}{\cdot}w_{i}{-}w_{i}^{2})e^{ib_{i}{\cdot}q_{i}}\\
 & \hspace{2cm}\,\,\,\,\,\,\,\,\,\,\,\,\times\hat{\delta}^{4}(w_{1}{+}w_{2}{+}r_{X}{-}k)\hat{\delta}^{4}(w_{1}{+}w_{2}{+}r_{X}{+}q_{1}{+}q_{2})\\
 & \hspace{2cm}\,\,\,\,\,\,\,\,\,\,\,\,\,\times\sum_{a_{1}=0}^{n-1-X}\mathcal{A}^{(a_{1})\,\mu}(p_{1},p_{2}{\rightarrow}p_{1}{-}w_{1},p_{2}{-}w_{2},r_{X},k)\\
 & \hspace{2cm}\,\,\,\,\,\,\,\,\,\,\,\,\,\times\mathcal{A}^{(n-a_{1}-X-1)\star}(p_{1}{-}w_{1},p_{2}{-}w_{2},r_{X}{\rightarrow}p_{1}-q_{1},p_{2}-q_{2})\,,
\end{split}
\end{equation}}
where the $\star$ in one  of the amplitude indicates complex conjugation.
We refer to the $\mathcal{C}^{(n)\,\mu}(k)$ term as the \textit{cut-box} contribution, to indicate that it is given by the cut of higher loop amplitudes.  In this expression, $r_X$ denotes the collection of  momenta $\{r_1,\cdots,r_X\}$ carried by additional particles propagating thorough the cut, whose momentum phase space integration has been explicitly indicated by $d\Phi(r_m)=\hat{d}^4r_m\hat{\delta}^{(+)}(r_m^2)$. For $n=0\,\,\rm{and}\,\,1$,  no additional photons propagate thought the cut, since they only appear   starting from  $\rm{N}^2$LO in the perturbative expansion (i.e.  two-loops). 

$\langle J^{(n)\,\mu} (k)\rangle $ can then  be interpreted as a classical source entering into the RHS of the field equations, and is computed directly from the scattering amplitudes. It is particularly  remarkable how the classical field is   controlled by  \textit{single} photon emission amplitudes, while the classical field should be composed from many photon. In \cite{Cristofoli:2021jas}, it was shown such amplitudes parametrize the high photon occupation number as expected for a classical field. An analogous expression  for the source in the gravitational case $\langle T^{(n)\,\mu\nu} (k)\rangle $ follows from the scattering amplitudes. The difference is in the double Lorentz index characterizing the graviton polarization tensor. 

In this thesis we will mostly be interested in the computation of the previous source in both, the electromagnetic and the gravitational case. We do not perform explicitly the photon/graviton phase space integration in \eqref{eq:rad_field_full} although a  simple proof can be found in the review \cite{Kosower:2022yvp}. Here we just mention the integration in $k$ can be made in an almost independent way from the amplitude. The result is then to just bring down a power of $R$ in the denominator which ratifies the radiative nature of the classical field. There is additional exponential factor from the retarded nature of the radiation. In $d=4$ one can show \eqref{eq:rad_field_full} becomes

\begin{equation}\label{eq:rad_field_full_n}
   \langle  \hat{A}^\mu(x)\rangle=\frac{1}{4\pi R}\text{Re}\int d\omega  e^{-i\omega T_R} \langle J^\mu(\omega,\hat{n})\rangle 
\end{equation}
where we have used $k=\omega(1,\hat{n})$, with  $\hat{n}=\vec{x}/R$, is the unit direction of emission of the radiation, and $T_R=t-R$ is the retarded time.

\section{A few worlds on the double copy}\label{sec:double-copy-preliminaires}
Let us now move to study some generalities of the so called double copy of scattering amplitudes. 
The program of the double copy originally started from  the observation by \ac{KLT} in \cite{Kawai:1985xq} that $n$-point tree-level closed string  scattering  amplitudes  can be computed from the sum of products of n-point open string  partial amplitudes, with coefficients
that depend on the kinematic variables. This program  has however  seen many incarnations, ranging from perturbative \ac{QFT}  realizations \cite{Bern:2008qj}, to the understanding the double copy structure of   non perturbative  solutions classical gravity \cite{Luna:2016hge,Monteiro:2013rya,Luna:2015paa,Cardoso:2016amd,Cardoso:2016ngt,Carrillo-Gonzalez:2017iyj,Luna:2018dpt,CarrilloGonzalez:2019gof,Arkani-Hamed:2019ymq} . The double copy  colloquially goes by the slogan \ac{GR}  = YM$^2$, which is the simple observation that amplitudes involving massless gravitons in \ac{GR}  can be directly obtained from products of amplitudes for the scattering of gluons in non abelian gauge theories. Currently,  we understand the double copy is much more general feature of \ac{QFT}  amplitudes \cite{Bern:2019prr,Cachazo:2014xea}, and is naturally realized in classical sectors as well. Indeed, in this thesis we will learn how to connect classical and quantum versions of the double copy, including spinning massive particles \cref{ch:double_copy}.   

In the remaining of this section we give a brief introduction to the computation of \ac{GR}  amplitudes from the double copy of their \ac{YM}  counterparts. For that, let us first recall the  color decomposition of \ac{YM}  amplitudes, it will be useful when studying the \ac{KLT}  formulation of the double copy below. 

\subsubsection{Color Decomposition}

n-gluon scattering amplitudes can be factorized into two pieces. The firs piece  contains the information of the color  structure, whereas the second one containing only kinematics information of the scattering process. This  factorization is known as \textit{color decomposition} of gauge theory amplitudes \cite{Cvitanovic:1980bu,Mangano:1987xk}. More precisely,  for $n$-external gluon legs, the tree level\footnote{Color decomposition can be generalized to higher loops, where 
there will be double and higher trace contributions to the color decomposition, see for instance \cite{Bern:1996je}.  
} $n$-point scattering  amplitude is written in terms of $(n-1)!$ single-trace color structures as follows:

\begin{equation}\label{eq:color_decomp}
    \mathcal{A}^{\text{tree}}(g_1,\cdots,g_n) = \sum_{\sigma\in S_{n-1}}\text{tr}(T^{a_1}T^{a_{\sigma_2}}\cdots T^{a_{\sigma_n}})A(1,\sigma_2,\cdots,\sigma_n)\,.
\end{equation}
The   sum here runs  over  non-cyclic permutations of the indices $\{1,2,\cdots ,n\}$, corresponding to the set of inequivalent traces, and $T^{a_n}$ are the gauge group generators in the adjoint representation.  We have use cyclic invariance  of the trace to fix one of the entries. $A(1,\sigma_2,\cdots,\sigma_n)$ are known as  \textit{partial amplitudes} or \textit{color ordered amplitudes }, and are gauge invariant \cite{Mangano:1990by} objects, depending only on the momenta and polarization vectors of the particles in the scattering process. They are computed from the Feynman diagrams that respect the order of the momentum labels (in other words, planar diagrams), using Feynman rules that respect such a  order \cite{Bern:1996je}.

Since there is only $(n-1)!$ independent color factors,  these partial amplitude basis is over completed. Indeed, partial amplitudes  satisfy  linearly constraints that allow us to reduce the number of independent elements to $(n-2)!$. These constraints are known as   Kleiss-Kuijf relations \cite{Kleiss:1988ne,DelDuca:1999rs}, the simplest of which is the  U($1$) decoupling identity 
\begin{equation}
    A(1,2,3,\cdots,n)  + A(1,3,2,\cdots,n)+\cdots +  A(1,3,\cdots,n,2)=0, 
\end{equation}
which follows from the $T^a\to1$ replacing of the generators in \eqref{eq:color_decomp}. 
See  \cite{Elvang:2013cua} for a discussion of the additional  relations. Kleiss-Kuijf relations are not the only constraints on the partial amplitudes, indeed, there are  additional relations  known as the \ac{BCJ}  relations \cite{Bern:2008qj}, which impose a  series of constraints that reduces the  number of independent partial amplitudes  to $(n-3)!$. 
Let us stress here the choice of partial amplitudes basis is not unique since we could have chosen any other pair of legs  in replacement of the reference legs $1,n$, in \eqref{eq:color_decomp}.

\subsection{\ac{KLT} representation of the double copy}
Now that we understand the concept of partial amplitudes, we are ready to present  a first form of the double copy of \ac{YM}  amplitudes, this is the \ac{KLT}  form of the double copy. It says  $n$-point  axio-dilaton-gravity scattering amplitudes can be obtained from the sum of two copies of $n$-point partial \ac{YM}  amplitudes. More precisely
\begin{equation}\label{eq:kltmassless_}
    A_n^{\text{GR}} = \sum_{\alpha\beta}K_{\alpha\beta}A^{\text{YM}}(1,\cdots,n)\bar{A}^{\text{YM}}(1,\cdots ,n)
\end{equation}
The sum over $\mbox{\ensuremath{\alpha}},\,\beta$ 
ranges over $(n-3)!$ orderings, corresponding to the number of independent partial amplitudes, and $K_{\alpha,\beta}$ is the
standard \ac{KLT}  kernel \cite{Kawai:1985xq,Bern:1998sv,BjerrumBohr:2010ta}. Let us emphasise  formula \eqref{eq:kltmassless_} is valid in general space-time dimensions. 
Notice in addition, as natural from string theory, a graviton state come accompanied by an antisymmetric tensor $B^{\mu\nu}$, and a scalar, the dilaton. Amplitudes computed using  \eqref{eq:kltmassless_} have therefore  these additional states in the spectrum. We will see a more detailed discussion of this fact in  \cref{sec:spin_dc}. 

Let us provide a simple example of how to use formula \eqref{eq:kltmassless_}. 
The simplest double copy amplitude is indeed given for the $n=3$ case. The partial amplitude for the scattering of 3-gluons, with momentum conservation $p_1-p_2+p_3=0$, and $p_i{\cdot}\epsilon_i=\epsilon_i^2=0$,  is simply given by 
\begin{equation}\label{eq:3-gluon}
    A^{\text{YM}}_3(1,2,3) = 2g(p_1{\cdot}\epsilon_3\epsilon_1{\cdot}\epsilon_2- p_1{\cdot}\epsilon_2 \epsilon_1{\cdot}\epsilon_3+p_3{\cdot}\epsilon_1 \epsilon_2{\cdot}\epsilon_3 )\,.
\end{equation}

Formula \eqref{eq:kltmassless_} allows us to compute the 3-graviton scattering amplitude in general dimensions, and is  given by the squaring  of this simple amplitude. The \ac{KLT}  kernel at three-points is simply $K_3 = \kappa/(4g^2)$. We can choose graviton polarization tensors to be given $\epsilon_i^{\mu\nu} = \epsilon_i^\mu\epsilon_i^\nu$. 

\begin{equation}
    A_3^{\text{GR}} = \kappa(p_1{\cdot}\epsilon_3\epsilon_1{\cdot}\epsilon_2- p_1{\cdot}\epsilon_2 \epsilon_1{\cdot}\epsilon_3+p_3{\cdot}\epsilon_1 \epsilon_2{\cdot}\epsilon_3 )^2\,.
\end{equation}
See \cref{ch:double_copy} for a discussion on how to obtain amplitudes for dilaton scattering.

As a further  example we can compute the 4-graviton scattering amplitude from the double copy of    the 4-gluon scattering amplitude. In this case there is also one independent partial \ac{YM}  amplitude, say $A^{\text{YM}}_4(1,2,3,4)$, which for momentum conservation $p_1+p_2=p_3+p_4$ is simply given by 
\begin{equation}\label{eq:ABhhB-1}
\begin{split}
A^{\text{YM}}_4(1,2,3,4)  =\frac{2g^2\epsilon_{1,\alpha}\epsilon_{2,\beta}}{p_{1}{\cdot}p_{3}\,p_{1}{\cdot}p_{4}}\big[p_{1}{\cdot}p_{3}F_{4}^{\mu\alpha}F_{3,\mu}^{\beta}{+}
  p_{1}{\cdot}p_{4}F_{3}^{\mu\alpha}F_{4,\mu}^{\beta}{+}F_{3}^{\alpha\beta}p_{1}{\cdot}F_{4}{\cdot}p_{2}{+}F_{4}^{\alpha\beta}p_{1}{\cdot}F_{3}{\cdot}p_{2}{+}p_{1}{\cdot}F_{3}{\cdot}F_{4}{\cdot}p_{1}\eta^{\alpha\beta}\big]\,.
\end{split}
\end{equation}
Here we have used $F_i^{\mu\nu} = 2p_i^{[\mu}\epsilon_i^{\nu]}$. In this case, the 4-graviton scattering amplitude in general dimension is simply 

\begin{equation}\label{eq:4-grav}
    A^{\text{GR}}_4= K_4 A^{\text{YM}}_4(1,2,3,4)^2\,,
\end{equation}
where the 4-point \ac{KLT}  kernel is simply $K_4 =\frac{ p_1{\cdot}p_3 p_1{\cdot}p_4}{8 g^4 p_3{\cdot}p_4}$.

\subsection{The \ac{BCJ}  representation of the double copy}
We have seen how the \ac{KLT}  formula \eqref{eq:kltmassless_} allows us to compute \ac{GR}  amplitudes in a straightforward manner.  The formula however becomes quite non-trivial to use when the number of external legs  become big, this because we will have, as seen, $(n-3)!$ independent partial amplitudes to compute. On the other hand, this formula is only valid for tree-level amplitudes.  In this subsection we introduce a different representation of the double copy that overcomes these problems. This is the \ac{BCJ} \cite{Bern:2008qj} double copy formulation, which is one of the main computational tools in the modern amplitudes program in gravity \cite{Bern:2019prr}. 

\subsubsection{The Color-Kinematic duality}

The \ac{BCJ}  form of the double copy was originated from the following observations: A given $n$-point  \ac{YM}  amplitude can always be  written in the following fashion 
\begin{equation}
    A_n^{\rm{YM}} =g^{n-2} \sum_{\Gamma} \frac{c_i n_i}{d_i}\,,
\end{equation}
where the sum run over trivalent graphs\footnote{ Contributions from any diagram which has  quartic or higher-point vertices can be introduced 
to these graphs  by multiplying and dividing by appropriate missing propagators},  $d_i$ are kinematic denominators contains physical poles, and are made of  ordinary scalar Feynman propagators,  $c_i$ encode the color structure and $n_i$ are kinematics numerators. For a given triplet $(i,j,k)$, the color factors satisfy the Jacobi identity 
\begin{equation}\label{eq:jacobi}
    c_i\pm c_j=\pm c_k\,,
\end{equation}
then the numerators can be arrange in such a way  they satisfy an analog kinematic relation 
\begin{equation}\label{eq:ckd}
    n_i\pm n_j=\pm n_k\,.
\end{equation}
This is relation is known as the color-kinematic duality. 

The \ac{BCJ}  proposal is then that gravitational amplitudes can be computed by replacing the color factors $c_i$ by a second copy of kinematics numerators $\tilde{n}_i$ as follows 
\begin{equation}\label{eq:dcbcj}
     A^{\rm{GR}} = \sum_{\Gamma} \frac{n_i \tilde{n}_i}{d_i}\,.
\end{equation}
The two gauge theories can in general be different, and only one of theme is required to satisfy the color-kinematic duality \eqref{eq:ckd} in order for the gravitational amplitude \eqref{eq:dcbcj} to be gauge invariant \cite{Bern:2010yg,Bern:2010ue}. 

Let us remark although originally this formulation was done in the massless \ac{YM}  sector, it has been  extended to include both massless and massive matter, including spin effects. We will revisit this formulation in \cref{ch:double_copy} in the context of spinning matter. Also, there is analogous formulation of the  \ac{BCJ}  double copy at higher orders in perturbation theory \cite{Bern:2019prr}.

\begin{figure}
\begin {center}
\includegraphics[width=12truecm]{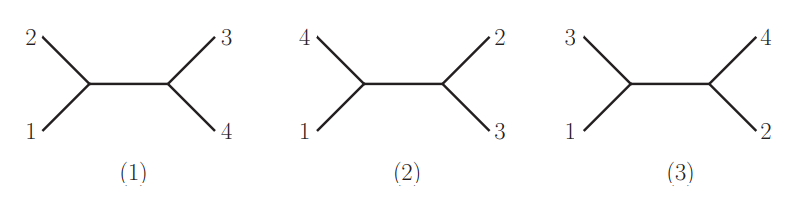} 
\end{center}
\caption{Feynman diagrams that contribute to the 4-gluon amplitude in the \ac{BCJ}  representation. Figure adapted from \cite{Bern:2022wqg}}
\label{fig:M4_gluon}
\end{figure}

Let us as an example recompute the 4-graviton scattering amplitude \eqref{eq:4-grav} using the \ac{BCJ}  double copy formula \eqref{eq:dcbcj}.
For this case, the \ac{YM}  amplitude has the contribution of 3 color structures, as associated to each of the graphs in Figure \ref{fig:M4_gluon}. 

The $s-$channel color factor is simply given by the contraction of the colour structure  constant associated to each 3-vertex
\begin{equation}
    c_s = f^{a_1 a_2 b}f^{b a_3 a_4}\,.
\end{equation}
The corresponding numerator is 
\begin{equation}
    n_s = \Big[ \epsilon_1{\cdot}\epsilon_2 p_1^\mu +2 \epsilon_1{\cdot}p_2\epsilon_2^\mu-(1\leftrightarrow2 ) \Big] \Big[\epsilon_3{\cdot}\epsilon_4 p_3^\mu +2 \epsilon_3{\cdot}p_4\epsilon_4^\mu-(3\leftrightarrow 4 ) \Big] + 
    s\Big[\epsilon_1{\cdot}\epsilon_3 \,  \epsilon_2{\cdot}\epsilon_4 - \epsilon_1{\cdot}\epsilon_4\, \epsilon_2{\cdot}\epsilon_3\Big]
\end{equation}
The additional numerators follow from index-relabeling as in Figure \eqref{fig:M4_gluon}. It is easy to show the color factors satisfy \eqref{eq:jacobi}, as it is just the usual Jacobi identity for the structure constants of the gauge group. Explicit computation also shows the numerators satisfy the analog relation \eqref{eq:ckd}, and can therefore be used in \eqref{eq:dcbcj} to compute the 4-graviton amplitude, which will agree with the \ac{KLT}  result \eqref{eq:4-grav}.

We have then two alternative constructions for the double copy, which we will explore further through the  body of  this thesis.


\section{The spinor-helicity formalism}\label{sec:spinor-helicity}
In this final section we introduce the spinor-helicity formalism, which is convenient to use  when dealing with observables in  $4$ spacetime dimensions\footnote{Extensions of this formalism to higher dimensions are also available, see for instance \cite{Cheung:2009dc,Jha:2018hag}.}. This formalism is based on the simple observation that  spin-1 vectors transforms as $(\frac{1}{2},\frac{1}{2})$ representations of the Lorentz group in $4$ spacetime dimensions and can therefore  be represented as a bi-spinors, where  each component acts  on its respective $\frac{1}{2}$ representation. Naturally,  particles momenta $p^\mu$  are Lorentz vectors, and can indeed  be represented in this spinorial form $p^{\alpha\dot{\alpha}}$. There is a distinction however between massive and massless momenta which we need to take into account. 

Recall under Wigner's classification \cite{Wigner1939},  particles correspond to irreducible, unitary representations of the Poincare group. In this sense, 
massless and massive particles are fundamentally different and need to be distinguished when written in their spinorial form. This is because they have associated different  little group. Remember the little group is defined as the  set of Lorentz transformations that leave  particles momenta invariant. Each particle has its own little group. 
For instance, for massless particles we can choose a frame in which the momentum vectors are of the form $p^\mu=\omega(1,0,0,1)$, and therefore the little group corresponds to the group of rotations in the x-y plane, SO$(2)=$U$(1)$. For massive particles on the other hand, one can choose   the particle's rest frame where $p^\mu = (m,0,0,0)$. This  allow us to identify  the little group as  the three-spatial rotations group SO$(3)\sim$SU$(2)$.

With this  distinction between massless and massive particles in mind, let use  introduce their correspondent spinor helicity formalism  in a separate way. For the former we follow the conventions of \cite{schwartz_2013} (see also \cite{Elvang:2013cua}), whereas for the latter we follow the seminal work in \cite{Arkani-Hamed:2017jhn}.

\subsection{Massless particles}
The transition from the vector to the spinorial representation of particle's momenta  is done thought the su($2$) sigma matrices $\sigma^\mu = (\mathbb{I},\sigma^i)$, via 
\begin{equation}\label{eq:mom}
   p^{\alpha\dot{\alpha}}=\sigma_{\mu}^{\alpha\dot{\alpha}}p^{\mu}=\left(\begin{array}{cc}
p^{0}-p^{3} & -p^{1}+ip^{2}\\
-p^{1}-ip^{2} & p^{0}+p^{3}
\end{array}\right)
\end{equation}
in this matrix notation, the    on-shell condition  becomes 
\begin{equation}
    p^2 = \det(p^{\alpha\dot{\alpha}}) = p_0^2 -p_1^2-p_2^2-p_3^2=0\,,
\end{equation}
The  two-dimensional momentum matrix $p^{\alpha\dot{\alpha}}$  has therefore rank 1 and can  be written as the outer product of two vectors 
\begin{equation}
    p^{\alpha\dot{\alpha}} = \lambda^\alpha \tilde{\lambda}^{\dot{\alpha}} \,.
\end{equation}
$\alpha$ and $\dot{\alpha}$ are SU($2$) indices, which can be  raised and lower with the invariant  $\epsilon$-tensor 
\begin{equation}
\epsilon^{\alpha\beta}=-\epsilon_{\alpha\beta}=\epsilon^{\dot{\alpha}\dot{\beta}}=-\epsilon_{\dot{\alpha}\dot{\beta}}=\left(\begin{array}{cc}
0 & 1\\
-1 & 0
\end{array}\right)\,.
\end{equation}

A possible parametrization of these spinors in terms of the momentum vector components  is 
\begin{equation}
\lambda^{\alpha}=\frac{z}{\sqrt{p^{0}-p^{3}}}\left(\begin{array}{c}
p^{0}-p^{3}\\
-p^{1}-ip^{2}
\end{array}\right)\,,\,\,\,\,\,\,\,\tilde{\lambda}^{\dot{\alpha}}=\frac{z^{-1}}{\sqrt{p^{0}-p^{3}}}\left(\begin{array}{cc}
p^{0}-p^{3} & -p^{1}+ip^{2}\end{array}\right)
\end{equation}
 with $p^0=\sqrt{p_1^2+p_2^2+p_3^2}$. Notice we have include a general scaling constant $z$. This is because   under the little group transformation, $\lambda\to z \lambda$ and $\tilde{\lambda}\to z^-1\lambda$, the matrix $p^{\alpha\dot{\alpha}}$ remains invariant. For real kinematics $\lambda^\alpha = (\tilde{\lambda}^{\dot{\alpha}})^\dagger$ and therefore the factor $z$ becomes a pure phase. $z=e^{i\phi}$. For Complex momenta, $\lambda$ and $\tilde{\lambda}$ are unrelated one to the other. 

We can analogously defined the conjugate matrix $\bar{p}_{\dot{\alpha}\alpha} = (\bar{\sigma}^\mu)_{\dot{\alpha}\alpha}p^\mu$, with $\bar{\sigma}^\mu =(\mathbb{I},-\sigma^i) $. It is furthermore convenient to introduce the bracket notation $\lambda_p^\alpha = \ket{p}$ and $\tilde{\lambda}_p^{\dot{\alpha}} = [p|$ to represent the spinor-helicity variables. In this way the momentum matrices become

\begin{equation}
    p^{\alpha\dot{\alpha}} = \ket{p} [p|\,,\qquad\bar{p}_{\dot{\alpha}\alpha} =|p]\bra{p}.  
\end{equation}
In this notation, the Lorentz product of two vectors, $p$ and $q$, becomes
\begin{equation}
    2p{\cdot}q =\text{tr}(\bar{p} q) =\text{tr}(\bar{q}p) = \text{tr}(|p]\bra{p}\ket{q}[q|) = \braket{pq}[qp]=\braket{qp}[pq]\,.
\end{equation}
Here we have denoted the contractions $\braket{pq} = \lambda_p^\alpha \lambda_q^\beta\epsilon_{\alpha\beta}$, and analogously $[pq] = \tilde{\lambda}_p^{\tilde{\alpha}}\tilde{\lambda}_q^{\dot{\beta}}\epsilon_{\dot{\alpha}\dot{\beta}}$. This  implies $\braket{pq}=-\braket{qp}$ and $[pq]=-[qp]$, and therefore, for any massless momentum $p$, we have $\braket{pp}=[pp] = 0$. 

Non surprisingly, massless polarization vectors (as well as Dirac spinors) can also be put into a spinorial form. Recall they satisfy the transversality condition $\epsilon{\cdot}p=0$. This  condition allows us to write positive and  negative helicity polarizations matrices in terms of the spinors for their associated massless momentum as follows
\begin{equation}\label{eq:polarizations}
    \epsilon_p^- = \sqrt{2} \frac{\ket{p}[r|}{[pr]}\,,\qquad \epsilon_p^+=\sqrt{2}\frac{\ket{r}[p|}{\braket{rp}}\,,
\end{equation}
here $\ket{r}$ and $[r|$ are two reference spinors (when computing amplitudes, they cancel out from the final answer), whose freedom to be chosen is the manifestation of the gauge freedom one has to shifting massless polarization vectors,  with vectors proportional to their correspondent momentum $\epsilon_\mu \to \epsilon_\mu+\frac{\sqrt{2}}{[pr]}p_\mu$, in a physical scattering amplitude. 
Notice  polarization vectors defined in \eqref{eq:polarizations} carry little group weights. That is, under a little group transformation, $\epsilon^-\to z^2 \epsilon^-$ and $\epsilon^+\to z^{-2}\epsilon^+$. 
Finally, polarization vectors satisfy the usual conditions $\epsilon_p^-{\cdot}\epsilon_p^+=-1$, and $\epsilon_p^{\pm}{\cdot}\epsilon_p^{\pm} =\epsilon_p^{\pm}{\cdot}p =0$. 

Any scattering amplitude in 4-dimensions can be written in terms of  inner products of spinor-helicity variables. Since reference spinors are little group invariant, the little group rescaling of an amplitude is fixed only by the external polarizations. This impose strong constrains on the permitted form of an amplitude, since arbitrary inner products of spinors must have the correct little group rescaling in order for the amplitude to describe the desired scattering process. Indeed, in \cite{Benincasa:2007xk} it was shown that for complex momenta\footnote{For real kinematics, 3-particle amplitudes vanish, see for discussion around \eqref{eq:np-3pt} },  on-shell 3-particle S-matrices
of massless particles of any spin can be uniquely determined from helicity arguments. This is where the power of the spinor helicity formalism overcomes the use of usual polarization tensors. That is, for a given particle of any spin, the spin structure is completely contained in the spinors $\{\ket{p}, |p]\}$. We will see this is also the case for massive spinning particles in \cref{sec:massive_spinor}. 

\subsubsection{Massless 3-point amplitude}
For the 3-point amplitude of massless particles with generic spin (or helicity $h=s$), and momentum conservation $p_1+p_2+p_3=0$, $p_i^2=0$, with $p_i{\cdot}p_j=0$,  Lorentz invariance impose the amplitude to be generic function of spinorial combinations $\braket{ij}$ and $[ij]$, with $i,j=1,2,3$. At 3-points, the most generic function is split into a holomorphic and anti-holomorphic contributions \cite{Benincasa:2007xk}:

\begin{equation}
    A_3(\{i,h_i\}) =\kappa_H \braket{12}^{d_3}\braket{23}^{d_1}\braket{31}^{d_2} +\kappa_A [12]^{-d_3}[23]^{-d_1}[31]^{-d_2} \,,
\end{equation}
where $\kappa_h,\kappa_A$ are constant coefficients and 
\begin{align}
    d_1 &=h_1-h_2-h_3\\
    d_2 &=h_2-h_3-h_1\\
    d_3 &= h_3-h_1-h_2
\end{align}
Imposing $ A_3(\{i,h_i\})$ has correct physical behaviour in the limit of real kinematics ($\braket{ij}=0$ and $[ij]=0$), implies that if 
$d_1+d_2+d_3>0$, one needs to set $\kappa_A=0$, and analogously if $d_1+d_2+d_3>0$ then $\kappa_H=0$. In this work we consider $d_1+d_2+d_3\neq0$ only, as for zero sum, the two contributions need to be kept. 

As an example, we can compute the 3-point amplitude  for a massless state of helicity $h$ emitting a massless particle of spin $h_3$:
\begin{equation}
A_{3}^{h_{3},h}\sim\left(\frac{\langle13\rangle}{\langle23\rangle}\right)^{2h}\left(\frac{\langle13\rangle\langle32\rangle}{\langle12\rangle}\right)^{h_{3}},\label{eq:massless3pths_n}
\end{equation}
We will come back to this amplitude in \cref{ch:double_copy} and \cref{ch:GW_scattering}.

\subsection{Massive particles}\label{sec:massive_spinor}
Let us now introduce the spinor helicity formalism for massive particles. We have learned that massless particles are labeled by their helicity weight, $h$. Massive particles on the other hand transform under some spin $S$ representation of SU$(2)$. The transition from vector to spinors notation can be done analogously to the massless case \eqref{eq:mom}. Since the on-shell condition for massive particles  is $p^2=m^2$,  the matrix $p^{\alpha\dot{\alpha}}$ is now of rank 2, instead of rank 1. This means, it can be written as the sum of two rank 1 matrices as follows

\begin{equation}
    p^{\alpha\dot{\alpha}} = \lambda^{\alpha\,I} \tilde{\lambda}_I^{\dot{\alpha}} \,,\qquad I=1,2,
\end{equation}
The massive on-shell condition now translates into 
\begin{equation}
    p^2=m^2\Rightarrow \det{\lambda}\times \det(\tilde{\lambda})=m^2\,.
\end{equation}

For massless particles, little group transformations were given by rescaling of the massless spinors. In the case of massive particles, the indices $I$, are SU($2$) indices (no to be confused with the spinorial $SL(2,\mathbb{C})$ indices $\alpha,\dot{\alpha}$), and the little group transformation correspond to $3$-dimensional rotations of these indices. The  transformation rules for the massive spinors are then $\lambda^{\alpha J}=W^{I}_J\lambda^{\alpha J}$, and $\tilde{\lambda}^{\dot{\alpha}}_I=(W^-1)_I^J \tilde{\lambda}_J^{\dot{\alpha}} $, with $W^{I}_J\in \text{su}(2)$. Of course $I$ indices can be raised and lowered with the $\epsilon^{IJ}\,\,, \epsilon_{IJ}$ invariant tensors of SU$(2)$.  Let us analogously to the massless case, introduce the bracket notation for massive spinors as follows $\lambda_p^{\alpha I} = \ket{p}^I$, and $\tilde{\lambda}^{\dot{\alpha}I}= [p|^I$, so that the massive momentum matrices now become
\begin{equation}
    p^{\alpha\dot{\alpha}} = \ket{p}^I[p|^J\epsilon_{IJ}\,.
\end{equation}
Angular and square brackets can be traded one to the other by means of the Dirac equation
\begin{equation}\label{eq:dirac}
    p^{\alpha\dot{\alpha}}|p]_{\dot{\alpha}}^I = m \ket{p}^{\alpha I}\,,\qquad \bar{p}_{\dot{\alpha    }\alpha}\ket{p}^{\alpha I} = m |p]_{\dot{\alpha}}^{I}\,.
\end{equation}

In addition to the helicity labels for massless particles, scattering amplitudes of massive spin-$S$ particles are given by totally symmetric tensors of rank $2S$. 
\begin{equation}
    M^{\{I_1\cdots I_{2S}\}} = \lambda_{\alpha_1}^{I_1}\cdots\lambda_{\alpha_{2S}}^{I_{2S}}M^{\{\alpha_1\cdots\alpha_{2S}\}}\,,
 \end{equation}
 where $M^{\{\alpha_1\cdots\alpha_{2S}\}}$ is totally symmetric in the $\alpha_i$ indices. Notice here we have chosen the angular as opposite to the square brackets basis to represent the scattering amplitudes. This is always possible since one can always convert from one basis to the other using \eqref{eq:dirac}. Let us also remark that massive spinor naturally recover their  massless counterparts  in the high energy ($m\to0$) limit. Let us not go into this discussion here but readers interested can see for instance \cite{Arkani-Hamed:2017jhn}.

\subsubsection{Minimal coupling massive 3-particle amplitude}
Let us analogously to the massless case, consider the 3-particle amplitude for a massive spin-$S$ state, emitting a helicity $h$ massless particle. We use the momentum conservation conventions $p_3 = p_1+k_2$.  This amplitude is completely fixed by the little group, and minimal coupling arguments \cite{Arkani-Hamed:2017jhn}\footnote{Minimal coupling in the sense of \cite{Arkani-Hamed:2017jhn}, is the statement that under the high energy limit, the 3-point amplitude \eqref{eq:3pt} reduces to the massless version \eqref{eq:massless3pths_n} }. 
Let us introduce the notation for the spin-s polarization states, using  totally-symmetric tensor products of spin $1/2$ spinors
\begin{align}
    |\varepsilon_1 \rangle =& \frac{1}{m^S} |1^{(a_1}\rangle \otimes \ldots \otimes |1^{a_{2S})}\rangle  \label{basi1} \,, \\
     |\varepsilon_1 ] =& \frac{1}{m^S} |1^{(a_1}] \otimes \ldots \otimes |1^{a_{2S})}] \,,
\end{align}
which are two different choices for a basis of $2S+1$ states. They can be mapped to each other using the operator \eqref{eq:dirac}. In this notation, the minimal coupling 3-point amplitude for the emission of a positive or negative helicity $h$ particle is  \cite{Arkani-Hamed:2017jhn}

\begin{equation}\label{eq:3pt}
    A_3^{+|h|,S} = (-1)^{2S+|h|}\frac{x^{|h|}}{m^S} \braket{\varepsilon_3\varepsilon_1}\,,\qquad 
    A_3^{-|h|,S}  = (-1)^{|h|}\frac{x^{-|h|}}{m^S}[\varepsilon_3\varepsilon_1]\,,
\end{equation}
where we have used the usual $x$-variable notation for massive spinors as follows:
\begin{equation}
    x = \frac{\bra{r}p_1|k_2]}{m\braket{r k_2}}\,,\qquad x^{-1} = \frac{[r|p_1\ket{k_2}}{m[r k_2]}\,,
\end{equation}
with $\ket{r}$ and $|r]$   reference spinors, associated to the massless particle polarization, as introduced  in \eqref{eq:polarizations}. In this sense, we can think of the $x$-variables as proportional to  the massless particles polarization tensors; or more precisely $x\sim \epsilon^+{\cdot}p_1 $ and $x^{-1}\sim \epsilon^-{\cdot}p_1$. 

Consider for instance the case $S=1/2$. $\ket{\epsilon_1} =\frac{1}{m^/2}\ket{1}$. In \cite{Guevara:2018wpp} (see also \cite{Aoude:2020onz}), it was shown that in terms of the spinorial realization of the spin $1/2$ Lorentz generators $J^{\mu\nu} = (\sigma^{\mu\nu}\otimes \mathbb{I}+\mathbb{I}\otimes \sigma^{\mu\nu}) $, where the angular momentum operator can be put in terms of the standard $SL(2,\mathbb{C})$ matrices, $\sigma^{\mu\nu} =  \sigma^{[\mu}\bar{\sigma}^{\nu]}/2$, the previous amplitudes can be written  in the following way (take for instance $|h|=2$):

\begin{equation}\label{eq:a3psin1half}
    A_3^{-2,S=1/2} \sim(\epsilon^-{\cdot}p_1)^2  \bra{3}\Big(1+ \frac{ k_{2\mu}\epsilon^-_\nu J^{\mu\nu}}{p_1{\cdot\epsilon^-}}\Big)\ket{1}\,.
\end{equation}
where one has to use 
\begin{equation}\label{eq:agmo}
 \frac{ k_{2\mu}\epsilon^-_\nu J^{\mu\nu}}{p_1{\cdot\epsilon^-}} = \frac{\ket{k_2}\bra{k_2}}{m x}\otimes\mathbb{I}+\mathbb{I}\otimes\frac{\ket{k_2}\bra{k_2}}{m x}\,.
\end{equation}
Analogously for the opposite polarization, we change angle to squared brackets, and do $k_2^\mu\to - k_2^\mu$. The trick to write the amplitudes \eqref{eq:3pt} in terms of the  spin operator is, for instance for the minus helicity one,  to change from the chiral (square brackets) to the inti-chiral basis (angular brackets) using the Dirac equation \eqref{eq:dirac}, and analogously for the other helicity.

The infinite spin generalization of \eqref{eq:a3psin1half} was also introduced in \cite{Guevara:2018wpp}. In this case, the spin-$j$ generalization of \eqref{eq:agmo} is
\begin{equation}
    \Big( \frac{ k_{2\mu}\epsilon^-_\nu J^{\mu\nu}}{p_1{\cdot\epsilon^-}}\Big)^{\odot j} = \begin{cases}
\frac{(2S)!}{(2S-j)!}\Big(\frac{\ket{k_2}\bra{k_2}}{mx}\Big)^{\otimes j}\odot\mathbb{I}^{\otimes2S-j}\,, & j\le2S\\
0\,, & j>2S
\end{cases}
\end{equation}
which leads immediately to an exponential representation of the 3-point amplitude 

\begin{equation}\label{eq:3pt}
    A^{-|h|,S}_{3}=A^{-|h|,S=0}_3 \times \langle \varepsilon_3| \exp\left(\frac{F_{2\mu \nu}J^{\mu\nu}}{2\epsilon^-_2\cdot p_1}\right) | \varepsilon_1 \rangle\,
\end{equation}
where we have used $F_2^{\mu\nu} = 2k_2^{[\mu}\epsilon_2^{-,\nu]}$, and $A^{-|h|,S=0}_3 = (\epsilon^-{\cdot}p_1)^{|h|}$. We will continue studding these 3-point amplitude in \cref{ch:GW_scattering} (see also \cref{ch:electromagnetism} and \cref{app_B}).  Naively we might think   this amplitude have unphysical poles for $S>h$, when one expands the exponential function.  In \cref{ch:GW_scattering} we will prove  this is not the case, and indeed, we will show how it's classical limit recovers the linearized effective metric for the Kerr BH, as originally shown in \cite{Guevara:2018wpp,Chung:2018kqs}.

\subsubsection{The Compton amplitude in spinor-helicity form}

Let us finalize this section by commenting on the spinor helicity form of the Gravitational Compton amplitude. As we will see in \cref{ch:electromagnetism}, this amplitude can be constructed from soft theorems, without the need of a Lagrangian. In \cite{Arkani-Hamed:2017jhn}, up to spin $S=2$,  it was also shown that it can be completely fixed using unitarity, and the 3-point amplitudes shown above, which are themselves fixed from little group, and minimal coupling arguments. In spinor helicity form, with momentum conservation $p_1+k_2=k_2+p_4$, and for incoming (outgoing) graviton helicity $+2$ ($-2$), the gravitational Compton amplitude reads

\begin{equation}\label{eq:s14d_}
A_{4}^{{\rm gr},s=2}\propto  \frac{\langle 2|1|3]^4}{p_1 \cdot k_2 \, p_1 \cdot k_3 \, k_2 \cdot k_3} \left([1^a 2]\langle 34^b\rangle + \langle 1^a 3 \rangle [4^b 2]\right)^4\,.
\end{equation}

Using arguments along the same lines above, the authors of \cite{Guevara:2018wpp} showed this amplitude can be written in terms of the spin generators in the form 

\begin{equation}\label{eq:compts}
 A^{+-,S}_{4}=A^{+-,S=0}_4 \times \langle \varepsilon_4 | \exp\left(\frac{F_{2,\mu \nu}J^{\mu\nu}}{2 \epsilon_2 \cdot p_1}\right) | \varepsilon_1 \rangle    
\end{equation}
where the scalar amplitude is simply
\begin{equation}\label{eq:scalar_compton}
    A^{+-,S=0}_4 =  \frac{\langle 2|1|3]^4}{p_1 \cdot k_2 \, p_1 \cdot k_3 \, k_2 \cdot k_3}\,.
\end{equation}
Unlike for the 3-point amplitude, \eqref{eq:compts} is valid only up to spin $S\le2$. For $S>2$, this amplitude has the unphysical pole $\epsilon_2{\cdot}p_1\sim \langle 2|1|3]$, which cancels from the scalar amplitude for lower spins. Up to $S=2$, this amplitude agrees with the Lagrangian derivation. We will comment on this in \cref{sec:new}. For the \ac{QCD} (single copy ) amplitude, the spinor-helicity amplitude
\begin{equation}\label{eq:s14qcd_}
A_{4}^{{\rm QCD},s=2}\propto  \frac{\langle 2|1|3]^2}{p_1 \cdot k_2 \, p_1 \cdot k_3} \left([1^a 2]\langle 34^b\rangle + \langle 1^a 3 \rangle [4^b 2]\right)^2\,.
\end{equation}
agrees with the $g=2$ form factor choice for $S=1/2,1$, but disagrees with the Lagrangian minimal coupling amplitude, where $g=1$ for $S=1$ Proca particles. In \cref{sec:new} we argue that  double copy criteria  fixes $g=2$ for \ac{QCD} for the named spin values. Furthermore, in \cref{ch:GW_scattering} we will show in the classical limit, amplitude \eqref{eq:compts} corresponds to an  effective description for the scattering of Gravitational waves off the Kerr BH in the low energy regime, traditionally studied using BHPT.

\section{Outlook of the chapter}\label{sec:outlook_preliminaries}
In this chapter we have introduced some modern amplitude techniques that will facilitate the  understanding for most of  the content of this thesis. These techniques are some of the cornerstones of the modern amplitudes program in classical gauge theories and gravity, and have shown remarkable simplifications at the moment of performing hard core computations. This chapter was intended as a short review but readers interested in a more pedagogical introduction can consult the reference cited through the chapter. 

%% file: Chapters/chapter_electromagnetic_radiation.tex
\chapter{Classical E\&M observables   from \ac{SQED}  amplitudes}\label{ch:electromagnetism}

\section{Introduction}
As  motivated in the Introduction and in \cref{sec:KMOC}, the main ingredients in the computation of two-body  classical observables  in gauge and gravity theories  are the  conservative 4-point ($M_4$) and radiative 5-point ($M_5$) scattering amplitudes   \eqref{eq:4_and_5_point}.  These  amplitudes have been  subject of exhaustive studies in the last decade,  including matter  with and without spin  in  gauge theories, as  well as  scalar and spinning sources in   gravitational scenarios.  Remarkable modern amplitudes techniques are used in the computation of these objects, aiming to simplify the  calculations and extract the relevant contributions needed for  classical physics at the earliest possible stage of the computation. Among some of these techniques we have  spinor helicity variables introduced in \cref{sec:spinor-helicity}, generalized unitarity , the double copy briefly introduced in \cref{sec:double-copy-preliminaires}, which we will expand in \cref{ch:double_copy},  as well the use of   integration techniques developed for the computation of \ac{QCD} cross sections, many of which  will be used in the body of this thesis. 
In this chapter we  provide a pedagogical introduction to the computation of these amplitudes in the simplest scenario, that is, for scalar particles minimally couple to the photon field, otherwise known as \ac{SQED}. This will allow us to introduce many of the ingredients needed in more complicated scenarios including spinning sources both in \ac{QED} (\ac{QCD}) and Gravity, while avoiding the complications introduced by the latter. We postpone the study of spin for both gauge and gravity theories  for  \cref{sec:spin_in_qed} . 

In the first part of this chapter we will concentrate on the computation of $M_4$ and $M_5$ at lowest orders in perturbation theory, that is, at 2\ac{PL} and 3\ac{PL} order respectively. We will show that as suggested by \eqref{cuts m4 m5}, these amplitudes can be obtained from  elementary building blocks given by the  3 and 4-point amplitude for one massive line emitting photons (gravitons) \footnote{ Let us stress here that factorization \eqref{cuts m4 m5} is in fact more general (including spinning particles) and holds for both gauge and gravity theories.}. We provide a Lagrangian derivation of these building blocks and give some of their simple  applications in classical physics: As first application we will discuss how no radiative content propagates to  future null infinity   from the 3-point amplitude. Secondly,  we show how  the  classical Thomson scattering of electromagnetic waves off  structure-less compact charge  objects can be obtained from the classical limit of the Compton amplitude. We further point out interesting properties of these building blocks as soft exponentiation and the orbit multipole decomposition. This then allows us to argue the same amplitudes can be constructed directly from soft theorems and Lorentz symmetry of the scattering matrix, without the need of a Lagrangian. 
Furthermore, we will check how the   soft exponentiation of the Compton amplitude induce an all order exponential soft decomposition of the classical 5-point amplitude. 
We proceed by illustrating  the computation of simple  two-body observables, including the   2\ac{PL} linear impulse, the 3\ac{PL} radiated photon field in a $2\to3$ scattering process in \ac{SQED} at leading order in the frequency of the radiated photon and show that it agrees with the well known Weinberg soft theorem \cite{PhysRev.140.B516}, whose universality  is a consequence of the spin  universality of the mentioned building blocks, which we study in more detail on  \cref{sec:spin_in_qed}. 

As advertised in previous sections,  one of the main subjects of this thesis is the computation of gravitational radiation from the classical limit of quantum scattering amplitudes. These gravitational  amplitudes can be computed  with the  help of the double copy, as we have stressed several times in previous sections. We have however introduced the  double copy in the context of Yang Mills theories with  the slogan $\text{GR}=\text{YM}^2$ in \cref{sec:double-copy-preliminaires}. 
The reason we chose to discuss electromagnetic radiation in this chapter as opposite to color radiation is that, as we will show in \cref{sec:spin_in_qed} and \cref{ch:double_copy}, the double copy of the  electromagnetic amplitudes discussed in this section will be  enough to compute the classical gravitational radiation in the two-body problem at \ac{LO} in perturbation theory, avoiding the complications arising from the non-Abelian nature of \ac{YM} theory. 
This is somehow a different approach to the one taken in the work of Goldberger and Ridgway in \cite{Goldberger:2016iau}, and Luna et al \cite{Luna:2017dtq}, where  the \ac{LO} gravitational 5-point amplitude  was computed from the \ac{BCJ} double copy of scalar-YM. Of course these two approaches  are equivalent as we will explicitly show in \cref{ch:double_copy}, with the reason behind this equivalence being the agreement of   $A_3$ and $A_4$ amplitudes, with  \ac{YM}  partial amplitude; we will expand on this in \cref{ch:double_copy}.

We finalize this chapter with the explicit computation of the 2\ac{PL} (1-loop) linear impulse for the scattering of two   structure-less, charged compact objects interacting through the exchange of electromagnetic waves, recovering the classical  results of Saketh et al \cite{Saketh:2021sri}.    For this we make use of the integral representation of the linear impulse derived in \cite{Kosower:2018adc}, and use integration technique introduced in \cref{sec:examplesa_tree_level} below.

This chapter takes elements of previous  work by the author  \cite{Bautista:2019tdr,Bautista:2021wfy,Bautista:2021llr}, and for completeness, the discussion in \cref{sec:radiantion3pt} is done along the lines of \cite{Guevara:2020xjx}.

\section{Scalar Electrodynamics}\label{sec:scalarqedlag}
As a warm up, let us study the  simple theory describing the minimal coupling  between  a charge scalar complex field and  the photon field. This will avoid all of the complication arising from spin, while capture many interesting   features of radiation, also present  for spinning bodies. 
For one matter line, the interaction is described by the scalar-QED Lagrangian 
\begin{equation}\label{eq:sqed_lagrangian}
    \mathcal{L}_{\text{SQED}} = -\frac{1}{4}F_{\mu\nu}F^{\mu\nu} +D_\mu\phi D^\mu\phi^* - m^2 |\phi|^2,
\end{equation}
where we have introduced the position space photon field strength tensor $F_{\mu\nu} = \partial_\mu A_\nu -\partial_\nu A_\mu $, whereas the covariant derivative is  $D_\mu=\partial_\mu + i Q e A_\mu$, with $Q$ the charge of the scalar field, and $e$ is the electron charge. It is a straightforward task to derive the Feynman rules from this Lagrangian. For instance,  the 3-point vertex and the seagull vertex are given  respectively by
\begin{eqnarray}\label{eq:3vertexSQED}
     \vcenter{\hbox{\includegraphics[width=20mm,height=18mm]{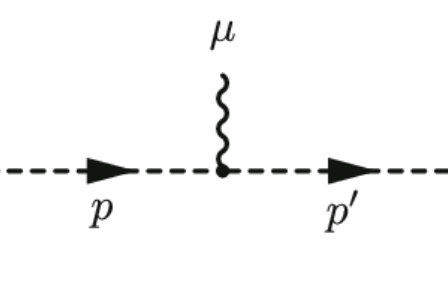}}} &= -i e Q(p+p')^\mu \\
     \vcenter{\hbox{\includegraphics[width=20mm,height=18mm]{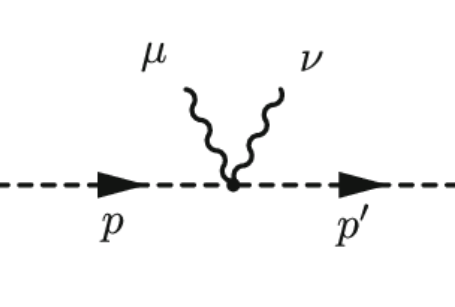}}} & = 2 e^2 Q^2 \eta^{\mu\nu}\label{eq:seagullSQED}
\end{eqnarray}

The simplest scattering amplitude one can compute with this Feynman rules is  the reduced 3-point amplitude $A_3$, for a massive scalar emitting a photon
\begin{equation}\label{eq:a3sqed}
    A_{3}^{\text{SQED}} = i 2 e\, Q\,  \epsilon^{\pm}{\cdot}p_1 \,,
\end{equation}
where the  momentum conservation condition reads $p_2= p_1-q$, and $\epsilon^\pm $ corresponds to the polarization vector for the emitted photon. 

We can also consider the  amplitude for the $2\to2$ scattering of our scalar particle with a photon. That is, the scalar Compton amplitude $A_4^{\text{SQED}}$, which for momentum conservation $p_1+k_2=k_3+p_4$, is simply given by 
\begin{equation}\label{eq:sqed_comtpon}
    A_4^{\text{SQED}} = 2 e^2 Q^2 \frac{ p_1{\cdot}F_2{\cdot}F_3^\star{\cdot}p_1}{p_1{\cdot}\,k_2 p_1{\cdot}k_3}\,,
\end{equation}
where, with  some abuse of notation, we have introduced the  momentum-space photon field strength $F_i^{\mu\nu} = 2 k_i^{[\mu}\epsilon_i^{\nu]} $. 

These will be the main building blocks in the computation of classical electromagnetic (gravitational) two-body observables, as we will shortly see. Before going into that, let us first  comment on two direct applications of this amplitudes in the computation of classical radiation. As  first example we will show how although the 3-point amplitude contains an external photons, it does not carry any radiative degrees of freedom (DoF) in Minkowski space-time. The second example will be the direct use of the classical limit of the Compton amplitude to describe the Thomson scattering process in classical electrodynamics.

\subsection{Radiation scalar and the 3-pt amplitude}\label{sec:radiantion3pt}

The radiative content in a classical scattering process is encoded in the so called Radiation Newman-Penrose scalars \cite{Newman1962}.  These correspond to  solutions of the classical field equations, which decay as $1/R$ at past and future null  infinity, with $R$  the distance from the position in which the scattering process took place, and   the position of the  detector. From a classical perspective, the momentum space scattering amplitude with external massless particles can be interpreted as the source entering into the right hand side of the field equations, and therefore are directly related to the radiation scalars. For instance, at the level of the 3-point amplitude,  the photon emission is capture by the Maxwell spinor \cite{Monteiro:2020plf}
\begin{equation}\label{eq:np-3pt}
    \phi(x) = \frac{\sqrt{2}}{m} \text{Re} \int \hat{d}\Phi \hat{\delta}(u{\cdot}q) \ket{q}\bra{q}e^{-q{\cdot}x}A_3^{\text{SQED}}\,.
\end{equation}
 Here we have introduced  the massive particle four-velocity $u^\mu = \frac{1}{m}p^\mu$, whereas  $\ket{q}\bra{q} = \sigma^\mu q_\mu$. 
We have also used  $\hat{\delta}(u{\cdot}q)$ to represent the  on-shell condition for the outgoing massive particle (Notice we have use the on-shell condition  $q^2=0$ for the emitted photon).  
  This integral is straightforward to evaluate in  the rest frame for the massive particle, where $u^\mu = (1,0,0,0)$ (or $x=(\tau,0,0,0)$), where the on-shell condition    $\hat{\delta}(u{\cdot}q)$ for the outgoing massive  particle becomes the zero energy condition $\omega = 0$, for the emitted photon,  which in Minkowski space time, $q^\mu = \omega (1,0,0,1)$  is solved for $q^\mu = 0$ identically. We conclude then that $\phi(x)=0$  and therefore 3-point amplitude contains no radiative modes in $(1,3)$ signature. This is an statement that holds to all orders in perturbation theory, and for generic massless emission at $3$-points. 
Interestingly, in split signature $(2,2)$, $\phi(x)$ is a non-vanishing object, containing radiative modes as shown in \cite{Guevara:2020xjx}, whose interesting properties are beyond the scope of  the present thesis. 

This is the reason a charged massive  particle cannot emit radiative \ac{DoF} towards future null infinity unless it is disturbed by an additional entity, for instance, an additional charged particle, or an electromagnetic wave. Let us however remark that although the amplitude \eqref{eq:a3sqed} does not provide radiative \ac{DoF}, it will be an important building block in the constriction of higher point amplitudes, which do carry radiate \ac{DoF}.  As a final remark, let us stress the Maxwell spinor \eqref{eq:np-3pt} can also be determined using the methods described earlier in \cref{sec:KMOC}.

\subsection{The classical electromagnetic  Compton amplitude and  Thomson scattering}

\begin{figure}
\begin {center}
\includegraphics[width=5truecm]{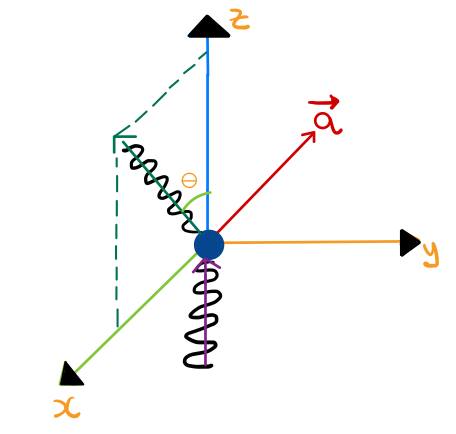} 
\end{center}
\caption{ Wave scattering in an Amplitudes setup. A incoming plane wave  traveling along the $z$-axis, hits a classical compact object at rest. The wave gets scattered with outgoing  momentum lying in the $x-z$ plane.
We have introduced a spin vector oriented in a generic direction,  preparing for the process of   gravitational wave scattering off Kerr BH,  treated in \cref{ch:GW_scattering}. 
}
\label{fig:ampl}
\end{figure}

Let us now proceed with a first direct application of our scalar Compton amplitude \eqref{eq:sqed_comtpon} in the low energy description of the scattering of light off a  charged particles in classical electrodynamics.  This is known as the Thomson process where the incoming wave hits the charge making it accelerate and therefore emitting a wave  with the same frequency of the incoming wave (see Figure \ref{fig:ampl} for our conventions used in \eqref{eq:kinematics}). The observable for  this process is the classical differential cross section which can be obtained from the classical (here equivalent to the low energy limit) limit of the Compton amplitude.

 In what follows we introduce some general notation that will be used not only for the scattering of electromagnetic waves in \ac{QED}, but also will be used in the context of the scattering of waves of general helicity $h$ off scalar and rotating Black holes, which will be studied in great detail in  \cref{ch:GW_scattering}. 

In order to define the classical piece of the \ac{QFT} $A_4$  amplitude and link it to a  wave scattering process, we proceed as follows. The null momenta of the massless particles, $k_i$, is to be identified with the classical wavenumbers, $\hat{k}_i$, as  dictate by the \ac{KMOC}  formalism  \cref{sec:KMOC}, and  corresponds  to the direction of wave propagation. Thus, this allows us to write
\begin{equation}\label{hscaling}
    k_i = \hbar \hat{k}_i  \,, [\hat{k_i}]=[1/L]\,,
\end{equation}
as $\hbar\to 0$. This scaling will be sufficient for  \ac{QFT}  amplitudes involving a single matter line (see also \cite{Kosower:2018adc,Bautista:2019evw,Bautista:2019tdr}). For such case, this also implies that internal massless momentum $q=\sum_i \pm k_i$ has the same $\hbar$-scaling, $q=\hbar \hat{q}$.

Consider now the Compton amplitude \eqref{eq:sqed_comtpon},  representing the  four-point scattering amplitude of two massive scalar legs of momenta $p_{1}$ and $p_{4}$ and two massless legs
of momenta $k_{2}$ and $k_{3}$. In the classical interpretation, the massive momenta will be associated to initial and final states of classical charged compact objects (or BHs in the gravitational case),  whereas the massless momenta $k_2$ and $k_3$ represent the incident and scattered wave respectively. The classical limit of \eqref{eq:sqed_comtpon} is the achieved by taking the  leading order term  in the $\hbar\rightarrow 0$ expansion of the amplitude,
\begin{equation}
\langle A_{4}\rangle :=\lim_{\hbar\rightarrow0}A_{4}\,.\label{eq:classical amplitude}
\end{equation}
We choose to evaluate our classical amplitude  in the reference frame for which the massive particle  is
initially at rest and the scattering process is restricted to the $x-z$ plane. By adopting the scaling given in \eqref{hscaling} and the rest frame for $p_1$, the momenta of the particles read explicitly  (see  Figure \ref{fig:ampl})
\begin{equation}
\label{eq:kinematics}
    \begin{split}
    p_{1}^{\mu} & =(M,0,0,0),\\
k_{2}^{\mu} & =\hbar \omega(1,0,0,1),\\
k_{3}^{\mu} & =\frac{\hbar \omega(1,\sin\theta,0,\cos\theta)}{1+2\frac{\hbar \omega}{M}\sin^{2}(\theta/2)},\\
p_{4}^{\mu} & =p_{1}^{\mu}+k_{2}^{\mu}-k_{3}^{\mu},  
    \end{split}
\end{equation}
 with the form of the energy for the outgoing wave of momentum $k_3$,  fixed by the on-shell condition for the outgoing massive particle. Here $\theta$ corresponds to the scattering angle.  The independent kinematic invariants are
\begin{equation}\label{eq:sandtclassical}
    \begin{split}
     s & =(p_1+k_2)^2=M^{2}\left(1+2\frac{\hbar \omega}{M}\right)\,,\\
t & =(k_3-k_2)^2=-\frac{4\hbar^2 \omega^{2}\sin^{2}\left(\theta/2\right)}{1+2\frac{\hbar \omega}{M}\sin^{2}(\theta/2)}\,, 
    \end{split}
\end{equation}
which for the case of electromagnetic scattering only receive contribution from the $s-$channel, from  the identity 
\begin{equation}\label{eq:ident}
s-M^2\approx M^2-u +\mathcal{O}(\hbar)    \,,
\end{equation}
hiding in  the classical limit, the latter can then be taken as the limit in which $\hbar \omega/M <<1$. This is equivalent to a multi-soft limit, for which the momenta of the incoming and outgoing photon  are much smaller than the mass of the  scalar particle.
 It will also be convenient to introduce the optical parameter \cite{Bautista:2021wfy}:
\begin{equation}\label{eq:xidef}
    \xi^{-1}:=-\frac{M^2 t}{(s-M^2)(u-M^2)}= \sin^2(\theta/2) \,.
\end{equation}
 We will make use of this  parametrization for particles momenta in \cref{ch:GW_scattering} when discussing the scattering of waves off rotating BHs. 

The next task is to relate the classical amplitude to the classical observable, in this case, the differential cross section. 
For that we use the well known formula for the differential cross section
in QFT, and then proceed to take its classical limit according to our prescription. Let us assume  that the incoming massless particles have fixed helicity $h$, whereas the outgoing massless particle can in general have a different helicity $h^\prime = \pm h$.
Then, the unpolarized differential cross section will be given by 
\begin{equation}\label{eq:crsec}
d\sigma=\sum_{h^\prime}\frac{|A_{4}(h\rightarrow h^\prime)|^{2}d{\rm LIPS}_{2}}{2E_{1}2E_{2}|\vec{v}_{1}-\vec{v}_{2}|}\,,
\end{equation}
where the sum runs over all the polarization states for the outgoing  massless particle, and the two particle Lorentz invariant phase space has the simple
form
\begin{equation}
    d{\rm LIPS}_{2}=\frac{s-M^{2}}{32\pi^{2}s}d\Omega\,.
\end{equation}
Noting that
in the classical limit $k_{3}^{\mu}\rightarrow  \omega(1,\sin\theta,0,\cos\theta)$, the differential cross section simply
becomes
\begin{equation}
\frac{d\langle\sigma\rangle}{d\Omega}=\sum_{h^\prime} \frac{|\langle A_{4}(h\rightarrow  h^\prime)\rangle|^{2}}{64\pi^{2}M^{2}}\,.\label{eq:classical cross section}
\end{equation}
The  impinging wave can also be unpolarized. In such case, the helicity states for both the incoming  and the  outgoing waves allow us to define the elements of the scattering matrix as follows 
\begin{equation}\label{eq:scattering-matrix-helicities}
A_{4}^{h}=\left[\begin{array}{cc}
A_{++}^{h} & A_{+-}^{h}\\
A_{-+}^{h} & A_{--}^{h}
\end{array}\right],
\end{equation}
where the sub-indices denote the polarization  of the incoming and outgoing wave respectively, and $h$ denotes the nature of the wave. We associate $+h$ ($-h$) states with circular left (right) wave polarizations. Motivated by the discussion of  wave spin induce polarization in  the next sections, specially in the context of spinning black holes in \cref{ch:GW_scattering}, we will refer to the diagonal elements $A^h_{++},A^h_{--}$ as \textit{helicity preserving} amplitudes, and to the off diagonals $A^h_{+-},A^h_{-+}$ as \textit{helicity reversing}. An important caveat here is that the helicity of particle $k_3$ appears flipped with respect to somewhat standard conventions: As $k_3$ is outgoing with helicity $h'$ it is equivalent to an incoming particle with helicity $-h'$.  

We are now in good position to evaluate the classical amplitude \eqref{eq:classical amplitude}, given by the classical limit of the Compton amplitude \eqref{eq:sqed_comtpon}. The corresponding polarization directions are 
\begin{equation}\label{eq:photon polarization1}
    \begin{split}
        \epsilon_{3}^{+}= & m=\frac{1}{\sqrt{2}}(0,\cos\theta, i,-\sin\theta)\, ,\\
\epsilon_{3}^{-}  =&
-\bar{m}=- \frac{1}{\sqrt{2}}(0,\cos\theta,- i,-\sin\theta)\,. 
    \end{split}
\end{equation}
Analogously for the  incoming wave
\begin{equation}\label{eq:photon polarization2}
    \begin{split}
        \epsilon_{2}^{+}= & \frac{1}{\sqrt{2}}(0,1, i,0) \,,\\
\epsilon_{2}^{-} =& - \frac{1}{\sqrt{2}}(0,1,- i,0)\,. 
    \end{split}
\end{equation}
Using previous prescription for the kinematics of the problem, one can easily show that the elements of the scattering matrix  \eqref{eq:scattering-matrix-helicities} for the scattering of a electromagnetic wave off a scalar charged massive particle read 
\begin{equation}
\langle A_{4,++}^{\text{SQED}}\rangle=\langle A_{4,--}^{\text{SQED}}\rangle=2\,e^2 \cos^{2}\left(\frac{\theta}{2}\right)\,,\label{eq:m+-sqed}
\end{equation}
whereas for the off-diagonal elements we have 
\begin{equation}
\langle A_{4,+-}^{h=2}\rangle=\langle A_{4,-+}^{h=2}\rangle=2\,e^2 \sin^{2}\left(\frac{\theta}{2}\right)\,.\label{eq:m4++sqed}
\end{equation}
One can immediately  obtain the unpolarized classical differential cross section 
\begin{equation}\label{eq:Thomson}
\frac{d\langle\sigma^{\rm{SQED}}\rangle }{d\Omega}= \left(\frac{e^2}{4\pi M}\right)^2\left[\cos^4\left(\frac{\theta}{2}\right)+\sin^4\left(\frac{\theta}{2}\right)\right]    
\end{equation}
which recovers the well known unpolarized differential cross section for the Thomson scattering \cite{Jackson:100964}.  A similar scattering amplitude approach  was taken in \cite{Cristofoli:2021vyo} reproducing the Thomson result analogously. 
Notice this differential cross section does  not diverges in the $\theta\to0$ limit, and is a consequence of  the form of the classical amplitude, which reduce to a contact term of the form $\langle A_4^{SQED}\rangle\sim 2\, e^2 \epsilon_2{\cdot} \epsilon^{\star}_3$. This is due to the fact classical electrodynamics is a Linear theory, unlike the case for  gravitational, where the non linearity nature allows to write non-contact diagrams with poles of the form $\frac{1}{\sin^2(\theta/2})$, these are basically t-channel poles, as we will see in detail in \cref{ch:GW_scattering}. 
As final observation, the cross section for the Thomson process 
\begin{equation}
    \sigma = \frac{8\pi}{3}r_Q^2\,,
\end{equation}
where $r_Q$ is the classical radius of the charged particle (see Table \ref{tab:grvsqed}), is independent of the energy of the incoming wave and only depends on the radius of the charge particle.  
\subsection{Soft exponentiation and orbit multipoles}\label{sec:soft_exponential_ph}

Another way one can  understand   why $A_3$ does not carries radiative \ac{DoF}  is through the orbit multipole moments. As we have seen above,  $A_{3}$ corresponds to a classical on-shell current entering into the r.h.s. of the classical field equations, and although it can be  used to evaluate conservative effects in the two-body problem,   it is not enough for the
computation of radiative effects \cite{Shen:2018ebu,Goldberger:2017ogt}.
This can be understood from the fact that it does not possess orbit multipoles, in contrast with $A_{4}$. We define the  orbit multipoles as each
of the terms appearing in the soft-expansion of $A_n$ for $n=3,4$, with respect to an external
photon (or graviton as we will illustrate in \cref{ch:double_copy})\footnote{The soft expansion is the  analog to the multipole expansion of a classical source \cite{Jackson:100964}, and therefore the name orbit multipoles.}. 
Such expansion is trivial for $A_{3}$ as seen from \eqref{eq:a3sqed}. 
For $A_{4}$, however, it
truncates at subleading order for photons \cite{PhysRev.96.1428,PhysRev.96.1433}. 
As a consequence, both amplitudes can be directly
constructed via Soft Theorems without the need for a Lagrangian.
The only seed is the three point  amplitude  \eqref{eq:a3sqed}
which is can be  fixed up to a constant using 3-point. kinematics arguments as we illustrated in \cref{sec:spinor-helicity}. Let us then write the soft expansion of $A_4$
with respect to $k_{3}\rightarrow0$ as
\begin{equation}
A_{4}^{{\rm ph}}=e\,Q\,\sum_{a{=}1,4}\frac{\epsilon_{2}{\cdot}p_{a}}{k_{3}{\cdot}p_{a}}e^{\frac{2F_{3}{\cdot} J_{a}}{\epsilon_{3}{\cdot} p_{a}}}A_{3}^{{\rm ph}} = 2e^2\,Q^2\,\left[\frac{p_{1}{\cdot}\epsilon_{2}F_{k}}{p_{1}{\cdot}k_{3}\,p_{4}{\cdot}k_{3}}{-}\frac{F_{\epsilon}}{p_{1}{\cdot}k_{3}}
\right]\label{eq:scphcompton},
\end{equation}
where $F_{3}{\cdot} J_{a}=F_{3}^{\mu\nu}J_{a\mu\nu}$, is the action
of the angular momentum operator $J_a^{\mu\nu}=  [p_a\wedge \partial_{p_a}]^{\mu\nu} $, on its corresponding massive particle  \cite{Cachazo:2014fwa}. We have also introduced the variables 
$F_{k}=p_{1}{\cdot}F_{3}{\cdot}k_{2}$, $F_{\epsilon}=p_{1}{\cdot}F_{3}{\cdot}\epsilon_{2}$.
This exponential representation of the four point amplitude will be use when we discuss radiation in the two-body problem, where the exponential expansion of $A_4$ induces and all order soft exponentiation of the 5-point amplitude  \eqref{cuts m4 m5}, for both \ac{QED} and Gravity. 

\section{1\ac{PL} linear impulse and 3\ac{PL} photon radiation in SQED}\label{sec:examplesa_tree_level}

With the previus building blocks at hand, we are now in position to compute simple classical observables in the scalar two-body problem in classical electrodynamics, for structure-less charged compact objects. In this section we will illustrate the computation of two main observable, the leading order linear impulse\eqref{eq:impulse_final}, and the radiated photon field in \eqref{eq:radi_field}, as provided by the \ac{KMOC}  formalism. The former was originally computed in \cite{Kosower:2018adc}, and we include it here for completeness, whereas the latter was computed by the author in \cite{Bautista:2019tdr,Bautista:2021llr} at leading order in the soft expansion. 

\subsubsection*{Leading order electromagnetic  impulse}

Let us start with the computation of the linear impulse at 1\ac{PL} order. For that we need the classical limit of the  amplitude $M_4^{\text{SQED}}$,  for the scattering of two massive scalar interchanging one photon. The quantum amplitude  can be easily computed using the Feynman rules \eqref{eq:3vertexSQED} and \eqref{eq:seagullSQED}, together with the photon propagator 
\begin{equation}
   \vcenter{\hbox{\includegraphics[width=22mm,height=18mm]{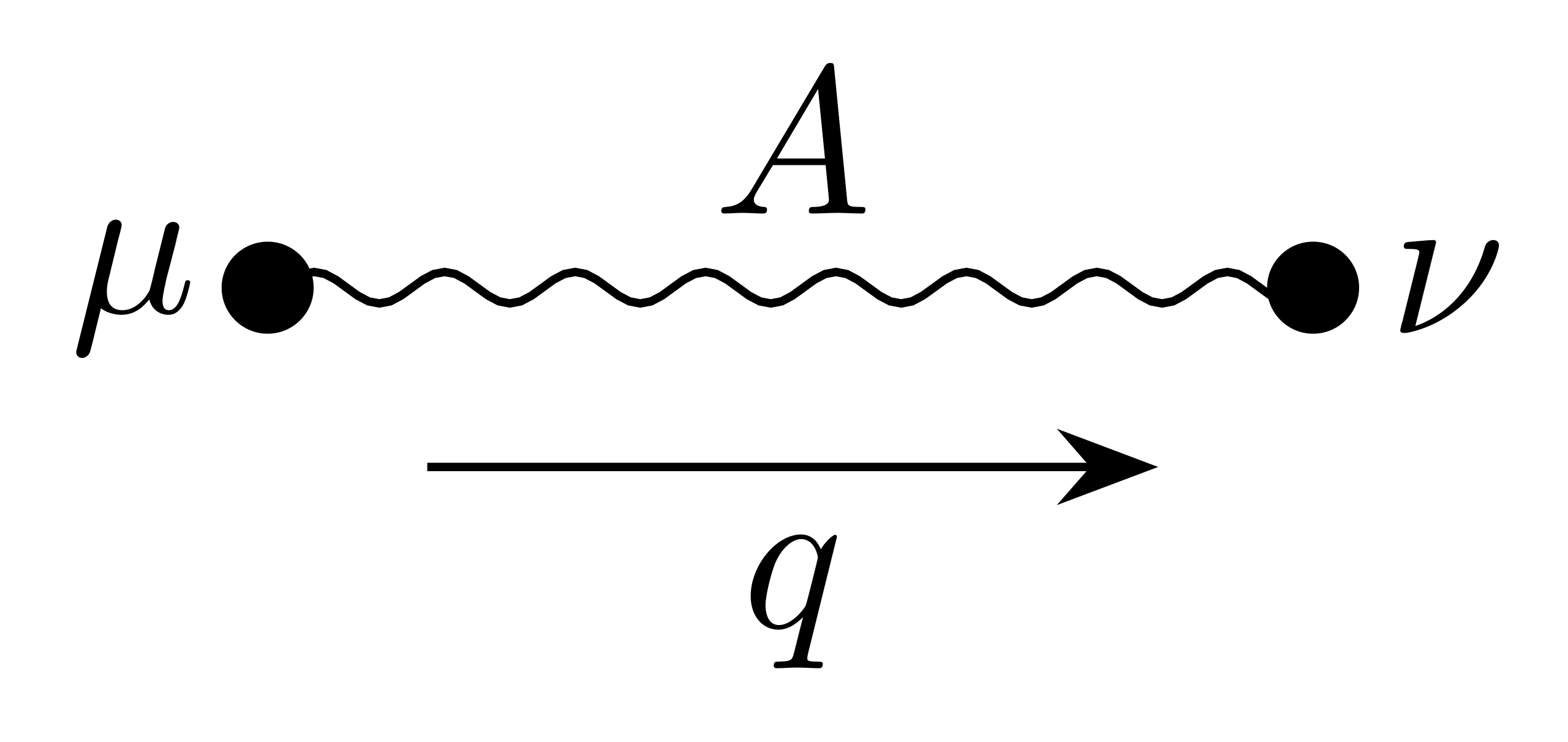}}}  = \frac{i\eta^{\mu\nu}}{q^2+i\epsilon}\,.
\end{equation}
Explicitly we have 
\begin{equation}
    M_4^{\text{SQED}} =4e^2 Q_1 Q_2 \frac{p_1{\cdot}p_2+q^2}{q^2+i\epsilon}\,.
\end{equation}
From the \ac{KMOC}  prescription (see \cref{sec:KMOC}), the  classical limit of this amplitude can be taken by recalling $q\to \hbar q$, and take the leading order as $\hbar\to0$. This amounts to simply drop the  term $q^2$ in the numerator of the previous expression (effectively removing contact terms, and therefore the classical expansion can be interpret as the large impact parameter expansion). 
Notice by doing so, the classical amplitude can be alternative computed from the unitary gluing of two three point amplitudes \eqref{eq:a3sqed}, where the internal photon is on-shell, as indicated in \eqref{cuts m4 m5}.  This is in general the usual  approach taken when computing higher multiplicity, as well as higher loop amplitudes, needed for the two-body problem; that is, amplitudes are computed   using generalized unitarity  from lower order building blocks  (see for instance \cite{Bjerrum-Bohr:2021vuf}).

Back to the 4-point amplitude, the classical piece simply reads 
\begin{equation}\label{eq:M4sqedclass}
    \langle M_4^{\text{SQED}}\rangle = 4e^2Q_1 Q_2 \frac{p_1{\cdot}p_2}{q^2+i\epsilon}\,.
\end{equation}

With this amplitude at hand, the  1\ac{PL} linear impulse can then be computed from the formula \eqref{eq:impulse_final}, and reads explicitly
\begin{equation}\label{eq:LO_impulse_integral}
    \Delta p^{(0)\,\mu}_{1} =  e^2\,Q_1\, Q_2\,p_1{\cdot}p_2 \int \hat{d}^4q \hat{\delta}(p_1{\cdot}q)\hat{\delta}(p_2{\cdot}q) \frac{iq^\mu}{q^2+i\epsilon} e^{-i q{\cdot}b}\,.
\end{equation}

Let us finish this example, by   the explicit  evaluating this integral,  although it has been evaluated in several previous works (see for instance \cite{Kosower:2018adc,Guevara:2018wpp}). We aim however to introduce  some notation  and  conventions that will be further used in \cref{ch:soft_constraints} and other parts of this work. 

We start by noticing that since there are two  delta functions inside the integral  \eqref{eq:LO_impulse_integral}, they  allow us to evaluate the integrals in the time and longitudinal directions of $q$. Let us now remember the role of the $i\epsilon$ prescription is to ensure that the energy integral, $q^0$,  does not diverges when $q^0$ hits any of the singular values (remember $q^0$ runs from $-\infty$ to $+\infty$).   
\begin{equation}
    \frac{1}{q^2+i\epsilon}=\frac{1}{(q^0+|\vec{q}| +i\epsilon)(q^0-|\vec{q}|-i\epsilon)}
\end{equation}
however, since we have at least one delta function, say $\hat{\delta}(p_2{\cdot}q)$, we can chose to evaluate the integral in the reference frame of particle 2, then, $\hat{\delta}(p_2{\cdot}q)\to \frac{1}{m_2}\hat{\delta}(q^0)$, which localizes the $q^0$ integral to $q^0=0$, in that case, the propagators evaluate to 

\begin{equation}
    \frac{1}{(|\vec{q}| +i\epsilon)(-|\vec{q}|-i\epsilon)}
\end{equation}
which leaves us with denominators no longer divergent an therefore we can drop the $i\epsilon$.  We have learned then that when there is at least one delta function $\hat{\delta}(p_i{\cdot}q)$, and a delta function,  one cal always ignore the $i\epsilon$ prescription for the massless ( radiation) poles,  which in turn implies that  the result for the impulse will be the same irrespective of whether we used the Feynman or the Retarded propagator \footnote{However, this will not be the case for all of the integrals that we will find in this work. }. This is of course expected in this example since this is a computation purely in the conservative sector.

Let us however take a more covariant approach for the explicit computation of the integral. For that we   decompose the momentum $q$ in terms of  the massive momenta $p_i$, and the transverse momentum $q_\perp$, as follows
\begin{equation}\label{eq:momentum-expansion-q}
    q^\mu=\alpha_{2}p_{1}^\mu+\alpha_{1}p_{2}^\mu+q_{\perp}^\mu,\quad p_{i}{\cdot}q_{\perp}=0,
\end{equation}
where 
\begin{equation}\label{eq:alphai_factors}
    \alpha_{1}=\frac{1}{\mathcal{D}}\left[p_{1}{\cdot}p_{2}x_1{-}m_{1}^{2}x_2\right],\quad \alpha_{2}=\frac{1}{\mathcal{D}}\left[p_{1}{\cdot}p_{2}x_2{-}m_{2}^{2}x_1\right].
\end{equation}
Here we have introduced the dimension-full quantities $x_i  = p_i{\cdot}q$, and the Jacobian factor  $\mathcal{D}$, given by
\begin{equation}\label{eq:jacobian}
    \mathcal{D} = \left(p_{1}{\cdot}p_{2}\right)^{2}-m_{1}^{2}m_{2}^{2}.
\end{equation}
Notice  the  decomposition \eqref{eq:momentum-expansion-q} is generic and does not assume any conditions on the $x_i$ variables.
With the change of variables \eqref{eq:momentum-expansion-q}, the integral measure in \eqref{eq:LO_impulse_integral} becomes simply $ \hat{d}^{4}q=\frac{1}{\sqrt{\mathcal{D}}}\hat{d}^{2}q_{\perp}\hat{d}x_1\hat{d}x_2$.

In general, in latter sections  we will have to evaluate integrals of the form 
\begin{equation}\label{eq:general_integral}
    \mathcal{I} = \frac{1}{\sqrt{\mathcal{D}}}\int \hat{d}^{2}q_{\perp}\hat{d}x_1\hat{d}x_2 \hat{\delta}^{(n)}(x_1)\hat{\delta}^{(m)}(x_2) f(x_1,x_2,q_\perp,\sigma),
\end{equation}
that is, with  a certain number of derivatives acting over  the on-shell delta functions. We can  use integration by part multiple times in order to remove the derivatives acting over the delta functions,  transporting them to act over the integrand function $f(x_1,x_2,q_\perp,\sigma)$ \footnote{ Here  we have use $\sigma$ to represent additional momenta, masses and impact parameter labels.}; once we have the on-shell delta functions free of derivatives, we can use the latter to evaluate the $x_i$-integrals. At that point, the calculation  would have been  reduced  to  evaluate the lower-dimensional integrals  of the form
\begin{equation}\label{eq:general_ibp}
    \mathcal{I} = (-1)^{m+n}\frac{1}{\sqrt{\mathcal{D}}}\int \hat{d}^{2}q_{\perp} \frac{\partial^n}{\partial x_1^n}\frac{\partial^m}{\partial x_2^m}
    f(x_1,x_2,q_\perp ,\sigma)\Bigg|_{x_1=x_2=0}.
\end{equation}

Going back to the computation of  the leading order impulse integral \eqref{eq:LO_impulse_integral}, for this case  the evaluation of the  integrals in the time and longitudinal directions simply reduces to  fixing $\alpha_1=\alpha_2=0$.  We are left then  with a two-dimensional integral 
\begin{equation}\label{eq:LO_impulse_interm}
\Delta p^{(0)\,\mu}_{1} =\frac{Q_{1}Q_{2}\,p_{1}{\cdot}p_{2}}{\sqrt{\mathcal{D}}}\int \hat{d}^{2} q_\perp e^{-iq_\perp{\cdot}b} \frac{iq_\perp^{\mu}}{q_\perp^{2}}\,,
\end{equation}
which can be evaluated by trading  the momentum $q_\perp$ in the numerator by a derivative w.r.t  the transverse impact parameter. Afterwards, the two dimensional  integral can be evaluated in polar coordinates as follows
\begin{align}
\Delta p^{(0)\,\mu}_{1}  & =\frac{Q_{1}Q_{2}\,p_{1}{\cdot}p_{2}}{\sqrt{\mathcal{D}}}\frac{1}{2\pi}\partial_{b^{\mu}}\lim_{\mu\to0}\int_{\mu}^{\infty}\frac{dq_{\perp}}{q_{\perp}}\int_{0}^{2\pi}\frac{d\theta}{2\pi}e^{iq_{\perp}b_{\perp}\cos\theta}\,,\\
 & =\frac{Q_{1}Q_{2}\,p_{1}{\cdot}p_{2}}{\sqrt{\mathcal{D}}}\frac{1}{2\pi}\partial_{b^{\mu}}\lim_{\mu\to0}\int_{\mu}^{\infty}dq_{\perp}\frac{\mathcal{J}_{0}(q_{\perp}b_{\perp})}{q_{\perp}}\,,\\
 & =-\frac{1}{4\pi}\frac{Q_{1}Q_{2}\,p_{1}{\cdot}p_{2}}{\sqrt{\mathcal{D}}}\lim_{\mu\to0}\partial_{b^{\mu}}\ln\left(-b^{2}\mu^{2}\right)\label{eq:IR_impulse_interm}\,.
\end{align}
In the second line $\mathcal{J}_0(x)$ corresponds to the order zero Bessel functions of the first kind.
Evaluating the remaining  derivative and trivially computing the  $\mu \to 0$ limit,  leads to the well know result for the leading order electromagnetic impulse, first computed by Westpfahl in  \cite{Westpfahl1985} by explicilty solving the classical particles' equations of motion (EoM) 
\begin{equation}\label{eq:LO_Impulse_explicit}
   \Delta p^{(0)\,\mu}_{1}=-e^2\frac{Q_{1}Q_{2}\,p_{1}{\cdot}p_{2}}{2\pi\sqrt{\mathcal{D}}}\frac{b^{\mu}}{b^{2}},\,\,\,\,\,b^{2}=-\vec{b}^{2},
\end{equation}
where $\vec{b}$ is the two dimensional impact parameter.

\subsection*{Leading order radiation}

\begin{figure}

\begin{equation*}
    \vcenter{\hbox{\includegraphics[width=65mm,height=42mm]{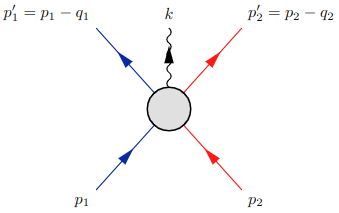}}} 
\end{equation*}

\caption{Bremsstrahlung radiation (outgoing photon) emitted during the scattering of two massive charge particles thought the exchanging electromagnetic waves.}
\label{fig:5pt-amplitude}
\end{figure}

The second example we provide in this section is the computation of the classical radiated photon field  at 3\ac{PL} order, for the scattering of two interacting classical compact charged objects.  At this order in perturbation theory,  only the   linear in amplitude part of the radiated field \eqref{eq:r-kernel} contributes, whereas the second term contributes to higher \ac{PL} orders as we will see explicitly in \cref{ch:soft_constraints}.  Analogous to the computation of the linear impulse, we first need to provide the relevant amplitude $M_5^{\text{SQED}}$, for which we will use the momentum conventions given in Figure \ref{fig:5pt-amplitude}, 
and then proceed to take its classical limit using the \ac{KMOC}  prescription. 

\begin{figure}
\begin {center}
\includegraphics[width=12truecm]{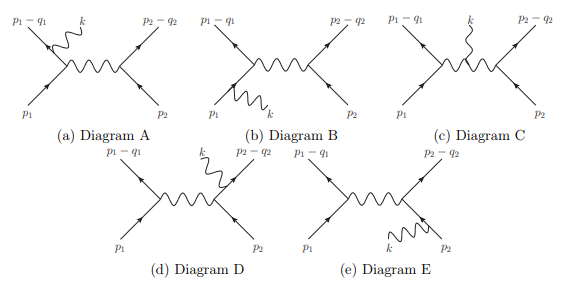} 
\end{center}
\caption{Feynman diagrams that contribute to the 5-point radiation amplitude in SQED. Figure adapted from \cite{Luna:2017dtq}}
\label{fig:M5_diags}
\end{figure}

The natural path for obtaining this amplitude  would  be by the  use of Feynman diagrams, there are five of them as shown in Figure \ref{fig:M5_diags}
, and for \ac{SQED} these are very easy to compute.
 Let us however take an alternative road which uses  what we have learned from the computation of the leading order impulse in the previous example. That is, in the classical limit, the amplitude can be obtained from the unitarity gluing of lower multiplicity amplitudes. This is nothing but the well known fact that up to contact terms (although in some cases they are not present), the scattering amplitudes can be reconstructed from unitary cuts, where the internal particles become on-shell,  and the amplitude factorizes into the product of two on-shell amplitudes \cite{Eden2002-hd}.

For the case of $M_4$, at leading order we in the \ac{PL} (PM) expansion, the classical piece of the amplitude can then be computed from the formula 
\begin{equation}\label{eq:M4_general}
\boxed{
    \langle M_4^h\rangle = \frac{n_h}{q^2}}\,,
\end{equation}
where $n_h$ is a local numerator. This form of the 4-point amplitude is general, and works for the electromagnetic and gravitational theory, including spin effects as we will see in \cref{sec:spin_in_qed}. For the case of the classical 4-point amplitude at leading \ac{PL} order, we can identify the scalar numerator  by caparison to \eqref{eq:M4sqedclass}. $n_{\text{ph}} =4e^2Q_1 Q_2 \,p_1{\cdot}p_2 $

For the case of the five point amplitude,   the relevant factorization channel that encapsulate the classical contribution, are those for which the   amplitudes factorizes into the product of  our favorite 3-point \eqref{eq:a3sqed} and the scalar Compton amplitude \eqref{eq:sqed_comtpon}, as follows

\begin{equation}\label{cutsm5}
    \vcenter{\hbox{\includegraphics[width=62mm,height=20mm]{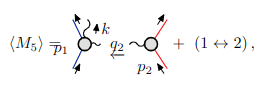}}} 
\end{equation}
\\
that is, the channels for which $q_1^2=0$ and $q_2^2=0$. 
We stress  this  factorization enclose the classical contribution  even in the presence of  spin as we will see in \cref{sec:spin_in_qed}. 
 
Let us now see how this factorization allow us to recover the Bremshrtralugh radiation formula for scalar objects. At the first factorization channel, $q_2^2=0$, the residues  can be computed via

\begin{equation}
    \text{Res}M_5^{\text{SQED}}\Big|_{q_2^2=0} = \Big(A_4^{\text{L}\,,\mu }\,\eta_{\mu\nu}\, A_3^{\text{R}\,,\nu}\Big)\Big|_{q_2^2=0}
\end{equation}
where
\begin{align}
     A_3^{\text{R}\,,\nu} &= 2e Q_2 \,p_2^\nu\,,\\
     A_4^{\text{L}\,,\mu } &=\frac{2e^2 Q_1^2}{(p_1{\cdot}k)(p_1{\cdot}q_2)}(p_1{\cdot}q_2\, F^{\mu\alpha}p_{1\alpha}-p_1^\mu q_{2}{\cdot}F{\cdot}p_1)\Big|_{q_2^2=0}\,.
\end{align}
Now, the on-shell condition for the outgoing massive particles imply  $p_1{\cdot}q_1 = q_1^2/2$ and $p_2{\cdot}q_2 = q_2^2/2$.  This shows  that, although  the products $p_i{\cdot}q_i$   naively scale as $\sim \hbar$, they in fact scale as $\sim \hbar^2$ and are therefore  subleading in the classical limit.  Notice also that   momentum conservation dictates $p_1{\cdot}q_2 = p_1{\cdot}k- q_1^2/2$, where the second factor can be dropped in the classical limit
The analog result follows for the factorization channel $q_1^2=0$. 

We have learned then that in the classical limit, the Bremsstrahlung amplitude has the following form: 
\begin{equation}\boxed{
\langle M_{5}^{h}\rangle{=}\frac{1}{(q{\cdot}k)^{h-1}}\left[\frac{n_h^{(a)}}{(q^{2}{-}q{\cdot}k)(p_{1}{\cdot}k)^{2}}+(-1)^{h+1}\frac{n_h^{(b)}}{(q^{2}{+}q{\cdot}k)(p_{2}{\cdot}k)^{2}}\right]}\,,\label{eq:newM5clas}
\end{equation}
where we have  further write the momentum transfer $q_i$ in the   symmetric variable $q$ via $q=\frac{q_1-q_2}{2}$. Here we have written $h=1,2$ to denote the photon and graviton emission, since, as we will argue in \cref{sec:spin_in_qed} and \cref{ch:double_copy}, this formula also holds for gravity, even in the presence of spin. For the time being we can just set $h=1$ as it is the case we are here interested in. 

The scalar numerators  for  photons emission, can be rearrange in the form
\begin{equation}
    n_{0,\rm{ph}}^{(a)}{=}4e^3 Q_1^2 Q_2 p_1{\cdot}R_3{\cdot}F{\cdot}p_1, \quad 
    n_{0,\rm{ph}}^{(b)}{=}4e^3 Q_1 Q_2^2 p_3{\cdot}R_1{\cdot}F{\cdot}p_3, \label{nphsc}
\end{equation}
where we have introduced the notation $R_i^{\mu\nu}{=}p_{i}^{[\mu}(\eta_i 2q{-}k)^{\nu]}$, with $\eta_1=1$ and $\eta_2=-1$.  We have to keep in mind this numerators  are valid  on the support of the on-shell condition for the outgoing massive particles. 

To be more precise, we can now write the radiated photon field for the scattering of two  scalar charged particles in SQED, as   given by our \ac{KMOC}  formula \eqref{eq:r-kernel} as follows

\begin{equation}\label{eq:r-kernel_a0}
    A^{\mu}(k)=ie^{k{\cdot}\tilde{b}}\int\hat{d}^{D}q\prod_{i=1}^{2}\hat{\delta}(2p_{i}{\cdot}q)e^{-iq{\cdot}b}\left[\frac{n_{0,\text{ph}}^{(a)}}{(q^{2}{-}q{\cdot}k)(p_{1}{\cdot}k)^{2}}+\frac{n_{0,\text{ph}}^{(b)}}{(q^{2}{+}q{\cdot}k)(p_{2}{\cdot}k)^{2}}\right]\,,
\end{equation}
where we have use $\tilde{b}=\frac{1}{2}(b_1+b_2)$, and we have defined the impact parameter $b=b_2-b_1$. One can show this formula recovers the classical result of  Goldberger and Ridgway \cite{Goldberger:2016iau} for colourless charges,  upon the change of variables $l_1 = q+k/2$ and $l_2=-q+k/2$, and set $b_2=0$.

\subsection*{Photon exponential soft theorem}\label{sec:esponential_soft}

We will not evaluate explicitly  integral \eqref{eq:r-kernel_a0} here, but will study a very interesting feature that arise if the photon emission is soft \footnote{By soft we means the limit in which the energy of the emitted photon is much smaller that the momentum of the  massive particles.}, namely,  the soft exponentiation of the scalar classical 5-point amplitude.  This will also be a feature for the gravitational case as we will show  in \cref{sec:spin_in_qed}. For this we make use of  the exponential form of the Compton amplitude \eqref{eq:scphcompton} to read exponential form of numerators entering in the 5-point amplitude \eqref{eq:newM5clas}.  This is, the numerators $n_{0,\text{ph}}^{(a)}$ can be read off directly from \eqref{eq:scphcompton} as follows: Replacing $\epsilon_{1}$ by $p_{2}$, powers of the orbit multipole $F_{\epsilon}$
translate to powers of $F_{p}{=}p_{1}{\cdot}F{\cdot}p_{2}$, whereas $F_{k}$ now
becomes $F_{iq}{=}\eta_i (p_i{\cdot}F{\cdot}q)$, with $\eta_1{=}{-}1,\eta_2=1$. The soft expansion \eqref{eq:scphcompton}
with respect to $k_{2}=k$ becomes
\begin{equation}
    n_{0,\text{ph}}^{(a)}=F_{1q}\,e\,Q_1\,e^{-\frac{F_{p}}{F_{1q}}(p_{1}{\cdot}k)\frac{\partial}{\partial(p_{1}{\cdot}p_{3})}}\left[4e^2 Q_1\,Q_2 p_{1}{\cdot}p_{2}\right].
\end{equation}
Further writing $\frac{1}{q^2\pm q\cdot k}=e^{\pm q\cdot k\frac{\partial}{\partial q^2}}\frac{1}{q^2}$
turns \eqref{eq:newM5clas} into
\begin{equation}
\boxed{
\langle M_{5}^{\rm SQED}\rangle=\sum_{i=1,2}\mathcal{S^{\text{SQED}}}_{i}e^{\eta_i \left(F_{p}\frac{p_{i}{\cdot}k}{F_{iq}}\frac{\partial}{\partial(p_{1}{\cdot}p_{3})}+q{\cdot}k\frac{\partial}{\partial q^2}\right)}\langle M_{4}^{\rm SQED}\rangle\,\label{eq:classicsoft}}
\end{equation}
where   is given in \eqref{eq:M4sqedclass}. We have defined  $\mathcal{S^{\text{SQED}}}_{i}{=} eQ_i\frac{F_{iq}}{(p_{i}{\cdot}k)^{2}}$ . This expression
can be used to obtain $\langle M^{\rm SQED}_{5}\rangle$ from $\langle M_{4}^{\rm SQED}\rangle$
as an expansion in the photon momentum $k^{\mu}$ to any desired
order in the soft expansion (sub-subleading orders were studied in \cite{Laddha:2018myi,Sahoo:2018lxl,Ciafaloni:2018uwe}).

One can check
explicitly that $\mathcal{S}_{1}+\mathcal{S}_{2}$ corresponds to
the $\hbar\to0$ limit of the Weinberg Soft Factor for the full $M_{5}$ \cite{Weinberg:1965nx}.
The first order of the exponential analogously corresponds to the
$\hbar\to0$ limit of the subleading soft factor of Low \cite{PhysRev.96.1428,PhysRev.110.974}.

Let us focus for simplicity on the leading order of \eqref{eq:classicsoft}. As we will see in \cref{ch:bounded}, for  bounded orbits   $\omega \sim \frac{v}{r}$ the wave frequency expansion becomes a non-relativistic expansion \cite{Goldberger:2017vcg}, where the Maxwell dipole emission formula (an analogously the Einstein quadrupole formula) can be derived from the Weinberg soft theorem.  For classical scattering we can use the leading soft term to obtain the Memory Effect as $R{\to}\infty$. Plugging \eqref{eq:classicsoft}
into \eqref{eq:r-kernel_a0} we get
\begin{equation*}
    \int \frac{d^{D}q}{(2\pi )^{D-2}}\delta(2q\cdot p_{1})\delta(2q\cdot p_{2})e^{iq\cdot b}\left(\sum_{i=1,2}\mathcal{S}_{i}\right)\langle M_{4}^{{\rm SQED}}\rangle
\end{equation*}
as $k\to 0$. Evaluating the sum and using \eqref{eq:LO_impulse_integral} as a definition of the linear impulse
$\Delta p_{1}=-\Delta p_{2}$ we obtain
\begin{equation}
    \epsilon_{\mu}A^{\mu}=\frac{1}{p_{1}{\cdot}k \,p_{2}{\cdot}k}\left(\frac{e\, Q_1\,p_{1}}{p_{1}{\cdot}k}{+}\frac{e\, Q_2\,p_{2}}{p_{2}{\cdot}k}\right){\cdot}F{\cdot}\Delta p{+}\mathcal{O}(k^{0}),
\end{equation}
which at leading order in $\Delta p$ (or $e$) becomes
\begin{equation}\label{eq:a_weinerg_tree}
  A^{\mu}(k)=\Delta\left[\frac{e\, Q_1\,p_{1}^{\mu}}{p_{1}{\cdot} k}+\frac{e\, Q_2\,p_{2}^{\mu}}{p_{2}{\cdot} k}\right]\, .
\end{equation}
This is nothing but the classical leading soft factor for two incoming and two  outgoing massive particles  \cite{Saha:2019tub}
\begin{flalign}\label{eq:weinber_LO}
A_{\mu}^{\omega^{-1}}(\hat{n})\, =\, \sum_{i=1}^{N_{in}}\, Q_{i}\, \frac{1}{p_{i} \cdot n}\, p_{i}^{\mu}\, -\, \sum_{i=1}^{N_{out}}\, Q_{i}\, \frac{1}{p_{i} \cdot n}\, p_{i}^{\mu}\,,
\end{flalign} 
once we write the outgoing momenta as a perturbative expansion 
  in powers of the impulse acquired by the massive particles   order by order in perturbation theory\footnote{We note that from the classical perspective, this expansion is convergent as final momenta are well defined.}
\begin{flalign}\label{eq:expansion_momentaa}
p_{i\, \text{out}}^{\mu}\, =\, p_{i\, \text{in}}^{\mu} + \sum_{L=0}\, e^{2(L+1)}\, (\triangle p^{(L)})_{i}^{\mu}\,,
\end{flalign}
and take the leading order in the coupling $e$ (i.e. $L=0$). We have then connected non-perturbative  results for classical soft theorems to the perturbative scattering amplitude approach to the computation of classical radiation. In general, we will show in \cref{ch:soft_constraints} that classical soft theorems impose and infinite tower of constraints on \ac{KMOC}  computations, where the tower arises from the loop expansion of the outgoing momenta \eqref{eq:expansion_momentaa}. 

As last comment, it is well known that the leading and subleading soft theorems \ac{QED} are universal,  independent of the details of the computation, as well as the matter content \cite{ Laddha:2018rle,Laddha:2018myi, Sahoo:2018lxl,Laddha:2019yaj, Saha:2019tub}. In this section we  have reproduced the leading soft theorem starting from the scattering of  scalar charge particles only. This means adding intrinsic structure to the particle such as spin, should not change the result \eqref{eq:weinber_LO} for the radiated photon field. This, as we will see in \cref{sec:spin_in_qed}, is a consequence of the spin universality of $A_3$ and $A_4$ amplitude, which is inherited by  the two-body radiative amplitude $M_5$. 
\section{2\ac{PL} linear impulse in SQED}\label{sec:2PLimpulse}

In the final part of this chapter we do the explicit evaluation of the 2\ac{PL} linear impulse integral for scalars particles. This  integral   was derived using the \ac{KMOC}  formalism in the original paper by the authors  \cite{Kosower:2018adc}, whose final result was left implicit, and can be obtained directly from the Feynman diagrams 1-loop diagrams in \ac{SQED}. We will use integration techniques outlined in \cref{sec:examplesa_tree_level} to show the final result agrees with the classical computation of  Saketh et al \cite{Saketh:2021sri}.

As shown by the authors \cite{Kosower:2018adc}, the classical impulse receives  contribution  from only of the triangle, boxes and cross-box diagrams 1-Loop diagrams. After carefully taken the classical limit of each contribution as described in \cref{sec:KMOC}, the authors show superclassical fragments cancel between the linear and quadratic in amplitude contributions to the linear impulse \eqref{eq:impulse_final}. As for the classical contribution, the authors arrive at the  following integral:

\begin{equation}
\Delta p^{(1)\,\mu}_{1}=\frac{i}{4}\int\hat{d}^{4}q\prod_{i}\hat{\delta}(p_{i}{\cdot}q)e^{-ib{\cdot}q}\left[\mathcal{I}_{1}^{\mu}+\mathcal{I}_{2}^{\mu}+\mathcal{I}_{3}^{\mu}\right],\label{eq:NLO impulse}
\end{equation}
 where the $\mathcal{I}_{i}^{\mu}$ integrals resemble the contributions
to the 4 point  amplitude from the different Feynman diagrams. The firs
one comes from the contribution from the triangle diagrams 
\begin{equation}
\mathcal{I}_{1}^{\mu}=2e^4\left(Q_{1}Q_{2}\right)^{2}q^{\mu}\sum_{i}\int\hat{d}^{4}l\frac{m_{i}^{2}\hat{\delta}(p_{i}{\cdot}l)}{l^{2}(l-q)^{2}}\,.\label{eq:I1}
\end{equation}
 Next we have the contribution coming from the Boxes, which once by canceling the term 
$Z$ in $(\ref{eq:4pt 1L classical explicit})$, using the cut-Box diagram, reads 
\begin{equation}
\mathcal{I}_{2}^{\mu}=2e^4\left(Q_{1}Q_{2}p_{1}{\cdot}p_{2}\right)^{2}q^{\mu}\sum_{i,j\,i\ne j}\int\hat{d}^{4}l\frac{l{\cdot}(l-q)}{l^{2}(l-q)^{2}}\frac{\hat{\delta}(p_{j}{\cdot}l)}{(p_{i}{\cdot}l+i\epsilon)^{2}},\label{eq:I2}
\end{equation}
 Finally, we have the 4 point cut-box contribution 
\begin{equation}
\mathcal{I}_{3}^{\mu}=-2ie^4\left(Q_{1}Q_{2}p_{1}{\cdot}p_{2}\right)^{2}\int\hat{d}^{4}l\frac{l{\cdot}(l-q)l^{\mu}}{l^{2}(l-q)^{2}}\left[\hat{\delta}'(p_{1}{\cdot}l)\hat{\delta}(p_{2}{\cdot}l)-\hat{\delta}(p_{1}{\cdot}l)\hat{\delta}'(p_{2}{\cdot}l)\right].\label{eq:I3}
\end{equation}

Let us start with the  computation of the triangle diagrams which corresponds to  the first term in  \eqref{eq:NLO impulse}. Using \eqref{eq:I1}, we have

\begin{equation}
    \tilde{\mathcal{I}}_1^\mu =\frac{ie^4}{2}(Q_1Q_2)^2\int \hat{d}^4q\hat{d}^4l\hat{\delta}(p_1{\cdot}q)\hat{\delta}(p_2{\cdot}q) \frac{q^\mu}{l^2(l-q)^2}  e^{-iq{\cdot}b}\left[m_1^2 \hat{\delta}(p_1{\cdot}l) +m_2^2 \hat{\delta}(p_2{\cdot}l) \right] \,.
\end{equation}
We can do the integral in $l^0$ by going to the rest frame of particle $1$ (or $2$) then getting $\hat{\delta}(l^0)$ as the zero energy condition. Notice that we can  also set $q^0=0$ by using one of the on-shell delta functions in $q$. With this in mind the previous integral takes the form
\begin{equation}
    \tilde{\mathcal{I}}_1^\mu =\frac{ie^4}{2}(Q_1Q_2)^2(m_1 +m_2 )\int \hat{d}^4q\hat{d}^3\vec{l}\hat{\delta}(p_1{\cdot}q)\hat{\delta}(p_2{\cdot}q) \frac{q^\mu}{\vec{l}^2(\vec{l}-\vec{q})^2}  e^{-iq{\cdot}b}\,.
\end{equation}
The integral in $\hat{d}^3\vec{l}$  is easy to evaluate using Schwinger parameters, see for instance eq. $(7.9)$ in \cite{Bern:2019crd} . Using those results we  get 
\begin{equation}
    \tilde{\mathcal{I}}_1^\mu =\frac{ie^4}{16\sqrt{D}}(Q_1Q_2)^2(m_1 +m_2 )\int \hat{d}^2q_\perp \frac{q_\perp^\mu}{q_\perp}  e^{-iq_\perp{\cdot}b},
\end{equation}
where we have further evaluated  two of the $\hat{d}^4q$ integrals using the expansions for the momenta \eqref{eq:expansion_momenta}. Evaluation of the remaining integral can be done in polar coordinates, upon trading  $q_\perp^\mu$ in the numerators by a derivative w.r.t. the impact parameter. The final answer will be 

\begin{equation}\label{eq:finali1}
      \tilde{\mathcal{I}}_1^\mu =-\frac{e^4}{32\pi}(m_1+m_2)\frac{(Q_1Q_2)^2}{\sqrt{D}}\frac{b^\mu}{|b|^3}.
\end{equation}

Next we move to the evaluation of  the last term in  \eqref{eq:NLO impulse} using \eqref{eq:I3},

\begin{equation}
    \tilde{\mathcal{I}}_3^\mu= \frac{e^4}{2}(Q_1Q_2p_q{\cdot}p_2)^2 \int \hat{d}^4q\hat{d}^4l \hat{\delta}(p_1{\cdot}q)\hat{\delta}(p_2{\cdot}q) \frac{l{\cdot}(l{-}q)}{l^2(l{-}q)^2}l^\mu e^{-iq{\cdot}b}[\hat{\delta}'(p_1{\cdot}l)\hat{\delta}(p_2{\cdot}l){-}\hat{\delta}(p_1{\cdot}l)\hat{\delta}'(p_2{\cdot}l)],
\end{equation}
where the tilde over $ \mathcal{I}_3$ indicates inclusion of the $q$ integration.
 To evaluate this integral  we can  expand the momentum $l$ in an analogous way to the $q$ momentum in (\ref{eq:momentum-expansion-q} - \ref{eq:alphai_factors}), with say $\alpha_i\to \beta_i$, and $x_i\to y_i= p_i\cdot l$. The resulting  integrand  takes the form   \eqref{eq:general_integral}, and therefore we can evaluate the time and longitudinal components using integrating by parts one time \eqref{eq:general_ibp}. That is, we can write 
 \begin{equation}\label{eq:i3some}
    \tilde{\mathcal{I}}_3^\mu = \frac{1}{\sqrt{\mathcal{D}}}\int \hat{d}^4q \hat{\delta}(p_1{\cdot}q)\hat{\delta}(p_2{\cdot}q)  \hat{d}^{2}l_{\perp}\hat{d}y_1\hat{d}y_2 e^{-iq{\cdot}b}\left[ \hat{\delta}^{(1)}(y_1)\hat{\delta}^{(0)}(y_2) - \hat{\delta}^{(0)}(y_1)\hat{\delta}^{(1)}(y_2)\right]  f_u^\alpha(y_1,y_2,l_\perp,\sigma),
 \end{equation}
 with the  identification of the  integrand function   
\begin{align}
f_u^\alpha(y_1,y_2,l_\perp,\sigma)=\frac{e^4}{2}\left(Q_{1}Q_{2}p_{1}{\cdot}p_{2}\right)^{2} \frac{(\beta_{2}p_{1}{+}\beta_{1}p_{2}+l_{\perp})^{\alpha}\left((\beta_{2}p_{1}{+}\beta_{1}p_{2})^{2}{+}l_{\perp}^{2}{-}l_{\perp}{\cdot}q_{\perp}\right)}{\left((\beta_{2}p_{1}{+}\beta_{1}p_{2})^{2}{+}l_{\perp}^{2}\right)\left((\beta_{2}p_{1}{+}\beta_{1}p_{2})^{2}{+}(l_{\perp}{-}q_{\perp})^{2}\right)}\nonumber \,.\label{eq:fu1}
\end{align}
where in addition to the $l$-expansion, we have used the expansion for the $q$-momentum \eqref{eq:momentum-expansion-q}, and set $x_i\to0$ using the support of the delta functions $\hat{\delta}(p_i{\cdot}q)$ . Next, to use \eqref{eq:general_ibp} after integration by parts we need to evaluate the derivatives of the form 
\begin{equation}\label{eq:derf}
\frac{\partial}{\partial y_{i}}f_{u,ij}^{\alpha}\Bigg|_{y_{i}=y_{j}=0}=\frac{4e^4}{\mathcal{D}}\left(Q_{i}Q_{j}p_{i}{\cdot}p_{j}\right)^{2}p_{j,\beta}p_{j}^{[\beta}p_{i}^{\alpha]}\frac{l_{\perp}{\cdot}(l_{\perp}-q_{\perp})}{l_{\perp}^{2}(l_{\perp}-q_{\perp})^{2}}\,,
\end{equation}
With all the tools at hand, it is then direct to show that the integral \eqref{eq:i3some} simplifies to
\begin{equation}
 \tilde{\mathcal{I}}_3^\mu =- e^4\frac{\left(Q_{1}Q_{2}p_{1}{\cdot}p_{2}\right)^{2}}{\mathcal{D}^2}\left[p_{2,\beta}p_{2}^{[\beta}p_{1}^{\alpha]}{-}p_{1,\beta}p_{1}^{[\beta}p_{2}^{\alpha]}\right]\int\hat{d}^{2}q_{\perp}\hat{d}^{2}l_{\perp}e^{-ib{\cdot}q_{\perp}}\frac{l_{\perp}{\cdot}(l_{\perp}{-}q_{\perp})}{l_{\perp}^{2}(l_{\perp}{-}q_{\perp})^{2}}.\label{eq:cut-box int2}
\end{equation}
 Next we do the usual change of variables $q_{\perp}=\bar{q}_{\perp}+l_\perp,$
so that 
\begin{equation}
\tilde{\mathcal{I}}_3^\mu=-\frac{e^4}{\mathcal{D}}\left[p_{2,\beta}p_{2}^{[\beta}p_{1}^{\alpha]}{-}p_{1,\beta}p_{1}^{[\beta}p_{2}^{\alpha]}\right]\left[\frac{Q_{1}Q_{2}p_{1}{\cdot}p_{2}}{\sqrt{\mathcal{D}}}\int\hat{d}^{2}q_{\perp}\hat{d}^{2}l_{\perp}e^{-ib{\cdot}q_{\perp}}i\frac{\bar{q}_{\perp}}{\bar{q}_{\perp}^{2}}\right]^{2}.\label{eq:cut-box int3}
\end{equation}
in the big bracket we recognize the Leading order impulse $(\ref{eq:LO_impulse_interm})$,
which in turn allow write the final result as
\begin{equation}\label{eq:finali3}
     \tilde{\mathcal{I}}_3^\mu=e^4 \frac{(Q_1Q_2p_1{\cdot}p_2)^2}{8\pi^2\mathcal{D}^2|b|^2}\left[\left(m_1^2+p_1{\cdot}p_2 \right)p_2^\mu -\left(m_2^2+p_1{\cdot}p_2 \right)p_1^\mu \right].
\end{equation}

The final  task the evaluation of the e box and cross-box diagrams from  integral \eqref{eq:I2}. We now show they provide vanishing contribution in the conservative sector. 
We can see this by first  dropping the term proportional to $l\cdot l $ in the numerator of \eqref{eq:I2} since it give rise  to non local contributions. Next, using the same philosophy of \cite{Kalin:2020fhe}, we can write $2 l\cdot q=l^2+q^2-(l-q)^2$, and discarding again non local contributions; the integral  \eqref{eq:I2} becomes 
\begin{equation}
\mathcal{I}_{2}^{\mu}=-e^4\left(Q_{1}Q_{2}p_{1}{\cdot}p_{2}\right)^{2}q^{\mu}\sum_{i,j\,i\ne j}q^2\int\hat{d}^{4}l\frac{\hat{\delta}(p_{j}{\cdot}l)}{l^{2}(l-q)^{2}(p_{i}{\cdot}l+i\epsilon)^{2}},\label{eq:I2n}
\end{equation}
using  the fact that at \ac{NLO}  no net four-momentum is radiated, radiation poles do not contribute to the integral, we can choose a contour in the opposite half of the plane were  \eqref{eq:I2} has the double poles $(p{\cdot}l+i\epsilon)^2$, and then getting a vanishing integral.  Indeed this was also done for the gravitational case in   \cite{Kalin:2020mvi} eq. (4.26).
In conclusion, the \ac{NLO}  electro-magnetic impulse is the sum of \eqref{eq:finali1} and \eqref{eq:finali3}, 
\begin{equation}
\Delta p_1^{(1)\,\mu} =  \tilde{\mathcal{I}}_1^\mu+ \tilde{\mathcal{I}}_3^\mu = -\frac{e^{4}}{32\pi^2|b|^{3}}\frac{(Q_{1}Q_{2})^2}{\mathcal{D}}\left[\pi\sqrt{\mathcal{D}}(m_{1}+m_{2}) b^{\mu}+4\frac{(p_{1}{\cdot}p_{2})^2(p_{1}{+}p_{2})^{2}|b|}{\mathcal{D}}p^{\mu}\right]\,,
\end{equation}
where we have introduced the center of mass momentum $p^\mu$ via 
\begin{equation}\label{eq:p-def_justin}
    p^\mu = \frac{m_1m_2}{(p_1+p_2)^2}\left[\left(\frac{m_2}{m_1}+ \frac{p_1\cdot p_2}{m_1 m_2}\right)p_1^\mu -  \left(\frac{m_1}{m_2}+ \frac{p_1\cdot p_2}{m_1 m_2}\right)p_2^\mu\right]\,,
\end{equation}
and therefore recovering the classical result of Saketh et al \cite{Saketh:2021sri}. 

\section{Outlook  of the chapter}\label{sec:outlook_sqed}
In this chapter we have introduced some of the main ideas in the computation of classical observables directly from the classical limit of scattering amplitudes, in company of the \ac{KMOC}  formalism.
In particular, we have focused in interactive,  structure-less compact charge objects, both,  tree, and 1-loop level. We have seen how the main ingredients $A_n$, $n=3.4$, in the computation of two body amplitudes $M_m$, $m=4,5$, have very interesting properties that are inherited by the latter. In addition, we have seen how soft theorems play a crucial role in the computation of low energy bremhstralugh radiation. 
 In \cref{ch:soft_constraints}, we will continue exploring some interesting properties  regarding the computation of classical soft radiation in \ac{SQED} to higher orders in perturbation theory.
In \cref{ch:bounded} we will show how soft theorems are actually also important in the computation of radiation for bounded orbit scenarios, where the soft expansion is closely connected to the source multipole moment expansion. 

 Many of the tools learned in this chapter will be of used in the remaining ones, specially when we discuss interacting spinning massive matter, and the covariant spin multipole double copy in \cref{sec:spin_in_qed}. The discussion regarding  the Thomson scattering will be generalized  for the scattering of waves off spinning black holes in \cref{ch:GW_scattering}.
 

%% file: Chapters/soft_constraints.tex
\chapter{Soft constraints on \ac{KMOC} for electromagnetic radiation}\label{ch:soft_constraints}

\section{Introduction}\label{sec:introduction_soft_constraints}
In \cref{ch:electromagnetism}  we have started the study of  classical radiation directly from the classical limit of \ac{QFT} scattering amplitudes through  the \ac{KMOC} formalism. In particular, we have seen that the radiated photon field in a classical $2\to3$ scattering process, at leading order in the frequency expansion of the emitted wave, is entirely capture by the classical limit of the so called Weinberg soft theorem \cite{Weinberg:1965nx}. In this chapter we extend the discussion of classical soft theorems, and in particular, we  will discuss the implication they have on the computation of classical soft radiation directly from perturbative amplitudes, to all orders in perturbation theory. We will show that to a given order in perturbation theory, the classical leading soft photon theorem impose an infinite tower of constraints on the expectation value of the product of monomials of exchange momenta in the \ac{KMOC} formula for radiation \eqref{eq:radi_field}. 

Before going into the main computation, let us in the remaining of this section, review some facts about classical soft theorems and summarize the main results of this chapter. This chapter is mostly  based on previous work by the author \cite{Bautista:2021llr}.

\subsubsection*{Facts from classical  soft theorems and summary of the results of the chapter}
Classical soft photon (graviton) theorems  \cite{ Laddha:2018rle,Laddha:2018myi, Sahoo:2018lxl,Laddha:2019yaj, Saha:2019tub}  are exact statements about  soft radiation emitted during a generic  electro-magnetic (gravitational) scattering process. As shown in the seminal works by Sahoo and Sen\cite{Sahoo:2018lxl}, Saha, Sahoo and Sen \cite{Saha:2019tub}, and Sahoo \cite{Sahoo:2020ryf}, in four dimensions if we expand the electro-magnetic (or gravitational) radiative field in the frequency of the emitted radiation, then the following terms in the expansion have a universal analytic form independent of the details of the scattering dynamics or even spins of the scattering particles
\begin{flalign}\label{one}
A_{\mu}(\omega, \hat{n})\, =\, \frac{1}{\omega}\, A_{\mu}^{\omega^{-1}}(\hat{n})\, +\, \sum_{I=1}^{2}\,  \omega^{I} (\ln\omega)^{I+1}\, A_{\mu}^{\ln^{I+1}}(\hat{n})\, +\, \cdots\,.
\end{flalign}
Here $\hat{n}$ is a unit vector pointing towards the direction of observation, 
and  $\cdots$ indicate sub-sub-leading terms in the soft expansion. It was conjectured in \cite{Sahoo:2020ryf} that even among the sub$^{n}$-leading  terms the coefficients of $\omega^{n}\, \ln^{n+1}\omega\ , n\, \geq\, 3$ are universal while other terms in the soft expansion are non-universal and depend on the details of the dynamics. In \cite{Ghosh:2021hsk}, first such non-universal soft factor proportional to $\omega\ln\omega$ was computed and was shown to depend on the spin of the scattering particles. 

Each coefficient in the above expansion is a function of incoming and outgoing momenta and charges of the scattering particles.  For example, the leading coefficient $A_{\mu}^{\omega^{-1}}(\hat{n})$ is simply the Weinberg soft photon factor \eqref{eq:weinber_LO}, which we reintroduce here for the reader's convenience
\begin{flalign}\label{eq:sphoton}
A_{\mu}^{\omega^{-1}}(\hat{n})\, =\, \sum_{i=1}^{N_{in}}\, Q_{i}\, \frac{1}{p_{i} \cdot n}\, p_{i}^{\mu}\, -\, \sum_{i=1}^{N_{out}}\, Q_{i}\, \frac{1}{p_{i} \cdot n}\, p_{i}^{\mu}\,.
\end{flalign} 
Here  $\{(Q_{1}, p_{1}),\, \cdots,\, (Q_{i}, p_{i})\, \}$ is the collection of  the charges and momenta of scattering particles, and   $n^{\mu}\, =\, (1, \hat{n})$.  Although the exact expressions for sub-leading and higher order log soft factors in eqn.(\ref{one}) are more complicated, they are all functions of asymptotic data, namely charges and momenta of scattering particles, which we do not show here explicitly\footnote{Readers interested are refer to the original works on soft theorems \cite{Sahoo:2018lxl, Laddha:2018rle,Laddha:2018myi, Sahoo:2018lxl,Laddha:2019yaj, Saha:2019tub,Saha:2019tub}.}. The form of \eqref{eq:sphoton} can be obtained by computing the   early and late time electromagnetic waveform emitted during the  scattering of the charged particles involved.  The computation requires then to solve the classical  \ac{EoM}  for all of the particles involved, obtaining $x_a(\sigma_\pm)$, with $\sigma_pm$ the incoming/outgoing particles proper time. Having these solutions at hand, we can then   compute a electromagnetic current of the form $j^\mu(x)\sim\sum_a \int d\sigma \delta^4(x-x_a(\sigma))\frac{dx_a}{d\sigma} $, which  enters  as a  source for electromagnetic waves, as given by  \eqref{eq:rad_field_full_n}, whose \ac{LO} behaviour in the low energy expansion  result into \eqref{eq:sphoton}.  For a detail computation see Section 4.1 in \cite{Saha:2019tub}. The key observation is this probe  assumes all particles momenta to be independent one from another, and in this regard, the solution is valid to all orders in perturbation theory.

From the perspective of scattering dynamics, these theorems are rather non-trivial as they are non perturbative in the coupling ans we just mentioned.\footnote{In an interesting recent work \cite{Freidel:2021qpz}, an attempt has been made to analyse the infinite set of constraints on the gravitational dynamics from  asymptotic symmetries which are in turn related to classical soft theorems.} If we consider a class of scattering processes which can be analysed perturbatively (such as large impact parameter scattering, as we have seen in previous section, with $N_{in} = N_{out} = N$) then every outgoing momenta  admits the perturbative expansion \eqref{eq:expansion_momentaa}, in terms of the incoming momenta \footnote{We note that from the classical perspective, this expansion is convergent as final momenta are well defined.} 
\begin{flalign}\label{eq:expansion_momenta}
p_{i\, +}^{\mu}\, =\, p_{i\, -}^{\mu} + \sum_{L=0}\, e^{2(L+1)}\, (\triangle p^{(L)})_{i}^{\mu}\,,
\end{flalign}
where $e$ is the coupling constant,  and  $(\triangle p^{(L)})^{\mu}$ is the linear impulse evaluated at   $L$-th order in the perturbation theory ($L=0$ being the \ac{LO} impulse which we explored in previous chapter). This expansion is just a fact that in a large impact parameter scattering process, particles'  outgoing momenta are determined by the incoming momenta plus the equations of motion governing the dynamics of the system. 
We thus see when expanded in the coupling, the Weinberg soft photon factor has a rather intricate structure
\begin{equation}\label{two} 
A_{\mu}^{\omega^{-1}}(\hat{n})\, =\, \sum_{s=1}^{N}\, e\,Q_{s} \sum_{n=0} e^{2(n+1)}\, \sum_{i=0}^{n}\, V^{\mu}_{s\, \alpha_{1}\, {\cdots}\, \alpha_{i+1}}\ \smashoperator{\sum_{\substack{L_{1}+\, \cdots+\, L_{i+1}=0 \\ (L_{1} + L_{i+1}) + (i+1)\, = n}}^{n +1 - i}}\ (\, (\triangle p^{(L_{1})})^{\alpha_{1}}\, {\cdots}\, (\triangle p^{(L_{i+1})})^{\alpha_{i+1}}\, )\,,
\end{equation}
where the sum $\sum_{L_{1},\, \cdots,\, L_{i+1}=0\, \vert\, (L_{1} + L_{i+1}) + (i+1)\, = n}^{n +1 - i}$ is over products of impulses at $L_{1},\, \cdots,\, L_{i+1}$ orders in the coupling respectively. In the above equation we have defined the tensors
\begin{flalign}\label{eq:projector_general_s}
V^{\mu}_{s\, \alpha_{1}\, \cdots\, \alpha_{i+1}}\, =\, \textcolor{black}{(-1)^i}\, \Big[\, \frac{ (i+1)\, !\,}{(p_{s} \cdot k)^{i+1}} \delta^{\mu}_{(\alpha_{1}}\, k_{\alpha_{2}}\, \cdots\, k_{\alpha_{i+1})}\, -\,  \frac{1}{(p_{s} \cdot k)^{i+2}}\, p_{s}^{\mu}\, k_{\alpha_{1}}\, \cdots\, k_{\alpha_{i+1}}\,\Big]\,,
\end{flalign}
which are the remainders  from doing the perturbative expansion of Weinberg soft factor. 

Classical Soft photon (or graviton) theorem are independent of the details of the hard scattering and are applicable to perturbative scattering at finite impact parameter as well as collisions.  However, to prove the classical soft theorems via perturbative analysis (even in the case where hard scattering can be treated perturbatively) is a highly non-trivial task as one has to resum the perturbation series. But the discussion above demonstrates that, 
due to their universality, classical soft theorems can serve as  powerful tool for any method which computes electro-magnetic (or gravitational) radiation using (perturbative) scattering amplitudes. For one, it can serve as a strong diagnostic for the perturbative results of radiation kernel and when used in conjunction with the perturbative results (such as analytic expressions for impulse in the \ac{PL} and \ac{PM}  expansions), it can produce interesting insights such as providing analytical formulae for classical radiation in terms of incoming kinematic data order by order in perturbation theory, an example of which we saw in previous section in the discussion around \eqref{eq:a_weinerg_tree}

One such methods aforementioned,  was  developed by \ac{KMOC} \cite{Kosower:2018adc}, as it is now familiar for us from \cref{sec:KMOC}, which allows to compute classical electromagnetic (gravitational) observables from the classical limit of quantum scattering amplitudes, as exemplified in previous chapter.  In this chapter we initiate a study of the implications of classical soft theorems for \ac{KMOC} formalism. As we will show, consistency with the leading classical soft theorem imposes an infinite hierarchy of  constraints on \ac{KMOC} observables. In order to state these constraints we introduce following conventions. 

The scattering process we consider is a $2\, \rightarrow\, 2$ scattering process in which two incoming charged particles with momenta $p_{1}, p_{2}$ and charges $Q_{1},\, Q_{2}$ scatter via electro-magnetic interactions as well as any other higher derivative interaction which is long range such that the \ac{KMOC} formalism applies to this scattering. As the classical soft theorems are universal and independent of the details of the scattering,  classical limit of the radiation kernel should generate the soft factors for \emph{any} perturbative amplitude involving charged particles in the external states and a photon. 

To each of the two massive particles we can associate certain classical observables defined as follows:\\
(1) Let  $V^{\mu}_{\alpha_{1}\, \cdots\, \alpha_{i}}$ be the  projection operator \eqref{eq:projector_general_s}, associated to particle $1$, that is
\begin{flalign}\label{eq:soft_projector_1}
V^{\mu}_{\alpha_{1}\, \cdots\, \alpha_{i}}\, =\, (-1)^{i+1}\, \Big[\, \frac{i!}{(p_{1} \cdot k)^{i}}\,\delta^{\mu}_{(\alpha_{1}}\, k_{\alpha_{2}}\, \cdots\, k_{\alpha_{i})}\, - \frac{1}{(p_{1} \cdot k)^{i+1}}\,p_{1}^{\mu}\, k_{\alpha_{1}}\, \cdots\, k_{\alpha_{i}}\, \Big]\,.
\end{flalign}
(2) Now consider certain  \textit{moments} of the exchange momenta 

\begin{flalign}\label{eq:moments}
\begin{array}{lll}
{\cal T}^{(n)\,\alpha_{1}\, \cdots\, \alpha_{i}}\, :=
 \,\hbar^{\frac{3}{2}}\Big[ i\int\, \hat{d}\mu_q\, q^{\alpha_{1}}\, \cdots\,  q^{\alpha_{i}}\,e^{-iq{\cdot}b} \, M^{(n)}(p_{1},p_2\, \rightarrow\, p_{1} - q, p_{2} + q)\\
\hspace*{3.5cm}+\, \sum_{X=0}^{n-1}\,\int\, \prod_{m=0}^X d\Phi(r_m) \,  \hat{d}\mu_q\, \hat{d}\mu_{w,X}  w^{\alpha_{1}}\, \cdots\, w^{\alpha_{i}}\,e^{-iq{\cdot}b} \\
\hspace{4.5cm} \times \sum_{a_{1}=0}^{n-1-X} M^{(a_{1}) }(p_{1},p_2\, \rightarrow\, p_{1} - w, p_{2} + w, \, r_{X})\,\\
\hspace*{4.5cm}\times M^{(n-a_{1} - X-1)\star}(p_{1} - w,p_2+w, \, r_{X}\, \rightarrow\, p_{1} - q,p_2+q)\Big]\, \,,
\end{array}
\end{flalign}
where $b$ is the impact parameter in the $2\to2$ scattering process. In the above equation we have introduced the following notations which will be used throughout this chapter. 
\begin{itemize}
\item $M^{(a)}( p_{1}^{i},\, p_{2}^{i}\, \rightarrow\, p_{1}^{f}, p_{2}^{f})$ is the (stripped) amplitude for a   4 point scattering at $a$-th  order in perturbation theory, and analogously for the other amplitudes with additional momentum labels. 
\item The  integral measure $\hat{d}\mu_q$ is defined via
\begin{equation}\label{eq:measure}
    \hat{d}\mu_q = \hat{d}^4q\hat{\delta}(2p_1{\cdot}q-q^2)\hat{\delta}(2p_2{\cdot}q+q^2),
\end{equation}
where $\hat{d}^4q =\frac{d^4q}{ (2\pi )^4}$, and the 
 hat on $\delta$-fn. indicates it is defined as
\begin{flalign}
\hat{\delta}(x)\, =\, -i\, [\, \frac{1}{x-i\epsilon}\, - \frac{1}{x + i\epsilon}\, ]\,,
\end{flalign}
and analogous  for $\hat{d}\mu_{w,X}$,
\begin{equation}\label{eq:measurex}
    \hat{d}\mu_{w,X} = \hat{d}^4q\hat{\delta}(2p_1{\cdot}w-w^2)\hat{\delta}(2p_2{\cdot}(w+r_X)+(w+r_X)^2)
\end{equation}
\item The sum over $X$ is a sum over number of intermediate photons with momenta $\{ r_{1}\, \cdots,\, r_{X}\, \}$. Even though integration over the momentum space of these photons is indicated explicitly by $d\Phi(r_m)=\hat{d}^4r_m\hat{\delta}^{(+)}(r_m^2)$, we assume that the sum over $X$ includes the sum over intermediate helicity states.
\item It is understood that for $n=0$, the second term in \eqref{eq:moments} vanishes. 
\end{itemize}

All of these conventions arise naturally from the \ac{KMOC} formalism of \cref{sec:KMOC}, and we will see how they naturally emerge in our computation below. 

As we will show in \cref{sec:weinberg_kernel}, consistency of \ac{KMOC} with the  classical leading soft photon theorem \cite{Saha:2019tub}, implies that at $n$-order in perturbative expansion  we have the following identities
\begin{itemize}
    \item $\forall\, n>0$ and $\forall\, 1\, \leq\, i\, \leq\, n$ :
\end{itemize}
\begin{equation}\label{hier1}
\begin{split}
 \lim_{\hbar \rightarrow\, 0}\, \hbar^{m}\, V^{\mu}_{\alpha_{1}\, \cdots\, \alpha_{i}}\, {\cal T}^{(n)\,\alpha_{1} \cdots \alpha_{i}}\, =\, 0\, \forall\, m\, \in\, \{1,\cdots,n+ 1-i\}\, ,
  \end{split}
\end{equation}
\begin{itemize}
    \item and  
$\forall\, n$ and $\forall\, 1\, \leq\, i\, \leq\, n\, +\, 1$ :
\end{itemize}
\begin{equation}\label{hier2}
    \lim_{\hbar\, \rightarrow\, 0}\, V^{\mu}_{\alpha_{1}\,{\cdots}\, \alpha_{i}}  {\cal T}^{\alpha_{1}\, \cdots\, \alpha_{i}}_{(n)}\, = e^{2(n+1)} \, V^{\mu}_{\alpha_{1}\, \cdots\, \alpha_{i}}\, \smashoperator{\sum_{L_{1}+\, {\cdots}+\, L_{i-1} =0}^{n}}\, (\triangle p_{1}^{L_{1}})^{\alpha_{1}}\, {\cdots}\, (\triangle p_{1}^{n {-} (L_{1} {+} \cdots {+} L_{i-1})})^{\alpha_{i}}  \,.
\end{equation}

The first set of identities  eqn.(\ref{hier1}), arise by demanding that  all the super-classical fragments in the radiated field vanish as mandated by consistency of \ac{KMOC} formalism. The second set  on the other hand (eqn. \eqref{hier2}),    relate the classical limit of the (expectation value) of the \textit{monomials} with perturbative coefficients of the classical soft factor. 
These constraints were shown to be satisfied at tree-level in the  earlier work of \cite{Bautista:2019tdr,Manu:2020zxl}, as we showed explicitly in \eqref{eq:a_weinerg_tree}, 
where at \ac{LO} in the coupling,  leading and sub-leading classical soft photon theorem was derived from \ac{KMOC} formula, which we will review in \cref{sec:tree-level-radiation} for the leading soft result.

Notice for $i=1$, and to any order in perturbation theory,  constraints \eqref{hier1} and \eqref{hier2} are trivial prove using \ac{KMOC} definition for the linear impulse \eqref{eq:impulse_final} .  That, is, for $i=1$, $\tau^{(n)\alpha}=\Delta p_1^{(n)\alpha}$, which is a well defined classical object, with no superclassical fragments in it. 
This also hints that identities \eqref{hier1} and \eqref{hier2}  might be valid removing the support of the projector tensors $V_{\alpha_i}$. We will check this explicitly to be true up to 1-loop below, but conjecture to be true to all orders in perturbation theory.  With the  removal of the $V$-projectors from these identities, they could  then be used to simplify complicate \ac{KMOC} expectation values integrals,   as we know the result is fixed by certain powers of linear impulse, at the desired perturbative order. 


This chapter  is organized as follows: In \cref{sec:classical_review} we  review perturbative results for classical soft photon theorem at leading and subleading orders in the soft expansion.  We then move to  the
derivation of identities \eqref{hier1} and \eqref{hier2} in \cref{sec:weinberg_kernel}.
In \cref{sec:lo-soft_costraitns} 
 we  show how the \ac{KMOC} formula indeed satisfies these constraints at leading \cref{sec:tree-level-radiation} and next to leading \cref{sec:1-loop_radiation} order in the coupling, by working with amplitudes in scalar QED.
That is, contribution of the tree-level and one loop amplitudes to  the $\frac{1}{\omega}$ coefficient of the radiative field indeed matches with eqn. \eqref{two}. Finally, in \cref{sec:discusion_} we provide some outlook of the chapter. 
 For computational details in \cref{sec:1-loop_radiation}, we refer the reader  to   appendix \ref{app:radiation_NLO}.

\section{Soft Radiation in Classical Scattering}\label{sec:classical_review}

In this section, we analyse the classical radiated soft  photon field at  leading order in soft expansion and \ac{NLO} in the coupling in terms of explicit expressions for the linear impulse. For that we  use the results in \cite{Saketh:2021sri}, in conjunction with classical soft theorem to write the radiative field at the desired order.  That is, we compute $A_{\mu}^{\omega^{-1}}(\hat{n})$ to \ac{NLO} in the coupling in a classical scattering involving two charged particles with masses $m_{1}, m_{2}$ which are interacting only via electro-magnetic interactions. 


\subsection{Leading soft factor}

\textit{Leading order  radiation}:

Let   $p_{i}\, \vert\, i = 1,2$ be the  momenta for in incoming massive particles, moving in the asymptotic free trajectories $b^\mu+v^\mu\tau$ in the far pass. If we denote the null vector $(1, \hat{n})$ as $n^{\mu}$ we can write the leading soft factor at tree level from formulas \eqref{two} and \eqref{eq:projector_general_s}, that is

\begin{equation}\label{eq:leading-soft-tree}
    A^{(0)\,\mu}_{\omega^{-1}}(\hat{n}) = e^3\sum_{i=1}^{2}Q_{i}\left[\frac{\Delta p_{i}^{(0)\,\mu}}{p_{i}{\cdot}n}-\frac{\Delta p_{i}^{(0)\,\mu}\cdot n}{(p_{i}{\cdot}n)^{2}}p_{i}^{\mu}\right]\,,
\end{equation}
where the leading order linear impulse we computed explicitly in previous section \eqref{eq:LO_Impulse_explicit}

 This  simple examples shows explicitly how the  radiated field to leading order in the soft expansion  is determined only from asymptotic data, and in  particular for perturbation theory, from only incoming data since the outgoing momenta are determined by the perturbative expansion \eqref{eq:expansion_momenta}, which we have truncated at leading order in the coupling. \\\\
\textit{Sub-Leading order radiation}

At  \ac{NLO}, the radiated field has a more interesting form, since as indicated in  \eqref{two} and \eqref{eq:projector_general_s}, both, the leading and subleading impulse enter into the field. Indeed, it explicitly reads 
\begin{equation}\label{eq:soft-one-loop}
    A_{\omega^{-1}}^{(1)\,\mu}(\hat{n}) =e^5 \sum_{i=1}^{2}Q_{i}\left[\frac{\Delta p_{i}^{(1)\,\mu}}{p_{i}{\cdot}n}-\frac{\Delta p_{i}^{(1)}\cdot n}{(p_{i}{\cdot}n)^{2}}p_{i}^{\mu}-\frac{\Delta p_{i}^{(0)}\cdot n}{(p_{i}{\cdot}n)^{2}}\Delta p_{i}^{(0)\,\mu}+\frac{(\Delta p_{i}^{(0)}\cdot n)^{2}}{(p_{i}{\cdot}n)^{3}}p_{i}^{\mu}\right]\,.
\end{equation}
At this order, it is still true that $\Delta p_{1}^{(1)\,\mu}=-\Delta p_{2}^{(1)\,\mu}$, where  the \ac{NLO} impulse was obtained similarly in the previous section \eqref{eq:p-def_justin}.

\section{KMOC  radiated photon field : A Soft Expansion}\label{sec:soft_constraints_deriv}
In this section we will study the $\mathcal{R}$ \eqref{eq:r-kernel} and $\mathcal{C}$ \eqref{eq:c-kernel} contributions to the radiated field in  the \ac{KMOC} formalism at leading and subleading order in the soft expansion in 4-dimensions. We will show that consistence of the \ac{KMOC} formalism with the soft theorems at the orders considered, generates a hierarchy of  constraints on the expectation value of several operators.

\subsection{Leading soft constraints }\label{sec:weinberg_kernel}

The aim of this part of the chapter is to  derive the set of identities  \eqref{hier1} and \eqref{hier2}. Our idea now is to use \ac{KMOC} formalism of \cref{sec:KMOC}, in conjunction with quantum soft theorems to obtain radiation kernel in the soft limit. In other words, we start with the exact formula for the radiated photon field in \ac{KMOC} form \eqref{eq:radi_field}. We then follow the theme from \cref{sec:esponential_soft}  where we showed that taking the soft limit before the classical limit generates the leading soft expansion of the radiated  field in \ac{KMOC} form at \ac{LO} in the perturbative expansion, we generalize this approach to all orders in perturbation theory.  That is, to a given order in soft expansion, we can apply quantum soft photon theorems to factorise the 5-point amplitude in terms of a 4 point amplitude and a soft factor. 

At higher orders in the loop expansion, one also has to take into account the order between loop integration and soft expansion.  If we first do a soft expansion and then loop integration, then one can use the tree-level soft theorems to factorise the loop integrand into a soft factor and a Four point integrand. However, as it was shown in a seminal paper by Sahoo and Sen \cite{Sahoo:2018lxl}, the two operations do not commute in Four dimensions beyond the leading order in soft expansion. That is, the soft expansion done after integrating over loop momenta results in $\ln$ soft factors  which are absent in the soft expansion of the loop integrand. 
At leading order however, this subtlety  does not enter as Weinberg soft photon theorem is a universal statement in all dimensions.

Let us then substitute the  soft photon theorem in eqns. (\ref{eq:r-kernel}, \ref{eq:c-kernel}), and use the momentum conserving delta functions to do the integrals in $q_2$ and $w_2$, we   get \footnote{We only consider the radiation emitted by the first particle, as the radiative field emitted by the outgoing particles is additive. We will denote this contribution as ${\cal R}^{\mu}_{1}(k)$.}  
\begin{equation}\label{eq:r-kerneln}
  \mathcal{R}_{1}^{(n)\,\mu}(k)=i\lim_{\hbar\to0}\hbar^{\frac{3}{2}}\int\hat{d}\mu_{q}e^{-ib{\cdot}q}S^{(0)\,\mu}(p_{1},q,k)M^{(n)}\left(p_{1},p_{2}{\rightarrow}p_{1}-q,p_{2}+q\right)\,.
\end{equation}
We have additionally  defined the impact parameter by  $b=b_2-b_1$, and used \eqref{eq:measure} to rewrite the momentum measure. Analogously, for the $\mathcal{C}$-term we have 
\begin{equation}\label{eq:c-kerneln}
\begin{split}\mathcal{C}_{1}^{(n)\,\mu}(k) & ={\color{black}}\lim_{\hbar\to0}\hbar^{\frac{3}{2}}\sum_{X=0}^{n-1}\int\prod_{m=0}^{X}d\Phi(r_{m})\hat{d}\mu_{q}\,\hat{d}\mu_{w,X}e^{-ib{\cdot}q}\\
 & \hspace{2cm}\,\,\,\,\,\,\,\,\,\,\,\,\times\sum_{a_{1}=0}^{n-1-X}S^{(0)\,\mu}(p_{1},w,k)M^{(a_{1})}(p_{1},p_{2}{\rightarrow}p_{1}{-}w,p_{2}{+}(w+r_{X}),r_{X})\\
 & \hspace{2cm}\,\,\,\,\,\,\,\,\,\,\,\,\,\times M^{(n-a_{1}-X-1)\star}(p_{1}{-}w,p_{2}{+}(w+r_{X}),r_{X}{\rightarrow}p_{1}-q_{1},p_{2}-q_{2})\,,
\end{split}
\end{equation}
with $\hat{d}\mu_{w,X}$ given in \eqref{eq:measurex}.

The Weinberg soft factor has the following ``quantum" expansion when expressed in terms of exchange momenta.
For the first particle (with charge and mass $Q_{1}, m_{1}$)
\begin{flalign}\label{hexpsoft}
S^{(0) \mu}(p_{1}, p_{1}{-}q, k)\, =\, Q_{1}\, \Big[\, \sum_{i=0}^{\infty}\, q^{\alpha_{1}}\, \cdots\, q^{\alpha_{i+1}}\, V_{\alpha_{1}\, \cdots\, \alpha_{i+1}}^{\mu}\, \Big]\,,
\end{flalign}
where $V^{\mu}_{\alpha_{1}\, \cdots\, \alpha_{i+1}}$ is defined in \eqref{eq:soft_projector_1}. 
We see that the $i$-th term inside the square bracket in \eqref{hexpsoft}, scales as $\hbar^{i}$ in the \ac{KMOC} sense.

We will now derive the constraints proposed in eqns. (\ref{hier1}) and \eqref{hier2} by associating the soft limit of the radiated field written in \ac{KMOC} form,  with the classical soft factor at all orders in the coupling.  The first contribution $ \mathcal{R}_{1}^{(n)\,\mu}(k)$ can be written as 
\begin{equation}\label{r-s-exp1}
\begin{split}
\mathcal{R}_{1}^{(n)\,\mu}(k)\, & =i\lim_{\hbar\to0}\hbar^{\frac{3}{2}}\,e\,Q_{1}\,e^{2(n+1)}\\
 & \,\,\,V_{1\,\alpha_{1}\,\cdots\,\alpha_{i+1}}^{\mu}\,\sum_{i=0}^{n  }\int\hat{d}\mu_{q}\,e^{-ib{\cdot}q}\,q^{\alpha_{1}}\,\cdots\,q^{\alpha_{i+1}}\,\bar{M}^{(n)}\left(p_{1},p_{2}{\rightarrow}p_{1}-q,p_{2}+q\right)\,,
\end{split}
\end{equation}
where the bar over the amplitude indicates that we have striped the coupling constant. Notice here we  have restricted the sum over $i$ at $n $ (where $n$ is the order of the loop expansion). This can be argued using $\hbar$ scaling arguments. The \ac{KMOC} scaling implies that
\begin{flalign}
\begin{array}{lll}
\hbar^{\frac{3}{2}}\, e^{2(n+1)}\, e\, \sim\, \frac{1}{\hbar^{n}}\,,\\
V^{\mu}_{1\,\alpha_{1}\, \cdots\, \alpha_{i+1}}\, q^{\alpha_{1}}\, \cdots\, q^{\alpha_{i+1}}\, \sim\, \hbar^{i}\,.
\end{array}
\end{flalign}
Additionally, we now notice that at $n$-th order in the loop expansion, the $\hbar$ scaling of the perturbative amplitude is quantified by \ac{KMOC} as follows 
\begin{equation}
    \begin{split}
\bar{M}^{(n)}\left(p_{1},p_{2}{\rightarrow}p_{1}-q,p_{2}+q\right)\, & =:\,{\cal I}^{(n)}_{4}(p_{1},p_{2}\,\rightarrow\,p_{1}-q,p_{2}+q)\\
 & \sim\,[\,\frac{1}{\hbar^{2}}\,+\,\frac{1}{\hbar}\,+\,O(\hbar^{0})\,]\,,\\
\hat{d}\mu_{q}\, & \sim\,\hbar^{4}[\frac{1}{\hbar^{2}}\,+O(\frac{1}{\hbar^{3}})\,]^{2}\,,\\
b{\cdot}q & \sim \hbar^0\,.
    \end{split}
\end{equation}

It can  be immediately verified that if  the sum $\sum_{i}$ in eqn. (\ref{r-s-exp1}) goes beyond $i = n$, the right hand side ($\hbar\, \rightarrow\, 0$ limit) vanishes. 
In fact, these scaling arguments can be used to immediately verify that the $\hbar$ expansion of the \textit{moments} are
\begin{equation}\label{eq:scaling_moments_R}
e^{2(n+1)}\, e\,Q_{1}\, \hbar^{\frac{3}{2}}\, V^{\mu}_{1\, \alpha_{1}\, \cdots\, \alpha_{i+1}}\, \int\, \hat{d}^{4}\mu_q\, e^{- i b \cdot q}\, (\, q^{\alpha_{1}}\, \cdots\, q^{\alpha_{i+1}}\, )\, \, \bar{M}^{(n)} =\, \sum_{\beta = 0}^{n -i}\, \frac{1}{\hbar^{\beta}}\, {\cal S}_{i\, \beta}^{\mu}\, + O(\hbar) \,.
\end{equation}
The contribution of $\mathcal{C}^{(n)\,\mu}(k)$ at leading order in  the soft limit can be analysed as in eqn.(\ref{r-s-exp1}). 
\begin{equation}\label{c-ker1}
\begin{split}\mathcal{C}_{1}^{(n)\,\mu}(k) & ={\color{black}}\lim_{\hbar\to0}\hbar^{\frac{3}{2}}\sum_{i=1}^{n}\, V^{\mu}_{\alpha_{1}\, \cdots\, \alpha_{i+1}}\, \sum_{X=0}^{n-1}\int\prod_{m=0}^{X}d\Phi(r_{m})\hat{d}\mu_{q}\,\hat{d}\mu_{w,X}e^{-ib{\cdot}q}\\
 & \hspace{2cm}\,\,\,\,\,\,\,\,\,\,\,\,\times\sum_{a_{1}=0}^{n-1-X}\, (w^{\alpha_{i_{1}}}\, \cdots\, w^{\alpha_{i+1}})\, M^{(a_{1})}(p_{1},p_{2}{\rightarrow}p_{1}{-}w,p_{2}{+}(w+r_{X}),r_{X})\\
 & \hspace{2cm}\,\,\,\,\,\,\,\,\,\,\,\,\,\times M^{(n-a_{1}-X-1)\star}(p_{1}{-}w,p_{2}{+}(w+r_{X}),r_{X}{\rightarrow}p_{1}-q_{1},p_{2}-q_{2})\,.
 \end{split}
 \end{equation}
 Once again, $\hbar$-scaling arguments can be used to immediately verify that the quadratic in amplitude  contribution to the moments at $e^{2n+3}$ order in the perturbative expansion can be written as
 \begin{flalign}
\begin{split}
e^{2(n+1)}\, g\,Q_{1}\, \hbar^{\frac{3}{2}}\, &V^{\mu}_{1\, \alpha_{1}\, \cdots\, \alpha_{i+1}}\, \sum_{X=0}^{n-1}\int\prod_{m=0}^{X}d\Phi(r_{m})\hat{d}\mu_{q}\,\hat{d}\mu_{w,X}e^{-ib{\cdot}q}\\
 & \hspace{2cm}\,\,\,\,\,\,\,\,\,\,\,\,\times\sum_{a_{1}=0}^{n-1-X}\, (w^{\alpha_{i_{1}}}\, \cdots\, w^{\alpha_{i+1}})\, \bar{M}^{(a_{1})}(p_{1},p_{2}{\rightarrow}p_{1}{-}w,p_{2}{+}(w+r_{X}),r_{X})\\
 & \hspace{2cm}\,\,\,\,\,\,\,\,\,\,\,\,\,\times\bar{M}^{(n-a_{1}-X-1)\star}(p_{1}{-}w,p_{2}{+}(w+r_{X}),r_{X}{\rightarrow}p_{1}-q_{1},p_{2}-q_{2})\\
&=\, \sum_{\beta=0}^{n -i}\, \frac{1}{\hbar^{\beta}}\, {\cal S}^{\prime\, \mu}_{i\, \beta}
 \end{split}
 \end{flalign}
 Here the sum over $X$ is constrained by the order (in the coupling) at which we are evaluating the $\mathcal{C}$-contribution.
 
We finally see that for each $i$, 
 \begin{flalign}
 V^{\mu}_{\alpha_{1}\, \cdots\, \alpha_{i+1}}\, {\cal T}^{(n)\,\alpha_{1}\, \cdots\, \alpha_{i+1}}\, =\, \sum_{\beta=0}^{n -i}\, \frac{1}{\hbar^{\beta}}\, ({\cal S}_{i\, \beta} + {\cal S}_{i\, \beta}^{\prime})^{\mu}\, +\, O(\hbar)\,.
 \end{flalign}
 
 Thus, at a given order in the perturbative expansion $V^{\mu}_{\alpha_{1}\, \cdots\, \alpha_{i+1}}\, {\cal T}^{(n)\,\alpha_{1}\, \cdots\, \alpha_{i+1}}$ has a hierarchy of  super-classical terms which scales as $\frac{1}{\hbar^{\beta}}\, \vert\, \beta \in\, \{1,\, \cdots\,, n  -i\}$.  As the classical limit in \ac{KMOC} formalism must be smooth, we thus conclude that to $n$-th order in the loop expansion and for each $i$, one has a tower of constraints which state that all the super-classical terms must vanish
\begin{flalign}
{\cal S}^{\mu}_{i\, \beta} + {\cal S}^{\prime}_{i\, \beta}\, =\, 0\, \forall\, \beta\, \in\, \{1,\, \cdots,\, n -i\}\,,\,\,\,\,\textcolor{black}{n>0 \, |\, 1\le i+1 \le n}.
\end{flalign}
This is precisely the first identity  \eqref{hier1}, written in a slightly different notation. 

We now analyse the classical $\beta = 0$ contribution explicitly. We can schematically write it in a form which makes the $\hbar$ scaling of various terms manifest. This can be done by isolating all the terms which do not have an $\hbar$ expansion. In particular:
(1) we separate the measure factor $\hat{d}\mu_{q}\, =\, \hat{d}^{4}q\, \hat{\delta}(q)$, and 
(2) we isolate all the measure factors over loop momenta and the $(n+1)$ massless propagators.
As can be checked, this implies that in the classical term, $\hat{\delta}(q)\, \mathcal{I}^{(n)}_4$ should scale as $\frac{1}{\hbar^{n+i-1}}$.

Let us illustrate this with ${\cal S}^{\mu}_{i\, \beta=0}\, =:\, {\cal S}^{\mu}_{i}$. 
\begin{flalign}
\begin{array}{lll}
{\cal S}^{\mu}_{i}\, =\\
e^{2(n+1)}\,g\, Q_{1}\, \hbar^{\frac{3}{2}}\, V^{\mu}_{1\, \alpha_{1}\, \cdots\, \alpha_{i+1}}\, \int\, \hat{d}^{4}q\, e^{- i b \cdot q}\, (\, q^{\alpha_{1}}\, \cdots\, q^{\alpha_{i+1}}\, )&\\  &\hspace*{-2.6in}\int\, \prod_{j=1}^{n}\hat{d}^{4}l_{j}\, \frac{1}{\prod_{m=1}^{n}l_{m}^{2}\, (\sum l_{m} - q)^{2}}\,  [\, \hat{\delta}(q)\,  {\cal I}^{(n)}_{4}(p_{1},\, p_{2}\, \rightarrow\, p_{1} - q,\, p_{2} + q)\, ]_{\frac{1}{\hbar^{n+2+i-1}}}\,.
\end{array}
\end{flalign}
One can write such a formal expression for ${\cal S}^{\mu \prime}_{i}$ analogously. 

The classical soft theorem is then a statement that $\forall\, n$ and $\forall\, 1\, \leq\, i\, \leq\, n + 1$, 
\begin{flalign}\label{csiid}
    {\cal S}^{(n)\,\mu}_{ i} + {\cal S}^{\prime (n)\, \mu}_{ i}\, =\, e^{2(n+1)}\, V^{\mu}_{s\, \alpha_{1}\, {\cdots}\, \alpha_{i+1}}\hspace{0.4in} \smashoperator{\sum_{\substack{L_{1}+\,\cdots+\,L_{i+1}=0\\
(L_{1}+L_{i+1})+(i+1)\,=n
}
}^{n+1-i}}\, (\, (\triangle p^{(L_{1})})^{\alpha_{1}}\, {\cdots}\, (\triangle p^{(L_{i+1})})^{\alpha_{i+1}}\, )\,,
\end{flalign}
which in turn recovers identity \eqref{hier2}.

In \cref{sec:tree-level-radiation} and \cref{sec:1-loop_radiation}, we verify identities \eqref{hier1} and \eqref{hier2}  up to subleading order in the perturbative expansion, i.e. $n=0$ and $n=1$. 

\subsubsection{Monomials of linear impulses}
In the previous section we expressed the soft radiation kernel as sum over certain classical \textit{moments}. Classical soft theorem implies that (expectation value) of each such \textit{moments} is sum over products of linear impulses. We can thus ask if ${\cal S}^{\mu}_{i} + {\cal S}^{\prime \mu}_{i}$ is an expectation value of certain observable. It is easy to see that the answer is indeed  affirmative. 
The tensor $V^{\mu}_{1\, \alpha_{1}\, \cdots\, \alpha_{i+1}}$ can be thought of as a map from symmetric rank $i+1$ tensor to a vector. It has a kernel spanned by $p_{1}^{\alpha_{1}}\, \cdots\, p_{1}^{\alpha_{i+1}}$. We can hence consider following quantum operators. 
Let 
\begin{flalign}
\Pi^{\mu}_{\alpha}\, =\, \delta^{\mu}_{\alpha} + \frac{1}{m_{1}^{2}}\, p_{1}^{\mu}\, p_{1 \alpha}\,.
\end{flalign}
Now consider a quantum operator, 
\begin{flalign}
{\cal P}_{1}^{\mu}\, =\, \Pi^{\mu}_{1\, \alpha}\, \hat{P}_{1}^{\alpha}\,,
\end{flalign}
with $\hat{P}_1$ the momentum operator for particle 1. 
The identities (given in eqn.(\ref{csiid}) implied by consistency with classical soft theorem is then a statement that 
\begin{flalign}
\begin{array}{lll}
\lim_{\hbar\rightarrow\, 0}\, V^{\mu}_{1\, \alpha_{1}\, \cdots\, \alpha_{i+1}}\langle\langle\, {\cal P}^{\alpha_{1}}\, \cdots\, {\cal P}^{\alpha_{i+1}}\, \rangle\rangle^{(n)}\ =
V^{\mu}_{s\, \alpha_{1}\, {\cdots}\, \alpha_{i+1}}\, \smashoperator{\sum_{\substack{L_{1}+\,\cdots+\,L_{i+1}=0\\
(L_{1}+L_{i+1})+(i+1)\,=n
}
}^{n+1-i}}\, (\, (\triangle p^{(L_{1})})^{\alpha_{1}}\, {\cdots}\, (\triangle p^{(L_{i+1})})^{\alpha_{i+1}}\, )
\end{array}\,.
\end{flalign}

\section{Leading Soft Constraints Verification up to \ac{NLO} }\label{sec:lo-soft_costraitns}
Let us in the remaining of this chapter to provide some specific tests for identities \eqref{hier1} and \eqref{hier2}, at leading ($n=0$) and subleading ($n=1$) orders in perturbation theory. 

\subsection{Tree-level leading soft \textit{moments}}\label{sec:tree-level-radiation}

We have already shown in \cref{sec:examplesa_tree_level} that at tree level and to leading order in the soft expansion, the \ac{KMOC} formula for radiation recovers the  result from the  classical leading  soft photon theorems. Let us however comment on this using the language of the soft constraints. 
At tree-level ($n=0$),  there is not superclassical term and therefore \eqref{hier1} does not impose any constrain. 
On the other hand, since the number of exchange momenta, $i$,  is bounded by $n$, via $i \le n+1$, at this order  there is only one 
 classical   \textit{moment} 
contributing, that is for $i=1$,  only  ${\cal T}^{(0)\,\alpha}$ survives in \eqref{eq:moments}. Proving the soft constraints means then  that in the classical limit  we just need to show  $\lim_{\hbar\to0}{\cal T}^{(0)\,\alpha}= e^2\,\Delta p_1^{(0),\alpha}$ as required by \eqref{hier2}.
We have already pointed out in the discussion around \eqref{hier2}, that for $i=1$, to any order in perturbation theory, the soft constraints is trivially to prove by means of the definition of the linear impulse \eqref{eq:impulse_final}. To be more precise, let us see how this emerge from our definition of the moments \eqref{eq:moments}. 

Since we are taking the classical limit, the following expansion for the momentum measure \eqref{eq:measure} will be useful for us
\begin{align}
\hat{d}\mu_q & =\hat{d}\mu_{1\,q}+\hat{d}\mu_{2\,q}+\cdots,\label{eq:full measure}\\
\hat{d}\mu_{1\, q} & =\hat{d}^{4}q\,\hat{\delta}(2p_{1}{\cdot}q)\hat{\delta}(2p_{2}{\cdot}q),\label{eq:hbar0 measure}\\
\hat{d}\mu_{2\,q }& =-\,\hat{d}^{4}q\,q^{2}\left[\hat{\delta}'(2p_{1}{\cdot}q)\hat{\delta}(2p_{2}{\cdot}q)-\hat{\delta}(2p_{1}{\cdot}q)\hat{\delta}'(2p_{2}{\cdot}q)\right].\label{eq:hbar1measure}
\end{align}
To compute ${\cal T}^{(0)\,\alpha}$, we will  need the classical piece of the  tree-level 4 point amplitude, given in \eqref{eq:M4sqedclass}. 
With all these ingredients at hand, the only non-vanishing contribution to the   \textit{moment} ${\cal T}^{(0)\,\alpha}$ in the classical limit,  can be obtained by replacing  \eqref{eq:hbar0 measure} and \eqref{eq:M4sqedclass} into \eqref{eq:moments}, after which it follows 
\begin{equation}\label{eq:moment0}
\begin{split}
  \lim_{\hbar\, \rightarrow\, 0}\, V^{\mu}_{\alpha}  
    {\cal T}^{(0)\,\alpha} &=  e^2V^{\mu}_{\alpha}   \int \hat{d}^4q \hat{\delta}(p_1{\cdot}q)\hat{\delta}(p_2{\cdot}q) \frac{iQ_1Q_2\,p_1{\cdot}p_2q^\alpha}{q^2+i\epsilon} e^{-i q{\cdot}b}\,,\\
    & =e^2V^{\mu}_{\alpha}   \Delta p_1^{(0)\,\alpha}\,,
   \end{split} 
\end{equation}
which indeed satisfies the identity \eqref{hier2} for $n=0$, as expected. In the second line we have used \eqref{eq:LO_impulse_integral} to identify the \ac{LO} linear impulse.

\subsection{One-loop leading soft \textit{moments}}\label{sec:1-loop_radiation}

At \ac{NLO} in the perturbative  expansion  the contributing 
\textit{moments} are  ${\cal T}^{(1)\,\alpha}$ and  ${\cal T}^{(1)\,\alpha\,\beta}$.  
We then need to show that  $\lim_{\hbar\to0}{\cal T}^{(1)\,\alpha} = e^4 \Delta p^{(1)\,\alpha}_{1} $, recovering  the \ac{NLO} impulse, whereas $ \lim_{\hbar\to0}{\cal T}^{(1)\,\alpha\,\beta} = e^4 \Delta p^{(0)\,\alpha}_{1} \Delta p^{(0)\,\beta}_{1}$, as suggested by the second identity  \eqref{hier2}. Combination of these two results allow us to recover the one loop contribution to the  radiated field given explicitly in 
\eqref{eq:soft-one-loop}.
Of course the verification of the soft constraints for the former are trivial as we have pointed out several times, whereas for the latter there is more work to do.

At  \ac{NLO},  the radiated field scales as $e^5$ and therefore
the \textit{moments} receive contributions from both the $\mathcal{R}$ and the $\mathcal{C}$ terms, given by the first and second line of \eqref{eq:moments}, respectively. However, at this order no  extra photons propagate through  the cut and we can simply set $X=0$  in \eqref{eq:moments}, which also implies that $\hat{d}\mu_{w,X}=\hat{d}\mu_{w}$ in \eqref{eq:measurex}. 
In addition, we  will  show that superclassical terms give vanishing contribution as suggested by the first identity  \eqref{hier1}. Indeed, this corresponds to  a cancellation between the  $\mathcal{R}$ and the $\mathcal{C}$ contributions to the aforementioned  \textit{moments}, which are consequence of the cancellations of the superclassical terms for  the computation of the   1-loop impulse  \cite{Kosower:2018adc}. Since only the  \textit{moment}  ${\cal T}^{(1)\,\alpha}$ will have potential superclassical contributions, coming from the superclassical piece of the 4 point amplitude at 1-loop \cite{Kosower:2018adc}, we only have to show that for $m=1$, $\lim_{\hbar\to0}\hbar^m\,{\cal T}^{(1)\,\alpha}=0$, as for higher values of $m$, this identity is trivially satisfied.

Let us split the computation as follows: For the potentially superclassical contributions we will compute

\begin{equation}\label{eq:super_class_total}
    \lim_{\hbar\to0}\hbar\,V_\alpha^\mu{\cal T}^{(1)\,\alpha} = \lim_{\hbar\to0}\,V_\alpha^\mu\Big[ {\cal T}^{(1)\,\alpha}_{\mathcal{R}_0} +   {\cal T}^{(1)\,\alpha}_{\mathcal{C}_0}\Big]\,,
\end{equation}
where
\begin{eqnarray}
{\cal T}^{(1)\,\alpha}_{\mathcal{R}_0}  & =& i \hbar^{5/2} \int\hat{d}\mu_{1\,q}\, e^{-ib{\cdot}q}\,q^\alpha\,M^{(1)}_{\rm{sc}}(q)\,,\label{eq:R0-1} \\
 {\cal T}^{(1)\,\alpha}_{\mathcal{C}_0} & =&  \hbar^{5/2}
 \int\hat{d}\mu_{1\,q}\,\hat{d}\mu_{1\,w}\,e^{-ib{\cdot}q}\,w^\alpha M^{(0)\,\star}(w-q)M^{(0)}(w)\,.\label{eq:C0 definition-2}
\end{eqnarray}
Here $M^{(1)}_{\rm{sc}}(q)$ is the superclassical piece of the 1-loop, 4 point amplitude, which we will write explicitly below. The tree level amplitudes in the second line are  given by \eqref{eq:M4sqedclass}, where we have removed the massive momenta labels to alleviate notation.

Next, we will have to compute the classical  contributions, from the one and two index moment. For the former we have 
\begin{equation}\label{eq:1_index_moment_one_loop}
    \lim_{\hbar\to0}\,V_\alpha^\mu{\cal T}^{(1)\,\alpha} = \lim_{\hbar\to0}\,V_\alpha^\mu\Big[ {\cal T}^{(1)\,\alpha}_{\mathcal{R}_1}+{\cal T}^{(1)\,\alpha}_{\mathcal{R}_2} +   {\cal T}^{(1)\,\alpha}_{\mathcal{C}_1}+  {\cal T}^{(1)\,\alpha}_{\mathcal{C}_2}\Big]\,,
\end{equation}
with each term computed as follows
\begin{eqnarray}
 {\cal T}^{(1)\,\alpha}_{\mathcal{R}_1} &= & i \hbar^{3/2} \int\hat{d}\mu_{1\,q}\, e^{-ib{\cdot}q}\,q^\alpha\,M^{(1)}_{\rm{c}}(q)\,,\label{eq:R1-1}\\
{\cal T}^{(1)\,\alpha}_{\mathcal{R}_2}  &=&i \hbar^{3/2}\, \int\hat{d}\mu_{2\,q}\,e^{-ib{\cdot}q} \,q^\alpha M^{(1)}_{\rm{sc}}(q)\,,\label{eq:R2-1}\\
 {\cal T}^{(1)\,\alpha}_{\mathcal{C}_1} &=&   \hbar^{3/2}
 \int\hat{d}\mu_{1\,q}\,\hat{d}\mu_{2\,w}\,e^{-ib{\cdot}q}\,w^\alpha M^{(0)\,\star}(w-q)M^{(0)}(w)\,,\label{eq:C1 definition-2} \\
 {\cal T}^{(1)\,\alpha}_{\mathcal{C}_2} &=&  \hbar^{3/2}
 \int\hat{d}\mu_{2\,q}\,\hat{d}\mu_{1\,w}\,e^{-ib{\cdot}q}\,w^\alpha M^{(0)\,\star}(w-q)M^{(0)}(w)\,.\label{eq:C2 definition-2} 
\end{eqnarray}
In the first line, $M^{(1)}_{\rm{c}}(q)$ is the classical part of the 1-loop 4 point amplitude, which we  will write  explicitly in a moment. 

Finally, the classical contribution from the  two-index \textit{moment} will be computed from
\begin{equation}\label{eq:two-index-moment-1-loop}
     \lim_{\hbar\to0}\,V_{\alpha\,\beta}^\mu{\cal T}^{(1)\,\alpha\,\beta} = \lim_{\hbar\to0}V_{\alpha\,\beta}^\mu\,\Big[ {\cal T}^{(1)\,\alpha\,\beta}_{\mathcal{R}_3}+   {\cal T}^{(1)\,\alpha\,\beta}_{\mathcal{C}_3} \Big]\,,
\end{equation}
with the respective terms evaluated via
\begin{eqnarray}
  {\cal T}^{(1)\,\alpha\,\beta}_{\mathcal{R}_3} &=&  i \hbar^{3/2} \int\hat{d}\mu_{1\,q}\, e^{-ib{\cdot}q}\,q^\alpha\,q^\beta \,M^{(1)}_{\rm{sc}}(q)\,,\label{eq:R3-1} \\
  {\cal T}^{(1)\,\alpha\,\beta}_{\mathcal{C}_3} &=&  \hbar^{3/2}
 \int\hat{d}\mu_{1\,q}\,\hat{d}\mu_{1\,w}\,e^{-ib{\cdot}q}\,w^\alpha\,w^\beta M^{(0)\,\star}(w-q)M^{(0)}(w)\,,\label{eq:C3 definition-2}
\end{eqnarray}

By explicit evaluation,  we will show that the actual terms contributing to  the radiated photon field are  \eqref{eq:R1-1}, \eqref{eq:R2-1} and  \eqref{eq:R3-1} -- as suggestively written  in \eqref{eq:scaling_moments_R} --  with the first two giving the \ac{NLO} impulse, and the last one giving the square of the leading order impulse. As for the   remaining contributions we show that they canceling among themselves. In what follows we will adventure  in this  computation.

\subsubsection{The superclassical fragments}
Let us start by computing the superclassical fragments   \eqref{eq:R0-1} and \eqref{eq:C0 definition-2}.  As we will see, these terms are IR divergent, in analogy to the IR divergent integrals appearing in the computation of the 2\ac{PM} two-body potential \cite{Cheung:2018wkq,Bern:2019crd}, and the cancellation here is the \ac{KMOC} analog of the cancellation for the \ac{EFT} and full theory amplitudes matching \cite{Cheung:2018wkq,Bern:2019crd}. Indeed, we will see that analogous comparisons follow for the different terms appearing in the 2\ac{PM} two-body potential as we will see below

The 4 point amplitude at 1-loop was computed in \cite{Kosower:2018adc}. The superclassical contribution $ M^{(1)}_{\rm{sc}}(q)$, arises from the addition of superclassical parts in the box $B_{-1}$, and cut-box $C_{-1}$ diagrams, given by  eq. $(5.31)$ in \cite{Kosower:2018adc}, 
\begin{equation}\label{eq:A4_super_classical}
   M^{(1)}_{\rm{sc}}(q) = \left(B_{-1}+C_{-1}\right)_{\hbar^{-1}}=2i\,e^4\left(Q_{1}Q_{2}\,p_{1}{\cdot}p_{2}\right)^{2}\int\hat{d}^{4}l\prod_{i}\hat{\delta}(p_{i}{\cdot}l)\frac{1}{l^{2}(l-q)^{2}}\,.
\end{equation}
Using  it into  \eqref{eq:R0-1}, together with the measure  \eqref{eq:hbar0 measure},  \eqref{eq:R0-1} becomes
\begin{equation}\label{eq:supcR}
{\cal T}^{(1)\,\alpha}_{\mathcal{R}_0} =  - \frac{1}{2}e^4\left(Q_{1}Q_{2}\,p_{1}{\cdot}p_{2}\right)^{2} \int\hat{d}^{4}l\hat{d}^{4}q\prod_{i}\hat{\delta}(p_{i}{\cdot}l)\hat{\delta}(p_{i}{\cdot}q)\frac{q^{\alpha}}{l^{2}(l-q)^{2}}e^{-ib{\cdot}q}\,,
\end{equation}
where we see that the explicit dependence in $\hbar$ drops away by using the \ac{KMOC} $\hbar$-rescaling mentioned in \cref{sec:KMOC}.  We can now do the change of variables $q=l+\bar{q}$. This in turn factorizes the integrals into two factors corresponding to a vector, and a scalar integrals; that is
\begin{equation}\begin{split}
{\cal T}^{(1)\,\alpha}_{\mathcal{R}_0} =ie^4\left(Q_{1}Q_{2}\,p_{1}{\cdot}p_{2}\right)^2S_{\alpha,\omega^{-1}}^{\mu}\left[\int\hat{d}^{4}\bar{q}\prod_{i}\hat{\delta}(p_{i}{\cdot}q)e^{-ib{\cdot}\bar{q}}\frac{i\bar{q}^{\alpha}}{\bar{q}^{2}}\right]
 \left[\int\hat{d}^{4}l\prod_{i}\hat{\delta}(p_{i}{\cdot}l)\frac{e^{-ib{\cdot}q}}{l^{2}}\right],
\end{split}
\label{eq:super classical intermediate}
\end{equation}
where  the change of variables has produced a factor of $2$ that
canceled the  $\frac{1}{2}$ overall factor in \eqref{eq:supcR} \footnote{Note that formally the change of variables implies that  we should
had changed $\hat{\delta}(p_{i}.q)\rightarrow\hat{\delta}(p_{i}.\bar{q}-p_{i}{\cdot}l),$
however, the delta functions $\hat{\delta}(p_1{\cdot}q)$ allow us to set
$p_{i}{\cdot}l\rightarrow0$.}. In the integral on the left, we recognize the leading order impulse $(\ref{eq:LO_impulse_integral})$, whereas  for the integral  on
the right, we obtain an IR-divergent expression, which can be evaluated along  similar steps used for the  computation of the leading order impulse \eqref{eq:IR_impulse_interm}, obtaining

\begin{equation}
I_{1}=\int\hat{d}^{4}l\prod_{i}\hat{\delta}(p_{i}{\cdot}l)\frac{e^{-ib{\cdot}l}}{l^{2}}=-\frac{1}{4\pi\sqrt{\mathcal{D}}}\ln\left(-\mu^{2}b^{2}\right)\,.\label{eq:scalar integral I1}
\end{equation}
Here we have   introduced the   IR-regulator $\mu$. Then,  the first superclassical contribution becomes 
\begin{equation}
{\cal T}^{(1)\,\alpha}_{\mathcal{R}_0}=-ie^4 \frac{Q_{1}Q_{2}\,p_{1}{\cdot}p_{2}}{4\pi\sqrt{\mathcal{D}}} \,\Delta p_{1}^{(0)\,\alpha}\,\ln\left(-\mu^{2}b^{2}\right),\label{eq:super classical final}
\end{equation}

Let us now evaluate the $\mathcal{C}$-contribution \eqref{eq:C0 definition-2}. For that we just need the tree-level 4 point amplitude \eqref{eq:M4sqedclass}, as well as the measure factors \eqref{eq:hbar0 measure}; we arrive at
\begin{equation}
   {\cal T}^{(1)\,\alpha}_{\mathcal{C}_0}  =16e^4\left(Q_{1}Q_{2}p_{1}{\cdot}p_{2}\right)^{2}\,\int\hat{d}^{4}q\hat{d}^{4}l\prod_i\hat{\delta}(2p_{i}{\cdot}q)\hat{\delta}(2p_{i}{\cdot}l)e^{-ib{\cdot}q}\frac{l^\alpha}{l^{2}(q-l)^{2}}\,.
\end{equation}
After doing the same change of variables  $q=l+\bar{q}$, we can analogously identify the leading order  impulse from the $l$-integral, whereas the $\bar{q}$-integral will result into the  IR-divergent expression  \eqref{eq:scalar integral I1}. We finally get
\begin{equation}
     {\cal T}^{(1)\,\alpha}_{\mathcal{C}_0}  =ie^4\frac{Q_{1}Q_{2}\,p_{1}{\cdot}p_{2}}{4\pi\sqrt{\mathcal{D}}} \,\Delta p_{1}^{(0)\,\alpha}\,\ln\left(-\mu^{2}b^{2}\right),
\end{equation}
which is equal to \eqref{eq:super classical final} but with opposite sign. This explicitly  shows that the r.h.s of \eqref{eq:super_class_total} evaluates to  zero,  as demanded from the first identity  \eqref{hier1}.

\subsubsection{Classical one-index \textit{moment} at 1-loop}\label{sec:one-index-1-loop}
Let us move to evaluate the classical contribution from the one-index \textit{moment} \eqref{eq:1_index_moment_one_loop}. We start from term \eqref{eq:R1-1}. For that,  we need the classical contribution to 4 point  amplitude at 1-loop. Likewise for the superclassical term, we obtain it from the sum  $(B_{0}+C_{0})+\left(B_{-1}+C_{-1}\right)+T_{12}+T_{21}$, where the different components where evaluated in eqs.  $(5.21)$ and $\,(5.34)$  in \cite{Kosower:2018adc}. This gives

\begin{align}
M_{\textrm{c}}^{(1)}(q) & =2e^4\left(Q_{1}Q_{2}\,p_{1}{\cdot}p_{2}\right)^{2}\int\frac{\hat{d}^{4}l}{l^{2}(l{-}q)^{2}}\Bigg\{ l{\cdot}(l{-}q)\left[\frac{\hat{\delta}(p_{2}{\cdot}l)}{(p_{1}{\cdot}l{+}i\epsilon)^{2}}{+}\frac{\hat{\delta}(p_{1}{\cdot}l)}{(p_{2}{\cdot}l{-}i\epsilon)^{2}}\right]\nonumber \\
 & +\frac{1}{\left(p_{1}{\cdot}p_{2}\right)^{2}}\left[m_{2}^{2}\hat{\delta}(p_{2}{\cdot}l){+}m_{1}^{2}\hat{\delta}(p_{1}{\cdot}l)\right]\Bigg\}+Z,\label{eq:4pt 1L classical explicit}
\end{align}
with
\begin{equation}
Z=ie^4\left(Q_{1}Q_{2}\,p_{1}{\cdot}p_{2}\right)^{2}\int\frac{\hat{d}^{4}l}{l^{2}(l{-}q)^{2}}(2l{\cdot}q{-}l^{2})\left[\hat{\delta}'(p_{1}{\cdot}l)\hat{\delta}(p_{2}{\cdot}q){-}\hat{\delta}(p_{1}{\cdot}l)\hat{\delta}'(p_{2}{\cdot}l)\right].\label{eq:Z mu contribution to boxes}
\end{equation}

By introducing  all these definitions we can check that the computation of $ {\cal T}^{\alpha}_{(1),\mathcal{R}_1}$ in 
 in \eqref{eq:R1-1}, toghether with the measure \eqref{eq:hbar0 measure}, can be rearrange to give exactly the \ac{NLO} impulse \eqref{eq:NLO impulse}  plus an additional contribution coming from adding and subtracting the 4-pt cut-Box diagram 
\begin{equation}
{\cal T}^{(1)\,\alpha}_{\mathcal{R}_1} = e^4\Delta p^{(1)\,\alpha}_{1} + {\rm [cut{-}box]^{(1)\,\alpha},}\label{eq:R1_final}
\end{equation}
with  the extra  contribution  ${\rm [cut{-}box]^{\mu}}$   given
by 
\begin{equation}
{\rm [cut{-}box]^{(1)\,\alpha}}=- e^4
     \left[p_{1,\beta}p_{1}^{[\beta}p_{2}^{\alpha]}-p_{2,\beta}p_{2}^{[\beta}p_{1}^{\alpha]}\right]\frac{\left(\Delta p^{(0)}_{1}\right)^{2}}{\mathcal{D}}\,,\label{eq:cut-box final}
\end{equation}
 The proof of this statement is lengthy  and we therefore  postpone it to be discussed  in   Appendix \ref{app:cut_box}.   
For the moment, let us notice that the first term of  eq. \eqref{eq:R1_final}
gives exactly the expected result from the second identity  \eqref{hier2}. Therefore, to conclude the proof we simply need to show that the remaining terms in \eqref{eq:1_index_moment_one_loop} together with \eqref{eq:cut-box final}, add up to zero. In fact, also in Appendix  \ref{app:cut_box} we will show that 
\begin{equation}\label{eq:cancel_r2_cb_c2}
   {\cal T}^{(1)\,\alpha}_{\mathcal{R}_2} + {\rm [cut{-}box]^{(1)\,\alpha}} = 0\,,
\end{equation}
whereas ${\cal T}^{(1)\,\alpha}_{\mathcal{C}_1}$ and ${\cal T}^{(1)\,\alpha}_{\mathcal{C}_2}$ evaluate to zero individually.

\subsubsection{Classical two-index \textit{moment} at 1-loop}\label{sec:r3}
The remaining task to complete the proof of identity \eqref{hier2} at 1-loop is to evaluate two-index \textit{moment}  \eqref{eq:two-index-moment-1-loop}. Similar to previous computation, we start  from its first term, given by  \eqref{eq:R3-1}, and  after inserting the measure \eqref{eq:hbar0 measure},  and the superclassical amplitude \eqref{eq:A4_super_classical}, we arrive at
\begin{equation}\label{eq:r3_start}
    {\cal T}^{(1)\,\alpha\,\beta}_{\mathcal{R}_3} =-\frac{1}{2}e^4\left(Q_{1}Q_{2}\,p_{1}{\cdot}p_{2}\right)^{2}   \int\hat{d}^{4}q\hat{d}^{4}l\prod_{i}\hat{\delta}(p_{i}{\cdot}q)\hat{\delta}(p_{i}{\cdot}l)e^{-ib{\cdot}q}\frac{q^{\alpha}q^{\beta}}{l^{2}(l-q)^{2}}\,.
\end{equation}
Next we can do our usual change of variables $q = l+\bar{q}$ 
\begin{equation}\label{eq:r3_start-1}
    {\cal T}^{(1)\,\alpha\,\beta}_{\mathcal{R}_3}=-\frac{1}{2}e^4\left(Q_{1}Q_{2}\,p_{1}{\cdot}p_{2}\right)^{2}  \int\hat{d}^{4}\bar{q}\hat{d}^{4}l\prod_{i}\hat{\delta}(p_{i}{\cdot}\bar{q})\hat{\delta}(p_{i}{\cdot}l)e^{-ib{\cdot}\bar{q}}e^{-ib{\cdot}l}\frac{\left(l^{\alpha}{+}\bar{q}^{\alpha}\right)\left(l^{\beta}{+}\bar{q}^{\beta}\right)}{l^{2}\bar{q}^{2}}\,.
\end{equation}
We  recognize the square of the leading order impulse \eqref{eq:LO_impulse_integral} coming from the crossed terms. On the other hand, the non-crossed terms give us the product of two integrals, one is them is the  usual IR-divergent
 integral  $I_{1}$ in $(\ref{eq:scalar integral I1})$, whereas the second one corresponds to the derivative 
of the leading order impulse  w.r.t. the impact parameter; notice  there is a factor of two for each case, which cancels the overall $1/2$ factor. That is 
\begin{equation}
   {\cal T}^{(1)\,\alpha\,\beta}_{\mathcal{R}_3} =  e^4{\color{black}} 
    \Delta p^{(0)\,\alpha}_{1}  \Delta p^{(0)\,\beta}_{1} 
   -e^4\left(Q_{1}Q_{2}\,p_{1}{\cdot}p_{2}\right)I_{1}\partial_{b^{\alpha}} \Delta p^{(0)\,\beta}_{1} \,.\label{eq:R3 interm2}
\end{equation}
The change of the  sign  for the first term comes from inserting a factor of $ i^2$ both, in the numerator and denominator, and absorb it  for the former, to complete the   square of the leading order impulse. 
Using $(\ref{eq:scalar integral I1})$ and the derivative of the
leading order impulse 
\begin{equation}
\partial_{b^{\alpha}} \Delta p^{(0)\,\beta}_{1} =-\frac{Q_{1}Q_{2}p_{1}{\cdot}p_{2}}{2\pi\sqrt{\mathcal{D}}}\left(b^{2}\eta^{\alpha\beta}-2b^{\alpha}b^{\beta}\right)\frac{1}{b^{4}}\,,\label{eq:derivative lo-impulse wrt b}
\end{equation}
 and  dropping the term proportional to $\eta^{\alpha\beta}$ , using
the on-shell condition for the photon momentum and gauge invariance,
we finally arrive at 
\begin{equation}
   {\cal T}^{(1)\,\alpha\,\beta}_{\mathcal{R}_3} =e^4 \Delta p_{1}^{(0)\,\alpha}\Delta p_{1}^{(0)\,\beta}+\mathcal{J}_{3}^{(1)\,\alpha\beta}\,,\label{eq:R3 final}
\end{equation}
 where 
\begin{equation}
\mathcal{J}_{3}^{(1)\,\alpha\,\beta}= -2e^4\,\Delta p_{1}^{(0)\,\alpha}\Delta p_{1}^{(0)\,\beta}\ln\left(-\mu^{2}b^{2}\right)\,. \label{eq:J3}
\end{equation}

Similar to the previous subsection, to complete the proof of the second identity for the two-index \textit{moment} at 1-loop, we simple need to show that the second term in \eqref{eq:two-index-moment-1-loop} added to \eqref{eq:J3} evaluates to zero

\begin{equation}\label{eq:final_cancelations}
{\cal T}^{(1)\,\alpha\,\beta}_{\mathcal{C}_3} +\mathcal{J}_{3}^{(1)\,\alpha\,\beta} = 0\,.
\end{equation}
We leave the proof of this equation for Appendix \ref{app:cancelations_nlo_radiation}.

With this we have  concluded the proof of identity \eqref{hier2} at \ac{NLO} in the perturbative expansion. 
Let us notice that the appearance of the square of the leading order impulse is a result of  $q$-expansion of the  Weinberg soft factor, iterated with the superclassical contributions from the box and cross box diagrams. However, remnants from the IR-divergent contributions as appearing in \eqref{eq:J3}, are nicely canceled by the $\mathcal{C}-$contribution to the radiated field, in analogy to the cancellation of IR-divergent integrals from the \ac{EFT} and full theory amplitudes matching \cite{Cheung:2018wkq,Bern:2019crd}.

\section{Outlook of the chapter}\label{sec:discusion_}

In this chapter, we have  analyse implications of classical soft theorems for \ac{KMOC} formalism through which radiative field can be computed using on-shell techniques. As we have argued, classical soft theorems impose a tower of an  infinite hierarchy of constraints  on expectation values of a class of composite operators in the \ac{KMOC} formalism. At leading order in the soft expansion, these operators are constructed from monomials of momentum operators. 

At leading order in perturbation theory, these constraints were verified in \cite{Bautista:2019tdr}, as we show explicitly in \cref{sec:examplesa_tree_level} and \cref{sec:tree-level-radiation},  and at sub-leading order in the soft expansion by \cite{Manu:2020zxl}. In this chapter,  we have  also verified them at \ac{NLO} in the coupling and at leading order in the soft expansion in scalar QED with no higher-derivative interactions.  We note that addition of other interactions will not change the structure of classical soft factor but will change the analytic expressions for the outgoing momenta in terms of incoming kinematics and impact parameter.  Verifying the sub-leading soft constraints at higher orders in the perturbative expansion requires a deeper investigation into the integration regions involving the loop momenta. This analysis out of the scope of this thesis. 

At  \ac{NLO}, the verification of the leading soft constraint is analogous to the \ac{EFT} and full theory amplitudes matching procedure for the computation of the 2\ac{PM} two-body potential \cite{Cheung:2018wkq,Bern:2019crd}. A difference between  the two computations is in the treatment of  super-classical terms. In the soft constraints derived from \ac{KMOC} formalism, the IR divergent terms  cancel by the addition of the $\mathcal{C}$-contributions to the radiative field \eqref{c-ker1}, in contrast to the  matching procedure. We have also seen that the powers of the leading order impulse were the analogs to the iterated tree-level amplitudes appearing in the 2\ac{PM} potential. Furthermore,  contribution to the \ac{NLO} impulse coming from the triangle  and cut-box integrals have the respective counterpart in the 2\ac{PM} potential. Viewed in this light, the classical soft theorems impose constraints in the conservative dynamics of the two-body problem. Indeed, once the frequency of the radiated photon (graviton) is fixed, soft-theorems become an statement on the conservative sector. At 3\ac{PM} for instance, the appearance of iterative 1-loop and tree-level contributions  to the potential \cite{Bern:2019crd,Bjerrum-Bohr:2021din, Kalin:2020fhe}, will be the analogs of products of the form $\Delta p^{(1)}{\cdot}\Delta p^{(0)}$, appearing at two loops in \eqref{two}, in addition to the cubic appearance of the tree-level amplitude, which will be the analog of $(\Delta p^{(0)})^3$, and analogously for the 4\ac{PM} result \cite{Bern:2021dqo,Dlapa:2021npj}

In this chapter we have solely focused on soft  electromagnetic radiation. We believe that the leading soft constraints can be generalised to gravitational interactions directly at \ac{NLO}. Beyond \ac{NLO} order, classical soft graviton factor will receive contribution from finite energy gravitational flux. On the other hand if we take classical limit after applying Weinberg soft theorem inside the radiation kernel, the result will be once again turn out to be in terms of monomials of linear impulses.  We believe that this result once again should be equated to the contribution to the classical soft graviton factor only from outgoing massive particles. However this remains to be shown. As \ac{KMOC} naturally takes into account the dissipative effects in computation of linear impulse, we expect this procedure to be consistent. \footnote{We thank Ashoke Sen for discussion on this issue.} \footnote{The generalisation of the sub-leading soft constraints may be even more subtle as the classical log soft factor in gravity has an additional contribution effect of space-time curvature on soft radiation. These terms may not simply arise from sub-leading soft graviton theorem for the integrands \cite{Sahoo:2018lxl}.}

It will be interesting to prove the leading and sub-leading soft constraints within \ac{KMOC} formalism  for perturbative scattering with large impact parameter. Universality of classical soft theorems imply that the proof is likely to involve ideas along the lines of the classical proof in  \cite{Saha:2019tub}, in which it was only assumed that the interactions outside a ``hard scattering region" (which can be parametrized as a space-time region bounded in spatial and temporal directions by $\pm\, t_{0}$ for some sufficiently large $t_{0}$) are simply the Coulombic interactions. However formulating the quantum dynamics in this fashion may require use of the time-ordered perturbation theory \cite{Sterman1993} which has in fact also been adopted to hard-soft factorisation in the seminal paper by Schwartz and Hannesdottir \cite{Hannesdottir:2019opa}.

%% file: Chapters/multiple_double_copy.tex
\chapter{Spinning particles and the  multipole double copy}\label{sec:spin_in_qed}
\section{Introduction}

In \cref{ch:electromagnetism} and \cref{ch:soft_constraints} we have shown in  great detail how to use  scattering amplitudes  to   compute  classical observable for interacting charged and  structure-less   compact objects in classical electrodynamics, at leading and subleading orders in perturbation theory. 
In  this chapter we aim to generalize that discussion for the case in which the classical objects have structure such as  classical spin  associated to them. This in turn  introduces a rich new set of structures not present for the scalar case.  In particular, in \cref{sec:soft_exponential_ph} we saw that the Compton amplitude can be fully determined from soft theorems, with the   
seed $A_{3}^{h,s}$ completely fixed by Lorentz invariant arguments. For the spin case, this seed  is not unique and contains a soft expansion
encoding corrections to $A_{3}^{h,0}$  \cite{osti_4073049,PhysRev.96.1428,PhysRev.110.974,Guevara:2018wpp}. This soft corrections are present as operators of the form $q_\mu\epsilon_\nu J^{\mu\nu}$, where $q$ is some massless momentum, with $\epsilon$ is corresponding polarization,  and $ J^{\mu\nu}$ corresponds to the angular momentum operator, which in 4-dimensions can be mapped  to the Pauli-Lubanski spin operator $s^\mu$. This is a general feature for electromagnetic $h=1$, and gravitational amplitudes $h=2$, as we shall see in this chapter. 

In this chapter we organize our favorite amplitudes $A_n$ for $n=3,4$ in a \textit{covariant spin multipole expansion}, where the spin multipole, are operators of the Lorentz group SO($D-1,1$). We then take the classical limit for amplitudes written in this fashion, and argue that in order to interpret our results in a classical context for compact objects with a given classical spin structure, the \textit{covariant spin multipole moments} need to be branched into the \textit{rotation multipole moments}, which are irreducible representations of the rotation subgroup SO($D-1$).

Furthermore, we will also start the study of the computation of classical observables in gravity through the \textit{spin multipole double copy}. In particular, we will show how to obtain a classical double copy formula for the two-body amplitudes $M_4$ and $M_5$, which follows as a consequence of the factorization \eqref{cuts m4 m5}, and  from the \ac{KLT} double copy of the $A_n$ amplitudes. This classical double copy formula  can be constructed directly from the classical limit of the  \ac{BCJ} double copy as we will show explicitly in \cref{ch:double_copy}. We provide explicit example of how to use this formula in the context of scalar, as well as spinning black holes.  In the scalar case, we show how a  soft exponentiation of the gravitational amplitude, analog to the electromagnetic case \eqref{eq:newM5clas} arises from the soft exponentiation of the scalar gravitational Compton amplitude. At leading order in the soft expansion, this amplitude allows us to recover the  burst memory waveform derived by Braginsky and Thorne \cite{Braginsky1987}. For the spinning case we obtain amplitudes up to quadrupole level both in terms of the \textit{covariant}, as well as the \textit{rotation} multipole moments. 
Amplitudes written in the latter fashion will be used in \cref{ch:bounded} to study radiation in the   two-body problem  for Kerr black holes  in bounded orbits.

This chapter combines elements introduced by the author  in \cite{Bautista:2019tdr} and further extended in \cite{Bautista:2021inx} and \cite{BGKV}.

\section{Amplitudes involving spinning particles in QED}\label{sec:spinin_matter}

In this chapter we will study a  richer sector for the  scattering amplitudes discussed in previous chapter, which arises from  allowing massive particles to have spin. At the level of the electromagnetic theory, in this section we will compute scattering amplitudes for spin $1/2$ and spin $1$ massive particles minimally coupled to the photon field. These amplitudes will be written in a  \textit{spin multipole} fashion, whose multipole structure will be kept unchanged when using the double copy. 
As we mentioned by the end of  \cref{ch:electromagnetism}, studying the electromagnetic sector will be enough for describing gravitational radiation in the two-body problem from low-multiplicity amplitudes, and at the lower orders in perturbation theory. In \cref{ch:double_copy} we will generalize to the non-abelian case, by studying the double copy for massive particles with spin, in more generality.

Let us start, in analogy to \eqref{eq:sqed_lagrangian},  introducing the Lagrangians we will use to compute amplitudes for each case. For spin $1/2$ particles we will use the standard \ac{QED} Lagrangian   
\begin{equation}\label{eq:qed_lagrangian}
    \mathcal{L}_{\text{QED}} = -\frac{1}{4}F_{\mu\nu}F^{\mu\nu} +\bar{\psi}(i\gamma^\mu D_\mu-m)\psi \,,
\end{equation}
whereas $\gamma^\mu$ correspond to the Dirac gamma matrices, and the covariant derivative is once again $D_\mu=\partial_\mu+i e Q A_\mu$. In the same way, for a spin $1$ particles we use the Maxwell-Proca Lagrangian 
\begin{equation}\label{eq:M-P-lagrangian}
    \mathcal{L}_{\text{MP}}=  -\frac{1}{4}F_{\mu\nu}F^{\mu\nu} -\frac{1}{2}\big(D_\mu B^*_\nu-D_\nu B^*_\mu \big)\big(D_\mu B^\nu-D_\nu B^\mu \big) - m^2 B^*_\mu B^\mu\,,
\end{equation}
which describes the interaction between charged complex vector field and the photon.
In \cref{ch:double_copy} we will promote these theories to their non-abelian analogs, i.e.   \ac{QCD}, and non-abelian Gluon-Proca theories, in order to study more in greater detail the double copy for massive spinning matter. 

Derivation of the Feynman rules from these Lagrangians is a straightforward task and we will not show them explicitly here ( readers interested can see for instance \cite{Holstein:2008sw} ).

\subsection{3-point amplitude and the spin multipole decomposition}
The simplest amplitude one can compute from these Lagrangians is   the    3-point amplitude for a  massive particle emitting a single photon. 
Starting with the \ac{QED} theory, this amplitude is simply given by
\begin{equation}\label{eq:A3qed}
    A_3^{\text{QED}} = i e m Q \epsilon_\mu\bar{u}_2\gamma^\mu u_1\,,
\end{equation}
where $u_1/\bar{u}_2$ are Dirac spinors associated to the incoming/outgoing massive particle, whereas $\epsilon$ is the polarization vector for the emitted photon, which satisfies the condition $\epsilon{\cdot}q=0$. Here we have use the momentum conservation condition $p_2=p_1-q$. 
In the same way, for the spin $1$ theory the 3-point amplitude reads
\begin{equation}\label{eq:A3proca}
     A_3^{\text{MP}} =  2i e\,Q\Big(p_1{\cdot}\epsilon \,\varepsilon_1{\cdot}\varepsilon_2^*+\frac{1}{2}( q{\cdot}\varepsilon_1\, \epsilon{\cdot}\varepsilon_2^*-p_1{\cdot}\varepsilon_2^*\, \epsilon{\cdot}\varepsilon_1 ) \Big)\,,
\end{equation}
with $\epsilon_1/\epsilon_2^*$  the massive polarization state, which satisfy the condition $\varepsilon_1{\cdot}p_1=\varepsilon_2^*{\cdot}p_2=0$.  Notice this formula is almost the same as the 3-gluon partial amplitude \eqref{eq:3-gluon} except for the $\frac{1}{2}$ factor. This factor has  very interesting consequences as we will discuss  below and more formally in \cref{sec:s1W}. 

By inspection of these two amplitudes, it is not really clear these amplitudes have anything in common, except  for the photon polarization vector. As it turns out, these two amplitudes  actually correspond to the same object, satisfying hidden properties when written in the usual \ac{QFT} language. We now aim to unravel these interesting properties, among of which we have the \textit{covariant multipole decomposition},  the \textit{ universality}   of the multipole expansion, and the \textit{spin exponentiation} amplitudes in the helicity basis.

\subsubsection{The spin multipole decomposition}
As  spin is the only quantum number available to characterize the massive state, it is natural to think   amplitudes \eqref{eq:A3qed} and \eqref{eq:A3proca} can be written as as a function of the  intrinsic angular-momentum
operator  $J^{\mu\nu}$.
Let us then be more general and propose this can be done for any particle multiplicity $n$, in both, the electromagnetic as well as the gravitational theory. That is,  at the operator level,  we can write amplitudes for one  spinning matter line emitting $n$-photons (or gravitons) in a \textit{covariant  spin multipole expansion} of the form \footnote{Formally, this can be argued via the generalized Wigner-Eckart theorem of e.g. \cite{Agrawala1980}, even if the group is non-compact.}
 \begin{equation}
\bar{A}_{n}^{h,s}(J)={\rm \mathcal{H}}_n\times\sum_{j=0}^{\infty}\omega_{n\,\mu_{1}\cdots\mu_{2j}}^{(2j)}J_s^{\mu_{1}\mu_{2}}\cdots J_s^{\mu_{2j-1}\mu_{2j}}\label{eq:multiexp},
\end{equation}
where the spin multipole moments are SO($D-1,1$) operators, $J_s^{\mu\nu}$, acting  on spin-$s$ states,  and $\omega_{n\,\mu_{1}\cdots\mu_{2j}}^{(2j)}$ correspond to multipole coefficients which are functions  of   particles kinematic quantities only. 
In this formula, products of $J_s^{\mu\nu}$ are understood to be 
symmetrized since, due to the Lorentz algebra,  $[J_s,J_s]\sim J_s$ can be put in terms of lower multipole moments. The sum is then guaranteed to truncate due to the Cayley-Hamilton theorem. The prefactors  $\mathcal{H}_n$ are functions encoding  the helicity structure of  the emitted  photons/gravitons (that is,  $h=1$ or $h=2$ respectively).

Let us explicitly  see how this works  for our previous amplitudes  \eqref{eq:A3qed} and \eqref{eq:A3proca} . The spin generator  corresponds  to nothing but to the Lorentz generator written in the spin-$s$ representation. For particles of spin $1/2$, they can be put in  terms of the Dirac Gamma matrices
$J^{\mu\nu}_{1/2}=\frac{\gamma^{\mu\nu}}{2}=\frac{1}{4}\gamma^{[\mu}\gamma^{\nu]}$, whereas for spin $1$ representation, we have $(J_1^{\mu\nu})^\alpha_{\,\,\beta} = \eta^{\mu\alpha}\eta^{\nu}_{\beta}-\eta^{\nu\alpha}\eta^{\mu}_{\beta} $.
 For the fermionic case, rewriting of the amplitude  \eqref{eq:A3qed} in terms of  $J_s^{\mu\nu}$ is  usually achieved by employing the support of the  Dirac equation and the momentum conservation condition, whereas for the vector case, no new manipulations are needed. 
 
With this considerations at hand, one can easily checked that in terms of the $J^{\mu\nu}$ operators, amplitudes \eqref{eq:A3qed} and \eqref{eq:A3proca} take  a unified form 
\begin{equation}\label{eq:A3_multipole}
     A_3^{\text{ph}} = \,_s\bra{p_2,\varepsilon_2} \bar{A}_3^{\text{ph}}\ket{\varepsilon_1}_s = \,_s\bra{\varepsilon_2} ie\,Q\left(2p_1{\cdot}\epsilon -g\epsilon_{[\mu} q_{\nu]} J_s^{\mu\nu}\right)\ket{p_1,\varepsilon_1}_s\,.
\end{equation}
Here we have used a Dirac bracket notation to represent the massive particle polarization states in the spin-$s$ representation, which for our case corresponds to Dirac spinors for the $s=1/2$ case, and to massive polarization vectors for the $s=1$ case. We will concentrate on $\bar{A}^{\text{ph}}_3$ as an operator acting on the polarization states.  
The  (tree-level) value for the  form  factor $g=2$ is  fixed  for a Dirac spinor coupled
to a photon/gluon. For the Proca field we have actually $g=1$. In \cref{ch:double_copy} we see that in order to set $g\to2$, as required from the double copy,  the Proca theory needs to be modified to become the W-boson model in QCD. We will then argue that it is this  double copy criteria  what fixes $g=2$ for generic spins, both, in the electromagnetic (or \ac{QCD} for the non abelian generalization), and the gravitational theory \cite{Bautista:2019tdr,Holstein:2006pq,Holstein:2006wi}. Let us for the moment assume we can set $g=2$ for both theories and come back to the 3-pt amplitude in \cref{sec:qcd-theories}.

By direct  comparison of  \eqref{eq:A3_multipole} to the general multipole expansion \eqref{eq:multiexp}, we can identify $\mathcal{H}_{3} = 2ie\,Q\,p_2{\cdot}\epsilon$, whereas the multipole coefficients ,  which  are \textit{ universal} for our minimal coupling theories \eqref{eq:qed_lagrangian} and \eqref{eq:M-P-lagrangian},  are given explicitly by $\omega_{3}^{(0)}=1$ and $\omega_{3\,\mu\nu}^{(2)} = -\frac{g}{2} \frac{\epsilon_{[\mu}q_{\nu]}}{p_1{\cdot}\epsilon}$. Notice then our 3-point  seeds in any dimension can be simply put as 
\begin{equation}
\bar{A}_{3}^{s,{\rm ph}}=2i\,e\,Q\,\epsilon\cdot p_{1}\left(\mathbb{I}+J_s\right)\,,\quad J_s=\frac{\epsilon_{\mu}q_{\nu}J_s^{\mu\nu}}{\epsilon\cdot p_{1}}\,,\label{eq:3ptshalf}
\end{equation} 
This indeed hints an \textit{exponential } structure for the 3-point amplitude $e^{J}$,  for higher spinning particles, and  truncates at the $2s$ order. We will return to this exponentiation in \cref{app_B} (see also \cref{ch:GW_scattering}) in the context of the massive  spinor helicity variables introduced in \cref{sec:spinor-helicity}, and in \cref{ch:GW_scattering} we will show this exponential is indeed achieved in a helicity basis, using the spinor helicity formalism introduced in \cref{sec:spinor-helicity}.

It is useful to introduce some diagrammatic notation to refer to the covariant SO($D-1,1$) multipole moments operators. This is done by  assigning each multipole operator  to the corresponding ${\rm SO}(D-1,1)$ Young
diagram, i.e.
\begin{equation}
    1=\mathbb{I}\,,\,\, \qquad\ytableausetup{mathmode,boxsize=0.5em}\ydiagram{1,1}=J_s^{\mu\nu}
\end{equation}

As mentioned above, the scalar 3-point seeds have been corrected by soft the soft operator $\frac{\epsilon_{\mu}q_{\nu}J_s^{\mu\nu}}{\epsilon\cdot p_{1}}$. This is indeed a property that holds also to higher multiplicity amplitudes as we shall see soon, and to higher spins. In fact, as an spoiler alert, we will see that the same exponential structure holds for the gravitational 3-point amplitude, where in the classical limit, the spin of the massive particle can be mapped to the spin of the Kerr \ac{BH}, whose linearized metric expansion encapsulates an infinite tower of classical soft multipole moments in it. We will come back to this discussion in \cref{ch:GW_scattering}.

\subsection{The spinning  Compton amplitude, a multipole decomposition}\label{sec:compton_qed_spin}

What is the meaning of the exponential $e^{J_s}$? It corresponds to
a finite Lorentz transformation induced by the massless emission.
That is, $p_{2}=e^{J_s}p_{1}$, hence for generic spin it maps the
state $|p_{1},\varepsilon_{1}\rangle$ into $|p_{2},\tilde{\varepsilon}_{2}\rangle$,
where $\tilde{\varepsilon}_{2}\neq\varepsilon_{2}$ is another polarization
for $p_{2}$. This means $e^{J}$ is composed both of a boost and
a ${\rm SO}(D-1)$ Wigner rotation. The boost can be removed in order
to match ${\rm SO}(D-1)$ multipoles in the classical theory, as we will see in \cref{sec:branching}. Also, as $e^{J}$ is a Lorentz transformation, $|\varepsilon_{2}\rangle$
must live in the same irrep as $|\varepsilon_{1}\rangle$. This means
that a projector is not needed when these objects are glued through unitarity. A corollary
of this is a simple formula for the full factorization in the massive poles of $A_{n}^{h,s}$,
e.g.

\begin{equation}\label{eq:fact}
    \vcenter{\hbox{\includegraphics[width=59mm,height=17mm]{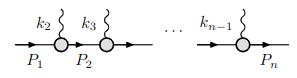}}} = \prod_{i}(2i\,e\,Q\,P_{i}{\cdot}\epsilon_{i})^{h}\langle\varepsilon_{2}|e^{J_{s,n-1}}{\cdots}e^{J_{s,2}}|\varepsilon_{1}\rangle=\prod_{i}(2i\,e\,Q\,P_{i}{\cdot}\epsilon_{i})^{h}\langle\varepsilon_{2}|\tilde{\varepsilon}_{2}\rangle,
\end{equation}

where $P_{i}=p_{1}+k_{2}+\ldots+k_{i}$ and $J_{s,i}=\frac{k_{i\mu}\epsilon_{i\nu}J_s^{\mu\nu}}{\epsilon_{i}\cdot P_{i}}$.
Each 3-pt. amplitude here maps $P_{i}$ to $P_{i+1}$ and their composition
maps $p_{1}$ to $p_{2}$. The state $|\tilde{\varepsilon}_{2}\rangle$
depends on all $\{k_{i},\epsilon_{i}\}_{i=2}^{n-1}$ as well as their
ordering. In particular, we can use  this factorization formula to reconstruct the spinning Compton amplitude in the \ac{QED} for spinning matter up to contact terms.  This can be seen  from  theories \eqref{eq:qed_lagrangian} and \eqref{eq:M-P-lagrangian}, where  the Compton amplitude only has two factorization channels  with  massive propagators only. 
To see this, we use the Baker-Campbell-Hausdorff formula in \eqref{eq:fact} and get the form
\begin{equation}\label{eq:sumofexp}
    \begin{split}
        \bar{A}_{4}^{{\rm ph},s}{=} 2e^2\,Q^2\,\left[ \frac{p_{1}{\cdot}\epsilon_{2}p_{4}{\cdot}\epsilon_{3}}{p_{1}{\cdot} k_{2}}\langle\varepsilon_{2}|e^{J_{1}+J_{2}-\frac{1}{2}[J_{1},J_{2}]+\ldots}|\varepsilon_{1}\rangle + 
        \frac{p_{4}{\cdot}\epsilon_{2}p_{1}{\cdot}\epsilon_{3}}{p_{4}{\cdot} k_{2}}\langle\varepsilon_{2}|e^{J'_{1}+J'_{2}+\frac{1}{2}[J'_{1},J'_{2}]+\ldots}|\varepsilon_{1}\rangle+{\rm c.t.}\right].
    \end{split}
\end{equation}
This is the spin analog of \eqref{eq:scphcompton}, where the exponential tracks the desired order. The role of the contact term in \eqref{eq:sumofexp} is to restore gauge invariance. 

Setting $\mathcal{H}_4{=}\frac{2e^2 Q^2}{p_1{\cdot}k_2 p_1{\cdot}k_3}$ in \eqref{eq:multiexp},  and by requiring the scalar piece ($J=0$) recovers the  result for the scalar Compton amplitude \eqref{eq:scphcompton}, one finds ${\rm c.t.}=\epsilon_{2}{\cdot}\epsilon_{3}$ and
\begin{equation}\label{eq:w0}
 \omega_4^{(0)}=p_1{\cdot}F_2 {\cdot}F_3{\cdot}p_1\,.
\end{equation}
This gives for $s{\leq}1$ the multipole decomposition for the Compton amplitude 
\begin{equation}
\bar{A}_{s,4}^{{\rm {QED}}}=\mathcal{H}_4\left[{\omega_4^{(0)}}J_s^{(0)} {+}\omega_{4\,\mu\nu}^{(1)}J_s^{(1)\mu\nu}{+}\omega_{4\,\mu\nu\rho\sigma}^{(2)}J_s^{(2)\mu\nu\rho\sigma}  \right]\,, \label{eq:comptom multipoles}
\end{equation}
where the spin multipoles are $J_s^{(0)} = \mathbb{I}_s$,   $J_s^{(1)\mu\nu}=J_s^{\mu\nu}$ and $J_s^{(2)\mu\nu\rho\sigma} = \{J_s^{\mu\nu},J_s^{\rho\sigma}\}$, as required from the symmetrization mentioned in \eqref{eq:multiexp}. The  remaining  multipole  coefficients $\omega^{(i)}$ read explicitly
\begin{align}
\omega_4^{(1)\,\mu\nu}{=} & \frac{p_{1}{\cdot}F_{2}{\cdot}p_{4}}{2}F_{3}^{\mu\nu}{+}\frac{p_{1}{\cdot}F_{3}{\cdot}p_{4}}{2}F_{2}^{\mu\nu}{+}\frac{p_{1}{\cdot}(k_{2}{+}k_{3})}{4}[F_{2}{,}F_{3}]^{\mu\nu} \,,\label{eq:w_2}\\
\omega^{(2)\,\mu\nu\rho\sigma}_4{=} & -\frac{k_{1}{\cdot}k_{2}}{16}\left(F_{2}^{\mu\nu}F_{3}^{\rho\sigma}+F_{3}^{\mu\nu}F_{2}^{\rho\sigma}\right)\,.\label{eq:w2}
\end{align}
These are universal for the theories \eqref{eq:qed_lagrangian} and the abelian sector of  \eqref{qcdW-boson}, the W-boson theory (correcting the Proca theory \eqref{eq:M-P-lagrangian}). 
Already for spin-$\frac{1}{2}$ it is clear that this decomposition
of the Compton amplitude is not evident at all from a Feynman-diagram
computation \cite{Bjerrum-Bohr:2013bxa,Ochirov:2018uyq}, 
whereas here it is direct. 
A key point of this splitting is that under the \textit{double}
soft deformation $k_{2}=\tau\hat{k}_{2},k_{3}=\tau\hat{k}_{3}$, the
multipole $\omega^{(2j)}$ is $\mathcal{O}(\tau^j)$, whose
leading order will be the classical contribution to the amplitude in the \ac{KMOC} sense. An explicit computation of the Compton amplitude for theories  \eqref{eq:qed_lagrangian} and the abelian sector of \eqref{qcdW-boson}, show exact agreement with the  result \eqref{eq:comptom multipoles}

Note now  that while $J_s^{(0)}$ and $J_s^{(1)\mu\nu}$ are  irreducible representations of the Lorentz group SO($D-1,1$),  the  operator $J_s^{(2)\mu\nu\rho\sigma}$ has the symmetries of the Riemann tensor and can be further decomposed into irreducible SO($D-1,1$) representations . This decomposition goes by the name of Ricci decomposition which we outline as follows:

\textit{Ricci decomposition:}
Let $R^{\mu\nu\rho\sigma}$ be a Riemannian tensor. The Ricci decomposition is the statement that $R^{\mu\nu\rho\sigma}$ can be split in the following form\footnote{See for instance the Wikipedia article \url{https://en.wikipedia.org/wiki/Ricci_decomposition} }:
\begin{equation}
    R^{\mu\nu\rho\sigma}=   S^{\mu\nu\rho\sigma}+ E^{\mu\nu\rho\sigma}+ C^{\mu\nu\rho\sigma}\,,
\end{equation}
where $S,E$ and $C$ corresponds to the scalar, symmetric and Weyl parts of the tensor, respectively defined by:
\begin{align}
       S^{\mu\nu\rho\sigma} &= \frac{R}{D(D-1)}(\eta^{\mu \sigma}\eta^{\nu\rho} -\eta^{\mu \rho}\eta^{\nu\sigma} )\,,\\
       E^{\mu\nu\rho\sigma} &= \frac{1}{(D-2)}(\eta^{\nu\rho}R^{\mu\sigma}-\eta^{\mu\rho}R^{\nu\sigma}+\eta^{\mu\sigma}R^{\nu\rho}-\eta^{\nu\sigma}R^{\mu\rho} )\,,\\
       C^{\mu\nu\rho\sigma} &=  R^{\mu\nu\rho\sigma}-  S^{\mu\nu\rho\sigma}- E^{\mu\nu\rho\sigma}\,.
\end{align}
Here $R^{\mu\nu} =  R^{\rho\mu\nu\sigma}\eta_{\rho\sigma}$ is the Ricci curvature tensor, whereas $R= R^{\mu\nu}\eta_{\mu\nu}$ corresponds to the scalar of  curvature. It is also useful to know that the Riemann of a tensor $T$,  can be computed by \cite{Bekaert:2006py}
\begin{equation}
\begin{split}
    \text{Riemann}(T)^{\mu\nu\rho\sigma}& = \frac{1}{12} (T^{\mu \nu \rho \sigma }-T^{\mu \nu \sigma \rho }-T^{\mu \rho \sigma \nu }+T^{\mu \sigma \rho \nu }-T^{\nu \mu \rho
\sigma }+T^{\nu \mu \sigma \rho }+T^{\nu \rho \sigma \mu }-T^{\nu \sigma \rho \mu }\\&\,\,\quad
-T^{\rho \mu \nu \sigma }+T^{\rho \nu \mu \sigma} +T^{\rho
\sigma \mu \nu }-T^{\rho \sigma \nu \mu }+T^{\sigma \mu \nu \rho }-T^{\sigma \nu \mu \rho }-T^{\sigma \rho \mu \nu }+T^{\sigma \rho \nu
\mu} )
\end{split}
\end{equation}

Let us  apply this decomposition to the quadratic in $J_s$ contribution to the Compton amplitude. For that we use the following notation   
\begin{equation}\label{eq:ricciA4qed}
\omega_{4\,\mu\nu\rho\sigma}^{(2)}J_s^{(2)\mu\nu\rho\sigma} =\left\{ \begin{matrix}\hat{1}_s[\omega_4^{(4)}]+[\omega_4^{(2)}]_{\mu\nu}Q_s^{\mu\nu},\quad s=1,\\
\hat{1}_s[\omega_4^{(2)}]+[\omega_4^{(2)}]_{\mu\nu\rho\sigma}\ell^{\mu\nu\rho\sigma},\,s=\frac{1}{2},
\end{matrix}\right.
\end{equation}
where $\ell_s^{\mu\nu\rho\sigma}=J_s^{(2)[\mu\nu\rho\sigma]}=\parbox{4pt}{\ytableausetup{mathmode,boxsize=0.2em}\ydiagram{1,1,1,1}},$
and
\begin{equation}
\hat{1}_s=\frac{J_{s,\mu\nu}J_s^{\mu\nu}}{2},\,\,Q_s^{\mu\nu}=\ytableausetup{mathmode,boxsize=0.4em}\ydiagram{2}=\{J_s^{\mu\rho},J_{s,\rho}^{\,\,\nu}\}+\frac{4}{D}\eta^{\mu\nu}\hat{1}_s\,.\label{eq:quadandl}
\end{equation}
We   then identify $Q_s^{\mu\nu}$ with  the traceless Ricci tensor, whereas $\hat{1}_s$ corresponds to the scalar curvature.  Notice remarkably, \eqref{eq:comptom multipoles} does not possess a Weyl contribution. This will be important when we discuss the double copy below.  In addition, for spin $1/2$ we get a totally antisymmetric contribution due to the non commutativity nature of the Dirac gamma matrices. 

We have also introduced the notation  $[\omega_4^{(2)}]$ ,  $[\omega_4^{(2)}]_{\mu\nu}$ and $[\omega_4^{(2)}]_{\mu\nu\rho\sigma}$ for the corresponding projections of the multipole coefficient $\omega_4^{(2)\mu\nu\rho\sigma}$ . The explicit form for the  first two read
\begin{equation}
    [\omega^{(2)}_4] = \frac{4}{D(D-1)}\omega_{4\,\mu\nu\rho\sigma}^{(2)} \eta^{\mu[\rho}\eta^{\sigma]\nu}=\frac{k_2{\cdot}k_3}{D(D-1)}F_{2,\mu\nu}F_3^{\mu\nu}
\end{equation}
\begin{equation}
    [\omega_{4}^{(2)}]_{\mu\nu}=\frac{k_2{\cdot}k_3}{D-2}F_{2\,(\mu|\rho}F_{3\,|\nu)}^{\rho}\,,
\end{equation}
whereas for the latter we simply have $[\omega_4^{(2)}]^{\mu\nu\rho\sigma}=\omega^{(2)\,[\mu\nu\rho\sigma]}_4$. 
We refer to the irreducible operators of  SO($D-1,1$) as the \textit{ covariant spin multipole moments}. 
This then allows us to identify the covariant traceless spin quadrupole moment $Q_s^{\mu\nu}$ existing only for spin $1$ particles in \ac{QED}.

\subsection{Bremsstrahlung  radiation for spinning sources}\label{sec:bremsstra_ph}
Now that we have studied our 3-point and 4-point amplitudes for spinning particles in great detail, it is natural to ask how the Bremsstrahlung radiation amplitude of Figure \ref{fig:5pt-amplitude} changes in the presence of spin. 
In \cref{sec:examplesa_tree_level} we have argued that formula \eqref{eq:newM5clas} is indeed more general and can be used to compute photon and gravitational  Bremsstrahlung  radiation, even in the presence of spin. To compute the numerators entering in \eqref{eq:newM5clas},   in the presence of spin, we use the same unitarity method of \eqref{cutsm5}, i.e. the numerators for the 5-point amplitude are given by 
the residues of $M_5$ at the  null momenta $q_i$ for $i=1,2$. This then corresponded to the unitary gluing of the Compton to the 3-point amplitude on the support of null momenta $q_i$ for the  respective factorization channel. For instance, at linear order in $J$,  and introducing the notation
\begin{align}\label{eq:notation}
 \hat{R}_{i}^{\mu\nu }	=2(2\eta_{i}q-k)^{[\mu}J_{s,i}^{\nu]\alpha}(2\eta_{i}q-k)_{\alpha},
\end{align}
recall $\eta_{1}=-1$ and $\eta_{2}=1$, the Bremsstrahlung radiation numerators have the form 
\begin{equation}
\label{eq:lineas-in-spin-numerators-photon}
\begin{split}
  n_{\frac{1}{2},\rm{ph}}^{(a)}	&=n_{0,\rm{ph}}^{(a)}{-}2e^{3}\left[p_{1}{\cdot}R_{2}{\cdot}kF{\cdot}J_{s,1}{-}F_{1q}R_{2}{\cdot}J_{s,1}{+}p_{1}{\cdot}k\,[F,R_{2}]{\cdot}J_{s,1}-p_{1}{\cdot}F{\cdot}\hat{R}_{2}{\cdot}p_{1}\right],\\
n_{\frac{1}{2},\rm{ph}}^{(b)}	&=n_{0,\rm{ph}}^{(b)}{-}2e^{3}\left[p_{2}{\cdot}R_{1}{\cdot}kF{\cdot}J_{s,2}{-}F_{2q}R_{1}{\cdot}J_{s,2}{+}p_{2}{\cdot}k\,[F,R_{1}]{\cdot}J_{s,2}-p_{2}{\cdot}F{\cdot}\hat{R}_{1}{\cdot}p_{2}\right],
\end{split}
\end{equation}
where the scalar numerators were obtained in \eqref{nphsc}, and follow naturally from the universality of $A_3$ and $A_4$. We  have also introduced the commutator notation $[F,R_{2}]{\cdot}J_{s,i} = (F^\mu{}_{\nu} R_{2}^{\nu\alpha}- R^\mu_{2,\nu} F^{\nu\alpha})(J_{s,i})_{\mu\alpha}  $. 
These numerators follow analogously  from the  gluing of the electromagnetic spin-$1/2$, 3-point \eqref{eq:3ptshalf} and 4-point \eqref{eq:comptom multipoles}  amplitudes. Using  variables \eqref{eq:notation} to rewrite the numerators trivializes the check for gauge invariance. Notice on the other hand, the spin contribution in the  $\hat{R}_i$ terms emerges purely  from the linear-in-spin piece of the  3-pt amplitude, whereas the linear-in-spin  Compton amplitude is responsible for   the remaining terms. 

Finally, the quadratic order in spin numerators are given in \cref{ch:quadratic_spin} by equations \eqref{eq:elepsin21} and \eqref{eq:elepsin22}. These, again, follow from the electromagnetic quadratic-in-spin 3-pt and 4-pt amplitude, the latter obtained from the single copy in \eqref{eq:a4CLASSSPIN2}. For simplicity, at this order we have  restricted to the case in which only one particle has spin, while the other is scalar.

\subsubsection{Classical Limit}

We have already seen that 
 by replacing  the numerators \eqref{nphsc}\footnote{ These numerators have the support of $\delta(p_i{\cdot}(\eta_i q-k))$, which imposes the on-shell condition for the outgoing massive particles in the classical limit.} into the general formula \eqref{eq:newM5clas}, we recover the classical photon  radiation amplitude for the scattering of two colorless scalar charges, as computed  by Goldberger and Ridgway in \cite{Goldberger:2016iau}. This is of course true since we have already taken the classical limit of the amplitude using the \ac{KMOC} prescription. In the presence of spin however,  additional considerations need to be taken  in order to: 1) Extract the correct classical contribution, and 2) Extract the spin multipole moments that correctly describe classical rotating objects. 
 
 To solve 1), notice that when scaling the scalar numerators  in \eqref{eq:lineas-in-spin-numerators-photon} using the \ac{KMOC} prescription, $q\to \hbar q$ and $k\to \hbar k$, the leading order contribution of  $n_{0,\rm{ph}}^{(a/b)}\sim \hbar^2 $. However, by doing the same scaling, the linear in spin terms in \eqref{eq:lineas-in-spin-numerators-photon} scale now as $\sim \hbar^3$. Naively one might think  this is a quantum contribution and can be discarded in the classical limit. This would however also discard any spin information, and we know classical rotating objects have associated a classical rotation tensor which is not present in the scalar contribution. To solve this discrepancy,  \cite{Guevara:2018wpp} (see also  \cite{Guevara:2018wpp,Chung:2018kqs, Bautista:2019evw,Maybee:2019jus}) proposed the Lorentz generators $J_s^{\mu\nu}$ need to also have an $\hbar$ scaling in such way there exist a classical spin structure extracted from the Quantum amplitude. Then, it is natural to take $J_s^{\mu\nu}\to \frac{J_s^{\mu\nu}}{\hbar}$, which will make the scaling of the spin contributions in \eqref{eq:lineas-in-spin-numerators-photon} to be of the same order in $\hbar$ as the scalar part. This scaling of the spin follows naturally from generalizing  \ac{KMOC} for spin \cite{Maybee:2019jus}, as dictated by the  correspondence principle mentioned in the general Introduction.
 
 To solve 2) one needs to recall that we have thought of  the spinning amplitudes $\bar{A}_n$, as a SO($D-1,1$) operator acting on spin-$s$ states of the form $\ket{p_i,\varepsilon_i}$. One needs to recall however these states transform under different representations of the little group, as indicated by the massive momentum labels. Then, in  order to interpret the results for the previously computed amplitude as 
 those for the interaction of a the same incoming and outgoing classical spinning object in electrodynamics, we need to choose a reference frame -- which can be fixed by choosing a time-like vector $u^\mu $ satisfying the \ac{SSC} -- so that the massive polarization states are aligned towards the same  canonical polarization states\footnote{These alignment of the polarization states also goes by the name of \textit{Hilbert space matching} \cite{Chung:2020rrz}}. 
 When doing so, the  SO$(D-1,1)$ generator $J_s^{\mu\nu}$, which consist of a  $SO(D-1)$  Wigner rotation plus a boost, $J_s^{\mu\nu} =S^{\mu\nu}-2u^{[\mu}K^{\nu]}$, can be interpret as a classical spin tensor for the rotating object, once the boost  component is removed. The \ac{SSC} to be satisfied is  simply  $u_\mu S^{\mu\nu}=0$.  After this is done, the  polarization states can be removed, leaving us with a classical object, which we will interpret as the classical amplitude.  This alignment   effectively  induces a map of the  $SO(D-1,1)$ \textit{covariant multipoles moments} towards the SO$(D-1)$ \textit{rotation multipole moments}, we will expand on this map in \cref{sec:branching}.

At the linear order in spin, the effect of choosing a reference frame results into simply setting $J_{s,i}^{\mu\nu} \to S_i^{\mu\nu}$, followed by the removal of the  polarisation states.  Only by then,  $S^{\mu\nu}$   can be  interpreted as a classical spin tensor characterizing the intrinsic rotation of the compact object. The reference vector $u_i$ can be taken to be either the incoming, outgoing or the average momentum for each matter line, in the classical limit they reduce to taking $u_i=p_i/m_i$. This in turn implies the  \ac{SSC} to satisfy is $p_{i,\mu}S_i^{\mu\nu}=0$. One can easily check that by  changing $J\to S$ in \eqref{eq:lineas-in-spin-numerators-photon}, and replacing the resulting numerators into  the general formula for radiation \eqref{eq:newM5clas}, we recover the classical result of Li and  Prabhu  \cite{Li:2018qap}, computed from classical Worldline theory arguments.

Let us finish this section by commenting on the resulting form of the radiated electromagnetic field in the soft  (large wavelength) limit. In \eqref{eq:a_weinerg_tree} we  showed at  leading order in the  soft, the waveform is entirely capture by the Weinberg soft factor, which  corresponds to a  universal, and in fact, non-perturbative result, as we  extensively studied  in \cref{ch:soft_constraints}. This universality translates into zero spin corrections to the low energy waveform. One then might wonder whether the spin structure in the numerators \eqref{eq:lineas-in-spin-numerators-photon} indeed provides  zero contribution to the   waveform \eqref{eq:r-kernel_a0}  in the soft  limit. As  can  be explicitly checked, the spin contribution in \eqref{eq:lineas-in-spin-numerators-photon} is indeed subleading in the soft expansion $k
\to\tau k,\,\, \tau \to 0$. This is nothing but a consequence of the universality of $A_3$ and $A_4$. That is, as we discussed  above, spin corrections of $A_3$ correspond to a tower of subleading$^n$ soft operators. This is indeed also the case for $A_4$ as can be seen from  \eqref{eq:comptom multipoles} and \eqref{eq:fact}, where each multipole coefficient scales  with an additional power of $\tau$, in for instance the soft expansion in the outgoing massless momentum $k_3\to\tau k_3$. That is $\bar{A}_{s,4}^{{\rm {QED}}}\sim\tilde {\mathcal{H}}_4\left[\frac{1}{\tau}{\tilde{\omega}_4^{(0)}}J_s^{(0)} {+}\tau^0\tilde{\omega}_{4\,\mu\nu}^{(1)}J_s^{(1)\mu\nu}{+}\tau^1\,\tilde{\omega}_{4\,\mu\nu\rho\sigma}^{(2)}J_s^{(2)\mu\nu\rho\sigma}  \right]$
where the tilde indicates we need to keep only the soft $\tau\to0$ contribution. This then show the leading soft contribution is exactly given by the scalar piece, which is present for for all of the  spin$-s$ amplitudes, above, and therefore, no spin-correction will be added to the leading soft waveform \eqref{eq:a_weinerg_tree}.

\section{The spin multipole double copy}\label{sec:spin_dc}

So far we have been concerned with the computation of observables for the electromagnetic theory. We have discussed in great detail the scalar case both, at tree and loop levels, emphasising the power soft theorems have in the computation of radiation at leading order in the soft expansion.  Furthermore, in the first part of this chapter we have introduced spin effects and argued that the main building blocks $A_3$ and $A_4$  have universality properties that are carried  over  $M_4$ and $M_5$. This discussion is indeed more general and can be extended to the gravitational case as we will see now. 
In the remaining of this chapter we will start the study of classical observables for the   gravitational theory, using  double copy arguments. In \cref{ch:double_copy} we will provide a more formal study of the double copy for massive spinning matter. 

We introduced the double copy of massless particles in  \cref{sec:double-copy-preliminaires}. In particular, we have seen that two copies of $S=1$ massless polarization tensors have the Clebsh-Gordon decomposition 
\begin{equation}\label{eq:CG_decomp}
    (D-1)\otimes(D-1) = \frac{(D+1)(D-2)}{2}\oplus 1\oplus\frac{(D-1)(D-2)}{2},
\end{equation}
where the first two terms in the r.h.s. correspond to $S=2$ (graviton ) and $S=0$ (dilaton) respectively, whereas the third piece is the antisymmetric piece (  Kalb-Ramond field), which for $D=4$ can be dualized to an $S=0$ pseudo-scalar, the axion. We
will indistinctly refer to this  two-form  as axion or Kalb-Ramond
field. 

Explicitly, if $\varepsilon^\mu$ and $\tilde{\varepsilon}^\mu$ correspond to two copies of an  $S=1$ representation, then the  Clebsh-Gordon decomposition reads
\begin{align}\label{eq:dc_s1}
    \epsilon^\mu \tilde{\epsilon}^\nu &=  \left(\frac{ \epsilon^\mu \tilde{\epsilon}^\nu+ \epsilon^\nu \tilde{\epsilon}^\mu}{2} -\frac{\varepsilon{\cdot}\tilde{\varepsilon}}{D-1}\bar{\eta}^{\mu\nu}\right)+\frac{\varepsilon{\cdot}\tilde{\varepsilon}}{D-1}\bar{\eta}^{\mu\nu}+\left(\frac{ \epsilon^\mu \tilde{\epsilon}^\nu- \epsilon^\nu \tilde{\epsilon}^\mu}{2} \right)\,.
\end{align}
For integer $S$, the representations are always isomorphic to transverse, traceless-symmetric
tensors, which dimension is given by $\dim_S = \frac{(D-3+2S)(D-4+S)!}{S!(D-3)!}$, (which reduces to the familiar expression $2S+1$ in $D=4$) \cite{Bekaert:2006py}.  For instance, for $S = 1$ we have the vector representation
$\dim_{S=1} = D -1$. The tensor representations are constructed from direct products of lower ones.  The simplest example is the $S = 2$ tensor which can be constructed
from two copies of $S = 1$ as given by \eqref{eq:dc_s1}.
This decomposition is the reason we can obtain the graviton amplitudes from $S=1$ amplitudes. 

The double copy for massive particles -- as we will formally discuss in \cref{ch:double_copy} -- can be constructed using analogous \ac{KLT} formulas for their massless counterparts. This is due to the fact that for a single matter line emitting gravitons/dilatons/axions,  massive double copies can be obtained from compactifications of their massless higher dimensional analogs. In this section we will concentrate on the double copy for our favorite amplitudes $A_n$, $n=3,4$, involving one spinning matter line emitting one and two gravitons, respectively. This will be sufficient to also induce a double copy formula for $M_4$ and $M_5$, due to the factorization \eqref{cuts m4 m5}.

We will use the $\odot$ symmetric product to denote the double copy of the amplitudes written in a spin-multipole expansion. To begin, let us consider then the photon emission amplitudes for $s\in\{0,\frac{1}{2},1\}$, as given in the previous sections,  and
define their double copy. From two multipole operators $X$ and $X'$
acting on spin-$s$ states, we introduce an operator $X\odot X'$ 
acting on spin-$2s$ as
\begin{equation}
X\odot X'=\left\{ \begin{matrix}X\times X',\,\,\,2s=0\\
2^{-\left\lfloor D/2\right\rfloor }{\rm tr}(X\slashed{\varepsilon}_{1}\tilde{X}'\slashed{\varepsilon}_{2})\,,\quad2s=1,\\
\phi_{1\mu_{1}\nu_{1}}\left(X_{\,\mu_{2}}^{\mu_{1}}X'{}_{\,\nu_{2}}^{\nu_{1}}\right)\phi_{2}^{\mu_{2}\nu_{2}}\,,\quad2s=2,
\end{matrix}\right.\label{eq:starprod}
\end{equation}
where $\varepsilon$ and $\phi$ are the respective massive polarizations
and $\tilde{X}$ denotes charge conjugation.
In \cref{ch:double_copy} we will prove this multipole  operation naturally arises from the  \ac{KLT} double copy, and show furthermore it can be used to obtain scattering amplitudes in a gravity theory of
a massive spin-$2s$ field. Let us for simplicity in this chapter assume it as a  valid double copy operation. 

For our favorite amplitudes, the double copy formula will read 
\begin{equation}
K^{-1}_{n}A_{n}^{{\rm gr},s+\tilde{s}}=A_{n}^{{\rm ph},s}\odot A_{n}^{{\rm ph},\tilde{s}}\,,\quad n=3,4\,.\label{eq:doublecopyspin}
\end{equation}
where $K_n$ corresponds to the massive   \ac{KLT} kernel for the n-point amplitude which is   simply given by  the inverse of the
 biadjoint amplitude involving two massive scalars of the same species ( see e.g. \cite{Naculich:2015zha}
 for details on this theory ).

The case $s=0$, $\tilde{s}\neq0$ was introduced by Holstein et
al. \cite{Holstein:2006pq,Bjerrum-Bohr:2013bxa}. It was used 
to argue that the gyromagnetic ratios
of both $A_{n}^{{\rm ph},1}$ and $A_{n}^{{\rm gr},1}$ must coincide, setting $g=2$ as a natural value \cite{Holstein:2006pq,Chung:2018kqs}.
We introduce the case $s,\tilde{s}\neq0$ as a further universality
condition, and find it imposes strong restrictions on $A_{n}^{h,s}$
for higher spins. More importantly, it can be used to directly
obtain multipoles in the classical gravitational theory.

For \eqref{eq:doublecopyspin} to hold we need to put $A_{n}^{h,s}$
into the form \eqref{eq:multiexp}, which we have done in the first part of the present chapter (although we will lift this restriction \cref{ch:double_copy}). The coefficients $\omega^{(2j)}$ are universal
once we consider minimal-coupling amplitudes, which are obtained from
 \ac{QED} at $s=\frac{1}{2}$ and from the $W^{\pm}$-boson model at $s=1$ \cite{Holstein:2006pq}, as seen above. In a diagramatic notation, the operation \eqref{eq:starprod}  gives  the rules
\begin{equation}
1_{s}\odot1_{s}=1_{2s}\,,\quad1_{s}\odot\ytableausetup{mathmode,boxsize=0.8em}\ydiagram{1,1}_{s}=\frac{1}{2}\ydiagram{1,1}_{2s}\,,\label{eq:rules}
\end{equation}
\begin{equation}
\ytableausetup{mathmode,boxsize=0.8em}\ydiagram{1,1}_{s}\odot\ydiagram{1,1}_{s}=\ydiagram{2,2}_{2s}+\ydiagram{2}_{2s}+\hat{1}_{2s}\, ,\label{eq:dcweyls2}
\end{equation}
which are a subset of the irreducible representations allowed
by the Clebsch-Gordan decomposition. Rule \eqref{eq:dcweyls2} is  nothing but the Ricci decomposition of the symmetric product of two multipole operators $J^{\mu\nu}_s$, and is analog to the decomposition \eqref{eq:ricciA4qed} above, but with non-vanishing Weyl component. Indeed, the  first term we denote by $\Sigma^{\mu\nu\rho\sigma}$
and has the symmetries of a Weyl tensor, i.e. is the traceless part
of $\{J^{\mu\nu},J^{\rho\sigma}\}$. For instance, the $s=2$, 3-point gravitational  amplitude
as obtained from \eqref{eq:doublecopyspin}, and using two copies of the 3-point seed \eqref{eq:3ptshalf}, and the double copy operation \eqref{eq:rules} and \eqref{eq:dcweyls2},  results into
\begin{equation}
A_{3}^{{\rm gr},2}=\kappa\left(\epsilon{\cdot}p_{1}\right)^{2}\phi_{2}{\cdot}\left(\mathbb{I}{+}\frac{\epsilon_{\mu}q_{\nu}J^{\mu\nu}}{\epsilon{\cdot} p_{1}}{+}\frac{W_{\mu\nu\alpha\beta}}{4(\epsilon{\cdot} p_{1})^{2}}\Sigma^{\mu\nu\alpha\beta}\right){\cdot}\phi_{1}\, ,\label{eq:3pts2}
\end{equation}
where $W_{\mu\nu\alpha\beta}:=q_{[\mu}\epsilon_{\nu]}q_{[\alpha}\epsilon_{\beta]}$
is the Weyl tensor of the graviton. We have also used $K_3 = \frac{\kappa}{4e^2}$ for the 3-point \ac{KLT} kernel, with  $\kappa=\sqrt{32\pi G}$ and $G$ the Newton Constant. $\phi_i^{\alpha\beta}$ are massive spin-2 polarization tensors, built from spin-1 polarization vectors; in $D=4$ they  have $5$-independent components.  One can show the same amplitude can be computed from the covariantization of the Fierz-Pauli Lagrangian \cite{Vaidya:2014kza}. 
The classical limit of this amplitude reproduces the
expected Weyl-quadrupole coupling \cite{Goldberger:2004jt,Porto:2005ac,Porto:2006bt,Levi:2015msa,Chung:2018kqs},  as we will discuss in \cref{sec:branching}.

 To deeper understand these results, let us demand $A_{3}^{{\rm gr},s}$
to be constructible from the double copy \eqref{eq:doublecopyspin}
for \textit{any} spin:
\begin{equation}
  A_{3}^{{\rm gr,}s+\tilde{s}}(J^{\mu\nu}\oplus\tilde{J}^{\mu\nu})=A_{3}^{{\rm ph},s}(J^{\mu\nu})\odot A_{3}^{{\rm ph},\tilde{s}}(\tilde{J}^{\mu\nu})\, ,\label{eq:group}
\end{equation}
where $J^{\mu\nu}\oplus\tilde{J}^{\mu\nu}$ is the generator acting on a
spin $s+\tilde{s}$ representation. This relation yields the condition $A_{3}^{1,s}A_{3}^{1,\tilde{s}}=A_{3}^{1,s+\tilde{s}}A_{3}^{1,0}$ on the $J^{\mu\nu}$ operators. Using that $[J,\tilde{J}]=0$ and assuming the coefficients in \eqref{eq:multiexp} to be independent of the spin leads to
\begin{equation}
\bar{A}_{3}^{h,s}(J)=g_h\left({\rm \epsilon\cdot}p_{1}\right)^{h}\times e^{\omega_{\mu\nu}J^{\mu\nu}}\,,\quad h=1,2\label{eq:3ptexp}
\end{equation}
with $\omega_{\mu\nu}=\frac{q_{[\mu}\epsilon_{\nu]}}{\epsilon\cdot p_{1}}$ and $\mathcal{H}_3=\left(\epsilon{\cdot}p_{1}\right)^h$
 fixed by the previous examples. Here $g_1= 4e^2$ and $g_2=\kappa$. This easily recovers such cases
and matches the Lagrangian derivation \cite{Vaidya:2014kza} 
for $s\in\{\frac{1}{2},1,2\}$
in any dimension $D$. After some algebra, \eqref{eq:3ptexp} leads to the $D{=}4$
photon-current derived in \cite{Lorce:2009br,Lorce:2009bs} 
for \textit{arbitrary}
spin via completely different arguments. On the gravity side, its classical limit
matches the Kerr stress-energy tensor derived in \cite{Vines:2017hyw}, as we will discuss in \cref{ch:GW_scattering},
together with its spinor-helicity form  found in \cite{Guevara:2018wpp}, as we show in \cref{app_B}. For $s>h$ and $D>4$, \eqref{eq:3ptexp} contains
a pole in $\epsilon{\cdot}p$ which reflects such interactions being
non elementary \cite{Arkani-Hamed:2017jhn}. In \cref{sec:branching} we show such pole cancels
for the classical multipoles and provide a local form of \eqref{eq:3ptexp}.

Now, the full quantum double copy amplitude $A_4^{\text{gr},2}$ for $s=2$ can be obtained by simply  squaring \eqref{eq:comptom multipoles} for $s=1$, and using the rule given in the third line of \eqref{eq:starprod}, which similar $A_3^{\text{gr},2}$, can be obtained from the covariant  Fierz-Pauli Lagrangian \cite{Vaidya:2014kza}. 
In a multipole decomposition, it will contain terms up to hexadecapole order. However, in general  there is not a known prescription to obtain the spin multipole expansion in terms of irreducible representations of SO($D-,1,1$), and we therefore write the full amplitude in terms of polarization vectors (i.e. not in an operator language), as follows
\begin{equation}
A_{4}^{\text{gr},2}=K_4^{-1}\mathcal{H}_4^2\left[{\omega_4^{(0)}}\varepsilon_1{\cdot}\varepsilon_4^* {+}\omega_{4\,\mu\nu}^{(1)} \varepsilon_1{\cdot} J_1^{(1)\mu\nu}{\cdot}\varepsilon_4^*{+}\omega_{4\,\mu\nu\rho\sigma}^{(2)}\varepsilon_1{\cdot} J_1^{(2)\mu\nu\rho\sigma}{\cdot}\varepsilon_4^*  \right]^2\,, \label{eq:comptonspin2}
\end{equation}
where we need to use the \ac{KLT} kernel 
\begin{equation}
  K_{4}{=}\frac{2e^2}{\kappa^2}\frac{k_2{\cdot}k_3}{p_{1}{\cdot}k_{2}\,p_{1}{\cdot}k_{3}}\,,
\end{equation}
and the polarization tensor for the massive spin $2$ states will be given by $\varepsilon_i^{\mu\nu} =\varepsilon_i^\mu\varepsilon_i^\nu  $, and satisfy $\varepsilon_i^{\mu\nu} p_{i,\mu}=0$, for $i=1,4$, and the traceless condition $\varepsilon_{i\,\mu}^{\mu}=0$. For the massless gravitons we have $\epsilon_i^{\mu\nu} = \epsilon_i^\mu\epsilon_i^\nu$, with analog properties to the massive polarizations. Here we use the contractions $\varepsilon_1{\cdot}J_1^{\mu\nu}{\cdot}\varepsilon_4^* =\varepsilon_{1\alpha}(J_1^{\mu\nu})^{\alpha}_\beta \varepsilon_4^{*\beta} $, and analog for $\varepsilon_1{\cdot} J_1^{(2)\mu\nu\rho\sigma}{\cdot}\varepsilon_4^* =\varepsilon_{1,\alpha} \{(J_1^{\mu\nu})^{\alpha}_\delta,(J_s^{\rho\sigma})^{\delta}_\beta\}\varepsilon_4^{*\beta}$. 

Notice when written in terms of \eqref{eq:w0}, \eqref{eq:w_2} and \eqref{eq:w2}, the spin 2 gravitational Compton amplitude  \eqref{eq:comptonspin2}, becomes dimension-independent. In fact, if we think of $p_1$ and $p_4$ as massless momenta, with  $\varepsilon_i^{\mu\nu}$ their correspondent massless polarization tensors, one recovers the 4-point amplitude for the scattering of 4 gravitons in General Relativity. This is of course not a coincidence, and is a consequence of the fact this amplitude can be written from a compactification of its massless higher dimensional counterpart, as we will show explicitly in \cref{ch:double_copy}.

In this chapter we are interested in computing the spin quadrupole radiation amplitude and for that we can simply take the double copy of two spin-$1/2$ amplitudes and apply the second line of \eqref{eq:dcweyls2}. In a multipolar expansion, and up to quadrupolar order,  the gravitational amplitude is given then by:

\begin{itemize}
    \item The scalar piece

\begin{equation}
    \bar{A}_4^{(0)\text{gr}} = \frac{\kappa^{2}}{8}\frac{\omega^{(0)}\omega^{(0)}}{k_{2}{\cdot}k_{3}\left(p_{1}{\cdot}k_{2}\right)\left(p_{1}{\cdot}k_{3}\right)}
\end{equation}

\item The spin dipole piece
\begin{equation}
\bar{A}_{4}^{(\frac{1}{2})\text{gr}}=\frac{\kappa^{2}}{8}\frac{\omega^{(0)}\omega_{\mu\nu}^{(1)}J_{1/2}^{\mu\nu}}{k_{2}{\cdot}k_{3}\left(p_{1}{\cdot}k_{1}\right)\left(p_{1}{\cdot}k_{2}\right)}
\end{equation}
\item And finally  quadrupolar piece 

\begin{equation}
\begin{split}\bar{A}_{4}^{(1)\text{gr}}= & \frac{\kappa^{2}}{8}\frac{\frac{1}{2}\omega^{(0)}[\omega^{(2)}]\hat{1}_{1}}{k_{2}{\cdot}k_{3}\left(p_{1}{\cdot}k_{2}\right)\left(p_{1}{\cdot}k_{3}\right)}\\
 & +\frac{\kappa^{2}}{8}\frac{\omega_{\mu\nu}^{(1)}\omega_{\rho\sigma}^{(1)}}{k_{2}{\cdot}k_{3}\left(p_{1}{\cdot}k_{2}\right)\left(p_{1}{\cdot}k_{3}\right)}\Big[\frac{1}{4}\Sigma_{1}^{\mu\nu\rho\sigma}{+}\frac{1}{D{-}2}\eta^{[\sigma[\nu}Q_{1}^{\mu]\rho]}{+}\frac{1}{2D(D{-}1)}\eta^{\sigma[\nu}\eta^{\mu]\rho}\hat{1}_{1}\Big]
\end{split}
\end{equation}
\end{itemize}
Here we have omitted the contribution from $\parbox{4pt}{\ytableausetup{mathmode,boxsize=0.2em}\ydiagram{1,1,1,1}}$, since it does not contribute to the classical amplitude. 

We have used that \eqref{eq:dcweyls2} reads
\begin{equation}
\begin{split}
      J_{s}^{\mu\nu}\odot J_{s}^{\rho\sigma} &{=}  \frac{1}{4}\Sigma_{2s}^{\mu\nu\rho\sigma}{+}\frac{\alpha_{D}}{D{-}2}\eta^{[\sigma[\nu}Q_{2s}^{\mu]\rho]}
  {+}\frac{\beta_{D}}{2D(D{-}1)}\eta^{\sigma[\nu}\eta^{\mu]\rho}\hat{1}_{2s} \,.
\end{split}
\label{eq:dc12}
\end{equation}
The normalizations $\alpha_{D},\beta_{D}$ depend solely on $D$.
However, it cancels out in the full computation  and hence we set $\alpha_{D}{=}\beta_{D}{=}1$
hereafter. Similarly, the condition $A_{4}^{{\rm ph},\frac{1}{2}}A_{4}^{{\rm ph},\frac{1}{2}}{=}A_{4}^{{\rm ph},0}A_{4}^{{\rm ph},1}$,
as implied by \eqref{eq:doublecopyspin}, for the quadrupole tensor $Q^{\mu\nu}_{2s}$, can be traced at this order 
to $[\omega_4^{(1)}\omega_4^{(1)}]_{\mu\nu}=[\omega_4^{(2)}]_{\mu\nu}\omega_4^{(0)}$,
which shows the universality of the quadrupole, and  holds up to terms subleading in the double soft limit.

Notice when doing the double copy of the massless spin-$1$ polarization vectors we have always projected the graviton component via $\epsilon_i^{\mu\nu} = \epsilon_i^\mu\epsilon_i^\nu$. 
Then, the removal of the dilaton and axion component is trivial and does not require additional subtraction schemes, or the introduction of ghost particles \cite{Johansson:2014zca}.  
This is a general feature of amplitudes for one matter line emitting $n$-massless particles, as we will see in \cref{ch:double_copy}. In addition, since $M_4^{\text{gr}}$ and $M_5^{\text{gr}}$ are built from these amplitudes via the factorization \eqref{cuts m4 m5}, the removal of the dilaton and axion states from these amplitudes is also straightforwards by using the graviton propagator. 
 
\subsection{ From $\rm{SO}(D-1,1)$ to $\rm{SO}(D-1)$ multipoles} \label{sec:branching}

As  mentioned in \cref{sec:bremsstra_ph},  in order to  compare spinning amplitudes 
with classical results for spinning bodies it is sometimes necessary
to choose a frame through the \ac{SSC}. Let us show how this arises from our setup, and make more formal the discussion introduced by the end of  \cref{sec:bremsstra_ph}.

We have shown that the spin multipoles correspond to finite $\rm{SO}(D-1,1)$
transformations which map $p_{1}$ $\rightarrow$ $p_{2}$. Such Lorentz
transformations are composed of both a boost and a $\rm{SO}(D-1)$ Wigner
rotation. Spin multipoles of a massive spinning body are defined with
respect to a reference time-like direction and form irreps. of $\rm{SO}(D-1)$
acting on the transverse directions \cite{Levi:2015msa,Levi:2018nxp}. Hence, it is natural
to identify such action with Wigner rotations of the massive states
entering our amplitude. A simple choice for the time-like direction
is the average momentum $u=\frac{p}{m}=\frac{p_{1}+p_{2}}{2m}$. In
this frame boosts are obtained as $K^{\nu}=u_{\nu}J^{\mu\nu}$ whereas
Wigner rotations read $S^{\mu\nu}=J^{\mu\nu}-2u^{[\mu}K^{\nu]}$.
Adopting $S^{\mu\nu}$ as classical spin tensor then corresponds to
the \textit{covariant} \ac{SSC}, i.e. $u_{\nu}S^{\nu\mu}=0$ \cite{Porto:2008tb,Porto:2008jj,Vines:2017hyw}
(another choice was used in \cite{Chung:2018kqs,Holstein:2008sx}).
The momenta
$p_{1}$ and $p_{2}$ can be aligned canonically to $p$ through the
boost,
\begin{equation}
p_{1}=e^{\frac{q}{2m}\cdot K}p\,,\quad p_{2}=e^{-\frac{q}{2m}\cdot K}p\,,\label{eq:boost}
\end{equation}
which defines canonical polarization vectors $\varepsilon$, $\tilde{\varepsilon}$
for $p$ through (recall $p_{2}$ is outgoing):
\begin{equation}
   \varepsilon_{1}{=}e^{\frac{q}{2m}\cdot K}\,\varepsilon\,\,,\quad\varepsilon_{2}{=}\tilde{\varepsilon}\,e^{\frac{q}{2m}\cdot K}\,.
\end{equation}
This replacement can then be applied to the multipole expansion \eqref{eq:multiexp},
yielding an extra power of $q$ for each power of $J$, hence preserving
the $\hbar$-scaling. We find
\begin{eqnarray}
\varepsilon_{1}{\cdot}\varepsilon_{2} & {=} &
\varepsilon{\cdot}\tilde{\varepsilon}{+}\frac{1}{m}q_{\mu}\varepsilon K^{\mu}\tilde{\varepsilon}{+}\mathcal{O}(K^{2})\,,\\
\varepsilon_{1}J^{\mu\nu}\varepsilon_{2} & {=}& 
\varepsilon S^{\mu\nu}\tilde{\varepsilon}{+}2u^{[\mu}\varepsilon K^{\nu]}\tilde{\varepsilon}{+}\nonumber \\
 & &
 \frac{q_{\alpha}}{m}\varepsilon\{K^{\alpha}{,}S^{\mu\nu}\}\tilde{\varepsilon}{+}\mathcal{O}(K^{2})\,,\\
\varepsilon_{1}\{J^{\mu\nu}{,}J^{\rho\sigma}\}\varepsilon_{2} &{ =}  & 
\varepsilon\{S^{\mu\nu}{,}S^{\rho\sigma}\}\tilde{\varepsilon}{+}\mathcal{O}(K)\,,\label{eq:quadK}
\end{eqnarray}
(for generic spin $K$ and $S$ are independent). In terms of irreducible
representations this decomposition can be thought of as branching
$\rm{SO}(D-1,1)$ into $\rm{SO}(D-1)$ \cite{Bekaert:2006py}. For instance, the dipole branches as
$\ytableausetup{mathmode,boxsize=0.6em}\ydiagram{1,1}\rightarrow\ydiagram{1,1}+\ydiagram{1}$,
which is a transverse dipole plus a transverse vector irrep, $K^{\mu}$.
In the same way, in general the $\ytableausetup{mathmode,boxsize=0.6em}\ydiagram{2,2}$
irrep. of $\rm{SO}(D-1,1)$ also contains a $\ytableausetup{mathmode,boxsize=0.6em}\ydiagram{2}$
piece for $\rm{SO}(D-1)$. This is the reason we can extract a quadrupole
from Weyl piece in \eqref{eq:dcweyls2}, namely by combining \eqref{eq:quadK} with
the replacement rule
\begin{equation}
\{S^{\mu\nu},S^{\rho\sigma}\}=\frac{2}{D{-}3}\left(\bar{\eta}^{\sigma[\mu}\bar{Q}^{\nu]\rho}{-}\bar{\eta}^{\rho[\mu}\bar{Q}^{\nu]\sigma}\right)+{\rm other}\,\,{\rm irreps}\,\label{eq:SStoQ}
\end{equation}
where $\bar{\eta}^{\mu\nu}=\eta^{\mu\nu}-u^{\mu}u^{\nu}$. Thus we
have the identity (c.f. \cite{Steinhoff:2012rw,Chen:2019hac})
\begin{equation}
    \begin{split}
      \omega_{\mu\nu\rho\sigma}\Sigma^{\mu\nu\rho\sigma}  &=  [\omega]_{\mu\nu\rho\sigma}^{\ytableausetup{mathmode,boxsize=0.15em}\ydiagram{2,2}}\langle\varepsilon_{1}|\{J^{\mu\nu}{,}J^{\rho\sigma}\}|\varepsilon_{2}\rangle\label{eq:weylproj}\,,\\
  &=  \frac{4}{D-3}[\omega]_{\mu\nu\rho\sigma}^{\ytableausetup{mathmode,boxsize=0.15em}\ydiagram{2,2}}u^{\nu}\bar{Q}^{\mu\rho}u^{\sigma}+O(K)\,.
    \end{split}
\end{equation}
For instance, we extract a quadrupole contribution from $A_{3}^{h,s}$ in \eqref{eq:3pts2}:
\begin{equation}
A_{3}^{h,s}|_{\bar{Q}} { =}  \frac{1}{4}\left(\epsilon\cdot p_{1}\right)^{h}\frac{q\cdot\bar{Q}\cdot q}{D-3}\label{eq:quad3pt2}\,.
\end{equation}

Of course, the $\rm{SO}(D-1,1)$ quadrupole present in $A_{4}^{h,s}$ also
contains a $\rm{SO}(D-1)$ quadrupole. It follows from \eqref{eq:quadK}. Similarly, the $\hat{1}$ piece in \eqref{eq:dcweyls2} also have a contribution proportional to $S^2$.    We can summarize the map in the following way
\begin{equation}\label{eq:branching-general-d}
    \begin{split}
        J_1^{\mu\nu}&\rightarrow S^{\mu\nu}\\
        Q_1^{\mu\nu}&\rightarrow \bar{Q}^{\mu\nu}+ \frac{1}{D-1}\bar{\eta}^{\mu\nu} \hat{1}_1 - \frac{1}{D}\hat{1}_1 \\
        \Sigma^{\mu\nu\rho\sigma}&\rightarrow\frac{4}{D-3}u^\nu \bar{Q}^{\mu\rho}u^{\sigma}\\
        \hat{1}_1&\to \frac{1}{2}S_{\mu\nu}S^{\mu\nu},
    \end{split}
\end{equation}
In general the $\rm{SO}(D-1)$ multipoles defined
through the covariant \ac{SSC} are given directly from the $\rm{SO}(D-1,1)$ ones,
up to $O(K)$ terms. Due to unitarity, one expects the latter to drop from the amplitude, at least for $A_3$. Let us show explicitly how this happens. Note that 3-pt. kinematics implies $[q{\cdot} K,q{\cdot} J{\cdot}\epsilon]=0$
and hence the spin piece of the 3-pt. amplitude \eqref{eq:3ptexp} reads
\begin{equation}
    \begin{split}
        \varepsilon_{1}e^{\frac{q\cdot J\cdot\epsilon}{\epsilon\cdot p}}\varepsilon_{2} &{ = } \tilde{\varepsilon}\exp\left(\frac{q_{\mu}\epsilon_{\nu}J^{\mu\nu}}{\epsilon\cdot p}{+}\frac{q_{\mu}K^{\mu}}{m}\right)\varepsilon\,=\tilde{\varepsilon}e^{\mathcal{S}}\varepsilon\, \\
 & =  \sum_{n=0}^{\infty}\frac{1}{n!}\tilde{\varepsilon}\left(\frac{q_{\mu}\epsilon_{\nu}S^{\mu\nu}}{\epsilon\cdot p}\right)^{n}\varepsilon\label{eq:expred-2}\,,
    \end{split}
\end{equation}
where one can check that the sum truncates at order $2s$. Thus the boost \eqref{eq:boost} is effectively subtracted from the
finite Lorentz transformation leading to the interpretation of the
3-point formula as a little-group rotation induced via photon/graviton
emission. We end with a comment on the case $s>h$ and $D>4$: Note
that the pole $\epsilon\cdot p$ cancels in \eqref{eq:quad3pt2} for
any dimension. This means we can provide a local form of the 3-pt.
amplitude which contains the same multipoles as the exponential. For instance,
\begin{equation}
    \begin{split}
       A_{3}^{{\rm ph},2}  {=}  \left(2e\epsilon{\cdot} p\right)\phi_{2}{\cdot}\left(\mathbb{I}{+}\frac{\epsilon_{\mu}q_{\nu}J^{\mu\nu}}{\epsilon{\cdot} p}{+}\frac{q_{\mu}q_{\rho}}{4m^{2}\,\epsilon{\cdot} p}\times\right.
   \left.\left[\epsilon_{\nu}p_{\sigma}{+}\epsilon_{\sigma}p_{\nu}{-}\frac{\eta_{\nu\sigma}\left(\epsilon{\cdot} p\right)}{D{-}3}\right]\{J^{\mu\nu}{,}J^{\rho\sigma}\}\right){\cdot}\phi_{1} \,,
    \end{split}
\end{equation}
also yields \eqref{eq:quad3pt2} and reduces to \eqref{eq:3ptexp} in $D=4$.(Recall $\phi_i^{\alpha\beta}$ are polarization tensors for the spin 2 matter fields.) In general the $2^{n}$-poles
\cite{Levi:2018nxp,Vines:2017hyw} of \eqref{eq:expred-2} are obtained by performing $\left\lfloor \frac{n}{2}\right\rfloor $
traces with the spatial metric $\bar{\eta}^{\alpha\beta}$ appearing
in \eqref{eq:SStoQ}. The result takes the local form
\begin{equation}
    \begin{split}
        \left.\bar{A}_{3}^{h,s}\right|_{2^{n}-{\rm poles}}  {=}g_h \left(\epsilon{\cdot} p\right)^{h}\sum_{n=0}^{\infty}\left(\alpha_{n}{+}\beta_{n}\frac{q_{\mu}\epsilon_{\nu}S^{\mu\nu}}{\epsilon{\cdot} p}\right)    \times\bar{Q}_{\mu_{1}{\ldots}\mu_{2n}}^{(n)}q^{\mu_{1}}{\cdots} q^{\mu_{2n}}\label{eq:localmultip}\,,
    \end{split}
\end{equation}
where $\alpha_{n}$, $\beta_{n}$ depend on the dimension $D$, and
$\bar{Q}{}^{(n)}$ are the transverse multipoles. In four dimensions
we find $\bar{Q}^{(n)}$ to be a tensor product of the Pauli-Lubanski
vector $S^{\mu}$ \cite{Levi:2018nxp,Chung:2018kqs},
and $\alpha_{n}=\frac{m^{-2n}}{(2n)!},$
$\beta_{n}=\frac{m^{-2n}}{(2n+1)!}$.

\textit{Spin-multipoles for D=4}. 

Let us finish this section by extracting the classical limit of the gravitational Compton amplitude, up to quadratic order in spin. 
Now, since we are interested in making contact with  the scattering of waves off the Kerr \ac{BH}, as we will see in \cref{ch:GW_scattering},   we specify the spin multipoles for the $D=4$ scenario.  As already pointed out, the  spin dipole can be written in terms of the  Pauli-Lubanski vector, via $S^{\mu\nu}=\epsilon^{\mu\nu\rho\sigma}p_{1\rho }a_{\sigma}$, where $a^\mu$ corresponds to the 
radius of the Kerr ring singularity. We will expand on this in \cref{ch:GW_scattering}, for the moment we can think of it as a classical spin tenor representing the intrinsic rotation of the K \ac{BH}. In the same way, the  spin quadrupole can be put in terms of $a^\mu$ via
\begin{equation}\label{eq:branching-4-d}
\begin{split}
    \bar{Q}^{\mu\nu}&=m^2 (a^\mu a^\nu-\frac{1}{3}\bar{\eta}^{\mu\nu}a^2),
    \end{split}
\end{equation}
where now the \ac{SSC} is satisfied by the  spin vector  $p_{1\mu} a^\mu=0$. Finally we have to do the usual $\hbar$ scaling of the massless momenta, $k_i\to \hbar k_1$, and in the same way for the spin vector  $a^\mu\to a^\mu/\hbar$. Where we have also identified  $u_\mu$  with the incoming massive object's four-velocity.   In that form, one can explicitly check that the final classical amplitude up to quadratic order in spin can be written as:
\begin{equation}\boxed{
    \langle A_4 ^{\text{gr}}\rangle = \frac{\kappa^2}{8}\frac{\langle\omega_4^{(0)}\rangle}{k_2{\cdot}k_3 (p_1{\cdot}k_2)^2}\left[\langle \omega^{(0)} \rangle + \langle \omega_4^{(1)\mu\nu}\rangle\epsilon_{\mu\nu\rho\sigma}p_1^{\rho}a^{\sigma} + \langle \overline{\omega}^{(2)}_{4\,\alpha\beta} \rangle a^{\alpha}a^{\beta}   \right]+ \mathcal{O}(a^3)\,,}\label{eq:a4CLASSSPIN2}
\end{equation}
where the angles indicate the classical limit of the corresponding multipole  coefficients given in \eqref{eq:w0} and \eqref{eq:w_2}. We have also identify the classical multipole coefficient for the quadratic in spin amplitude in  classical electromagnetism as 
\begin{align}
  \langle \overline{\omega}_4^{(2)\alpha\beta} \rangle &=\Big[p_{1}{\cdot}F_{2}{\cdot}F_{3}{\cdot}p_{1}(k_{1}{-}k_{2})_{\mu}\bar{\mathcal{P}}^{\mu\nu\alpha\beta}(k_{1}{-}k_{2})_{\nu}+\frac{k_{1}{\cdot}k_{2}m^{2}}{2}\Big(\bar{\mathcal{P}}^{\mu\nu\alpha\beta}+\frac{\eta^{\mu\nu}\eta^{\alpha\beta}}{2}\Big)F_{2}^{(\mu|\delta}F_{3}^{\gamma|\nu)}\eta_{\gamma\delta}\Big]\,,\label{eq:quad_class}\\
   \bar{\mathcal{P}}^{\mu\nu\alpha\beta}&=\frac{\eta^{\mu\alpha}\eta^{\nu\beta}+\eta^{\mu\alpha}\eta^{\nu\beta}}{2}-\eta^{\mu\nu}\eta^{\alpha\beta}\,.\label{eq:varp}
\end{align}
Notice remarkably that after summing up all the rotation spin multiple contributions,  the classical \ac{GR} amplitude has the factorization form $A^{\text{gr}} = \langle A_0^{\text{ph}}\rangle \times \langle A_s^{\text{ph}}\rangle $, as suggested by the authors  in \cite{Bern:2020buy}, and reflecting the universality of the coupling of matter to the graviton. The quadratic in spin contribution and its unitarity gluing with the 3-point amplitude, recovers the quadratic in spin  two-body radiation amplitude obtained in \cite{Jakobsen:2021lvp}, as we will explicitly show in \cref{sec:M4M5_spin}.

\subsection{Spin multipolar expansion of $M_4$ and $ M_5$}\label{sec:M4M5_spin}

Let us in the remaining of this chapter compute the gravitational $M_4$ and $M_5$ amplitudes up to the spin quadrupole level from the double copy. Let us recall at leading order in the perturbative expansion,  the $\hbar {\to} 0$ limit of the amplitudes is captured by the cuts of $M_{4}$ and $M_{5}$ given in (\ref{cuts m4 m5}).  $M_4$ can then be computed from the formula \eqref{eq:M4_general}, whereas for $M_5$, the key point is to introduce the average momentum transfer $q=\frac{q_1-q_2}{2}$, as done for the electromagnetic case \cref{sec:examplesa_tree_level}, after which one expects the same construction to apply. This  leave us then  with the formula \eqref{eq:newM5clas}, which we now use in the gravitational context. 

\subsubsection{Classical Double Copy}
As the numerators in \cref{eq:M4_general,eq:newM5clas}          
correspond
to $A_{n}^{h,s}$ amplitudes, the multipole double copy can be directly
promoted to $\langle M_{4}\rangle$ and $\langle M_{5}\rangle$. From
a classical perspective, the factorization of (\ref{cuts m4 m5}) implies that
the photon numerators can always be written as $n_{{\rm ph}}=t_{a\mu}t_{b}^{\mu}$
where $t_{a}$ and $t_{b}$ \textit{only} depend on kinematics for  particle 1 and
$2$ respectively. The simplest example is the scalar piece in $\langle M_{4}^{{\rm ph}}\rangle$,
where $t_{a}=p_{1}$ and $t_{b}=p_{2}$. The \ac{KLT} formula \eqref{eq:doublecopyspin} translates
to
\begin{equation}
\boxed{
n_{{\rm gr}}=n_{{\rm ph}}\odot n_{{\rm ph}}-{\rm tr}(n_{{\rm ph}}\odot n_{{\rm ph}})\,\label{eq:numdc}}
\end{equation}
where we defined the trace operation as
${\rm tr}(n\odot n)=\frac{(t_{a\mu}\odot t_{a}^{\mu})(t_{b\mu}\odot t_{b}^{\mu})}{D-2}$. By combining \eqref{eq:numdc} with \cref{eq:M4_general,eq:newM5clas}, 
this establishes for the first time a classical double-copy
formula that can be directly proved from the standard \ac{BCJ} construction
as we will see in \cref{ch:double_copy}.  Moreover, up to this order it only requires as input Maxwell
radiation as opposed to gluon color-radiation \cite{Goldberger:2016iau,Goldberger:2017ogt} and contains no Dilaton/Axion states \cite{Johansson:2014zca,Luna:2017dtq,Goldberger:2017ogt}, the latter of which are removed by the trace subtraction in \eqref{eq:numdc}. Since the 
formula for the  radiative amplitude for the gravitational case \eqref{eq:newM5clas} follows from the gravitational Compton amplitude,  the spurious  pole  $\frac{1}{q{\cdot}k}$ arises from the $t$-channel of the Compton amplitude, and its cancellation from the final result provides an strong check of our double copy formula. 

Let us in the following provide  some examples of how to use this double copy formula \eqref{eq:numdc}. Starting with  $\langle M_{4}\rangle$: The simplest example is the scalar amplitude, for which, as we already mention $n_{0}^{\text{ph}} = p_1{\cdot}p_2$, then  $t_{a}=p_{1}$ and $t_{b}=p_{3}$, so that $tr(n_{\text{ph}}\odot n_{\text{ph}}) =\frac{m_1^2 m_2^2}{D-2} $. This then leads to the scalar gravitational amplitude 

\begin{equation}
\langle M_{4}^{{(0),\rm gr}}\rangle=\frac{n_{\rm gr}}{q^{2}}=\frac{32\pi G}{q^{2}}\left[(p_{1}{\cdot} p_{2})^{2}-\frac{m_{1}^{2}m_{2}^{2}}{D-2}\right]\label{eq:grnum}, 
\end{equation}
where the factor of $D-2$ arises from the graviton propagator. In
$D=4$ we can evaluate  \eqref{eq:impulse_final} to recover the 1\ac{PM} scattering
angle as in \cite{Bjerrum-Bohr:2018xdl}, 
first derived in the classical context
by Portilla \cite{Portilla_1979,Portilla_1980}. 

Next we can consider the quadratic in spin contribution. 
To keep notation simple consider only particle $a$ to have spin. From \eqref{eq:3ptshalf}
we find that at the dipole level the numerator for $\langle M_{4}^{{\rm ph}}\rangle$
is $n_{\frac{1}{2}}^{{\rm ph}}=n_{{\rm 0}}^{{\rm ph}}+p_{3}{\cdot}J_{a}{\cdot}q$.
The gravity result follows from \eqref{eq:numdc} by dropping contact
terms in $q^2$. The rules \eqref{eq:rules} readily scalar as we just saw, as well as the dipole
parts. let us compute the more interesting  quadrupole part; rule \eqref{eq:dc12}
gives
\begin{equation}
   \frac{(p_{3}{\cdot}J_{a}{\cdot}q)\odot(p_{3}{\cdot}J_{a}{\cdot}q)-{\rm tr(\cdots)}}{q^{2}}=\frac{1}{4}\frac{p_{3\mu}q_{\nu}p_{3\alpha}q_{\beta}\Sigma_a^{\mu\nu\alpha\beta}}{q^{2}}\,, 
\end{equation}
Using  \eqref{eq:weylproj}, the $\rm{SO}(D-1)$ quadrupole \cite{Porto:2005ac,Levi:2014gsa,Levi:2015msa} reads 
\begin{equation}
\frac{1}{4}\frac{p_{3\mu}q_{\nu}p_{3\alpha}q_{\beta}\Sigma_a^{\mu\nu\alpha\beta}}{q^{2}}\rightarrow\left((p_1{\cdot}p_2)^2{-}\frac{m^2_1 m^2_2}{D{-}2}\right)\frac{q\cdot\bar{Q}_a\cdot q}{2(D{-}3)q^2 m_a^2}.\label{weyl4}
\end{equation}
Up to this order this agrees with the $D=4$ computation \cite{Vines:2017hyw,Guevara:2018wpp,o'connell-vines}. Agreement to all orders in spin is obtained from the formula
\eqref{eq:localmultip}.

Let us now move to compute the classical numerators for the gravitational $\langle M_5\rangle$ amplitude. We start with the scalar case, for which the gravitational numerators can be computed from putting the photon numerators \eqref{nphsc} into the double copy formula \eqref{eq:numdc} 
\begin{equation}
\label{eq:scalar-numerators-gr}
    \begin{split}
        n^{(a)}_{0,\rm{gr}}&=\frac{\kappa^3}{4}\left[\left(p_{1}{\cdot}p_{2}F_{1q}-p_{1}{\cdot}k F_{p}\right)^{2}- \frac{m_{1}^{2}m_{2}^{2}}{D-2}F_{1q}^{2} \right],\\
         n^{(b)}_{0,\rm{gr}}&= -\frac{\kappa^3}{4}\left[\left(p_{1}{\cdot}p_{2}F_{2q}+p_{2}{\cdot}k F_{p}\right)^{2}-\frac{m_{1}^{2}m_{2}^{2}}{D-2}F_{2q}^{2}\right],
    \end{split}
\end{equation}
where we have used  \textcolor{black}{ $F_{iq}=\eta_{i}(p_{i}{\cdot}F{\cdot}q)$, and $F_p= p_1 {\cdot} F{\cdot} p_2$}. Analogous to the electromagnetic numerators, this can also be obtained from the gluing of the scalar, 3-pt and 4-pt amplitudes through the graviton propagator.    These numerators  can  be introduced in the general formula \eqref{eq:newM5clas}, to  recover the result for the classical limit of the gravitational  amplitude for scalar particles \cite{Goldberger:2016iau,Luna:2017dtq,Bautista:2019evw}. 

In analogy to the electromagnetic case, the   gravitational 5-point amplitude for scalar sources can also be put in an all order exponential soft expansion.  This is a consequence of the  soft exponentiation  of the gravitational Compton amplitude, in analogy  to its electromagnetic counterpart \eqref{eq:scphcompton},
\begin{equation}
    \begin{split}
     A_{4}^{{\rm gr}}{=}\sum_{p_{a}{=}p_{1},p_{2},k_{2}}\frac{1}{2}\frac{(\epsilon_{3}\cdot p_{a})^{2}}{k_{3}\cdot p_{a}}e^{\frac{2F_{3}\cdot J_{a}}{\epsilon_{3}\cdot p_{a}}}A_{3}^{{\rm gr}}=\frac{1}{2k_{2}{\cdot}k_{3}}\times\left[\frac{(p_{1}{\cdot}\epsilon_{2})^{2}}{p_{1}{\cdot}k_{3}p_{4}{\cdot}k_{3}}F_{k}^{2}-2\frac{p_{1}{\cdot}\epsilon_{2}}{p_{1}{\cdot}k_{3}}F_{k}F_{\epsilon}+\frac{p_{4}{\cdot}k_{3}}{p_{1}{\cdot}k_{3}}F_{\epsilon}^{2}\right]
\label{eq:scgrcompton}.  
    \end{split}
\end{equation}
which induces the exponentiation for the scalar numerators \eqref{eq:scalar-numerators-gr} 
\begin{equation}
   n_{{\rm gr}}^{(a)}=\frac{F_{1q}^{2}}{2}e^{-\frac{F_{p}}{F_{1q}}(p_{1}{\cdot}k)\frac{\partial}{\partial(p_{1}{\cdot}p_{2})}}\left[(p_{1}{\cdot}p_{2})^{2}-\frac{m_{1}^{2}m_{2}^{2}}{D-2}\right]\,.
\end{equation}
Further writing $\frac{1}{q^2\\ac{PM} q\cdot k}=e^{\\ac{PM} q\cdot k\frac{\partial}{\partial q^2}}\frac{1}{q^2}$
turns \eqref{eq:newM5clas} into
\begin{equation}
\boxed{
\langle M_{5}^{\rm gr}\rangle=\sum_{i=1,3}\mathcal{S}^{\text{gr}}_{i}e^{\eta_i \left(F_{p}\frac{p_{i}{\cdot}k}{F_{iq}}\frac{\partial}{\partial(p_{1}{\cdot}p_{2})}+q{\cdot}k\frac{\partial}{\partial q^2}\right)}\langle M_{4}^{\rm gr}\rangle\,\label{eq:classicsoftGR}}
\end{equation}
where now the soft factor are  $\mathcal{S}^{\text{gr}}_{i}{=}\frac{\eta_i}{2} \frac{F_{iq}^2}{(p_{i}{\cdot}k)^{2} q{\cdot}k}$. This is the gravitational analog of   \eqref{eq:classicsoft}. Similarly to the electromagnetic case \eqref{eq:a_weinerg_tree}, we can show to leading order in the soft expansion, amplitude \eqref{eq:classicsoftGR}, when used into \eqref{eq:radi_field} leads to the memory waveform. That is
\begin{equation*}
    \int \frac{d^{D}q}{(2\pi )^{D-2}}\delta(2q\cdot p_{1})\delta(2q\cdot p_{2})e^{iq\cdot(b_{1}-b_{2})}\left(\sum_{i=1,3}\mathcal{S}_{i}\right)\langle M_{4}^{{\rm gr}}\rangle
\end{equation*}
as $k\to 0$. Evaluating the sum using  
$\Delta p_{1}=-\Delta p_{2}$ we obtain
\begin{equation}
    \epsilon_{\mu\nu}T^{\mu\nu}=\frac{F_{p}/2}{p_{1}{\cdot}kp_{2}{\cdot}k}\left(\frac{p_{1}}{p_{1}{\cdot}k}{+}\frac{p_{2}}{p_{2}{\cdot}k}\right){\cdot}F{\cdot}\Delta p{+}\mathcal{O}(k^{0}),
\end{equation}
which at leading order in $\Delta p$ (or $G$, if restored) becomes
\begin{equation}\label{eq:classicsoftGRwe}
  T^{\mu\nu}(k)=\sqrt{8\pi G}\times\Delta\left[\frac{p_{1}^{\mu}p_{1}^{\nu}}{p_{1}{\cdot} k}+\frac{p_{2}^{\mu}p_{2}^{\nu}}{p_{2}{\cdot} k}\right]^{{\rm TT}}\, .
\end{equation}
In position space this gives
the burst memory wave derived by Braginsky and Thorne \cite{Braginsky1987} in $D=4$ (a $\frac{1}{4\pi R}$ factor arises from the ret. propagator as $R{\to}\infty$, see \cref{sec:KMOC} and \cite{Goldberger:2016iau,Hamada:2018cjj,Kosower:2022yvp}), see also \cite{Pate:2017fgt,Mao:2017wvx,Satishchandran:2017pek} for $D>4$. Here we have provided a direct connection with the Soft Theorem in the gravitational case, alternative to the expectation-value arguments of \cite{Strominger:2014pwa,Strominger:2017zoo}. This can also be seen as the Black Hole Bremsstrahlung of \cite{Luna:2016due,Luna:2016hge} generalized to consistently include the dynamics of the sources.

Next, the gravitational numerators to linear-order  in spin  can analogously be computed. For that we use a copy of the scalar numerators  \eqref{nphsc}, and one of the linear in spin numerators 
\eqref{eq:lineas-in-spin-numerators-photon}, into our double copy formula  \eqref{eq:numdc}. We  get 
\begin{equation}\label{eq:linear-spin-amplitude-gr}
\begin{split}
    n^{(a)}_{\frac{1}{2},\rm{gr}}& = \frac{\kappa^3}{8}\Big\{\left(p_{1}{\cdot}p_{2}F_{1q}-p_{1}{\cdot}k F_{p}\right) \left[\left(  p_1{\cdot} p_2\,q{\cdot} k+p_1{\cdot} k\,p_2{\cdot} k   \right)   F{\cdot}J_{2s,1}{-}F_{1q}R_{2}{\cdot}J_{2s,1}{+}p_{1}{\cdot}k\,[F,R_{2}]{\cdot}J_{2s,1}\right]\\
     &\hspace{1.5cm}+\frac{m_{2}^{2}F_{1q}}{D-2}\left[F_{1q}(2q{-}k){\cdot}J_{2s,1}{\cdot}p_{1}{-}m_{1}^{2}q{\cdot}k\,F{\cdot}J_{2s,1}+p_{1}{\cdot}k(2q{-}k){\cdot}F{\cdot}J_{2s,1}{\cdot}p_{1}\right]\Big\},\\
      n^{(b)}_{\frac{1}{2},\rm{gr}} & =- \frac{\kappa^3}{8}\Big\{ \left(p_{1}{\cdot}p_{2}F_{2q}+p_{2}{\cdot}k F_{p}\right)
       \left(F_{2q} (2q+k){\cdot}J_{2s,1}{\cdot}p_2 -  p_2{\cdot}k\,p_2{\cdot}F{\cdot}J_{2s,1}{\cdot}(2q+k)\right) \\
     &\hspace{1.5cm} +\frac{m_{2}^{2}F_{2q}^{2}}{D-2}(2q{+}k){\cdot}J_{2s,1}{\cdot}p_{1}\Big\}.
    \end{split} 
\end{equation}
Here we remark  the generators $J_{2s,1}$ act in the gravitational theory rather than in the electromagnetic counterpart.  Similarly to the scalar case, these numerators can be placed in \eqref{eq:newM5clas} to recover the corresponding gravitational amplitude. To obtain the full amplitude for both particles with spin, we utilize the symmetrization mappings 
\begin{align}
m_1\leftrightarrow m_2, & & p_1\leftrightarrow p_2, & & q\rightarrow-q, & & J_{2s,1}\rightarrow J_{2s,2},
\end{align}
in the final formula. The resulting amplitude  recovers the spinning amplitude in dilaton gravity computed in \cite{Li:2018qap} for classical spinning sources, once we remove the terms proportional to $m_i$ in the  numerators in \eqref{eq:linear-spin-amplitude-gr}, which arise  from the graviton projection, and branch $J^{\mu\nu}\to S^{\mu\nu}$.
 This provides a strong cross-check of our method.

 Using \eqref{eq:dc12} we can also compute the quadrupolar order. For instance, the $Q^{\mu\nu}$ piece can be computed from two copies of the linear in spin numerators \eqref{eq:lineas-in-spin-numerators-photon}. Let us again for simplicity consider only particle $a$ with spin. 
\begin{equation}\label{eq:quqdcov}
     \begin{split}
          \frac{n^{(a)}|_Q}{q{\cdot}k}{=}&\frac{(32\pi G)^{\frac{3}{2}}}{8(D{-}2)}\biggl[\left(p_{1}{\cdot} p_{2}F_{1q}{-}p_{1}{\cdot}k\,F_{p}\right)\{R_2{,}F\}{\cdot}Q_{1}+ 
 \frac{m_{2}^{2}}{(D{-}2)}\left(F_{1q}\{F{,}Y\}{\cdot}Q_{1}{-}2p_{1}{\cdot}k\,p_{1}{\cdot}F{\cdot}Q_{1}{\cdot}F{\cdot}q\right)\biggr],
     \end{split}
 \end{equation}
with $Y^{\mu \nu}=p_1^{[\mu}(2q{-}k)^{\nu]}$, whereas $n^{(b)}|_Q=0$. As before, we have dropped contact terms in $q^2$ and used the support of $\delta(p_i{\cdot}q_i)$. This result  can be shown to agree with a much more lengthy computation of
the full $M_{5}^{{\rm gr}}$ using Feynman diagrams. At this order,
$M_{5}^{{\rm gr}}$ contains classical quadrupole pieces and quantum
scalar and dipole pieces. Interestingly, while the scalar part is trivial to identify, we have found that the dipole part can be cancelled by adding the spin-1 spin-0 interaction $(B_{\mu}\partial^{\mu}\phi)^{2}$ to the
Lagrangian, which signals its quantum nature.

Let us stress the numerators \eqref{eq:linear-spin-amplitude-gr} and \eqref{eq:quqdcov} are written in terms of the SO($D-1,1$)  multipole operators. Furthermore, for the quadrupolar contribution, the full amplitude  includes the additional irreps. ( Weyl and trace pieces in \eqref{eq:dcweyls2}). In order to use the radiative amplitude in a more realistic context, as for instance for two   coalescing  KBHs, as we will 
study in \cref{ch:bounded}, 
the amplitude needs to be computed in terms of the spin multipoles of the rotation group SO($3$), in $D=4$, where all the irreps. can be written in powers of the Pauli-Lubanski vector $s^\mu=m\times a^\mu$. This is achieved through the map \eqref{eq:branching-general-d}, where the rotation quadrupole moment in 4-dimensions is given in \eqref{eq:branching-4-d}. 

In $D=4$, at the quadrupole level we can however take an alternative route in the computation of the full $M_5^{\text{gr}}$. Since we already have the classical gravitational Compton amplitude written in terms of $a^\mu$ as given in \eqref{eq:branching-4-d},  we can glue it to the classical  3-point amplitude \eqref{eq:localmultip} as given by the unitary prescription of eq. \eqref{cuts m4 m5}. This in turn will allow us to identify the gravitational numerators entering in \eqref{eq:newM5clas} in a simpler way. We arrive at the following numerators given in \eqref{eq:num_quad_class} and \eqref{eq:num_quad_class2} in  \cref{ch:quadratic_spin} 
They agree with the more involving   computation of the irreps in \eqref{eq:dcweyls2}. In \cref{ch:bounded}, we will use numerators \eqref{eq:scalar-numerators-gr}, \eqref{eq:linear-spin-amplitude-gr}, \eqref{eq:num_quad_class} and \eqref{eq:num_quad_class2} to compute the gravitational waveform at leading and subleading order in the velocity expansion for the coalescing of two KBHs in general closed orbits, whose spins are aligned  with the angular momentum of the system. We will then specialize to circular orbits where the waveform can be computed to all orders in the BHs' spins.

\section{Outlook of the chapter}\label{sec:outlook_spin}
We have shown that key techniques of Scattering Amplitudes such as soft
theorems and double copy can be promoted directly to study classical phenomena arising in \ac{GWs} . These
techniques drastically streamline the computation of radiation and spin
effects; both are phenomenologically important for Black Holes, which are believed to be extremely spinning in nature \cite{Risaliti2013,Reis2014}. In that direction, one could for instance apply our formalism to derive the hexadecapole ($s=2$) order in radiation \cite{Marsat:2014xea,Siemonsen:2017yux} to \ac{LO} in $G$ but all orders in $1/c$. 

\textit{Soft Theorem/Memory Effect:} It would be interesting to understand the meaning of the higher orders of \eqref{eq:classicsoft}, considering for instance the Spin Memory Effect \cite{Pasterski:2015tva,Nichols:2017rqr}. Motivated by the infinite soft theorems of \cite{Hamada:2018vrw,Campiglia:2018dyi} one could expect the corrections are related to a hierarchy of symmetries. One may also incorporate spin contributions and study their interplay with such orders \cite{Hamada:2018cjj}. In the applications side, it is desirable to further investigate \eqref{eq:classicsoft} at loop level \cite{Bern:2014oka,He:2014bga}, which could lead to a simple way of obtaining $\langle M_5\rangle$ from $\langle M_4\rangle$.

\textit{Generic Orbits:}  
For orbits more general than scattering $\mathcal{J}(k)$ does not have the support of $\delta(2 p_i{\cdot}q_i)$, as will become clear in \cref{ch:bounded} \cite{Goldberger:2017vcg,Shen:2018ebu}. In fact, for bounded orbits it contains the subleading terms $p_i{\cdot}q_i \sim \omega$. Very nicely, by keeping such terms in the classical calculation we have checked they match with eqs. \eqref{nphsc},\eqref{eq:lineas-in-spin-numerators-photon}, which in turn arise from the form in \eqref{eq:w2} via a natural "$F{\to}R$ replacement". As we will show in \cref{ch:bounded}, one can use the amplitudes computed in the present chapter to approach the two-body problem in General Relativity at lower orders in the velovoty expansion, where the terms removed by the on-shell conditions do not contribute to the waveform.

%% file: Chapters/bounded.tex
\chapter{Bounded systems and waveforms from Spinning
 Amplitudes}\label{ch:bounded}

\section{Introduction}\label{sec:intro_bounded}

So far we have been concerned with the computation of classical  observables for bodies moving in scattering orbits. However, more realistic scenarios, as for instance, the coalescing of compact objects observed by the   LIGO/Virgo Collaboration \cite{LIGOScientific:2016aoc},   require the study of observables for bodies moving in general closed orbits. The first approach  to this problem was done in  the early days of general relativity, by Einstein  predicting the existence of gravitational waves \cite{1918SPAW154E} and cast the emission from a compact system into the, now famous, Quadrupole formula for gravitational radiation.
A little while later, in a spectacular breakthrough the LIGO/Virgo Collaboration \cite{LIGOScientific:2016aoc} confirmed Einstein's prediction by directly detecting the gravitational waves emitted from a \ac{BBH}. Higher order corrections to Einstein's Quadrupole formula in the context of the quasi-circular orbit general relativistic two-body problem -- needed to enable such detections -- have traditionally been obtained in the \ac{PN} \cite{Blanchet:2013haa,Futamase:2007zz} formalism, within numerical relativity \cite{Pretorius:2005gq} and black hole perturbation theory \cite{1973Teukolsky,Kokkotas:1999bd}, as well as models combining these approaches \cite{PhysRevD.59.084006,Buonanno:2000ef,Santamaria:2010yb}. More recently, however, efforts have been focused on the \ac{BBH} scattering problem, in order to connect classical computations performed in the context of the \ac{PM} theory \cite{Damour_2016,Porto:2016pyg,Goldberger:2017vcg,Goldberger:2017ogt,Vines:2018gqi,Damour:2019lcq,Damour:2020tta,Kalin:2019rwq,Kalin:2019inp,Kalin:2020mvi,Kalin:2020fhe,Goldberger:2020fot,Brandhuber:2021kpo,Brandhuber:2021eyq}, with those approaches based on the classical limit of \ac{QFT} scattering amplitudes \cite{Cheung:2020gyp,Cheung:2018wkq,Bern:2019nnu,Bern:2019crd,Bern:2021dqo,Bjerrum-Bohr:2018xdl,Cristofoli:2019neg,Bjerrum-Bohr:2019kec,Bjerrum-Bohr:2021vuf,DiVecchia:2020ymx,DiVecchia:2021ndb,Bern:2020buy,Chung:2019duq,Chung:2020rrz,Cachazo:2017jef,Guevara:2017csg,Guevara:2018wpp,Guevara:2019fsj,Aoude:2020onz,Bautista:2019evw, Blumlein:2020pyo, Blumlein:2020znm}.

Until recently, the scattering amplitudes approach to the two-body scattering problem had mostly focused it's efforts in the conservative sector, although in this work we have shown how the radiative sector to leading orders in perturbation theory can be similarly approach from scattering amplitudes.  In addition, soft theorems \cite{Laddha:2018rle,Saha:2019tub,Ghosh:2021hsk} suggest that the full radiative sector can be approached from the classical limit of a 5-point scattering amplitude, as we have seen in previous chapters.
The introduction of the \ac{KMOC} formalism \cite{Kosower:2018adc}, enabling the computation of classical observables directly from the scattering amplitude, proved to be extremely useful in determining radiative observables as extensively exemplified in during the body of this thesis, ranging from the leading in $G$  memory waveform from hyperbolic, soft encounters was presented \cref{sec:spin_in_qed} , to the prediction of the  waveform to all orders in perturbation theory, but to leading order in the soft expansion. 
 In this same formalism, the computation of the full leading \ac{PM} order radiated four-momentum was recently presented in \cite{Herrmann:2021lqe,Herrmann:2021tct}; these results were subsequently confirmed by other methods in \cite{DiVecchia:2021bdo,Bini:2021gat,Riva:2021vnj}. Simultaneously, using a worldline-QFT formalism \cite{Mogull:2020sak}, the computation of the  gravitational waveform valid for all values for the momentum of the  emitted graviton, was computed  in \cite{Jakobsen:2021smu} (see also \cite{Mougiakakos:2021ckm}), and extended to include spin effects in \cite{Jakobsen:2021lvp}. Analogously, the scattering amplitudes approach has been employed to study radiation scattering off of a single massive source \cite{Cristofoli:2021vyo,Kol:2021jjc}, where a novel connection between  scattering amplitudes and black hole perturbation theory has emerged  \cite{Bautista:2021wfy}, shedding light on how to obtaining the higher-spin gravitational Compton amplitude \cite{BCGV}, as we will expand in \cref{ch:GW_scattering}
 (see also \cite{Falkowski:2020aso,Chiodaroli:2021eug}). 

Even with the powerful scattering amplitudes techniques at hand, so far, radiative information from bodies moving on \textit{bounded} orbits has been obtained only via analytic continuation \cite{Kalin:2019inp, Kalin:2019rwq} of radiation observables of scattering bodies \cite{Herrmann:2021lqe,Dlapa:2021npj,Saketh:2021sri} (applying mainly in the large eccentricity limit). However, the almost $40$ year old derivation of the Einstein quadrupole formula from a Feynman diagrammatic perspective by Hari Dass and Soni \cite{HariDass:1980tq}, and the more recent derivation by Goldberger and Ridgway using the classical double copy \cite{Goldberger:2017vcg}, suggest that scattering amplitudes can indeed be used to derive gravitational radiation emitted from objects moving on general \textit{closed} orbits (including the zero eccentricity limit, i.e., quasi-circular orbits). In this chapter we follow this philosophy to compute the gravitational waveform emitted from an aligned spin \ac{BBH} on general and quasi-circular orbits, up to quadratic order in the constituents spin at the leading order in the velocity expansion and to sub-leading order in the no-spin limit, from the classical  5-point scattering amplitude derived in \cref{sec:spin_in_qed}. We contrast and compare these results to the analogous classical derivation of the corrections to the Einstein quadrupole formula using the well-established multipolar post-Minkowskian formalism  \cite{Thorne:1980ru, Blanchet:1985sp, Blanchet:1995fg, Blanchet:1998in, Blanchet:2013haa}. 

We find perfect agreement between the classical and the scattering amplitudes derivation of all radiative observables we consider, to the respective orders in the spin and velocity expansions. Furthermore, we show that at leading order in the \ac{BBH} velocities, there is a one-to-one correspondence between the \ac{BBH} source's mass and current multipole moments, and the scalar and linear-in-spin 5-point scattering amplitude, respectively. At quadratic order in the spin of the black holes, we demonstrate explicitly that the corresponding contribution from the quadratic-in-spin scattering amplitude does not provide additional spin information at the level of the waveform; hence, we conjecture this to hold for higher-spin amplitudes as well, based on the aforementioned correspondence. Then, the leading in velocity, all orders-in-spin waveform, is obtained purely through the solutions to the \ac{EoM} of the conservative sector of the \ac{BBH}. Furthermore, the gauge dependence of gravitational radiation information at future null infinity is a potential source of difficulty when comparing results obtained by different approaches. In this work, we provide evidence that gauge freedom partially manifests itself in the integration procedure appearing in the computation of the waveform directly from the scattering amplitude. For quasi-circular orbits, the orbit's kinematic variables are subject to certain relations, such that the gravitational waveform can take different forms without affecting the gauge invariant information contained in the total instantaneous gravitational wave energy flux.

This chapter is organized as follows: In \cref{sec:classical_derivation}, we begin by reviewing the classical derivation of the conservative sector of the spinning \ac{BBH} to all orders in the spins at leading \ac{PN}. In \cref{sec:RadDyn}, we derive the associated gravitational wave emission from this system to all order in the \ac{BH}s' spins. We then proceed with the scattering amplitudes derivation of the waveform in  \cref{sec:scatteringampl}, with the general formalism outlined in \cref{sec:general_approach_amp}. In  \cref{sec:radiated_field_amplitudes} we us the classical spinning amplitudes $M_4$ and $M_5$ obtained in \cref{sec:spin_in_qed}
for   explicitly determining  the waveform. In section \cref{sec:results}, we briefly discuss the computation of the gauge invariant energy flux, and comment on the manifestation of the gauge freedom. We conclude with a outlook  of the chapter in  \cref{sec:discusion}. In this chapter we use Greek letters $\alpha, \beta \dots$ for spacetime indices and Latin letters $i,j \dots$ for purely spatial indices. Furthermore, we use $G=c=1$ units throughout, assume $\varepsilon_{0123}=1$, and use the $2\nabla_{[\alpha}\nabla_{\beta]}\omega_\mu = R_{\alpha\beta\mu}{}^\nu \omega_\nu$ Riemann tensor sign convention. 

This chapter is based on the work by the author \cite{Bautista:2021inx}.

\section{Classical derivation}\label{sec:classical_derivation}

In order to approach the bound orbit from a classical point of view, we utilize an effective worldline action \cite{Porto:2005ac, Porto:2006bt, Porto:2008tb, Levi:2015msa, Levi:2014gsa, Porto:2016pyg, Levi:2018nxp}, parametrizing the complete set of spin-induced interactions of the two spinning  \ac{BH}s  in the weak-field regime, at linear order in the gravitational constant, i.e. at \ac{PM} order. As we are interested in \textit{bound}, as opposed to \textit{unbound}, orbits, we will be focusing on the leading \ac{PN} contribution to the 1\ac{PM} conservative sector at each order in the \ac{BH}s' spins. In the following, we first briefly summarize the necessary conservative results established in Refs.~\cite{Levi:2016ofk, Tulczyjew:1959b, Barker:1970zr, Barker:1975ae, D'Eath:1975vw,Thorne:1984mz, Poisson:1997ha, Damour:2001tu,Hergt:2007ha, Hergt:2008jn, Levi:2014gsa, Vaidya:2014kza, Marsat:2014xea, Vines:2016qwa, Siemonsen:2017yux}. Using these results, we then tackle the radiative sector, utilizing the multipolar post-Minkowskian formalism \cite{Thorne:1980ru, Blanchet:1985sp, Blanchet:1995fg, Blanchet:1998in, Futamase:2007zz} (see also Ref.~\cite{Blanchet:2013haa} and references therein). We derive the transverse-traceless (TT) pieces of the linear metric perturbations, $h_{\mu\nu}^\text{TT}$, and the total instantaneous gravitational wave power, $\mathcal{F}$, radiated by this source to future null infinity. We achieve this, considering all orders in the spins, both for an aligned spin system on general orbits at leading order in velocities, as well as specialize to quasi-circular orbits at leading and first sub-leading orders in velocities. In this section we work in the $-+++$ signature for the flat metric.


\subsection{Classical spinning Binary black hole} \label{sec:linKerr}

Let us begin by  briefly reviewing the approach to the conservative sector of the \ac{BBH} dynamics at the respective orders in the weak-field and low-velocity regimes using an effective worldline action. We start by presenting the  necessary spin-interactions to describe a rotating \ac{BH}, and then move to  review how an effective spinning \ac{BBH} action, needed for the computation of the radiation field, can be derived.

\subsubsection{Effective binary black hole action} \label{sec:effBBHaction}

An effective description of a rotating \ac{BH}, obeying the no-hair theorems, as a point particle with suitable multipolar structure in the weak-field regime rests solely on its worldline and spin degrees of freedom \cite{Porto:2016pyg, Porto:2006bt, Steinhoff:2014kwa, Marsat:2014xea, Levi:2015msa}. The former are given by a worldline $z^\mu(\lambda)$ of mass $m$, with 4-velocity  $u^\mu=dz^\mu/d\lambda$, while the latter are encoded in the BH's (mass-rescaled) angular momentum vector\footnote{This angular momentum vector $a^\mu =\varepsilon^\mu {}_{\nu \alpha\beta}u^\nu S^{\alpha\beta}/(2m)$ emerges from the spin tensor $S^{\alpha\beta}$ assuming the covariant spin supplementary condition, $p_\mu S^{\mu\nu}=0$, and a local body-fixed frame $e_A^\mu(\lambda)$. See, for instance, Ref.~\cite{Steinhoff:2014kwa} for details.} $a^\mu$, and local frame $e_A^\mu(\lambda)$. An effective worldline action, $S$, that entails the dynamics of such a \ac{BH} (or, more generally, a compact object) in the weak-field regime was developed in Refs.~\cite{Porto:2005ac, Porto:2006bt, Porto:2008tb, Levi:2015msa, Levi:2014gsa}; see Ref.~\cite{Levi:2018nxp} for further details. This action $S[h,\mathcal{K}]$, describing a rotating compact object, is built considering all possible couplings of gravitational, $h=\{h_{\mu\nu}\}$, and object specific degrees of freedom, $\mathcal{K}=\{z^\mu,u^\mu,a^\mu,e^\mu_A\}$, requiring covariance, as well as reparameterization and parity invariance \cite{Goldberger:2009qd, Ross:2012fc, Levi:2015msa, Levi:2018nxp}. At  the 1\ac{PM} level, a matching procedure between the linearized Kerr metric \cite{Harte:2016vwo, Vines:2017hyw, Vines:2016qwa} and the gravitational field $h_{\mu\nu}$, emanating from a generic compact object described by $S[h,\mathcal{K}]$, leads to a \textit{unique} set of non-minimal couplings between $h$ and $\mathcal{K}$. This ultimately results in an effective 1\ac{PM} \ac{BH} worldline action $S_\text{BH}[h,\mathcal{K}]$. This action can be extended to higher orders in $G$ in spins (see for instance Refs.~\cite{Levi:2020kvb, Levi:2020uwu, Levi:2020lfn,Siemonsen:2019dsu}).

It was shown in Ref.~\cite{Vines:2017hyw} that for a harmonic gauge linearized Kerr \ac{BH} the infinite set of spin-couplings present in the 1\ac{PM} effective worldline action $S_\text{BH}[h,\mathcal{K}]$ can be resummed into an exponential function. In a linear setup, a \ac{BH} of mass $m$ traveling along the worldline $z^\mu(\lambda)$, sources the gravitational field, $g_{\mu\nu}=\eta_{\mu\nu}+h_{\mu\nu}^\text{Kerr}+\mathcal{O}(h^2)$, with \cite{Vines:2017hyw}
\begin{align}
h_{\mu\nu}^\text{Kerr}=4\mathcal{P}_{\mu\nu}{}^{\alpha\beta}\hat{\mathcal{T}}^\text{Kerr}_{\alpha\beta} \frac{1}{\hat{r}}, & & \hat{\mathcal{T}}_{\mu\nu}^\text{Kerr}=m \exp(a * \partial)_{(\mu}{}^\rho u_{\nu)}u_\rho.
\label{eq:LinKerr}
\end{align}
Here we define $(a * \partial)^\mu {}_\nu = \epsilon^\mu {}_{\nu\alpha\beta} a^\alpha \partial^\beta$ and introduced the trace reverser  $\mathcal{P}_{\mu\nu\alpha\beta}=(\eta_{\mu \alpha}\eta_{\nu \beta}+\eta_{\nu\alpha}\eta_{\mu\beta}-\eta_{\mu\nu}\eta_{\alpha\beta})/2$
\footnote{This is not to be confused with the $\bar{\mathcal{P}}^{\mu\nu\alpha\beta}$ tensor defined in \eqref{eq:varp} .}. Additionally, $\hat{r}$ labels the proper distance between the spacetime point $x$ and the worldline $z^\mu(\lambda)$, within the slice orthogonal to $u^\mu$ \cite{Vines:2017hyw}. In the following, we restrict ourselves to the leading \ac{PN} part of the 1\ac{PM} ansatz, since this is the natural setting for \textit{closed} orbits in the weak field regime. However, while we are expanding in $\epsilon_\text{PN}\sim v^2/c^2\sim GM/rc^2$, we consider all orders in the spins, i.e., consider $\epsilon_\text{spin}\sim \chi GM/rc^2$ non-perturbatively (here, $\chi$ the black hole's dimensionless spin parameter). To that end, we choose the Minkowski coordinate time $t$ to parameterize the worldline $z^\mu$, i.e., $\lambda\rightarrow t$, and expand the 4-velocity $u^\mu=(1,\boldsymbol{v})^\mu+\mathcal{O}(v^2)$, with $\dot{z}^i=dz^i/dt=v^i$. Given this and utilizing the three-dimensional product $(a\times \partial)_i=\varepsilon_{ijk}a^j\partial^k$, the metric \eqref{eq:LinKerr} reduces to its leading \ac{PN} form:
\begin{align}
\begin{aligned}
h^\text{Kerr}_{00} & \ =\left(2 \cosh (a\times \partial)-4 v_i \sinh (a\times \partial)^i\right)\frac{m}{\hat{r}} + \mathcal{O}(v^2),\\
h^\text{Kerr}_{0i} & \ =\left(4 v_i \cosh (a\times \partial)-2 \sinh (a\times \partial)_i\right)\frac{m}{\hat{r}} + \mathcal{O}(v^2),\\
h^\text{Kerr}_{ij} & \ =\left(2 \delta_{ij}\cosh (a\times \partial)-4 v_{(i}\sinh (a\times \partial)_{j)}\right)\frac{m}{\hat{r}} + \mathcal{O}(v^2).
\end{aligned}
\label{eq:hmunuPN}
\end{align}

Note that even at zeroth order in velocity, the solution contains non-trivial gravito-magnetic contributions, $h^\text{Kerr}_{0i}$, due to the presence of the \ac{BH} spin. Conversely, an effective stress-energy distribution $T_{\mu\nu}$ can be derived that yields \eqref{eq:hmunuPN} via the linearized Einstein equations\footnote{At leading \ac{PN} order, the spacetime effectively decomposes into space and time parts, yielding a simplification of the linearized Einstein equations: $\square_\text{ret.}^{-1}T_{\mu\nu}\rightarrow \Delta^{-1}T_{\mu\nu}$ (see Ref.~\cite{Blanchet:2013haa} for details).} $\square h_{\mu\nu}^\text{Kerr}=-16\pi \mathcal{P}_{\mu\nu}{}^{\alpha\beta}T_{\alpha\beta}$. This distribution has support only on the worldline $z^i(t)$ and, with the above parameterization, is given by
\begin{align}
T_{\mu\nu}(t,x^i) = \hat{\mathcal{T}}_{\mu\nu}^\text{Kerr}\delta^3(\boldsymbol{x}-\hat{\boldsymbol{z}}(t))+\mathcal{O}(\hat{v}^2).
\label{eq:KerrStressEn}
\end{align}

Collecting these within the worldline action, we can construct an effective binary \ac{BBH} $S_\text{BBH}$ that encodes the conservative dynamics with the complete spin information at the leading \ac{PM} level \cite{Vines:2017hyw} or leading \ac{PN} level \cite{Vines:2016qwa, Siemonsen:2017yux}. That is, given two worldlines $z_{1,2}^\mu$, with velocities $u^\mu_{1,2}$, masses $m_{1,2}$, and two spin vectors $a^\mu_{1,2}$ -- conveniently collected in the sets $\mathcal{K}_{1,2}$ -- the spin interactions within the binary are obtained by integrating out the gravitational field in a Fokker-type approach \cite{Bernard:2015njp}. Following \cite{Vines:2017hyw, Siemonsen:2017yux}, in practice, the effective action for the second \ac{BH} $S_\text{BH}[h,\mathcal{K}_2]$ (containing this BH's degrees of freedom $\mathcal{K}_2$) is evaluated at the metric of the first \ac{BH} $h\rightarrow h_1$, such that $S_\text{BH}[h,\mathcal{K}_2]\rightarrow S_\text{BH}[h_1,\mathcal{K}_2]$. However, since the metric $h_1$, explicitly given in \eqref{eq:LinKerr}, is effectively a map from the gravitational degrees for freedom into that \ac{BH}s' degrees of freedom, i.e., $h_{1}\rightarrow\mathcal{K}_{1}$, the \ac{BBH} action $S_\text{BH}[h_1\rightarrow \mathcal{K}_1,\mathcal{K}_2]\rightarrow S_\text{BBH}[\mathcal{K}_1,\mathcal{K}_2]$, solely depends on the \ac{BH}s' degrees of freedom.

\subsubsection{Conservative dynamics} \label{sec:consvdyn}

In order to write out the effective \ac{BBH} action $S_\text{BBH}[\mathcal{K}_1,\mathcal{K}_2]$ explicitly, let us define the spatial separation $r^i=z^i_{1}-z^i_{2}$, with $r=|\boldsymbol{r}|$, between the two worldlines, as well as the spin sums $a^i_+=a_1^i+a_2^i$ and $a^i_-=a_1^i-a_2^i$. The angular velocity\footnote{The angular velocity tensor $\Omega^{\mu\nu}=e^\mu \cdot D e^\nu/d\lambda$ is defined by means of the body fixed frame $e_A^\mu(\lambda)$ along the worldline. The corresponding angular velocity vector is then given by $\Omega^i=\varepsilon^i{}_{jk}\Omega^{jk}/2$. See, for instance, Ref.~\cite{Steinhoff:2014kwa} for details.} 3-vectors $\Omega^i_{1,2}$ are introduced for completeness, however, the aligned-spin dynamics are independent of $\Omega_{1,2}^i$. Finally, we define the center of mass frame velocity $v^i=\dot{r}^i=v_1^i-v_2^i$. In Refs.~\cite{Vines:2016qwa, Siemonsen:2017yux} it was shown that after integrating out the gravitational degrees of freedom, as described in the previous section, the effective \ac{BBH} action $S_\text{BBH}$ reduces to the two-body Lagrangian
\begin{align}
\begin{aligned}
\mathcal L_\text{BBH}= \bigg[\frac{m_1}{2}v_1^2  + \frac{m_1}{2}\varepsilon_{ijk} a_1^i v^j_1
\dot{v}_1^k + m_1 a^i_1\Omega_{1,i} + (1\leftrightarrow 2) \bigg] 
 + \bigg[ \cosh (a_+\times \partial)+ 2v_i\sinh (a_+\times \partial)^i \bigg]\frac{m_1 m_2}{r},
\end{aligned}
\label{eq:Leff}
\end{align}
at the leading \ac{PN} level. Note that here and in the remainder of this section $\partial_i r^{-1}=\partial r^{-1}/\partial z^i_1=-\partial r^{-1}/\partial z_2^i$. So far, we have assumed a leading \ac{PN} treatment at each order in spin, but kept the dynamics unrestricted. In the following we assume that the spin degrees of freedom are fixed, i.e., the spin vectors are independent of time, $\dot{a}_{1,2}=0$, and aligned with the orbital angular momentum of the system: $a_{1,2}^i\propto L^i$; hence, the motion is confined to the plane orthogonal to $L^i$. For later convenience, we define the unit vector $\ell^i$, such that $L^i=|\boldsymbol{L}|\ell^i$. Varying this action with respect to the worldline $z_1^i$, the classical \ac{EoM} of the system are\footnote{The corresponding equation for $-\dot{v}^i_2$ emerges from the right hand side of \eqref{eq:EoM} under the replacement $a_1^i\leftrightarrow a_2^i$.} \cite{Siemonsen:2017yux, Vines:2016qwa}
\begin{align}
\begin{aligned}
\dot{v}_1^i= \Big(\partial^i-  & \ \varepsilon^i{}_{jk}a_1^kv^l\partial_l\partial^j\Big)\cosh(a_+\times\partial)\frac{ m_2}{r}
 \ +2 \left( v_j\partial^i-\delta^i_jv^k\partial_k\right)\sinh(a_+\times \partial)^j\frac{ m_2}{r} + \mathcal{O}(v^2).
\end{aligned}
\label{eq:EoM}
\end{align}
A geometric approach using oblate spheroidal coordinates \cite{Vines:2016qwa, Harte:2016vwo, Vines:2017hyw} or an algebraic approach, exploiting properties of the Legendre polynomials \cite{Siemonsen:2017yux}, under the assumption that the motion takes place in the plane orthogonal to the spin vectors, can be used to resum the series of differential operators in \eqref{eq:EoM}.

In order to present the contribution of the conservative sector needed for the radiative dynamics, we specialize to the center of mass frame for the rest of this section. The transformation into the center of mass variables $r^i$ based on \eqref{eq:Leff} (and using the total mass $M=m_1+m_2$), is corrected by the presence of the spins only at sub-leading orders in velocities:
\begin{align}
\begin{aligned}
\begin{split}
z^i_1=\frac{m_2}{M}r^i-b^i, \qquad z^i_2=-\frac{m_1}{M}r^i-b^i, \quad b^i:=\frac{1}{M}\varepsilon^i{}_{jk}(v^j_1 S^k_1+v^j_2 S^k_2).
\end{split}
\label{COMframe}
\end{aligned}
\end{align}
In this center of mass frame, the \ac{EoM} are readily solved for quasi-circular motion. In that scenario, the separation $r^i$ is related to its acceleration $\ddot{r}^i$ by $r^i=-\ddot{r}^i/\omega^2$, where $\omega$ is the system's orbital frequency. This ansatz picks out the quasi-circular orbits allowed by the \ac{BBH} \ac{EoM} \eqref{eq:EoM} and is equivalent to finding a relation between the frequency $x=(M \omega)^{2/3}$, the  \ac{BH}s  spins $a^i_{1,2}$, and the separation of the binary $r$. This relation, at the leading \ac{PN} order at each order in the \ac{BH}s' spins, is given by \cite{Siemonsen:2017yux}
\begin{align}
r(x)=\sqrt{\frac{M^2}{x^2}+\bar{a}_+^2}\left(1-\frac{x^{3/2}M}{3}\frac{\bar{\sigma}^*+2\bar{a}_+}{M^2+x^2\bar{a}_+^2}\right),
\label{eq:consvsol}
\end{align}
where $\bar{\sigma}^*=(m_2 \bar{a}_1+m_1 \bar{a}_2)/M$ and we defined $\bar{a}_{1,2}=a_{1,2}^i\ell_i$. It should be emphasized that the even-in-spin part of \eqref{eq:consvsol} contains only $\mathcal{O}(v^0)$ information, while the odd-in-spin pieces are non-zero only at first sub-leading order in velocities, at $\mathcal{O}(v^1)$. This solution can then be used to compute gauge invariant quantities of the conservative sector, such as the total binding energy and angular momentum \cite{Siemonsen:2017yux}.

\subsection{Linearized metric perturbations at null infinity} \label{sec:RadDyn}

With the conservative results in hand, in this subsection, we compute the gravitational waves from the \ac{BBH} system at future null infinity. We start by  briefly reviewing the general approach of mapping the source's multipole moments into the radiation field, and then move to the derivation of the TT part of the linear metric perturbations (the gravitational waves) at null infinity utilizing this mapping. 

\subsubsection{General approach} \label{sec:psi4generalapproach}

A natural choice of gauge invariant quantity capturing the radiative dynamics at null infinity is the Newman-Penrose Weyl scalar $\Psi_4$. This contains both polarization states, $h_+$ and $h_\times$, of the emitted waves, which are the observables measured by gravitational wave detectors. Upon choosing a suitable null tetrad, the TT part of the gravitational field, $h_{\mu\nu}^\text{TT}$, can be related to $\Psi_4$:
\begin{align}\label{eq:NP scalar}
\Psi_4\sim\ddot{h}_+-i\ddot{h}_\times=\bar{m}^\mu \bar{m}^\nu \ddot{h}^\text{TT}_{\mu\nu}.
\end{align}
The complex conjugate pair $\{m^\alpha, \bar{m}^\alpha\}$ is typically defined with respect to the flat spherically symmetric angular coordinate directions $m =(\boldsymbol{\Theta}+i \boldsymbol{\Phi})/\sqrt{2}$. With this choice in place, we restrict our attention to the spatial components $h^\text{TT}_{ij}$, as these contain the full information of $\Psi_4$, i.e., the radiative, non-stationary, degrees of freedom\footnote{As we will see below, this choice of purely spatial $m^\alpha$ is equivalent to choosing a gauge, in which the graviton polarization tensor is also purely spatial.}. 

In the previous section, we summarized the leading \ac{PN} conservative dynamics of a spinning \ac{BBH} to all orders in their spins. Given this, the well-established multipolar post-Minkowskian formalism \cite{Thorne:1980ru, Blanchet:1985sp, Blanchet:1995fg, Blanchet:1998in, Blanchet:2013haa} is ideally suited to determine the time-dependent metric perturbations at null infinity. Within this framework, the stress energy distribution of the source, $T_{\mu\nu}^\text{source}$, is mapped into a set of mass and current symmetric and trace free (STF) source multipole moments $\mathcal{I}_{i_1\dots i_\ell}(t)$ and $\mathcal{J}_{i_1\dots i_\ell}(t)$. We denote $\langle i_1\dots i_\ell\rangle$ as the STF projections  of the indices $i_1\dots i_\ell$. Then the STF multipole moments evaluated at the retarded time $T_R=t-R$ are defined by \cite{Blanchet:2013haa}
\begin{align}
\begin{aligned}
\mathcal{I}_{i_1\dots i_\ell}= & \ \int d\mu\left(\delta_\ell x_{\langle i_1\dots i_\ell\rangle} \Sigma - f_{1,\ell}\delta_{\ell+1}x_{\langle i i_1\dots i_\ell\rangle} \dot{\Sigma}^i + f_{2,\ell}\delta_{\ell+2}x_{\langle ij i_1\dots i_\ell\rangle}\ddot{\Sigma}^{ij} \right)(\boldsymbol{x},T_R+zr),\\
\mathcal{J}_{i_1\dots i_\ell}= & \ \int d\mu \ \varepsilon_{ab\langle i_\ell}\left(\delta_\ell x_{i_1\dots i_{\ell-1}\rangle}{}^a\Sigma^b- g_{1,\ell}\delta_{\ell+1} x_{i_1\dots i_{\ell-1}\rangle c}{}^a \dot{\Sigma}^{bc} \right)(\boldsymbol{x},T_R+zr),
\label{eq:generalsourcemultipoles}
\end{aligned}
\end{align}
where $x_{i_1\dots i_\ell}=x_{i_1}\dots x_{i_\ell}$,
\begin{align}
f_{1,\ell}=\frac{4(2\ell+1)}{(\ell+1)(2\ell+3)}, & & 
f_{2,\ell}=\frac{2(2\ell+1)}{(\ell+1)(\ell+2)(2\ell+5)}, & & g_{1,\ell}=\frac{2\ell+1}{(\ell+2)(2\ell+3)},
\end{align}
and the integration measure $\int d\mu=\text{FP}\int d^3 \boldsymbol{x} \int_{-1}^1 dz$. The source energy-momentum distribution enters in $\Sigma$, via (valid only at leading \ac{PN} orders)\footnote{At sub-leading \ac{PN} orders, the stress energy of the emitted gravitational waves contributes to $\Sigma$.}
\begin{align}
\Sigma=T^{00}+T^{ij}\delta_{ij}, & & \Sigma^i = T^{0i}, & & \Sigma^{ij}=T^{ij}.
\end{align}
The source's finite-size retardation effects are contained in the $z$-integral with $\delta_\ell=\delta_\ell(z)$ in \eqref{eq:generalsourcemultipoles}, which are given explicitly in eq.~(120) of Ref.~\cite{Blanchet:2013haa}. At the orders considered in this work, at the leading \ac{PN} orders, finite size-retardation effects vanish and the $z$-integral trivializes: $\int_{-1}^1 dz \  \delta_\ell(z)f(\boldsymbol{x},T_R+rz)=f(\boldsymbol{x},T_R)+\mathcal{O}(v^2)$. We discuss in \cref{sec:general_approach_amp}, how a similar structure as in \eqref{eq:generalsourcemultipoles} appears in the scattering amplitudes approach, as well as what precisely encapsulates the ``finite size" retardation effects in that context. The lowest order moments $\mathcal{I}$, $\mathcal{I}_i$, and $\mathcal{J}_i$ are constants of motion representing the total conserved energy, center of mass position and total angular momentum, respectively. Only for $\ell\geq 2$, do the multipoles contribute non-trivially.

A matching scheme enables to directly relate these functionals for the source's stress-energy distribution, to the radiation field at null infinity (at 1\ac{PM} order)\footnote{Beyond linear theory, corrections to these multipole moments are necessary \cite{Blanchet:2013haa}.} \cite{Blanchet:2013haa}
\begin{align}
h_{ij}=-4 \sum_{\ell= 2}^\infty \frac{(-1)^\ell}{\ell!}\left[\partial_{i_1\dots i_{\ell-2}}\ddot{\mathcal{I}}_{ij}{}^{i_1\dots i_{\ell-2}}R^{-1}+\frac{2\ell}{\ell+1}\partial_{a i_1 \dots i_{\ell-2}}\varepsilon^{ab}{}_{(i}\dot{\mathcal{J}}_{j)b}{}^{i_1\dots i_{\ell-2}}R^{-1} \right].
\label{eq:generalhij}
\end{align}
 Here $\partial_a R^{-1}=-N_a/R^2$ is to be understood as the derivative in the background Minkowski spacetime, where $N_a$ is radially outwards pointing from the source to spatial infinity, with $N_a N^a=1$. To solely focus on the radiation at null infinity, we work to leading order in the expansion in $R^{-1}$. Therefore, the spatial derivatives in \eqref{eq:generalhij} act purely on the source multipole moments, and there, can be traded for time derivatives: $\partial_a f(t-R)=-\dot{f}N_a$. Similarly, the total instantaneous gravitational wave energy flux $\mathcal{F}$ can be derived directly from the source multipole moments \cite{Blanchet:2013haa}.

\subsubsection{Gravitational radiation from spinning binary black hole} \label{sec:hijclassical}

At the 1\ac{PM} level, non-linear effects vanish such that the energy-momentum of the \ac{BBH} is simply the superposition of two linearized Kerr \ac{BH}s' energy momentum distributions \eqref{eq:KerrStressEn},  $T_{\mu\nu}^\text{source}=T_{\mu\nu}^{\text{Kerr},1}+T_{\mu\nu}^{\text{Kerr},2}$. This superposition holds in the conservative sector, while the radiative dynamics are derived directly from derivatives acting on $T_{\mu\nu}^\text{source}$ in the manner described in the previous section. From the scattering amplitudes perspective, this superposition is reflected in the \textit{only }two-channel factorization of the  classical 5-point amplitude,  into the product of a 3-pt amplitude and the gravitational Compton amplitude as given in \eqref{cuts m4 m5}. 

\textit{Leading order in velocities} -- As the radiative quantities $h_{ij}^\text{TT}$ and $\mathcal{F}$ depend on time derivatives of the source multipole moments, we focus on time-dependent terms after fixing the angular momentum dynamics. For the case of the above spinning \ac{BBH} with aligned spins, at the leading \ac{PN} order, we expand the source $T_{\mu\nu}^\text{source}$ analogously to \eqref{eq:hmunuPN}. Given this, the resulting leading-in-velocity contributions to the source multipole moments, utilizing \eqref{eq:generalsourcemultipoles}, are \cite{Buonanno:2012rv, Marsat:2014xea, Siemonsen:2017yux}
\begin{align}
\mathcal{I}^{ij}_{(0)} = m_1 z_1^{\langle ij\rangle}+(1\leftrightarrow 2), & & \mathcal{J}^{ij}_{(0)} = \frac{3}{2}S_1^{\langle i}z_1^{j\rangle} +(1\leftrightarrow 2),
\label{eq:multipolemomentv0}
\end{align}
where $(0)$ indicates the order in velocities. It should be stressed that these are \textit{all} the multipoles needed for the gravitational waveform to \textit{all} orders in the \ac{BH}s' spins, at leading order in velocity \cite{Siemonsen:2017yux}. From the amplitudes perspective, this will be reflected in the need for only the scalar and linear-in-spin scattering amplitudes at the leading orders in velocities. While all higher-order spin terms in the source multipole moments vanish identically, spin contributions to the waveform at arbitrary order in the spin expansion could enter through the solution to the \ac{EoM} \eqref{eq:consvsol}. We see below that this solution to the classical \ac{EoM} \eqref{eq:EoM} introduces non-zero contributions at arbitrary orders in the \ac{BH}s' spins for quasi-circular orbits.

Given \eqref{eq:multipolemomentv0}, the metric perturbation at null infinity, for general orbits at zeroth order in velocities, assuming aligned spins, is
\begin{align}
h^{(0)ij}_{S^\infty}(T_R,R,\boldsymbol{N},\boldsymbol{z}_1,\boldsymbol{z}_2) = \frac{2 m_1}{R}\bigg\{\frac{d^2}{dt^2}\left[z_1^i z_1^j\right] + \varepsilon_{pq}{}^{(i}\left(a_1^{j)}\dot{v}_1^p+\dot{v}_1^{j)}a_1^p\right)N^q\bigg\}\bigg|_{t=T_R}+(1\leftrightarrow 2),
\label{eq:classicalhijgeneralorbit}
\end{align}
i.e. the Einstein Quadrupole formula with spinning corrections for a binary system. We specialize to quasi-circular orbits by introducing the orthogonal unit vectors
\begin{align}
n^i=r^i/r = (\cos \omega t, \sin\omega t,0)^i, & & \lambda^i=v^i/v=(-\sin\omega t, \cos\omega t, 0)^i,
\label{eq:circorbits}
\end{align}
in the center of mass frame that rotate with frequency $\omega$ in the orbital plane. The spin vectors $a_{1,2}^i\propto \ell^i$ are aligned orthogonal to the orbital plane, $n^i\lambda^j \varepsilon_{ij}{}^k=\ell^k$, such that $\ell^i=(0,0,1)^i$. Furthermore, the TT projector
\begin{align}
\Pi^{ij}{}_{kl}=P^i{}_k P^j{}_l-\frac{1}{2}P^{ij}P_{kl}
\label{eq:TTprojector}
\end{align}
is defined relative to $N_a$, where $P_{ij}=\delta_{ij}-N_i N_j$. Utilizing \eqref{COMframe}, together with the solution \eqref{eq:consvsol} to the \ac{EoM}, as well as \eqref{eq:multipolemomentv0}, the gravitational waves emitted by the spinning \ac{BBH} to all orders in the \ac{BH}s' spins is conveniently written as
\begin{align}
h^{\text{TT}}_{ij}(T_R) =\frac{2\mu}{R} \ \Pi_{ij}{}^{ab} \hat{h}_{ab}\Big|_{t=T_R},
\label{eq:hijeasier}
\end{align}
where at leading order in velocities, we have $\hat{h}^{(0)}_{ab}=\hat{h}^{(0),\text{I2}}_{ab}+\hat{h}^{(0),\text{J2}}_{ab}$, with
\begin{align}
\label{eq:hijcircularv0}
\begin{aligned}
\hat{h}^{(0),\text{I2}}_{ab} = & \ -2x \left(1+\frac{\bar{a}_+^2x^2}{M^2}\right)(n_an_b-\lambda_a\lambda_b) \\
\hat{h}^{(0),\text{J2}}_{ab} = & \ -\frac{x^2 \bar{a}_-}{M}\sqrt{1+\frac{\bar{a}_+^2x^2}{M^2}},\ \varepsilon_{kl(a}(\ell_{b)}n^k+n_{b)}\ell^k)N^l.
\end{aligned}
\end{align}

Notice that here, the odd-in-spin contribution, $\hat{h}^{(0),\text{J2}}_{ab}$, is a series that has non-zero coefficients at arbitrary orders in spin, arising from the odd part of the solution  \eqref{eq:consvsol},  while, on the other hand, the even-in-spin part, $\hat{h}^{(0),\text{I2}}_{ab}$, provides coefficients that vanish for $\mathcal{O}(a^{\ell\geq 3})$. This is analogous to the cancellations observed in the conservative and radiative sectors reported in Ref.~\cite{Siemonsen:2017yux}. We find agreement with  the results reported in Refs.~\cite{Kidder:1992fr, Kidder:1995zr, Buonanno:2012rv} to the respective finite order in spin.  To check for consistency to all orders in spin, the gravitational wave modes are extracted from the spatial part of the metric perturbations, in \eqref{eq:hijcircularv0}, by projecting onto a suitably defined basis of spin-weighted spherical harmonics, ${}_{-2}Y_{\ell m}(\Theta, \Phi)$. Explicitly, the gravitational wave modes $h^{\ell m}$ are defined to be $h^{\ell m}=\int d\Omega \ {}_{-2}\bar{Y}_{\ell m}(\Theta, \Phi) \bar{m}^\mu \bar{m}^\nu h^\text{TT}_{\mu\nu}$. These modes, obtained from \eqref{eq:hijcircularv0} in conjunction with the above defined polarization tensor $\bar{m}^\alpha\bar{m}^\beta$, agree with the results in Ref.~\cite{Siemonsen:2017yux} to all orders in the \ac{BH}s' spins at leading order in their velocities.

\textit{Sub-leading order in velocities} -- The sub-leading corrections to the above radiation field are obtained in much the same way. The additional contributions to the source multipole moments, beyond the leading pieces \eqref{eq:multipolemomentv0}, at sub-leading orders in velocities are \cite{Buonanno:2012rv, Marsat:2014xea, Siemonsen:2017yux}
\begin{align}
\begin{aligned}
\mathcal{I}^{ij}_{(1)} = & \ \frac{4}{3}\left(2v_1^a S^b_1\varepsilon_{ab}{}^{\langle i}z_1^{j \rangle}-z_1^aS_1^b\varepsilon_{ab}{}^{\langle i} v^{j\rangle} \right) +(1\leftrightarrow 2), \\
\mathcal{I}^{ijk}_{(1)} = & \ m_1 z_1^{\langle ijk\rangle}-\frac{3}{m_1}S_1^{\langle i}S_1^j z_1^{k\rangle} +(1\leftrightarrow 2), \\
\mathcal{J}^{ij}_{(1)} = & \ m_1 z_1^av_1^b\varepsilon_{ab}{}^{\langle i}z_1^{j\rangle} + \frac{1}{m_1}v_1^a S_1^b\varepsilon_{ab}{}^{\langle i}S_1^{j\rangle}+(1\leftrightarrow 2), \\
\mathcal{J}^{ijk}_{(1)} = & \ 2S_1^{\langle i}z_1^{jk\rangle}+(1\leftrightarrow 2).
\label{eq:multipolemomentv1}
\end{aligned}
\end{align}
Also here, we focused only on those pieces that are time-dependent, i.e., that will contribute non-vanishing terms in $h_{ij}^{(1)\text{TT}}$. Additionally, as pointed out above, these are \textit{all} necessary contributions for the full all orders-in-spin information at sub-leading orders in velocities (at leading \ac{PN} order) \cite{Siemonsen:2017yux}. Using this, together with the mapping \eqref{eq:generalhij}, the decomposition \eqref{eq:hijeasier}, and $\hat{h}^{(1)}_{ab}=\hat{h}^{(1),\text{I2}}_{ab}+\hat{h}^{(1),\text{J2}}_{ab}+\hat{h}^{(1),\text{I3}}_{ab}+\hat{h}^{(1),\text{J3}}_{ab}$, the sub-leading contribution $h_{ij}^{(1)\text{TT}}$ to all orders in spin from a spinning binary black hole on quasi-circular orbits are
\begin{align}
\label{eq:hijcircularv1}
\begin{aligned}
\hat{h}^{(1),\text{I2}}_{ab}= & \   \frac{4 x^{5/2}}{3M^3}\left(2\bar{a}_+M^2+\left(M^2-2 r_e^2 x^2\right)\bar{\sigma}^*\right)\left(n_a n_b-\lambda_a \lambda_b\right), \\
\hat{h}^{(1),\text{J2}}_{ab}= & \  \frac{ x^{5/2}}{3M^4 r_e}\left[2r_e^4 x^2 \delta m+\bar{a}_- M\left(2\bar{a}_+(M^2-r_e^2x^2)+3r_e^2 x^2 \bar{\sigma}+M^2\bar{\sigma}^*\right)\right] \\ 
& \ \times\varepsilon_{pq(a}N^q \left(n^p\ell_{b)}+n_{b)}\ell^p\right),\\
\hat{h}^{(1),\text{I3}}_{ab}= & \  \frac{r_e x^{9/2}}{15M^4}\left[15 \bar{a}_+ \bar{a}_- M \ell_{\langle a}\ell_b \lambda_{k\rangle}-r_e^2\delta m\left(30 \lambda_{\langle a}\lambda_b\lambda_{k\rangle}-105 n_{\langle a}n_b \lambda_{k\rangle}\right)\right]N^k, \\
\hat{h}^{(1),\text{J3}}_{ab}= & \ -\frac{48 r_e^2 x^{9/2} \bar{\sigma}^*}{6M^3}\epsilon_{pq(a} \delta_{b)k} n^{(k}\lambda^p\ell^{e)}N^q N_e.
\end{aligned}
\end{align}
Here $r_e=(\bar{a}_+^2+M^2/x^2)^{1/2}$, which is just the leading-in-velocities (even-in-spin) solution to the classical \ac{EoM} \eqref{eq:consvsol} for quasi-circular orbits. We check the gravitational wave modes obtained from \eqref{eq:hijcircularv1} with those presented in Ref.~\cite{Siemonsen:2017yux} and find agreement to all orders in spin. Additionally, we compute the gauge invariant gravitational wave energy flux with the above result together with the leading-in-velocities radiation field and find agreement with results reported in \cite{Marsat:2014xea, Siemonsen:2017yux} (see also a detailed discussion in \cref{sec:results} below). Finally, in order to compare to the scalar amplitude at first sub-leading orders in the \ac{BH}s' velocities in \cref{sec:subleading-v}, we also present the radiation field of a non-spinning \ac{BBH} system on general orbits, to sub-leading order in velocities:
\begin{align}\label{eq:h1_v1_scalar}
h^{(1),ij}_{S^0,\text{TT}}=\frac{2m_1}{3R}\Pi^{ij}{}_{ab}\left[4\varepsilon_{pq}{}^{(a}\Big\{ \partial_t^2 (\varepsilon_{cd}{}^{e)}z_1^c v_1^d\delta_e{}^{\langle b}z_1^{p\rangle})\Big\}N^q +N_k\partial_t^3 (z_1^{\langle a}z_1^b z_1^{k\rangle})\right]+(1\leftrightarrow 2).
\end{align}

\section{Scattering Amplitudes derivation} \label{sec:scatteringampl}

In the previous sections, we obtained the form of the  gravitational waves emitted from a spinning \ac{BBH} on general \textit{closed} orbits with aligned spins, to leading order in the \ac{BH}s'  velocities  \eqref{eq:classicalhijgeneralorbit}   [and on quasi-circular orbits given in \eqref{eq:hijcircularv0}], whereas  at sub-leading  order in $v$,  and for quasi-circular orbits, we derived  \eqref{eq:hijcircularv1}, at each order (and to \textit{all} orders) in the \ac{BH}s' spins. In the following, we show that these results follow directly from the classical limit of the spinning 5-point  \textit{scattering} amplitudes derived in \cref{sec:spin_in_qed}.
More precisely, at leading order in velocity there is a one-to-one correspondence between the source's mass and current  multipole moments \eqref{eq:multipolemomentv0}, and the scalar and linear-in-spin  contribution to the scattering amplitude, respectively. This correspondence allow us to derive the linear in spin, general orbit result for the radiated gravitational field \eqref{eq:classicalhijgeneralorbit}, from an amplitudes perspective. At  quadratic order in the \ac{BH}s' spins, and for quasi-circular orbits, we demonstrate that the contribution from the quadratic in spin amplitude is canceled by the contribution of the  scalar amplitude in conjunction with the $\mc{O}(S^2)$-piece of the \ac{EoM} \eqref{eq:EoM}. This leaves only the quadrupole field, \eqref{eq:einstein-quadrupole}, supplemented with the solution to the \ac{EoM} \eqref{eq:consvsol}, to enter at quadratic order in spin. Although we explicitly demonstrate the cancellation for quasi-circular orbits and up to quadratic order in spin only, we expect this theme to continue to hold for more complicated bound orbits, as well as to higher spin orders in the 5-point scattering amplitude, as suggested by the classical multipole moments \eqref{eq:multipolemomentv0}. At sub-leading orders in the \ac{BH}s' velocities the situation becomes more complicated; there, we demonstrate the matching of the amplitudes to the classical computation in the spin-less limit for quasi-circular orbits, and briefly comment on extensions to higher orders in spin. In this section, we use the mostly minus signature convention for the flat metric $\eta_{\mu\nu}=\rm{diag}(1,-1,-1,-1)$.\\

\subsection{General approach}\label{sec:general_approach_amp}

To compute the radiated field at future null infinity from the \ac{BBH} system we follow the approach used by Goldberger and Ridgway in  \cite{Goldberger:2017vcg} to derive the Quadrupole formula, and extend it to include relativistic and spin effects. This approach is  based on the classical \ac{EoM} for the orbiting objects in combination with the corresponding 5-point (spinning) scattering amplitude (see Figure \ref{fig:5pt-amplitude}). It is valid for \ac{BBH}s whose components have Schwarzschild radii $r_{1,2}=2m_{1,2}$ much smaller than their spatial separation $r$, i.e., $r_{1,2}\ll r$, while the radiation field wavelength is much bigger than the size of the individual components $\lambda\gg m_{1,2}$, as well as the size of the system $\lambda\gg r$\footnote{In the long distance separation regime, radiation reaction effects can be neglected, since they become important only when the separation of the two bodies is comparable to the system's gravitational radius \cite{Misner1973} eq. $36.11$.}.  Therefore, we expect our results to be situated in the \ac{PN} regime of the binary inspiral\footnote{We stress that even though we concentrate mostly in the computation of gravitational waveform, an analogous derivation follows for  electromagnetic radiation, as   already pointed out in \cite{Goldberger:2017vcg}.}.

Let us start by noting that in the limit in which $R\rightarrow\infty$, where $R$ is the distance from the source to the observer (i.e., the radial coordinate in Bondi-Sachs gauge) as defined above, the time-domain waveform  at retarded time $T_{R}$, has the asymptotic form \cite{Maggiore:1900zz}  (see also \eqref{eq:rad_field_full_n})
\begin{equation}\label{eq:radiated field}
h_\text{TT}^{ij}(T_{R},R,\boldsymbol{N},\boldsymbol{z}_1,\boldsymbol{z}_2)=\frac{\kappa}{16\pi R }\Pi^{ij}{}_{ab}\int d\bar{\omega}e^{-i\bar{\omega}\,T_{R}}T^{ab}(\bar{\omega},\boldsymbol{N},\boldsymbol{z}_1,\boldsymbol{z}_2).
\end{equation}
Here $\kappa^2=32\pi$ (recall we set $G=1$),  $\bar{\omega}$ is the frequency of the radiated wave with four  momentum $k^\mu=\bar{\omega}N^\mu=\bar{\omega}(1,\boldsymbol{N})^\mu$, and $\Pi^{ij}{}_{ab}$ is the TT-projector defined in \eqref{eq:TTprojector}.
As above, the locations of the binary's components are denoted by $z^i_{1,2}$. Analogous to the previous section,  we focus only  on the spatial components of $h^{\mu\nu}$, which contain all the radiative degrees of freedom. In what follows we also simplify the notation for the source $T^{ab}(\bar{\omega},\boldsymbol{N},\boldsymbol{z}_1,\boldsymbol{z}_2)\to T^{ab}(k,\boldsymbol{z}_1,\boldsymbol{z}_2) $, where it is understood that $k^\mu$ has implicit the dependence in both, $\bar{\omega}$ and $\boldsymbol{N}$. 

The source $T^{ab}(k,\boldsymbol{z}_1,\boldsymbol{z}_2)$, is related directly to the 5-point scattering amplitude in Figure \ref{fig:5pt-amplitude}; therefore, in order to focus on the spatial components, it is sufficient to work in a gauge in  which the graviton polarization tensor $\epsilon^{\mu\nu} =\epsilon^\mu\epsilon^\nu$, is the tensor product of two purely spatial  polarization vectors $\epsilon^\nu$. From the classical perspective, this choice of gauge is analogous to the conjugate pair $\{m^\alpha,\bar{m}^\alpha\}$ (defined in \cref{sec:psi4generalapproach}) to be purely spatial. Notice, however, the radiation field computed from a 5-point scattering amplitude, and the corresponding field computed classically in the previous section, can in general differ by a time independent constant, for which, gravitational observables such as the gravitational wave energy flux, or the radiation scalar will be insensitive to, since they are computed from one or two time derivatives of the waveform. 
As shown below, this is directly related to a freedom in choice of an integration by parts (IBP) prescription in \eqref{eq:radiated field}.

We proceed by writing the explicit form of the source $T^{ij}(k,\boldsymbol{z}_1,\boldsymbol{z}_2) $ in terms of the  classical 5-point scattering amplitude. In the classical computation, $T^{ij}$ corresponds to the source entering on the right hand side of  field equations, at a given order in perturbation theory. To leading order, for scalar particles, it was shown in \cite{Goldberger:2017vcg}  that the source can be rearranged in such a way, so that the scalar  5-point   amplitude can be identified as the main kinematic object entering the graviton phase space integration, as well as the integration over the  particles proper times (which account for the particles history). In this thesis we propose that formula to also hold for spinning particles.  That is,
\begin{equation}\label{eq:source_some1}
T^{ij}(k,\boldsymbol{z}_1,\boldsymbol{z}_2) = \frac{i}{m_1 m_2}\int d\tau_{1}d\tau_{2}\hat{d}^{4}q_{1}\hat{d}^{4}q_{2}\hat{\delta}^4\left(k-q_{1}-q_{2}\right)e^{iq_{1}{\cdot}z_{1}}e^{iq_{2}{\cdot}z_{2}}\langle M_{5}^{ij}(q_{1},q_{2},k)\rangle.
\end{equation}
Here $\langle M_{5}^{ij}\rangle$ is the classical 5-point amplitude . Conventions for the particles' momenta  and the spins are shown in  Figure \ref{fig:5pt-amplitude}, with the condition for  momentum conservation     $q_{1}+q_{2}=k$. We have used the notation $\hat{d}^4q_i = \frac{d^4q_i}{(2\pi )^4} $, and similarly for the momentum-conserving delta function  $\hat{\delta}^4(p) = (2\pi)^4\delta^4(p)$, in analogy to the notation used in \cref{sec:KMOC}. 

This notation was in fact selected for a good reason. To motivate this formula, although this is by no means a formal derivation,  as already observed in \cite{Goldberger:2017vcg} for the tree-level amplitude, we can take the expression for the linear in amplitude contribution to the radiation field in \ac{KMOC} form for scattering scenarios \eqref{eq:r-kernel}  for the gravitational case
\begin{equation}
    \mathcal{J}=\lim_{\hbar\to 0}\frac{1}{m_1m_2}\Big\langle
    \int \prod_{i=1}^2\left[
    \hat{d}^4q_i \hat{\delta}(v_i{\cdot}q_i-q_i^2/(2m_i))e^{ib_i{\cdot}q_i} \right]\hat{\delta}^4(k-q_1-q_2)M_5
   \Big\rangle,
\end{equation}
and use the integral representation for the on-shell delta functions $\delta(x)\sim\int dy e^{ix y}$. Identifying the   asymptotic trajectories for the particles $z_i(\tau_i) = b_i^\mu +v_i^\mu \tau_i$, plus a quantum correction  $z_{Q}^\mu(\tau_i) = -\frac{ q_i}{2m_i}\tau_i $, and upon restoring the $\hbar$-counting in the exponential,   the radiation kernel can be rewritten as 
\begin{equation}\label{eq:radiation_kernell_gen}
    \mathcal{J}=\lim_{\hbar\to 0}\frac{1}{m_1m_2}\Big\langle
    \int \prod_{i=1}^2\left[
    d\tau_i\hat{d}^4q_i  e^{i q_i{\cdot}(z_i(\tau_i)+  \hbar z_{Q}(\tau_i))} \right]\hat{\delta}^4(k-q_1-q_2)M_5
   \Big\rangle.
\end{equation}
In the  classical limit, and to leading order in perturbation theory,   we can simply drop quantum correction to the particles trajectories    $z_{Q}^\mu(\tau_i)  $, and recover the formula \eqref{eq:source_some1} upon  promoting   $z_i(\tau_i)$ to be valid for generic time dependent  orbits. A similar argument can be given to derive the \ac{BH}s' \ac{EoM} directly from the amplitude, in this case, an instantaneus impulse, starting from the linear in amplitude contribution to the linear impulse \eqref{eq:impulse_final}. 

\begin{equation}\label{eq:instantaneus_impulse}
    \Delta p_1^{\mu} = \frac{1}{4m_1 m_2}\int \hat{d}^4q\,d\tau_1\,d\tau_2 iq^{\mu}e^{-i(z_2(\tau_2)-z_1(\tau_1)){\cdot}q}\langle M_4 \rangle\,.
\end{equation}
Below we will show how to use this formula to obtain the  \ac{BH}s  \ac{EoM} to leading order in the velocity expansion. 

The position vectors are $z^\mu_A= (\tau_A,\boldsymbol{z}_A)^\mu$, with $A=1,2$, as described in \cref{sec:effBBHaction}, where the proper times $\tau_A$, parametrize the \ac{BH}s' trajectories. Here the product of the exponential functions,  $\prod_Ae^{i q_A\cdot z_A}$, represents the two-particles initial state where each  particle is taken to be in a plane-wave state. This is nothing but the Born approximation in Quantum Mechanics (See also \cite{HariDass:1980tq}). Here we emphasize formulas \eqref{eq:source_some1} and \eqref{eq:instantaneus_impulse} are valid up to subleading order in the velocity expansion, as we are dropping the quantum corrections $z_Q^\mu(\tau_i)$. In addition, the notice the delta functions $ \hat{\delta}(v_i{\cdot}q_i-q_i^2/(2m_i))$ effectively impose the on-shell condition for particles in a scattering scenario. For general trajectories, outgoing particles are no longer on-shell and therefore the on-shell condition cannot  be imposed in the amplitude. 

We have striped away the graviton polarization tensor in \eqref{eq:source_some1}, assuming there exists the aforementioned gauge fixing for which the graviton polarization tensor is purely spatial. We can further  rewrite the source  using the  symmetric variable $q=(q_{1}-q_{2})/2$, as well as exploiting the  momentum conserving delta function to remove one of the $q_{i}$-integrals. The result reduces to
\begin{equation}\label{eq:source_some2}
T^{ij}(k,\boldsymbol{z}_1,\boldsymbol{z}_2)=\frac{i}{m_1 m_2}\int d\tau_{1}d\tau_{2}\hat{d}^{4}qe^{ik \cdot\tilde{z}}e^{-iq\cdot z_{21}}\langle M_{5}^{ij}(q,k)\rangle,
\end{equation}
where $\tilde{z}=(z_{1}+z_{2})/2$ and $z_{BA}=z_{B}-z_{A}$. Since we are interested in the bound-orbit problem, we take the slow-motion limit. Therefore, we can write the momenta of the  \ac{BH}s  moving on \textit{closed}  orbits in the form $p_{1,2}^{\mu}=m_{1,2}v^{\mu}_{1,2}$. As noted above, we choose the frame in which $v_{1,2}^{\mu}=(1,\boldsymbol{v}_{1,2})^\mu+\mathcal{O}(v_{1,2}^{2})$, where $v^i_{1,2}=  dz^i_{1,2}/dt$, i.e. with the   proper times $\tau_{1,2}$  replaced by the coordinate time (see details below). On the other hand, in the closed orbits scenario the typical frequency of the orbit $\omega$, scales with $v$ as $\omega\sim v/r$, where $\omega=v/r$ for quasi-circular orbits (see also \eqref{eq:circorbits}). In this bound-orbits case, the integration in $q$ is restricted to the potential region (technically, as an expansion in powers of $q^0/|\mathbf{q}|$), where the internal graviton momentum has the scaling $q\sim(v/r,1/r)$, while the radiated graviton momentum scaling is   $k\sim(v/r,v/r)=\bar{\omega}(1,\boldsymbol{N})$ (with $\bar{\omega}\sim \omega$). Integration in the potential region ensures that from the retarded propagators,
\begin{equation}
    \frac{1}{(q_0+i0)^2-\boldsymbol{q}^2}\rightarrow \frac{1}{v^2(q_0+i0)^2-\boldsymbol{q}^2} \approx -\frac{1}{\boldsymbol{q}^2}+\mathcal{O}(v^2),
\end{equation}
entering in the scattering amplitude, retardation effects only  become important  at order  $\mathcal{O}(v^2)$, which we do not consider here. At subleading  orders in velocities, the amplitude  $\langle M_{5}^{ij}(q,k)\rangle$ has no explicit dependence on $q^{0}$. This takes care of the $q^0$-integration in \eqref{eq:source_some2}, which results in the delta function $\delta(t_{2}-t_{1})$; this can be used to trivialize one of the time integrals\footnote{As a connection with the classical computation, the source multipole moments [given in \eqref{eq:multipolemomentv0}] contain the finite size and retardation effects of the binary, though, at leading and sub-leading orders in velocities, these effects vanish (see e.g., \cite{Blanchet:2013haa}), which is equivalent to the replacement $\tau_{1,2}\rightarrow t$ above.}. With all these simplifications in hand,   the source \eqref{eq:source_some2} becomes
\begin{equation}\label{source}
T^{(0)\,ij}(k,\boldsymbol{z}_1,\boldsymbol{z}_2){=}\frac{i}{m_1 m_2}\int dt \hat{d}^3\boldsymbol{q}e^{i\bar{\omega}\,t{-}i\boldsymbol{q}{\cdot}\boldsymbol{z}_{21}}\langle M_{5,{\rm S}^0}^{(0)\,ij}(\boldsymbol{q},\bar{\omega}){+}M_{5,{\rm S}^1}^{(0)\,ij}(\boldsymbol{q},\bar{\omega}){+}M_{5,{\rm S}^2}^{(0)\,ij}(\boldsymbol{q},\bar{\omega})\rangle {+}\cdots,
\end{equation}
where the amplitude was written in a  spin-multipole decomposition. The superscript $^{(0)}$ indicates that we restrict these to the leading-in-$v$ contribution to the scattering amplitude (See \cref{sec:subleading-v} for the computation at the first sub-leading order in velocities contribution, for spinless  \ac{BH}s ). 

\subsubsection{Instantaneous impulse and particles EoM}\label{sec:amplitude_double_copy}

In the seminal work of Dass and Soni \cite{HariDass:1980tq}, it was claimed the conservative 4-point amplitude can be used to reproduce the particles \ac{EoM} for scalar sources. In this section we  will  show that indeed, the instantaneous impulse formula \eqref{eq:instantaneus_impulse} can be used to reproduce the particles \ac{EoM} \eqref{eq:EoM}, to leading order in velocity, but to all spin orders from the conservative amplitude. 
For that we will first have to compute the  classical conservative $\langle M_4^{\text{gr}}\rangle$ amplitude to all orders in spin. At leading order in perturbation theory this is just given by the t-channel cut as indicated in \eqref{cuts m4 m5}. We will then need the classical limit of the 3-point amplitud to all orders in spin. 
In \cref{ch:GW_scattering} we will show the spin exponentiation of the classical  gravitational 3-point amplitude \eqref{eq:3ptexp}, in 4-dimensions in terms of the Kerr \ac{BH} spin vector $a^\mu$ takes the form 
\begin{equation}
  \langle  A_3^{a\,\text{gr}}(p_1,q^{\pm}) \rangle =  \kappa (p_1{\cdot}\epsilon^{\pm})^2 e^{\pm a {\cdot}q }\,,
\end{equation}
where we have included explicitly the helicity of the emitted graviton. The classical 4-point amplitude can be computed then following the theme of \cite{Guevara:2017csg,Guevara:2018wpp}. 
\begin{align}
    \langle M_4^{\text{gr}}\rangle 
    &= \frac{1}{q^2} \Big[\langle  A_3^{a_1\,\text{gr}}(p_1,q^-)  \rangle\times \langle  A_3^{a_2\,\text{gr}}(p_2,-q^{+})  \rangle + \langle  A_3^{a_1\,\text{gr}}(p_1,q^{+})  \rangle\times \langle  A_3^{a_2\,\text{gr}}(p_2,-q^{-})  \rangle\Big]\\
    &= \frac{m_1 m_2}{q^2}\Big[\frac{x_1^2}{x_2^2}e^{q{\cdot}a_+}+ \frac{x_2^2}{x_1^2}e^{-q{\cdot}a_+}\Big]
\end{align}
where we have summed over the helicities of the exchanged graviton,  and set $a_+=a_1+a_2$. We have also used the well known $x_i$-helicity variables from the spinor helicity formalism \cite{Arkani-Hamed:2017jhn}, by fixing the little group rescaling of the internal graviton as follows \cite{Guevara:2017csg}
\begin{equation}
     x_2=\sqrt{2}\frac{p_2{\cdot}\epsilon^-(-q)}{m_2}=-\sqrt{2}\frac{p_2{\cdot}\epsilon^-}{m_2}=1
\end{equation}
which implies 
\begin{equation}
    x_1^{-1} = -\sqrt{2}\frac{p_1{\cdot}\epsilon^+}{m_1} =\gamma(1-v)\,,\quad x_1 =  -\sqrt{2}\frac{p_1{\cdot}\epsilon^-}{m_1}=\gamma(1+v),\,,
\end{equation}
where $\gamma =\frac{1}{\sqrt{1-v^2}}=\frac{p_1{\cdot}p_2}{m_1 m_2}$. Using the on-shell identity $i\epsilon_{\mu\nu\rho\sigma}p_1^\mu p_2^\nu q^\rho a^\sigma = m_1 m_2\sqrt{\gamma^2-1}q{\cdot}a$, and defining $\hat{\boldsymbol{p}}$ as a unit vector in the direction of the relative momentum in the ac{CoM}, the classical amplitude simply becomes
\begin{align}\label{eq:M4_treespin}
    \langle M_4^{\text{gr}}\rangle = \frac{\kappa^2 m_1^2 m_2^2}{2q^2}\gamma^2\sum_{\pm}(1\pm v)^2 e^{\pm i\boldsymbol{q}\times \boldsymbol{p}{\cdot}\boldsymbol{a}_+}\,,\\
    \langle M_4^{\text{gr}}\rangle =\frac{\kappa^2 m_1^2 m_2^2}{2q^2}\gamma^2\sum_{\pm}(1\pm v)^2 e^{\pm i q_i\times a^i_{+}}\,.
\end{align}
where in the second line we have specialized to the aligned spin scenario. In the \ac{CoM} frame, $q$ is purely spatial. We can then use \eqref{eq:M4_treespin} into \eqref{eq:instantaneus_impulse}, the $q^0$ integral results into the delta function $\delta(t_1-t_2) $, reflecting non-retardation effects in the conservative sector at this order in perturbation theory. Finally, the particles \ac{EoM} result from $\boldsymbol{q}$-integration, and to leading order in the velocity expansion,  and after using the fundamental theorem of calculus\footnote{Here we assume boundary terms do not contribute. At leading order in the velocity expansion seems to be a valid assumption since the amplitudes recover the correct results for the \ac{EoM}.}, with $\frac{\Delta p_1}{\Delta t} = m_1 \dot{v}_1$, results into 
\begin{equation}
    \dot{v}^i_1 = \partial^i\cosh(a_+\times\partial)\frac{m_2}{r}\,
\end{equation}
where we used $|z_{21}|=r$. This recovers the leading in velocity contribution to the particles \ac{EoM} \eqref{eq:EoM}, directly from the scattering amplitude and the instantaneous impulse.

\subsection{Computation of the radiated field}\label{sec:radiated_field_amplitudes}

In the previous sections, we built up the 5-point gravitational spinning scattering amplitude up to quadratic order in the \ac{BH}s' spins. With this, we can now return to \eqref{eq:radiated field} to successively construct the emitted classical gravitational radiation from the spinning \ac{BBH} at increasing \ac{PN} order. First, we compute the gravitational waveform to leading order in velocity up to quadratic order in the \ac{BH}s' spins, while turning to the  computation of the waveform  at sub-leading order in the \ac{BH}s' velocities in the spin-less limit in  \cref{sec:subleading-v}.

\subsubsection{Scalar waveform}\label{sec:scalar_waveform}

The derivation of the Einstein quadrupole formula from scattering amplitudes was first done by Hari Dass and Soni in \cite{HariDass:1980tq}; more  recently, it was derived by Goldberger's and Ridgway's classical double copy approach \cite{Goldberger:2017vcg}. In the following, we re-derive the scalar term of the  waveform in the Goldberger and Ridgway setup, for completeness. This, in turn, will outline the formalism used throughout the remaining sections to arrive at the corrections to the quadrupole formula.  Expanding the scalar amplitude -- obtained by replacing the scalar numerators \eqref{eq:scalar-numerators-gr} into the general formula \eqref{eq:newM5clas} -- to leading order in velocities $v$, we find  (one can check that actually the non-relativistic limit of the leading order in the soft expansion produce the same result)
\begin{equation}\label{eq:scalar-amplitude}
\langle M_{5,{\rm {S}^0}}^{(0)\,ab}(\boldsymbol{q},\bar{\omega})\rangle=-i\frac{m_{1}^{2}m_{2}^{2}}{4}\kappa^{3}\left[2\frac{q^{a}q^{b}}{\boldsymbol{q}^4}+\frac{1}{\bar{\omega}\boldsymbol{q}^2}\left(q^{a}v_{12}^{b}+q^{b}v_{12}^{a}\right)\right],
\end{equation}
where $v_{AB}=v_A-v_B$. Substituting this amplitude into the  scalar source \eqref{source}, and integrating over $\boldsymbol{q}$ using \eqref{eq:integrals}, the non-spinning source reduces to
\begin{equation}\label{eq:scalar-source-befor-eom}
T_{\rm{S}^0}^{(0)\,ab}(k,\boldsymbol{z}_1,\boldsymbol{z}_2){=}{ -}\int dte^{i\bar{\omega}\,t}\frac{\kappa^{3}}{32\pi}\sum_{A,B}\frac{m_{A}m_{B}}{r^{3}}\left[\left(z_{AB}^{a}z_{AB}^{a}  {-}r^2\delta^{ab} \right)+\frac{2i}{\bar{\omega}}\left(z_{AB}^{a}v_{A}^{b}{+}z_{AB}^{a}v_{A}^{a}\right)\right].
\end{equation} 
Here, and in the following, single label sums are understood to run over the two massive particle labels, $\sum_A:=\sum_{A=1}^2$, while the double sum  is performed imposing the constraint $A\neq B$: $\sum_{A,B}:=\sum_{A\neq B;A,B=1}^2$.

Notice that the term proportional to $\delta^{ab}$ in \eqref{eq:scalar-source-befor-eom} vanishes under the action of the TT-projector in \eqref{eq:radiated field}. Therefore, in the following, we remove this term from the source and focus only on those parts contributing non-trivially to the TT radiated field. Now, we use the non-spinning part of the \ac{EoM} \eqref{eq:EoM} to rewrite the second term in the square bracket of \eqref{eq:scalar-source-befor-eom}:
\begin{equation}\label{eq:scalar-source-intermediate}
T_{\rm{S}^0}^{(0)\,ab}(k,\boldsymbol{z}_1,\boldsymbol{z}_2)=-\kappa\int dte^{i\bar{\omega}\,t} \left[\sum_{A,B}\frac{\kappa^{2}m_{A}m_{B}}{32\pi}\frac{z_{AB}^{a}z_{AB}^{a}}{r^{3}}-\frac{2i}{\bar{\omega}}\sum_{A}m_{A}\left(v_{A}^{b}\dot{v}_{A}^{a}+v_{A}^{a}\dot{v}_{A}^{b}\right)\right].
\end{equation}
The second term of this expression can be further integrated, since $v_{A}^{b}\dot{v}_{A}^{a}+v_{A}^{a}\dot{v}_{A}^{b}=\frac{d}{dt}\left(v_{A}^{a}v_{A}^{b}\right)$. As for the first term, this can be rewritten using
\begin{equation}\label{eq:identity_eom}
\frac{\kappa^{2}}{32\pi}\sum_{A,B}m_{A}m_{B}\frac{z_{AB}^{a}z_{AB}^{a}}{r^{3}}=-\sum_{A}m_{A}\left(\ddot{z}_{A}^{a}z_{A}^{b}+z_{A}^{a}\ddot{z}_{A}^{b}\right),
\end{equation}
derived from the scalar EoM. Putting these ingredients together into \eqref{eq:scalar-source-intermediate}, we find the scalar source to be
\begin{equation}
T_{\rm{S}^0}^{(0)\,ab}(k,\boldsymbol{z}_1,\boldsymbol{z}_2)=\kappa\int dte^{i\bar{\omega}\,t}\sum_{A} m_{A}\left(\ddot{z}_{A}^{a}z_{A}^{b}+z_{A}^{a}\ddot{z}_{A}^{b}+2v_{A}^{a}v_{A}^{b}\right).
\end{equation}
Using the relation $2v_{A}^{a}v_{A}^{b}=\frac{d^{2}}{dt^{2}}\left(z_{A}^{a}z_{A}^{b}\right)-\left(\ddot{z}_{A}^{a}z_{A}^{b}+z_{A}^{a}\ddot{z}_{A}^{b}\right)$, the above expression can be put into the more compact form
\begin{equation}
T_{\rm{S}^0}^{(0)\,ab}(k,\boldsymbol{z}_1,\boldsymbol{z}_2)= \kappa \int dte^{i\bar{\omega}\,t}\sum_{A}m_{A}\frac{d^{2}}{dt^{2}}\left(z_{A}^{a}z_{A}^{b}\right),
\label{eq:scalarsource}
\end{equation}
which in turn implies that the radiated field \eqref{eq:radiated field} for a non-spinning \ac{BBH} takes the familiar Einstein quadrupolar form:
\begin{equation}\label{eq:einstein-quadrupole}
\boxed{
h_{TT,\,\rm{S}^0}^{(0)\,ij}(T_{R},R,\boldsymbol{N} ,\boldsymbol{z}_1,\boldsymbol{z}_2)= \frac{\kappa^2}{16\pi R}\Pi^{ij}{}_{ab}\sum_{A} m_{A}\left[\frac{d^{2}}{dt^{2}}\left(z_{A}^{a}z_{A}^{b}\right)\right]_{t=T_{R}}.}
\end{equation}
The sequence of Fourier transforms in the source \eqref{eq:scalarsource} and \eqref{eq:radiated field} leads to the evaluation of the emitted gravitational radiation at retarded time $T_R$, therefore, recovering the classical result \eqref{eq:classicalhijgeneralorbit} in the no-spin-limit. As a quick remark,  notice when restoring Newton's constant $G$ the quadrupole radiation is linear in $G$, as opposed to gravitational Bremsstrahlung, which is  quadratic \cite{Peters:1970mx,1985Konradin,1977KT,1978KT}. This is of course just a  feature of using the \ac{EoM} to rewrite the source. 

\subsubsection{Linear-in-spin waveform}\label{sec:linear_spin_waveform}

In the previous section, the main components of the derivation of the gravitational waveform from a compact binary system were outlined. In particular, we have seen that the classical \ac{EoM} play an important role in recovering the quadrupole formula. Going beyond this, at linear order in the \ac{BH}s' spins, there are two contributions to the waveform. First, the scalar amplitude could be iterated with the linear-in-spin part of the classical \ac{EoM} \eqref{eq:EoM}; this contribution, however, is sub-leading in velocity as made explicit in \eqref{eq:EoM}. Secondly, the linear-in-spin amplitude, in conjunction with the non-spinning part of the \ac{EoM} gives rise to a leading in \ac{BH}s' velocities and linear-in-their spins contribution to the waveform. To determine the latter, we start from the linear-in-spin amplitude obtained by replacing the linear in spin numerators  \eqref{eq:linear-spin-amplitude-gr} into the general formula \eqref{eq:newM5clas}, setting $J^{\mu\nu}\to S^{\mu\nu}=\frac{1}{2m}\epsilon^{\mu\nu\rho\sigma}p_1^{\rho}S_1^\sigma $, where the leading-in-$v$ contribution is given by
\begin{equation}\label{eq:linear_spin_v0}
   \langle M_{5,{\rm S}^{1}}^{(0)\,ab}(\boldsymbol{q},\bar{\omega})\rangle=-\frac{m_{1}m_{2}\kappa^{3}}{8}\varepsilon_{efk}\left(m_{2}S_{1}^{k}{-}m_{1}S_{2}^{k}\right)N^{[e}\left(\delta^{f]a}\delta^{bc}{+}\delta^{f]b}\delta^{ac}\right)\frac{q^{c}}{\boldsymbol{q}^{2}}.
\end{equation}
Analogous to the scalar case, we can substitute this amplitude into  \eqref{source} to get the  linear-in-spin source $T_{{\rm {S}^{1}}}^{(0)\,ab}$. After integrating over $\bs{q}$, utilizing \eqref{eq:integrals}, this source simplifies to
\begin{equation}
    T_{{\rm {S}^{1}}}^{(0)\,ab}(k,\boldsymbol{z}_{1},\boldsymbol{z}_{2})=\frac{\kappa^{3}}{32\pi}\varepsilon_{efk}\left(m_{2}S_{1}^{k}{-}m_{1}S_{2}^{k}\right)N^{[e}\left(\delta^{f]a}\delta^{bc}{+}\delta^{f]b}\delta^{ac}\right)\int dte^{i\bar{\omega}\,t}\frac{z_{21}^{c}}{r^{3}}.
\end{equation}
Powers of $r$ in the denominator can be removed by using the scalar limit of the classical \ac{EoM} \eqref{eq:EoM}. Then, analogous to the scalar computation, the linear-in-spin source is
\begin{equation}
T_{{\rm {S}^{1}}}^{(0)\,ab}(k,\boldsymbol{z}_{1},\boldsymbol{z}_{2})=\kappa\varepsilon_{efk}S_{1}^{k}N^{[e}\left(\delta^{f]a}\delta^{bc}{+}\delta^{f]b}\delta^{ac}\right)\int dte^{i\bar{\omega}\,t}\dot{v}_{1}^{c}+(1\leftrightarrow2).
\label{eq:spin1source}
\end{equation}
Finally, the linear in spin corrections to the Einstein quadrupole formula, derived from the above amplitude, obtained from \eqref{eq:spin1source}, together with \eqref{eq:radiated field}, are
\begin{equation}\boxed{h_{TT,\,{\rm {S}^{1}}}^{(0)\,ij}(T_{R},R,\boldsymbol{N},\boldsymbol{z}_1,\boldsymbol{z}_2)=\frac{\kappa^{2}}{16\pi R}\Pi^{ij}{}_{ab}\varepsilon_{efk}\sum_{A}S_{A}^{k}\left[N^{[e}\left(\delta^{f]a}\delta^{bc}{+}\delta^{f]b}\delta^{ac}\right)\dot{v}_{A}^{c}\right]\Big|_{T_{R}}.}
\label{eq:amphijS1}
\end{equation}
At this stage, this correction is valid, similar to the quadrupole formula, for general closed orbits. We find a perfect match of these spinning corrections at linear order in the objects' spins, with the classical derivation, \eqref{eq:classicalhijgeneralorbit}, using the identity \eqref{eq:identityTT}. The linear-in-spin scattering amplitude is universal \cite{Bjerrum-Bohr:2013bxa,Bautista:2019tdr}, therefore, so is the radiated gravitational field \eqref{eq:amphijS1}. Equivalently, the classical spin dipole of a point particle is universal, describing any spinning compact object at leading order. Therefore, non-universality of the waveform at higher spin orders may enter only through a solution to the classical \ac{EoM} for a particular compact binary system. We showed in \cref{sec:hijclassical} that the closed orbits waveform \eqref{eq:classicalhijgeneralorbit} contains all possible spin effects at leading order in the \ac{BH}s' velocities, \textit{before} specializing the constituents' trajectories; i.e., $h^{(0),ij}_{TT,S^{\ell\geq 2}}=0$. Therefore, we expect to find cancellations at higher orders in spins at the level of the scattering amplitude for $\ell> 1$. Finally, as claimed above, there exists a one-to-one correspondence between source multipole moments and spinning scattering amplitudes: $\mc{I}_{ij}\leftrightarrow \langle M_{5,{\rm S}^{0}}^{(0)\,ab}\rangle$ and $\mc{J}_{ij}\leftrightarrow \langle M_{5,{\rm S}^{1}}^{(0)\,ab}\rangle$. This holds in the sense that both $\mc{I}_{ij}$ and $\langle M_{5,{\rm S}^{0}}^{(0)\,ab}\rangle$ produce the quadrupole formula (and similarly for the linear-in-spin waveform).

\subsubsection{Cancellations at quadratic order in spin}\label{sec:quadratic in spin}

In the previous section, we showed that the gravitational waveform emitted from a spinning \ac{BBH} at leading order in its velocities is entirely contained in the linear-in-spin radiation field \eqref{eq:classicalhijgeneralorbit}. Equivalently, this waveform is obtained only using the scalar and linear-in-spin amplitude. The remaining all orders in spin result \eqref{eq:hijcircularv0} emerges solely from the solution \eqref{eq:consvsol} for quasi-circular orbits. To confirm this from the scattering amplitudes perspective, we are left to show that higher spin amplitudes do not provide additional non-trivial contributions to the general closed orbit results presented above. In this section, we demonstrate the cancellation at the quadratic order in the \ac{BH}s' spins, by specializing to circular orbits and by focusing on the $S_1\neq 0, S_2\rightarrow0$ limit.

At leading order in the \ac{BH}s' velocities, there are two distinct contributions to the radiated field from our approach. There is the quadratic-in-spin part of the amplitude on the one hand -- obtained from replacing the classical numerators \eqref{eq:num_quad_class} into the general formula \eqref{eq:newM5clas}  -- leading to $T_{1,{\rm {S}^{2}}}^{(0)\,ij}$, and the scalar part \eqref{eq:scalar-amplitude} in conjunction with the quadratic-in-spin part of the classical \ac{EoM} \eqref{eq:EoM}, yielding $T_{2,{\rm {S}^{2}}}^{(0)\,ij}$, on the other hand\footnote{Notice, the linear-in-spin part of the \ac{EoM} is sub-leading in $v$, and therefore, when convoluted with the linear-in-spin amplitude, the resulting quadratic in spin contribution is pushed to sub-leading order in velocities.}; both combine as
\begin{align}
    T_{\rm {S}^{2}}^{(0)\,ij}(k,\boldsymbol{z}_{1},\boldsymbol{z}_{2})=T_{1,{\rm {S}^{2}}}^{(0)\,ij}(k,\boldsymbol{z}_{1},\boldsymbol{z}_{2})+T_{2,{\rm {S}^{2}}}^{(0)\,ij}(k,\boldsymbol{z}_{1},\boldsymbol{z}_{2}).
\end{align}
Focusing first on the contribution from the quadratic-in-spin part of the  amplitude, to leading order in $v$ it reads 
\begin{equation}\label{eq:amplitude spin 2 nonrel}
    \langle M_{5,{\rm S}^{2}}^{(0)\,ab}(\boldsymbol{q},\bar{\omega})\rangle= \textcolor{black}{\frac{1}{4}\,}i m_{2}^{2}\kappa^{3}S_{1}^{k}S_{1}^{l}\left[V_{kl,df}^{ab}\frac{q^{d}q^{f}}{\boldsymbol{q}^{2}}+C_{kl}^{ab}\right],
\end{equation}
where we have defined the tensor 
$ V_{kl,df}^{ab} = \delta_{kl}\delta_{d}^{a}\delta_{f}^{b}-\frac{1}{2}\delta_{kd}\left(\delta_{f}^{a}\delta_{l}^{b}+\delta^{fb}\delta_{l}^{a}\right)$, and $C_{kl}^{ab}$ is a contact term, which we discard, as it is irrelevant for the gravitational waveform. As before, we insert this amplitude into the source \eqref{source}, and perform the $\bs{q}$-integrals aided by \eqref{eq:integrals}. The first contribution to the source $T_{\rm {S}^{2}}^{(0)\,ij}$ is then
\begin{equation}
    T_{1,{\rm {S}^{2}}}^{(0)\,ab}(k,\boldsymbol{z}_{1},\boldsymbol{z}_{2})=-\textcolor{black}{\frac{1}{4}\,}\frac{m_{2}\kappa^{3}}{m_14\pi}S_{1}^{k}S_{1}^{l}V_{kl,df}^{ab}\int dte^{i\bar{\omega}\,t}\frac{1}{r^{5}}\left[r^{2}\delta^{df}-3z_{21}^{d}z_{21}^{f}\right].
\end{equation}
Using the scalar part of the  \ac{EoM} \eqref{eq:EoM} to remove three powers of $r$ in the denominator, the above reduces to
\begin{equation}\label{eq:quad source quad ampl}
    T_{{1,\rm {S}^{2}}}^{(0)\,ab}(k,\boldsymbol{z}_{1},\boldsymbol{z}_{2})=-3\frac{m_{2}}{m_1}\text{\ensuremath{\kappa}}S_{1}^{k}S_{1}^{l}V_{kl,df}^{ab}\int dte^{i\bar{\omega}\,t}\frac{1}{r^{2}}\left[\left(\frac{\dot{v}_{2}{\cdot}z_{12}}{m_{1}}{+}\frac{\dot{v}_{1}{\cdot}z_{21}}{m_{2}}\right)\frac{\delta^{df}}{3}{-}\left(\frac{\dot{v}_{2}^{(d}z_{12}^{f)}}{m_{1}}{+}\frac{\dot{v}_{1}^{(d}z_{21}^{f)}}{m_{2}}\right)\right], 
\end{equation}
which, for quasi-circular orbits \eqref{eq:circorbits}, reads
\begin{equation}\label{eq:cirspin2amp2}
     T_{1,{\rm {S}^{2}}}^{(0)\,ab}(k,\boldsymbol{z}_{1},\boldsymbol{z}_{2})\Big|_{\rm{circular}} = \textcolor{black}{-2}\kappa\bar{\omega}^2\mu \bar{a}_1^2\int dte^{i\bar{\omega}t}\left[2n^a n^b - \lambda^a \lambda^b\right].
\end{equation}
Recall the definition for the symmetric mass ratio $\mu = m_1 m_2/M$, and  $\bar{a}_1 = S_1^i \ell_i/m_1$, with $\ell^i$  perpendicular to both $n^i$ and $\lambda^i$. Note, the solution to the classical \ac{EoM}, $r(x)$, in the numerator, cancels with the two powers of $r$ in the denominator.

We now turn to the second contribution to the source: $T_{2,\rm {S}^{2}}^{(0)\,ij}$. To that end, we first rewrite \eqref{eq:scalar-source-befor-eom} by expanding the sums and removing those terms that vanish under the TT projection:
\begin{equation}
    T_{{2,\rm {S}^{2}}}^{(0)\,ab}(k,\boldsymbol{z}_{1},\boldsymbol{z}_{2})= -\kappa^3 \int dt e^{i\bar \omega t}\frac{m_1 m_2 z_{12}^c}{16\pi r^3}\left[  \delta ^{c(a} z_{12}^{b)}+\frac{2i}{\bar \omega }\delta^{c(a}v_{12}^{b)}\right].
\label{eq:t2quadratic}
\end{equation}
Next we use the classical \ac{EoM} to quadratic order in spin, which can be written in the following form (see  \cref{sec: eom quad spin})
\begin{equation}\label{eq:eom_wuad_spin}
     \dot{v}_1^l = {-}\frac{m_2\kappa^2}{32\pi}  \frac{z_{12}^l}{r^3}{\textcolor{black}{+}}\frac{3}{4}\frac{S_1^iS_1^j}{m_1^2r^2}\left[\left(\delta_{ij}{-}\frac{5z_{12,i}z_{12,j}}{r^2}\right)\left(\dot{v}_1^l {-}\frac{m_2}{m_1}\dot{v}_2^l  \right){+}2 \delta^l_{(i}\left(\dot{v}_{1,j)}{-}\frac{m_2}{m_1} \dot{v}_{2,j)}  \right) \right].
\end{equation}
Combining this with \eqref{eq:t2quadratic}, the scalar part will recover the Einstein quadrupole radiation formula \eqref{eq:einstein-quadrupole}. We stress that although the quadrupole formula appears to be spin-independent for general orbits, spin information arises through a specific solution to the \ac{EoM}, as pointed out above. In particular, for quasi-circular orbits the Einstein quadrupole formula provides the quadratic-in-spin result \eqref{eq:hijcircularv0}. Let us, therefore, focus in the remaining contribution of \eqref{eq:eom_wuad_spin}, which is 
\begin{equation}
\begin{split}
T_{{2,{\rm {S}^{2}}}}^{(0)\,ab}(k,\boldsymbol{z}_{1},\boldsymbol{z}_{2}) & =-\frac{\textcolor{black}{3}}{4}\kappa S_{1}^{k}S_{1}^{l}\int dte^{i\bar{\omega}t}\left[\delta^{c(a}z_{12}^{b)}+\frac{2i}{\bar{\omega}}\delta^{c(a}v_{12}^{b)}\right]\times\frac{1}{m_{1}r^{2}}\\
 &  \left[\left(\delta_{kl}{-}\frac{5z_{12,k}z_{12,l}}{r^{2}}\right)\left(\dot{v}_{1}^{c}{-}\frac{m_{2}}{m_{1}}\dot{v}_{2}^{c}\right){+}2\delta_{(k}^{c}\left(\dot{v}_{1,l)}{-}\frac{m_{2}}{m_{1}}\dot{v}_{2,l)}\right){+}\frac{m_{2}}{m_{1}}(1\leftrightarrow2)\right].
\end{split}
\end{equation}
Using the center of mass parametrization\footnote{Note, the linear-in-spin corrections of this parametrization is sub-leading in velocities.} \eqref{COMframe}, the quasi-circular orbits condition $\ddot{r} = -\bar{\omega}r$, and the unit vectors \eqref{eq:circorbits}, the source reduces to
\begin{equation}
    T_{{2,{\rm {S}^{2}}}}^{(0)\,ab}(k,\boldsymbol{z}_{1},\boldsymbol{z}_{2})\Big|_{\rm{circular}} =\textcolor{black}{3}\kappa\bar{\omega}^2 \mu \bar{a}_1^2\int dte^{i\bar{\omega}t}\left[n^a n^b + i (\lambda^a n^b+\lambda^b n^a)\right],
\end{equation}
In order to remove the imaginary part of the source, we proceed as before and use an IBP prescription. Notice, since $(\lambda^a n^b+\lambda^b n^a) = -\frac{1}{\omega}\frac{d}{dt}(\lambda^a\lambda^b)$, the IBP yields
\begin{equation}\label{eq:source quad spin scalar ampl}
    T_{{2,{\rm {S}^{2}}}}^{(0)\,ab}(k,\boldsymbol{z}_{1},\boldsymbol{z}_{2})\Big|_{\rm{circular}} =\textcolor{black}{3}\kappa\bar{\omega}^2\mu \bar{a}_1^2\int dte^{i\bar{\omega}t}\left[n^a n^b - \lambda^a \lambda^b\right].
\end{equation}
This has the familiar form found in \eqref{eq:hijcircularv0}. Unlike this form, in \eqref{eq:cirspin2amp2} an extra factor of two appears in the $n^an^b$ term. This obscures the desired cancellation between \eqref{eq:source quad spin scalar ampl} and \eqref{eq:cirspin2amp2} in $T_{\rm {S}^{2}}^{(0)\,ij}$. To address this subtlety, we emphasize the degeneracy in choice of the IBP prescription. For instance, the relations of the kinematic variables in the center of mass frame results in $-\frac{d}{dt}(\lambda^a\lambda^b) = \frac{d}{dt}(n^a n^b)=\omega(\lambda^a n^b+\lambda^b n^a)$. Using the latter equality, the IBP performed in \eqref{eq:source quad spin scalar ampl} results in $2n^a n^b$, instead of $n^a n^b - \lambda^a \lambda^b$. A priori, neither of these two choices are preferred. The solution is to notice that the freedom in the choice of the IBP prescription is a manifestation of the gauge redundancy of the gravitational waveform at null infinity. That is, below in \cref{sec:results} we show that either choice yields the same result for the gauge invariant gravitational wave energy flux. For now, we note only that at the level of the gauge invariant energy flux, one factor of $n^a n^b $ in  \eqref{eq:cirspin2amp2} is equivalent to $n^a n^b\to \frac{1}{2}(n^a n^b - \lambda^a \lambda^b)$, and postpone the justification to \cref{sec:results}. Therefore, both \eqref{eq:cirspin2amp2} and \eqref{eq:source quad spin scalar ampl} yield the same result, but with opposite sign. This implies the desired cancellation of the waveform contributions at the quadratic order in \ac{BH}s' spins. Equivalently, using the waveform derived from \eqref{eq:cirspin2amp2} and \eqref{eq:source quad spin scalar ampl} to determine the energy flux from each contribution, we see that both contributions are identical up to an overall sign, hence, cancelling at the level of the gauge invariant gravitational wave energy flux as well (more on this below).

\subsubsection{Scalar waveform at sub-leading order in velocities} \label{sec:subleading-v}

So far we have dealt with leading in \ac{BH}s' velocities spinning corrections to the Einstein quadrupole formula \eqref{eq:einstein-quadrupole}. In this section, we go beyond this restriction and consider a non-spinning \ac{BBH} at the first sub-leading order in velocities, therefore, demonstrating the applicability of our approach \eqref{eq:radiated field} to determine the radiated gravitational waves also in this regime. At this order, the scalar 5-point amplitude is also independent of $q^0$, therefore, arguments made above in \cref{sec:general_approach_amp} concerning the time integration still holds. In this case, however, the first relativistic correction, in our    Born approximation, as coming from the product of the plane wave functions in \eqref{eq:source_some2}, appears in the source through the     kinematic exponential $e^{-i\boldsymbol{ k}\cdot\tilde{\bs{z}}}$, and therefore contributes to the sub-leading source $T_{S^0}^{(1)\,ab}$, due to the scaling $\bar{\omega}\sim v$. That is, after time integration, the exponential function reduces as $e^{i \bs{k}\cdot\tilde{\bs{z}}}\rightarrow 1-i\bar{\omega} \boldsymbol{N}{\cdot} \tilde{\boldsymbol{z}}+\mathcal{O}(v^2)= 1-\frac{i}{2}\bar{\omega} \boldsymbol{N}{\cdot}( \boldsymbol{z}_1+\boldsymbol{z}_2)+\mathcal{O}(v^2)$; hence, the source is built from the order-$v^0$ non-spinning scattering amplitude, $T_{2,S^0}^{(1)\,ab}$, as well as from the $v^1$-amplitude, $T_{1,S^0}^{(1)\,ab}$. More concretely, the sub-leading source decomposes as\footnote{In principle, the classical \ac{EoM} \eqref{eq:EoM} also contain higher-order-in-$v$ corrections, which could be used in an iterative manner, starting purely from the leading in $v$-scalar amplitude. However, these velocity corrections vanish in the no-spin limit considered in this section.}
\begin{equation}\label{eq:sourcev1gen}
    T_{S^0}^{(1)\,ab}(k,\boldsymbol{z}_1,\boldsymbol{z}_2) = T_{1,S^0}^{(1)\,ab}(k,\boldsymbol{z}_1,\boldsymbol{z}_2)+T_{2,S^0}^{(1)\,ab}(k,\boldsymbol{z}_1,\boldsymbol{z}_2),
\end{equation}
where
\begin{equation}\label{eq:soruce1}
T_{1,S^0}^{(1)\,ab}(k,\boldsymbol{z}_1,\boldsymbol{z}_2)=\frac{i}{m_1 m_2}\int dte^{i\bar{\omega}\,t}\int\frac{d^{3}\boldsymbol{q}}{\left(2\pi\right)^{3}}e^{-i\boldsymbol{q}{\cdot}\boldsymbol{z}_{21}}\langle M_{5,{\rm S}^0}^{(1)\,ab}(\boldsymbol{q},\bar{\omega})\rangle ,
\end{equation}
and
\begin{equation}\label{eq:soruce2}
T_{2,S^0}^{(1)\,ab}(k,\boldsymbol{z}_1,\boldsymbol{z}_2)=\frac{1}{m_1 m_2}\int dte^{i\bar{\omega}\,t}\int\frac{d^{3}\boldsymbol{q}}{\left(2\pi\right)^{3}}e^{-i\boldsymbol{q}{\cdot}\boldsymbol{z}_{21}}\frac{\bar{\omega}}{2} \boldsymbol{N}{\cdot}( \boldsymbol{z}_1+\boldsymbol{z}_2)\langle M_{5,{\rm S}^0}^{(0)\,ab}(\boldsymbol{q},\bar{\omega})\rangle.
\end{equation}
Notice the superscripts in the amplitude. First, we focus on the relativistically corrected scalar amplitude. Analogous to before, we insert \eqref{eq:scalar-numerators-gr} into \eqref{eq:newM5clas}, but now keep the non-trivial order $\mathcal{O}(v^1)$ contributions to the 5-point amplitude:
\begin{equation}
\begin{split}
\langle M_{5,{\rm S^0}}^{(1)\,ab}(\boldsymbol{q},\bar{\omega})\rangle & =-\frac{i m_{1}^{2}m_{2}^{2}\kappa^{3}}{2}N_{l}\Bigg[\frac{q^{l}q^{m}}{\boldsymbol{q}^{4}}\delta_{m}^{(a}(v_{1}{+}v_{2})^{b)}+\frac{q^{l}}{2\bar{\omega}\boldsymbol{q}^{2}}\left(v_{1}^{a}v_{1}^{b}-v_{2}^{a}v_{2}^{b}\right)\\
 & \hspace{3cm}+\frac{q^{m}}{\bar{\omega}\boldsymbol{q}^{2}}\left(v_{1}^{l}v_{1}^{(a}\delta_{m}^{b)}-v_{2}^{l}v_{2}^{(a}\delta_{m}^{b)}\right)\Bigg].
 \end{split}
\end{equation}
Subsequently, the source \eqref{eq:soruce1}, after the $\bs{q}$-integration, takes the form
\begin{equation}
\begin{split}
    T_{1,S^0}^{(1)\,ab}(k,\boldsymbol{z}_1,\boldsymbol{z}_2) & =\frac{\kappa^{3}}{16\pi}N_{l}\int dte^{i\bar{\omega}\,t}\sum_{A,B}\frac{m_{A}m_{B}}{r^{3}}\Bigg[\frac{1}{2}\left(\boldsymbol{z}_{AB}^{2}\delta^{lm}{-}z_{AB}^{l}z_{AB}^{m}\right)\delta_{m}^{(a}(v_{A}{+}v_{B})^{b)}\\
 & \hspace{3cm}{-}\frac{i}{\bar{\omega}}z_{AB}^{n}\left(\delta_{n}^{l}v_{A}^{a}v_{A}^{b}{+}2\delta_{n}^{m}v_{A}^{l}v_{A}^{(a}\delta_{m}^{b)}\right)\Bigg].
 \end{split}
\end{equation}
In order to remove the powers of $z_{AB}\sim r$ in the denominator, we use the scalar part of the \ac{EoM} \eqref{eq:EoM}, to obtain
\begin{align}
\begin{split}
T_{1,S^0}^{(1)\,ab}(k,\boldsymbol{z}_1,\boldsymbol{z}_2) & =2\kappa N_{l}\int dte^{i\bar{\omega}\,t}\Bigg[-\frac{1}{2}\sum_{A,B}m_{A}\left(\dot{\boldsymbol{v}}_{A}{\cdot}\boldsymbol{z}_{AB}\delta^{lm}{-}\dot{v}_{A}^{m}z_{AB}^{l}\right)\delta_{m}^{(a}(v_{A}{+}v_{B})^{b)}\\
 & \hspace{3cm}{+}\frac{i}{\bar{\omega}}\sum_{A} m_{A}\dot{v}_{A}^{n}\left(\delta_{n}^{l}v_{A}^{a}v_{A}^{b}{+}2\delta_{n}^{m}v_{A}^{l}v_{A}^{(a}\delta_{m}^{b)}\right)\Bigg].
\end{split}
\end{align}
The term in the second line can be integrated utilizing the relation $\dot{v}_{A}^{n}\left(\delta_{n}^{l}v_{A}^{i}v_{A}^{j}{+}2\delta_{n}^{m}v_{A}^{l}v_{A}^{(i}\delta_{m}^{j)}\right)= \frac{d}{dt}(v_{A}^i v_{A}^j v_{A}^l)$. With this, this piece of the sub-leading scalar source simplifies to
\begin{align}
\begin{split}
  T_{1,S^0 }^{(1)\,ab}(k,\boldsymbol{z}_1,\boldsymbol{z}_2) & =-2\kappa N_{l}\int dte^{i\bar{\omega}\,t}\Bigg[\sum_{A,B}\frac{m_{A}}{2}\left(\dot{\boldsymbol{v}}_{A}{\cdot}\boldsymbol{ z}_{AB}\delta^{lm}{-}\dot{v}_{A}^{m}z_{AB}^{l}\right)\delta_{m}^{(a}(v_{A}{+}v_{B})^{b)}
  {-}\sum_{A} m_{A}v_{A}^{l}v_{A}^{a}v_{A}^{b}\Bigg].
\end{split}
\end{align}
We now address the second term in \eqref{eq:sourcev1gen} -- the computation of the second contribution \eqref{eq:soruce2} to the sub-leading scalar source. The $\bs{q}$-integration is identical to the one used leading up to \eqref{eq:scalar-source-intermediate}. Starting from the latter, using the relation \eqref{eq:identity_eom}, and multiplying the sub-leading prefactor $-\frac{i}{2}\bar{\omega}\boldsymbol{N}{\cdot}(\boldsymbol{z}_1+\boldsymbol{z}_2)$ we arrive at
\begin{equation}
T_{2,S^0}^{(1)\,ab}(k,\boldsymbol{z}_1,\boldsymbol{z}_2)
=-\kappa\int dte^{i\bar{\omega}\,t}\boldsymbol{N}{\cdot}(\boldsymbol{z}_1+\boldsymbol{z}_2)\sum_A m_{A}\left[i\bar{\omega}\ddot{z}_{A}^{(a}z_{A}^{b)}-2v_{A}^{(a}\dot{v}_{A}^{b)}\right].
\end{equation}
Lastly, with the replacement $\bar{\omega}\rightarrow i\frac{d}{dt}$ the first term is integrated. The gravitational radiation field is then determined by putting the two sources together in \eqref{eq:sourcev1gen}, and substituting this into \eqref{eq:radiated field}, to end up at the first sub-leading in \ac{BH} velocities non-spinning correction to the Einstein quadrupole formula:
\begin{equation}\label{eq:amplitudes_scalar_v^1}
\begin{split}
h_{\text{TT},S^0}^{(1)\,ij}(T_{R},R,\boldsymbol{N},\boldsymbol{z}_1,\boldsymbol{z}_2)&=-\frac{\kappa^2 m_1}{8\pi R }\Pi^{ij}{}_{ab} N_l\Bigg[\frac{1}{2}
\left(\dot{\boldsymbol{v}}_{1}{\cdot}\boldsymbol{z}_{12}\delta^{lm}{-}\dot{v}_{1}^{m}z_{12}^{l}\right)\delta_{m}^{(a}(v_{1}+v_2)^{b)} {-}v_{1}^{l}v_{1}^{a}v_{1}^{b}
\\& 
-\frac{1}{2} \left(\frac{d}{dt}\left(\ddot{z}_{1}^{(a}z_{1}^{b)}(z_1+z_2)^l\right)+2(z_1+z_2)^lv_{1}^{(a}\dot{v}_{1}^{b)}\right)
\Bigg]+(1\leftrightarrow 2).
\end{split}
\end{equation}
Based on our derivation, this result is valid for generic \textit{closed} orbits, provided the corresponding EoM. However, the form of this waveform is different from the compact classical result in \eqref{eq:h1_v1_scalar}. This is not surprising since, as illustrated above, there is always the freedom of choice of IBP prescription, which casts the waveform into different forms. Finding the prescription, for which both the amplitude's and the classical waveforms match, could be cumbersome for generic closed orbits. Therefore, we specialize to the quasi-circular setting \eqref{eq:circorbits}; In the latter, we find perfect agreement between   \eqref{eq:amplitudes_scalar_v^1} and \eqref{eq:h1_v1_scalar}. We close with a remark on the correspondence between the classical source multipole moments leading to the gravitational radiation via the multipolar post-Minkowskian approach, and our ansatz to compute the associate gravitational waves using spinning scattering amplitudes. We saw in \cref{sec:hijclassical}, the sub-leading order result \eqref{eq:hijcircularv1} is built from both $\mathcal{I}_{ijk}$ and $\mathcal{J}_{ij}$. While at leading order in the \ac{BH}s' velocities (see \cref{sec:linear_spin_waveform}), there exists a certain one-to-one correspondence between the source multipole moments, at sub-leading orders in velocities, no trivial correspondence can be extracted from our results.

\subsection{Radiated gravitational wave energy flux} \label{sec:results}

In the previous sections, we showed explicitly that the radiated gravitational field, $h^\text{TT}_{ij}$, computed using a classical approach and utilizing a 5-point spinning scattering amplitude, agree in the aligned spin, general (and quasi-circular) orbit setup at the considered orders in the velocity and spin expansions. These are the gravitational waves emitted at an instant in the binary's evolution. Information about the frequency dynamics of the radiation is contained in the emitted gauge invariant gravitational wave energy flux. The latter is ultimately responsible for the inspiral of the two  \ac{BH}s  and for the characteristic increase in gravitational wave frequency towards the merger, therefore, a crucial ingredient for gravitational wave search strategies.

In this section, we derive the instantaneous gravitational wave energy flux $\mathcal{F}$ using the TT metric perturbations at null infinity computed in the previous subsections to the respective orders in the spin and velocity expansions. In general, the total instantaneous energy loss $\mathcal{F}$ can be obtained with
\begin{align}
    \mathcal{F}=\frac{R^2}{32\pi}\int_{S^2} d\Omega \ \dot{h}^\text{TT}_{ij} \dot{h}^{\text{TT},ij}.
    \label{eq:generalfluxexpression}
\end{align}
Let us   return here to the justification for the replacements and claims made in \cref{sec:quadratic in spin}. The time dependence of $h^{\text{TT}}_{ij}$ is solely contained in the center of mass variables $n^a$ and $\lambda^a$, which, in the center of mass frame and for circular orbits, are related by $\frac{d}{dt}(n^a n^b)=\frac{1}{2}\frac{d}{dt}(n^a n^b-\lambda^a \lambda^b)$. Since only the time derivative of the radiated field, $\dot{h}^\text{TT}_{ij}$, enters in \eqref{eq:generalfluxexpression}, this justifies the replacement $n^a n^b\rightarrow\frac{1}{2}(n^an^b-\lambda^a\lambda^b)$ made in \cref{sec:quadratic in spin} at the level of the radiated field. Furthermore, this also shows that the gauge invariant energy flux is, in fact, independent of the IBP prescription discussed in \cref{sec:quadratic in spin}. Therefore, the latter can be viewed as a manifestation of the gauge freedom in the emitted waveform. Indeed, this extends to the Newman-Penrose scalar $\Psi_4\sim \bar{m}^\mu \bar{m}^\nu \ddot{h}^\text{TT}_{\mu\nu}$ in an identical fashion. Exploiting this, the gravitational wave energy flux is obtained by combining the scalar, \eqref{eq:einstein-quadrupole}, and linear-in-spin, \eqref{eq:amphijS1}, metric perturbations $h^{\text{TT}}_{ij}$ at leading order in the \ac{BH}s' velocities, in \eqref{eq:generalfluxexpression}. For quasi-circular  orbits \eqref{eq:consvsol}, together with \eqref{eq:identityOmegaInt}, we find the energy loss
\begin{align}
\mathcal{F}^{(0)}_\text{circular}=\frac{32}{5}\frac{\mu^2x^5}{M^2}+\frac{2}{5}\frac{\mu^2 x^7}{M^2}(32a_+^2+a_-^2)+ \mathcal{O}(a^3_{1,2}, a_1 a_2).
\label{eq:circorbflux}
\end{align}
Recall from above that $x=(M\omega)^{2/3}$. This matches perfectly with the results reported in Refs.~\cite{Poisson:1997ha, Kidder:1992fr, Kidder:1995zr, Buonanno:2012rv, Marsat:2014xea, Siemonsen:2017yux} to the respective orders in spin. In addition to this match at leading order in the black holes velocities, the metric perturbations computed in \eqref{eq:amplitudes_scalar_v^1} and \eqref{eq:h1_v1_scalar} specialized to circular orbits reproduce the correct no-spin gravitational wave energy flux $\mathcal{F}^{(1)}_\text{circular}=0$ at the first sub-leading order in velocities; this is, again, consistent with the leading no-spin \ac{PN} gravitational wave power (see, e.g., \cite{Kidder:1995zr}). Notice, we explicitly computed the quadratic-in-spin contributions only for one \ac{BH} with spin: $S_1\neq 0, S_2\rightarrow 0$. However, as noted above, the classical derivation in \cref{sec:hijclassical} revealed that the high-order-in-spin contributions to the circular orbit $h^\text{TT}_{ij}$ emerge solely from the solution to the \ac{EoM}, indicating that \eqref{eq:circorbflux} already contains the $a_1 a_2$-type interactions; this is the case, as can be seen in, for instance, \cite{Marsat:2014xea, Siemonsen:2017yux}, or from using \eqref{eq:hijeasier} together with \eqref{eq:generalfluxexpression}. At the level of the transverse traceless metric perturbations $h^\text{TT}_{ij}$, the classical derivation showed that \eqref{eq:classicalhijgeneralorbit} contains the complete all orders-in-spin information at leading order in velocities, since the remaining contributions to the radiation field -- i.e. $h_{ij,S^{\ell\ge 2}}^{(0)\text{TT}}=0$ -- vanish, \textit{without} a specific solution to the \ac{EoM}. In the scattering amplitudes setting, we confirmed this explicitly up to $\ell=1$, since \eqref{eq:einstein-quadrupole} and \eqref{eq:amphijS1} agree with \eqref{eq:classicalhijgeneralorbit} (exploiting \eqref{eq:identityTT}), and we showed the necessary cancellation for $\ell=2$ in \cref{sec:quadratic in spin}. Therefore, we conjecture such cancellations to occur at arbitrary order in spin, such that the solution to the \ac{EoM} provides the remaining spin-information, at leading order in velocities. The complete all-orders in spin gravitational power result partially presented in \eqref{eq:circorbflux} was determined in \cite{Siemonsen:2017yux}.

\section{Outlook of the chapter }\label{sec:discusion}

In this chapter we studied the relationship between the radiative dynamics of an aligned-spin spinning binary black hole from both, a classical, and a scattering amplitude perspective. For the former we employed the  multipolar post-Minkowskian formalism, whereas for the latter we proposed a dictionary  built from the 5-point \ac{QFT} scattering amplitude extensively studied in \cref{sec:spin_in_qed}. More precisely, the dictionary maps the  classical limit of the 5-pt scattering amplitude of two massive spinning particles exchanging and emitting a graviton, to   the source entering in Einstein's equation.  Furthermore, we included information of the conservative dynamics using the classical equations of motion, which we obtained from the instantaneous impulse formula which takes as main input the conservative two-body amplitude. 
We worked in linearized gravity, i.e., at tree-level, and to leading order in the black holes' velocities, but to all orders in their spin, as well as present preliminary results at sub-leading orders in velocities (in the no-spin limit). To leading order in the system's velocities, we showed that there exists a one-to-one correspondence between the source's multipole moments, and the scattering amplitudes. That is, the mass quadrupole in \eqref{eq:multipolemomentv0} corresponds to the scalar amplitude  \eqref{eq:scalar-amplitude}, while similarly, the current quadrupole in \eqref{eq:multipolemomentv0} is associated with the linear-in-spin amplitude \eqref{eq:linear_spin_v0}. This correspondence was made explicit  in the computation  of the  transverse-traceless part of the linear metric perturbations emitted to null infinity, as well as on the  gauge invariant gravitational wave energy flux. The latter agrees for quasi-circular orbits with the existing literature \cite{Kidder:1992fr, Kidder:1995zr, Marsat:2014xea, Siemonsen:2017yux}, both at the considered leading and sub-leading orders in the black holes' velocities. Therefore, gravitational waveforms and gauge invariant powers needed for detecting gravitational waves from \textit{inspiraling} black holes can be consistently computed from the classical limit of quantum \textit{scattering} amplitudes. 

The gravitational waveform is, in general, a gauge-dependent object, which makes a comparison between the classical and the scattering amplitude's derivations potentially difficult. In particular, and especially for general orbits and with spin effects, finding the corresponding gauge to undertake such comparisons can become cumbersome. In this chapter, we found evidence that such gauge freedom is related to the integration procedure used in the source for Einstein's equation, within the scattering amplitudes derivation. We demonstrated this explicitly for quasi-circular orbits, as this restriction simplifies the problem drastically. Importantly, we find that while the form of the gravitational radiation field is dependent upon the integration procedure used, the gauge-invariant gravitational wave power is independent of such a prescription -- as desired.

In this chapter, we focused entirely on the derivation of radiative degrees of freedom from the 5-point scattering amplitude, and show  classical \ac{EoM} for the system follow directly from the conservative amplitude, providing then a self-contained amplitudes derivation for the radiated field at leading and subleading order in the velocity expansion, but to all orders in spin. This expand the claims for scalar sources made by Dass and Soni in \cite{HariDass:1980tq}.

The amplitudes-based construction of the radiated field  \eqref{eq:radiated field}, provided in this chapter, has implicitly  used  the on-shell condition for the outgoing massive particles $\delta(p_i{\cdot}q_i)$, which discards terms quadratic in the velocities as indicated by the quantum corrections to the particles trajectories  $z_{Q}(\tau_i)$ in  \eqref{eq:radiation_kernell_gen}. These corrections can become important if convoluted with superclassical terms coming from loop amplitudes. This then   hints that at higher orders in perturbation theory, a  subtraction scheme would be needed to cancel those superclassical contributions  at the level of the gauge invariant observable, which in  this case corresponds to  the radiated energy flux  $\mathcal{F} \sim \int d\omega \dot{h}^{ij}\dot{h}_{ij}$; in addition, it would be desirable to study the connection of our approach and that of analytic continuation methods of scattering observables \cite{Kalin:2019inp, Kalin:2019rwq,Herrmann:2021lqe} .

 Besides,  exploring gauge fixing procedures that allow to match the general orbit result \eqref{eq:amplitudes_scalar_v^1} to the  classical result \eqref{eq:h1_v1_scalar}, as well as the inclusion of spin effects at sub-leading order in velocity is left for future work. Furthermore, in the context of scattering amplitudes, higher orders in velocities are naturally included. However, for closed orbits, these corrections are consistent only -- by virtue of the  virial theorem -- when also higher orders in the gravitational constant $G$ are considered. For instance, at quadratic order in the \ac{BH}s' velocities, the radiated field could contain contributions from both the tree-level and the one-loop 5-pt scattering amplitudes. One might wonder whether the amplitudes approach could reproduce the higher-order corrections to the energy flux for non-spinning binary black holes \cite{Blanchet:2013haa}.  

Finally, the source \eqref{eq:source_some1} was written in the Born approximation, where the initial state consists of two particles in their plane-wave states. However, the long-range nature of the gravitational  interactions renders the Born approximation to be invalid in this setting. Although this is expected to be a higher-$G$-effect (or equivalently a higher-$v$-effect in the closed orbit case), it plays an important role in the determination of the correct gravitational waveform. A modification to the Born approximation was proposed in \cite{HariDass:1980tq}, and claimed to contain all non-perturbative aspects of the S-matrix elements. We leave the exploration of this proposal for future work.

%% file: Chapters/chapter_double_copy.tex
\chapter{The double copy for massive spinning matter}\label{ch:double_copy}
\section{Introduction}\label{sec:intro_dc}

In \cref{sec:double-copy-preliminaires} we have briefly  introduced the \ac{BCJ} double copy program \cite{Bern:2008qj}, and show how certain gravitational
quantities can be obtained as a square of gauge-theory ones. This was done  in the context of massless particles, where the slogan was \ac{GR}= \ac{YM}$^2$. However, to test the extent of the double copy, and  to  study phenomenologically
relevant setups, it is desirable to introduce fundamental matter in
the construction. This has already been explored in the context of standard \ac{QCD} \cite{Johansson:2014zca,Johansson:2015oia, delaCruz:2015dpa,delaCruz:2016wbr,Brown:2018wss,Plefka:2018zwm}. Also a number of other interesting cases
has been considered,\footnote{For matter-coupled \ac{YM} theory the gravitational $D=4$ Lagrangians were first obtained from double copy in \cite{Chiodaroli:2015wal}, see also \cite{Anastasiou:2017nsz}.} including quiver theories with bifundamental matter \cite{Chiodaroli:2013upa,Huang:2012wr,Huang:2013kca} and theories with
spontaneously broken symmetries \cite{Chiodaroli:2015rdg,Chiodaroli:2017ehv}. On the other hand the classical double copy, in its many realizations, inherently contains massive matter and hence it is important to clarify the connection between the quantum and classical approaches.

In  \cref{sec:spin_in_qed} we have taken several steps in this direction, where  we introduced a classical double copy prescription for fundamental matter with spin,  which connects gravitational wave phenomena 
 with the spin-multipole expansion and  soft theorems, whose classical amplitudes were used in  \cref{ch:bounded} to study gravitational radiation for systems in bounded orbits. 
 In this chapter we will thoroughly expand on the spin multipole  double copy, and show how it arises in a purely \ac{QFT} framework. We will consider tree-level double copy of massive particles with generic spins and explore several interesting cases.

One of the main results of \cref{sec:spin_in_qed} was to obtain graviton-matter amplitudes from double copy at low
multiplicities but generic spin quantum number $s$. For a single matter line, the double copy was summarized in the operation \eqref{eq:doublecopyspin}. Amplitudes constricted in this way,  and their higher multiplicity extensions
are relevant for a number of reasons as we have pointed out in \cref{sec:spin_in_qed}, and we recall here: First, they have been recently
pinpointed to control the classical limit where the massive lines correspond to compact objects \cite{Neill:2013wsa,Bern:2019nnu,Bautista:2019tdr}. Second,
they  have an exponential form in accord with their multipole expansion \cite{Guevara:2018wpp,Chung:2018kqs,Bautista:2019tdr,Guevara:2019fsj,Arkani-Hamed:2019ymq} (see also \cref{app_B}). Third, they are dimension-independent and are
not polluted with additional states arising from the double copy such as dilaton and axion fields \cite{Bautista:2019tdr}. This will become even more evident once we provide the corresponding Lagrangians. Fourth, they are the building blocks in the two body problem, whose double copy properties are inherited by the two-body amplitudes via \eqref{eq:numdc}.

In this chapter we will rederive and extend \eqref{eq:doublecopyspin}, mainly focusing
on the simplest cases with $s,\tilde{s}\leq1$. For this spin values,  interactions are fundamental in the sense that their amplitudes  have a healthy high-energy behaviour \cite{Arkani-Hamed:2017jhn}. By promoting \ac{QED} to
\ac{QCD}\footnote{For the lower multiplicity cases $n=3,4$, one can choose \ac{QCD} partial amplitudes to coincide with \ac{QED} amplitudes.}, studying higher multiplicity amplitudes and the relevant cases
for two massive lines, we will identify the gravitational
theories obtained by this construction. In order to do this we must observe
that formula \eqref{eq:doublecopyspin} has implicit a rather strong assumption,
namely the fact that the \ac{LHS}   only depends on the quantum number $s+\tilde{s}$
and not on $s,\tilde{s}$ individually. For instance, this means that
for gravitons coupled to a spin-1 field, it should hold that

\begin{equation}
A_{n}^{{\rm {\rm GR}},1}\sim A_{n}^{{\rm QCD},\frac{1}{2}}\odot A_{n}^{{\rm QCD},\frac{1}{2}}=A_{n}^{{\rm QCD},0}\odot A_{n}^{{\rm QCD},1}\, , \label{eq:dcintro-1}
\end{equation}
(we have changed \ac{QED} to \ac{QCD} in preparation for $n>4$). This means $A_{n}^{{\rm {\rm gr}},1}$ not only realizes the equivalence principle in the sense of Weinberg \cite{PhysRev.135.B1049} but extends it to deeper orders in the soft expansion \cite{Bautista:2019tdr,Guevara:2019fsj}. In the classical limit, the $A^{\rm gr,s}_n$ amplitudes so constructed will reproduce a well defined
compact object irrespective of its double copy factorization. In \cref{sec:spin_in_qed} 
we exploited condition \eqref{eq:dcintro-1} at \textit{arbitrary} spin to argue that the 3-point
amplitude should indeed take an exponential structure, which has recently
been identified as a characteristic feature of the Kerr black hole
in the sense of \cite{Vines:2017hyw}, we will expand on this in \cref{ch:GW_scattering}. Here we will argue that despite having arbitrary spin, this 3-pt. amplitude can still be considered fundamental as it is essentially equal to its high-energy limit, which in fact implies \eqref{eq:doublecopyspin}-\eqref{eq:dcintro-1}.

A simple instance of (\ref{eq:doublecopyspin}) for gravitons was verified
explicitly by Holstein \cite{Holstein:2006pq,Holstein:2006wi} (see also \cite{Bjerrum-Bohr:2013bxa}) for $s=0\,,\tilde{s}\leq1$. He observed that as gravitational amplitudes have an intrinsic gravitomagnetic
ratio $g=2$, the double copy (\ref{eq:doublecopyspin}) can only hold by
modifying $A_{3}^{{\rm QED},1}$ away from its ``minimal-coupling''
value of $g=1$. This modification yields the gyromagnetic ratio $g=2$
characteristic of the electroweak model and was proposed as natural by Weinberg \cite{osti_4073049}. As observed long ago by Ferrara, Porrati and Telegdi \cite{PhysRevD.46.3529} this modification precisely cancels all powers of $1/m^{2}$ in $A_{4}^{{\rm QED},1}$ (see \eqref{eq:hlqcd}), which otherwise prevented the Compton amplitude to have a smooth high-energy limit.
This is a crucial feature, as it hints that the theories with a natural
value $g=2$ have a simple massless limit, and indeed can be obtained
conversely by compactifying pure massless amplitudes at any multiplicity. Furthermore,
it was pointed out in \cite{Cucchieri:1994tx} (and recently from a modern perspective \cite{Arkani-Hamed:2017jhn}) that the appearance of $1/m^{2}$ can
be avoided up to $s=2$ in the gravitational Compton amplitude $A_{4}^{{\rm GR},s}$
since it corresponds to fundamental interactions. By working on general dimensions, we will see that
indeed all such fundamental amplitudes follow from dimensional reduction of
massless amplitudes, and ultimately from a compactification of a pure graviton/gluon master amplitude. This is the underlying reason
they can be arranged to satisfy (\ref{eq:doublecopyspin}), which in turn simplifies the multipole expansion we exploited in \cref{sec:spin_in_qed}.

On a different front, as we argued in \cref{sec:spin_in_qed},  the squaring relations in the massless sector
yield additional degrees of freedom corresponding to a dilaton $\phi$
and 2-form potential $B_{\mu\nu}$, which is a consequence of the Clebsh-Gordon decomposition \eqref{eq:CG_decomp}. Their classical counterparts also arise in classical solutions (e.g. string theory backgrounds \cite{CAMPBELL1992199,CAMPBELL1993137,PhysRevD.43.3140,Gibbons:1982ih}) and therefore emerge naturally (and perhaps inevitably) in the classical double copy \cite{Goldberger:2016iau,Luna:2015paa,Luna:2016hge,Luna:2017dtq}. It is therefore natural to ask
whether the condition (\ref{eq:dcintro-1}) also holds when the massless
states involve such fields. As we have explained this is a non-trivial
constraint, and in fact, it only holds for graviton states! To exhibit this phenomena we are led to identify two different gravitational theories, which we refer
to as \12x12 and \0x1 theories for brevity. The corresponding
tree amplitudes will be constructed as
\begin{equation}
A_{n}^{\frac{1}{2}\otimes\frac{1}{2}}\sim A_{n}^{{\rm QCD},\frac{1}{2}}\otimes A_{n}^{{\rm QCD},\frac{1}{2}}\,,\quad A_{n}^{0\otimes1}\sim A_{n}^{{\rm QCD},0}\otimes A_{n}^{{\rm QCD},1}\label{eq:2dcs}
\end{equation}
We conjecture that at all orders in $\kappa=\sqrt{32\pi G}$ such tree-level interactions
follow from the more general Lagrangians,
\begin{equation}
\frac{\mathcal{L}^{\frac{1}{2}\otimes\frac{1}{2}}}{\sqrt{g}}=-\frac{2}{\kappa^{2}}R+\frac{(d-2)}{2}(\partial\phi)^{2}-\frac{1}{4}e^{\fb{\frac{\kappa}{2}}(d-4)\phi}F^I_{\mu\nu} F_I^{\mu\nu}+\frac{m_I^{2}}{2}e^{\fb{\frac{\kappa}{2}}(d-2)\phi}A_{\mu}^I A^{\mu}_I\,,\label{eq:1212dc}
\end{equation}
and 
\begin{align}
\frac{\mathcal{L}^{0\otimes1}}{\sqrt{g}}=&-\frac{2}{\kappa^{2}}R+\frac{(d-2)}{2}(\partial\phi)^{2}-\frac{e^{-2\kappa\phi}}{6}H_{\mu\nu\rho}(H^{\mu\nu\rho}+\frac{3\kappa}{2}A_I^{\mu}F_I^{\nu\rho}) \nonumber \\ 
&-\frac{1}{4}e^{-\kappa\phi}F^I_{\mu\nu}F_I^{\mu\nu}+\frac{m_I^{2}}{2}A^I_{\mu}A_I^{\mu}+\rm{quartic\,\, terms}\,,\label{eq:01dc}
\end{align}
where $H=dB$ is the field strength of a two-form $B$. Here a sum over $I=1,2$, the flavour index, is implicit and "quartic terms" denote contact interactions between two matter lines that we will identify in  \cref{two matter lines}. These actions will be constructed in general dimensions from simple considerations such as 1) classical behaviour and 2) massless limit/compactification in the
string frame. These methods were  cross-check  against the corresponding \ac{QFT} amplitudes in \cite{Bautista:2019evw} using modern tools such as massive versions of CHY  \cite{Cachazo:2013hca,Cachazo:2013iea,Cachazo:2014nsa,Cachazo:2014xea} and the connected formalism \cite{Cachazo:2018hqa,Geyer:2018xgb,Schwarz:2019aat}. In the massless limit, the \12x12 Lagrangian
is known as the Brans-Dicke-Maxwell (BDM) model with unit coupling
\cite{Cai:1996pj}. This theory is simpler than \0x1 in
many features, for instance in that the $B$-field is not sourced by
the matter line and it does not feature quartic interactions. Not surprisingly, in $d=4$ and in the massless
limit the \0x1 theory reproduces the bosonic interactions of
$\mathcal{N}=4$ Supergravity \cite{CREMMER197861,PhysRevD.15.2805}, which is known to arise
as the double copy between $\mathcal{N}=4$ \ac{SYM} and pure \ac{YM} theories  \cite{Bern:2010ue,Bern:2011rj,BoucherVeronneau:2011qv}. In general dimension we will see that the \0x1 theory is precisely the \ac{QFT} version
of the worldline model constructed by Goldberger and Ridgway in \cite{Goldberger:2016iau,Goldberger:2017ogt} and later extended in \cite{Goldberger:2017vcg,Li:2018qap} to exhibit a classical double copy construction with spin. This explains their findings on the fact that the \textit{classical} double copy not only fixes $g=2$ on the \ac{YM} side, but also precisely sets the dilaton/axion-matter coupling on the gravity side.

As explained in \cref{sec:KMOC}, and extensively exemplified in previous chapters,  the long-range radiation of a two-body system, emerging in the classical double copy, has been directly linked to a 5-point amplitude at leading order
\cite{Luna:2017dtq,Kosower:2018adc,Bautista:2019tdr,PV:2019uuv}. We show that
by implementing generalized gauge transformations \cite{Bern:2008qj} one can define
a \ac{BCJ} gauge in which the $\hbar\to0$ limit is smooth, i.e. there are
no "superclassical" $\sim \frac{1}{\hbar}$ contributions to cancel \cite{Kosower:2018adc}. The result precisely takes the
form derived in \eqref{eq:newM5clas}. This then  allows us to translate between the \ac{QFT} version of the double copy and
a classical version of it. We employ this formulae to test double copy in several cases, including the computation of dilaton-axion-graviton radiation with spin \cite{Goldberger:2017vcg,Li:2018qap}.

This chapter is organized as follows. In  \cref{sec:sec2} we introduce the double copy for one matter line by studying its massless origin, focusing on the \12x12 theory and later extending it in more generality. In  \cref{constructing the lagrangians} we construct the Lagrangians for both \ac{QCD} and Gravity from simple arguments, which are then checked against the previous amplitudes. In  \cref{two matter lines} we extend both the amplitudes and the Lagrangian construction to two matter lines and define the classical limit to make contact with previous results. In the appendices we provide some further details on the constructions, and  perform checks such as tree-level unitarity, and consistency with the $d=4$ formulation of the $\frac{1}{2}\otimes\frac{1}{2}$ double copy in \cite{Johansson:2019dnu}.
This chapter is mostly based on previous work by the author \cite{Bautista:2019evw}.

\section{Double Copy from Dimensional Reduction}\label{sec:sec2}

In this section we will introduce the double copy construction by considering
a single massive line. In this case one should expect the double copy to hold for massive
scalars as their amplitudes can be obtained via compactification of
higher dimensional amplitudes \cite{Bjerrum-Bohr:2013bxa,Bern:2019crd,Bern:2019nnu}. Here we will explicitly demonstrate how
this holds even for the case of spinning matter as long as such particles
are \textit{elementary}. This means we consider particles of spin
$s\leq2$ coupled to \ac{GR}  and particles of spin $s\leq1$ coupled to
\ac{QCD}, in accordance with the notion of \cite{Arkani-Hamed:2017jhn}, see also \cite{Cucchieri:1994tx,Deser:2001dt}. The
fact that these amplitudes can be chosen to have a smooth high-energy
limit can be used backwards to construct them directly from their
massless counterparts. On the other hand, once the double copy form of gravitational-matter amplitudes is achieved one may use it to manifest properties such as the multipole expansion of \cref{sec:spin_in_qed},\fb{ we will expand on this in Sec \ref{sec:multipoles}.}

\subsection{The \texorpdfstring{$\frac{1}{2}\otimes\frac{1}{2}$}{1212} construction}\label{sec:new}

Let us consider first the case $s=\tilde{s}=\frac{1}{2}$ in (\ref{eq:doublecopyspin}) and
relegate the other configurations for the next section. For $D=4$
massless \ac{QCD}, the double copy procedure was first studied by Johansson
and Ochirov \cite{Johansson:2014zca}. In particular they observed that Weyl-spinors
in \ac{QCD} can be double copied according to the rule $2\otimes2=2\oplus1\oplus1$,
where the two new states correspond to a photon $\gamma^{\pm}$ and
the remaining ones to axion and dilaton scalars. This implies that
we can obtain amplitudes in a certain Einstein-Maxwell theory directly
from massless \ac{QCD}.  More precisely, for two massive particles we can
write (see also \eqref{eq:kltmassless})

\begin{equation}
A_{n}^{\frac{1}{2}\otimes\frac{1}{2}}(\gamma_{1}^{-}H_{3}{\cdots}H_{n}\gamma_{2}^{+})=\sum_{\alpha\beta}K_{\alpha\beta}[2|A_{n,\alpha}^{{\rm QCD}}(g{\cdots}g)|1\rangle \langle 1|\bar{A}_{n,\beta}^{{\rm QCD}}(g{\cdots}g)|2],\label{eq:dconeline}
\end{equation}
here we have used the massless Weyl spinors  $v_{1}^{-}=|1\rangle$
and $\bar{u}_{2}^{+}=[2|$ for   matter particles  (See \cref{sec:spinor-helicity}). $\bar{A}$ here denotes charge conjugation, which will be relevant
in the massive case. In the gravitational amplitude the two photon
states $\gamma_{1}^{+}$,$\gamma_{2}^{-}$ make a matter line while
interacting with the ``fat'' states $H_{i}$. The latter are obtained
from the double copy of the gluons $g_{i}$, and can be taken to be
either a Kalb-Ramond field\footnote{In $D=4$ this field can be dualized to an axion pseudoscalar. We
will indistinctly refer to the two-form $B_{\mu\nu}$ as axion or Kalb-Ramond
field.}, a dilaton or a graviton by projecting the product representation
into the respective irreps. as dictated by the Clebsh-Gordon decomposition \eqref{eq:CG_decomp}:

\begin{equation}
  H_{i}^{\mu\nu}\to\epsilon_{i}^{\mu}\tilde{\epsilon}_{i}^{\nu}=\underbrace{\epsilon_{i}^{[\mu}\tilde{\epsilon}_{i}^{\nu]}}_{B^{\mu\nu}}+\underbrace{\frac{\eta^{\mu\nu}}{D-2}\epsilon_{i}\cdot\tilde{\epsilon}_{i}}_{\eta^{\mu\nu}\frac{\phi}{\sqrt{D-2}}}+\underbrace{\left(\epsilon_{i}^{(\mu}\tilde{\epsilon}_{i}^{\nu)}-\frac{\eta^{\mu\nu}}{D-2}\epsilon_{i}\cdot\tilde{\epsilon}_{i}\right)}_{h^{\mu\nu}}\,.
\end{equation}\label{eq:fatgraviton}
The sum over $\mbox{\ensuremath{\alpha}},\,\beta$ in (\ref{eq:dconeline})
ranges over $(n-3)!$ orderings, where $K_{\alpha,\beta}$ is the
standard \ac{KLT} kernel \cite{Kawai:1985xq,Bern:1998sv,BjerrumBohr:2010ta}.\footnote{We define the \ac{KLT} kernel with no coupling constants and  absorb the gauge theory coupling $\tilde{g}$ into the generators $\tilde{ T}^a$ as $\tilde{g}\tilde{T}^a\to T^a$. We also absorb the overall factors of $i$ in the definition of the amplitudes and  use the conventions for the metric to be in the mostly minus signature.} This construction can be implemented
because for a single matter line we can take the matter particles
to be either in the fundamental or in the adjoint representation and
the basis of partial amplitudes will be identical \cite{Johansson:2014zca}. In  \cref{two matter lines}  we will switch to a more natural prescription for the case of two
matter lines.

The \ac{RHS}   of (\ref{eq:dconeline}) exhibits explicitly the helicity
weight $\pm\frac{1}{2}$ associated to the Weyl spinors. This
means the operators $A^{{\rm QCD}}$ and $\bar{A}^{{\rm QCD}}$ ,
defined as the amplitude with such spinors stripped, do not carry
helicity weight. They can be written as products of Pauli matrices
$\sigma^{\mu}$, $\bar{\sigma}^{\mu}$ where the free Lorentz index
is contracted with momenta $p_{i}^{\mu}$ or gluon polarizations $\epsilon_{i}^{\mu}$,
as we will see in the examples of the next section. We can alternatively
write them in terms of the corresponding spinor-helicity variables
as in \cite{Johansson:2015oia}.

Quite generally, the \ac{LHS}   of (\ref{eq:dconeline}) defines a gauge
invariant quantity due to the fact that it is constructed from partial
gauge-theory amplitudes. It also has the correct factorization properties
(see e.g. \cite{Bern:2010yg,Chiodaroli:2017ngp}). Furthermore, by providing the Lagrangian it
will become evident that when the states $H_{i}$ are chosen to be
gravitons the amplitude we get for a single matter-line is that of
\textit{pure} Einstein-Maxwell theory, where the dilatons and axions
simply decouple. This decoupling is one of the key properties of these
objects, which we have exploited in \cref{sec:spin_in_qed}. Similarly, the decoupling of further matter particles will be treated in  \cref{residues at 4 points}.

In order to extend (\ref{eq:dconeline}) to the massive case we rewrite
it in a way in which it is not sensitive to the dimension, and then
use dimensional reduction. This can be done by introducing polarization
vectors for the photons $\gamma^{\pm}$. Recall from the spinor helicity section \cref{sec:spinor-helicity}, that a photon polarization
vector can be taken to be $\epsilon_{\mu}^{+}\sigma^{\mu}=\sqrt{2}\frac{|\mu\rangle[p|}{\langle\mu p\rangle}$
where $[\mu|$ is a reference spinor carrying the gauge freedom, and
analogously $\epsilon_{\mu}^{-}\bar{\sigma}^{\mu}=\sqrt{2}\frac{|\mu]\langle p|}{[\mu p]}$.
We then have the identity

\begin{eqnarray}
[2|X|1\rangle \langle 1 |\bar{Y}|2] & = & \frac{{\rm Tr}(X|1\rangle [1\mu_1]  \langle 1|\bar{Y}|2]\langle 2\mu_{2}\rangle |2])}{[1\mu_{1}]\langle 2\mu_{2}\rangle}, \\
 & = & \frac{1}{2}{\rm Tr}\left(X\bar{p}_{1}\epsilon_{1}\bar{Y}p_{2}\bar{\epsilon}_{2}\right),\label{eq:massless}
\end{eqnarray}
where the bottom line now can be naturally extended to higher dimensions.\footnote{We represent the Dirac algebra in terms of the $2^{D/2}\times2^{D/2}$
matrices $\Gamma_{D}^{\mu}=\left(\begin{array}{cc}
0 & \sigma_{D}^{\mu}\\
\bar{\sigma}_{D}^{\mu} & 0
\end{array}\right)$ and define $X=X_{\mu}\sigma_{D}^{\mu},\,\bar{X}=X_{\mu}\bar{\sigma}_{D}^{\mu}$
etc. The extension of (\ref{eq:massless}) to general dimension simply
states that linear combinations $c_{ab}u_{i}^{a}\bar{v}_{i}^{b}$
of the Weyl spinors can be replaced as $c_{ab}v_{i}^{a}\bar{u}_{i}^{b}=p_{i}\bar{\epsilon}_{i}$
for some particular choice of $\epsilon_{i}^{\mu}$ depending on $c_{ab}$.
A formula for general dimension is of course obtained by replacing
$\sigma^{\mu},\bar{\sigma}^{\mu}\to\Gamma^{\mu}$, which in $D=4$
also reduces to (\ref{eq:massless}).} It is manifestly gauge invariant since the shift  $\epsilon_{i}\to \epsilon_{i}+p_{i}$
is projected out due to the on-shell condition for massless particles $p_{i}\bar{p}_{i}=0$. 

Using this identity, the double copy (\ref{eq:dconeline}) can be uplifted
to dimension $D=2m$ as
\begin{equation}
A_{n}^{\frac{1}{2}\otimes\frac{1}{2}}(\gamma_{1}H_{3}\cdots H_{n}\gamma_{2}^{*})=\frac{1}{2}\sum_{\alpha\beta}K_{\alpha\beta}{\rm Tr(}A_{n,\alpha}^{{\rm QCD}}(g{\cdots}g)\bar{p}_{1}\epsilon_{1}\bar{A}_{n,\beta}^{{\rm QCD}}(g{\cdots}g)p_{2}\bar{\epsilon}_{2}).\label{eq:kltmassless}
\end{equation}
Note that the operators $A_{n}^{{\rm QCD}}$, $\bar{A}_{n}^{{\rm QCD}}$
in (\ref{eq:dconeline}) are defined under the support of the Dirac
equation. This means that they can be shifted by operators proportional
to $p_{1}$ or $p_{2}$. The insertion of $p_{1},p_{2}$ in (\ref{eq:kltmassless})
certainly projects out these contributions by using the on-shell condition
$p\bar{p}=\bar{p}p=0$. For instance, if the matrix operator $A_{n}^{{\rm QCD}}$
is shifted by $p_{2\mu}\sigma^{\mu}$ the \ac{QCD} amplitude $\bar{u}_{2}A_{n}^{{\rm QCD}}v_{1}$
is invariant, and consistently (\ref{eq:kltmassless}) picks
up no extra contribution, i.e.
\begin{equation}
    {\rm Tr(}p_{2}\bar{p}_{1}\epsilon_{1}\bar{A}_{n,\beta}^{{\rm QCD}}(g\cdots g)p_{2}\bar{\epsilon}_{2})=-{\rm Tr(}p_{2}p_{2}\bar{p}_{1}\epsilon_{1}\bar{A}_{n,\beta}^{{\rm QCD}}(g\cdots g)\epsilon_{2})=0,
\end{equation}
where we used $p_{2}\bar{\epsilon}_{2}={-}\epsilon_{2}\bar{p}_{2}$.
This kind of manipulations are usual when bringing the \ac{QCD} amplitude
into multipole form as explored in \cref{sec:spin_in_qed} to make explicit the corresponding form
factors. 

 We now proceed to dimensionally reduce our formulae in order to
obtain a \ac{KLT} expression for massive spin-$\frac{1}{2}$ particles.
This follows from a standard \ac{KK} compactification on a torus, as we
explain in the next section. In terms of momenta, we can define the
$d=D-1$ components $p_{1}$ and $p_{2}$ via
\begin{equation}\begin{split}
P_{1} & =  (m,p_{1}),\\
P_{2} & =  (-m,p_{2}),\\
P_{i} & =  (0,k_{i}),\qquad i\in\{3,\ldots,n\}    
\end{split}
\end{equation}
which trivially satisfies momentum conservation in the \ac{KK} component, which we take with minus signature. We also take all momenta outgoing. In terms of Feynman diagrams,
the reduction induces the flow of \ac{KK} momentum through the only path
that connects particles $p_{1}$ and $p_{2}$. The propagators in
this line are deformed to massive propagators as

\begin{equation}
    \frac{1}{P_{I}^{2}}=\frac{1}{p_{I}^{2}{-}m^{2}},
\end{equation}
where $P_{I}=(m,p_{I})$ is the internal momentum. The procedure works
straightforwardly when compactifying more particles as long as the
KK lines do not cross (i.e. we will not allow interactions between
massive particles), as we will explain in the case of two matter lines.\\
 By applying these rules to (\ref{eq:kltmassless}) the amplitudes
$A^{{\rm gr}}$, $A^{{\rm QCD}}$ now contain massive lines and lead
to a (gravitational) Proca theory and the massive \ac{QCD} theory in $d=D-1$
dimensions, respectively. This can be observed easily by applying
the dimensional reduction to the Lagrangian as we do in  \cref{constructing the lagrangians}.
In the case of the spin-1 theory we choose the polarization vectors
$\epsilon_{1}$,$\epsilon_{2}$ to be $d$-dimensional, i.e.
$\epsilon\to(0,\varepsilon)$, so that the transverse condition
$\epsilon{\cdot}P=0$ now imposes $\varepsilon{\cdot}p=0$. In
the \ac{QCD} case we note that the Dirac equation now becomes
\begin{equation}\begin{split}
(p_{\mu}\Gamma_{d}^{\mu})u & =  m\,u,\\
(p_{\mu}\Gamma_{d}^{\mu})v & =  -m\,v,
\end{split}
\end{equation}
where we have used
\begin{equation}\label{sigmad}
\sigma_{D}=(\mathbb{I},\Gamma_{d})\,,\qquad \bar{\sigma}_{D}=(-\mathbb{I},\Gamma_{d})\,,
\end{equation}
in the chiral representation. Denoting by $W$ and $W^*$ the Proca fields obtained from the photons, the construction (\ref{eq:kltmassless})
now reads 
\begin{equation}
A_{n}^{\frac{1}{2}\otimes\frac{1}{2}}(W_{1}H_{3}{\cdots}H_{n}W_{2}^{*}){=}\sum_{\alpha\beta}\frac{K_{\alpha\beta}}{2^{\left\lfloor d/\text{2}\right\rfloor {-}1}} {\rm Tr(}A_{n,\alpha}^{{\rm QCD}}(g{\cdots}g)(\slashed p_{1}{-}m)\slashed\varepsilon_{1}\bar{A}_{n,\beta}^{{\rm QCD}}(g{\cdots}g)(\slashed p_{2}{-}m)\slashed\varepsilon_{2}),\label{eq:massiveklt}
\end{equation}
where the normalization factor follows from the Dirac trace ${\rm tr}(\mathbb{I})=2^{\left\lfloor D/\text{2}\right\rfloor }$. Even though our derivation used that $d=2m-1$ for the reduction procedure, our final result is written explicitly in terms of $d$-dimensional Dirac matrices so we assume it to be valid in generic dimensions. To confirm this we will indeed compute both sides of \eqref{eq:massiveklt} from generic-dimensional Lagrangians and find a precise agreement. 

From now on we refer to the double-copy theory as the \textit{$\frac{1}{2}\otimes \frac{1}{2}$ theory} because it is constructed from two (conjugated) copies of massive \ac{QCD}.  As in the massless case, the
role of the projectors $\slashed p_{i}\pm m$ is to put the \ac{QCD} amplitudes
on the support of the massive Dirac equation. With a slight abuse
of notation, we have left here the symbol $K_{\alpha\beta}$ for the
massive \ac{KLT} kernel, which we used in \cref{sec:spin_in_qed} for the 3 and 4-point amplitudes.

We have thus derived an explicit \ac{KLT} relation for massive amplitudes
of one matter line, (\ref{eq:massiveklt}) as a direct consequence
of the massless counterpart. The resulting theory will be extended to two matter lines in Section \ref{two matter lines}. The partial
amplitudes $A_{n,\alpha}^{{\rm QCD}}$ are associated to Dirac spinors
in general dimension, as opposed to Majorana ones, and hence the resulting
spin-1 field is a complex 
Proca state coupled to gravity. Moreover, it follows from the massless
case that when all the gravitational states $H_{i}$ are chosen as
gravitons, the dilaton and axion field decouple and the theory simply
corresponds to Einstein-Hilbert gravity plus a covariantized (minimally
coupled) spin-1 Lagrangian. We will see that this holds quite generally
and is consistent with the observations made around eq. \eqref{eq:doublecopyspin} for generic
spin.

In our formula the states $H_{i}$ denote the fat gravitons \eqref{eq:fatgraviton} characteristic of the double copy construction. However,
a particular feature arises in that amplitudes with an odd number
of axion fields vanish. This can be traced back to the symmetry in
the two \ac{QCD} factors of the \12x12 construction.
To see this, let us slightly rewrite (\ref{eq:massiveklt}) as
\begin{equation}
A_{n}^{\frac{1}{2}\otimes\frac{1}{2}}(W_{1}H_{1}^{\mu_{1}\nu_{1}}{\cdots}H_{n{-}2}^{\mu_{n-2}\nu_{n-2}}W_{2}^{*})=\sum_{\alpha\beta}K_{\alpha\beta}(A_{n,\alpha}^{{\rm QCD}})^{\mu_{1}\cdots\mu_{n-2}}\otimes(A_{n,\beta}^{{\rm QCD}})^{\nu_{1}\cdots\nu_{n-2}}\,,\label{eq:abklt}
\end{equation}
where
\begin{equation}\label{tensor product }
X\otimes Y=\frac{1}{2^{\left\lfloor d/\text{2}\right\rfloor -1}}{\rm Tr(}X(\slashed p_{1}{-}m)\slashed\varepsilon_{1}\bar{Y}(\slashed p_{2}{-}m)\slashed\varepsilon_{2}).
\end{equation}
It is not hard to check that (see for instance the explicit form in
(\ref{eq:dconeline}))
\begin{equation}
    (A_{n,\alpha}^{{\rm QCD}})^{\mu_{1}\cdots\mu_{n-2}}\otimes(A_{n,\beta}^{{\rm QCD}})^{\nu_{1}\cdots\nu_{n-2}}=(A_{n,\beta}^{{\rm QCD}})^{\nu_{1}\cdots\nu_{n-2}}\otimes(A_{n,\alpha}^{{\rm QCD}})^{\mu_{1}\cdots\mu_{n-2}}\,.
\end{equation}
Now, since the Kernel $K_{\alpha\beta}$ in (\ref{eq:abklt}) can
be arranged to be symmetric in $\alpha\leftrightarrow\beta$, this
implies that the \ac{RHS}   of (\ref{eq:abklt}) is symmetric under the exchange
of \textit{all} $\mu_{i}\leftrightarrow\nu_{i}$ at the same time,
namely $(\mu_{1},\mu_{2}\ldots)\leftrightarrow(\nu_{1},\nu_{2}\ldots)$.
However, if we antisymmetrize an odd number of pairs $\{\mu_{k},\nu_{k}\}$,
i.e. compute the amplitude for an odd number of axions, and symmetrize
the rest of the pairs, we obtain an expression which is antisymmetric
under the full exchange $(\mu_{1},\mu_{2}\ldots)\leftrightarrow(\nu_{1},\nu_{2}\ldots)$.
Hence amplitudes with an odd number of axions must vanish. 

The above considerations imply that the axion field is pair-produced
and cannot be sourced by the Proca field. This is also true for amplitudes with no matter (i.e. the massless double copy) and even for amplitudes with
more matter lines: For e.g. two matter lines, provided a double copy formula as in  \cref{two matter lines}, we can test axion propagation by examining all possible factorization channels. Since the factorization always contains amplitudes with either one or none matter lines we conclude that the axion will not emerge in the cut unless introduced also as an external state. The argument carries over for an arbitrary number of matter lines. This is the reason we were able to remove axionic states from the double copy amplitudes in \cref{sec:spin_in_qed}.

The previous fact is surprising from the gravitational perspective since it is known that the axion couples naturally to the spin of matter particles. We interpret this fact as an avatar of the spin-$\frac{1}{2}$ origin of the construction. In  \cref{4 d double copy} and \cref{residues at 4 points} we will specialize the construction to $d=4$: In particular we will show that being a pseudoscalar,
the axion can only be sourced when the Proca field decays into a massive
pseudoscalar as well, as considered  in \cite{Johansson:2019dnu}. In the massless theory such field is obtained by selecting anticorrelated fermion helicities
in the \ac{RHS}   of (\ref{eq:dconeline}) which leads to massless (pseudo)scalars
instead of photons $\gamma^{\pm}$ \cite{Johansson:2014zca}. The analysis becomes
more involved in higher dimensions. For our purposes here we can neglect
these processes and simply keep the theory containing a Proca field,
a graviton and a dilaton as a consistent tree-level truncation of the spectrum
in arbitrary number of dimensions.

A further clarification is needed regarding the compactification and
the dilaton states. In the massless case these are obtained via the
replacement
\begin{equation}\label{eq:pr1}
   \epsilon_{i}^{\bar{\mu}}\tilde{\epsilon}_{i}^{\bar{\nu}}\to\frac{\eta^{\bar{\mu}\bar{\nu}}}{\sqrt{D-2}},
\end{equation}
where we have denoted the indices as $\bar{\mu},\mathbf{\bar{\nu}}$
to emphasize that the trace is taken in $D=d+1$ dimensions. However,
after dimensional reduction we have $\epsilon^{\bar{\mu}}\to\epsilon^{\mu}$,
and we extract the corresponding dilaton via
\begin{equation}\label{eq:pr2}
    \epsilon_{i}^{\mu}\tilde{\epsilon}_{i}^{\nu}\to\frac{\eta^{\mu\nu}}{\sqrt{d-2}}.
\end{equation}
This means that taking the dimensional reduction does not commute
with extracting dilaton states, as e.g. terms of the form $P_{1}\cdot\epsilon\,P_{2}\cdot\tilde{\epsilon}$
are projected to $P_{1}\cdot P_{2}=p_{1}\cdot p_{2}+m^{2}$ in the
first case and to $p_{1}\cdot p_{2}$ in the second case. In order to match certain results in the literature (e.g. \cite{Goldberger:2016iau}) we find that we need to adopt the second construction: first implement dimensional reduction
on the fat states, and then project onto either dilatons or gravitons. 

Let us close this section by providing some key examples of this procedure
for $n=3,4$. The 3-point. dilaton amplitude from (\ref{eq:massiveklt}), using \eqref{eq:fatgraviton},
gives 
\begin{eqnarray}
A_{3}^{\frac{1}{2}\otimes\frac{1}{2}}(W{}_{1}\phi W_{2}^{*}) & = & \frac{2K_{3}}{2^{\left\lfloor d/2\right\rfloor }\sqrt{d-2}}{\rm Tr(}A_{3}^{{\rm \mu}}\slashed\varepsilon_{1}(\slashed p_{1}{-}m)\bar{A}_{3\mu}\slashed\varepsilon_{2}(\slashed p_{2}{-}m)), \nonumber \\
 & = & \frac{\kappa}{2\sqrt{d-2}}{\rm (}2m^{2}\varepsilon_{1}{\cdot}\varepsilon_{2}{+}(d-4)k_{3}{\cdot}\varepsilon_{1}k_{3}{\cdot}\varepsilon_{2}),\label{eq:dila3pt}
\end{eqnarray}
\textcolor{black}{where we have use the momentum conservation $p_{1}+p_{2}+k_{3}=0$,
and the dilaton projection $\epsilon_{3}^{\mu}\tilde{\epsilon}_{3}^{\nu}\to\frac{\eta^{\mu\nu}}{\sqrt{d-2}}$}.
This example will exhibit one of the main differences between the \12x12
construction and the other cases, namely that the dilaton (and the
axion) fields couple differently to matter in each case, as opposed
to gravitons which couple universally, as we will see when obtaining explicitly the Lagrangians  \eqref{eq:2dcs} and\eqref{eq:1212dc} below. 


Now we can move on to $n=4$. The only independent \ac{QCD} amplitude can be computed from the Feynman rules derived from the \ac{QED}\footnote{Since, as already mentioned, at  4-points the \ac{QED} and the partial order  \ac{QCD} amplitude coincide, we can use the Feynman rules derived from the \ac{QED}  Lagrangian.  } Lagrangian  \eqref{eq:qed_lagrangian}, and  reads 

\begin{equation}
A_{4,1324}^{\mu_{3}\mu_{4}}=-\frac{1}{4}\frac{\gamma^{\mu_{4}}(\slashed{p}_{1}{+}\slashed{k}_{3}{-}m)\gamma^{\mu_{3}}}{(p_1+k_3)^2-m^2}-\frac{1}{4}\frac{\gamma^{\mu_{3}}(\slashed{p}_{1}{+}\slashed{k}_{4}{-}m)\gamma^{\mu_{4}}}{(p_1+k_4)^2-m^2}\,. \label{eq:A41324}
\end{equation}
Analogously, the charge conjugated amplitude is 
\begin{equation}
\bar{A}_{4,1324}^{\mu_{3}\mu_{4}}=-\frac{1}{4}\frac{\gamma^{\mu_{3}}(\slashed{p}_{1}{+}\slashed{k}_{3}{+}m)\gamma^{\mu_{4}}}{(p_1+k_3)^2-m^2}-\frac{1}{4}\frac{\gamma^{\mu_{4}}(\slashed{p}_{1}{+}\slashed{k}_{4}{+}m)\gamma^{\mu_{3}}}{(p_1+k_4)^2-m^2},\label{eq:A41324bar}
\end{equation}
where the conjugated amplitude is obtained by inverting the direction of the massive line. Note that this ordering corresponds to the Compton amplitude in  \ac{QED}  \eqref{eq:comptom multipoles} for spin $1/2$.

The full Compton amplitude for fat gravitons can be computed from the double
copy $(\ref{eq:massiveklt})$,
\begin{equation}
A_{4}^{\frac{1}{2}\otimes\frac{1}{2}}(W_{1}H_{3}^{\mu_{3}\nu_{3}}H_{4}^{\mu_{4}\nu_{4}}W_{2}^{*})=\frac{1}{2^{\left\lfloor d/2\right\rfloor -1}}K_{1324,1324}\,\text{tr} \left[A_{4}^{\mu_{3}\mu_{4}}\slashed{\varepsilon}_{1}(\slashed{p}_{1}{+}m)\bar{{A}}_{4}^{\nu_{3}\nu_{4}}\slashed{\varepsilon}_{2}(\slashed{p}_{2}{+}m)\right],
\end{equation}\label{eq:4pt1212}
where the massive \ac{KLT} kernel takes the compact form
\begin{equation}
K_{1324,1324}=\frac{2p_{1}{\cdot}k_{3}\,p_{1}{\cdot}k_{4}}{k_{3}{\cdot}k_{4}}.\label{eq:klt 1324}
\end{equation}
For instance, the two-dilaton emission amplitude reads

\begin{equation}\label{eq:ABphiphiB-2}
\begin{split}
A_{4}^{\frac{1}{2}\otimes\frac{1}{2}}(W_{1}\phi_{3}\phi_{4}W_{2}^{*}) & {=}\frac{\kappa^{2}\varepsilon_{1,\alpha}\varepsilon_{2,\beta}^{*}}{32(d-2)p_{1}{\cdot}k_{3}\,p_{1}{\cdot}k_{4}k_{3}{\cdot}k_{4}}\left\{ \left[(d{-}4)^{2}s_{34}^{2}{-}16(d{-}2)p_{1}{\cdot}k_{3}p_{2}{\cdot}k_{3}\right]\times\right.\\
&
\left[p_{1}{\cdot}k_{3}k_{4}^{\alpha}k_{3}^{\beta}{+}p_{2}{\cdot}k_{3}(k_{3}^{\alpha}k_{4}^{\beta}{+}p_{1}{\cdot}k_{3}\eta^{\alpha\beta})\right] 
 {+}2m^{2}s_{34}\left[4p_{1}{\cdot}k_{3}\left(k_{4}^{\alpha}k_{3}^{\beta}{-}k_{3}^{\alpha}k_{4}^{\beta}\right.\right.
\\& \left.\left.\left.{+}2p_{2}{\cdot}k_{3}\eta^{\alpha\beta}\right)
 {+}s_{34}\left((d{-}4)(k_{3}^{\alpha}k_{3}^{\beta}{+}k_{4}^{\alpha}k_{4}^{\beta})-2(k_{3}^{\alpha}k_{4}^{\beta}{+}m^{2}\eta^{\alpha\beta})\right)\right]\right\},
 \end{split}
\end{equation}
which again exhibits explicit mass dependence in accord with our discussion. On the other hand, extracting the pure graviton emission from \eqref{eq:4pt1212} gives 

\begin{equation}\label{eq:ABhhB-1}
\begin{split}
A_{4}^{\frac{1}{2}\otimes\frac{1}{2}}(W_{1}h_{3}h_{4}W_{2}^{*}) & =\frac{\kappa^{2}\varepsilon_{1,\alpha}\varepsilon_{2,\beta}^{*}}{2p_{1}{\cdot}k_{3}\,p_{1}{\cdot}k_{4}\,k_{3}{\cdot}k_{4}}p_{1}{\cdot}F_{3}{\cdot}F_{4}{\cdot}p_{1}\big[p_{1}{\cdot}p_{3}F_{4}^{\mu\alpha}F_{3,\mu}^{\beta}{+}\\
 &\quad p_{1}{\cdot}k_{4}F_{3}^{\mu\alpha}F_{4,\mu}^{\beta}{+}F_{3}^{\alpha\beta}p_{1}{\cdot}F_{4}{\cdot}p_{2}{+}F_{4}^{\alpha\beta}p_{1}{\cdot}F_{3}{\cdot}p_{2}{+}p_{1}{\cdot}F_{3}{\cdot}F_{4}{\cdot}p_{1}\eta^{\alpha\beta}\big]\,.
\end{split}
\end{equation}
 Quite non-trivially, we find that the Dirac trace leads to a factorized formula. The underlying reason is of course that the graviton amplitudes are universal as announced in \eqref{eq:1212dc} and \eqref{eq:01dc}. This means these results can also be obtained via the $0 \otimes 1$ factorization that we will introduce in the next subsection, but we can already guess it is  given by the double copy of the $s=0$  and $s=1$ part of  \eqref{eq:comptom multipoles}.

\subsubsection{Exempli Gratia: The Multipole double copy revisited}\label{sec:multipoles}

We have introduced the operation (\ref{eq:massiveklt}) with a slight
modification in \eqref{eq:starprod}. This is because the main utility of this
construction is \textit{not} the fact that we can build gravitational
amplitudes by squaring those of \ac{QCD} (we have just seen that the
former follow from a dimensional reduction of the Einsten-Maxwell
system), but the fact that by rearranging the massive \ac{QCD} amplitudes
in a multipole form we obtain a multipole expansion on the gravitational
side \cite{Porto:2005ac,Porto:2006bt,Levi:2015msa,Levi:2014gsa,Levi:2018nxp,Porto:2008tb,Porto:2008jj}.

Consider two spin $1/2$ multipole  operators $X,Y$ of order $p,q$ respectively, namely $X\sim(\gamma^{\mu\nu})^{p}$ and $Y\sim(\gamma^{\mu\nu})^{q}$ acting on Dirac spinors.
As they involve an even number of gamma matrices, and the Dirac trace
vanishes for an odd number of such, we have
\begin{equation}
{\rm Tr(}X\fb{(g{\cdots}g)}(\slashed p_{1}{-}m)\slashed\varepsilon_{1}\bar{Y}(g{\cdots}g)(\slashed p_{2}{-}m)\slashed\varepsilon_{2})={\rm Tr}(Xp_{1}\varepsilon_{1}\bar{Y}p_{2}\varepsilon_{2})+m^{2}{\rm Tr}(X\varepsilon_{1}\bar{Y}\varepsilon_{2}),\label{eq:expdc}
\end{equation}
where the conjugated operator $\bar{Y}$ is obtained by $\gamma^{\mu\nu}\to-\gamma^{\mu\nu}$.
In the cases studied in \cref{sec:spin_in_qed}  (for $n=3,4$) both terms in the
RHS of previous equation coincide and hence we defined the double copy product simply as
\begin{equation}
X\odot Y=\frac{1}{2^{\left\lfloor D/\text{2}\right\rfloor }}{\rm Tr}(X\varepsilon_{1}\bar{Y}\varepsilon_{2}),
\end{equation}
i.e. using twice the second term. At $s=\frac{1}{2}$ we explicitly tested
this definition for operators up to the quadratic order in $\gamma^{\mu\nu}.$
Let us here just recall the example of $A_{3}$, given in \eqref{eq:3ptshalf},  which exhibits an
explicit exponential form and we write we for the reader's convenience
\begin{equation}\label{eq:3ptextr}
\bar{u}_{2}A_{3}^{{\rm QCD}}v_{1}\propto\epsilon\cdot p_{1}\times\bar{u}_{2}e^{J}v_{1}\,,
\end{equation}
where $J$ is a Lorentz generator that reads
\begin{equation}
J=-\frac{k_{3\mu}\epsilon_{3\nu}}{\epsilon_{3}\cdot p_{1}}J^{\mu\nu}=-\frac{k_{3\mu}\epsilon_{3\nu}}{\epsilon_{3}\cdot p_{1}}\frac{\gamma^{\mu\nu}}{2}.\label{eq:gens12}
\end{equation}
The exponential form for $s=\frac{1}{2}$ generators is only linear in this case since higher multipoles vanish. Note now that while the second equality holds for $s=\frac{1}{2}$, the generator $J$
itself makes sense in any representation  \cite{Bautista:2019tdr}. In the representation
$(J^{\mu\nu})_{\beta}^{\alpha}=\eta^{\alpha[\mu}\delta_{\beta}^{\nu]}$
we can check that $(e^{J})_{\alpha}^{\beta}p_{1}^{\alpha}=(p_{1}+k)^{\beta}=-p_{2}^{\beta}$
and hence the generator acts as a boost $p_{1}\to-p_{2}$. Now we
can plug the operator \eqref{eq:3ptextr} and its conjugate in (\ref{eq:expdc}) and check that in fact
both terms yield the same contribution:

\begin{eqnarray}
A_{3}^{{\rm QCD}}\otimes A_{3}^{{\rm QCD}} & \propto & {\rm Tr(}e^{J}(\slashed p_{1}{-}m)\slashed\varepsilon_{1}e^{-J}(\slashed p_{2}{-}m)\slashed\varepsilon_{2}),\nonumber \\
 & = & {\rm Tr}(e^{J}p_{1}e^{-J}e^{J}\varepsilon_{1}e^{-J}p_{2}\varepsilon_{2})+m^{2}{\rm Tr}(e^{J}\varepsilon_{1}e^{-J}\varepsilon_{2}), \nonumber\\
 & = &- {\rm Tr}(p_{2}\tilde{\varepsilon}_{2}p_{2}\varepsilon_{2})+m^{2}{\rm Tr}(\tilde{\varepsilon}_{2}\varepsilon_{2})=2m^{2}{\rm {\rm Tr(\mathbb{I})}}\tilde{\varepsilon}_{2}\cdot\varepsilon_{2}, \label{eq:expdc-1}
\end{eqnarray}
where $\tilde{\varepsilon}_{2}^{\alpha}=(e^{J})_{\beta}^{\alpha}\varepsilon_{1}^{\beta}$
is a new polarization state for $p_{2}$, that is, it satisfies $p_{2}\cdot\tilde{\varepsilon}_{2}=0$.
Thus we obtain the gravitational (Proca) amplitude as
\begin{equation}\label{12123ptcopy}
A_{3}^{\frac{1}{2}\otimes\frac{1}{2}}\propto\epsilon_{3}\cdot p_{1}\times\varepsilon_{2}\cdot e^{J}\cdot\varepsilon_{1}=\epsilon_{3}\cdot p_{1}\varepsilon_{2}\cdot\varepsilon_{1}\fb{-}k_{3\mu}\epsilon_{3\nu}\varepsilon_{2}^{\alpha}(J^{\mu\nu})_{\alpha}^{\beta}\varepsilon_{1\beta},
\end{equation}
where higher multipoles also vanish for $s=1$, in contrast with higher spins (see \eqref{eq:mass3pths}).This simple example shows that the exponential form is preserved under
double copy (this is particular of $n=3$), but more importantly it
shows the general fact that, as observed in \cref{sec:spin_in_qed}, the gravitational
amplitude is obtained in multipole form as well. For $n=3,4$, the multipole operators
can be double copied via the general rules \eqref{eq:rules}, and in turn the resulting
multipole expansion can be used to decode the classical information
contained in the amplitude by comparison to either one body observables such as the linearized Kerr metric and the scattering of gravitational waves off the Kerr black hole, as we will do in \cref{ch:GW_scattering}, or by computing two-body observables for unbounded \cref{ch:electromagnetism,ch:soft_constraints,sec:spin_in_qed,}, or bounded scenarios as in \cref{ch:bounded}.

\subsubsection*{Detour: Arbitrary spin at $n=3$}

The massless origin of all these constructions should be by now clear. Let us take a brief detour to emphasize some remarkable properties at $n=3$.
In $D=4$, in \cref{sec:spinor-helicity} we learned  the massless three-point amplitude is fixed from helicity
weights as in \eqref{eq:massless3pths_n},

\begin{equation}
A_{3}^{h_{3},h}\sim\left(\frac{\langle13\rangle}{\langle23\rangle}\right)^{2h}\left(\frac{\langle13\rangle\langle32\rangle}{\langle12\rangle}\right)^{h_{3}},\label{eq:massless3pths}
\end{equation}
for a state of arbitrary helicity $h$ emitting a gluon ($h_{3}=1$) or a
graviton $(h_{3}=2$). Consequently, it directly satisfies the double
copy relation
\begin{equation}
A_{3}^{{\rm gr},h+\bar{h}}=K_{3}A_{3}^{{\rm QCD},h}A_{3}^{{\rm QCD},\bar{h}}.\label{eq:3ptgen}
\end{equation}
On the other hand, by implementing the multipole expansion, in \cref{sec:spin_in_qed}  
we have found that the same relation can be imposed for massive
amplitudes of arbitrary spin, and fixes their full form as
\begin{equation}
A_{3}^{h_{3},s}\sim(\epsilon_{3}\cdot p_{1})^{h_{3}}\varepsilon_{2}{\cdot}\exp\left(-\frac{k_{3\mu}\epsilon_{3\nu}}{\epsilon_{3}\cdot p_{1}}J_{s}^{\mu\nu}\right){\cdot}\varepsilon_{1},\label{eq:mass3pths}
\end{equation}
where $J_{s}^{\mu\nu}$ is the generator in e.g. (\ref{eq:gens12})
naturally adapted to higher spin $s$.\footnote{A local form of this amplitude can be found in \cite{Lorce:2009br,Lorce:2009bs,Bautista:2019tdr}, which however features $1/m$ divergences.}
Observe that this form does not depend explicitly on the mass and,
as noted in \cite{Guevara:2017csg}, reduces to (\ref{eq:massless3pths}) when
written in terms of the $D=4$ spinor helicity variables.\footnote{For a quick derivation of this fact write the polarization tensors
as $\varepsilon_{1}\propto\left(\frac{|1\rangle[\mu_{1}|}{[1\mu_{1}]}\right)^{h}$
and $\varepsilon_{2}\propto\left(\frac{|2]\langle\mu_{2}|}{\langle2\mu_{2}\rangle}\right)^{h}$, together with $\frac{k_{3\mu}\epsilon_{3\nu}}{\epsilon_{3}\cdot p_{1}}J^{\mu\nu}=\frac{\langle12\rangle}{\langle32\rangle}\langle3\frac{\partial}{\partial\lambda_{1}}\rangle$
as in e.g. \cite{Cachazo:2014fwa}. Then,
\begin{align*}
\varepsilon_{2}\cdot e^{\fb{-}\frac{\langle12\rangle}{\langle32\rangle}\langle3\frac{\partial}{\partial\lambda_{1}}\rangle}\cdot\varepsilon_{1} & =\langle\mu_{2}|e^{\frac{\langle21\rangle}{\langle32\rangle}\langle3\frac{\partial}{\partial\lambda_{1}}\rangle}|1\rangle^{h}\left(\frac{[\mu_{1}2]}{[1\mu_{1}]\langle\mu_{2}2\rangle}\right)^{h}\\
 & =\left(\langle\mu_{2}1\rangle-\frac{\langle12\rangle\langle\mu_{2}3\rangle}{\langle32\rangle}\right)^{h}\left(\frac{[\mu_{1}2]}{[1\mu_{1}]\langle\mu_{2}2\rangle}\right)^{h}=\left(\frac{\langle31\rangle}{\langle32\rangle}\right)^{2h}\,,
\end{align*}
where we have used that $e^{-\frac{\langle12\rangle}{\langle32\rangle}\langle3\frac{\partial}{\partial\lambda_{1}}\rangle}$
acts as a Lorentz boost on $|1\rangle$, see \cref{app_B}. Finally, the $h_3$ dependence is also the same in \eqref{eq:massless3pths} and \eqref{eq:mass3pths}.} Hence (\ref{eq:mass3pths}) is nothing but the natural
extension of (\ref{eq:massless3pths}) to generic dimension and helicities, whose
dimensional reduction in the sense of the previous section is trivial.
Curiously, when interpreted as a $D=4$ massless amplitude this object is
known to be inconsistent with locality for $|h|>1$ (or analogously
$s>1$) whereas in the massive case it has the physical interpretation
given in \cite{Guevara:2018wpp,Chung:2018kqs,Arkani-Hamed:2019ymq}. On the other hand, these
inconsistencies will only appear in the ``four-point test''  \cite{Arkani-Hamed:2017jhn,Benincasa:2007xk}, namely by computing $A_{4}^{{\rm QCD}}$ or $A_{4}^{\text{{\rm gr}}}$.
In the massive case they can be cured by including
contact interactions \cite{Chung:2018kqs}, as we will see in \cref{ch:GW_scattering}. In the same chapter, we will see how to take the classical limit of \eqref{eq:mass3pths} recovering which recovers the linearized metric for the Kerr \ac{BH}.

\subsubsection*{Arbitrary multiplicity at low spins}

From the above discussion we see that at least at low spins we can extend
the relation (\ref{eq:3ptgen}) and its compactification to arbitrary
multiplicity, since the massless theory is healthy. The starting \ac{QCD} theories for scalars, Dirac fermions and gluons are standard and catalogued in the next section. Let us then write
\begin{equation}
A_{n}^{h+\bar{h}}(\varphi_{1}^{h+\bar{h}}H_{3}\cdots H_{n}\varphi_{2}^{-h-\bar{h}}):=\frac{1}{2}\sum_{\alpha\beta}K_{\alpha\beta}A_{n,\alpha}^{{\rm QCD}}(\varphi_{1}^{h}g_3{\cdots}g_n\varphi_{2}^{-h})A_{n,\beta}^{{\rm QCD}}(\varphi_{1}^{\bar{h}}g_3{\cdots}g_n\varphi_{2}^{-\bar{h}}),\label{eq:def}
\end{equation}
where we have denoted by $\varphi_{i}^{h}$ the state of helicity
$h$ and particle label $i$. This extends the relation (\ref{eq:dconeline})
for the cases $h,\bar{h}\leq1$. We can also uplift it to arbitrary
dimensions. Following the previous section we first rewrite the amplitudes
in terms of the corresponding polarization vectors/spinors and the
implement the tensor products $\otimes$ between representations.
For simplicity of the argument we regard (\ref{eq:def}) as a \textit{definition}
of the object $A_{n}^{{\rm gr}}$, and we claim that it corresponds
to a tree-level amplitude in a certain \ac{QFT} coupled to gravity. We recall from the previous section that this
is because 1) diffeomorphism (gauge) invariance and crossing-symmetry
are manifest and 2) tree-level unitarity follows from general arguments \cite{Bern:2010yg,Chiodaroli:2017ngp}. This means that we just need to construct
a corresponding Lagrangian to identify the theory, which we will do
for most cases in Section \ref{constructing the lagrangians}.

We have already explained how under the dimensional reduction $D=d+1\to d$
we obtain massive momenta and the corresponding propagators. We have
also shown how the $D$-dimensional polarization vectors/spinors of
the compactified particles, $\varepsilon^{\mu}$ and $u^{\alpha}$,
can now be regarded as satisfying the corresponding massive wave equations.
The result of (\ref{eq:def}) after this procedure leads to the general
formula for one-massive line

\begin{equation}
\boxed{A_{n}^{s+\tilde{s}}(\varphi_{1}^{s+\tilde{s}}H_{3}\cdots H_{n}\varphi_{2}^{s+\tilde{s}}):=\frac{1}{2}\sum_{\alpha\beta}K_{\alpha\beta}A_{n,\alpha}^{{\rm QCD}}(\varphi_{1}^{s}g_3{\cdots}g_n\varphi_{2}^{s})\otimes A_{n,\beta}^{{\rm QCD}}(\varphi_{1}^{\tilde{s}}g_3{\cdots}g_n\varphi_{2}^{\tilde{s}}).}\label{eq:genklt}
\end{equation}
which holds for $s,\tilde{s}\leq1$ and has a smooth high-energy limit
by construction. Thus, this gives a double-copy formula for the minimally-coupled
partial amplitudes defined in the sense of \cite{Arkani-Hamed:2017jhn}. 

Even though
we have not yet specified the theory, let us momentarily restrict
the states $H_{i}$ to gravitons. We have explicitly checked, by inserting
massive spinor-helicity variables, that in $D=4$ we can obtain the
gravitational and \ac{QCD} amplitudes given in \cite{Arkani-Hamed:2017jhn} for $n=3,4$, see \eqref{eq:s14d} below. This establishes a $D=4$ double-copy formula between these amplitudes, analogous to the one studied in Appendix \ref{4 d double copy}. In general
dimensions, we have also checked that this agrees with the amplitudes and
double copy for $s=0,\tilde{s}\neq0$ pointed out in \cite{Bjerrum-Bohr:2013bxa}. We remark that these
are precisely the gravitational amplitudes used to obtain perturbative
black hole observables in \cite{Vaidya:2014kza,Guevara:2017csg,Guevara:2018wpp,Maybee:2019jus,Guevara:2019fsj}, and that for the all-graviton
case the \ac{LHS}   of (\ref{eq:genklt}) is unique given the sum $s+\tilde{s}$.

We now provide simple examples to illustrate these points. In the rest of this section we shall indistinctly use $\varepsilon_2$ or $\varepsilon^{*}_2$ to refer to the (conjugated) polarization of the outgoing massive state.

\subsubsection{Non-universality of Dilaton Couplings}\label{sec221}
As opposed to gravitons, we have anticipated that the dilaton field
couples differently in the $0\otimes 1$ than in the \12x12
case. So let us compute the amplitude $A_{3}(W_{1}\phi W_{2}^{*})$
via double copy of $s=0$ and $s=1$. This is to say, we take the
trace of

\begin{equation}
A_{3}^{{\rm 0\otimes1}}(W_{1}H^{\mu\nu}W_{2}^{*})=A_{3}^{{\rm QCD},s=0}(\varphi_{1}g^{\mu}\varphi_{2})A_{3}^{{\rm QCD},s=1}(W_{1}g^{\nu}W_{2}^{*})\label{eq:01dc-2}
\end{equation}
i.e. the \0x1 double copy, and contrast it with (\ref{eq:dila3pt})
from the \12x12 double copy. The spin-1 QCD
factor arising from dimensional reduction is equivalent to a covariantized Proca action plus a correction
on the gyromagnetic ratio $g$, see next section. Explicitly,

\begin{equation}\label{eq:prevqcd}
A_{3}^{{\rm QCD},s=1}(W_{1}g^{\mu}W_{2}^{*})= p_{1}^{\mu}\varepsilon_{1}\cdot\varepsilon_{2}- \varepsilon_{1}^{\alpha}(J^{\mu\nu})_{\alpha}^{\beta}\varepsilon_{2\beta} k_{3\nu}
\end{equation}
where we used that $(J^{\mu\nu})_{\beta}^{\alpha}=\eta^{\alpha[\mu}\delta_{\beta}^{\nu]}$ according to our conventions in \eqref{12123ptcopy}. Recalling that for spin-0 $A_3^{\mu}\propto p_1^{\mu }$ , the trace of \eqref{eq:01dc-2} gives 
\begin{equation}\label{eq:01dila}
{\color{black}A_{3}^{{\rm 0\otimes1}}(W_{1}\phi W_{2}^{*})=\frac{\kappa}{\sqrt{d-2}}\left(m^{2}\varepsilon_{1}{\cdot}\varepsilon_{2}+k_{3}{\cdot}\varepsilon_{1}\,k_{3}{\cdot}\varepsilon_{2}\right)},
\end{equation}
where we restored the coupling $\kappa$ in order to be more precise.
We now observe that this differs from (\ref{eq:dila3pt}) in a term proportional
to $\varepsilon_{1}{\cdot}k_{3}\,\varepsilon_{2}{\cdot}k_{3}$, controlled
by a coupling $\phi F^{2}$ with the matter field that we derive in the next section. At first this may look like a contradiction given that we pinpointed the massless origin of this double copy, namely eq. \eqref{eq:3ptgen}. Here $A_{3}^{{\rm }}(W_{1}\phi W_{2}^{*})$ should be uniquely fixed by little-group as happened for the graviton case \eqref{eq:mass3pths}. The difference however lies in the coupling constant, which vanishes in the $d{\to} 4\,, m{\to}0$ limit for  $A_{3}^{{\rm \frac{1}{2}\otimes\frac{1}{2}}}(W_{1}\phi W_{2}^{*})$ but not for $A_{3}^{{ 0\otimes 1}}(W_{1}\phi W_{2}^{*})$. Hence \textit{the reason why graviton amplitudes are the same in both \12x12 and \0x1  double-copies is not only because of its massless form \eqref{eq:massless3pths}, but also because the coupling $\kappa$ is fixed by the equivalence principle.}

A final and crucial remark is as follows. From general considerations it is known that the dilaton cannot couple linearly to the spin of a matter line \cite{Goldberger:2017ogt,Li:2018qap}. This is consistent, as we will see that \eqref{eq:01dila} contains only a quadrupole $\sim J^2$ term, but appears in contradiction with the fact that $A^{s=1}$ in \eqref{eq:prevqcd}, which carries the spin dependence, seems to have a dipole and no quadrupole. The resolution of this puzzle comes from distinguishing two types of multipoles. The first type are the covariant multipoles carrying the action of the full Lorentz group $\text{SO}(d-1,1)$, as generated by $J^{\mu\nu}$. The second type are the rotation multipoles defined by the condition $p_\mu S^{\mu\nu}=0$ with respect to e.g. the average momentum $p=\frac{p_1 + p_2}{2}$. They generate the $\text{SO}(d-1)$ rotation subgroup and in the classical limit represent the classical spin-tensor of compact objects. The relation between the two multipoles is the decomposition $\text{SO}(d-1,1)\to \text{SO}(d-1)$ explained in \cref{sec:branching}, such that one can write $J_{\mu \nu}=S_{\mu\nu}+\textrm{boost terms}.$ Using this, \eqref{eq:prevqcd} can be written as

\begin{equation}
A_{3}^{{\rm QCD},s=1,\mu}= p^{\mu}\left(1+\frac{k_{3\mu}S^{\mu \alpha}S_{\alpha}^{\,\,\nu}k_{3\nu} }{m^2 (d-3)}\right)- S^{\mu\nu} k_{3\nu}\,,
\end{equation}
where the quadrupole term $S^{\mu \alpha}S_{\alpha}^{\,\,\nu}$ is obtained precisely from the boost piece and we have stripped polarization states.\footnote{Here the massive polarization vectors have been removed and the quantum amplitude is  understood to be  an operator acting on them. On the other hand, in the classical context, $S^{\mu\nu}$ is interpreted as a spin tensor (c-number) describing the intrinsic rotation of the classical object. } The double copy now gives
\begin{align}
A_{3}^{{\rm 0\otimes1}}(W_{1}\phi W_{2}^{*}) =& \frac{\kappa}{\sqrt{d-2}}p_\mu \left[p^{\mu}\left(1+\frac{k_{3\mu}S^{\mu \alpha}S_{\alpha}^{\,\,\nu}k_{3\nu} }{m^2 (d-3)}\right)- S^{\mu\nu} k_{3\nu}\right] \nonumber \\
 =& \frac{\kappa\,m^2}{\sqrt{d-2}}\left(1+\frac{k_{3\mu}S^{\mu \alpha}S_{\alpha}^{\,\,\nu}k_{3\nu} }{m^2 (d-3)}\right)\,.
\end{align}
Comparing this to our previous result, it is clear that the term $k_3{\cdot} \varepsilon_1 k_3 {\cdot} \varepsilon_2$ in \eqref{eq:01dila} is in direct correspondence with the quadrupole operator. A similar argument holds for the \12x12 theory: In this case there is genuinely no quadrupole contribution in the \ac{QCD} factor,

\begin{equation}
A_{3}^{{\rm QCD},s=\frac{1}{2},\mu}= p^{\mu}- S^{\mu\nu} k_{3\nu}\,,
\end{equation}
whereas in the double copy $A^{\frac{1}{2}\otimes \frac{1}{2}}$ the linear-in-spin terms again cancel due to $p_\mu S^{\mu \nu}=0$. We are left again with a quadrupole term $\sim S^2$, as can be also seen from \eqref{eq:dila3pt}. We conclude that the \12x12 and \0x1 theories differ in the dilaton coupling only at the level of the matter quadrupole. We come back to this point during  \cref{GGT} in the context of classical double copy.

\subsubsection{Compton Amplitude and the $g$ factor}

Moving on to $n=4$, we can explore the interplay between the double copy and the multipole expansion. Let us first quote here the spin-1 \ac{QCD} result for general
gyromagnetic factor $g$ computed  by Holstein in \cite{Holstein:2006pq}

\begin{equation}
\begin{aligned}A_{4}^{{\rm QCD},s=1}(1324)= & \frac{1}{4}\bigg\{-2\varepsilon_{1}\cdot\varepsilon_{2}\left[\frac{\epsilon_{3}\cdot p_{1}\epsilon_{4}\cdot p_{2}}{p_{1}\cdot k_{3}}+\frac{\epsilon_{3}\cdot p_{2}\epsilon_{4}\cdot p_{1}}{p_{1}\cdot k_{4}}+\epsilon_{3}\cdot\epsilon_{4}\right]\\
 & -g\left[\varepsilon_{1}\cdot F_{4}\cdot\varepsilon_{2}\left(\frac{\epsilon_{3}\cdot p_{1}}{p_{1}\cdot k_{3}}-\frac{\epsilon_{3}\cdot p_{2}}{p_{1}\cdot k_{4}}\right)+\varepsilon_{1}\cdot F_{3}\cdot\varepsilon_{2}\left(\frac{\epsilon_{4}\cdot p_{2}}{p_{1}\cdot k_{3}}-\frac{\epsilon_{4}\cdot p_{1}}{p_{1}\cdot k_{4}}\right)\right]\\
 & +g^{2}\left[\frac{1}{2p_{1}\cdot k_{3}}\varepsilon_{1}\cdot F_{3}\cdot F_{4}\cdot\varepsilon_{2}-\frac{1}{2p_{1}\cdot k_{4}}\varepsilon_{1}\cdot F_{4}\cdot F_{3}\cdot\varepsilon_{2}\right]\\
 & -\frac{(g-2)^{2}}{m^{2}}\bigg[\frac{1}{2p_{1}\cdot k_{3}}\varepsilon_{1}\cdot F_{3}\cdot p_{1}\varepsilon_{2}\cdot F_{4}\cdot p_{2}\\
 & -\frac{1}{2p_{1}\cdot k_{4}}\varepsilon_{1}\cdot F_{4}\cdot p_{1}\varepsilon_{2}\cdot F_{3}\cdot p_{1}\bigg]\,\bigg\} ,
\end{aligned}
\label{eq:hlqcd}
\end{equation}
where $F_{i}^{\mu\nu}=2k_{i}^{[\mu}\epsilon_{i}^{\nu]}$. Here all momenta are outgoing and satisfy the on-shell conditions $p_{1}^{2}=p_{2}^{2}=m^{2}$
and $k_{3}^{2}=k_{4}^{2}=0$. The covariantized Proca theory is obtained
by setting $g=1$ and hence contains a $1/m$ divergence. On the other
hand, if the Proca field is identified with a $W^{\pm}$ boson of
the electroweak model we obtain $g=2$ and completely cancel the $1/m$ term. This is a general feature of the $g=2$ theory at any multiplicity \cite{PhysRevD.46.3529}.  Moreover, in this case we observe not only a well behaved high
energy limit, but also not apparent dependence on $m$ at all! This
means that the amplitude is essentially equal to its massless limit,
which corresponds to a $n=4$ color-ordered gluon amplitude. This is essentially the single copy amplitude we refereed to in the discussion around  \eqref{eq:comptonspin2} above.

From the above we find that for this amplitude setting $g=2$ will automatically yield to the double copy relation \eqref{eq:genklt}. This is the underlying reason for the result found in \cite{Holstein:2006pq,Holstein:2006wi} \fb{for the natural value of $g$}. The converse is also true as gravitational amplitudes always have $g=2$, thus imposing the same value on its \ac{QCD} factors. The universality of $g$ is a feature of the gravitational Lagrangians, independently of the covariantization or the couplings considered. It was checked explicitly in  \cite{Chung:2018kqs} and is a direct
consequence of the universal subleading soft theorem in gravity \cite{Bautista:2019tdr}.
This contrasts to \ac{QCD} in that only the leading soft factor is universal
there and hence $g$ becomes a parameter. Finally, it can also be understood
from the fact that both rotating black hole or neutron stars also
yield $g=2$ indistinctly in classical \ac{GR}  \cite{Pfister_2002}.

Let us elaborate on the relation between \eqref{eq:hlqcd} and the 4-gluon amplitude. Pretend
that (\ref{eq:hlqcd}) (with $g=2$) is indeed the massless amplitude.
As we compactify we must send $p_{i}\to P_{i}=(p_{i},\pm m)$ and
$k_{i}\to(k_{i},0)$, while setting the polarizations $\varepsilon_{i},\epsilon_{i}$
to lie also in $D-1$ dimensions. As the amplitude itself only depends
on $p_{i}$ through $P_{i}\cdot k_{j}$ and $P_{i}\cdot\epsilon_{j}$
the extra dimensional component of $P_{i}$ drops and the mass $m$
simply does not appear. More generally, the reader can convince themselves that the only appearances of $m$ are through 1) $P_{1}\cdot P_{2}=p_{1}\cdot p_{2}+m^{2}$
or 2) $P_{1}\cdot\epsilon_{i}P_{2}\cdot\epsilon_{j}$, which we have
seen lead to $p_{1}\cdot p_{2}$ after dilaton projection. In the
first case we can use momentum conservation to write $P_{1}\cdot P_{2}=\sum_{i<j}k_{i}\cdot k_{j}$
and effectively cancel the mass dependence. Hence, if we choose a basis
of kinematic invariants that excludes $P_{1}\cdot P_{2}$ the compactification
will be trivial: The amplitudes $A_{n}$ will essentially be identical
to their massless limit \textit{except} in the cases of dilaton amplitudes,
since they contain terms like $p_{1}\cdot p_{2}=-m^{2}+\sum_{i<j}k_{i}\cdot k_{j}$.
The same observation applies to the \ac{KLT} construction (\ref{eq:genklt})
and the \ac{KLT} kernel introduced in the previous section. We will extend
these observations to more matter lines in  \cref{two matter lines}.

Note also that the explicit mass dependence can as well be hidden by means of $d=4$ massive spinor-helicity variables.\footnote{See  \cite{Arkani-Hamed:2017jhn}
for the details on this formalism and \cite{Guevara:2018wpp,Bautista:2019tdr}  for a construction
of these amplitudes via soft factors.} For instance, using these variables eq. (\ref{eq:hlqcd})
with $g=2$ reads

\begin{equation}\label{eq:s14dqcd}
A_{4}^{{\rm QCD},s=1}(1324)\propto \frac{\langle 3|1|4]^2}{p_1 \cdot k_3 \, p_1 \cdot k_4 } \left([1^a 3]\langle 42^b\rangle + \langle 1^a4 \rangle [2^b3]\right)^2
\end{equation}
In this form the double copy can be performed as in  \cref{4 d double copy}. For instance, from two copies of the previous spin-1 amplitude we obtain the following spin-2 amplitude:

\begin{equation}\label{eq:s14d}
A_{4}^{{\rm gr},s=2}\propto  \frac{\langle 3|1|4]^4}{p_1 \cdot k_3 \, p_1 \cdot k_4 \, k_3 \cdot k_4} \left([1^a 3]\langle 42^b\rangle + \langle 1^a 4 \rangle [2^b 3]\right)^4
\end{equation}
This result has been used to construct observables associated to the Kerr \ac{BH} in \cite{Guevara:2018wpp,Chung:2018kqs}, in fact, as reviewed in \cref{sec:spinor-helicity}, the Compton amplitude can be written in an exponential form, we will use such formula in \cref{ch:GW_scattering} to show it matches the classical solutions of the Teukolsky equitation for the scattering of gravitational waves off the Kerr \ac{BH}. Here we can conclude that such amplitude is nothing but the 4-graviton amplitude in higher dimensions.  Again, since there are no massless higher spin particles in flat space, this framework provides a natural explanation for the fact that $A_{4}^{{\rm gr},s>2}$ and $A_{4}^{{\rm QCD},s>1}$ must contain $\frac{1}{m}$ divergences. 

\section{Constructing the Lagrangians} \label{constructing the lagrangians}

In this section we will provide the Lagrangians associated to the previous constructions, covering all the \ac{QCD} theories and mainly focusing on the
\12x12 and \0x1 gravitational cases. This will allow us to gain further insight in the
corresponding amplitudes. On the \ac{QCD} side we will employ the compactification method to obtain the actions. On the gravity side we will construct them from simple considerations in the string frame, including classical regime.
For two matter lines some of these Lagrangians acquire contact terms
which we further study in   \cref{two matter lines}.

\subsection{QCD Theories}\label{sec:qcd-theories}

We start by considering the \ac{QCD} factors associated to the double copy.
The cases of spin-0 and spin-$\frac{1}{2}$ are standard and we can
provide the Lagrangian for more than one matter line straight away.
The case of the \ac{QCD} theory of spin-1 \cite{Holstein:2006wi,Holstein:2006pq} is more interesting
and will be treated in a separate subsection.

\subsubsection{Spins \texorpdfstring{ $s=0,\frac{1}{2}$}{s012}}\label{sec:qcdspin0}

We have explained in the previous section how the scalar theory coupled
to \ac{QCD} arises from a particular compactification both in momenta and
polarization vectors. The compactification in polarization vectors
is obtained by considering a pure gluon amplitude and setting $\varepsilon_{i}=(0,\ldots0|1)$
where the non-zero component explores an ``internal space''. We
can immediately ask what happens if the internal space is enlarged
to $N$ slots, namely the scalars are obtained by setting

\begin{equation}
\varepsilon_{i}=(\underbrace{0,\ldots,0}_{D}|\underbrace{0,\ldots1,\ldots,0}_{N})\,.\label{eq:compeps}
\end{equation}
This construction is well known from string theory and the resulting
amplitudes correspond to $N$ scalars in \ac{QCD}.   
In other words, letting $I,J=1,\ldots,N$ the resulting amplitudes
for any number of scalar lines are given by the aforementioned ``special''
Yang-Mills scalar theory:

\begin{equation}
\mathcal{L}_{D}^{s=0}=-\frac{1}{4}{\rm tr}(F_{\mu\nu}F^{\mu\nu})+\frac{1}{2}{\rm tr}(D_{\mu}\varphi_{I}D^{\mu}\varphi^{I})\textcolor{black}{-}\frac{1}{4}{\rm tr}([\varphi^{I},\varphi^{J}][\varphi_{I},\varphi_{J}]).\label{eq:syms}
\end{equation}
The proof of this compactification is very simple and illustrative so we briefly outline it here. It follows from
decomposing the gluon polarization in $D+N$ dimensions as

\begin{equation}\label{eq:dim_red_YM}
A_{\mu}\to(A_{\mu}|\varphi_{1},\ldots,\varphi_{N}),
\end{equation}
which implies 
\begin{equation}
F_{\mu I}=D_{\mu}\varphi_{I}\,,\quad F_{IJ}=[\varphi_{I},\varphi_{J}],
\end{equation}
together with the $D$ dimensional $F_{\mu\nu}$ components. Then,
the resulting Lagrangian \eqref{eq:syms}, just follows from expanding ${\rm tr}(F^2)$. 
Note that the fields only depend on $D$ coordinates (see e.g. \cite{Cheung:2017yef}). Two key remarks which will be useful later are as follows: First, the extra dimensional (scalar) modes are always pair-produced and hence will assemble into matter lines in the Feynman diagrams. In particular this means that even after dimensional reduction the pure gluon amplitudes coincide with the ones of \ac{YM} theory.
Second, as already pointed out in the original construction \cite{Brink:1976bc} of the compactified Yang-Mills action, the action (\ref{eq:syms}) indeed corresponds to
the bosonic sector of $\mathcal{N}=4$ Super Yang-Mills theory (in that case $D=4$ and $N=6$). 

Let us now provide masses to the scalars in the Lagrangian (\ref{eq:syms}). This requires to consider complex fields as is standard in \ac{KK} reductions. There are a number of ways to achieve this. For instance, still following \cite{Brink:1976bc}, we can consider an even number of compact dimensions $N$ after which the scalars can be grouped as $\psi=\varphi_I+i\varphi_{I+1}$. 

Here we will instead take an alternative route that connects more directly to our previous amplitudes discussion, and therefore extends to particles with spin. Recall that so far we have constructed the double-copy formula for a single matter line \eqref{eq:genklt}. We can also consider scattering amplitudes for more matter lines as long as they have different flavors, a restriction that we impose throughout this chapter. Now, for a given flavor $I$, the Lagrangian \eqref{eq:syms} takes the form $\mathcal{L_D}\supset  \textcolor{black}{\frac{1}{2}} \varphi_I \mathcal{D} \varphi^I$ (without summation) where $\mathcal{D}$ is a Hermitian operator that can depend on other fields. This Lagrangian generates the same Feynman rules than $\varphi_I^* \mathcal{D} \varphi^I$, which is the previous statement that the scalar fields are pair-produced. Repeating the argument for $I,J=1,\ldots,N$, we conclude that we can replace

\begin{equation}
\mathcal{L}_{D}^{s=0} \to -\frac{1}{4}{\rm tr}(F_{\mu\nu}F^{\mu\nu})+{\rm tr}(D_{\mu}\varphi^*_{I}D^{\mu}\varphi^{I})\textcolor{black}{-}{\rm tr}([\varphi^{*I},\varphi^{*J}][\varphi_{I},\varphi_{J}]).\label{eq:symsc}
\end{equation}
carrying a $U(1)^{N}$ flavour. After providing  masses to the complex fields, they can be turned into real fields again via the same argument. We will use this procedure in the remaining compactifications presented in this chapter.  

We now proceed then  via \ac{KK} reduction on a torus, $M_{D}=\mathbb{R}^{d}\times T^{N}$,
and we let each of $N$  scalars to have a non-zero momentum in one
of the circles $S^{1}$,
\begin{equation}
\varphi_{I}(x,\theta)=e^{im_{I}\theta_{I}}\varphi_{I}(x),
\end{equation}
where $0<\theta_{I}\leq\frac{2\pi}{m_{I}}$. The gluon field has no
momenta on $T^{N}$, i.e.  is $\theta$-independent, and its only non-zero
components are $A_{\mu}(x)$, where now $\mu=0,\ldots,d-1$. By acting
with the derivative $\partial_{\bar{\mu}}$, where $\bar{\mu}=0,\ldots,D-1=d+N-1$ ,
we can read off the momentum of the flavour $\varphi_{I}$:

\begin{equation}
p_{i\bar{\mu}}^{(I)}=(\underbrace{p_{i\mu}}_{d}|\underbrace{0,\ldots,m_{I},\ldots,0}_{N}).\label{eq:masscom}
\end{equation}
Thus the on-shell condition becomes $(p_{i}^{I})^{2}=p_{i}^{2}-m_{I}^{2}=0$
and, for $N=1$, this procedure is equivalent to the one described
in the previous section. It generalizes it to more massive lines by
imposing that the momenta of scalars of different flavour are orthogonal
in the \ac{KK} directions, i.e. $p_{i}^{(I)}\cdot p_{j}^{(J)}=p_{i}\cdot p_{j}$
for $I\neq J$. By integration on $T^{N}$ we find the corresponding
massive action:

\begin{equation}\label{scalarqcd}
 \int d^{d}xd^{N}\theta\mathcal{L}_{D}^{s=0}\propto\int d^{d}x{\rm tr}(-\frac{1}{4}F_{\mu\nu}F^{\mu\nu}+\fb{\frac{1}{2}}D_{\mu}\varphi_{I}D^{\mu}\varphi^{I}+\fb{\frac{1}{2}}m_{I}^{2}\varphi_{I}\varphi^{I}\textcolor{black}{-}\fb{\frac{1}{4}}[\varphi^{I},\varphi^{J}][\varphi_{I},\varphi_{J}]),   
\end{equation}
which corresponds to a scalar \ac{QCD} theory, with a sum over flavours $I$ implicit. Here the scalars inherit the adjoint representation from the higher-dimensional gluons. For one matter line we can nevertheless take them in the fundamental representation (see sec. \ref{sec:s1W} below) and also drop the quartic term from the Lagrangian: The double copy of the resulting theory has been studied in \cite{Luna:2017dtq} and we will come back to it in  \cref{two matter lines}. On the other hand, by keeping the last term we have a non-trivial contact interaction between flavours. In the massless case the double copy of this theory with itself corresponds to Einstein-YM as first observed in \cite{Chiodaroli:2014xia}. In our case we will be interested in the double copy of \eqref{scalarqcd} with the spin-1 theory constructed in the next subsection, leading to the $0\otimes 1$ gravitational theory. In the classical regime we also anticipate that this distinction is irrelevant and both cases can be regarded as equivalent.
\\

Finally, we note that we can also apply the reduction procedure to massless \ac{QCD} in order to get the massive theory, as discussed previously from the amplitudes perspective. Using the splitting \eqref{sigmad} we obtain, after dropping some irrelevant \ac{KK} modes,

\begin{equation}\label{qcdfermions}
  \int d^{d}xd^{N}\theta\mathcal{L}_{D}^{s=\frac{1}{2}}\propto\int d^{d}x{\rm tr}\left(-\frac{1}{4}F_{\mu\nu}F^{\mu\nu}+i\bar{\psi}_{I}\Gamma_{\mu}D^{\mu}\psi^{I}-m \bar{\psi}_{I}\psi^{I}\right).
\end{equation}
In $d=4$ and for a single fermion line, we note that this reproduces
the fermion amplitudes of $\mathcal{N}=4$ \ac{SYM} in the Coulomb branch. 

\subsubsection{Spin  \texorpdfstring{$s=1$}{s1}}\label{sec:s1W}

We now consider in detail the case of spin-1, that is, a complex Proca field
coupled to \ac{QCD}.  In order to motivate this theory we will reproduce
here the argument given by Holstein in \cite{Holstein:2006pq} regarding the natural value of $g$,  which we  used  in
\cref{sec:spin_in_qed} to derive  the three-point amplitude for spinning particles in \ac{QED}, but here we consider a slightly more general setup by promoting \ac{QED} to \ac{QCD} 
amplitudes.

Consider first the (complex) Proca theory minimally coupled to $SU(N)$ Yang-Mills
theory,
\begin{equation}
\mathcal{L}=-\frac{1}{4}F_{\mu\nu}^{a}F_{a}^{\mu\nu}-\frac{1}{4}W_{\mu\nu}^{\bar{I}}W_{I}^{\mu\nu}+\frac{m^{2}}{2}W_{\bar{I}}^{\mu}W_{\mu}^{I},\label{eq:wym}
\end{equation}
where we have distinguished color indices $I,\bar{I}$ to emphasize
that ($W^{\bar{I}}$) $W^{I}$ transforms in the (anti)fundamental
representation. This is just a formal feature since for now we will
only consider one matter line (note also that the mass does not depend
on $I$). Here
\begin{equation}
\begin{split}
W_{\mu\nu}^{I} & = D_{\mu}W_{\nu}^{I}-D_{\nu}W_{\mu}^{I}\label{eq:defwuv},\\
D_{\mu}W_{\nu}^{I} & =  \partial_{\mu}W_{\nu}^{I}+A_{\mu}^{a}T_{a}^{I\bar{J}}W_{\nu\bar{J}}.
\end{split}
\end{equation}
Now consider the three point amplitude obtained from (\ref{eq:wym}),
\begin{equation}
A_{3}^{{\rm QCD,1}}(W_{1}^{I}A_{3}^{a}W_{2}^{\bar{J}}) =  2T^{aI\bar{J}}\times(\epsilon_{3}\cdot p_{1}\varepsilon_{1}\cdot\varepsilon_{2}^{*}-\,\epsilon_{3\mu}k_{3\nu}\varepsilon_{1}^{[\mu}\varepsilon_{2}^{*\nu]})\,,\label{eq:3ptqcd1}
\end{equation}
which is equivalent to \eqref{eq:A3proca} with the additional colour generators from the non-abelian structure of \ac{QCD}.   
By recalling the example of \eqref{eq:A3_multipole} we can easily identify
the scalar and dipole pieces in these two terms. Note that $\varepsilon_{1}\cdot J^{\mu\nu}\cdot\varepsilon_{2}^{*}=2 \varepsilon_{1}^{[\mu}\varepsilon_{2}^{*\nu]}$
and hence we obtain $g=1$. This is consistent with the value of $g=\frac{1}{s}$ obtained for minimally covariantized Lagrangians as conjectured by Belinfante \cite{Belinfante1953}. We
then proceed to modify the value of $g$ by adding the interaction
\begin{equation}
\mathcal{L}_{int}=\beta\,F_{\mu\nu}^{a}T_{a}^{I\bar{J}}W_{I}^{\mu}W_{\bar{J}}^{\nu}.\label{eq:lint}
\end{equation}
This interaction was studied in e.g. \cite{Holstein:2006pq} restricted to the context of QED. In such case we can take $T_{a}^{I\bar{J}}\to\delta^{+-}$ and $\mathcal{L}_{int}$ arises
from the spontaneous symmetry breaking in the $W^{\pm}$-boson model (with
$\beta=1$). In our case we need to promote this to \ac{QCD} so that we
can perform the double copy at higher multiplicity. In any case, this
term precisely deforms the value of the dipole interaction to $g=1+\beta$,
because
\begin{equation}
\mathcal{L}_{int}\to -2\beta\,T_{a}^{I\bar{J}}\times\epsilon_{3\mu}k_{3\nu}\varepsilon_{1}^{[\mu}\varepsilon_{2}^{*\nu]}\,.\label{eq:bint}
\end{equation}
Now, we claim that in order for $A_{3}^{{\rm QCD}}$ to be consistent
with the double copy for the graviton states we will need to set $g=2$,
i.e. $\beta=1$ as in the electroweak model. This is because only
in such case we find\footnote{This is a slight simplification of the argument, which is what we
used in \cite{Bautista:2019tdr} at $n=3$, arbitrary spin. Actually, Holstein \cite{Holstein:2006wi} studied the double copy of $A_{4}^{{\rm QED}}$ with the purpose
of showing the $1/m$ cancellations which are equivalent to $g=2$
as we saw in \eqref{eq:hlqcd}. Of course, the amplitude $A_{4}^{{\rm gr}}$
did not feature any such divergences.}
\begin{eqnarray}
A_{3}^{{\rm QCD},0}\times A_{3}^{{\rm QCD},1} & \sim & A_{3}^{{\rm gr},1}(W_{1}h_{3}W_{2}), \nonumber \\
 & \sim & \epsilon_{3}\cdot p_{1}\times(\epsilon_{3}\cdot p_{1}\varepsilon_{1}\cdot\varepsilon_{2}^{*}-2\,\epsilon_{3\mu}k_{3\nu}\varepsilon_{1}^{[\mu}\varepsilon_{2}^{*\nu]})\label{eq:013pts}
\end{eqnarray}
Here we have stripped the coupling constants to make the comparison
direct and written the graviton polarization as $\epsilon_{3}^{\mu\nu}=\epsilon_{3}^{\mu}\epsilon_{3}^{\nu}$
for simplicity, which can then be promoted to a general polarization
$\epsilon_{3}^{\mu\nu}$. The fixing of $g=2$ follows then from the
fact that gravitational amplitudes for any spin will always lead to
$g=2$ as we outlined in the Compton example in \eqref{eq:hlqcd}.

The fact that the double copy is satisfied for the $W$-boson model
but not for the ``minimally coupled'' Proca action is not a coincidence.
As we have explained, the concept of minimal coupling that we attain
here does not necessarily agree with the covariantization of derivatives
in (\ref{eq:wym}). Our condition for minimal coupling, and that of
\cite{Arkani-Hamed:2017jhn}, is that the $m\to0$ limit of $A_{n}^{{\rm QCD}}$ is
well defined at any multiplicity $n$. The $W$-boson model arises
from spontaneous symmetry breaking in $SU(2)_{L}\times U(1)_{Y}$
gauge theory, and as such, will be deformed back to Yang-Mills as
we take $m\to0$. This will precisely fix $\beta=1$ in (\ref{eq:bint})
and we now show how.

From a Feynman diagram perspective, we have already explained how
the \ac{QCD} amplitudes we are after can be obtained from massive compactification
of \ac{YM} amplitudes. In the case of spin-1 and a single matter line,
we interpret the cubic Feynman diagrams of $A_{n}^{{\rm YM}}$ as
associated to a color factor made of fundamental and adjoint structure
constants, following  \cite{Johansson:2014zca}. As an example, for partial amplitudes
in the half ladder (DDM) basis, we will consider the color factor
associated to the ordering $\alpha=(1\beta_{1}\ldots\beta_{n-2}2)$
as
\begin{equation}
f^{a_{1}a_{\beta_{1}}b_{1}}f^{b_{1}a_{\beta_{2}}b_{2}}\ldots f^{b_{n-3}a_{\beta_{n-2}}a_{2}}\to T_{a_{\beta_{1}}}^{I_{1}\bar{J}_{1}}T_{a_{\beta_{2}}}^{J_{1}\bar{J}_{2}}\ldots T_{a_{\beta_{n-2}}}^{J_{n-3}\bar{I}_{2}},
\end{equation}
where particles in $\{\beta_{1},\ldots,\beta_{n}\}$ are gluons and
particles 1 and 2 are bosons $W^{I_{1}},W^{\bar{I}_{2}}$ respectively.
The same operation can be repeated in any cubic color numerator of
YM, which in general means to replace $f^{abc}\to T_{a}^{I\bar{J}}$
for matter vertices or just leave them as $f^{abc}$ for the 3-gluon
vertices. This means we identify three types of color indices: $A=(a,I,\bar{I})$.\footnote{Formally we take $T_{a}^{I\bar{J}}=-T_{a}^{\bar{J}I}$ as in \cite{Johansson:2014zca}.
One must also be careful in that the structure constants $\{T_{a}^{I\bar{J}},f^{abc}\,\}$
do not form a Lie algebra (except in the $SU(2)$ case) and hence cannot be used as an input to construct a pure \ac{YM} action. However, the inconsistency
appears in the Jacobi relation $T_{a}^{I\bar{J}}T_{a}^{K\bar{L}}+\ldots$
which is associated to two matter lines, which we are not interested
here: We drop such interactions in our resulting
Lagrangian.} After relabelling the structure constants and the fields accordingly,
the field strength $\mathcal{F}_{\mu\nu}^{A}$ can be split into the
components

\begin{equation}
\mathcal{F}_{\mu\nu}^{a}=F_{\mu\nu}^{a}+2\,T_{I\bar{J}}^{a}W_{[\mu}^{I}W_{\nu]}^{\bar{J}}\,,\quad\mathcal{F}_{\mu\nu}^{I}=W_{\mu\nu}^{I}\,,\quad\mathcal{F}_{\mu\nu}^{\bar{I}}=W_{\mu\nu}^{\bar{I}},
\end{equation}
where $W_{\mu\nu}$ is defined in (\ref{eq:defwuv}). Now consider
the \ac{YM} action after relabelling
\begin{equation}
\frac{1}{4}\mathcal{F}_{\mu\nu}^{A}\mathcal{F}_{A}^{\mu\nu}=\frac{1}{4}F_{\mu\nu}^{a}F_{a}^{\mu\nu}+\frac{1}{4}W_{\mu\nu}^{\bar{I}}W_{I}^{\mu\nu}+F_{a}^{\mu\nu}T_{I\bar{J}}^{a}W_{\mu}^{I}W_{\nu}^{\bar{J}}+\ldots,
\end{equation}
where we have dropped the term with four $W$-bosons. Repeating the
compactification procedure, this time on a single circle $S^{1}$,
gives
\begin{equation}\label{qcdW-boson}
\mathcal{L}^{s=1}=-\frac{1}{4}F_{\mu\nu}^{a}F_{a}^{\mu\nu}-\frac{1}{4}W_{\mu\nu}^{\bar{I}}W_{I}^{\mu\nu}+\frac{m^{2}}{2}W_{\bar{I}}^{\mu}W_{\mu}^{I}-F_{a}^{\mu\nu}T_{I\bar{J}}^{a}W_{\mu}^{I}W_{\nu}^{\bar{J}},
\end{equation}
which is indeed the deformation of (\ref{eq:wym}) by the ``spin-dipole''
coupling (\ref{eq:lint}). Thus, we have shown that the massive spin-1
theory yielding the $g=2$ interaction when coupled to \ac{QCD} is precisely
the compactification of Yang-Mills theory for a single matter line,
as described in Section \ref{sec:sec2}.

\subsection{Proposal for Gravitational Theories}\label{sec:grav_theories}

Let us now introduce the gravitational Lagrangians. We begin by a construction of both $0\otimes 1$ and \12x12 theories in the string
frame, following some simple guidelines. First, let us assume momentarily
that the base massless theory, leading to the amplitudes $A_{n}^{{\rm gr}}(\gamma^{-}h_{3}{\cdots}h_{n}\gamma^{+})$
is indeed Einstein-Maxwell in both \12x12 and
\0x1 cases,
\begin{equation}
\mathcal{L}_{{\rm base}}=-\sqrt{g}\left[\frac{2}{\kappa^{2}}R+\frac{1}{2}F_{\mu\nu}^{*}F^{\mu\nu}\right].
\end{equation}
This allow us to signal the crucial difference between the \12x12
and \0x1 theories in the dilaton coupling. Following \cite{9780511816079},
in the string frame this can be generated by adding the kinetic term
and promoting $\sqrt{g}\to\sqrt{g}e^{-\frac{\kappa}{2}\phi}$. Thus
we propose
\begin{eqnarray}
\mathcal{L}_{{\rm base}}^{0\otimes1} & = & \sqrt{g}e^{-\frac{\kappa}{2}\phi}\left[-\frac{2}{\kappa^{2}}R+\frac{1}{2}(\partial\phi)^{2}-\frac{1}{2}F_{\mu\nu}^{*}F^{\mu\nu}\right],\\
\mathcal{L}_{{\rm base}}^{\frac{1}{2}\otimes\frac{1}{2}} & = & \sqrt{g}e^{-\frac{\kappa}{2}\phi}\left[-\frac{2}{\kappa^{2}}R+\frac{1}{2}(\partial\phi)^{2}\right]-\sqrt{g}\times\frac{1}{2}F_{\mu\nu}^{*}F^{\mu\nu}.
\end{eqnarray}
We now see the difference lies in the fact that the Maxwell term has
been added before and after incorporating the dilaton, respectively.
The coupling of the dilaton is simpler and in a sense trivial in the
\12x12 theory, which is characteristic of the
Brans-Dicke-Maxwell action \cite{Frau:1997mq}. In fact, we can take such
theory into the so-called Jordan frame by setting
\begin{equation}
\phi=-\frac{2}{\kappa}\ln\Phi,
\end{equation}
which leads to the standard Brans-Dicke theory \cite{Sheykhi:2008tt}
\begin{equation}
\mathcal{L}_{{\rm base}}^{\frac{1}{2}\otimes\frac{1}{2}}=\frac{2}{\kappa^{2}}\sqrt{g}\left[-\Phi R+\frac{(\partial\Phi)^{2}}{\Phi}-\frac{\kappa^{2}}{2}\times\frac{1}{2}F_{\mu\nu}^{*}F^{\mu\nu}\right]\,.
\end{equation}
On the other hand, our proposal that the \0x1 action involves
a non-trivial coupling to the dilaton arises from a careful consideration
of the classical results of \cite{Goldberger:2017ogt}, which construction we
further realize in  \cref{two matter lines} as a double copy of a spinning source (e.g.
$s=1$) in \ac{QCD} with a scalar theory ($s=0$).

At this point we can generate a mass term by performing the compactification
on a circle, $M_{D}=\mathbb{R}^{d}\times S^{1}$, letting the Proca
field to have a non-zero (quantized) momentum on $S^{1}$
\begin{equation}
A_{\mu}(x,\theta)=e^{im\theta}A_{\mu}(x)\,,
\end{equation}
whereas the remaining fields have not, i.e. $h_{\mu\nu}(x)$ and $\phi(x)$
are $\theta$-independent. Notice we have also implicitly restricted
the polarizations to lie in $d=D-1$ dimensions. For instance, the
full metric reads
\begin{equation}
g_{\bar{\mu}\bar{\nu}}=\eta_{\bar{\mu}\bar{\nu}}+\frac{\kappa}{2}h_{\bar{\mu}\bar{\nu}}\,,
\end{equation}
but $h_{\bar{\mu}\bar{\nu}}$ only has non-zero components $h_{\mu\nu}$. This relies on the assumption, exemplified in section \ref{sec:qcdspin0}, that additional \ac{KK} components will assemble into matter lines and hence can be decoupled. The only exception is the dilaton field, which would in principle obtain a contribution from the extra component $h_{DD}$ in $h_{\bar{\mu}\bar{\nu}}$. The reason we set this component to zero beforehand is precisely to reproduce our prescription \eqref{eq:pr2} as opposed to \eqref{eq:pr1} (which would lead to the standard dimensional reduction of the dilaton amplitudes).

After this clarification we can now readily perform the integration of the action
over the compact direction, leading to
\begin{equation}
\frac{1}{2\pi}\int d^{d}xd\theta\,\mathcal{L}_{{\rm base}}{=}\int d^{d}x\,\sqrt{g}\begin{cases}
e^{{-}\frac{\kappa}{2}\phi}\left[-\frac{2}{\kappa^{2}}R{-}\frac{1}{2}(\partial\phi)^{2}{-}\frac{1}{2}F_{\mu\nu}^{*}F^{\mu\nu}{+}m^{2}A_{\mu}^{*}A^{\mu}\right] & ,\,\text{for }0\otimes1\\
e^{{-}\frac{\kappa}{2}\phi}\left[-\frac{2}{\kappa^{2}}R{-}\frac{1}{2}(\partial\phi)^{2}\right]{-}\frac{1}{2}F_{\mu\nu}^{*}F^{\mu\nu}{+}m^{2}A_{\mu}^{*}A^{\mu} & ,\,\text{for }\frac{1}{2}\otimes\frac{1}{2}
\end{cases}
\end{equation}
The key point here is that we have performed the compactification
in the string frame, where the dilaton coupling is trivial. We can
move to the Einstein frame by setting $g_{\mu\nu}\to e^{-\frac{\kappa\phi}{d-2}}g_{\mu\nu}$.
Perturbatively, this is equivalent to a change of basis in the asymptotic
states, given by 
\begin{equation}
h_{\mu\nu}\to h_{\mu\nu}-\frac{\phi}{d-2}\eta_{\mu\nu}+\mathcal{O}(\kappa),
\end{equation}
which means the amplitudes in this frame can be computed as linear
combinations of the string frame ones. Returning to the Lagrangian,
we use
\begin{eqnarray}
R & \to & e^{-\frac{\kappa\phi}{d-2}}(R-\kappa\frac{d-1}{d-2}D^{2}\phi-\frac{d-1}{d-2}\frac{\kappa^{2}}{4}\partial_{\mu}\phi\partial^{\mu}\phi)
\end{eqnarray}
after which we perform a trivial rescaling ($\phi\to(d-2)\phi$) to
get
\begin{equation}
\mathcal{L}^{\frac{1}{2}\otimes\frac{1}{2}}=\sqrt{g}\left[-\frac{2}{\kappa^{2}}R+\frac{(d-2)}{2}(\partial\phi)^{2}-\frac{1}{2}e^{\frac{\kappa}{2}(d-4)\phi}F_{\mu\nu}^{*}F^{\mu\nu}+m^{2}e^{\frac{\kappa}{2}(d-2)\phi}A_{\mu}^{*}A^{\mu}\right],\label{eq:1212dc-1}
\end{equation}
and
\begin{equation}
\mathcal{L}^{0\otimes1}=\sqrt{g}\left[-\frac{2}{\kappa^{2}}R+\frac{(d-2)}{2}(\partial\phi)^{2}-\frac{1}{2}e^{-\kappa\phi}F_{\mu\nu}^{*}F^{\mu\nu}+m^{2}A_{\mu}^{*}A^{\mu}\right].\label{eq:01dc-1}
\end{equation}
Note that only in $d=4$ the dilaton is not sourced by matter in the
\12x12 theory. Indeed, consider momentarily
the massless limit $m=0$. A general Einstein-Maxwell-Dilaton theory
in four dimensions can be classified in the Einstein frame from the
coupling $e^{-\kappa\alpha\phi}F^{2}$, with $0\leq\alpha\leq\sqrt{3}$ \cite{PhysRevD.46.1340,Pacilio:2018gom}. The Brans-Dicke
theory corresponds to $\alpha=0$ whereas the low-energy limit of
string theory yields $\alpha=1$. This is not surprising as we will
soon identify the \0x1 with a dimensional extension of $\mathcal{N}=4$
Supergravity. We should mention that the $\alpha=\sqrt{3}$ case is
characteristic of the well-known five dimensional \ac{KK} theory, whose
double copy structure was considered in \cite{Chiodaroli:2014xia}.

These actions would be enough for amplitudes involving only gravitons,
dilatons and two Proca fields as external states. However, in the
case of the \0x1 theory we have seen that axions can be sourced
by matter. Keeping the classical application in mind, this means that
for two matter lines we will need to compute such contributions, as
they will appear as virtual states. We begin by constructing the interaction
that reproduces single matter-line amplitudes with external axions.

In order to introduce the axion coupling in the \0x1 theory we
again resort to the classical results of \cite{Goldberger:2017ogt}, which found
that in the string-frame the axion couples to the matter through 
\begin{equation}
\kappa\int d\tau\,H_{\mu\nu\rho}v^{\mu}S^{\nu\rho}.\label{eq:3ptwl}
\end{equation}
Here $S^{\mu \nu}$ is the spin operator as introduced in section \ref{sec221}. This coupling can be reproduced in \ac{QFT} by computing a ``three-point''
amplitude between the dipole and the axion,
\begin{equation}\label{eq:target}
A_{3}^{\mu\nu}\propto\kappa p^{[\mu}\times S^{\nu]\rho}q_{\rho},
\end{equation}
where $q^{\mu}$ and $p^{\mu}$ are the momentum of the axion and
the matter line respectively. As predicted, we identify the first
factor as the scalar 3pt. amplitude $A_{3}^{\mu,s=0}\propto p^{\mu}$
and the second factor as the dipole of the spin-1 amplitude $\left.A_{3}^{\mu,s=1}\right|_{J}\propto S^{\mu\rho}q_{\rho}$ \cite{Bautista:2019tdr}, which signals this corresponds to the \0x1 theory. The
overall proportionality factor can be adjusted accordingly. The \ac{QFT} 3-pt. vertex leading to \eqref{eq:target} is then the direct analog of (\ref{eq:3ptwl}), That is, after identifying $S^{\mu \nu} \to J^{\mu \nu}$ up to longitudinal terms, (\ref{eq:3ptwl}) becomes
\begin{equation}
- B_{ \mu\nu}(q)\times \kappa p_2^{\mu}A^{*[\nu}(p_2)A^{\rho]}(p_1)q_{\rho}\rightarrow \frac{\kappa}{2}H_{\mu\nu\rho}\partial^{\mu}A^{*[\nu}A^{\rho]}=\frac{\kappa}{4}H_{\mu\nu\rho}A^{*\mu}F^{\nu\rho}.
\end{equation}
Attaching then the canonically normalized kinetic term $\frac{1}{6}H_{\mu\nu\rho}H^{\mu\nu\rho}$
we can readily take this vertex into the Einstein frame (also applying
the aforementioned rescaling to $\phi$), 
\begin{equation}
\sqrt{g}e^{-\frac{\kappa}{2}\phi}\times\frac{1}{6}H_{\mu\nu\rho}(H^{\mu\nu\rho}+\frac{3\kappa}{2}\left(A^{*\mu}F^{\nu\rho}{+}\rm{c.c.}\right))\to\sqrt{g}e^{-2\kappa\phi}\times\frac{1}{6}H_{\mu\nu\rho}(H^{\mu\nu\rho}+\frac{3\kappa}{2}\left(A^{*\mu}F^{\nu\rho}{+}\rm{c.c.}\right)).
\end{equation}
Note that this term is not deformed by the massive compactification
since the derivatives in $F^{\mu \nu}$ are contracted with $H_{\mu\nu\rho}$ living
in $d=D-1$ dimensions. We note that the complex character of the fields is important for the following compactification. However, once the compactification is done we are left with a quadratic action in the Proca field, which can then be turned into real invoking the argument above \eqref{eq:symsc}. 
Thus we finally arrive at the action principle presented
in the introduction for one matter line:
\begin{equation}
\mathcal{L}^{0\otimes1}{=}\sqrt{g}\left[{-}\frac{2R}{\kappa^{2}}{+}\frac{(d{-}2)}{2}(\partial\phi)^{2}{-}\frac{e^{{-}2\kappa\phi}}{6}H_{\mu\nu\rho}(H^{\mu\nu\rho}{+}\frac{3\kappa}{2}A^{\mu}F^{\nu\rho}){-}\frac{1}{4}e^{{-}\kappa\phi}F_{\mu\nu}F^{\mu\nu}{+}\frac{m^{2}}{2}A_{\mu}A^{\mu}\right]\label{eq:final011line}
\end{equation}
Note that the massless sector corresponds to $\mathcal{N}=0$ Supergravity \cite{9780511816079}
as seen also in \cite{Goldberger:2017ogt}. We will rederive
this result from a pure on-shell point of view in the following subsubsection,
and extend it to two-matter lines. We will also perform various checks
in our proposals for both \12x12 and \0x1
actions. We can also already draw some conclusion regarding the interactions:
Even though the axion is sourced by the Proca field, it is pair produced
in the massless sector. This means that the axion is projected
out in amplitudes involving external gravitons and dilatons with a
single matter line, just as in the \12x12 theory. More importantly, an analogous
reasoning can be applied to dilatons to show that in both \0x1
and \12x12 theories the graviton emission
amplitudes are precisely the same, as we observed first in \cite{Bautista:2019tdr}.  Now, as we have mentioned, when the dilaton is included as an external state its coupling
differs in both theories: In particular, it follows from \eqref{eq:1212dc-1} that in the massless four dimensional
case the dilaton is not sourced by the photon in the \12x12 theory,
see e.g. the 4-pt. example in \cite{Johansson:2014zca}.

\section{Two matter lines from the  \ac{BCJ} construction}\label{two matter lines}

So far we have used the \ac{KLT} double copy mostly to compute the
amplitudes $A_{n}$, i.e. those involving one matter line. To test the extent of the double copy it is important to include more matter lines transforming in the fundamental representation. In our case it will be enough to consider two matter lines of different flavours in order to make contact with the classical results of previous chapters. The full quantum  amplitudes lose many nice features of the $A_n$ amplitudes: For instance we cannot trivially remove the dilaton-axion propagation nor write the multipole expansion of \cref{sec:spin_in_qed} directly. We shall anyhow conclude that the relevant classical information is already contained in the $A_n$ amplitudes, as pointed out in e.g. \cite{Neill:2013wsa}, which we have used to remove the dilaton/axion from the classical perspective in \cref{sec:spin_in_qed}.

For more than one matter line a basis of amplitudes based on Dyck words was introduced by Melia \cite{Melia:2013bta,Melia:2013epa} and later refined by Johansson and Ochirov \cite{Johansson:2015oia,Melia:2015ika}.\footnote{The amplitudes in Melia basis satisfy a restricted set of \ac{BCJ} relations \cite{Johansson:2015oia,delaCruz:2015dpa}, and consequently a \textit{generalized} \ac{KLT} construction has been recently introduced in \cite{Brown:2018wss,Johansson:2019dnu}, see also \cite{delaCruz:2016wbr}. For loop level extensions of colour-kinematics duality in this context see \cite{Ochirov:2019mtf,Kalin:2018thp,Johansson:2017bfl,Kalin:2017oqr}.} Since we only consider here two matter lines we choose to resort instead to the \ac{BCJ} representation we introduced in \cref{sec:double-copy-preliminaires}, thereby extending the approach of \cite{Luna:2017dtq}. The equivalence between the approaches has been detailed, including spin-$\frac{1}{2}$ applications, in e.g. \cite{delaCruz:2016wbr}.

Consider the two matter  lines to have mass $m_a$  and $m_b$, and spin $s_a$ and $s_b$. 
For \ac{QCD} scattering, the two  massive particles have different flavours, and  we restrict their spins to  lie 
in $\left\{0,\frac{1}{2},1\right\}$. These amplitudes are defined by the Lagrangians provided in Section \ref{constructing the lagrangians}: For the spin-0 case we use the scalar \ac{QCD} Lagrangian $(\ref{scalarqcd})$ with the removed quartic term as per our previous discussion; for spin-1 we use the $W-$boson model $(\ref{qcdW-boson})$
and for spin-$1/2$ we use the standard \ac{QCD} Lagrangian for massive Dirac
fermions $(\ref{qcdfermions})$. \\
Following the \ac{BCJ} prescription 
we arrange the \ac{QCD} amplitudes into a sum of the form
\begin{equation}
M_{n}^{\rm{QCD}}=\sum_{i\in\Gamma}\frac{c_{i}n_{i}^{(s_{a},s_{b})}}{d_{i}},\label{eq:qcd bcj}
\end{equation}
running over the set $\Gamma$ of all cubic diagrams, with denominators $d_i$. The superscript $(s_{a},s_{b})$ here denotes the spin of the lines and may be omitted. For a given triplet $(i,j,k)$, if the color factors satisfy the Jacobi identity
\begin{equation}\label{jacobi}
c_{i}\pm c_{j}=\pm c_{k},
\end{equation}
then colour kinematics duality requires there is a choice of numerators $n_i$ such that
\begin{equation}\label{color-kinematic}
  n_{i}\pm n_{j}=\pm n_{k}.  
\end{equation}
The gravitational amplitudes can be computed starting  from $(\ref{eq:qcd bcj})$ by replacing the color factors with further kinematic factors, which can be associated to a different \ac{QCD} theory. In this section
we will explore some of the choices for \ac{QCD} theories, and write the explicit form of the resulting gravitational Lagrangians. With this in mind, the $n-$point
gravitational amplitude, where now the massive lines have spins $s_a + \tilde{s}_a$ and $s_b+\tilde{s}_b$ respectively, reads
\begin{equation}
M_{n}^{(s_a \otimes \tilde{s}_a, s_b \otimes \tilde{s}_b)}=\sum_{i\in\Gamma}\frac{n^{(s_a,s_b)}_{i}\otimes \tilde{n}^{(\tilde{s}_a,\tilde{s}_b)}_{i}}{d_{i}},\label{eq:BCJ gravity}
\end{equation}
 where the product $\otimes$  depends on the spin of the massive particles in the \ac{QCD} theory.
For instance, 
for $s_a=\tilde{s}_a=s_b=\tilde{s}_b=1/2$ we define it in an analogous way to the case of only  one matter line  $(\ref{tensor product })$; that is: consider the spin $\frac{1}{2}$ operators $\mathcal{X}_{i}$ and $\mathcal{Y}_{i}$, entering in a
QCD numerator $n^{\rm{QCD}}$ with four external fermions whose momenta we choose to be all
 outgoing as follows
\begin{equation}
n^{(\frac{1}{2},\frac{1}{2})}=\bar{u}_{2}\mathcal{X}_{i}v_{1}\bar{u}_{4}\mathcal{Y}_{i}v_{3},
\end{equation}
analogously, the charge conjugated numerator reads 
\begin{equation}
\bar{n}^{(\frac{1}{2},\frac{1}{2})}=\bar{u}_{1}\bar{\mathcal{X}}_{i}v_{2}\bar{u}_{3}\bar{\mathcal{Y}}_{i}v_{4}.
\end{equation}
 We define the spin-1 gravitational numerator as the tensor product of the two \ac{QCD} numerators as follows:
\begin{equation}
n^{(\frac{1}{2},\frac{1}{2})}\otimes\bar{n}^{(\frac{1}{2},\frac{1}{2})}=\frac{1}{2^{2\left\lfloor d/2\right\rfloor -2}}\text{tr}\left[\mathcal{X}_{i}\slashed{\varepsilon}_{1}(\slashed{p}_{1}{+}m_{a})\bar{\mathcal{X}}_{i}\slashed{\varepsilon}_{i}(\slashed{p}_{2}{+}m_{a})\right]\text{tr}\left[\mathcal{Y}_{i}\slashed{\varepsilon}_{3}(\slashed{p}_{3}{+}m_{b})\bar{\mathcal{Y}}_{i}\slashed{\varepsilon}_{4}(\slashed{p}_{4}{+}m_{b})\right],\label{eq:doublestar-1-1}
\end{equation}
This is the analog double copy numerators of the multipole double copy in \eqref{eq:numdc}.
 Notice that the generalization
of $(\ref{eq:doublestar-1-1})$ to an arbitrary number of massive lines could be done analogously by introducing one Dirac trace for each matter line.

 In this section we focus on elastic scattering, given by $M_4$, and inelastic scattering, given by $M_5$, firstly from a \ac{QFT} perspective and then from a classical perspective. Nevertheless, we propose Lagrangians for arbitrary multiplicity as long as we keep two matter lines.
 
 Setting conventions, the momenta of the particles are taken as follows: For the $2\rightarrow2$ elastic scattering, the two incoming momenta are $p_1$ and $p_3$, and the outgoing momenta are $p_2=p_1-q$ and $p_4=p_3+q$, for $q$ the momentum transfer. For the $2\rightarrow3$ inelastic scattering, again the two incoming momenta are  $p_1$ and $p_3$, whereas the momenta for the two outgoing massive particles are $p_2=p_1-q_1$ and $p_4=p_3-q_3$, and the outgoing gluon or graviton has momentum $k$.

\subsection{Elastic scattering}

The simplest example of the scattering of two massive particles of
mass $m_{a}$ and $m_{b}$ , and spin $s_{a}$ and $s_{b}$, is the
elastic scattering, which we call $M_{4}^{(s_{a},s_{b})}$ amplitudes.
Let us illustrate how the double copy works for some choices of $s_{a}$
and $s_{b}$.

\subsubsection{Case  \texorpdfstring{$s_a=s_b=0+1$}{sasb01}}\label{sec:casee}

The gravitational scattering amplitude $\eqref{eq:BCJ gravity}$ at four points  can be obtained from the double copy of the scalar numerators  $n^{(0,0)}$ and the spin-$1$ numerator $n^{(1,1)}$. This numerator  can be computed from the gluon exchange between two massive spin-0,1 fields, each described by the matter part of \eqref{qcdW-boson}, and results into  
\begin{equation}\label{numM4scalar}
 n^{(0,0)}=-e^{2}\left(4p_{1}{\cdot}p_{3}+q^{2}\right),\qquad d_{4}=q^{2}.
\end{equation}
\begin{equation}\begin{split}
     n^{(1,1)}&=-4e^2\bigg[\frac{1}{4}\left(4p_1{\cdot}p_3{+}q^2\right)\varepsilon_1{\cdot}\varepsilon_2\,\varepsilon_3{\cdot}\varepsilon_4-\left(p_1{\cdot}\varepsilon_3\,p_3{\cdot}\varepsilon_4{+}p_1{\cdot}\varepsilon_3\,q{\cdot}\varepsilon_4\right)\varepsilon_1{\cdot}\varepsilon_2 \\
     &\qquad\qquad-\left(p_1{\cdot}\varepsilon_2\,p_3{\cdot}\varepsilon_1{-}p_3{\cdot}\varepsilon_2\,q{\cdot}\varepsilon_1\right)\varepsilon_3{\cdot}\varepsilon_4
     +p_1{\cdot}\varepsilon_2\,p_3{\cdot}\varepsilon_4\,\varepsilon_1{\cdot}\varepsilon_3\\
     &\qquad\qquad
    -q{\cdot}\varepsilon_1\,q{\cdot}\varepsilon_3\,\varepsilon_2{\cdot}\varepsilon_4 -p_3{\cdot}\varepsilon_4\,q{\cdot}\varepsilon_1\,\varepsilon_2{\cdot}\varepsilon_3+p_1{\cdot}\varepsilon_2\,q{\cdot}\varepsilon_3\,\varepsilon_1{\cdot}\varepsilon_4\bigg].
\end{split}
\end{equation}
The gravitational Lagrangian for this theory has a more intricate structure than the one for a single matter line, which is natural due to additional propagation of the axion coupling to the spin of the matter lines. It can be shown that the Lagrangian is given by  
\begin{equation}\label{lagrangian01x01}
\begin{split}
    \mathcal{L}^{(0\otimes1,0\otimes1)} &=\mathcal{L}_{ct}+\sqrt{g}\bigg[{-}\frac{2}{\kappa^{2}}R{+}\frac{2(d{-}2)}{\kappa^{2}}(\partial\phi)^{2}-\frac{e^{-4\phi}}{6\kappa^{2}}H_{\mu\nu\rho}H^{\mu\nu\rho}\\&\qquad -\frac{e^{-4\phi}}{6\kappa^{2}}H_{\mu\nu\rho}A_I^{\mu}F^{I\nu\rho}-\frac{1}{4}e^{-2\phi}F_{I,\mu\nu}F^{I\mu\nu}{+}\frac{m_{I}^{2}}{2}A_{I,\mu}A^{I\mu}\bigg] ,
\end{split}
\end{equation}
where the flavour index  $I\in\{1,2\}$, and once again the masses $m_1=m_a$ and $m_2=m_b$. The contact interaction Lagrangian for this case  has the form
\begin{equation}\label{contact term 01}
    \begin{split}
        \mathcal{L}_{ct}&\sim \sqrt{g}\big[ 2A_{1}{\cdot} A_{2}\,(\partial_{\mu}A_{1,\nu}{-}3\partial_{\nu}A_{1,\mu})\partial^{\mu}A_{2}^{\nu}-2A_{2}{\cdot} F_{1}{\cdot}F_{2}{\cdot} A_{1}  \\
        &-2A_{2}^{\mu}\partial_{\mu}A_{1}^{\alpha}A_{2}^{\nu}\partial_{\nu}A_{1,\alpha} {-}A_{1}^{\mu}\partial_{\mu}A_{2}^{\alpha}A_{1}^{\nu}\partial_{\nu}A_{2\alpha}{-}A_{1}^{\mu}\partial_{\alpha}A_{1,\mu}A_{2}^{\nu}\partial^{\alpha}A_{2,\nu}\big],
    \end{split}
\end{equation}
where the  product of field strength tensors reads explicitly
\begin{equation}\begin{split}
     A_{2}{\cdot} F_{1}{\cdot} F_{2}{\cdot} A_{1}&=A_{2}^{\mu}\partial_{\mu}A_{1}^{\alpha}\partial_{\alpha}A_{2,\nu}A_{1}^{\nu}{-}A_{2}^{\mu}\partial^{\alpha}A_{1,\mu}\partial_{\alpha}A_{2,\nu}A_{1}^{\nu}\\
     &{-}A_{2}^{\mu}\partial_{\mu}A_{1}^{\alpha}\partial_{\nu}A_{2,\alpha}A_{1}^{\nu}{-}A_{2}^{\mu}\partial^{\alpha}A_{1,\mu}\partial_{\nu}A_{2,\alpha}A_{1}^{\nu} .
\end{split}
\end{equation}
Thus in this case, for two particles including spin, we have found an elevated level of complexity even for the four-point terms in the Lagrangian, not present in the single matter line case. 

\subsubsection{Case  \texorpdfstring{$s_{a}=s_{b}=\frac{1}{2}+\frac{1}{2}$}{sasb12p12}}

We  finish the discussion  for the  elastic scattering considering the simplest gravitational theory for both of the massive lines with spin-1. As we mentioned previously, this theory is  dictated  by the factorization $s_{a}=s_{b}=\frac{1}{2}{+}\frac{1}{2}$. The gravity amplitude \eqref{eq:BCJ gravity} at $4$ pt. is computed 
from the double copy of the \ac{QCD} spin $\frac{1}{2}$ numerator $n^{(\frac{1}{2},\frac{1}{2})}$, and its charge conjugated pair. They have a simple form
\begin{equation}\begin{split}
      n^{(\frac{1}{2},\frac{1}{2})} & =e^{2}\bar{u}_{2}\gamma^{\mu}u_{1}\bar{u}_{4}\gamma_{\mu}u_{3},\\
\bar{n}^{(\frac{1}{2},\frac{1}{2})} & =e^{2}\bar{v}_{1}\gamma^{\mu}v_{2}\bar{v}_{3}\gamma_{\mu}v_{4},
\end{split}
\end{equation}
where we use the condition for momentum conservation  $p_{2}=p_{1}{-}q$ and $p_{4}=p_{3}{+}q$.
Now, using the double copy operation for two matter lines $(\ref{eq:doublestar-1-1})$,
the gravitational amplitude takes the compact form 
\begin{equation}
\begin{split}
M_{4}^{(\frac{1}{2}\otimes\frac{1}{2},\frac{1}{2}\otimes\frac{1}{2})}&=\frac{4}{2^{2\left\lfloor D/2\right\rfloor }}\frac{\kappa^{2}}{q^{2}}\text{tr}\big[\gamma^{\mu}\slashed{\varepsilon}_{1}(\slashed{p}_{1}{-}m_{a})\gamma^{\nu}\slashed{\varepsilon}_{2}(\slashed{p}_{2}{+}m_{a})\big]\text{tr}\big[\gamma_{\mu}\slashed{\varepsilon}_{3}(\slashed{p}_{3}{-}m_{b})\gamma_{\nu}\slashed{\varepsilon}_{4}(\slashed{p}_{4}{+}m_{b})\big],
\end{split}
\end{equation}
 Notice the momenta $p_1$ and $p_3$  are incoming, therefore the sign in the projector  changes. After taking  the traces the amplitude reads
\begin{equation}
\begin{split}
M_{4}^{(\frac{1}{2}\otimes\frac{1}{2},\frac{1}{2}\otimes\frac{1}{2})} & =\frac{4\kappa^{2}}{q^{2}}\bigg\{\big[\varepsilon_{1}{\cdot}\varepsilon_{2}\left((d{-}6)p_{1}^{\nu}p_{2}^{\mu}{+}(d{-}2)p_{1}^{\mu}p_{2}^{\nu}\right){-}p_{1}{\cdot}\varepsilon_{2}\left((d{-}6)\varepsilon_{1}^{\nu}p_{2}^{\mu}{+}(d{-}2)\varepsilon_{1}^{\mu}p_{2}^{\nu}\right){-}\\
 & p_{2}{\cdot}\varepsilon_{1}\left((d{-}6)p_{1}^{\nu}\varepsilon_{2}^{\mu}{+}(d{-}2)p_{1}^{\mu}\varepsilon_{2}^{\nu}\right){+}\left((d{-}6)p_{1}{\cdot}p_{2}{+}(d{-}4)m_{a}^{2}\right)\left(\varepsilon_{1}^{\mu}\varepsilon_{2}^{\nu}{-}\varepsilon_{1}{\cdot}\varepsilon_{2}\eta^{\mu\nu}\right)\\
 & +\left((d{-}2)p_{1}{\cdot}p_{2}{+}d\,m_{a}^{2}\right)\varepsilon_{1}^{\mu}\varepsilon_{2}^{\nu}{+}(d{-}6)p_{1}{\cdot}\varepsilon_{2}\,p_{2}{\cdot}\varepsilon_{1}\eta^{\mu\nu}\big]\times\big[{\rm {line}}\,a\rightarrow{\rm {line}}\,b\big]_{\mu \nu}\bigg\},\\
\end{split}
\end{equation}
where the change $\big[{\rm {line}}\,a\rightarrow{\rm {line}}\,b\big]$
means to do $\big[1\rightarrow3,\,2\rightarrow4,\,a\rightarrow b\big].$
Likewise for the two previous cases, we can write the gravitational
Lagrangian for this theory, surprisingly it has a very simple form 
\begin{equation}
\mathcal{L}^{(\frac{1}{2}\otimes\frac{1}{2},\frac{1}{2}\otimes\frac{1}{2})}=\sqrt{g}\bigg[{-}\frac{2}{\kappa^{2}}R{+}\frac{2(d{-}2)}{\kappa^{2}}(\partial\phi)^{2}{-}\frac{1}{4}e^{(d{-}4)\phi}F_{I,\mu\nu}F_{I,}^{\mu\nu}{+}\frac{1}{2}e^{(d{-}2)\phi}m_{I}^{2}A_{I\mu}A_{I}^{\mu}\bigg]\, ,\label{eq:1212LAGRANGIAN}
\end{equation}
We say  that  this is the simplest theory for spinning particles coupled to gravity in two senses:
First, even thought the two massive lines have spin, there is no propagation of the axion. This  confirms that in the \12x12 double copy setup the spin-$1$ field does not source the axion. Second and more importantly, there is no need for adding a contact interaction between matter lines, a feature we will confirm also in $M_5$. This is the only gravitational theory we have found for which this happens and reflects its underlying fermionic origin.

\subsection{Inelastic Scattering}
Moving on to the inelastic scattering, we consider the emission of a gluon or a (fat) graviton in the final state. The relevance of this amplitude is that it allows us to make contact with classical double copy introduced in \cref{sec:spin_in_qed}.

The \ac{QCD} amplitude obtained from Feynman diagrams can be arranged into the color decomposition \eqref{eq:qcd bcj} with only five terms as shown in Figure \ref{fig:M5_diags}. The color factors and  denominators given by 

\begin{equation}\label{color and denominators}
\begin{split}
c_{1} & =(T_{1}^{a}.T_{1}^{b})T_{3}^{b},\qquad d_{1}=q_{3}^{2}\left(2p_{1}{\cal \cdot}k-q_{1}^{2}+q_{3}^{2}\right),\\
c_{2} & =(T_{1}^{b}.T_{1}^{a})T_{3}^{b},\qquad d_{2}=-2\left(p_{1}{\cdot}k\right)q_{3}^{2},\\
c_{3} & =f^{abc}T_{1}^{b}T_{3}^{c},\qquad\,\,\, d_{3}=q_{1}^{2}q_{3}^{2},\\
c_{4} & =(T_{3}^{a}.T_{3}^{b})T_{1}^{b},\qquad d_{4}=q_{1}^{2}\left(2p_{3}{\cal \cdot}k+q_{1}^{2}-q_{3}^{2}\right),\\
c_{5} & =(T_{3}^{b}.T_{3}^{a})T_{1}^{b}.\qquad\, d_{5}=-2\left(p_{3}{\cdot}k\right)q_{1}^{2}\,,
\end{split}
\end{equation}
they satisfy the Jacobi relations 
\begin{equation}
c_{1}-c_{2}=-c_{3},\,\,\,c_{4}-c_{5}=c_{3},\label{eq:color identities-1}
\end{equation}
and in the same way, the numerators can be arrange to satisfy the same algebra
\begin{equation}
n_{1}-n_{2}=-n_{3},\,\,\:n_{4}-n_{5}=n_{3}.\label{eq:color-kinematics duality}
\end{equation}
The gravitational amplitude will be given again by $(\ref{eq:BCJ gravity})$, with the sum running from $1$ to $5$.
The product of polarization vectors of the external gluon $\epsilon_{\mu}\tilde{\epsilon}_{\nu}$
corresponds to a fat graviton state $H_5$. To extract the graviton amplitude we replace $\epsilon_{\mu}\tilde{\epsilon}_{\nu}\rightarrow\epsilon_{\mu\nu}^{\rm{TT}}$
i.e. the symmetric, transverse and traceless polarization tensor for the graviton.
If on the other hand we want to compute the dilaton amplitude,  we replace
$\epsilon_{\mu}\tilde{\epsilon}_{\nu}\rightarrow\frac{\eta_{\mu\nu}}{\sqrt{D{-}2}}.$ Finally, in the case of the $0\otimes1$ theory, there will be also the existence of axion radiation which can be obtained by taking the antisymmetric part,   $\epsilon_{[\mu}\tilde{\epsilon}_{\nu]}$. 

In order to make direct contact with the classical double copy introduce in \cref{sec:spin_in_qed}, we choose however to write the 5-point, and therefore the numerators entering into the amplitude for the different theories in  more convenient \textit{generalized gauge}.

\subsection{Generalized Gauge Transformations and Classical Radiation}\label{GGT}
As we have seen in previous chapters, the 5-poitn amplitude encodes information regarding the  classical radiated momentum in a 2-3 scattering process,  which is carried by long range fields (photons, gravitons, dilatons and axions) to null infinity \cite{Goldberger:2016iau,Kosower:2018adc}. This momentum is determined by a phase space integral, 
\begin{equation}
     K^{\mu} = \int \text{dLIPS}(k)\, k^{\mu} \,|J(k)|^2 \,,
\end{equation}
as outlined in \cref{sec:KMOC}, 
where $J(k)$ is the radiative piece of the stress energy tensor (or current) related to the amplitude via the LSZ formula. This also requires to implement a prescription for the classical limit, $J(k)=\lim_{\hbar \to 0 } M_5 $ a la KMOC. In light of the promising  developments of \cite{Luna:2016due,PV:2019uuv,Chester:2017vcz,Shen:2018ebu} it is desirable to understand how a double copy structure turns out to be realized in classical radiation, and more specifically, how it follows from the \ac{BCJ} construction in \ac{QFT}. 

We would like to extract the classical piece of the amplitude in such a way that the double copy structure is preserved untouched in the final result.  Taking the classical limit of  \eqref{eq:BCJ gravity} however  does not show explicitly the double copy form of the classical amplitude in \eqref{eq:newM5clas}, as we will see in a moment. This was first observed for scalar sources in \cite{Luna:2017dtq}, but is also true for the spinning case.  We find that the problem can be fixed if we  write the double copy for inelastic scattering in a more convenient \textit{generalized} gauge.

\subsubsection{Classical radiation from the standard \ac{BCJ} double copy }

Here we will use the usual \ac{KMOC} approach to take the classical limit. For that, it is convenient to introduce the   average momentum transfer $q=\frac{q_1{-}q_2}{2}$ as we did in previous chapters.  The re-scaled momenta
can be interpreted as a classical wave vector $q\rightarrow\hbar\bar{q}$.
Notice that momentum conservation implies that the radiated on-shell momenta
needs to be re-scaled as well $k\rightarrow\hbar\bar{k}.$ For spinning radiation the classical limit was outlined in \cref{sec:spin_in_qed} and requires to introduce the angular momentum operator, performing the multipole expansion as we have described in the previous sections. We then scale such operator as $J\rightarrow\hbar^{-1}\bar{J}$  \cite{Guevara:2018wpp,Bautista:2019tdr} and strip the respective polarization states \cite{Guevara:2019fsj}. Finally, for the
case of \ac{QCD} amplitudes, one further scaling needs to be done in order
to correctly extract the classical piece. In reminiscence of the color-kinematics duality, we find that the generators of the color group
$T^a$ must also scale as those of angular momentum, i.e. $T^a\rightarrow\hbar^{-1}T^a$.

In order to motivate our procedure let us first consider the 5-pt. amplitudes for both \ac{QCD} and gravity in the standard \ac{BCJ} form we have provided. In other words, we want to see how the ingredients in \eqref{eq:qcd bcj} and \eqref{eq:BCJ gravity} behave in the
$\hbar$-expansion. By inspection, the leading order of the numerators $n_{i}$ goes
as $\hbar^{0},$ and the sub-leading correction is of order $\hbar$. 
Let us denote the expansion of the numerators as $n_{i}=\langle n_{i}\rangle +\delta n_{i}\hbar+{\cdots}$.
The denominators can also be expanded as $d_{i}= \langle d_{i} \rangle\hbar^{3}+\delta d_{i}\hbar^{4}+\cdots$.
At leading order, it is easy to check that $\langle n_{3}\rangle $=0, $\langle n_{1}\rangle=\langle n_{2}\rangle$
and $\langle n_{4}\rangle =\langle n_{5}\rangle$, whereas for  the denominators we have $\langle d_{1}\rangle=-\langle d_{2}\rangle$
and $\langle d_{4}\rangle=-\langle d_{5}\rangle .$ At sub-leading order $\delta d_{2}=\delta d_{5}=0.$
Finally, for the color factors we have $c_{i}\rightarrow\hbar^{-3}c_{i}$
for $i=1,2,4,5$ and $c_{3}\rightarrow\hbar^{-2}c_{3}$.\\

 With this in mind, the classical piece of the \ac{QCD} amplitude for gluon radiation reads
\begin{equation}\label{M5qcd classical}
\langle M_{5}^{\rm{QCD}}\rangle
  ={-}c_{1}\left[\frac{\langle n_{1}\rangle \delta d_{1}}{\langle d_{1}\rangle^{2}}{-}\frac{\delta n_{1}{-}\delta n_{2}}{\langle d_{1}\rangle}\right]{-}c_{3}\left[\frac{\langle n_{1}\rangle}{\langle d_{1}\rangle}{-}\frac{\delta n_{3}}{\delta d_{3}}{-}\frac{\langle n_{4}\rangle}{\langle d_{4}\rangle}\right]{-}c_{4}\left[\frac{\langle n_{4}\rangle \delta d_{4}}{\langle d_{4}\rangle^{2}}{-}\frac{\delta n_{4}{-}\delta n_{5}}{\langle d_{4}\rangle}\right],
\end{equation}
where  $\langle M_{n}\rangle:{=}\lim_{\hbar\rightarrow0}M_{n}$.  A similar expansion can be done for the gravitational amplitude
given by the double copy \eqref{eq:BCJ gravity}
\begin{equation}\label{eq:m5gr classical}
\begin{split}
\langle M_{5}^{gr}\rangle & ={-}\frac{\langle n_{1}\rangle \otimes\langle \tilde{n}_{1}\rangle}{\langle d_{1,0}\rangle^{2}}\delta d_{1}{+}\frac{\langle n_{1}\rangle\otimes\left(\delta\tilde{n}_{1}{-}\delta\tilde{n}_{2}\right)+\left(\delta n_{1}{-}\delta n_{2}\right)\otimes\langle \tilde{n}_{1}\rangle}{\langle d_{1}\rangle}{+}\frac{\delta n_{3}\otimes\delta\tilde{n}_{3}}{\langle d_{3}\rangle}\\
 &\qquad-\frac{\langle n_{4}\rangle \otimes\langle \tilde{n}_{4}\rangle}{\langle d_{4,0}\rangle^{2}}\delta d_{4}{+}\frac{\langle n_{4}\rangle \otimes\left(\delta\tilde{n}_{4}{-}\delta\tilde{n}_{5}\right)+\left(\delta n_{4}{-}\delta n_{5}\right)\otimes \langle \tilde{n}_{4}\rangle}{\langle d_{4}\rangle}
 \end{split}
\end{equation}

Hence, we find that the classical piece of the gravitational amplitude \eqref{eq:m5gr classical} does not reflect
the \ac{BCJ} double copy structure as expected. This can be traced back to the presence of $\frac{1}{\hbar}$ terms which will still contribute to the expansion even though the overall leading order (as $\hbar{\to}0$) cancels. We shall find a way to make such limit smooth and preserve the double copy structure.

\subsubsection{Generalized gauge transformation}

In order to rewrite the quantum amplitudes \eqref{eq:qcd bcj}  and \eqref{eq:BCJ gravity} in a convenient
gauge we proceed as follows.  Observe that the non-abelian contribution to the \ac{QCD}  amplitude
\eqref{eq:qcd bcj} comes from the diagram whose color factor  \eqref{color and denominators} is $c_{3}$,
which is proportional to the structure constants of the gauge group.
We can however gauge away this non-abelian piece of the amplitude
using a \textit{Generalized Gauge Transformation} (GGT) \cite{Bern:2008qj}. Recall that a  GGT is
a transformation on the kinematic numerators that leaves the amplitude invariant. This transformation allow us to move terms
between diagrams. For the case of the inelastic scattering, consider
the following shift on the numerators entering in \eqref{eq:qcd bcj}
\begin{equation}\label{eq:shifted numeratos}
\begin{split}
n_{1}^{\prime} & =n_{1}-\alpha d_{1},\\
n_{2}^{\prime} & =n_{2}+\alpha d_{2}, \\
n_{3}^{\prime} & =n_{3}-\alpha d_{3}+\gamma d_{3},\\
n_{4}^{\prime} & =n_{4}-\gamma d_{4}, \\
n_{5}^{\prime} & =n_{5}+\gamma d_{5}. 
\end{split}
\end{equation}
This shift leaves invariant the amplitude \eqref{eq:qcd bcj} since under it, 
\begin{equation}
\Delta M_{5}^{\rm{QCD}} =-\alpha(c_{1}-c_{2}+c_{3})-\gamma(c_{4}-c_{5}-c_{3})=0,
\end{equation}
where we have use the color identities $(\ref{eq:color identities-1})$
in the last equality. We can now solve for the values of $\alpha$ and
$\gamma$ that allow to impose $n_{3}^{\prime}=0$ , while satisfying
the color-kinematic duality for the shifted numerators
\begin{equation}
n_{1}^{\prime}-n_{2}^{\prime}=-n_{3}^{\prime}=0,\,\,\:n_{4}^{\prime}-n_{5}^{\prime}=n_{3}^{\prime}=0.
\end{equation}
The solution can be written as
\begin{align}
\alpha & =-\frac{n_{3}}{d_{1}+d_{2}},\,\,\,\,\,\gamma=-\frac{d_{1}+d_{2}+d_{3}}{d_{1}+d_{2}}\frac{n_{3}}{d_{3}}.\label{eq:alpha gamma parameters}
\end{align}
Explicitly these  parameters 
take the simple form
\begin{equation}
  \alpha=\frac{n_{3}}{2q{\cdot}k\left(q^{2}{-}q{\cdot}k\right)},\qquad\gamma=\frac{n_{3}}{2q{\cdot}k\left(q^{2}{+}q{\cdot}k\right)},  
\end{equation}
Importantly, this solution is general and independent of the spin of scattered particles as we wish to make contact with the classical formula \eqref{eq:newM5clas}.

The new numerators \eqref{eq:shifted numeratos} will be non-local since they have absorbed $n_3$. However, they exhibit nice features: They are independent, gauge invariant, and in the classical limit they will lead to a remarkably simple (and local!) form. Indeed, the \ac{QCD} amplitude \eqref{eq:qcd bcj} for inelastic scattering takes already a more compact form 
\begin{equation}\label{m5qcd GGT}
M_{5}^{\rm{QCD}}=\left[\frac{c_{1}}{d_{1}}+\frac{c_{2}}{d_{2}}\right]n_{1}^{\prime}+\left[\frac{c_{4}}{d_{4}}+\frac{c_{5}}{d_{5}}\right]n_{4}^{\prime}.
\end{equation}
The gravitational amplitude \eqref{eq:BCJ gravity} then is given by the double copy of \eqref{m5qcd GGT}  as follows 
\begin{equation}
M_{5}^{\rm {gr}}=\frac{n_{1}^{\prime}\otimes\tilde{n}_{1}^{\prime}}{d_{1}^{\prime}}+\frac{n_{4}^{\prime}\otimes\tilde{n}_{4}^{\prime}}{d_{4}^{\prime}},\label{m5 gr CBCJ}
\end{equation}
 where 
\begin{equation}
d_{1}^{\prime}=\frac{d_{1}d_{2}}{d_{1}+d_{2}},\qquad d_{4}^{\prime}=\frac{d_{4}d_{5}}{d_{4}+d_{5}}.
\end{equation}
Explicitly, this gives
\begin{equation}\label{denominators explicit}
\frac{1}{d_{1}^{\prime}}{=}-\frac{q{\cdot}k}{p_{1}{\cdot}k\,q{\cdot}(q-k)\left(2q{\cdot}k{-}2p_{1}{\cdot}k\right)},\,\,\,\,\frac{1}{d_{4}^{\prime}}{=}-\frac{q{\cdot}k}{p_{3}{\cdot}k\,q{\cdot}(q+k)\left(2q{\cdot}k{+}2p_{3}{\cdot}k\right)},
\end{equation}
 When performing the double copy, there will in principle be a
 pole in $q{\cdot}k$  both in $(\ref{m5 gr CBCJ})$ and in the classical formula  \eqref{M5 classical CBCJ} below, which is nevertheless spurious and cancels out
in the final result. This is the spurious pole we saw in \eqref{eq:newM5clas}, arising from the t-channel of the gravitational Compton amplitude. 
 Notice we have reduced the problem of doing the double copy of five
numerators to do the double copy of  just two (the dimension of the \ac{BCJ} basis). Indeed, now we can take $c_3\to 0$, setting $c_2 \to c_1$ and $c_5 \to c_4$. Further fixing $c_1=c_4=1$ we obtain the \ac{QED} case (see \eqref{color and denominators}) with

\begin{equation}\label{m5qedd}
M_{5}^{\rm {QED}}=\frac{n_{1}^{\prime}}{d_{1}^{\prime}}+\frac{n_{4}^{\prime}}{d_{4}^{\prime}},
\end{equation}

The double copy formula \eqref{m5 gr CBCJ} agrees with \eqref{eq:BCJ gravity}. Remarkably, we can use \eqref{m5qedd} as a starting point for the (classical) double copy since the numerators $n_1'$ and $n_4'$ can be read off from $M_5^{\rm QED}$ from its pole structure. This has the advantage that the
classical limit of the amplitude will be smooth and will preserve the double copy form.

\subsubsection{Classical limit and Compton Residue}

In the gauge $(\ref{eq:shifted numeratos})$, extracting the classical
piece of the gravitational amplitude \eqref{m5 gr CBCJ} is straightforward. The shifted numerators
scale as $n_{i}^{\prime}=\langle n_{i}^{\prime}\rangle+\delta n_{i}^{\prime}\hbar $,
whereas the denominators scale as $d_{i}^{\prime}=\langle d_{i}^{\prime}\rangle\hbar^{2}+\delta d_{i}^{\prime}\hbar^3$. With this
in mind, the classical piece of the gravitational amplitude \eqref{m5 gr CBCJ} is simply 
\begin{equation}\label{M5 classical CBCJ}
\boxed{
\langle M_{5}^{(s_a\otimes\tilde{s}_a,s_b\otimes \tilde{s}_b)}\rangle =\frac{\langle n_{1}^{\prime\,(s_a,s_b)}\rangle \otimes\langle \tilde{n}_{1}^{\prime
\,(\tilde{s}_a,\tilde{s}_b)}\rangle}{\langle d_{1}^{\prime}\rangle}+\frac{\langle n_{4}^{\prime\,(s_a,s_b)}\rangle \otimes\langle\tilde{n}_{4}^{\prime\,(\tilde{s}_a,\tilde{s}_b)}\rangle}{\langle d_{4}^{\prime}\rangle},}
\end{equation}
 which shows explicitly the double copy structure. Indeed, the classical limit of the \ac{QED} amplitude is naturally identified as the single copy in this gauge:
 
 \begin{equation}\label{qedclas}
 \boxed{
\langle M_{5}^{\text{QED},(s_a,s_b)}\rangle =\frac{\langle n_{1}^{\prime\, (s_a,s_b)}\rangle }{\langle d_{1}^{\prime}\rangle}+\frac{\langle n_{4}^{\prime \,(s_a,s_b)}\rangle }{\langle d_{4}^{\prime}\rangle}.}
\end{equation}
Taking the classical piece of the denominators \eqref{denominators explicit} leads to   
\begin{equation}\label{classical denominators}
\frac{1}{\langle d_{1}^{\prime}\rangle}{=}\frac{q{\cdot}k}{2\left(p_{1}{\cdot}k\right)^{2}(q^2-q{\cdot}k)},\,\,\,\,\frac{1}{\langle d_{4}^{\prime}\rangle}{=}-\frac{q{\cdot}k}{2\left(p_{3}{\cdot}k\right)^{2}(q^2+q{\cdot}k)}.
\end{equation}
 As a whole, the formulas \eqref{M5 classical CBCJ}, \eqref{qedclas} and \eqref{classical denominators} correspond to the construction given in \cref{ch:electromagnetism} and \cref{sec:spin_in_qed}. The conversion can be done via $\langle n_i^{\prime} \rangle=\frac{2}{q\cdot k}n_i^{\rm{there}} $, where $n_i^{\rm{there}}$ is a local numerator in the classical limit. We have thus found here an alternative derivation which follows directly from the standard \ac{BCJ} double copy of $M_5$, up to certain details we now describe.
 
Suppose first that the numerators $\langle n_i^{\prime} \rangle$ do not depend on $q^2$. Then we find they can be read off from the \ac{QED} Compton residues at $q^2 \to \pm q\cdot k$. Indeed, using that \eqref{qedclas}-\eqref{classical denominators} should factor into the Compton amplitude $A_4$ together with a 3-pt. amplitude $A_3$, we get
 \begin{equation}
     \langle n_{i}^{\prime(s_{a},s_{b})}\rangle=\frac{2(p{\cdot}k)^2}{q{\cdot}k}\langle A_{4}^{{\rm QED},s_{a},\mu}\rangle\times\langle A_{3}^{{\rm QED},s_{b},\mu},\rangle
 \end{equation}
where the contraction in $\mu$ denotes propagation of photons. This guarantees the same is true for the gravitational numerators in \eqref{M5 classical CBCJ}, that is

\begin{align}
    \langle n_{i}^{\prime(s_{a},s_{b})}\rangle\otimes\langle n_{i}^{'(s_{a},s_{b})}\rangle&=\frac{4(p{\cdot}k)^4}{(q{\cdot}k)^2}\langle A_{4}^{{\rm QED},s_{a},\mu}\rangle\otimes\langle A_{4}^{{\rm QED},\tilde{s}_{a},\nu}\rangle\times\langle A_{3}^{{\rm QED},s_{b},\mu}\rangle\otimes\langle A_{3}^{{\rm QED},\tilde{s}_{b},\nu}\rangle, \nonumber \\&=\frac{4(p{\cdot}k)^4}{(q{\cdot}k)^2}\langle A_{4}^{s_{a}\otimes\tilde{s}_{a},\mu\nu}\rangle\times\langle A_{3}^{s_{b}\otimes\tilde{s}_{b},\mu\nu}\rangle,
\end{align}
where the contracted indices denote propagation of fat states. Thus we conclude that \textit{the classical limit is controlled by $A_4$ and $A_3$ via the Compton residues} provided the numerators do not depend on $q^2$. Considering the scaling of the multipoles $J\rightarrow\hbar^{-1} \bar{J}$ and that $q\rightarrow \hbar \bar{q}$, we see that this is true up to dipole ${\sim} J$ order. We will confirm this explicitly in the cases below. 

At quadrupole order ${\sim} J^2$, associated to spin-1 particles, we will find explicit dependence on $q^2$ in the numerators. Nevertheless, it is still true that the classical multipoles are given by the  Compton residues as we have extensively  exemplified previous chapters, specially \cref{sec:spin_in_qed}. Indeed, as a quick analysis shows, the $q^2$ dependence in $M_5$ that is not captured by them can only arise from 1) contact terms in $M_5$ or 2) contact terms in $M_4$ entering through the residues at $p\cdot k \to 0$. Both contributions can be canceled by adding appropriate (quantum) interactions between the matter particles. Note that canceling such contributions in the \ac{QCD} side will automatically imply their cancellation on the gravity side.

Let us now provide  some specific examples of how to write the amplitudes \eqref{m5 gr CBCJ} and their classical pieces \eqref{M5 classical CBCJ}-\eqref{qedclas}, in the gauge \eqref{eq:shifted numeratos}, for both, the $\frac{1}{2}\otimes \frac{1}{2}$ and the $0\otimes1$ theories.

\subsubsection{Case  \texorpdfstring{$s_{a}=s_b=0+1$}{s10}}\label{sec:sasb01}
We want to compute  the gravitational amplitude for  inelastic scattering $M_5^{(0\otimes1,0\otimes1)}$ using \eqref{m5 gr CBCJ}. The scalar numerators are given by 
\begin{align}
n_{1}^{\prime(0,0)} & =e^{3}\frac{8p_{1}{\cdot}k\left(p_{1}{\cdot}F{\cdot}p_{3}{-}q{\cdot}F{\cdot}p_{3}\right){+}2\left(4p_{3}{\cdot}k{-}4p_{1}{\cdot}p_{3}{-}q{\cdot}(q-k)\right)q{\cdot}F{\cdot}p_{1}}{q{\cdot}k},\label{eq:numeratorBCJ1}\\
n_{4}^{\prime(0,0)} & =e^{3}\frac{8p_{3}{\cdot}k\left(p_{1}{\cdot}F{\cdot}p_{3}{-}q{\cdot}F{\cdot}p_{1}\right){+}2\left(4p_{1}{\cdot}k{-}4p_{1}{\cdot}p_{3}{-}q{\cdot}(q+k)\right)q{\cdot}F{\cdot}p_{3}}{q{\cdot}k}.\label{eq:numerator BCJ2}
\end{align}
Observe these numerators contain $q^2$ dependence. Nevertheless it is completely quantum as the only classical piece is the leading order in $q$
where $R_i^{\mu\nu}{=}p_{i}^{[\mu}(\eta_i 2q{-}k)^{\nu]}$, and  $\eta_1{=}{-}1,\eta_3=1$.
 The numerators for the spinning case are constructed following the considerations of Sec. \ref{sec:casee} and give
\begin{equation}\label{n1p11}
\begin{split}
n_{1}^{\prime(1,1)} &{=}\frac{2e^{3}}{q{\cdot}k}\bigg\{\left[\left(q^{2}{-}q{\cdot}k{+}4p_{1}{\cdot}p_{3}\right)q{\cdot}F{\cdot}p_{1}{+}4(q{-}p_{1}){\cdot}k\,p_{1}{\cdot}F{\cdot}p_{3}\right]\varepsilon_{1}{\cdot}\varepsilon_{2}\,\varepsilon_{3}{\cdot}\varepsilon_{4}+\\
 & \bigg[8\left(q{-}p_{1}\right){\cdot}k\,q{\cdot}\epsilon_{2}\,q_{\mu}\epsilon_{1}^{\alpha}F_{\alpha\nu}+4\left[q{\cdot}k\left(2p_{1\mu}q_{\nu}{+}(q{-}p_{1})_{\mu}k_{\nu}\right){+}p_{1}{\cdot}k\,k_{\mu}q_{\nu}\right]\varepsilon_{1}{\cdot}F{\cdot}\varepsilon_{2}+\\
 & 4q_{\mu}\left(2q{\cdot}\varepsilon_{1}\,p_{1}{\cdot}k{-}k{\cdot}\varepsilon_{1}\,q{\cdot}k\right)\varepsilon_{2}^{\alpha}F_{\alpha\nu}{-}\left[4p_{1}{\cdot}k\,q_{\alpha}k_{\beta}\varepsilon_{1}^{[\alpha}\varepsilon_{2}^{\beta]}{+}q{\cdot}k\,k{\cdot}\varepsilon_{1}\left(2q{-}k\right){\cdot}\varepsilon_{2}\right]F_{\mu\nu}+\\
 & \left[2\left(q{-}p_{1}\right){\cdot}k\left(4q_{\mu}p_{1}^{\alpha}F_{\nu\alpha}{+}p_{1}{\cdot}k\,F_{\mu\nu}\right){+}4p_{1\mu}\left(2q-k\right)_{\nu}q{\cdot}F{\cdot}p_{1}\right]\varepsilon_{1}{\cdot}\varepsilon_{2}-\\
 & 4\left(p_{1}{\cdot}k\,q_{\rho}F^{\rho\sigma}{+}q{\cdot}k\,p_{1\rho}F^{\rho\sigma}{+}2q^{\sigma}\,q{\cdot}F{\cdot}p_{1}\right)\varepsilon_{1[\mu}\varepsilon_{2\sigma]}\left(2q{-}k\right)_{\nu}{-}4q{\cdot}k\,q{\cdot}F{\cdot}\varepsilon_{1}\varepsilon_{2\mu}\left(2q{-}k\right)_{\nu}\bigg]\\
 &\times \varepsilon_{3}^{[\mu}\varepsilon_{4}^{\nu]}{+} \bigg[4q{\cdot}\varepsilon_{2}\left(q{-}p_{1}\right){\cdot}k\,p_{3}{\cdot}F{\cdot}\varepsilon_{1}{+}2\left(2q{\cdot}\varepsilon_{1}\,p_{1}{\cdot}k{-}q{\cdot}k\,k{\cdot}\varepsilon_{1}\right)p_{3}{\cdot}F{\cdot}\varepsilon_{2}\\&{-}8p_{3\mu}q_{\nu}q{\cdot}F{\cdot}p_{1}\varepsilon_{1}^{[\mu}\varepsilon_{2}^{\nu]}-
  2p_{1}{\cdot}k\,p_{3}{\cdot}\varepsilon_{1}q{\cdot}F{\cdot}\varepsilon_{2}{-}4q{\cdot}k\,p_{3\mu}p_{1}^{\alpha}F_{\alpha\nu}\varepsilon_{1}^{[\mu}\varepsilon_{2}^{\nu]}{-}\\
 & 2\left(2q{-}p_{1}\right){\cdot}k\,p_{3}{\cdot}\varepsilon_{2}q{\cdot}F{\cdot}\varepsilon_{1}+ \left(q{\cdot}k\left(q^{2}{-}q{\cdot}k{+}4p_{1}{\cdot}p_{3}\right)-2\left(q{-}p_{1}\right){\cdot}k\,p_{3}{\cdot}k\right)\varepsilon_{1}{\cdot}F{\cdot}\varepsilon_{2}\bigg]\varepsilon_{3}{\cdot}\varepsilon_{4}\bigg\}.
 \end{split}
\end{equation}
The numerator $n_{4}^{\prime(1,1)}$ is given by exchanging particles $ a\leftrightarrow b$ in $n_{1}^{\prime(1,1)}$, with $q\rightarrow-q$. The result expressed in terms of these numerators is far more compact than the Feynman diagram expansion obtained from the covariantized Lagrangian \eqref{lagrangian01x01}.

Now, by taking the classical limit of the numerators \eqref{n1p11} we  can compute the amplitude $\langle M_5^{(0\otimes1,0\otimes1)}\rangle$ via \eqref{M5 classical CBCJ}, using also \eqref{n1primeclassical} and \eqref{classical denominators}. In the multipole form of the previous section, the numerators read, up to dipole order,

\begin{equation}\label{spin1classical numerators}
    \begin{split}
\langle n_{1}^{\prime(1,1)}\rangle & =\langle n_{1}^{\prime(0,0)}\rangle{-}4e^{3}\left[p_{1}{\cdot}R_{3}{\cdot}kF{\cdot}J_{1}{-}F_{1q}R_{3}{\cdot}J_{1}{+}p_{1}{\cdot}k\,[F,R_{3}]{\cdot}J_{1}-p_{1}{\cdot}F{\cdot}\hat{R}_{3}{\cdot}p_{1}\right],\\
\langle n_{4}^{\prime(1,1)}\rangle & =\langle n_{4}^{\prime(0,0)}\rangle{-}4e^{3}\left[p_{3}{\cdot}R_{1}{\cdot}kF{\cdot}J_{3}{-}F_{3q}R_{1}{\cdot}J_{3}{+}p_{3}{\cdot}k\,[F,R_{1}]{\cdot}J_{3}-p_{3}{\cdot}F{\cdot}\hat{R}_{1}{\cdot}p_{3}\right]. 
\end{split}
\end{equation}
where 
\begin{equation}
   \langle n_{1}^{\prime(0,0)}\rangle{=}\frac{8e^{3}}{q{\cdot}k} p_1{\cdot}R_3{\cdot}F{\cdot}p_1, \quad 
   \langle n_{4}^{\prime(0,0)}\rangle{=}\frac{8e^{3}}{q{\cdot}k} p_3{\cdot}R_1{\cdot}F{\cdot}p_3, \label{n1primeclassical}
\end{equation}
Notice up to the spurious pole $q{\cdot}k$, which cancel in formulas \eqref{qedclas} and \eqref{M5 classical CBCJ} via \eqref{classical denominators}, these numerators agree with the classical ones provided in \eqref{nphsc} and \eqref{eq:lineas-in-spin-numerators-photon}. This indeed provides the direct connection for the derivation of classical radiation  using the \ac{BCJ} or the Compton residues of \cref{ch:electromagnetism}.

\subsubsection{Case  \texorpdfstring{$s_{a}=s_b=\frac{1}{2}+\frac{1}{2}$}{s1212}}
The final case for inelastic scattering in the gauge \eqref{eq:shifted numeratos} is given by the factorization of the  gravitational amplitude \eqref{m5 gr CBCJ} as $s_{a}=s_b=\frac{1}{2}+\frac{1}{2}$. For the \ac{QCD} theory, the shifted numerators entering in \eqref{m5qcd GGT} are
\begin{align}
n_{1}^{\prime(\frac{1}{2},\frac{1}{2})} & =\frac{4e^{3}F_{\alpha\beta}}{q{\cdot}k}\big[q^{[\alpha}p_{1}^{\beta]}\bar{u}_{2}\gamma^{\mu}u_{1}\bar{u}_{4}\gamma_{\mu}u_{3}{+}(q{-}p_{1}){\cdot}k\,\bar{u}_{2}\gamma^{[\alpha}u_{1}\bar{u}_{4}\gamma^{\beta]}u_{3}{-}\frac{q{\cdot}k}{4}\,\bar{u}_{2}\gamma^{[\alpha}\gamma^{\beta]}\gamma^{\mu}u_{1}\bar{u}_{4}\gamma_{\mu}u_{3}\big],\\
n_{4}^{\prime(\frac{1}{2},\frac{1}{2})} & =\frac{4e^{3}F_{\alpha\beta}}{q{\cdot}k}\big[q^{[\alpha}p_{3}^{\beta]}\bar{u}_{4}\gamma^{\mu}u_{3}\bar{u}_{2}\gamma_{\mu}u_{1}{+}(q{+}p_{3}){\cdot}k\,\bar{u}_{4}\gamma^{[\alpha}u_{3}\bar{u}_{2}\gamma^{\beta]}u_{1}{-}\frac{q{\cdot}k}{4}\,\bar{u}_{4}\gamma^{[\alpha}\gamma^{\beta]}\gamma^{\mu}u_{3}\bar{u}_{2}\gamma_{\mu}u_{1}\big].
\end{align}
Analogously, their  charge conjugated pairs read
\begin{align}
\bar{n}_{1}^{\prime(\frac{1}{2},\frac{1}{2})} & =\frac{4e^{3}}{q{\cdot}k}F_{\alpha\beta}\big[q^{[\alpha}p_{1}^{\beta]}\bar{v}_{1}\gamma^{\mu}v_{2}\bar{v}_{3}\gamma_{\mu}v_{4}{+}(q{-}p_{1}){\cdot}k\,\bar{v}_{1}\gamma^{[\alpha}v_{2}\bar{v}_{3}\gamma^{\beta]}v_{4}{+}\frac{q{\cdot}k}{4}\,\bar{v}_{1}\gamma^{\mu}\gamma^{[\alpha}\gamma^{\beta]}v_{2}\bar{v}_{3}\gamma_{\mu}v_{4}\big],\\
\bar{n}_{4}^{\prime(\frac{1}{2},\frac{1}{2})} & =\frac{4e^{3}}{q{\cdot}k}F_{\alpha\beta}\big[q^{[\alpha}p_{3}^{\beta]}\bar{v}_{3}\gamma^{\mu}v_{4}\bar{v}_{1}\gamma_{\mu}v_{2}{+}(q{+}p_{3}){\cdot}k\,\bar{v}_{3}\gamma^{[\alpha}v_{4}\bar{v}_{1}\gamma^{\beta]}v_{2}{+}\frac{q{\cdot}k}{4}\,\bar{v}_{3}\gamma^{\mu}\gamma^{[\alpha}\gamma^{\beta]}v_{4}\bar{v}_{1}\gamma_{\mu}v_{2}\big].
\end{align}

The  gravitational amplitude $M_5^{(\frac{1}{2}\otimes\frac{1}{2},\frac{1}{2}\otimes\frac{1}{2})}$ can be computed from the double copy of the above  numerators with their charge conjugated pairs, using  the operation defined in  \eqref{eq:doublestar-1-1}. The result  is in complete agreement with the Feynman diagrammatic computation from the  Lagrangian  \eqref{eq:1212LAGRANGIAN}. 

On the classical side, although the classical limit of these \ac{QCD} numerators agrees with \eqref{spin1classical numerators} (with appropriate conjugated numerators and up to dipole order), it is clear that the double copy  $\langle M_5^{(\frac{1}{2}\otimes\frac{1}{2},\frac{1}{2}\otimes\frac{1}{2})}\rangle$ differs  from  $\langle M_5^{(0\otimes1,0\otimes1)}\rangle$. For instance, as the double copy for the former is symmetric in the numerators the axion field has no radiative amplitude, whereas for the latter it is unavoidably present.

We do not provide the explicit result for  $\langle M_5^{(\frac{1}{2}\otimes\frac{1}{2},\frac{1}{2}\otimes\frac{1}{2})}\rangle$, but let us mention that it is naturally computed using the symmetric double copy product defined in \cref{sec:spin_in_qed} which preserves the multipole structure of the amplitude, and recovers the results of previous chapters.

\section{Outlook of the chapter}\label{sec:discusion_dc}

Based on the analysis performed along refs. \cite{Porrati:2010hm,osti_4073049,Deser:2000dz,Holstein:2006pq,Pfister_2002} and in the current work we can draw an equivalence for lower spins between the following three statements:
\begin{enumerate}
\item The cancellation of $\frac{1}{m}$ divergences in the tree-level high-energy
limit of single matter lines.
\item The ``natural value'' of the gyromagnetic ratio
$g=2$.
\item The double copy construction for the single matter line ($A_{n}$) amplitudes.
\end{enumerate}

Let us remark that this equivalence not only seems to show up in \ac{QFT} amplitudes but also in classical solutions \cite{Pfister_2002}. One instance of this is the so-called $\sqrt{\text{Kerr}}$ solution in electrodynamics which has been the focus of recent studies \cite{Arkani-Hamed:2019ymq,Guevara:2019fsj}. This EM solution can be double-copied into the Kerr metric via the Kerr-Schild ansatz \cite{Monteiro:2014cda}, and also features $g=2$. Since these classical solutions contain the full tower of spin-multipoles, and so do higher spin particles in QFT, a natural question that arises is: \textit{How much of the above equivalence can be promoted to higher spins?} 

A hint of the answer may come from the 3-pt amplitudes first derived in \cite{Arkani-Hamed:2017jhn} which are directly related to the aforementioned classical solutions \cite{Guevara:2017csg,Guevara:2018wpp,Guevara:2019fsj,Chung:2018kqs,Chung:2019duq,Arkani-Hamed:2019ymq}, at least at leading order in the coupling. In \cref{ch:electromagnetism} and \cref{sec:spin_in_qed} we have emphasized their double copy structure, which fixes not only $g=2$ but also the full tower of multipoles in both gravity and \ac{QCD} side. Here we have pointed out that these objects are in correspondence with higher spins massless amplitudes, thereby providing an underlying reason for double copy. Quite paradoxically, the latter are known to be inconsistent \cite{Benincasa:2007xk} whereas the former have an striking physical realization. To elucidate this contradiction we recall that massless higher spin amplitudes only fail at the level of the "4-particle" test \cite{Benincasa:2007xk,Arkani-Hamed:2017jhn}. 

Indeed, the higher spin 4-point (Compton) $A_4$ amplitudes suffer from ambiguities in the form of contact terms and from $\frac{1}{m}$ divergences, although recent progress to understand these has been made in \cite{Guevara:2018wpp,Chung:2018kqs,Chung:2019duq,Bautista:2019tdr}, we will go back to this point in \cref{ch:GW_scattering}. The importance of this object at higher spins was emphasized  in \cite{Guevara:2017csg} and proposed to control the subleading order associated to gravitational and EM classical potentials. These potentials emerge in the two-body problem \cite{Chung:2018kqs,Chung:2019duq,Damgaard:2019lfh,Maybee:2019jus,Neill:2013wsa,Holstein:2008sw,Holstein:2008sx}  (particularly outside the test body limit) and hence their understanding could have not only theoretical but practical implications. In fact, the relevance of the full tower of $A_n$ amplitudes lies in that they have been proposed to control the classical piece of conservative potentials at deeper orders in the coupling \cite{Neill:2013wsa,Cheung:2018wkq,Bern:2019crd,Bern:2019nnu,Bjerrum-Bohr:2019nws}. 

In \cref{ch:electromagnetism} and \cref{sec:spin_in_qed} we demonstrated the latter fact is true also for radiation: At least at order ${\sim} \kappa^3 $ and at spins $s\leq 2$ the non-conservative observables are controlled by $A_4$ and $A_3$ instead of the full $M_5$ amplitude. Here we have rederived this construction from a \ac{BCJ} double-copy perspective and use it to make contact with the results of Goldberger et al. \cite{Goldberger:2016iau,Goldberger:2017frp,Goldberger:2017vcg,Goldberger:2017ogt,Li:2018qap} for the full massless spectrum including dilatons, axions and gravitons. As we have mentioned it is remarkable how via \ac{QFT} double copy we have found the precise couplings of these fields to matter, besides the aforementioned $g=2$ condition. On the practical side it is important to evaluate the relevance of these additional fields, as well as string theory corrections, from the perspective of effective classical potentials arising from amplitudes, see e.g. \cite{Emond:2019crr,Brandhuber:2019qpg} for recent related results.

%% file: Chapters/chapter_higher_spins.tex
\chapter{Spinning amplitudes and the Kerr Black Hole}\label{ch:GW_scattering}
\section{Introduction}\label{sec:intro_gwscatt}

In previous chapters we have studied classical electromagnetic and gravitational observables directly from the classical limit of spinning quantum amplitudes. We have learnt how to approach the conservative and radiative sectors for both, unbounded and bounded scenarios to leading and subleading orders in the perturbative expansion, but keeping spin effects. In particular, \cref{sec:spin_in_qed} we have learned how the burst memory waveform for the hyperbolic scattering of classical astrophysical objects is controlled by the universality of the gravitational Weinberg soft factor. In the same way for the bounded scenario, in \cref{ch:bounded} we have shown how the spinning 5-point amplitude encapsulates the radiative dynamics of a coalescing \ac{BBH} with Kerr components, whose  spins are aligned with the direction of the angular momentum of the binary.  These amplitude description of classical processes hints a strong correspondence  between the  SO$(3)$ spin multipole moments of the minimal coupling classical gravitational amplitudes, and the  spin multipole moments of the Kerr ac{BH}. 

Currently, it is in general widely accepted that minimal coupled spinning amplitudes indeed encode vast part  of the information encoded in the Kerr \ac{BH}. In particular, and  since as already mentioned, the exponential structure of the gravitational 3-point amplitude can be mapped  to the exponential structure of the linearized effective metric for the Kerr \ac{BH} in momentum space, as shown in the seminal works \cite{Guevara:2018wpp,Chung:2018kqs,Arkani-Hamed:2019ymq}, which were at the same time inspired by previous work by \cite{Holstein:2008sw,Holstein:2008sx,Vaidya:2014kza}. This fact   was  then  used to construct two-body observables  for the conservative sector up to 2\ac{PM}    and to quartic order in spin \cite{Guevara:2018wpp,Chen:2021qkk,Bautista:2021inx}, whose results are in agreement with other approaches to the two-body problem such as the worldline \ac{EFT}    approaches (see for instance \cite{Kim:2021rfj,Cho:2021mqw,Jakobsen:2022fcj,Siemonsen:2019dsu}  ) and \ac{EFT}    approaches \cite{Bern:2020buy,Kosmopoulos:2021zoq} . In \cite{Siemonsen:2019dsu} it was shown how the predictions for the aligned spin 2\ac{PM}    scattering function of \cite{Guevara:2018wpp} agree up to third order in the \ac{BH}'s spins result  expected from  self-force computation\footnote{In this approximation, the two-body problem is assumed to have one black hole of mass $M$ and the other with mass $m$, so that $m/M\ll1$. }. 
However, a deeper understanding of the agreement of this different approaches to the two-body problem is needed. If it is true we understand very well how the 3-point amplitude contains all the spin structure of the Kerr \ac{BH}, we have learned that it is not the only building block in the construction of two-body amplitudes, but instead we have a full tower of $A_n$ amplitudes which can be used to construct unitarity cuts to higher orders in perturbation theory. 

\begin{figure}
    \centering
    \includegraphics[width=50mm]{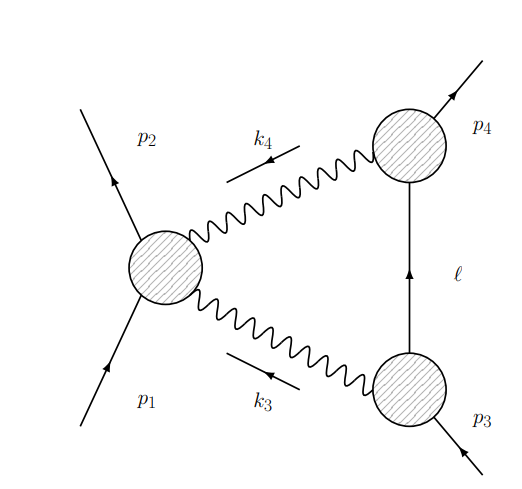}
    \caption{Triangle leading-singularity configuration \cite{Cachazo:2017jef}. The gravitational Compton amplitude is glued to two 3 point amplitudes, where the internal lines are on-shell. The Leading singularity corresponds to the loop integration of this amplitude, where internal gravitons are soft as compared to the external massive lines.}
    \label{leading-singularity}
\end{figure}

This then calls for a study of these $A_n$ amplitudes and their direct correspondence to the Kerr \ac{BH}. In this chapter we aim to initiate such study for the  simple cases $n=3,4$. In particular, we will review how for $n=3$ we reproduce the expected linearized Kerr metric solution, whereas for $n=4$, we will argue $A_4$ is an effective description of the low energy regime for the  scattering of gravitational waves off the KBH, the latter of which is traditional studied using \ac{BHPT}. At 2\ac{PM}, these amplitudes are sufficient to obtain the aligned spin scattering function thought the  triangle Leading-Singularity  \cite{Cachazo:2017jef,Guevara:2018wpp}, of Figure \ref{leading-singularity}

This chapter is organized as follows: In \cref{sec:esponential34} we study in more detail amplitudes $A_n$  for $n=3.4$ in spinor-helicity variables, and show in the infinite spin limit, they can be arranged in an exponential form, as originally proposed by \cite{Guevara:2018wpp,Aoude:2020onz}. In \cref{sec:class_l} we show how to extract the classical information of such amplitudes. For $A_3$ we indeed recover the exponential form of the linearized Kerr effective metric, whereas for $A_4$ we recover the classical exponential form suggested by the on-shell heavy particle \ac{EFT}    of \cite{Aoude:2020onz} . In \cref{sec:gw_Scatt} we use the classical Compton amplitude to study the low energy limit for the scattering of a gravitational wave off the Kerr \ac{BH},     up to fourth order in the spin of the \ac{BH}.     Spin induced polarization of the wave after the scattering process is discussed. In \cref{ap:hcl} we show the classical $A_4$ can indeed be used to reproduce the 2\ac{PM}    scattering angle computation of \cite{Guevara:2018wpp}. We leave for \cref{ch:teukolsky} Teukolsky  formulation for the scattering of the gravitational wave off Kerr, and argue it agrees with the amplitudes derivation of the present chapter up to fourth order in spin.

This chapter is mainly based on work \cite{Bautista:2021wfy}, as well as  work in progress \cite{BGKV}.

\section{Exponential 3 and 4 point spinning amplitude}\label{sec:esponential34}

Following \cite{Guevara:2018wpp}, and as reviewed in \cref{sec:spinor-helicity}, 
in terms of the angular momentum operator in spinor helicity variables, the gravitational 3-point and 4-point amplitudes  can be put into the exponential form \eqref{eq:3pt} and \eqref{eq:compts} respectively. For $A_3$, in \cref{sec:spin_in_qed} we saw how an analogous formula holds in covariant notation up to quadratic order in spin, and in \cref{sec:branching} we provided a local form of the 3-point amplitude to all orders in spin in $D=4$. For the 4-point amplitude in covariant form, such exponentiation is not evident, whereas in spinor-helicity variables it is immediate to get as given in  \eqref{eq:s14d} for the opposite helicity configuration\footnote{An analogous formula can be found for the same helicity configuration as we will discuss below. }.

In this section we aim to extract the classical limit of amplitudes  \eqref{eq:3pt} and \eqref{eq:compts}, and show they agree with the classical description of the Kerr \ac{BH}.     
Let us rewrite explicitly the amplitudes \eqref{eq:3pt} and \eqref{eq:compts} for the reader's convenience
\begin{equation}\label{ampli}
        A^{S}_{3}=A^{0}_3 \times \langle \varepsilon_3 | \exp\left(\frac{F_{2\mu \nu}J^{\mu\nu}}{2i\epsilon_2\cdot p_1}\right) | \varepsilon_1 \rangle\,, A^{S}_{4}=A^{0}_4 \times \langle \varepsilon_4 | \exp\left(\frac{F_{2,\mu \nu}J^{\mu\nu}}{2i \epsilon_2 \cdot p_1}\right) | \varepsilon_1 \rangle 
\end{equation}
 Recall that the graviton polarization vector is given by $\epsilon_{\mu \nu}=\epsilon_\mu \epsilon_\nu$ and we have defined $F_{\mu \nu}=2k_{[\mu}\epsilon_{\nu]}$. Let us also use  momentum conservation, respectively for the 3-point and 4-point amplitudes as follows:
\begin{eqnarray}\label{34kins}
p_3&=& p_1+k_2 \nonumber \\
p_4 &=& p_1 + k_2 + k_3  \,.
\end{eqnarray}
In both cases we assume the graviton associated to $k_2$ to have negative helicity, and for $n=4$ the graviton $k_3$ has positive helicity\footnote{This is somehow opposite to the conventions used in \cref{ch:electromagnetism}, where the $k_3$ graviton had opposite momentum as compared to conventions here, to connect to previous section, we simply take $k_3\to-k_3$ here,  which also flips its helicity from positive to negative. }. The gauge is fixed, in spinor-helicity variables, as \cite{Guevara:2018wpp}

\begin{equation}\label{eq:chkgauge}
    \epsilon_2=\frac{\sqrt{2} |3]\langle 2|}{[32]} \propto \tilde{\epsilon}_3 = \frac{\sqrt{2} |3]\langle 2|}{\langle32\rangle }\,.
\end{equation}
The operator $J_{\mu\nu}$ in \eqref{ampli} is a Lorentz generator in the spin-$s$ representation. In this case we will realize it as a fully quantum operator acting linearly on the representation $|\varepsilon\rangle$. The exponential series truncates at order $2s$ in the expansion of the exponential. The pole $\epsilon\cdot p$ will cancel in the cases treated here. As anticipated, this effectively restricts $S\leq 2$ in the Compton amplitude $A_4$, since as mentioned in \cref{sec:spinor-helicity}, unphysical poles, which we shall not discuss in this thesis, arise from the exponent in the 4-point case which cannot be canceled by the scalar amplitude.  In previous amplitudes we have introduced a factor of $i$ in the exponent and defined the Lorentz generators with that extra factor.

As explained in \cref{sec:branching}, in order to extract the classical piece of the spinning amplitude, we need to align the polarization states towards the same little group. 
We therefore introduce a four-velocity vector $u^\mu$ together with a generic mass scale $m$. They will be mapped to the four-velocity and rest frame mass of the classical object, respectively. However, the identification with the kinematic momenta in the Compton amplitude is ambiguous, some choices are  $u=\frac{p_1}{m}, \frac{p_4}{m}, \frac{p}{m}$ (with $m=M$ in the former cases, and $m^2=p^2,p=\frac{p_1+p_4}{2}$ in the latter), which will all coincide after we take the classical limit. Now, in order to parametrize the degrees of freedom associated with spin in four dimensions we introduce the Pauli-Lubanski operator
\begin{equation}\label{spinvec}
    a^{\mu} :=\frac{1}{2m} \epsilon^{\mu \nu \rho \sigma } u_{\nu} J_{\rho \sigma}
\end{equation}
This will play the role of the spin vector introduced in previous sections. However, this gives us only a  classical  relation between  $J^{\mu \nu}$ and the  spin vector $a^\mu$. Using spinor-helicity variables we can find an exact quantum relation between operators. For this, note that in  \eqref{ampli} the field strength $F_2^{\mu\nu}$ is self-dual since the graviton $k_2$ has negative helicity. Consequently, the generator $J_{\mu \nu}$ is also self-dual and it is associated with the chiral basis \eqref{basi1}, i.e. $J^{\mu \nu}=\frac{i}{2} \epsilon^{\mu \nu \rho \sigma}J_{\rho \sigma}$. \footnote{More precisely, we have \cite{Guevara:2018wpp}

\begin{equation}
     \langle \varepsilon_4 | \exp\left(\frac{F_{2,\mu \nu}J^{\mu\nu}}{2i \epsilon_2 \cdot p_1}\right) | \varepsilon_1 \rangle =  [ \varepsilon_4 | \exp\left(\frac{F_{3,\mu \nu}\tilde{J}^{\mu\nu}}{2i \epsilon_3 \cdot p_1}\right) | \varepsilon_1 ]\,, \nonumber
\end{equation}

i.e. using the negative helicity graviton also changes the chirality of the Lorentz generator.
} We use this property to rewrite the exponents of \eqref{ampli} in terms of the spin vector \eqref{spinvec} as follows. Following the discussion of  \cref{sec:branching} , for a given 4-velocity $u^{\mu}$ we decompose the full Lorentz generator $J^{\mu \nu}$ into a spin and a boost operator:

\begin{equation}\label{eq:BandS}
B^\mu:=J^{\mu \nu} u_\nu\, ,\qquad S^{\mu \nu}:=J^{\mu \nu} - 2 u^{[\mu} B^{\nu]} \,.
\end{equation}
 One can easily check that $u_{\mu}S^{\mu \nu}=0$, hence $S^{\mu\nu}$ generates little group transformations on states $|\varepsilon\rangle$ and shall be related to the Pauli-Lubanski vector $a^{\mu}$. Indeed, from \eqref{spinvec} one easily finds
\begin{equation}\label{aprop}
     a^{\mu} =\frac{1}{2m} \epsilon^{\mu \nu \rho \sigma } u_{\nu} S_{\rho \sigma} \Leftrightarrow S^{\mu \nu} = - m \epsilon^{\mu \nu \rho \sigma} u_{\rho} a_{\sigma} \,.
\end{equation}
Furthermore, due to the self-dual condition on $J^{\mu \nu}$, it turns out that the boost and spin parts are indeed related. From \eqref{spinvec} and \eqref{eq:BandS} we find:

\begin{equation}\label{bprop}
    B^\mu = i m a^\mu \,.
\end{equation}
We can now decompose the exponent of \eqref{ampli}. We proceed for both $n=3,4$ at the same time, introducing the generic field strength $F_{\mu \nu}=2 k_{[\mu} \epsilon_{\nu]}$. Using \eqref{aprop} and \eqref{bprop} we have
\begin{eqnarray}
F_{\mu \nu} J^{\mu \nu} &=& F_{\mu \nu} S^{\mu \nu} + 2 u_\mu F^{\mu \nu} B_\nu \nonumber \\
&=& -m \, \epsilon^{\mu \nu \rho \sigma} F_{\mu \nu} u_\rho a_\sigma + 2i m\, u_\mu F^{\mu \nu} a_\nu \,.
\end{eqnarray}
Regarding $F_{\mu \nu}$ as self dual, which follows from the contraction with $J^{\mu \nu}$ on the \ac{LHS}, we finally get
\begin{equation}
  F_{\mu \nu} J^{\mu \nu}   = \\ac{PM}    4im\,  u_\mu F^{\mu \nu} a_\nu \,.
\end{equation}
The $\pm$ sign accounts for self-duality or anti self-duality of the Lorentz generator $J_{\mu \nu}$, or equivalently, the helicity associated to $F_{\mu \nu}$. We remark that the classical limit has not yet been applied. Note that the LHS does not depend on the four-vector $u^\mu$, which we are free to choose. In any case, for $u=\frac{p_1}{M}, \frac{p_4}{M}, \frac{p}{m}$ we can now rewrite \eqref{ampli} as

\begin{equation}\label{ampli2}
    A^{S}_{3}=A^{0}_3 \times \langle \varepsilon_3 | \exp\left(2\frac{u\cdot F_2 \cdot a}{u\cdot \epsilon_2}\right) | \varepsilon_1 \rangle\,, A^{S}_{4}=A^{0}_4 \times \langle \varepsilon_4 | \exp\left(2\frac{u\cdot F_2 \cdot a}{u\cdot \epsilon_2}\right) | \varepsilon_1 \rangle 
\end{equation}
For $n=3$ we have $u\cdot k_2=0$ from the on-shell conditions. This automatically implies that the pole $u\cdot \epsilon_2$ cancels and we have

\begin{equation}\label{3ptex}
    A^{S}_{3}=A^{0}_3 \times \langle \varepsilon_3 | e^{-2 k_2 \cdot a}  | \varepsilon_1 \rangle \,.
\end{equation}
For $n=4$, the pole does not cancel in the exponential, as $u\cdot k_2\neq 0$ generically. Since the prefactor $A_4^0$ contains a term $(u\cdot \epsilon_2)^4$, the form \eqref{ampli2} is valid only up quartic order in the expansion of the exponential, i.e. up to spin $S=2$. We can encode the unphysical pole in the vector 

\begin{equation}\label{eq:wdef}
    w^\mu := \frac{u\cdot k_2}{u\cdot \epsilon_2} \epsilon^\mu_2\,,
\end{equation}
so that
\begin{equation}\label{eq:4ptexpdef}
    A^{S}_{4}=A^{0}_4 \times \langle \varepsilon_4 | e^{2(w\cdot a - k_2\cdot a)} | \varepsilon_1 \rangle \,.
\end{equation}

The polarization states $|\varepsilon_1\rangle,\langle\varepsilon_4|$ are associated with initial and final momentum, $p_1,p_4$ respectively. It will be convenient to rewrite them as associated to the 4-velocity $u^\mu$. For instance, taking $u=\frac{p_1}{M}$, we can write
\begin{equation}
    p_4= e^{i\mu M  (k_2+k_3)\cdot B } p_1=  e^{-\mu M^2 (k_2+k_3)\cdot a } p_1\,,
\end{equation}
Here $\mu$ is a scalar which explicit expression we do not need, but which is given explicitly in \cite{Guevara:2019fsj}. The analogous formula holds for $n=3$; in this case three-particle kinematics yields $\mu M^2 =1$, hence
\begin{equation}
    p_3=  e^{- k_2\cdot a } p_1\,,
\end{equation}
This implies that we can write
\begin{eqnarray}
  |\varepsilon_3\rangle   &=&   e^{-k_2\cdot a } |\varepsilon'_1\rangle \,,\qquad \,\,\qquad n=3 \\
  |\varepsilon_4\rangle   &=&   e^{-\mu M^2 (k_2+k_3)\cdot a } |\varepsilon'_1\rangle \,,\quad n=4
\end{eqnarray}
where $|\varepsilon'_1\rangle$ is a polarization state associated to $p_1=M u$. Thus we have the following \ac{QFT}     amplitudes

\begin{eqnarray}\label{eq:qftamp3}
    A^{S}_{3}= A^{0}_3 \times \langle \varepsilon'_1| e^{k_2\cdot a } e^{-2 k_2 \cdot a}  | \varepsilon_1 \rangle = A^{0}_3 \times \langle \varepsilon'_1| e^{-k_2\cdot a }  | \varepsilon_1 \rangle \,,
    \end{eqnarray}
and
\begin{eqnarray}\label{eq:qftamp4}
    A^{S}_{4}=A^{0}_4 \times \langle \varepsilon'_1 | e^{\mu M^2 (k_2+k_3)\cdot a } e^{2(w - k_2)\cdot a} | \varepsilon_1 \rangle \,.
\end{eqnarray}
The constraint $u\cdot a=0$ implies that the Pauli-Lubanski vector $a^\mu$ only yields three independent operators. In the rest frame of $u^\mu$ they satisfy $[a^i,a^j]=\epsilon^{ijk}a_k$, or covariantly

\begin{equation}\label{eq:a-commutator}
    [a^\mu,a^\nu]=M^{-1} S^{\mu \nu}=\epsilon^{\mu \nu \rho \sigma}a_\rho u_\sigma \,.
\end{equation}
In eq. \eqref{eq:qftamp3} only the combination $k_2 \cdot a$ appears. Furthermore, note that in this case the boost component $e^{k_2 \cdot a}$ commutes with the amplitude $e^{-2 k_2\cdot a}$. This is not the case for eq. \eqref{eq:qftamp4} where indeed all three combinations $k_2\cdot a,k_3\cdot a, w\cdot a$ appear and do not commute. As the spin is the only quantum number available, we assume that in general these combinations span a basis of operators in the space of states associated to $u^\mu$, namely $|\varepsilon_1\rangle, \langle \varepsilon'_1|$.

\subsection{Classical Limit}\label{sec:class_l}

As argued in the previous section, the operator $O$ in the contraction $\langle \varepsilon'_1 |O|\varepsilon_1\rangle$ can be attributed a classical nature. We note that the three-point amplitude \eqref{eq:qftamp3} is invariant under such limit

\begin{equation}\boxed{
\label{eq:classical3pt}
    \langle A_3^S\rangle  = \langle A_3^0\rangle e^{-k_2\cdot a},}
\end{equation}
whereas for the four-point \eqref{eq:qftamp4} we obtain 
\begin{equation}
    w^\mu,k_2^\mu,k_3^\mu\sim \hbar \,, \quad a^\mu \sim 1/\hbar \,.
\end{equation}
where the scaling of $w^\mu$ follows from its definition \eqref{eq:wdef}. Together \eqref{eq:a-commutator} this implies

\begin{equation}
    [(k_2+k_3)\cdot a , (w - k_2)\cdot a] \sim \hbar 
\end{equation}
i.e. the exponents of \eqref{eq:qftamp4} commute in the classical limit. Furthermore, from the explicit expression in \cite{Guevara:2019fsj} we see that $\mu M^2 = 1+\mathcal{O}(\hbar)$, hence the limit of \eqref{eq:qftamp4} becomes

\begin{eqnarray}\label{eq:qftamp4clas}
 A^{S}_{4} =A^{0}_4 \times    \langle \varepsilon'_1|e^{(2w +k_3 - k_2)\cdot a}|\varepsilon_1\rangle +\mathcal{O}(\hbar)  \Longrightarrow \boxed{ \langle A^{S}_{4} \rangle =\langle A^{0}_4  \rangle \times  e^{(2w +k_3 - k_2)\cdot a}  \,.}
\end{eqnarray}
This result agrees with the one obtained in \cite{Aoude:2020onz} from Heavy Particle \ac{EFT}.      This is expected since we have argued in  \cite{Bautista:2021wfy} that the limits $\hbar \to 0$ and $M\to \infty$ are equivalent.
Note that in the last step of \eqref{eq:qftamp4clas} we have stripped off the polarization states $|\varepsilon_1\rangle, |\varepsilon'_1\rangle$. In this case, $a^\mu$ is interpret as a classical spin-vector and not an operator. 

In an analogous way, one can show that the classical limit for the same helicity configuration of the gravitational Compton amplitude is simply given by 

\begin{equation}\label{eq:samehcompton}\boxed{
    \langle \tilde{A}_4^S\rangle = \langle\tilde{A}_4^0\rangle e^{(k_3+k_2){\cdot}a}\,.}
\end{equation}  
which agrees with the result of \cite{Aoude:2020onz} from Heavy Particle \ac{EFT}.      

\subsection{Gravitational wave scattering}\label{sec:gw_Scatt}

Amplitude \eqref{eq:classical3pt} precisely agrees with the exponentiated spin structure of the linearized Kerr metric as shown in the seminal work \cite{Guevara:2018wpp}. The natural question to ask  is, what does $A_4$ have to do with the Kerr \ac{BH}?
In the follows, and in \cref{ch:teukolsky} we will argue that  \eqref{eq:qftamp4clas}  corresponds to a effective description of   the low energy regime of the a gravitational  wave scattering  off the Kerr \ac{BH}.     

\begin{figure}

\begin{equation*}
    \vcenter{\hbox{\includegraphics[width=140mm,height=42mm]{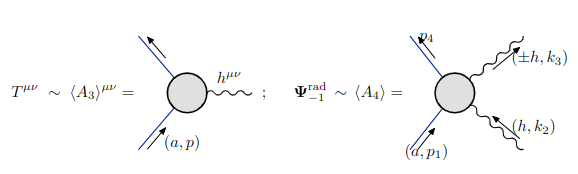}}} 
\end{equation*}

\caption{Left: Schematic representation of the correspondence between the spin multiple moments of the  Kerr \ac{BH} and the minimal coupling $3$ pt amplitude. Right: Graphic representation for the scattering of a plane wave of helicity $h$, off the Kerr \ac{BH}.  }
\label{eq:A4-figure}
\end{figure}

For that let us consider the gravitational analog of the Thomson scattering process in \ac{QED} as reviewed in \cref{ch:electromagnetism}. In Figure \ref{eq:A4-figure} (right) we do a schematic representation of the process from an amplitudes perspective (see also Figure \ref{fig:ampl}): A wave of helicity $h=2$ is scattered off the Kerr \ac{BH}.

In terms of the kinematics \eqref{eq:kinematics}\footnote{Here we take $k_3\to-k_3$ to use the conventions $p_1+k_2=k_3+p_4$.} in the classical limit, and using a generic orientation for the spin vector $a^\mu= (0,a_x,a_y,a_z)$, in the \ac{BH}      rest frame,  amplitudes \eqref{eq:qftamp4clas} and \eqref{eq:samehcompton} become
\begin{align}
   \langle A_{4}^{++}\rangle &= \frac{\kappa^2 M^2\cos^4(\theta/2)}{4\sin^2(\theta/2)}\big[1+\mathcal{F}(\omega,a,\theta)+\frac{1}{2!}\mathcal{F}(\omega,a,\theta)^2 +\frac{1}{3!}\mathcal{F}(\omega,a,\theta)^3+\frac{1}{4!}\mathcal{F}(\omega,a,\theta)^4 \big]\label{eq:App-gr}\,,\\
     \langle A_{4}^{- -}\rangle&=  \left[\langle A_{4}^{++}\rangle^*\right]_{\omega\rightarrow-\omega}\,,\\
    \langle A_{4}^{+ -}\rangle&= \frac{\kappa^2 M^2\sin^4(\theta/2)}{4\sin^2(\theta/2)}\big[\mathcal{G}(\omega,a,\theta) +\frac{1}{2!}\mathcal{G}(\omega,a,\theta)^2+\frac{1}{3!}\mathcal{G}(\omega,a,\theta)^3+\frac{1}{4!}\mathcal{G}(\omega,a,\theta)^4\big]\,\label{eq:Amp-gr}\,,\\
     \langle A_{4}^{- +}\rangle&=  \left[\langle A_{4}^{+ -}\rangle^*\right]_{\omega\rightarrow-\omega}\label{eq:Amp-grm}\,,
\end{align}
where we have truncated the expansion at $a^4$ ($S=2$), where the Compton amplitude has physical meaning, and we  have further used 
\begin{eqnarray}
     \mathcal{F} (\omega,a,\theta) &=& -2a_z\omega\sin^2(\theta/2) +a_x\omega\sin\theta-2(a_x-ia_y)\omega\tan(\theta/2)\,,\\
     \mathcal{G}(\omega,a,\theta) &= &2a_z\omega\sin^2(\theta/2) -a_x\omega\sin\theta \,.
\end{eqnarray}
which come naturally from rewriting  the exponent in \eqref{eq:qftamp4clas} and \eqref{eq:samehcompton}  respectively,  in terms of the scattering angle, using kinematics \eqref{eq:kinematics}. We have in addition  written all of the helicity configurations entering into the scattering matrix \eqref{eq:scattering-matrix-helicities}, which follow from changing the direction of the massless momenta. 
Up to $a^2$, one can easily show that the same result can be obtained starting from the quadratic in spin classical Compton amplitude derived from the covariant spin multipole double copy, and  written in vector notation in \eqref{eq:a4CLASSSPIN2}, once we use  \eqref{eq:photon polarization1} and \eqref{eq:photon polarization2} as the polarization states for the incoming and outgoing wave, respectively, and the kinematics \eqref{eq:kinematics} in the classical limit. This provides a strong consistency check for the amplitudes written in both, vector and spinor notation, and the validity of the spin multipole double copy introduced  in \cref{sec:spin_in_qed}.

Using \eqref{eq:classical cross section} we can then compute the unpolarized differential for the scattering of gravitational waves off  the Kerr \ac{BH}.     Up to quartic order in spin it is simply given by 
\begin{equation}\label{eq:sigmaspin2}
\begin{split}
     \frac{d\braket{\sigma}}{d\Omega} &= \frac{G^2M^2}{\sin^4(\theta/2)}\Big[\cos^8(\theta/2)\Big(1+2\tilde{\mathcal{F}}+\frac{(2\tilde{\mathcal{F}})^2}{2!}+\frac{(2\tilde{\mathcal{F}})^3}{3!}+\frac{(2\tilde{\mathcal{F}})^4}{4!}
     \Big)\\
     &\qquad\qquad\qquad+\sin^8(\theta/2)\Big(1+2\mathcal{G}+   \frac{(2\mathcal{G})^2}{2!}+ \frac{(2\mathcal{G})^3}{3!}+ \frac{(2\mathcal{G})^4}{4!}
     \Big)\Big)\Big]+\mathcal{O}(a^5)\,,
\end{split}
\end{equation}
where $\tilde{\mathcal{F}}=\mathcal{F}\big|_{a_y=0}$, and we have used $\kappa^2=32\pi G$. We then see the $a_y$ component of the spin corresponds to just a phase in the amplitude, unimportant for the cross section, as one could have guessed from the exponential structure of the amplitude.  Interestingly, the only spin components contributing to the actual observables are those with non zero projection on the scattering plane. 

The first difference we notice when comparison to the Thomson differential cross section for the scattering of electromagnetic waves off charge compact objects  \eqref{eq:Thomson} (set $f,g\to0$ for the moment) is that  unlike  for the latter, \eqref{eq:sigmaspin2} does diverge in the $\theta\to0$ limit. This is a forward divergence and is due to the long range nature of the gravitational potential. There is a second difference when spin is included, which manifests in a spin induced polarization of the incoming wave as we now discuss.

\subsection{Spin-induced Polarization}\label{sec:polapp}

In general, incoming waves can be linearly polarized, that is, they can be written as a superposition of  circularly polarized waves. When impinging on the black hole, waves of  different circular polarization can scatter by a different angle. This in turn will induce a polarization of the wave after  scattering, which  will  be reflected in the difference between elements of the scattering matrix \eqref{eq:scattering-matrix-helicities}. To see this explicitly we compare the scattering cross-sections for a left ($+$) and right ($-$) circularly polarized incoming wave:
\begin{equation}\label{eq:crsepm}
    \begin{split}
         64\pi^2 M^2 \frac{d\langle \sigma_+\rangle}{d\Omega}&=\langle A_{++} \rangle \langle A_{++} \rangle^* + \langle A_{+-} \rangle \langle A_{+-} \rangle ^*  \\
   64\pi^2 M^2 \frac{d\langle \sigma_-\rangle}{d\Omega}&=\langle A_{--} \rangle \langle A_{--} \rangle^* + \langle A_{-+} \rangle \langle A_{-+} \rangle ^* 
    \end{split}
\end{equation}
We have found from (\ref{eq:App-gr}-\ref{eq:Amp-grm}) that opposite helicity amplitudes are related via $ \langle A\rangle \to \langle A \rangle^*$, accompanied by the expected time reversal $\omega\to -\omega$, map. This is more transparent in the  spinor-helicity formalism, and can be seen as a consequence of CPT/crossing symmetry: Opposite helicities are related by chiral (i.e. complex) conjugation in the amplitude. This induces a parity transformation which flips the sign of $a^\mu$, which corresponds to a pseudovector as it describes the orientation of the rotating black hole. Due to the fact that spin only enters through the combination $a\omega$ the map $a^\mu \to -a^\mu$ is of course equivalent to $\omega\to -\omega$.

From the above discussion, using \eqref{eq:crsepm}, we easily conclude that

\begin{equation}\label{eq:symcr}
    \frac{d\langle \sigma_+\rangle}{d\Omega}=   \left[ \frac{d\langle \sigma_-\rangle}{d\Omega}\right]_{(\omega\to-\omega)}
\end{equation}
Following \cite{PhysRevD.16.237,Dolan:2008kf,Barbieri:2005kp} we also introduce the polarization measurement

\begin{equation}\label{eq:polarization-formula}
    \mathcal{P}=\frac{\frac{d\langle\sigma_+\rangle}{d\Omega}-\frac{d\langle\sigma_-\rangle}{d\Omega}}{\frac{d\langle\sigma_+\rangle}{d\Omega}+\frac{d\langle\sigma_-\rangle}{d\Omega}}\,.
\end{equation}
According to \eqref{eq:symcr} the numerator of the polarization depends only on odd powers of $a\omega$ in the cross-section, and in particular vanishes for the Schwarzschild case. Let us for simplicity restrict here   to  the  \textit{polar scattering} case, where the impinging wave moves along the direction of the spin of the \ac{BH},     i.e. $a_x=a_y=0$. Let us also consider the linear in spin term. Extension to general spin orientation is straightforwards to compute using the full expression for the differential cross section \eqref{eq:sigmaspin2}. In this case, the spin induced polarization simply reads 

\begin{equation}\label{eq:polariation}
    \mathcal{P} =  -\left(4a_z|\omega|\sin^2(\theta/2)\right) \frac{\cos^8(\theta/2)-\sin^8(\theta/2)}{\cos^8(\theta/2)+\sin^8(\theta/2)}\,,
\end{equation}
which for $\theta\rightarrow 0$ becomes $ \mathcal{P} =  -a_z|\omega|\theta^2$ and thus
recovers the classical result using  \ac{BHPT}  \cite{Dolan:2008kf} (see also \cref{ch:teukolsky}). It however disagrees with the prediction of  \cite{Guadagnini:2008ha,Barbieri:2005kp} to linear order in  spin. The reason for this mismatch is that  \cite{Guadagnini:2008ha,Barbieri:2005kp} only considered the graviton exchange diagram between the wave and the \ac{BH},     whereas in here we have shown in order to recover the classical result computed \ac{BHPT}, one needs to consider the full classical gravitational Compton amplitude. Indeed, in \cref{ch:teukolsky} we argue the cross section \eqref{eq:sigmaspin2} indeed matches in a  spectacular way the classical result obtained by solving the Teukolsky equation. This then allows us to conclude confidently the minimal gravitational Compton amplitude is indeed a equivalent description of  the scattering of gravitational waves off the Kerr \ac{BH}.     Let us stress this is the first time a direct connection between the classical piece of the  gravitational Compton amplitude and the Kerr \ac{BH}      is made.  Up to linear order in spin, the wave scattering process is independent whether the compact object corresponds to  a Kerr \ac{BH}      or any other spinning  of object; by the fact the  results for the spin monopole and dipole are universal. However, at quadratic and higher order in spins, the minimal coupling Compton amplitude uniquely describes Kerr \ac{BH}      and not any other compact object, due to its  unique spin multipole structure.   

\section{2\ac{PM}    Scattering Angle and the Holomorphic Classical Limit}\label{ap:hcl}
Let us in the remaining of this chapter show the classical result for the gravitational Compton amplitude can indeed be used to derive the aligned spin scattering angle for the scattering of two Kerr \ac{BH}s,
at order $G^2$ and up to $a^4$. This will recover the result obtained from the \ac{HCL} computation in \cite{Guevara:2018wpp}. 

Our aim is to compute the triangle leading singularity of Figure \ref{leading-singularity}  
\begin{equation}\label{LS}
\mathcal{M}_4=
\frac{i}{8m_b\sqrt{-t}} \int_{\Gamma_{\rm{LS}}} \frac{dy}{2\pi y} \langle A_4^{(s_a)}(P_1,-P_2,k_3^+,k_4^-)\rangle\langle A_3^{(s_b)}(P_3,-l,-k_3^-) \rangle \langle A_3^{(s_b)}(-P_4,l,-k_4^+)\rangle .
\end{equation}
where the brackets indicate that we take the classical limit of  the indicated amplitudes. It gives the scattering angle via the two dimensional Fourier transform from momentum space $k_\perp=(k_3-k_4)_\perp$ to impact parameter space $b_\perp$ (see eq. $(1.11)$ in \cite{Guevara:2018wpp}). 
The reason  we can reproduce the \ac{HCL} computation starting from the classical amplitudes is that as it turns out, the \ac{HCL} (i.e. $\beta\rightarrow1$ below) is contained in the classical limit, and it corresponds to the special case where the momenta for the two internal gravitons become proportional to each other\footnote{  \textcolor{black}{Effectively,   the \ac{HCL} is a complexification of  the classical limit which allow us to align $k_2$ and $k_3$ without having to take the scat angle  $\langle \xi \rangle$ in \eqref{eq:xihcl} to zero, ass opposite to the case in which the we enforce $k_2\propto k_3$ in the \textit{standard} classical limit. } }; this then implies that $k_2\,\alpha\, k_3\,\alpha \,w $, which in turn makes the basis for the spin directions with $3$ elements to get degenerated to include only one element, which corresponds to the direction in which the  spins of the two \ac{BH}s are aligned.  

Let us start our computation by considering that the two incoming black holes have momenta $P_1$ and $P_3$ and spin $a_a$ and $a_b$, respectively; likewise, the outgoing \ac{BH}s will have momentum $P_2$ and $P_4$, with their spins unchanged. The  \ac{HCL} parametrization  for the momenta of the massive particles in the center of mass frame is \cite{Guevara:2017csg,Chung:2018kqs} 

\begin{equation}\label{eq:kinematics HCL}
    \begin{split}
     P_1 &=|\hat{\eta}]\bra{\hat{\lambda}}+|\hat{\lambda}]\bra{\hat{\eta}},\\
     P_2 &=  \beta^\prime|\hat{\eta}]\bra{\hat{\lambda}}+\frac{1}{\beta ^\prime}|\hat{\lambda}]\bra{\hat{\eta}} +|\hat{\lambda}]\bra{\hat{\lambda}},\\
     P_3 &=|\eta]\bra{\lambda}+|\lambda]\bra{\eta},\\
     P_4 &=  \beta|\eta]\bra{\lambda}+\frac{1}{\beta}|\lambda]\bra{\eta} +|\lambda]\bra{\lambda},\\
     K &=  -|\hat{\lambda}]\bra{\hat{\lambda}}+\mathcal{O}(\beta-1)=  |\lambda]\bra{\lambda}+\mathcal{O}(\beta^\prime-1),
    \end{split}
\end{equation}
where $K$ is the complex momentum transfer. The on-shell conditions $P_1^2=P_2^2=m_a^2$ and $P_3^2=P_4^2=m_b^2$, impose the normalization for the spinors $\braket{\hat{\lambda}\hat{\eta}}=[\hat{\lambda}\hat{\eta}]=m_a$ and $\braket{\lambda\eta}=[\lambda\eta]=m_b$. 
For the internal gravitons the spinor helicity variables read

\begin{equation}\label{eq:kinematics HCL gravitons}
    \begin{split}
        \ket{k_2} &= \frac{1}{\beta+1}\left(\left(\beta^2-1 \right)\ket{\eta} -\frac{1+\beta y}{y}\ket{\lambda}\right),\,\,\,|k_2] = \frac{1}{\beta+1}\left(\left(\beta^2-1\right)y|\eta]+(1+\beta y)|\lambda]\right),\\
        \ket{k_3} &= \frac{1}{\beta+1}\left(\frac{\beta^2-1}{\beta} \ket{\eta} +\frac{1-y}{y}\ket{\lambda}\right),\,\,\,|k_3] = \frac{1}{\beta+1}\left(-\beta\left(\beta^2-1\right)y|\eta]+(1-\beta^2 y)|\lambda]\right).
    \end{split}
\end{equation}
We define the variables $U$, $V$ and $\gamma$ from the massive momenta as follows
\begin{equation}
    \begin{split}
        U&= [\lambda|P_1\ket{\eta},\\
        V &=[\eta|P_1\ket{\lambda},\\
        \gamma &= \frac{P_1\cdot P_3}{m_a m_b}=\frac{1}{\sqrt{1-v^2}},
    \end{split}
\end{equation}
which satisfy the useful identities \cite{Chung:2018kqs} :
\begin{equation}
\begin{split}
   [\lambda|P_1\ket{\lambda}&=-\frac{(\beta-1)^2}{\beta }m_b^2+(1-\beta)V+\frac{\beta-1}{\beta}U,\\
      [\eta|P_1\ket{\eta}[\lambda|P_1\ket{\lambda} &= U V-m_a^2m_b^2 .
\end{split}
\end{equation}
The \ac{HCL}  is achieved  by taking $\beta\rightarrow 1$ (or equivalently $\beta^\prime\rightarrow 1$). In this limit, the variables $U$  and $V$  are related to $v$ via
\begin{equation}
    \begin{split}
        U &= m_a m_b(1-v)\gamma,\\
        V&=m_a m_b(1+v)\gamma.
    \end{split}
\end{equation}
To evaluate the LS \eqref{LS}, let us consider first the non spinning piece of the amplitudes entering in the integral.
The classical piece for the scalar Compton amplitude   \eqref{eq:scalar_compton}, in the gauge \eqref{eq:chkgauge}, is simply given by %
\begin{equation}
     \langle A_4^{0}(P_1,-P_2,k_3^+,k_4^-)\rangle= 32\pi G m_a^2\frac{(\epsilon_2\cdot u)^2(\tilde{\epsilon}_3\cdot u)^2 }{\langle \xi \rangle},\,\,\,\langle \xi \rangle = \frac{(s_c-m_a^2)^2}{m_a^2t} 
\end{equation}
where $s_c= (P_1+k_2)^2$, and $t=(P_1-P_2)^2$, i.e. the t-channel for the Compton amplitude coincides with that for the massive 4-pt amplitude. 
We have taken the classical limit of the optical parameter $\xi$ as given in \eqref{eq:xidef} by means of the classical identity \eqref{eq:ident}.
In the \ac{HCL} the Compton Mandelstam invariants read
\begin{equation}
    \begin{split}
        s_c &= m_a^2-m_a  m_b \gamma  \frac{y^2-1}{2y}(\beta-1)\,,\\
        t & = m_b^2\frac{(\beta-1)^2}{\beta}\,.
    \end{split}
\end{equation}
Then, we can easily check that in the \ac{HCL}  parametrization \eqref{eq:kinematics HCL} and \eqref{eq:kinematics HCL gravitons}:
\begin{equation}\label{eq:xihcl}
    \langle\xi\rangle \rightarrow - \gamma^2 v^2 \frac{(1-y^2)^2}{4y},
\end{equation}
and in the gauge \eqref{eq:chkgauge}, the scalar classical Compton  amplitude becomes
\begin{equation}
      \langle A_4^{0}(P_1,-P_2,k_3^+,k_4^-)\rangle=32\pi G m_a^2\frac{(2y-v(1+y^2))^4\gamma^2}{16v^2y^2(1-y^2)^2}.
\end{equation}
In the same way, it is straightforwards to see that     in the HCL, the scalar piece of the 3-pt amplitudes evaluates to
\begin{equation}
    \langle A_3^{0}(P_3,-l,-k_3^-) \rangle \langle A_3^{0}(-P_4,l,-k_4^+)\rangle = 8\pi G m_b^4.
\end{equation}
We now turn our attention to the spinning pieces of the amplitudes. For that, it is useful to write the classical spin vector for the \ac{BH}s in terms of the $SL(2,C)$ sigma matrices
\begin{equation}\label{eq:spin SL2c}
    (S^\mu)_\alpha^{\,\,\beta} =\frac{1}{4}\left[(\sigma^\mu )_{\alpha\dot{\alpha}}u^{\dot{\alpha}\beta}-u_{\alpha\dot{\alpha}}(\bar{\sigma}^\mu)^{\dot{\alpha}\beta}\right]+\mathcal{O}(\hbar),
\end{equation}
where as mentioned before, $u$ can be chosen to be the velocity of the initial or the final \ac{BH},     or the average velocity. Quantum mechanical corrections arise if we  include contributions to $u$ coming from the  gravitons momenta. 

Let us choose for instance  $u=\frac{P_1}{m_a}$. Then, using  \eqref{eq:wdef} as the definition for $\omega^\mu$, and the kinematics in the \ac{HCL} (\ref{eq:kinematics HCL}\,-\,\ref{eq:kinematics HCL gravitons}), together with \eqref{eq:spin SL2c}, it is straightforward to  show that to leading order in $1-\beta$, the classical basis of spin $\{\omega\cdot a,k_2\cdot a,k_3\cdot a\}$ maps to

\begin{equation}
    \begin{split}
        w &\rightarrow i K\frac{v\left(1-y^2\right)^2}{4 y \left(v-2 y + v y^2 \right)}\\
        k_2  &\rightarrow i K \frac{(1+y)^2}{4 y},\\
        k_3 & \rightarrow - i K \frac{(1-y)^2}{4 y},
    \end{split}
\end{equation}
note that in the \ac{HCL} $k_3-k_2=K=|\lambda\rangle [\lambda|$, which  turns take the exponential piece in \eqref{eq:qftamp4clas} into
\begin{equation}
   e^{(2w+k3-k2)\cdot a_a}\rightarrow e^{-i \frac{1+y^2-2vy}{2y-v(1+y^2)} K\cdot a_a},
\end{equation}
and analogous the product of two exponential for the 3-pt amplitudes \eqref{eq:classical3pt} becomes 
\begin{equation}
    e^{-2k_2\cdot a_b}\rightarrow e^{-i\frac{(1+y^2)}{2y}K\cdot a_b}.
\end{equation}
Putting all these ingredients together, the LS  \eqref{LS} evaluates to
\begin{equation}\label{LSN}
\frac{i2\pi^2 G^2m_a^2m_b^3\gamma^2}{v^2\sqrt{-t}} \int_{\Gamma_{\rm{LS}}} \frac{dy[2y-v(1+y^2)]^4}{2\pi y^3(1-y^2)^2}  \exp\left( -i \frac{1+y^2-2vy}{2y-v(1+y^2)}K{\cdot} a_a-i\frac{(1+y^2)}{2y}K{\cdot} a_b\right),
\end{equation}
which recovers the result of \cite{Guevara:2018wpp} up to $a_a^4$. The evaluation of the integral was done in the same reference. 

As last comment, in principle in the computation of the leading singularity  in Figure \ref{leading-singularity}, a sum over the exchange graviton polarization is needed. This means, the same helicity Compton amplitude \eqref{eq:samehcompton} needs to be included in the computation. One can however check explicitly such configuration produces vanishing contribution to the scattering angle in the aligned spin setup.

\section{Outlook of the chapter}\label{sec:outlook_gw}
In this chapter we have argued the minimal coupling classical amplitudes $A_n$ for $n=3,4$ are indeed very related to the Kerr \ac{BH}     by means of the unique spin multipole structure fixed for minimal coupling amplitudes, which map directly to the   spin multipole moments of the Kerr \ac{BH}.     This matching further clarifies the mismatch between the Feynman diagram prediction for the spin induced polarization made by Barbieri and Guadagnini  \cite{Guadagnini:2008ha,Barbieri:2005kp} to those of \ac{BHPT} by Dolarn \cite{Dolan:2008kf} at linear order in spin.
In particular, the spin induced polarization,  had the simple pattern for haves of helicity $h<2$
\begin{equation}\label{eq:polariation_n}
    \mathcal{P}^{(h)} =  -h \sinh{\left(4a_z|\omega|\sin^2(\theta/2)\right)} \left[\cosh^2{\left(2a_z|\omega|\sin^2(\theta/2)\right)} +h^2\sinh^2{\left(2a_z|\omega|\sin^2(\theta/2)\right)}\right]^{-1},
\end{equation}
as computed by the author of this thesis and collaborators in \cite{Bautista:2021wfy}, which  for  $\theta\rightarrow 0$  is simply  $ \mathcal{P}^{(h)} =  -ha_z|\omega|\theta^2$,   disagrees with the \ac{BHPT} result \eqref{eq:polariation} if extrapolated to $h=2$. The solution to this disagreement is of course to include the additional diagrams that restore gauge invariance of the amplitude. Result \eqref{eq:polariation_n}  recovers those in \cite{PhysRevD.16.237,Barbieri:2005kp} for  $h<2$ at linear order in spin, which needed only from the one graviton exchange diagram.

In \cref{ch:teukolsky} we check explicitly amplitudes \eqref{eq:App-gr} and \eqref{eq:Amp-gr} match exactly the classical result for the scattering of a gravitational wave off the Kerr \ac{BH},     using   \ac{BHPT}. Amplitudes computed from this \ac{QFT}     approach are written in a closed form which in turn re-sums the infinite sums appearing from the partial wave expansion as required from usual classical wave physics. 

For spin $S>2$, amplitude \eqref{eq:qftamp4clas}   needs to be modified. In particular, the unphysical poles arising from $w{\cdot}a$, can be removed by adding contact terms with unfixed coefficients.
In \cite{BGKV}  it was  shown that by matching to higher spin solutions of the Teukolsky equation, these coefficients can be fixed for the Kerr \ac{BH}.     Explicit results  up to spin $S=3$ are provided, but their interpretation are beyond the scope of this thesis. 
The \ac{BHPT} solutions however  disagree with the recent proposal in \cite{Aoude:2022trd}, where the free coefficients of an  ansatz  for the higher spin gravitational Compton amplitude were fixed by imposing that for   the conservative  2\ac{PM}    binary black hole amplitude,   certain spin structure observed at lower spins, is conserved (see also \cite{Bern:2022kto}).

%% file: Chapters/conclusions.tex
\chapter{General Discussion}\label{conclusions}

In this thesis we have presented a study addressing the 
computation of classical observables in classical gauge theories and gravity directly from the classical limit of \ac{QFT} amplitudes, using modern amplitudes techniques such as  double copy \cite{Kawai:1985xq,Bern:2008qj,Bern:2019prr}, spinor-helicity variables \cite{schwartz_2013,Elvang:2013cua,Arkani-Hamed:2017jhn}, leading singularities and the \ac{HCL}  \cite{Guevara:2017csg,Guevara:2018wpp}, the \ac{KMOC} formalism \cite{Kosower:2018adc,Kosower:2018adc,Maybee:2019jus,Guevara:2019fsj,Cristofoli:2021vyo,Aoude:2021oqj,Herrmann:2021lqe}, the amplitudes to the Kerr \ac{BH} correspondence \cite{Chung:2018kqs,Guevara:2018wpp,Arkani-Hamed:2019ymq,Bautista:2021wfy,BGKV}. We have presented a detail study of amplitudes for a massive spinning line emitting photons/graviton, $A_n$, both in (S)QED (QCD) and Gravity as  they are the main building blocks to compute two-body observables at leading  and subleading orders in perturbation theory. Such building blocks possess many remarkable properties such as soft exponentiation, universal covariant spin multipole expansion,   multipole preserving double copy, healthy high energy limit due to the fact they can be constructed from dimensional reduction and compactifications arguments, and they are not polluted with additional massless degrees of freedom (dilaton, axion) from the  double copy. In addition, we have shown the classical limit of  $A_n$ amplitudes can be directly associated to represent classical processes in both  electrodynamics and general relativity. For $n=3$, and in the  (electromagnetic) gravitational case, this amplitude encodes the same spin multipole structure of the (root) Kerr \ac{BH}, as shown by the seminal work of \cite{Chung:2018kqs,Guevara:2018wpp,Arkani-Hamed:2019ymq}. The $n=4$ case on the other hand, describes the low energy limit for the scattering of waves off classical compact objects with and without spin structures. For instance, in the electromagnetic case, $A_4 $ recovers results for the differential cross section for the well known Thompson process, whereas for gravity, $A_4$ encodes the information for the scattering of gravitational waves off the Kerr \ac{BH}. This in turn open a new amplitudes-\ac{BHPT} correspondence \cite{Bautista:2021wfy,BGKV} , allowing to study complicated gravitational scattering problems from simple \ac{QFT} derivations. 

We presented a detailed study of amplitudes in the electromagnetic theory, and show how soft theorems provide an infinite tower of constraints on the \ac{KMOC} formula for the computation of radiation.  These constraints can naturally be generalized to the gravitational theory and extended to subleading orders in the soft expansion  \cite{BLZ}. Understanding soft radiation is  important since it encodes the so called \ac{GME}, as dictated by the infrared triangle \cite{Strominger:2017zoo}. This is a   strong gravity effect and although it has not yet been  detected in the current \ac{GW} observatories, preliminary analysis  \cite{Hubner:2019sly} of available data from the LIGO/VIRGO \ac{GW} catalog suggest that \ac{GME} is very likely to be observed in the era of the advance LIGO/VIRGO detector, where an order of $2000$ events will be needed to say something conclusive. In addition, positive LISA  prospects for measuring of the \ac{GME} from the coalescing of super massive \ac{BH} even at red-shift values of $\sim 5$  for \ac{BH}s masses of order $10^5- 10^6 M_\odot$  are expected    \cite{Favata:2010zu}. 
Furthermore, soft theorems being non perturbative  can give information about radiation, and radiation reaction to higher orders in perturbation theory 
\cite{Laddha:2018rle,Laddha:2018myi, Sahoo:2018lxl,Laddha:2019yaj,Saha:2019tub,DiVecchia:2021ndb,DiVecchia:2021bdo,DiVecchia:2022owy,DiVecchia:2022nna,BLZ}.
On the other hand, studding electromagnetic radiation rather than color radiation, simplifies the computations as it avoids the complications introduced by the non abelian character the colour group, and permits to obtain two-body gravitational radiation from \ac{KLT} double copy properties of $A_n$ theories at lower orders in perturbation theory, which at the same time permits to easily remove extra degrees of freedom product of  the double copy. 

Inspired by the early work of Holstein \cite{Holstein:2008sw,Holstein:2008sx},  Vaidya \cite{Vaidya:2014kza}, Cachazo and Guevara \cite{Guevara:2017csg,Cachazo:2017jef} for the study  of spin effects in the   conservative sector of the two body problem from minimal coupling amplitudes (see also \cite{Chung:2018kqs,Guevara:2018wpp} ), we have  introduced a detailed study for inclusion of spin effects in the electromagnetic and gravitational theories for the radiative sector,   as well as for one-body wave scattering processes through the spin multipole double copy. Results for lower   orders in the \textit{covariant} spin multipolar expansion naturally recover results from  classical worldline computations \cite{Goldberger:2016iau,Goldberger:2016iau,Luna:2017dtq,Li:2018qap}, and extend them  quadrupolar order\footnote{In principle expansion up to hexadecapole order are possibles using the methods described in \cref{sec:spin_in_qed}.}. 
Extraction of the spin structure from \ac{QFT} amplitudes in a vector notation in general is a non-trivial task, and it is therefore more suitable to use spinor-helicity variables for such endeavors,  as we  have exemplified through several parts of the body of this thesis, in particular, in the identification of Compton amplitude with low energy solutions of the Teukoslky equation in \cref{ch:GW_scattering} and \cref{ch:teukolsky}. Additional approaches to include spin effects in vector notation such as the \ac{EFT}  formulation of spin \cite{Bern:2020buy} or the worldline theory \cite{Goldberger:2004jt,Maia:2017gxn,Maia:2017yok,Kim:2021rfj,Levi:2020lfn,Jakobsen:2021lvp,Jakobsen:2021zvh,Jakobsen:2022fcj}  are also techniques of current use. Studying  spin effects is important since they encode information regarding the formation mechanism of the binary system (see for instance \cite{Gerosa:2013laa,Vitale:2015tea, Biscoveanu:2021nvg}), and for nearly extremal Kerr \ac{BH}s, the individual spins of the binary's components   are expected to be measured with great precision by LISA  \cite{Burke:2020vvk}.

Motivated by the non-universality of the bounded-unbounded character of the two-body problem \cite{Cho:2021arx,Buonanno:2022pgc}, we initiated the program of computing classical observables for systems in bounded orbits directly from  scattering amplitudes. Formulas for the radiated field were motivated by the classical worldline computation \cite{Goldberger:2017vcg} and the Feynman diagrammatic approach of \cite{HariDass:1980tq}, but we argued they can be seen as a promotion of \ac{KMOC} formalism to include generic particles trajectories. It is desirable to probe these formulas from first principles as they could naturally account for radiation reaction forces captured by the conservative amplitude (and therefore by the bodies  \ac{EoM}), which are added from balance equations arguments to the objects'  \ac{EoM} in the \ac{EOB} formalism when addressing the radiative sector  \cite{Khalil:2021txt}. Guidance from the Worldline \ac{QFT} \cite{Mogull:2020sak}, or the in-in formalism \cite{Bakshi1963,Bakshi1963,Schwinger1960,Keldysh:1964ud}, might be  of great use in this endeavor. In addition, it would be  interesting to establish a precise dictionary between the soft expansion  and the source multipole decomposition for  the bounded orbit problem. The  reason for this is that soft theorems are non perturbative and therefore encode all  orders in perturbation theory, and  higher velocity corrections to the waveform,  as dictated by the virial theorem.  

On the formal side of the double copy for spinning matter, we have shown how graviton coupling to matter produce universal term in the Lagrangian description, whereas coupling of additional double copy states such as dilaton and two-form field have non universal coupling. This led us to identify two independent double copy theories producing spin 1 massive matter couple to gravity, we refereed to them as the $\frac{1}{2}\otimes\frac{1}{2}$ and the $0\otimes 1$ theories, as introduced explicitly in \cref{ch:double_copy}. The former is a simpler theories since it does not need quartic terms for the two matter line case, whereas for the latter, identification of such contact terms was shown explicitly. The double copy spectrum of the $\frac{1}{2}\otimes \frac{1}{2}$ theory was consistently truncate, and shown to agree with the 4-dimensional double copy of \cite{Johansson:2019dnu}. It is nevertheless desirable to complete the double copy spectrum for the case of general dimensions, and surpass the complications introduced by higher Dirac structures, when allowing matter coupling to the two-form potential (See \cref{residues at 4 points}).  In addition, we have shown that double copy arguments fixes the gyromagnetic factor $g=2$, for spin 1-particles, in both, \ac{QCD} and gravity. This in turn allows to identify  spin 1 particles in the \ac{QCD} theory as W-bosons and not Proca fields \cite{Bautista:2019evw,Johansson:2019dnu,Holstein:2006pq}, allowing in addition to remove divergences in the amplitudes  in the massless limit. A well-defined high-energy behaviour is then explained from the fact massive amplitudes with a direct double copy application can be obtained from dimensional reduction and compactification arguments. 

Finally, in this thesis we have presented the first direct  connection of the minimal coupling gravitational Compton amplitude and the Kerr \ac{BH}. We have shown how it perfectly describes the low energy regime for the scattering of gravitational waves off the Kerr \ac{BH} up to quartic order in spin. This provides a non-trivial connection of minimal coupling amplitudes and Kerr, and provides the basis of the amplitudes-\ac{BHPT} correspondence \cite{Bautista:2021wfy,BGKV}.    This is however just the beginning of this correspondence as computing higher order spin corrections \cite{Bern:2022kto,Aoude:2022trd,BGKV}, extracting  higher multiplicity amplitudes from \ac{BHPT},  understanding \ac{BH} horizon dissipative effects \cite{Goldberger:2019sya,Goldberger:2020geb,Goldberger:2020fot},  as well as the  description of \ac{BH} quasi-normal modes \cite{Berti:2009kk} from a \ac{QFT} perspective remains to be  open problems. Of particular importance are the extraction of higher spinning amplitudes from Kerr, since this provides with guides for approaching the problem interactive higher spin particles in \ac{QFT} \cite{Camanho:2014apa}, and the extraction of higher multiplicity amplitudes from \ac{BHPT}, these are expected to be contained in higher $G$ solutions to the Teukolsky equation, which can be classified into loop corrections to the gravitational Compton amplitude, and higher multiplicity tree-level amplitudes, both of which are building blocks for computing two-body observables involving spinning black holes. 

%% file: appendices/soft_constraints.tex
\chapter{Computational details NLO radiation}\label{app:radiation_NLO}
In this appendix we walk through the computational details of several integrals given in section \cref{sec:1-loop_radiation}
\section{Cancellations in the classical one-index \textit{moment} at 1-loop
}\label{app:cut_box}
In this section we fill in  the computational details for the cancellations announced by the end of  \cref{sec:one-index-1-loop}.
\begin{itemize}
    \item \textit{Proof of eq. \eqref{eq:cut-box final}}
\end{itemize}
Let us begin  by walking through the proof of the statement of equation \eqref{eq:cut-box final}. We start from 
\begin{equation}\label{eq:cut_bot_start}
  {\rm [cut{-}box]^{(1)\,\alpha}}=\frac{1}{4}\int\hat{d}^{4}q\prod_{i}\hat{\delta}(p_{i}{\cdot}q)e^{-ib{\cdot}q}\left(\mathcal{I}_{3}^{\alpha}-q^\alpha Z=U^{\alpha}\right)\,,
\end{equation}
where  the integrand has the explicit form
\begin{equation}\label{eq:u_integrand_in_cut-box}
U^{\alpha}=-2g^4\left(Q_{1}Q_{2}p_{1}{\cdot}p_{2}\right)^{2}\int\frac{\hat{d}^{4}l}{l^{2}(l{-}q)^{2}}\left[\hat{\delta}'(p_{1}{\cdot}l)\hat{\delta}(p_{2}{\cdot}l){-}\hat{\delta}(p_{1}{\cdot}l)\hat{\delta}'(p_{2}{\cdot}l)\right]\left[l{\cdot}(l-q)l^{\alpha}{+}\frac{q^{\alpha}}{2}(2l{\cdot}q{-}l^{2})\right]\,.
\end{equation}
 To evaluate this integral  we can  expand the momentum $l$ in an analogous way to the $q$ momentum in (\ref{eq:momentum-expansion-q} - \ref{eq:alphai_factors}), with say $\alpha_i\to \beta_i$, and $x_i\to y_i= p_i\cdot l$. The resulting  integrand  takes the form   \eqref{eq:general_integral}, and therefore we can evaluate the time and longitudinal components using integrating by parts one time \eqref{eq:general_ibp}. That is, we can write 
 \begin{equation}
     U^\alpha = \frac{1}{\sqrt{\mathcal{D}}}\int  \hat{d}^{2}l_{\perp}\hat{d}y_1\hat{d}y_2\left[ \hat{\delta}^{(1)}(y_1)\hat{\delta}^{(0)}(y_2) - \hat{\delta}^{(0)}(y_1)\hat{\delta}^{(1)}(y_2)\right]  f_u^\alpha(y_1,y_2,l_\perp,\sigma),
 \end{equation}
 with the  identification of the  integrand function   
\begin{align}
f_u^\alpha(y_1,y_2,l_\perp,\sigma)=-2g^4\left(Q_{1}Q_{2}p_{1}{\cdot}p_{2}\right)^{2}\Bigg[ & \frac{(\beta_{2}p_{1}{+}\beta_{1}p_{2}+l_{\perp})^{\alpha}\left((\beta_{2}p_{1}{+}\beta_{1}p_{2})^{2}{+}l_{\perp}^{2}{-}l_{\perp}{\cdot}q_{\perp}\right)}{\left((\beta_{2}p_{1}{+}\beta_{1}p_{2})^{2}{+}l_{\perp}^{2}\right)\left((\beta_{2}p_{1}{+}\beta_{1}p_{2})^{2}{+}(l_{\perp}{-}q_{\perp})^{2}\right)}\nonumber \\
 & {-}\frac{\left((\beta_{2}p_{1}{+}\beta_{1}p_{2})^{2}{+}l_{\perp}^{2}-2l_{\perp}{\cdot}q_{\perp}\right)\frac{q_{\perp}^{\alpha}}{2}}{\left((\beta_{2}p_{1}{+}\beta_{1}p_{2})^{2}{+}l_{\perp}^{2}\right)\left((\beta_{2}p_{1}{+}\beta_{1}p_{2})^{2}{+}(l_{\perp}{-}q_{\perp})^{2}\right)}\Bigg]\,.\label{eq:fu1}
\end{align}
where in addition to the $l$-expansion, we have used the expansion for the $q$-momentum \eqref{eq:momentum-expansion-q}, and set $x_i\to0$ using the support of the delta function $\hat{\delta}(p_i{\cdot}q)$ in \eqref{eq:cut_bot_start}. Next, to use \eqref{eq:general_ibp} after integration by parts we need to evaluate the derivatives of the form 
\begin{equation}\label{eq:derf}
\frac{\partial}{\partial y_{i}}f_{u,ij}^{\alpha}\Bigg|_{y_{i}=y_{j}=0}=\frac{4g^4}{\mathcal{D}}\left(Q_{i}Q_{j}p_{i}{\cdot}p_{j}\right)^{2}p_{j,\beta}p_{j}^{[\beta}p_{i}^{\alpha]}\frac{l_{\perp}{\cdot}(l_{\perp}-q_{\perp})}{l_{\perp}^{2}(l_{\perp}-q_{\perp})^{2}}\,,
\end{equation}
where one can check that only the first line of \eqref{eq:fu1} contributed to \eqref{eq:derf}.
With all the tools at hand, it is then direct to show that the integral \eqref{eq:cut_bot_start} simplifies to
\begin{equation}
{\rm [cut{-}box]^{(1)\,\alpha}}=g^4\frac{\left(Q_{1}Q_{2}p_{1}{\cdot}p_{2}\right)^{2}}{\mathcal{D}^2}\left[p_{2,\beta}p_{2}^{[\beta}p_{1}^{\alpha]}{-}p_{1,\beta}p_{1}^{[\beta}p_{2}^{\alpha]}\right]\int\hat{d}^{2}q_{\perp}\hat{d}^{2}l_{\perp}e^{-ib{\cdot}q_{\perp}}\frac{l_{\perp}{\cdot}(l_{\perp}{-}q_{\perp})}{l_{\perp}^{2}(l_{\perp}{-}q_{\perp})^{2}}.\label{eq:cut-box int2}
\end{equation}
Next we do the usual change of variables $q_{\perp}=\bar{q}_{\perp}+l_\perp,$
so that 
\begin{equation}
{\rm [cut{-}box]^{(1)\,\alpha}}=\frac{g^4}{\mathcal{D}}\left[p_{2,\beta}p_{2}^{[\beta}p_{1}^{\alpha]}{-}p_{1,\beta}p_{1}^{[\beta}p_{2}^{\alpha]}\right]\left[\frac{Q_{1}Q_{2}p_{1}{\cdot}p_{2}}{\sqrt{\mathcal{D}}}\int\hat{d}^{2}q_{\perp}\hat{d}^{2}l_{\perp}e^{-ib{\cdot}q_{\perp}}i\frac{\bar{q}_{\perp}}{\bar{q}_{\perp}^{2}}\right]^{2}.\label{eq:cut-box int3}
\end{equation}
in the big bracket we recognize the Leading order impulse $(\ref{eq:LO_impulse_interm})$,
which in turn allow us to recover the announced result $(\ref{eq:cut-box final}).$
\begin{equation}
{\rm [cut{-}box]^{(1)\,\alpha}}= - g^4
     \left[p_{1,\beta}p_{1}^{[\beta}p_{2}^{\alpha]}-p_{2,\beta}p_{2}^{[\beta}p_{1}^{\alpha]}\right]\frac{\left(\Delta p^{(0)}_{1}\right)^{2}}{\mathcal{D}}\,,\label{eq:cut-box final_app}
\end{equation}
\begin{itemize}
    \item \textit{Proof of eq. \eqref{eq:cancel_r2_cb_c2}}
\end{itemize}
Next we move to prove the cancellation \eqref{eq:cancel_r2_cb_c2}. For that we still need to compute $ {\cal T}^{\alpha}_{(1),\mathcal{R}_2}$ starting from \eqref{eq:R2-1}, and using  \eqref{eq:hbar1measure} for the integral  measure, and \eqref{eq:M4sqedclass} for the 4 point amplitude. We get

\begin{equation}\begin{split}
     {\cal T}^{(1)\,\alpha}_{\mathcal{R}_2} =
\frac{1}{4}g^4\left(Q_{1}Q_{2}\,p_{1}{\cdot}p_{2}\right)^{2}&\int\hat{d}^{4}l\prod_{i}\hat{\delta}(p_{i}{\cdot}l)\hat{d}^{4}q e^{-ib{\cdot}q}\,\\
&\times\left[\hat{\delta}'(p_{1}{\cdot}q)\hat{\delta}(p_{2}{\cdot}q){-}\hat{\delta}(p_{1}{\cdot}q)\hat{\delta}'(p_{2}{\cdot}q)\right]\frac{q^{2}q^{\alpha}}{l^{2}(l-q)^{2}}\,,
\end{split}
\end{equation}
where we have  used $\delta'(2p_{i}{\cdot}q)=\frac{1}{4}\delta'(p_{i}{\cdot}q)$.
To proceed in the calculation, we follow the philosophy of the previous subsection for the computation of integrals involving derivatives of the Dirac delta function, i.e. using integration by parts. Doing the change of variables (\ref{eq:momentum-expansion-q}) (and the analogous change for  $q\to l$ and $x_i\to y_i$ ), and evaluating the integrals in $(p_{i}{\cdot}l) =  y_i $, using the corresponding delta function, we arrive at 
\begin{equation}\label{eq:R2-int}
    {\cal T}^{(1)\,\alpha}_{\mathcal{R}_2}= \frac{1}{\mathcal{D}}\int\hat{d}^{2}q_{\perp}\frac{\hat{d}^{2}l_{\perp}}{l_{\perp}^{2}}\prod_{i}\hat{d}x_i\left[\hat{\delta}'(x_i)\hat{\delta}(x_2){-}\hat{\delta}(x_1)\hat{\delta}'(x_2)\right]Y^{\alpha}\,,
\end{equation}
where we have defined
\begin{equation}
Y^{\alpha}=\frac{1}{4}g^4\left(Q_{1}Q_{2}\,p_{1}{\cdot}p_{2}\right)^{2}e^{-iq_{\perp}{\cdot}b}\frac{((\alpha_{2}p_{1}{+}\alpha_{1}p_{2})^{2}+q_{\perp}^{2})(\alpha_{2}p_{1}{+}\alpha_{1}p_{2}{+}q_{\perp})^{\alpha}}{(\alpha_{2}p_{1}{+}\alpha_{1}p_{2})^{2}{+}(l_{\perp}-q_{\perp})^{2}}\,,\label{eq:Y function}
\end{equation}
recalling that    $\alpha_i$ are  function of $x_i$. We then get an integral of the form \eqref{eq:general_integral}. Using 
\begin{equation}
\frac{\partial}{\partial x_i}Y_{ij}^{\alpha}\Bigg|_{x_1=x_2=0}=-\frac{1}{2}g^4\left(Q_{1}Q_{2}\,p_{1}{\cdot}p_{2}\right)^{2}\frac{q_{\perp}^{2}}{\mathcal{D}(l_{\perp}-q_{\perp})^{2}}p_{j,\nu}p_{j}^{[\nu}p_{i}^{\alpha]}e^{-iq_{\perp}{\cdot}b}\label{eq:derivative Y},
\end{equation}
for doing the integration  by parts procedure, we can write \eqref{eq:R2-int} as follows
\begin{equation}
     {\cal T}^{(1)\,\alpha}_{\mathcal{R}_2} ={-}\frac{1}{2}\frac{g^4\left(Q_{1}Q_{2}\,p_{1}{\cdot}p_{2}\right)^{2}}{\mathcal{D}^2}\left(p_{1,\beta}p_{1}^{[\beta}p_{2}^{\alpha]}{-}p_{2,\beta}p_{2}^{[\beta}p_{1}^{\alpha]}\right)
\left[\int\hat{d}^{2}q_{\perp}\hat{d}^{2}l_{\perp}\frac{q_{\perp}^{2}e^{-iq_{\perp}{\cdot}b}}{l_{\perp}^{2}(l_{\perp}-q_{\perp})^{2}}{=}\mathcal{J}_{2}\right]\,.\label{eq:R2 interm 2}
\end{equation}
Let us now  evaluate the integral $\mathcal{J}_{2}$ in the square brackets.  Doing the usual shift
$q_{\perp}=\bar{q}_{\perp}+l_{\perp},$ so that
\begin{equation}
\mathcal{J}_{2}=\int\hat{d}^{2}\bar{q}_{\perp}\hat{d}^{2}l_{\perp}\frac{(\bar{q}_{\perp}^{2}+l_{\perp}^{2}+2l_{\perp}{\cdot}\bar{q}_{\perp})e^{-i\bar{q}_{\perp}{\cdot}b}e^{-il_{\perp}{\cdot}b}}{l_{\perp}^{2}\bar{q}_{\perp}^{2}}\,.\label{eq:J2 definition}
\end{equation}
 From the crossed terms we identify the integral representation for the  leading order impulse, whereas the remaining terms are contact integrals which we drop assuming $b\ne0$. We finally get
\begin{equation}
\mathcal{J}_{2}=-2\frac{\left(\Delta p^{(0)}\right)^{2}\mathcal{D}}{g^4\left(Q_{1}Q_{2}\,p_{1}{\cdot}p_{2}\right)^{2}}\,,\label{eq:J2 final}
\end{equation}
which can be replace back  into \eqref{eq:R2 interm 2} to finally give
\begin{equation}
 {\cal T}^{(1)\,\alpha}_{\mathcal{R}_2} = g^4 \left[p_{1,\beta}p_{1}^{[\beta}p_{2}^{\alpha]}{-}p_{2,\beta}p_{2}^{[\beta}p_{1}^{\alpha]}\right]\frac{\left(\Delta p^{(0)}\right)^{2}}{\mathcal{D}}\,.\label{eq:R2 final}
\end{equation}
We note that this simply gives  ${\cal T}^{(1)\,\alpha}_{\mathcal{R}_2} =- {\rm [cut{-}box]^{(1)\,\alpha}}$, and therefore this concludes the proof of  \eqref{eq:cancel_r2_cb_c2}.
\begin{itemize}
    \item \textit{Vanishing of ${\cal T}^{(1)\,\alpha}_{\mathcal{C}_1}$}
\end{itemize}
This is a very simple proof since this term give us a contact integral. Our  starting point is  the definition \eqref{eq:C1 definition-2}, and using \eqref{eq:hbar0 measure}  and \eqref{eq:hbar1measure} for the integral measure in $q$ and $l$ respectively, and \eqref{eq:M4sqedclass} for the 4 point amplitude, we get  
\begin{equation}
\begin{split}
{\cal T}^{(1)\,\alpha}_{\mathcal{C}_1}  =  16 g^4\left(Q_{1}Q_{2}p_{1}{\cdot}p_{2}\right)^{2}&\int\hat{d}^{4}q\hat{d}^{4}l e^{-ib{\cdot}q}\prod_{i}\hat{\delta}(2p_{i}{\cdot}q)\\
 &\times \left[\hat{\delta}'(2p_{1}{\cdot}l)\hat{\delta}(2p_{2}{\cdot}l){-}\hat{\delta}(2p_{1}{\cdot}l)\hat{\delta}'(2p_{2}{\cdot}l)\right]
 \frac{\cancel{l^{2}}l^{\alpha}}{\cancel{l^{2}}(q-l)^{2}}
\end{split}\,,
\end{equation}
which gives us indeed a  contact integral for the $l$-variable, unimportant for long classical scattering, since $b\ne0$. 
\begin{itemize}
    \item \textit{Vanishing of ${\cal T}^{(1)\,\alpha}_{\mathcal{C}_2}$}
\end{itemize}
Here we carry out the final piece of the computation for the one-index \textit{moment}. 
As usual, we start from the definition \eqref{eq:C2 definition-2}, and use \eqref{eq:hbar1measure} and \eqref{eq:hbar0 measure} for the integral measure in $q$ and $l$ respectively,  and \eqref{eq:M4sqedclass} for the 4 point amplitude. This gives

\begin{equation}
{\cal T}^{(1)\,\alpha}_{\mathcal{C}_2}= \int\hat{d}^{4}q\hat{d}^{4}l\prod_{i}\hat{\delta}(p_{i}{\cdot}l)\left[\hat{\delta}'(p_{1}{\cdot}q)\hat{\delta}(p_{2}{\cdot}q){-}\hat{\delta}(p_{1}{\cdot}q)\hat{\delta}'(p_{2}{\cdot}q)\right]X^{\alpha}\,,
\end{equation}
where we have defined 
\begin{equation}
X^{\alpha}=-\frac{1}{2}g^4\left(Q_{1}Q_{2}p_{1}{\cdot}p_{2}\right)^{2}\frac{q^{2}l^{\alpha}}{l^{2}(q{-}l)^{2}}e^{-ib{\cdot}q}\,.
\end{equation}
Showing that this integral gives zero contribution is a straightforward task. We do the usual change of variables 
 \eqref{eq:momentum-expansion-q}, and doing the integrals in $y_i$ using the delta functions. However, since we will use integration by parts, we need to evaluate the derivative of $X^{\alpha}$ w.r.t. $x_1$ or $x_2$, which  after evaluating $x_1=x_2=0$, vanish identically. We therefore conclude that  ${\cal T}^{\alpha}_{(1),\mathcal{C}_2}=0$, as announced in \cref{sec:one-index-1-loop}

\section{Cancellations in the classical two-index \textit{moment} at 1-loop}\label{app:cancelations_nlo_radiation}
In the final part of this appendix we proof the cancellations announced  in 
\eqref{eq:final_cancelations}.  Recall we already obtained $\mathcal{J}_3$ in \eqref{eq:J3}. All that is left is to compute explicitly ${\cal T}^{\alpha\,\beta}_{(1),\mathcal{C}_3} $. As usual we start  from the definition \eqref{eq:C3 definition-2}. Use \eqref{eq:hbar0 measure} for the integral measure in both $q$ and $l$ variables, and \eqref{eq:M4sqedclass} for the 4 point  amplitude, to get
\begin{equation}
{\cal T}^{(1)\,\alpha\,\beta}_{\mathcal{C}_3}= g^4\left(Q_{1}Q_{2}p_{1}{\cdot}p_{2}\right)^{2}\int\hat{d}^{4}q\hat{d}^{4}l\prod_{i}\hat{\delta}(p_{i}{\cdot}q)\hat{\delta}(p_{i}{\cdot}l)\frac{l^{\alpha}l^{\beta}}{l^{2}(q{-}l)^{2}}e^{-ib{\cdot}q}\,.\label{C3 start}
\end{equation}
 Doing our usual  shift $q=\bar{q}+l,$ allows us to factorize out the  IR-divergent integral - for the $l$ variable - \eqref{eq:scalar integral I1}. This becomes
\begin{equation}
{\cal T}^{(1)\,\alpha\,\beta}_{\mathcal{C}_3} = g^4\left(Q_{1}Q_{2}p_{1}{\cdot}p_{2}\right)^{2}I_{1} \int\hat{d}^{4}\bar{q}\hat{\delta}(p_{1}{\cdot}\bar{q})\hat{\delta}(p_{2}{\cdot}\bar{q})\frac{l^{\alpha}l^{\beta}}{\bar{q}^{2}}e^{-ib{\cdot}\bar{q}}\,,\label{eq:C3 interm 2}
\end{equation}
 which can be further rewritten as 
\begin{equation}
{\cal T}^{(1)\,\alpha\,\beta}_{\mathcal{C}_3}  =g^4\left(Q_{1}Q_{2}p_{1}{\cdot}p_{2}\right)I_{1} \partial_{b^{\alpha}}\Delta p^{(0)\,\beta}_{1}\,.\label{eq:C3 interm 4}
\end{equation}
The computation of  the derivative of the leading order impulse   w.r.t. the impact parameter was given in  \eqref{eq:derivative lo-impulse wrt b}. Using it leads to 
\begin{equation}
    {\cal T}^{\alpha\,\beta}_{(1),\mathcal{C}_3} =  2 g^4 \Delta p^{(0)\,\alpha}_{1} \Delta p^{(0)\,\beta}_{1}\ln\left(-\mu^{2}b^{2}\right),
\end{equation}
which is nothing but $\mathcal{J}_3^{(1)\,\alpha\,\beta}$ as given in 
\eqref{eq:J3} but with opposite sign. Thus we simply conclude that 
$
{\cal T}^{(1)\,\alpha\,\beta}_{\mathcal{C}_3} +\mathcal{J}_{3}^{(1)\,\alpha\,\beta} = 0\,,$ as required from \eqref{eq:final_cancelations}.

%% file: appendices/appendix_multipoles.tex
\chapter{Spinor-Helicity Formulae}\label{app_B}

Here we show the exponential forms presented here for spin-multipoles
contain as particular cases the ones of \cite{Guevara:2018wpp}, which implemented
massive spinor-helicity variables in $D=4$ \cite{Arkani-Hamed:2017jhn}. Consider
first $A_{3}^{{\rm gr},s}$: For plus helicity of the graviton, the
expression derived in \cite{Guevara:2018wpp} reads
\begin{equation}
A_{3,+}^{{\rm gr},s}=\frac{(p\cdot\epsilon)^{2}}{m^{2s}}\langle2|^{2s}e^{\frac{k_{\mu}\epsilon_{\nu}J^{\mu\nu}}{p\cdot\epsilon}}|1\rangle^{2s}\label{eq:4dsetup}\,,
\end{equation}
where $\epsilon{=}\epsilon^{+}$ carries the graviton helicity and $|\lambda\rangle^{2s}$
stands for the product $|\lambda^{(a_{1}}\rangle_{\alpha_{1}}\cdots|\lambda^{a_{2s})}\rangle_{\alpha_{2s}}$
of ${\rm SL}(2,\mathbb{C})$ spinors associated to each massive particle.
The generator $J^{\mu\nu}$ in the exponent acts on such chiral
representation. The labels $a_{i}$ are completely symmetrized little-group
indices. The explicit construction of the massive spinors is not needed
here (c.f. \cite{Arkani-Hamed:2017jhn}), but solely the fact that spin-$s$ polarization
tensors can be expressed compactly as
\begin{equation}
\varepsilon_{1} { =}  \frac{1}{m^{s}}|1\rangle^{s}|1]^{s}\,,\quad\varepsilon_{2}{=}\frac{1}{m^{s}}|2\rangle^{s}|2]^{s}\,,
\end{equation}
where $|1^{a}]_{\dot{\alpha}}$ and $|2^{a}]_{\dot{\alpha}}$ live
in the antichiral representation of ${\rm SL}(2,\mathbb{C})$. Inserting
them  into \eqref{eq:3ptexp} we obtain 
\begin{equation}
\langle\varepsilon_{2}|A_{3}^{{\rm gr},s}|\varepsilon_{1}\rangle{=}\frac{(p\cdot\epsilon)^{2}}{m^{2s}}\langle2|^{s}e^{\frac{k_{\mu}\epsilon_{\nu}J^{\mu\nu}}{p\cdot\epsilon}}|1\rangle^{s}[2|^{s}e^{\frac{k_{\mu}\epsilon_{\nu}\tilde{J}^{\mu\nu}}{p\cdot\epsilon}}|1]^{s}\,,\label{eq:3ptredux}
\end{equation}
where $J^{\mu\nu}$ and $\tilde{J}^{\mu\nu}$ are given by
\begin{eqnarray}
J^{\mu\nu} & {=} & \frac{1}{2}\mbox{\ensuremath{\sigma}}^{\mu\nu}\otimes\mathbb{I}^{\otimes(s-1)}{+}\mathbb{I}\otimes\frac{1}{2}\mbox{\ensuremath{\sigma}}^{\mu\nu}\otimes\mathbb{I}^{\otimes(s-2)}{+}{\cdots}\,,\quad\\
\tilde{J}^{\mu\nu} & {=} & \frac{1}{2}\mbox{\ensuremath{\tilde{\sigma}}}^{\mu\nu}\otimes\mathbb{I}^{\otimes(s-1)}{+}\mathbb{I}\otimes\frac{1}{2}\mbox{\ensuremath{\tilde{\sigma}}}^{\mu\nu}\otimes\mathbb{I}^{\otimes(s-2)}{+}{\cdots}\,,\quad
\end{eqnarray}
with $\sigma^{\mu\nu}=\sigma^{[\mu}\tilde{\sigma}^{\nu]}$
and $\tilde{\sigma}^{\mu\nu}=\tilde{\sigma}^{[\mu}\sigma^{\nu]}$.
They satisfy the self-duality conditions
\begin{equation}
    J^{\mu\nu} =  \frac{i}{2}\epsilon^{\mu\nu\rho\sigma}J_{\rho\sigma}\,\,,\quad
\tilde{J}^{\mu\nu} = -\frac{i}{2}\epsilon^{\mu\nu\rho\sigma}\tilde{J}_{\rho\sigma}\,.
\end{equation}
As it is well known, choosing the graviton to have plus helicity leads
to a self-dual field strength tensor, which in turn implies that $k_{[\mu}\epsilon_{\nu]}^{+}\tilde{J}^{\mu\nu}=0$.
Then \eqref{eq:3ptredux} reads

\begin{equation}
\langle\varepsilon_{2}|A_{3}^{{\rm gr},s}|\varepsilon_{1}\rangle=\frac{(p\cdot\epsilon)^{2}}{m^{2s}}\langle2|^{s}e^{\frac{k_{\mu}\epsilon_{\nu}J^{\mu\nu}}{p\cdot\epsilon}}|1\rangle^{s}[21]^{s}\,.
\end{equation}
We can now plug the identity
$
[21]^{s}{=}\langle2|^{s}e^{\frac{k_{\mu}\epsilon_{\nu}J^{\mu\nu}}{p\cdot\epsilon}}|1\rangle^{s}
$ from \cite{Guevara:2018wpp}
to obtain:
\begin{equation}
  \langle\varepsilon_{2}|A_{3}^{{\rm {\rm gr}},s}|\varepsilon_{1}\rangle {=}  \frac{(p{\cdot}\epsilon)^{2}}{m^{2s}}\langle2|^{s}e^{\frac{k_{\mu}\epsilon_{\nu}J^{\mu\nu}}{p\cdot\epsilon}}|1\rangle^{s} \langle2|^{s}e^{\frac{k_{\mu}\epsilon_{\nu}J^{\mu\nu}}{p\cdot\epsilon}}|1\rangle^{s}\label{eq:prematch} .
\end{equation}
which has the structure of our formula \eqref{eq:group}, now in "spinor space". Extending the generators $J^{\mu\nu}$ to act on $2s$ slots, i.e.
$J^{\mu\nu}\otimes\mathbb{I}^{s}+\mathbb{I}^{s}\otimes J^{\mu\nu}\rightarrow J^{\mu\nu}$,
then recovers \eqref{eq:4dsetup}. Consider now $A_{4,{+}{-}}^{{\rm gr},s}$
for $s\leq2$ as given in \cite{Guevara:2018wpp}, where $(+-)$ denotes the helicity
of the gravitons $k_{1}=|\hat{1}]\langle\hat{1}|$ and $k_{2}=|\hat{2}]\langle\hat{2}|$,

\begin{equation}
A_{4,+-}^{{\rm gr},s}=\frac{\langle\hat{1}|P_{1}|\hat{2}]^{4}m^{-2s}}{p_1{\cdot}k_1\,p_1{\cdot}k_2\, k_1{\cdot}k_2}\langle2|^{2s}e^{\frac{k_{1\mu}\epsilon_{1\nu}J^{\mu\nu}}{p\cdot\epsilon_{1}}}|1\rangle^{2s}\label{eq:spinorcomp}\,.
\end{equation}
In order to match this we double copy our formula \eqref{eq:sumofexp}.
The sum in \eqref{eq:sumofexp} exponentiates if we impose $[J_{1},J_{2}]=0$,
which in turn is only possible if the polarizations are aligned,
i.e. $\epsilon_{1}\propto\epsilon_{2}$. When the states have opposite
helicity this can be achieved via a gauge choice. This yields
\begin{equation}
\frac{k_{1\mu}\epsilon_{1\nu}J^{\mu\nu}}{p_{1}\cdot\epsilon_{1}}+\frac{k_{2\mu}\epsilon_{2\nu}J^{\mu\nu}}{p_{2}\cdot\epsilon_{2}}=\frac{k_{\mu}\epsilon_{1\nu}J^{\mu\nu}}{p\cdot\epsilon_{1}},
\end{equation}
where $k=k_{1}+k_{2}$. Expression \eqref{eq:sumofexp} thus becomes
\begin{equation}
\left.A_{4}^{{\rm ph},s}\right|_{\epsilon_{1}\propto\epsilon_{2}}{=}\frac{p_{1}{\cdot}\epsilon_{1}\,p_{2}{\cdot}\epsilon_{2}\,k_1{\cdot}k_2}{p_{1}{\cdot}k_{1}p_{1}{\cdot}k_{2}}\langle\varepsilon_{1}|e^{\frac{k_{\mu}\epsilon_{1\nu}J^{\mu\nu}}{p{\cdot}\epsilon_{1}}}|\varepsilon_{2}\rangle\,.
\end{equation}
(note that ${\rm ct}=\epsilon_{1}\cdot\epsilon_{2}$ drops out). The
formula \eqref{eq:doublecopyspin} gives
\begin{equation}
\left.A_{4}^{{\rm gr},s}\right|_{\epsilon_{1}\propto\epsilon_{2}}=\frac{(p_{1}{\cdot}\epsilon_{1})^{2}(p_{2}{\cdot}\epsilon_{2})^{2}\,k_1{\cdot}k_2}{p_{1}{\cdot}k_{1}\,p_{1}{\cdot}k_{2}}\langle\varepsilon_{1}|e^{\frac{k_{\mu}\epsilon_{1\nu}J^{\mu\nu}}{p\cdot\epsilon_{1}}}|\varepsilon_{2}\rangle\label{eq:gr4dcomp}\,,
\end{equation}
for $s\leq2$. This can be shown to match \eqref{eq:spinorcomp} following
the same derivation as before and fixing  $\epsilon_{1}=\frac{|\hat{1}\rangle[\hat{2}|}{[\hat{1}\hat{2}]}$,
$\epsilon_{2}=\frac{|\hat{1}\rangle[\hat{2}|}{\langle\hat{1}\hat{2}\rangle}$.
Note finally that, even though in any dimension $D$ there is an helicity
choice such that \eqref{eq:sumofexp} becomes \eqref{eq:gr4dcomp},
the factorization of (\ref{cuts m4 m5}) requires to sum over all helicities of
internal gravitons.

%% file: appendices/spin_2.tex
\chapter{Quadratic in spin results}\label{ch:quadratic_spin}

In this appendix we write explicit formulae for some quadratic in spin numerators  written in vector notation which where used in \cref{sec:spin_in_qed} when writing the radiation amplitude \eqref{eq:newM5clas}. 
\section{Electromagnetic case}
Let us start with the quadratic in spin  numerators for the  electromagnetic case. The first numerator, consider only particle 1 has spin,  is given by
\begin{equation}\label{eq:elepsin21}
\begin{split}
 n_{1,\text{ph}}^{(a)}&=\frac{e^3}{m_1^2}\Big[-\frac{1}{2}  k.p_1 k.p_2 \left(k.S_1\right){}^2 \epsilon .p_1+\frac{1}{2}  \left(k.p_1\right){}^2 \left(k.S_1\right){}^2 \epsilon
.p_2+\frac{1}{2}  k.p_1 \left(k.S_1\right){}^2 q.\epsilon  p_1.p_2\\&
-\frac{1}{2}  k.q \left(k.S_1\right){}^2 \epsilon .p_1 p_1.p_2+2  \left(q.S_1\right){}^2 \left(-k.p_1 k.p_2 \epsilon .p_1+\left(k.p_1\right){}^2
\epsilon .p_2+k.p_1 q.\epsilon  p_1.p_2-k.q \epsilon .p_1 p_1.p_2\right)\\&
-2  k.q \left(k.S_1\right){}^2 \epsilon .p_2 m_1^2+2  k.q k.p_2 k.S_1 \epsilon .S_1 m_1^2-4  k.q k.S_1 q.\epsilon  p_2.S_1 m_1^2+4
 (k.q)^2 \epsilon .S_1 p_2.S_1 m_1^2\\&
+2  q.S_1 \left(-k.p_1 k.p_2 k.S_1 \epsilon .p_1+\left(k.p_1\right){}^2 k.S_1 \epsilon .p_2+k.p_1 k.S_1 q.\epsilon  p_1.p_2-k.q k.S_1 \epsilon
.p_1 p_1.p_2\right.\\&
\left.+2 k.q k.S_1 \epsilon .p_2 m_1^2-2 k.q k.p_2 \epsilon .S_1 m_1^2\right)-4  k.q \left(-k.p_1 k.p_2 \epsilon .p_1+\left(k.p_1\right){}^2
\epsilon .p_2+k.p_1 q.\epsilon  p_1.p_2\right.\\&
\left.-k.q \epsilon .p_1 p_1.p_2-k.p_2 q.\epsilon  m_1^2+k.q \epsilon .p_2 m_1^2\right) S_1^2\Big]
\end{split}
\end{equation}
and for the other numerator we analogously  have
\begin{equation}\label{eq:elepsin22}
 n_{1,\text{ph}}^{(b)}=
\frac{ e^3}{2 m_1^2}\left(k.S_1+2 q.S_1\right)^2 \left(-\left(k.p_2\right)^2 \epsilon .p_1+k.p_1 k.p_2 \epsilon .p_2+k.p_2 q.\epsilon
 p_1.p_2-k.q \epsilon .p_2 p_1.p_2\right)
\end{equation}

\subsubsection{Gravitational case}  
In the gravitational case, once again considering only particle 1 has spin, the first numerator reads
\begin{equation}\label{eq:num_quad_class}
\begin{split}
n_{1,\text{gr}}^{(a)}&=\frac{\kappa^3}{32 m_1^2} \left(8 k.q k.p_1 k.p_2 k.S_1 \left(-k.p_2 \epsilon .p_1+k.p_1 \epsilon .p_2\right) \epsilon .S_1 m_1^2-8 m_2^2
(k.q)^2 \left(k.p_1\right){}^2 \left(\epsilon .S_1\right){}^2 m_1^2\right.\\&
-16 (k.q)^2 k.p_1 k.p_2 \epsilon .p_1 \epsilon .S_1 p_2.S_1 m_1^2+16 (k.q)^2 \left(k.p_1\right){}^2 \epsilon .p_2 \epsilon .S_1 p_2.S_1 m_1^2\\&
-16 k.q k.p_1 k.S_1 (q.\epsilon )^2 p_1.p_2 p_2.S_1 m_1^2+16 (k.q)^2 k.p_1 q.\epsilon  \epsilon .S_1 p_1.p_2 p_2.S_1 m_1^2\\&
-16 (k.q)^3 \epsilon .p_1 \epsilon .S_1 p_1.p_2 p_2.S_1 m_1^2+4 k.q k.p_1 k.S_1 q.\epsilon  \left(m_2^2 k.p_1 \epsilon .S_1+2 k.p_2 \epsilon
.S_1 p_1.p_2\right.\\&
\left.+4 k.p_2 \epsilon .p_1 p_2.S_1-4 k.p_1 \epsilon .p_2 p_2.S_1\right) m_1^2-4 (k.q)^2 k.S_1 \epsilon .p_1 \left(-3 m_2^2 k.p_1 \epsilon
.S_1\right.\\&
\left.+2 k.p_2 \epsilon .S_1 p_1.p_2-4 q.\epsilon  p_1.p_2 p_2.S_1\right) m_1^2+4 (k.q)^2 \left(q.S_1\right){}^2 \left(\epsilon .p_1\right){}^2
\left(2 \left(p_1.p_2\right){}^2-m_2^2 m_1^2\right)\\&
-8 k.q k.p_1 \left(q.S_1\right){}^2 \epsilon .p_1 \left(-2 k.p_2 \epsilon .p_1 p_1.p_2+2 k.p_1 \epsilon .p_2 p_1.p_2+2 q.\epsilon  \left(p_1.p_2\right){}^2-m_2^2
q.\epsilon  m_1^2\right)\\&
+\left(k.p_1\right){}^2 \left(k.S_1+2 q.S_1\right){}^2 \left(2 \left(k.p_2\right){}^2 \left(\epsilon .p_1\right){}^2-4 k.p_1 k.p_2 \epsilon
.p_1 \epsilon .p_2+2 \left(k.p_1\right){}^2 \left(\epsilon .p_2\right){}^2\right.\\&
\left.-4 k.p_2 q.\epsilon  \epsilon .p_1 p_1.p_2+4 k.p_1 q.\epsilon  \epsilon .p_2 p_1.p_2+2 (q.\epsilon )^2 \left(p_1.p_2\right){}^2-m_2^2
(q.\epsilon )^2 m_1^2\right)\\&
-4 k.q k.p_1 \left(k.S_1\right){}^2 \left(-k.p_2 \epsilon .p_1+k.p_1 \epsilon .p_2\right) \left(\epsilon .p_1 p_1.p_2+2 \epsilon .p_2 m_1^2\right)\\&
-2 k.q k.p_1 \left(k.S_1\right){}^2 q.\epsilon  \left(2 \epsilon .p_1 \left(p_1.p_2\right){}^2+m_2^2 \epsilon .p_1 m_1^2+4 \epsilon .p_2
p_1.p_2 m_1^2\right)\\&
+(k.q)^2 \left(k.S_1\right){}^2 \epsilon .p_1 \left(2 \epsilon .p_1 \left(p_1.p_2\right){}^2-5 m_2^2 \epsilon .p_1 m_1^2+8 \epsilon .p_2
p_1.p_2 m_1^2\right)\\&
+4 (k.q)^2 q.S_1 \epsilon .p_1 \left(2 k.S_1 \epsilon .p_1 \left(p_1.p_2\right){}^2+m_2^2 k.S_1 \epsilon .p_1 m_1^2-2 m_2^2 k.p_1
\epsilon .S_1 m_1^2\right.\\&
\left.-4 k.S_1 \epsilon .p_2 p_1.p_2 m_1^2+4 k.p_2 \epsilon .S_1 p_1.p_2 m_1^2\right)-8 k.q k.p_1 q.S_1 \left(-2 k.p_2 k.S_1 \left(\epsilon
.p_1\right){}^2 p_1.p_2\right.\\&
+2 k.p_1 k.S_1 \epsilon .p_1 \epsilon .p_2 p_1.p_2+2 k.S_1 q.\epsilon  \epsilon .p_1 \left(p_1.p_2\right){}^2+2 k.p_2 k.S_1 \epsilon .p_1 \epsilon
.p_2 m_1^2\\&
-2 k.p_1 k.S_1 \left(\epsilon .p_2\right){}^2 m_1^2-m_2^2 k.p_1 q.\epsilon  \epsilon .S_1 m_1^2-2 \left(k.p_2\right){}^2 \epsilon .p_1
\epsilon .S_1 m_1^2+2 k.p_1 k.p_2 \epsilon .p_2 \epsilon .S_1 m_1^2\\&
\left.-2 k.S_1 q.\epsilon  \epsilon .p_2 p_1.p_2 m_1^2+2 k.p_2 q.\epsilon  \epsilon .S_1 p_1.p_2 m_1^2\right)+\left(-16 k.q \left(k.p_1\right){}^2
\left(-k.p_2 \epsilon .p_1+k.p_1 \epsilon .p_2\right){}^2\right.\\&
-32 (k.q)^2 k.p_1 k.p_2 \left(\epsilon .p_1\right){}^2 p_1.p_2+32 (k.q)^2 \left(k.p_1\right){}^2 \epsilon .p_1 \epsilon .p_2 p_1.p_2\\&
-16 k.q \left(k.p_1\right){}^2 (q.\epsilon )^2 \left(p_1.p_2\right){}^2+32 (k.q)^2 k.p_1 q.\epsilon  \epsilon .p_1 \left(p_1.p_2\right){}^2-16
(k.q)^3 \left(\epsilon .p_1\right){}^2 \left(p_1.p_2\right){}^2\\&
+16 (k.q)^2 k.p_1 k.p_2 \epsilon .p_1 \epsilon .p_2 m_1^2-16 (k.q)^2 \left(k.p_1\right){}^2 \left(\epsilon .p_2\right){}^2 m_1^2\\&
+16 k.q k.p_1 k.p_2 (q.\epsilon )^2 p_1.p_2 m_1^2-16 (k.q)^2 k.p_2 q.\epsilon  \epsilon .p_1 p_1.p_2 m_1^2\\&
-16 (k.q)^2 k.p_1 q.\epsilon  \epsilon .p_2 p_1.p_2 m_1^2+16 (k.q)^3 \epsilon .p_1 \epsilon .p_2 p_1.p_2 m_1^2\\&
\left.\left.-16 k.q k.p_1 q.\epsilon  \left(-k.p_2 \epsilon .p_1+k.p_1 \epsilon .p_2\right) \left(2 k.p_1 p_1.p_2-k.p_2 m_1^2\right)\right)
S_1^2\right)
\end{split}
\end{equation}
and similarly for the second numerator 
\begin{equation}\label{eq:num_quad_class2}
\begin{split}
n_{1,\text{gr}}^{(b)}&=\frac{\kappa^3}{32 m_1^2} \left(k.S_1+2 q.S_1\right){}^2 \left(-2 \left(\left(k.p_2\right){}^2 \epsilon .p_1-k.p_1 k.p_2 \epsilon .p_2-k.p_2
q.\epsilon  p_1.p_2+k.q \epsilon .p_2 p_1.p_2\right){}^2 \right.\\
&\qquad \qquad \qquad \qquad \qquad \qquad  \left.+\left(-k.p_2 q.\epsilon +k.q \epsilon .p_2\right){}^2 m_1^2 m_2^2\right)
\end{split}
\end{equation}

%% file: appendices/app_bounded.tex
\chapter{Tools for bounded systems}\label{sec:app_bounded}
In this appendix we provide some useful tools in the computation of bounded orbits radiation. 

\section{Useful integrals and identities}\label{ap:integrals}
Here we write out the identity used in \cref{sec:linear_spin_waveform} for the comparison of the gravitational waveforms at linear order in spin. That is, given two vectors $a^i$ and $b^i$, and the TT projector defined in  \eqref{eq:TTprojector}, we have  \cite{Kidder:1995zr, Buonanno:2012rv}
\begin{align}
\Pi^{ab}{}_{ij}b^j\varepsilon^{i}{}_{k\ell}a^k N^\ell  =\Pi^{ab}{}_{ij} a^{i}\varepsilon^{j}{}_{k\ell}b^k N^\ell .
\label{eq:identityTT}
\end{align}
Furthermore, the following identity  \cite{Maggiore:1900zz}, was used in the computation of the energy flux in \cref{sec:results}
\begin{align}
\int_{S^2} d\Omega \ N_{i_1 \dots i_{2\ell}} =\frac{4\pi}{(2\ell+1)!!}(\delta_{i_1 i_2}\delta_{i_3 i_4}\dots \delta_{i_{2\ell-1} i_{2\ell}}+\dots).
\label{eq:identityOmegaInt}
\end{align}
In addition, the following integrals were used during the computation of gravitational radiation from the amplitudes perspective:
\begin{equation}
\begin{split}
\int\frac{d^{3}q}{(2\pi)^{3}}e^{i\bs{q}{\cdot}\bs{z}}\frac{1}{\bs{q}^{2}}&=\frac{1}{4\pi|\bs{z}|},\\
\int\frac{d^{3}q}{(2\pi)^{3}}e^{i\bs{q}{\cdot}\bs{z}}\frac{q^{i}}{\bs{q}^{2}}&=\frac{iz^{i}}{4\pi|\bs{z}|^{3}},\\
\int\frac{d^{3}q}{(2\pi)^{3}}e^{i\bs{q}{\cdot}\bs{z}}\frac{q^{i}q^{j}}{\bs{q}^{2}}&=\frac{1}{4\pi|\bs{z}|^{5}}\left[|\bs{z}|^{2}\delta^{ij}-3z^{i}z^{j}\right],\\
\int\frac{d^{3}q}{(2\pi)^{3}}e^{i\bs{q}{\cdot}\bs{z}}\frac{q^{i}q^{j}}{\bs{q}^{4}}&=\frac{1}{8\pi|\bs{z}|^{3}}\left[|\bs{z}|^{2}\delta^{ij}-z^{i}z^{j}\right],\\
\int\frac{d^{3}q}{(2\pi)^{3}}e^{i\bs{q}{\cdot}\bs{z}}\frac{q^{i}q^{j}q^{k}}{\bs{q}^{4}}&=-\frac{i}{8\pi|\bs{z}|^{5}}\left[|\bs{z}|^{2}\left(z^{i}\delta^{jk}+z^{j}\delta^{ik}+z^{k}\delta^{ij}\right)-3z^{i}z^{j}z^{k}\right].
\end{split}\label{eq:integrals}
\end{equation}

\section{The quadratic in spin EoM}\label{sec: eom quad spin}

In this appendix, we expand the classical equations of motion in \eqref{eq:EoM} to quadratic order in the black holes' spins (used in  \cref{sec:quadratic in spin}). After setting $S_2=0$, and expanding to second  order in $S_1$,  as well as  taking the leading order in velocity, the equation of motion reduce to
\begin{equation}
    \dot{v}_1^l =  \frac{-m_2\kappa^2}{32\pi }\left[  \frac{z_{12}^l}{r^3}+\frac{1}{2m_1^2}S_1^iS_1^j\partial^l\partial_i\partial_j\frac{1}{r}\right].
    \label{eq:eom quad spin}
\end{equation}
The spatial derivatives acting on $1/r$ result in contractions of a symmetric trace-free tensor
\begin{equation}
\partial^l\partial_i\partial_j\frac{1}{r} = \frac{3}{r^5}\left[\delta_{ij}z_{12}^l+2\delta^l_{(i}z_{12,j)}-5\frac{z_{12,i}z_{12,j}z_{12}^l}{r^2}\right].
\label{eq:spatialderivatives}
\end{equation}
Furthermore, we use these equations recursively, to remove powers of $1/r$\footnote{By restoring Newton's constant $G$, the equations of motion can be used to remove powers of $G$ in the numerator.}. Since we are interested in the quadratic-in-spin contribution only, we consider only the scalar part of  \eqref{eq:eom quad spin} (as well as the analogous equation for $v_2^i$) to rewrite \eqref{eq:spatialderivatives} as follows
\begin{equation}
    \partial^l\partial_i\partial_j\frac{1}{r} \rightarrow 
    -\frac{32\pi }{\kappa^2}\frac{3}{2m_2r^2}\left[\left(\delta_{ij}{-}\frac{5z_{12,i}z_{12,j}}{r^2}\right)\left(\dot{v}_1^l {-}\frac{m_2}{m_1}\dot{v}_2^l  \right)+2 \delta^l_{(i}\left(\dot{v}_{1,j)}{-}\frac{m_2}{m_1} \dot{v}_{2,j)}  \right) \right] {+}\mathcal{O}(S_1^2),
\end{equation}
Notice a factor of $1/3$ arises from symmetrization. 
This, then finally allows us to write the quadratic-in-spin equations of motion as
\begin{equation}
     \dot{v}_1^l = {-}\frac{m_2\kappa^2}{32\pi}  \frac{z_{12}^l}{r^3}{+}\frac{3}{4}\frac{S_1^iS_1^j}{m_1^2r^2}\left[\left(\delta_{ij}{-}\frac{5z_{12,i}z_{12,j}}{r^2}\right)\left(\dot{v}_1^l {-}\frac{m_2}{m_1}\dot{v}_2^l  \right){+}2 \delta^l_{(i}\left(\dot{v}_{1,j)}{-}\frac{m_2}{m_1} \dot{v}_{2,j)}  \right) \right].
\end{equation}
And analogously we also find
\begin{equation}
     \dot{v}_2^l = \frac{m_1\kappa^2}{32\pi }  \frac{z_{12}^l}{r^3}+\frac{3}{4}\frac{S_1^iS_1^j}{m_1^2r^2}\left[\left(\delta_{ij}-\frac{5z_{12,i}z_{12,j}}{r^2}\right)\left(\dot{v}_2^l -\frac{m_1}{m_2}\dot{v}_1^l  \right)+2 \delta^l_{(i}\left(\dot{v}_{2,j)}-\frac{m_1}{m_2} \dot{v}_{1,j)}  \right) \right].
\end{equation}

%% file: appendices/app_dc.tex
\chapter{Double copy for spinning particles }
\section{Double Copy in  \texorpdfstring{$d=4$}{d4}}\label{4 d double copy}

In this appendix we outline the \12x12 construction in $d=4$. 
It is interesting to make connection with the spinor formalism
for massive particles introduced in \cref{sec:spinor-helicity} ( see also \cite{Arkani-Hamed:2017jhn}), recently implemented for obtaining a massive double copy in \cite{Johansson:2019dnu}. Let us briefly sketch how our operation will read in such variables. For this, observe that we can write

\begin{equation}
E_{\mu}^{ab}\sigma^{\mu}={\color{black}\frac{\sqrt{2}}{m}}|p^{(a}]\langle p^{b)}|\,\quad E_{\mu}^{ab}\tilde{\sigma}^{\mu}={\color{black}\frac{\sqrt{2}}{m}}|p^{(a}\rangle [ p^{b)}|.\label{eq:pols1}
\end{equation}
where $E_{\mu}^{ab}$ is a spin-1 polarization vector, $\tilde{E}^{ab}{\cdot}P=0$,
with the little group indices $\{a,b\}=\{1,1\},\{2,2\},\{1,2\}$. Note its spinors satisfy
the Dirac equation

\begin{equation}
P|p^{a}\rangle=m|p^{a}],\,\quad\tilde{P}|p^{a}]=m|p^{a}\rangle,\label{eq:Diraceq}
\end{equation}
where $P=P_{\mu}\sigma^{\mu}$ and $\tilde{P}=P_{\mu}\tilde{\sigma}^{\mu}$. Then it is true that $[1^a,1^b]=-m\epsilon^{ab}$, and $\langle 1^a,1^b\rangle=m\epsilon^{ab}$
Now, in terms of the Dirac matrices note that

\begin{eqnarray}
(\slashed P+m\mathbb{I}_{4\times4})\slashed{E}^{ab} & = & {\color{black}\frac{\sqrt{2}}{m}}\left(\begin{array}{cc}
m\mathbb{I}_{2\times2} & P\\
\tilde{P} & m\mathbb{I}_{2\times2}
\end{array}\right)\left(\begin{array}{cc}
0 & |1^{(a}]\langle 1^{b)}|\\
|1^{(a}\rangle [1^{b)}| & 0
\end{array}\right),\label{projector}\\
 & = & {\color{black}\sqrt{2}}\left(\begin{array}{cc}
 |1^{(a}][1^{b)}|&|1^{(a}]\langle1^{b)}|\\
|1^{(a}\rangle [1^{b)}| &|1^{(a}\rangle \langle1^{b)}| 
\end{array}\right),\\
 & = & {\color{black}\sqrt{2}}\left(\begin{array}{c}
|1^{(a}]\\
|1^{(a}\rangle
\end{array}\right)([1^{b)}|\,\,\langle1^{b)}|),\\
 & = & {\color{black}\sqrt{2}}\,u^{(a}\bar{v}^{b)}\label{product projector},
\end{eqnarray}
where $u$ and $v$ are Dirac spinors satisfying $\slashed{P}u=m u$, $\slashed{P}v=-m v$,
as follows from (\ref{eq:Diraceq}). Note that the spin-1 polarization
can be recovered from \eqref{product projector} via
\begin{equation}
E_{\mu}^{ab}=\frac{1}{{\color{black}\sqrt{2}}m}\bar{v}^{(a}\gamma_{\mu}u^{b)}.
\end{equation}
In this sense the spin-1 polarization vector is constructed out of spin-$\frac{1}{2}$ polarizations.
We see that in $d=4$ the choice of polarizations given by
(\ref{eq:pols1}) turns the product \eqref{tensor product } into
\begin{equation}
X\otimes Y=\bar{v}_{2}^{(b_{2}}Xu_{1}^{(a_{1}}\times\bar{v}_{1}^{b_{1})}\bar{Y}u_{2}^{a_{2})},
\end{equation}
which is simple multiplication together with symmetrization over the
spin-$\frac{1}{2}$ states. Since this operation coincides with the one given in \cite{Johansson:2019dnu} we conclude that the amplitudes for a spin-1 field will agree in $d=4$.

For instance, for one matter line we will write
\begin{equation}\label{eq:a9}
A_{n}^{{\rm gr}}(E_{1}^{a_{1}b_{1}},E_{2}^{a_{2}b_{2}})=\sum_{\alpha,\beta}K_{\alpha\beta}\left(A_{n,\alpha}^{\text{QCD}}\right){}^{(a_{1}(b_{2}}\left(A_{n,\beta}^{\text{QCD}}\right){}^{a_{2})b_{1})}.
\end{equation}
which exhibits the symmetry properties of the indices explicitly. In particular it can be used to streamline the argument given in Section \ref{sec:sec2} for axion pair-production.

In an analogous way to \eqref{projector}, in the representation where $\gamma^{5}=\left(\begin{array}{cc}
-\mathbb{I}_{2\times2} & 0\\
0 & \mathbb{I}_{2\times2}
\end{array}\right)$,  we have 
\begin{align}
(\slashed P+m\mathbb{I}_{4\times4})\gamma^{5} & =\left(\begin{array}{cc}
-m\mathbb{I}_{2\times2} & |1]^{a}\langle 1|^{b}\epsilon_{ab}\\
|1\rangle^{a}[1|^{b}\epsilon_{ab} & m\mathbb{I}_{2\times2}
\end{array}\right)\\
 & =u^{[a}\bar{{v}}^{b]}\epsilon_{ab}\,.
\end{align}
By inserting the projector on the LHS instead of \eqref{projector} into our double copy, we find that antisymmetrizing little group indices from the Dirac spinors leads to a pseudoscalar. This antisymmetrization will necessarily require an odd number of axion fields in \eqref{eq:a9}. Hence the axion can be sourced by matter if the Proca field decays to a pseudoscalar, which is again consistent with the Lagrangian of \cite{Johansson:2019dnu}. Further analysis in general dimensions is done in the next Appendix.\\

\subsection*{Lagrangian comparison with \cite{Johansson:2019dnu}}

The results  of \cite{Johansson:2019dnu} consider the full spectrum of the $1/2 \otimes 1/2$ double copy restricted to four-dimensions. In contrast, our work shows that there exists a truncated spectrum in general dimensions. It is interesting  to analyze the overlap by comparing the interactions in our Lagrangian \eqref{eq:1212dc} with a truncated version of the one in \cite{Johansson:2019dnu}. Note that in principle the matching at the level of amplitudes does not guarantee such an off-shell agreement due to diverse field redefinitions. However, in our case it is possible since 1) both actions are written on the  Einstein frame for the graviton-dilaton couplings and 2) It can be shown that the axion and massive pseudoscalar fields decouple in the amplitudes of \cite{Johansson:2019dnu}, hence the corresponding interaction terms can be ignored in their Lagrangian.

With the previous considerations the Lagrangian of \cite{Johansson:2019dnu} leads to the following explicit couplings of the dilaton to the Proca field at $\mathcal{O}(\kappa^2)$
\begin{equation}\label{eq:lagrangian_qcds}
    \mathcal{L}_{\rm{QCD^2}} = -\frac{2}{\kappa^2}R + \frac{\partial_\mu \bar{Z}\partial^\mu Z}{\left(1-\frac{\kappa^2}{4}\bar{Z}Z\right)}-\frac{1}{2}F_{\mu\nu }^* F^{\mu\nu}+m^2A_\mu ^*A^\mu \left(1-\frac{\kappa}{2}(\bar{Z}+Z)+\frac{\kappa^2}{2}\bar{Z}Z+\mathcal{O}(\kappa^3)\right)
\end{equation}
The kinetic term for $Z$ can be cast into the standard form when we identify the dilaton component. Indeed, recall the field $Z$ was defined by

\begin{equation}
    Z=\frac{2a+i(e^{-2\phi}-1)}{2a+i(e^{-2\phi}+1)}.
\end{equation}
Where the axion corresponds to the parity-odd piece, i.e. the field $a$. Setting $a\to0$ implies $\bar{Z}=Z=-\tanh{\phi}$. Doing the further field redefinition $Z\to \frac{\kappa}{2}Z $, the Lagrangian \eqref{eq:lagrangian_qcds} becomes

\begin{equation}
     \mathcal{L}_{\rm{QCD^2}} =  -\frac{2}{\kappa^2}R+ \frac{4}{\kappa^2}(\partial\phi)^2-\frac{1}{2}F_{\mu\nu }^* F^{\mu\nu} +m^2A_\mu ^*A^\mu \left(1+2\tanh{\phi} +2(\tanh{\phi})^2+\mathcal{O}(\kappa^3) \right).
\end{equation}
Finally, we do the field re-definition  $\phi\to \frac{\kappa}{2}\phi $,  expanding up to second order in $\phi$, which is the order of the validity of the Lagrangian \eqref{eq:lagrangian_qcds}; in addition, we  turn $A^\mu$ into a real  field   using the argument made above \eqref{eq:symsc},
to arrive at
\begin{equation}
 \mathcal{L}_{\rm{QCD^2}} =  -\frac{2}{\kappa^2}R+ (\partial\phi)^2-\frac{1}{4}F_{\mu\nu } F^{\mu\nu} +\frac{m^2}{2}A_\mu A^\mu \left(1+ \kappa\phi +\frac{\kappa^2}{2}\phi^2 +\mathcal{O}(\kappa^3) \right),
\end{equation}
which precisely agrees with the Lagrangian \eqref{eq:1212dc} for $d=4$ if we truncate at  $\mathcal{O}(\kappa^2)$.

 
\section{Tree-level Unitarity at $n=4$}\label{residues at 4 points}

In this appendix we compute the residues of the gravitational
amplitude $A_4^{\frac{1}{2}\otimes\frac{1}{2}}$.
The aim of this is twofold. On the one hand this checks explicitly that the operation $(\ref{eq:massiveklt})$ defines a \ac{QFT} amplitude for $n=4$ and outlines the argument for general $n$. On the other hand, we want to find the matter fields that propagate in a given factorization channel.
For two dilaton emissions we find only the propagation of the Proca field, which is consistent with our Lagrangian \eqref{eq:1212dc-1}. For two axion emissions we find the propagation of tensor structures of rank four and five. The former can be interpreted as a pseudoscalar in $d=4$. In general dimension, the propagation of these structures makes it more
involved to write the Lagrangian of the full \12x12 theory including axions. \\
Consider then the Compton amplitude
from the \12x12 theory $(\ref{eq:massiveklt})$
\begin{equation}
A_{4}^{\frac{1}{2}\otimes\frac{1}{2}}(W_{1}H_{3}^{\mu_{3}\nu_{3}}H_{4}^{\mu_{4}\nu_{4}}W_{2}^{*})=\frac{K_{1324,1324}}{2^{\left\lfloor d/2\right\rfloor -1}}\text{tr}\left[A_{4,1324}^{\rm{QCD},\mu_{3}\mu_{4}}\slashed{\varepsilon}_{1}\left(\slashed{p_{1}}{+}m\right)\bar{A}_{4,1324}^{\rm{QCD},\nu_{3}\nu_{4}}\slashed{\varepsilon}_{2}\left(\slashed{p_{2}}{+}m\right)\right],\label{A4 grklt}
\end{equation}
 where the $4$ pt. \ac{QCD} partial amplitudes are given in $(\ref{eq:A41324})$,
and the massive  \ac{KLT} kernel at four points was given in $(\ref{eq:klt 1324})$. \\
We have claimed that $(\ref{A4 grklt})$ defines a tree-level amplitude. First, from the standard argument it is clear that the \ac{RHS} is local. Let us then argue that unitarity of the gravitational amplitude follows from
the unitarity of the \ac{QCD} amplitudes.
Consider for instance the factorization channel $2p_{1}{\cdot}k_{3}\rightarrow0$. We know that in such case
the \ac{QCD} amplitude factorizes as
\begin{equation}
A_{4,1324}^{\rm{QCD},\mu_{3}\mu_{4}}\rightarrow\frac{1}{2p_{1}{\cdot}k_{3}}A_{3,\text{R}}^{\rm{QCD},\mu_{3}}\left(\slashed{p}_{13}-m\right)A_{3,\text{L}}^{\rm{QCD},\mu_{4}}{+}\cdots,
\end{equation}
 Analogously, the  charge conjugated amplitude  factorizes as 
\begin{equation}
\bar{A}_{4,1324}^{\rm{QCD},\nu_{3}\nu_{4}}\rightarrow\frac{1}{2p_{1}{\cdot}k_{3}}\bar{A}_{3,\text{L}}^{\rm{QCD},\nu_{4}}\left(\slashed{p}_{13}+m\right)\bar{A}_{3,\text{R}}^{\rm{QCD},\nu_{3}}{+}\cdots,
\end{equation}
This implies that $(\ref{A4 grklt})$ behaves as
\begin{equation}\label{res1}
\begin{split}
A_{4}^{\frac{1}{2}\otimes\frac{1}{2}}(W_{1}H_{3}^{\mu_{3}\nu_{3}}H_{4}^{\mu_{4}\nu_{4}}W_{2}^{*})\rightarrow & -\frac{1}{2p_{1}{\cdot}k_{3}2^{\left\lfloor d/2\right\rfloor -1}}\text{tr}\bigg[A_{3,\text{L}}^{\rm{QCD},\mu_{4}}\slashed{\varepsilon}_{1}\left(\slashed{p_{1}}{+}m\right)\bar{A}_{3,\text{L}}^{\rm{QCD},\nu_{4}}\left(\slashed{p}_{13}+m\right)\\
 & \qquad\qquad\bar{A}_{3,\text{R}{}}^{\rm{QCD},\nu_{3}}\slashed{\varepsilon}_{2}\left(\slashed{p_{2}}{+}m\right)A_{3,\text{R}}^{\rm{QCD},\mu_{3}}\left(\slashed{p}_{13}-m\right)\bigg]{+}\cdots,
\end{split}
\end{equation}
We can examine the inner spectrum in the factorization channel by using the Fierz relations for the product of two
matrices $M$ and $N$ \cite{9781139026833},
\begin{equation}
\text{tr}[M\times N]=\frac{1}{2^{\left\lfloor d/2\right\rfloor }}\sum_{J}^{[d]}\frac{(-1)^{|J|}}{|J|!}\text{tr}\left[M\Gamma_{J}\right]\text{tr}\left[N\Gamma^{J}\right],\qquad[d]=\begin{cases}
d & {\rm {for\,\,even}\,}\,d\\
\frac{d-1}{2} & {\rm {for\,\,odd}\,\,}d
\end{cases}\label{eq:fierrz relations-1}
\end{equation}
where $\{\Gamma^{J}=\mathbb{I},\gamma^{\alpha},\gamma^{\alpha_{1}\alpha_{2}},\cdots,\gamma^{\alpha_{1}\cdots\alpha_{d}}\}$
is the Clifford algebra basis, with $\alpha_{1}<\alpha_{2}<\cdots<\alpha_{r}$. The gravitational amplitude \eqref{res1} then takes the form 
\begin{equation}\label{sumJ}
\begin{split}
&-\frac{1}{4p_{1}{\cdot}k_{3}2^{2\left\lfloor d/2\right\rfloor -2}}\sum_{J}^{[d]}\frac{(-1)^{|J|}}{|J|!}\text{tr}\left[A_{3,\text{L}}^{\rm{QCD},\mu_{4}}\slashed{\varepsilon}_{1}\left(\slashed{p_{1}}{+}m\right)\bar{A}_{3,\text{L}}^{\rm{QCD},\nu_{4}}\left(\slashed{p}_{13}+m\right)\Gamma_{J}\right]\times\\
 & \qquad\qquad \text{tr}\left[\bar{A}_{3\text{R}}^{\rm{QCD},\nu_{3}}\slashed{\varepsilon}_{2}\left(\slashed{p_{2}}{+}m\right)A_{3,\text{R}}^{\rm{QCD},\mu_{3}}\left(\slashed{p}_{13}-m\right)\Gamma^{J}\right]{+}\cdots,
\end{split}
\end{equation}
Now it is clear that each trace corresponds to the double copy for
the $3$pt amplitudes, therefore we have
\begin{equation}
A_{4}^{\frac{1}{2}\otimes\frac{1}{2}}(W_{1}H_{3}^{\mu_{3}\nu_{3}}H_{4}^{\mu_{4}\nu_{4}}W_{2}^{*})\rightarrow-\frac{1}{4p_{1}{\cdot}k_{3}}\sum_{J}^{[d]}\frac{(-1)^{|J|}}{|J|!}A_{3,\text{L}}^{\frac{1}{2}\otimes\frac{1}{2}}(W_{1}H_{3}^{\mu_{3}\nu_{3}}\varPhi_{J})\times A_{3,\text{R}}^{\frac{1}{2}\otimes\frac{1}{2}}(\varPhi^{J}H_{4}^{\mu_{4}\nu_{4}}W_{2}^{*}).\label{eq:resp1k3 general}
\end{equation}
Hence, we have shown that the gravitational 4-pt. amplitude factorizes
into the product of two 3-pt. amplitudes. Moreover, $\varPhi_{J}$
indicates all possible Lorentz structure propagating in the given
factorization channel. We can expand the sum to see the explicit form
of some of these structures propagating in this channel. To do so, first  notice
that since $\left(\slashed{p}_{13}+m\right)\mathbb{I}=\frac{p_{13,\alpha}}{m}\left(\slashed{p}_{13}+m\right)\gamma^{\alpha},$
we can identify the contribution from the terms $|J|=0$ and $|J|=1$ with the transverse and longitudinal modes of the spin-1 field. With this consideration \eqref{eq:resp1k3 general} takes the form
\begin{equation}
\begin{split}
 & -\frac{1}{p_{1}{\cdot}k_{3}2^{2\left\lfloor d/2\right\rfloor }}\bigg\{ \text{tr}\left[A_{3,\text{L}}^{\rm{QCD},\mu_{4}}\slashed{\varepsilon}_{1}\left(\slashed{p_{1}}{+}m\right)\bar{A}_{3,\text{L}}^{\rm{QCD},\nu_{4}}\left(\slashed{p}_{13}+m\right)\gamma^{\alpha}\right]D_{W,\alpha\beta}\\
 & \qquad\quad \text{tr}\left[\bar{A}_{3,\text{R}}^{\rm{QCD},\nu_{3}}\slashed{\varepsilon}_{2}\left(\slashed{p_{2}}{+}m\right)A_{3,\text{R}}^{\rm{QCD},\mu_{3}}\left(\slashed{p}_{13}-m\right)\gamma^{\beta}\right]\\
&\qquad +\frac{1}{2}\text{tr}\left[A_{3,\text{L}}^{\rm{QCD},\mu_{4}}\slashed{\varepsilon}_{1}\left(\slashed{p_{1}}{+}m\right)\bar{A}_{3,\text{L}}^{\rm{QCD},\nu_{4}}\left(\slashed{p}_{13}+m\right)\gamma^{\mu\nu}\right]\\
 & \qquad \eta_{[\mu\alpha}\eta_{\nu]\beta}\text{tr}\left[\bar{A,\text{R}}_{3}^{\rm{QCD},\nu_{3}}\slashed{\varepsilon}_{2}\left(\slashed{p_{2}}{+}m\right)A_{3,\text{R}}^{\rm{QCD},\mu_{3}}\left(\slashed{p}_{13}-m\right)\gamma^{\alpha\beta}\right]+\cdots\bigg\},
\end{split}
\end{equation}
 where 
\begin{equation}
D_{W,\alpha\beta}=\eta_{\alpha\beta}-\frac{p_{13,\alpha}p_{13,\beta}}{m^{2}},
\end{equation}
 and the $\cdots$ indicate the terms with higher value of $|J|$.
 
A similar analysis can be made at higher multiplicity starting from \eqref{eq:massiveklt}. The additional complication is that we have to deal with the factorization of the \ac{KLT} kernel $K_{\alpha \beta}$, which is however standard. Once the dust settles we obtain

\begin{equation}
\begin{split}
&-\frac{1}{2(p_I^2 -m^2)2^{2\left\lfloor d/2\right\rfloor -2}}\sum_{J}^{[d]}\frac{(-1)^{|J|}}{|J|!}K_{\alpha_L \beta_L}\text{tr}\left[A_{n_L,\alpha_L}^{\rm{QCD}}\slashed{\varepsilon}_{1}\left(\slashed{p_{1}}{+}m\right)\bar{A}_{n_L,\beta_L}^{\rm{QCD}}\left(\slashed{p}_{I}+m\right)\Gamma_{J}\right]\times\\
 & \qquad\qquad K_{\alpha_R \beta_R}\text{tr}\left[\bar{A}_{n_R,\beta_R}^{\rm{QCD}}\slashed{\varepsilon}_{2}\left(\slashed{p_{2}}{+}m\right)A_{n_R,\beta_R}^{\rm{QCD}}\left(\slashed{p}_{I}-m\right)\Gamma^{J}\right]{+}\cdots,
\end{split}
\end{equation}
as $p_I^2\to m^2$, for $p_I$ any internal massive momenta. This means that unitarity of $A_n^{\frac{1}{2}\otimes \frac{1}{2}}$ should follow from that of $A_n^{\rm{QCD}}$ provided we correcltly account for the tensor structures $\Gamma^J$ as particles propagating in this channel.

Let us leave the analysis for general multiplicity for future work, and here instead focus in the internal spectrum at $n=4$. Next we consider two such cases and determine the fields propagating
in this channel. The first is the gravitational amplitude
for a massive line emitting two dilatons, whereas the second one corresponds
to the amplitude for the emission of two axions. 

\subsection*{Dilaton emission}

For this explicit
example the sum truncates at $|J|=3$.
Moreover, it can be checked that the terms $|J|=2$ and $|J|=3$
add up exactly to the contributions given by the $|J|=0$ and $|J|=1$
terms, namely, they account for a propagating spin-1 field. With this in mind,   $(\ref{eq:resp1k3 general})$ gives 
 
\begin{equation}
A_{4}^{\frac{1}{2}\otimes\frac{1}{2}}(W_{1}\phi_{3}\phi_{4}W_{2}^{*})\rightarrow\frac{\kappa^{2}}{32p_{1}{\cdot}k_{3}(2{-}d)}\left[\left(d{-}4\right)p_{1}^{\alpha}\,p_{3}{\cdot}\varepsilon_{1}{+}2m^{2}\varepsilon_{1}^{\alpha}\right]D_{W,\alpha\beta}\left[\left(d{-}4\right)p_{2}^{\beta}\,p_{4}{\cdot}\varepsilon_{2}{+}2m^{2}\varepsilon_{2}^{\beta}\right].\label{eq:residue final}
\end{equation}
 It can be also checked that the same residue is computed starting
from $(\ref{eq:ABphiphiB-2})$.

\subsection*{Axion emission}

Let us move on to the slightly more complicated example corresponding to the emission of two axions by a massive line. As we mentioned, the matter spectrum of the \12x12
double copy can be truncated to massive
vector fields once we consider
the emission of gravitons or dilatons, but not axions. On the other hand, via double copy we showed that the matter line can only produce axions in pairs.
An example of this is the four point amplitude for two axions:
\[
A_{4}^{\frac{1}{2}\otimes\frac{1}{2}}(W_{1}B_{3}B_{4}W_{2}^{*})=\frac{1}{2^{\left\lfloor d/2\right\rfloor -1}}K_{1324,1324}\left(A_{4,1324\,[\mu_{4}}^{\rm{QCD},[\mu_{3}}\bar{A}_{4,1324\,\nu_{4}]}^{\rm{QCD},\nu_{3}]}\right)\epsilon_{B_{3},\mu_{3}\nu_{3}}\epsilon_{B_{4}}^{\mu_{4}\nu_{4}}.
\]

Studying tree-level unitarity in this object leads to consider additional matter fields. For instance, consider  the channel $2p_{1}{\cdot}k_{3}\rightarrow0$ given by \eqref{eq:resp1k3 general}.
For two axion emissions,  the sum truncates at $|J|=5.$ The sum of the contributions
for $|J|=0$ and $|J|=1$ cancels out, therefore no Proca field will
propagate in this channel, as expected since $A_{3}^{\frac{1}{2}\otimes \frac{1}{2}}(W_{1}BW_{2}^{*})=0$. We can check that
the sum of the contributions for $|J|=2$ and $|J|=3$ equals the sum
of the contributions for $|J|=4$ and $|J|=5.$ Therefore, in this factorization
channel there is the propagation of particles associated to the structures $\{\gamma^{\mu_{1},\mu_{2}},\gamma^{\mu_{1}\mu_{2}\mu_{3}}\}$
or equivalently $\{\gamma^{\mu_{1}\mu_{2}\mu_{3}\mu_{4}},\gamma^{\mu_{1}\mu_{2}\mu_{3}\mu_{4}\mu_{5}}\}$. The propagation of these structures is what makes more
involved to write down a Lagrangian including the additional fields in general dimension. We leave this task for future work.
In $d=4$ however there is a simplification since the form $\gamma^{\mu_{1}\mu_{2}\mu_{3}\mu_{4}}$
can be dualized to a pseudoscalar, whereas the form $\gamma^{\mu_{1}\mu_{2}\mu_{3}\mu_{4}\mu_{5}}$
vanishes. The propagation of this pseudoscalar (as obtained in \cite{Johansson:2019dnu}) was pointed out
in the previous Appendix, as obtained from antisymmetrization of spinors in $d=4$. 

%% file: appendices/teukolsky.tex
\chapter{Gravitational wave scattering, Teukolsky formulation}\label{ch:teukolsky}

In this appendix we approach the classical problem of the scattering of a gravitational wave off the Kerr \ac{BH} from  \ac{BHPT}. We aim to show the classical solutions of Teukolsky equation,  indeed agree with the amplitudes derivation of the differential cross section \eqref{eq:sigmaspin2}. 
For the spinless problem, the problem of the   scattering of waves off the Schwarzschild \ac{BH} is approached  by means of solutions to   the so called  Regge-Wheeler (RW) equation 
\cite{Regge:1957td} (see also \cite{Chandrasekhar:579245}), which  shows how the Schwarzschild black hole was stable under small perturbations caused by the  wave. In the case of the Kerr \ac{BH}, an analogous equation was derived by Teukolsky  \cite{Teukolsky:1972my}, by applying the Newman-Penrose formalism \cite{Newman:1961qr}, to the problem of perturbations of of Kerr. This formalism allows to write separation of variables solutions for the equation for the perturbation in terms of the  radial and angular part, while keeping all orders in the Newton's constant and the \ac{BH}'s spin. 
The Teukolsky equation is the cornerstone of modern \ac{BHPT}, used to approach  problems for both one and  two-body processes in general relativity. 

\begin{figure}
     \centering
     \begin{subfigure}[b]{0.49\textwidth}
         \centering
         \includegraphics[width=\textwidth]{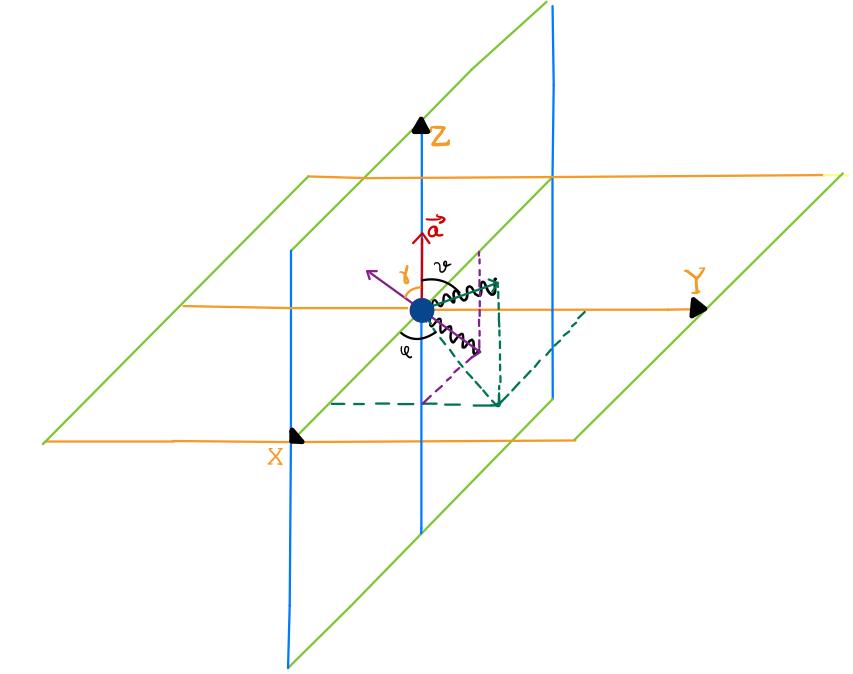}
         \caption{Initial \ac{BHPT} setup}
         \label{fig:Initial}
     \end{subfigure}
     \hfill
     \begin{subfigure}[b]{0.49\textwidth}
         \centering
         \includegraphics[width=\textwidth]{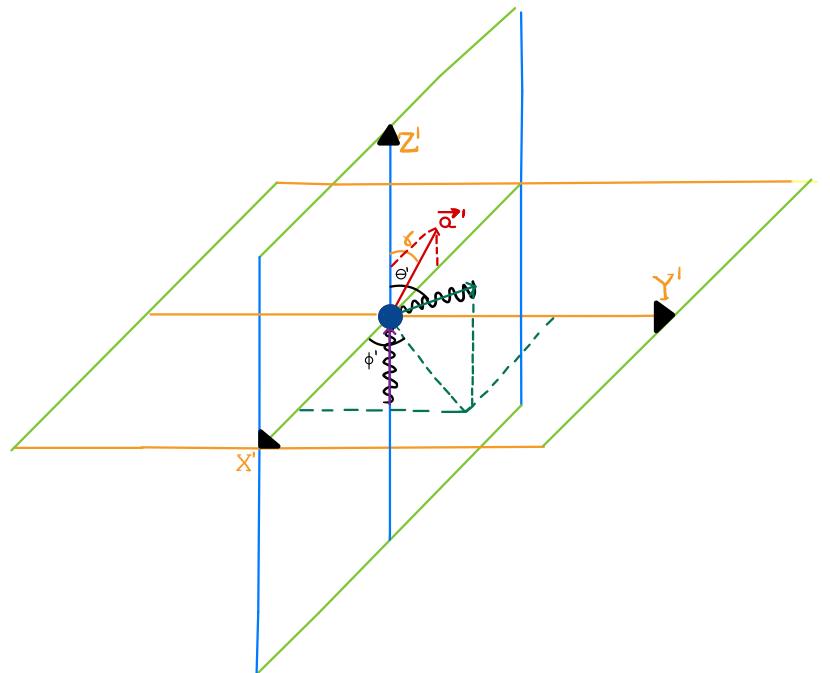}
         \caption{Rotated \ac{BHPT} setup}
         \label{fig:rotated}
     \end{subfigure}
     \caption{Gravitational wave scattering in the  \ac{BHPT} setup. (a) An incoming \ac{PW} (purple) impinges on the \ac{BH} at an angle $\gamma$ with respect to the direction of the \ac{BH} spin ($\vec{a} = a \hat{z}$). The outgoing scatter wave (green) moves in a general direction with angles $\vartheta$ and $\varphi$ with respect to eh $Z$ and $X$ axis respectively. (b) Rotated frame. In this frame, the incoming \ac{PW} (purple) moves along the vertical axis, whereas the spin of the \ac{BH} is rotated at an angle $\gamma$ with respect to $Z'$. The outgoing scatter wave (green) now moves in the general direction $\theta',\phi'$ with respect to the $Z'$ and $X'$ axis respectively. }
\end{figure}

The setup for this part is to  consider a \ac{PW} incoming  into the Kerr \ac{BH}, and subsequently  scatter into a  wave (S), which can be written as the superposition between  the incoming plane wave and the outgoing spherical wave. In this computation we use the conventions for the wave and black hole momenta as indicated in Figure \ref{fig:Initial}. 
For vacuum perturbations, the Teukolsky scalar (radiation scalar) ${}_{-2}\psi=\varrho^{4}\psi_4$, contains all the information of the radiative dynamics and will be  the main objects of study in this section. Here
 $\varrho=\frac{1}{r-i a \cos \vartheta}$, with $r$- and $\vartheta$ the spatial coordinates of the scattering problem. 
 The radiation scalar 
satisfies the homogeneous Teukolsky equation \cite{Teukolsky:1972my}
\begin{align}
&\left[\frac{(r^2+a^2)^2}{\Delta}-a^2\sin^2\vartheta\right]\spdiff{\psi}{t}+\frac{4 M a r}{\Delta}\frac{\partial^2 \psi}{\partial t \partial \varphi}+
\left[\frac{a^2}{\Delta}-\frac{1}{\sin^2\vartheta}\right]\spdiff{\psi}{\varphi} 
-\Delta^{-s}\pdiff{}{r}\left(\Delta^{s+1}\pdiff{\psi}{r}\right) \nonumber\\
&-\frac{1}{\sin \vartheta}\pdiff{}{\vartheta}\left(\sin\vartheta \pdiff{\psi}{\vartheta}\right)  
-2s\left[\frac{a(r-M)}{\Delta}+\frac{i \cos\vartheta}{\sin^2\vartheta}\right]\pdiff{\psi}{\varphi}
-2 s \left[\frac{M(r^2-a^2)}{\Delta}-r -i a \cos\vartheta\right]\pdiff{\psi}{t}\nonumber \\
&\qquad\qquad+s(s\cot^2\vartheta-1)\psi = 0\,,
\label{Eq:TeukMaster}
\end{align} 
where $s$ corresponds to the spin of the perturbation, which for  gravitational wave scattering   simply fixes  $s=-2$.
This equation is separable in the frequency domain via 
\begin{align}
	{}_{-2}\psi(t,r,\vartheta,\varphi)=\sum_{\ell m}\int d\omega e^{-i \omega t}{}_{-2}Z_{\ell m \omega}\Rlm{-2}{\ell}{m}{\omega}(r)\sSlm{-2}{\ell}{m}(\vartheta,\varphi,a \omega).
\end{align}
Here, ${}_{-2}Z_{l m \omega}$ are normalization coefficients.  $\Rlm{s}{\ell}{m}{\omega}(r)$ are solutions to the homogeneous radial Teukolsky equation, and $\sSlm{s}{\ell}{m}(\vartheta,\varphi,a \omega)$ are the spin-weighted spheroidal harmonics with respective defining equations
\begin{align}
\bigg[\Delta^{-s}\diff{}{r}\left(\Delta^{s+1}\diff{}{r}\right)+&\frac{K^2-2 i s (r-M)K}{\Delta}
+4 i s \omega r -{}_s \lambda_{lm}\bigg]{}_s R_{\ell m \omega}(r)
=0\,, \label{Eq:RadTeuk} 
\end{align}
and
\begin{align}
\bigg[\frac{1}{\sin \vartheta}\diff{}{\vartheta}\left(\sin\vartheta \diff{  }{\vartheta}\right)-&a^2 \omega^2 \sin^2\vartheta
-\frac{(m+s\cos\vartheta)^2}{\sin^2\vartheta} 
-2 a \omega s \cos\vartheta+s+2 m a \omega + {}_s \lambda_{lm}\bigg]\sSlm s l m (\vartheta, \varphi; a \omega)=0\,, \label{Eq:Slmeq} 
\end{align}
where $K = (r^2+a^2)\omega - a m$, and 
${}_s \lambda_{lm}$ is the spheroidal eigenvalue.
We will come back to the Teukolsky equation in a moment. Let us in the mean time  provide the classical definition of the differential cross section for the gravitational wave scattering process, which will be the observable to compare with the \ac{QFT} computation.  It corresponds to the  outward energy 
flux from the scattered wave (S) normalised by the energy per unit area in the incoming
plane wave (\ac{PW}). 
In Kerr spacetime, the differential cross-section for the scattering of a plane gravitational wave can be expressed in the following form
\begin{align}\label{eq:csecgr}
\frac{d\sigma}{d\Omega}=|f(\vartheta,\varphi)|^2+|g(\vartheta,\varphi|^2,
\end{align}
where $f$ and $g$ are respectively the complex helicity preserving and helicity reversing scattering amplitudes. Using a partial wave expansion they are given by the expressions
\begin{align}
     f(\vartheta,\varphi)&=\sum_{l=2}^{\infty} \sum_{m=-\infty}^{\infty}{}_{-2}S_{lm}(\gamma,0;a\omega){}_{-2}S_{lm}(\vartheta,\varphi;a\omega)f_{lm\omega}, \label{Eq:fKerrGeneric}\\
     g(\vartheta,\varphi)&=\sum_{l=2}^{\infty} \sum_{m=-\infty}^{\infty}{}_{-2}S_{lm}(\gamma,0;a\omega){}_{-2}S_{lm}(\pi-\vartheta,\varphi;a\omega)g_{lm\omega} \label{Eq:gKerrGeneric}
\end{align}
where $\gamma$ is the angle between the incoming wave vector and the axis of rotation of the Kerr \ac{BH} (see Figure \ref{fig:Initial}). The   amplitude modes can be obtained   by
\begin{align}
    f_{lm\omega}&=\frac{2\pi}{i\omega}\sum_{P=\pm1}\left(e^{2i\delta^{P}_{lm}}-1\right),\label{eq:fmodes} \\
    g_{lm\omega}&=\frac{2\pi}{i\omega}\sum_{P=\pm1}P(-1)^{l+m+2}\left(e^{2i\delta^{P}_{lm}}-1\right)\label{eq:gmodes}
\end{align}
where $\delta^{P}_{lm}$ are the phase shifts in a standard scattering process. These are computed  by solving the radial Teukolsky equation \eqref{Eq:RadTeuk}  \cite{Bautista:2021wfy,Dolan:2008kf}.  
Let us see. For our needs we will require a vacuum Teukolsky solution, typically labelled $\Rin{s}{\ell}{m}{\omega}$, which satisfies the physical boundary condition of purely ingoing waves at the horizon, namely
\begin{align}
	\Rin{-2}{\ell}{m}{\omega}(r)=\Btrans{\ell}{m}{\omega} \Delta^2e^{-i\tilde{\omega} \rs}, \qquad r\rightarrow r_+,
\end{align}
where $r_+=M+\sqrt{M^2-a^2}$ is the location of the outer horizon of Kerr, $\tilde{\omega}=\omega-\frac{m a}{2 M r_+}$, and $\Btrans{\ell}{m}{\omega}$ is the so called transmission coefficient.
Imposing this boundary condition fixes the asymptotic form at radial infinity for each $\ell,m$ mode to be 
\begin{align}
	\Rin{-2}{\ell}{m}{\omega}(r)=\Binc{\ell}{m}{\omega} r^{-1}e^{-i\omega \rs}+\Bref{\ell}{m}{\omega} r^3 e^{i\omega \rs} , \qquad r\rightarrow \infty, \label{Eq:InfFormTeuk}
\end{align}
where $\Binc{\ell}{m}{\omega}$ and $\Bref{\ell}{m}{\omega}$ are the incident and reflection coefficients.
Solutions to the radial Teukosky equation can  be written as infinite series of hypergeometric functions or confluent hypergeometric functions, depending on the required asymptotic boundary conditions \cite{Leaver86_1,Leaver86_2,Mino:1996nk,Sasaki:2003xr}. Investigation of the asymptotic behaviour of these infinite series yields expressions for the incident and reflection coefficients:
\begin{align}
\Binc{\ell}{m}{\omega}
=&\omega^{-1}\left[{K}_{\nu}-
ie^{-i\pi\nu} \frac{\sin \pi(\nu-s+i\epsilon)}
{\sin \pi(\nu+s-i\epsilon)}
{K}_{-\nu-1}\right]A_{+}^{\nu} e^{-i(\epsilon\ln\epsilon -\frac{1-\kappa}{2}\epsilon)},\label{Eq:Binc}
\\
\Bref{\ell}{m}{\omega}
=&\omega^{-1-2s}\left[{K}_{\nu}
+ie^{i\pi\nu} {K}_{-\nu-1}\right]A_{-}^{\nu}
e^{i(\epsilon\ln\epsilon -\frac{1-\kappa}{2}\epsilon)},\label{Eq:Bref}
\end{align}
with
\begin{align}
&A_{+}^\nu=e^{-{\pi\over 2}\epsilon}e^{{\pi\over 2}i(\nu+1-s)}
2^{-1+s-i\epsilon}{\Gamma(\nu+1-s+i\epsilon)\over 
\Gamma(\nu+1+s-i\epsilon)}\sum_{n=-\infty}^{+\infty}a_n^\nu,\\
&A_{-}^\nu=2^{-1-s+i\epsilon}e^{-{\pi\over 2}i(\nu+1+s)}e^{-{\pi\over 2}\epsilon}
\sum_{n=-\infty}^{+\infty}(-1)^n{(\nu+1+s-i\epsilon)_n\over 
(\nu+1-s+i\epsilon)_n}a_n^\nu, 
\end{align}
and 
\begin{align}
K_{\nu}=&	\frac{e^{i\epsilon\kappa}(2\epsilon \kappa )^{s-\nu-r}2^{-s}i^{r}
	\Gamma(1-s-2i\epsilon_+)\Gamma(r+2\nu+2)}
	{\Gamma(r+\nu+1-s+i\epsilon)
	\Gamma(r+\nu+1+i\tau)\Gamma(r+\nu+1+s+i\epsilon)}
	\nonumber\\
	&\times \left ( \sum_{n=r}^{\infty}
	(-1)^n\, \frac{\Gamma(n+r+2\nu+1)}{(n-r)!}
	\frac{\Gamma(n+\nu+1+s+i\epsilon)}{\Gamma(n+\nu+1-s-i\epsilon)}
	\frac{\Gamma(n+\nu+1+i\tau)}{\Gamma(n+\nu+1-i\tau)}
	\,a_n^{\nu}\right)
	\nonumber\\
	&\times \left(\sum_{n=-\infty}^{r}
	\frac{(-1)^n}{(r-n)!
	(r+2\nu+2)_n}\frac{(\nu+1+s-i\epsilon)_n}{(\nu+1-s+i\epsilon)_n}
	a_n^{\nu}\right)^{-1}.
	\label{eq:Knu}
\end{align}
Here $r$ is a free parameter (not to be confused with the radial coordinate) we set to be $0$, $\epsilon=2 G M \omega$, $\kappa=\sqrt{1-a^{\star 2}}$, $a^\star=\frac{a}{GM}$, $\tau=\frac{\epsilon-m q}{\kappa}$ and $ \epsilon_{\pm}=\frac{\epsilon\pm\tau}{2}$. In these expressions the series coefficients $a_n^{\nu}$ satisfy 3 term recurrence relations and the `renormalised angular momentum' $\nu$ is determined by insisting the series all converge. 
The phase shifts are then simply given by 
\begin{equation}\label{eq:phase_shift}
    e^{2i\delta_{\ell m}^P} = (-1)^{l+1}\frac{B^{\rm ref}_{lm\omega}}{B^{\rm inc}_{lm\omega}}\,.
\end{equation}

Solving the Teukolsky equation then means to solve for $\Binc{\ell}{m}{\omega}$ and $\Binc{\ell}{m}{\omega}$, and therefore for the phase shift \eqref{eq:phase_shift} which can be use to compute  the scattering amplitudes modes \eqref{eq:fmodes} and \eqref{eq:gmodes}. The differential cross section can be obtained by performing the infinite sums \eqref{Eq:fKerrGeneric} and \eqref{Eq:gKerrGeneric}, which can then be replaced in \eqref{eq:csecgr}, and compare to the amplitudes result \eqref{eq:sigmaspin2}. In general, solving for this infinite sums is a very non-trivial task, and we will have to engineer a method to compare to the closed form solutions from the \ac{QFT} computation. We will come bakc to this in \cref{sec:matching_BHPT_QFT}.

In general, and as mentioned above,  solutions of the Teukolsky equation encapsulate all orders in perturbation theory (all orders in $G$), and all orders in the \ac{BH}'s spin. 
In practice solving for this conditions,  this can become a non trivial task, however calculating the low frequency expansions of $\Binc{\ell}{m}{\omega}$ and $\Binc{\ell}{m}{\omega}$ ultimately come down to determining low frequency expansions of $a_n^{\nu}$ and $\nu$. These have been extensively studied (see e.g. \cite{Sasaki:2003xr,Kavanagh:2016idg}), and so we will not discuss this problem here. The relevant results will soon be available in the Black Hole Perturbation Toolkit \cite{BHPToolkit}. Comparison to the \ac{QFT} computation requres a further expansion of the results in powers of the \ac{BH}'s. This is a simple  task up to $a^4$ which is the order  we are interested in this thesis \footnote{Starting at order $a^5$ and higher, a more careful treatment of the spin expansion is needed since  terms of the form  $\sqrt{1-a^{\star 2}}$ needs to be analytically continue from $a\le a^\star <1$, to $a^\star>1$. This then introduces a  branch pick. In addition, a this spin order, there is the presence of abortive terms (entering as complex components to the phase shift) that are not present at $a^4$ and lower orders. These issues will be  discussed in more detail in \cite{BGKV}. }.

\section{Low energy expansion}
The matching to the \ac{QFT} computation can actually be done at the level of the scattering amplitude, which up to a phase should coincide with the \ac{BHPT} result. 
We calculate the partial wave amplitudes in a long wavelength limit $GM\omega\ll1$. For this, it is  crucially important that $0\leq\frac{a}{GM}<1$ when constructing the long wavelength expansion, so that we can use the tools of black hole perturbation theory. When  $\frac{a}{GM}>1$ the \ac{BH} ceases to have an horizon and standard methods for solving the Teukolsky equation are not clearly defined. In this thesis, we will be interested to compute the partial waves up to order $a^4$, where the final result is independent of whether $0\leq\frac{a}{GM}<1$ or $\frac{a}{GM}>1$. 

We first construct the low frequency expansion of the harmonic modes of the amplitude functions, holding $a^\star=\frac{a}{GM}<1$ fixed. This is essentially a two step process. \\

\noindent $1.$ Calculate $f_{lm\omega}$ and $g_{lm\omega}$ as a Taylor expansion in $\epsilon=2G M \omega$. This can be given order-by-order in closed form as a function of $a^\star$. Further, for all $l$ greater than some value $l_{\rm min}$, the solutions can also be written as a function of $l$ and $m$. $l_{\rm min}$ is determined by the order in $\epsilon$ to which one is working. For example, up to $\epsilon^5$ only $l=2$ differs from the closed form general $lm$ expressions. At higher orders, successively higher values of $l$ will disagree with the general forms.\\
\noindent $2.$ Project the spin-weighted spheroidal-harmonic representation on to a basis of spin-weighted spherical harmonics. This is fairly straightforwardly done since in a low frequency limit one can write 
\begin{align}\label{eq:spinwe}
    \sSlm{-2}{\ell}{m}(\vartheta,\varphi,a \omega)=\sum_{\pm i}d^i_{lm}\sYlm{-2}{\ell+i,}{m}(\vartheta,\varphi)(a\omega)^i
\end{align}\\
where the coefficients $d^i_{lm}$ are well known. See e.g. Appendix B of \cite{Kavanagh:2016idg}.

As a matter of computational complexity, this procedure is more or less independent of whether one is dealing with polar ($\gamma=0$) or non-polar  ($\gamma\ne 0$) scattering; for non-polar we simply need to keep all $m$-modes.

We now give some explicit details of the calculations of $f)$. Significant detail of such a calculation, and for $g$, can also be found in \cite{Dolan:2008kf} where the author computed the first correction in $a \omega$ for the polar ($\gamma=0$) case. We will compute up to and including $(a\omega)^4$  the relevant $lm$ modes.
As was noted in \cite{Dolan:2008kf}, when calculating the expansion of $\frac{\Bref{\ell}{m}{\omega}}{\Binc{\ell}{m}{\omega}}$ in small $\epsilon=2G M \omega$, the complicated function $K_\nu$ as given above, appears only in the schematic form $1+\frac{K_{-\nu-1}}{K_\nu}$. Explicit computation shows that $\frac{K_{-\nu-1}}{K_\nu}\sim\epsilon^{2\ell-1}$. This implies for our calculation it can only be relevant for $\ell=2,3$. As we will see below this leads to a correction to the partial wave series at $a^{\star 5}$ only for $\ell=2$, which are not relevant for the present thesis \footnote{ To correctly account for this we would need to treat the amplitudes in two parts, a generic-$l$ expansion which ignores all contributions from $\frac{K_{-\nu-1}}{K_\nu}$, and a specific $l$ piece which includes it. See \cite{BGKV}}.

Omitting the cumbersome intermediate expansions we arrive at the following expression for the amplitude modes
\begin{align}
f_{\ell m \omega}=\frac{\Gamma(\ell-1-i\epsilon)}{\Gamma(\ell+3+i\epsilon)}\frac{\Gamma(\ell+3)}{\Gamma(\ell-1)}\beta_{\ell m\omega} \label{Eq:Ftobeta}
\end{align}
where $\beta_{\ell m\omega}$ has the form
\begin{align}
	\beta_{\ell m}=1+\sum_{i=2}^\infty\beta^{(i)}_{\ell m}\epsilon^{i}.
\end{align}
The $\Gamma$-function prefactors in \eqref{Eq:Ftobeta} absorb a significant amount of complicated structure in the low frequency expansion of $f_{\ell m \omega}$, so that the $\beta^{(i)}_{\ell m}$ are relatively simple. We find explicitly that for $i\leq5$ the $\beta^{(i)}_{\ell m}$ are polynomials in $a^\star$ of order $i-1$.

The downside of the $\Gamma$-function prefactors is that the projection onto spherical harmonics is contains some subtlety. Schematically, writing the low frequency expansion of the harmonics as
\begin{align}
    \sSlm s l m (\vartheta,\varphi,a\omega)&=\sYlm s l m+\sSlm s l m ^{(1)}q\epsilon+\sSlm s l m ^{(2)}q^2\epsilon^2+\ldots,\\
    \sSlm s l m (0,0,a\omega)&=N^{(0)}_{lm}+N_{lm}^{(1)}q\epsilon+N_{lm}^{(1)}q^2\epsilon^2+\ldots,
\end{align}
where the $N^{(i)}_{lm}$ are constants. 

Both of the above expansions are available open source in the  \texttt{SpinWeightedSpheroidalHarmonics} package of the Black hole perturbation toolkit \cite{BHPToolkit}.
The full amplitude function is then
\begin{align}
    f(\vartheta,\varphi)&=\frac{1}{i\omega}\sum_{lm}\frac{\Gamma(\ell-1-i\epsilon)}{\Gamma(\ell+3+i\epsilon)}\frac{\Gamma(\ell+3)}{\Gamma(\ell-1)}\left(N^{(0)}_{lm}+N_{lm}^{(1)}a^\star\epsilon+N_{lm}^{(1)}a^{\star 2}\epsilon^2+\ldots\right)\\
  &  \times\left(\sYlm s l m+\sSlm s l m ^{(1)}a^{\star }\epsilon+\sSlm s l m ^{(2)}a^{\star 2}\epsilon^2+\ldots\right)\left(1+\beta^{(2)}_{\ell m}\epsilon^{2}+\ldots\right) \\
  &=\frac{1}{i\omega}\sum_{lm}\frac{\Gamma(\ell-1-i\epsilon)}{\Gamma(\ell+3+i\epsilon)}\frac{\Gamma(\ell+3)}{\Gamma(\ell-1)}\bigg\{N^{(0)}_{lm}\sYlm s l m+\big[N^{(0)}_{lm}\sSlm s l m ^{(1)}+\sYlm s l m N^{(1)}_{lm}\big]a^{\star }\epsilon\\
  &+\bigg[\beta^{(2)}_{\ell m}N^{(0)}_{lm}\sYlm s l m+(N^{(0)}_{lm}\sSlm s l m ^{(2)}+N^{(2)}_{lm}\sYlm s l m +N^{(1)}_{lm}\sSlm s l m ^{(1)})a^{\star 2}\bigg]\epsilon^2+\ldots\bigg\}.
\end{align}
We we will be investigating the large-$a^{\star }$ limit of this expression. Knowing that $\beta^{(i)}_{\ell m}$ is at most $\mathcal{O}(a^{\star (i-1))}$ in this limit, one might naively conclude that the dominant, and relevant contribution comes entirely from the cross terms of the expansions of the spheroidal harmonics. However, when we project onto the spherical harmonics, an explicit calculation reveals that
\begin{align}
  \sum_{lm}  \frac{\Gamma(\ell-1-i\epsilon)}{\Gamma(\ell+3+i\epsilon)}\frac{\Gamma(\ell+3)}{\Gamma(\ell-1)}\int \big[N^{(0)}_{lm}\sSlm s l m ^{(1)}+\sYlm s l m N^{(1)}_{lm}\big] \sYlm s l m^*d\Omega\sim\mathcal{O}(\epsilon),\\
  \sum_{lm}  \frac{\Gamma(\ell-1-i\epsilon)}{\Gamma(\ell+3+i\epsilon)}\frac{\Gamma(\ell+3)}{\Gamma(\ell-1)}\int (N^{(0)}_{lm}\sSlm s l m ^{(2)}+N^{(2)}_{lm}\sYlm s l m +N^{(1)}_{lm}\sSlm s l m ^{(1)}) \sYlm s l m^*d\Omega\sim\mathcal{O}(\epsilon),
  \label{eq:partial-wves}
\end{align}
so that the pure cross terms get an `order bump' in the frequency expansion upon projection to spherical harmonics. This pattern continues as far as we have checked. The end result is that the relevant terms in the large-$a^{\star }$ expansion will be these cross terms \emph{and} the leading order behaviour of the $\beta^{(i)}_{\ell m}$ functions.

For the generic-$l$ contribution we find explicitly that the $\beta^{(i)}_{\ell m}$ are polynomials in $a^{\star }$. In particular  $\beta^{(i)}_{\ell m}$ is an $(i-1)$th order polynomial in $a^{\star }$. Thus the highest power of $a^{\star }$ in each $\beta^{(i)}_{\ell m}$ gives the $\mathcal{O}(G)$ contribution we seek; all other terms are higher order in $G$. Focusing purely on these terms we will write
\begin{align}
	\beta^G_{\ell m}=1+\sum_{i=2}^\infty\beta^{G,(i)}_{\ell m}a^{\star (i-1)}\epsilon^{i}.
\end{align}
For example, for $m=2$ we find
\begin{align}
\beta^{G,(2)}_{\ell 2}&=-\frac{2 i}{l (l+1)}, \\
\beta^{G,(3)}_{\ell 2}&=-\frac{i \left(l^6+3 l^5+3 l^4+l^3-80 l^2-80 l-48\right)}{2 l^3 (l+1)^3 (2 l-1) (2
   l+3)}, \\
\beta^{G,(4)}_{\ell 2}&=\frac{i \left(l^{10}+5 l^9+3 l^8-18 l^7+75 l^6+309 l^5-151 l^4-848 l^3-1696 l^2-1216
   l-384\right)}{(l-1) l^5 (l+1)^5 (l+2) (2 l-1) (2 l+3)}, \\
\beta^{G,(5)}_{\ell 2}&=\frac{i}{8 (l-1) l^7 (l+1)^7 (l+2)
   (2 l-3) (2 l-1)^3 (2 l+3)^3 (2 l+5)} \big(12 l^{20}+120 l^{19}+263 l^{18}\nonumber\\
   &-1053 l^{17}-20767 l^{16}
   -126764
   l^{15}-122488 l^{14}+1199896 l^{13}+2612040 l^{12}-5081558 l^{11}\nonumber\\
   &-22234775
   l^{10}-22582443 l^9+29249651 l^8+123462810 l^7+142507808 l^6+33491616 l^5\nonumber\\
   &-65123264
   l^4-63746304 l^3-10990080 l^2+9262080 l+4147200\big),\nonumber \\
\end{align}
and analogously for other harmonics.

\section{Matching procedure: \ac{BHPT} and \ac{QFT} amplitudes }
\label{sec:matching_BHPT_QFT}

In order to match our \ac{QFT} amplitudes with the results from \ac{BHPT} it is convenient to now project the previous amplitude function \eqref{Eq:fKerrGeneric} onto spin weighted \textit{spherical} harmonics as in \eqref{eq:spinwe}
\begin{equation}
f(\vartheta,\varphi)=\sum_{lm}{}_{-2}Y_{lm}(\vartheta,\varphi)f_{lm}(\gamma),
\label{Eq:fSpher}
\end{equation}
where $f_{lm}=f^N+f^{(1)}_{lm}z+f^{(2)}_{lm}z^2+\ldots$ and $z=a^{\star}\epsilon={\textcolor{black}{2}}a\omega$. While there are some subtleties in the projection as discussed in Sec. 4.3.1. of \cite{Dolan:2008kf}, we will omit such details here.

An important feature emerges for \textit{polar scattering}, which is obtained by setting $\gamma=0$. One finds that only the modes $f^{(i)}_{l0}$ are non-trivial and
\begin{equation}
f^{(i)}_{l0}(0)=0\,,\,\,\textrm{ for  }\,\, i=2k+1\,\, \textrm{  or for  } i\leq l. 
\end{equation}
This means that, other than the Newtonian term, for  \textit{polar scattering} the infinite sum over the spherical harmonics reduces to a finite sum for each power of $z$. This has been noted in \cite{Dolan:2008kf} for the case of GW scattering in Kerr. 

In the off-axis case where $\gamma\neq0$, no such  simplification occurs, and we find it convenient to compare with the 4-pt amplitudes \eqref{eq:App-gr}
, by working mode-by-mode. To do this we first need to align our coordinates by a rotation. 

Let us see how this works. The amplitude function \eqref{Eq:fSpher} is written as a sum over the spin weighted spherical harmonics ${}_{-2}Y_{lm}(\vartheta,\varphi)$,
where $(\vartheta,\varphi)$ is the direction of the outgoing wave in a coordinate system where the spin direction is $\vartheta=0$, i.e. the $+Z$ direction, and the incoming wave is in the direction $(\vartheta,\varphi)=(\gamma,0)$ (in the $X$-$Z$ plane), so that $\gamma$ is the angle between the spin and the incoming wave.

Now let us rotate our $(\vartheta,\varphi)$--$(X,Y,Z)$ coordinate system about the $Y$ axis (the same as the new $Y'$-axis) by an angle $\gamma$, to bring the incoming wave direction to the new $+Z'$ direction, and call the new coordinates  $(\theta',\phi')$--$(X',Y',Z')$ (see Figure \ref{fig:rotated}).  This is still not the same as the $(\theta,\phi)$--$(x,y,z)$ coordinate system used in \cref{eq:App-gr}, but now the $z$-axes are the same. 
The rotation of the spin weighted spherical harmonics is known to be accomplished by
\begin{equation}
{}_{-2}Y_{lm}(\theta',\phi')=\sum_{m'}D^{l*}_{mm'}(\gamma)\;{}_{-2}Y_{lm'}(\vartheta,\varphi)\,,
\end{equation}
where $D^{l*}_{mm'}$ is the (complex conjugate) Wigner $D$-matrix with Euler angles $(0,\gamma,0)$,
\begin{equation}
D^{l*}_{mm'}(\gamma)=D^{l*}_{mm'}(0,\gamma,0)=(-1)^{m'}\sqrt{\frac{4\pi}{2l+1}}{}_{-m'}Y_{lm}(\gamma,0).
\end{equation}
Now the amplitude \eqref{Eq:fSpher} takes the form
\begin{equation}\label{Flm}
f=\sum_{lm}{}_{-2}Y_{lm}(\theta',\phi')f'_{lm}(\gamma),
\end{equation}
with
\begin{equation}\label{eq:modesfBHPT}\boxed{
f'_{lm}(\gamma)=\sum_{m'}D^{l*}_{m'm}(\gamma)\; f_{lm'}(\gamma)}\,,
\end{equation}
where we relabeled $m\leftrightarrow m'$ after substituting.
Now, in this new $(\theta',\phi')$ coordinate system, with corresponding $(X',Y',Z')$, the spin vector is $\vec a=a(-\sin\gamma,0,\cos\gamma)$, the incoming wave vector is $\vec k_2=(0,0,\omega)$, and the outgoing wave vector is $\vec k_3=\omega(\sin\theta'\cos\phi',\sin\theta'\sin\phi',\cos\theta')$.

Finally, we have the third $(\theta,\phi)$--$(x,y,z)$ coordinate system used in \eqref{eq:App-gr}, where $\vec k_2$ is in the $z$ direction (same as $Z'$ direction) and $\vec k_3$ is in the $x$-$z$ plane at an angle $\theta$ from $\vec k_2$, and this is the same $\theta=\theta'$ from the second coordinates.  To translate the result \eqref{eq:App-gr} into the $(\theta',\phi')$ coordinates, we use $\theta=\theta'$ and do a rotation by an angle of $\phi'$ around the $x$-axis. This simply amounts to take 
\begin{equation}\label{eq:axayaz}
a_z=a\cos\gamma, \quad
 a_x=-a\sin\gamma\cos\phi', \quad
 a_y=- a\sin\gamma\sin\phi',
\end{equation}
as can be confirmed by comparing the values of $\vec a\cdot\vec k_2$, $\vec a\cdot\vec k_3$ and $\vec k_2\cdot\vec k_3$.
  
 It is most convenient to compare our amplitudes results from \eqref{eq:App-gr} using $(\theta,\phi)$ to our \ac{BHPT} results using $(\vartheta,\varphi)$ by comparing the amplitudes of their spin weighted spherical harmonic modes in the intermediate $(\theta',\phi')$ coordinates as in \eqref{Flm}.

We recall that the amplitude function $f$ at the leading order in $\epsilon$ (at fixed $a\omega=a^\star\epsilon/2$), coming from the tree-level scattering amplitude \eqref{eq:App-gr}, is
 
 \begin{equation}\label{eq:f0Kerr}
    f = \frac{\kappa^2 M^2\cos^4(\theta/2)}{4\sin^2(\theta/2)}\big[1+\mathcal{F}(\omega,a,\theta)+\frac{1}{2!}\mathcal{F}(\omega,a,\theta)^2 +\frac{1}{3!}\mathcal{F}(\omega,a,\theta)^3+\frac{1}{4!}\mathcal{F}(\omega,a,\theta)^4 \big]  
 \end{equation}
  which is expressed in terms of $(\theta',\phi')$ by using $\theta=\theta'$ and \eqref{eq:axayaz}.  Its mode amplitudes $f'_{lm}$ from \eqref{Flm} are given by integrals over the 2-sphere,
\begin{equation}\boxed{
f'_{lm}(\gamma)=\int d\Omega'\,{}_{-2}Y^*_{lm}(\theta',\phi') f(\gamma,\theta',\phi')}\label{eq:modesfqft}\,,
\end{equation}
and depend only on the angle $\gamma$ between the incoming momentum and the spin, and on the parameters $\epsilon$ and $a^\star$. Then, matching of the \ac{BHPT} results to the \ac{QFT} results translates  to show  \eqref{eq:modesfqft} and \eqref{eq:modesfBHPT} agree for all $\ell, m$.

In general, let us write these as an expansion in $\epsilon$, focusing on the leading order in the large $a^\star$ expansion at each order in $\epsilon$,
\begin{equation}
f'_{lm}=\sum_{n=0}^\infty \epsilon^n \Big[f'_{lm,n}a^\star{}^n+\mathcal O(a^\star{}^{(n-1)})\Big].
\end{equation}
We find that this pattern also holds for the analytically continued \ac{BHPT} amplitudes in the large $a^\star$ expansion for the orders considered here, i.e. up to $a^4$.

At linear order in spin, from \eqref{eq:f0Kerr}, we find
\begin{alignat}{3}
\begin{aligned}
f'_{00,1}&=0,
\\
\{f'_{1m,1}\}&=\{0,0,0\},
\\
\{f'_{2m,1}\}&=\sqrt{\frac{\pi}{5}}\{0,0,0,3\sin\gamma,-2\cos\gamma\},
\\
\{f'_{3m,1}\}&=\sin\gamma\{0,0,0,0,\sqrt{\frac{7\pi}{10}},0,\sqrt{\frac{7\pi}{6}}\},
\end{aligned}
\end{alignat}
and so on, with $m=\{-l,\ldots,l\}$.  From \eqref{eq:f0Kerr} at quadratic order in spin, we find
\begin{alignat}{3}
\begin{aligned}
f'_{00,2}&=0,
\\
\{f'_{1m,2}\}&=\{0,0,0\},
\\
\{f'_{2m,2}\}&=\{0,0,\frac{1}{4}\sqrt{\frac{5\pi}{6}}\sin^{2}\gamma,-\frac{2}{3}\sqrt{\frac{\pi}{5}}\cos\gamma\sin\gamma,\frac{1}{24}\sqrt{\frac{\pi}{5}}(9\cos(2\gamma)-5)\},
\\
\{f'_{3m,2}\}&=\{0,0,0,\frac{1}{4}\sqrt{\frac{3\pi}{70}}\sin^{2}\gamma,\frac{1}{6}\sqrt{\frac{\pi}{70}}\sin(2\gamma),-\frac{1}{24}\sqrt{\frac{\pi}{7}}(3\cos(2\gamma)+1),-\sqrt{\frac{\pi}{42}}\sin\gamma\cos\gamma\}
,
\\
\{f'_{4m,2}\}&=\sin^{2}\gamma\{0,0,0,0,\frac{1}{8}\sqrt{\frac{\pi}{10}},0,0,0,\frac{3}{16}\sqrt{\frac{\pi}{7}}\}
,
\end{aligned}
\end{alignat}
and so on. At cubic order in spin, it follows

\begin{alignat}{3}
\begin{aligned}
f'_{00,3}&=0,
\\
\{f'_{1m,3}\}&=\{0,0,0\},
\\
\{f'_{2m,3}\}&=\{0,\frac{5}{168}\sqrt{5\pi}\sin^{3}\gamma,-\frac{2}{7}\sqrt{\frac{2\pi}{15}}\cos\gamma\sin^{2}\gamma,\frac{1}{336}\sqrt{\frac{\pi}{5}}(1+39\cos2\gamma)\sin\gamma,\\
&\qquad\qquad\frac{1}{504}\sqrt{\frac{\pi}{5}}\cos\gamma(23-31\cos2\gamma)\}
,
\\
\{f'_{3m,3}\}&=\{0,0,-\frac{1}{96}\sqrt{\frac{\pi}{70}}\sin^{3}\gamma,\frac{1}{8}\sqrt{\frac{3\pi}{70}}\cos\gamma\sin^{2}\gamma,-\frac{1}{192}\sqrt{\frac{\pi}{70}}(17+39\cos2\gamma)\sin\gamma,\\
&\qquad\qquad\frac{1}{144}\sqrt{\frac{\pi}{7}}(11\cos2\gamma-7)\cos\gamma,\frac{1}{64}\sqrt{\frac{\pi}{42}}(9\cos2\gamma-1)\sin\gamma\}
,
\end{aligned}
\end{alignat}
and so on. Finally, at quartic order in spin we have

\begin{alignat}{3}
\begin{aligned}
f'_{00,4}&=0,
\\
\{f'_{1m,4}\}&=\{0,0,0\},
\\
\{f'_{2m,4}\}&=\{\frac{769}{10752}\sqrt{\frac{\pi}{5}}\sin^{4}\gamma,-\frac{67}{672}\sqrt{\frac{\pi}{5}}\cos\gamma\sin^{3}\gamma,\frac{1}{896}\sqrt{\frac{\pi}{30}}(37+103\cos2\gamma)\sin^{2}\gamma,\\
&\qquad\qquad\frac{1}{2688}\sqrt{\frac{\pi}{5}}(17-41\cos2\gamma)\sin2\gamma,\frac{1}{43008}\sqrt{\frac{\pi}{5}}(77-156\cos2\gamma+143\cos4\gamma)\},
\\
\{f'_{3m,4}\}&=\{0,-\frac{233}{23040}\sqrt{\frac{\pi}{7}}\sin^{4}\gamma,\frac{1}{9}\sqrt{\frac{\pi}{70}}\cos\gamma\sin^{3}\gamma,-\frac{1}{1152}\sqrt{\frac{\pi}{210}}(97+179\cos2\gamma)\sin^{2}\gamma,\\
&\qquad\frac{1}{144}\sqrt{\frac{\pi}{70}}(7\cos2\gamma-2)\sin2\gamma,-\frac{1}{18432}\sqrt{\frac{\pi}{7}}(33-76\cos2\gamma+107\cos4\gamma),\frac{1}{144}\sqrt{\frac{\pi}{42}}(\sin2\gamma-\sin4\gamma)\},
\end{aligned}
\end{alignat}
and so on. 
Remarkably, also continuing to large values of $l$, we find that all of these $(\epsilon a^\star)^i,$ for $i=1,2,3,4$ terms in the mode amplitudes from the (minimal) tree-level scattering amplitude \eqref{eq:f0Kerr} precisely match those computed from the analytically continued \ac{BHPT} theory amplitudes as described above. 
Amplitude  \eqref{eq:f0Kerr} in turn provides a closed form for the infinite partial waves from \ac{BHPT}. An analogous mode expansion can be made for the amplitude  \eqref{eq:Amp-gr} and show it agrees with the \ac{BHPT} computation up to quartic order in spin.